\documentclass[12pt]{yalethesis} 

\usepackage[T1]{fontenc} 
\usepackage[utf8]{inputenc}
\usepackage[main=english,bulgarian]{babel}
\usepackage{array}   
\usepackage{amsmath}
\usepackage{amssymb}  
\usepackage{amsfonts}  
\usepackage{mathdots}  
\usepackage{mathtools} 
\usepackage{graphicx}
\usepackage{xcolor}       
\usepackage{epigraph} 
\usepackage{setspace} 
\usepackage{lettrine} 
\pdfoutput=1 					     

\global\long\def\ket#1{\left|#1\right\rangle }

\global\long\def\bra#1{\left\langle #1\right|}

\global\long\def\braket#1#2{\left\langle #1\middle|#2\right\rangle }

\global\long\def\ketbra#1#2{\left|#1\right\rangle \left\langle #2\right|}

\global\long\def\kb#1#2{\ketbra{#1}{#2}}

\global\long\def\braOket#1#2#3{\left\langle #1\middle|#2\middle|#3\right\rangle }

\global\long\def\half{\frac{1}{2}}

\global\long\def\o{\omega}

\global\long\def\Tr#1{\textrm{Tr}\left[#1\right]}

\global\long\def\ddt{\frac{\mathrm{d}}{\mathrm{d}t}}

\global\long\def\sign{\operatorname{sign}}

\let\Pr\undefined

\global\long\def\Pr#1{\textrm{Pr}\left[#1\right]}

\global\long\def\E#1{\textrm{E}\left[#1\right]}

\global\long\def\dt{\mathrm{d}t}

\global\long\def\dN{\mathrm{d}N}

\global\long\def\dW{\mathrm{d}W}

\global\long\def\dZ{\mathrm{d}Z}

\global\long\def\G{\ket{\mathrm{G}}}

\global\long\def\D{\ket{\mathrm{D}}}

\global\long\def\B{\ket{\mathrm{B}}}

\global\long\def\cg{c_{\mathrm{G}}}

\global\long\def\cd{c_{\mathrm{D}}}

\global\long\def\cb{c_{\mathrm{B}}}

\global\long\def\cgn{C_{\mathrm{G}}}

\global\long\def\cdn{C_{\mathrm{D}}}

\global\long\def\cbn{C_{\mathrm{B}}}

\global\long\def\cbnb{\bar{C}_{\mathrm{B}}}

\global\long\def\Obg{\Omega_{\mathrm{BG}}}

\global\long\def\Odg{\Omega_{\mathrm{DG}}}

\global\long\def\deltatcatch{\Delta t_{\mathrm{catch}}}

\def\zrem#1{}  

\usepackage{textcomp} 
\usepackage[numbers,round]{natbib} 
\setcitestyle{authoryear}

\usepackage[hidelinks]{hyperref} 

\usepackage[top=1.5in, bottom=1in, left=1.5in, right=1in]{geometry}

\usepackage{fancyhdr} 
\usepackage{chngcntr}
\counterwithin{figure}{chapter}
\counterwithin{table}{chapter}
\counterwithin{equation}{chapter}

\let\Chaptermark\chaptermark   
\def\chaptermark#1{\def\Chaptername{#1}\Chaptermark{#1}}  

\newcommand{\lofpost}{\vspace{1.1ex}} 

\usepackage[explicit]{titlesec} %

\definecolor{yaleBlue}{HTML}{00356B}
\colorlet{chapnumcol}{yaleBlue}        

\newcommand{\chapPrefont}{
	\usefont{T1}{pnc}{b}{sc}
	\fontsize{35}{10}
	\selectfont
}

\newcommand{\chapnumfont}{
	\usefont{T1}{pnc}{b}{n}
	\fontsize{80}{10}
	\selectfont
}

\titleformat{\chapter}[display]
{\filcenter\bfseries} 
{\filcenter
	\chapPrefont\textcolor{chapnumcol}{} 
	\chapnumfont\textcolor{chapnumcol}{\thechapter}}
{-2pt}
{\Huge \textcolor{chapnumcol}{#1} 
}

\AtBeginDocument{\setlength{\DefaultFindent}{0.25em}}
\usepackage{nomencl} 
\newcommand{\nomunit}[1]{\renewcommand{\nomentryend}{\hspace*{\fill}#1}}
\usepackage{ifthen}
\newcommand{\spacePre}{		\vspace{20pt} }
\newcommand{\spacePst}{		\vspace{0pt} }
\newcommand{\itemGrp}[1]{	\item[\textbf{\Large #1}] }
\renewcommand{\nomgroup}[1]{%
	\ifthenelse{\equal{#1}{A}}{\itemGrp{Acronyms}\spacePst}{%
	\ifthenelse{\equal{#1}{C}}{\spacePre\itemGrp{Physical constants}\spacePst}{%
	\ifthenelse{\equal{#1}{D}}{\spacePre\itemGrp{Classical stochastic differential equations}\spacePst}{%
	\ifthenelse{\equal{#1}{F}}{\spacePre\itemGrp{Quantum trajectory theory}\spacePst}{%
	\ifthenelse{\equal{#1}{H}}{\spacePre\itemGrp{Quantum jumps in the three-level atom}\spacePst}{%
	\ifthenelse{\equal{#1}{J}}{\spacePre\itemGrp{Quantum electromagnetic circuit design}\spacePst}{%
						}	
	}}}}}
	\hspace*{-\leftmargin}\vspace{5pt}%
}

\makenomenclature

\usepackage{caption}
\DeclareCaptionLabelSeparator{line}{  \scalebox{2.0}[1.0]{$\vert$} }
\usepackage[labelsep=line, labelfont=bf]{caption} 

\titlespacing*{\paragraph} {0pt}{1.2ex plus 1ex minus .2ex}{1em}

\title{Catching and Reversing a Quantum Jump Mid-Flight}
\author{Zlatko Kristev Minev}
\advisor{Michel H. Devoret}
\date{2018}

\begin{document}

\doublespacing 
\begin{abstract}
\begin{changemargin}{-0.25cm}{-0.25cm}  
A quantum system driven by a weak deterministic force while under strong continuous energy measurement exhibits quantum jumps between its energy levels \citep{Nagourney1986, Sauter1986, Bergquist1986}. This celebrated phenomenon is emblematic of the special nature of randomness in quantum physics. The times at which the jumps occur are reputed to be fundamentally unpredictable. However, certain classical phenomena, like tsunamis, while unpredictable in the long term, may possess a degree of predictability in the short term, and in some cases it may be possible to prevent a disaster by detecting an advance warning signal. 
Can there be, despite the indeterminism of quantum physics, a possibility to know if a quantum jump is about to occur or not?
In this dissertation, we answer this question affirmatively by experimentally demonstrating that the completed jump from the ground to an excited state of a superconducting artificial atom can be tracked, as it follows its predictable ``flight,'' by monitoring the population of an auxiliary level coupled to the ground state. Furthermore, the experimental results demonstrate that the jump when completed is continuous, coherent, and deterministic. Exploiting these features, we catch and reverse a quantum jump mid-flight, thus deterministically preventing its completion. 
This real-time intervention is based on a particular lull period in the population of the auxiliary level, which serves as our advance warning signal. 
Our results, which agree with theoretical predictions essentially without adjustable parameters,  support the modern quantum trajectory theory 
and  provide new ground for the exploration of real-time intervention techniques in the control of quantum systems, such  as early detection of error syndromes.
\end{changemargin}
\end{abstract}
\singlespacing

\maketitle
\pagestyle{plain} 

\makecopyright
\newpage

\hbox{\hfil}\vspace{4in}\begin{center}
	For those who gave the most but could see the least. \\ 
	\ \\
	\foreignlanguage{bulgarian}{Цветко Ликов Цветков}\\
	(April 11, 1925 - December 11, 2017)\\ 
	\ \\
	\& \\
	\ \\
	\foreignlanguage{bulgarian}{Ангелина Борисова Цветкова}\\ 
	(December 9, 1928 - February 2, 2016)
\end{center}
\clearpage
\newpage



\singlespacing
\phantomsection 
\addcontentsline{toc}{chapter}{Contents} 
\tableofcontents

\phantomsection
\addcontentsline{toc}{chapter}{\listfigurename}
\listoffigures

\phantomsection
\addcontentsline{toc}{chapter}{\listtablename}
\listoftables

\phantomsection
\addcontentsline{toc}{chapter}{List of Symbols}



\nomenclature[ACqed]{CQED}{Cavity quantum electrodynamics}
\nomenclature[Acqed]{cQED}{Circuit quantum electrodynamics}
\nomenclature[An]{NMR}{Nuclear magnetic resonance}
\nomenclature[Aqnd]{QND}{Quantum non-demolition}
\nomenclature[Aqec]{QEC}{Quantum error correction}

\nomenclature[Abbq]{BBQ}{Black box quantization}
\nomenclature[Ahfss]{HFSS}{High-frequency electromagnetic simulation}
\nomenclature[Aepr]{EPR}{Energy participation ratio}
\nomenclature[Aljc]{LJC}{Linearized Josephson circuit}

\nomenclature[Adac]{DAC}{Digital-to-analog converter}
\nomenclature[Aacd]{ADC}{Analog-to-digital converter}
\nomenclature[Afpga]{FPGA}{Field-programmable gate array}
\nomenclature[Aawg]{AWG}{Arbitrary waveform generator}
\nomenclature[Asnr]{SNR}{Signal-to-noise ratio}

\nomenclature[Acw]{CW}{Continuous wave}
\nomenclature[Arf]{RF}{Radio frequency}
\nomenclature[Aif]{IF}{Intermediate frequency}
\nomenclature[Alo]{LO}{Local oscillator}
\nomenclature[Asma]{SMA}{SubMiniature version A RF connector}
\nomenclature[Avna]{VNA}{Vector network analyzer}
\nomenclature[Aiq]{IQ}{In-phase/quadrature}

\nomenclature[Ahemt]{HEMT}{High electron mobility transistor}
\nomenclature[Asquid]{SQUID}{Superconducting quantum interference device}
\nomenclature[Ajpc]{JPC}{Josephson parametric converter}

\nomenclature[Adrag]{DRAG}{Derivative removal by adiabatic gate}
\nomenclature[Aspam]{SPAM}{State preparation and measurement}

\nomenclature[Apovm]{POVM}{Positive operator valued measure}
\nomenclature[Acptp]{CPTP}{Completely positive and trace preserving}
\nomenclature[Arwa]{RWA}{Rotating wave approximation}
\nomenclature[Ahc]{$\mathrm{H.c.}$}{Hermitian conjugate}
\nomenclature[Asnr]{cNOT}{Controlled-NOT gate}

\nomenclature[Asse]{SSE}{Stochastic Schr\"{o}dinger equation}
\nomenclature[Asme]{SME}{Stochastic master equation}
\nomenclature[Asde]{SDE}{Stochastic differential equation}
\nomenclature[Aqsde]{QSDE}{Quantum stochastic differential equation}

\nomenclature[Apmma]{PMMA}{Polymethyl methacrylate}
\nomenclature[Anmp]{NMP}{$N$-Methyl-2-pyrrolidone}
\nomenclature[Aipa]{IPA}{Isopropyl alcohol}


\nomenclature[C1]{$h$}{Planck's constant
	\nomunit{$6.63\times 10^{-34}~\mathrm{J\,Hz}$} }
\nomenclature[C2]{$\hbar$}{Reduced Planck's constant ($=h/2\pi$)
	\nomunit{$1.05\times 10^{-34}~\mathrm{J\,Hz}$} }
\nomenclature[C3]{$e$}{Electron charge
	\nomunit{$1.60\times 10^{-19}~\mathrm{C}$} }
\nomenclature[C6]{$\Phi_0$}{Magnetic flux quantum ($=h/2e$)
	\nomunit{$2.07\times 10^{-15}~\mathrm{Wb}$} }
\nomenclature[C7]{$\phi_0$}{Reduced magnetic flux quantum ($=\hbar/2e$)
	\nomunit{$3.29\times 10^{-16}~\mathrm{Wb}$} }
\nomenclature[C8]{$R_Q$}{Resistance quantum ($=\hbar/\left(2e\right)^2$)
	\nomunit{$2.58\times 10^{4}~\mathrm{\Omega}$} }
\nomenclature[C9]{$k_B$}{Boltzmann constant
	\nomunit{$1.38\times 10^{-23}~\mathrm{J\, K^{-1}}$} }


\nomenclature[Dd3]{$\Pr{X=x}$}{Probability that classical stochastic variable $X$ has value $x$}
\nomenclature[Dd4]{$\wp(x)$}{For discrete event outcome $x$: probability of the event $\wp(x) = \Pr{X=x}$; for continuous event outcome $x$: probability density of the event $\wp(x)\mathrm{d}x = \Pr{X\in\left(x,x+\mathrm{d}x\right)}$}
\nomenclature[Dd5]{$\E{X}$}{Expectation value of classical stochastic variable $X$}

\nomenclature[Df1]{$N(t)$}{Continuous-time stochastic point process whose value corresponds to the number of detection events (e.g., photodetections) up to time $t$}
\nomenclature[Df2]{$W(t)$}{Continuous-time stochastic Wiener process (often called standard Brownian motion process)}
\nomenclature[Df3]{$\xi(t)$}{Gaussian white noise process, completely characterized by the two moments $\E{\xi(t)\xi(t')}=\delta(t-t')$ and $\E{\xi(t)}=0$; typically found in Stratonovich SDEs}
\nomenclature[Df4]{$\delta(t)$}{Dirac delta function}
\nomenclature[Df5]{$\delta_{ij}$}{Kronecker delta}

\nomenclature[Dg1]{$\dt$}{Deterministic time increment}
\nomenclature[Dg2]{$\dN(t)$}{Stochastic point-process increment. Defined by $\dN^2 = \dN$ and  $E\left[\dN\right] = \lambda\left(X\right) \dt$, where $\lambda\left(X\right)$ is a positive function of the random variable X;  typically found in It\^{o} SDEs}

\nomenclature[Dg3]{$\dW(t)$}{Stochastic Wiener increment, which satisfies $\dW(t)^2 = \dt$ and $\E{dW(t)}=0$; typically found in It\^{o} SDEs}
\nomenclature[Dg4]{$\dZ(t)$}{Complex Wiener increment, defined by $\dZ(t) \equiv \left( \dW_x(t)+ i \dW_y(t) \right)/\sqrt{2} $, which satisfies $\dZ(t)^* \dZ(t) = \dt$ and $\dZ(t)^2 = 0$; typically found in It\^{o} SDEs}

\nomenclature[Di]{$\mathbb{I}\int, \mathbb{S}\int  $}{Integral in the It\^{o} and Stratonovich sense, respectively; in later chapters, we relax the blackboard notation to avoid undue notational complexity}



\nomenclature[Fa1]{$\ket{\Psi}$}{Pure quantum state of the system \textit{and} environment (in Schr\"{o}dinger picture)}
\nomenclature[Fa2]{$\ket{\psi}$}{Unconditioned pure quantum state of the system}
\nomenclature[Fa3]{$\rho$}{Unconditioned density matrix of the system}
\nomenclature[Fa6]{$\ket{\psi_r}$}{Pure quantum state of system conditioned on the measurement result $r$}
\nomenclature[Fa7]{$\rho_r$}{Density matrix of system conditioned on the measurement result $r$}
\nomenclature[Fa8]{$\ket{\alpha}$}{Coherent state at complex displacement $\alpha$ from the vacuum}
\nomenclature[Fa9]{$\ket{\beta}$}{Coherent state at complex displacement $\beta$ from the vacuum}

\nomenclature[Fc,1]{$\hat{I}$}{Identity operator}
\nomenclature[Fc,4]{$\hat{H}$}{Hamiltonian operator}
\nomenclature[Fc,5]{$\hat{H}_S$}{System Hamiltonian operator}
\nomenclature[Fc,6]{$\hat{H}_E$}{Environment Hamiltonian operator}
\nomenclature[Fc,7]{$\hat{H}_{SE}$}{Hamiltonian operator corresponding to the interaction between the system and environment}
\nomenclature[Fc,9]{$\hat{U}$}{Unitary operator}
\nomenclature[Fc,~a1]{$\hat{M}_r$}{Kraus map operator corresponding to measurement result $r$}

\nomenclature[Fd1]{$\hat \sigma_{x,y,z}$}{Pauli $X$, $Y$, $Z$ operators}
\nomenclature[Fd2]{$\hat a$,$\hat a^\dag$}{Creation and annihilation operators}
\nomenclature[Fd3]{$\hat c$}{Arbitrary system operator coupled to the environment, e.g., $\hat{c} = \sqrt{\Gamma} \hat{a}$, where $\Gamma$ is the coupling rate; also, arbitrary system operator subjected to monitoring}

\nomenclature[Fe1]{$\mathcal{L}$}{Liouvillian superoperator}
\nomenclature[Fe2]{$\mathcal{D}\left[\hat{c}\right]$}{Lindblad superoperator $\mathcal{D}\left[\hat{c}\right]\rho
	\equiv
	\hat{c}^\dagger\rho\hat{c}-\frac{1}{2}[\hat{c}^{\dagger}\hat{c},\rho]_{+}$}
\nomenclature[Fe3]{$\mathcal{G}\left[\hat{c}\right]$}{Photodetection superoperator 
	$\mathcal{G}\left[\hat{c}\right]\rho \equiv \frac{\hat{c}\rho\hat{c}^{\dagger}}{\Tr{\hat{c}\rho\hat{c}^{\dagger}}}-\rho$}
\nomenclature[Fe4]{$\mathcal{H}\left[\hat{c}\right]$}{Dyne detection superoperator 
$	\mathcal{H}\left[\hat{c}\right]\rho\equiv\hat{c}\rho+\rho\hat{c}^{\dagger}-\Tr{\hat{c}\rho+\rho\hat{c}^{\dagger}}\rho$}

\nomenclature[Fg1]{$r$}{Outcome resulting from the measurement}
\nomenclature[Fg2]{$N(t)$}{Total number of photodetections up to time $t$}
\nomenclature[Fg2b]{$I(t)$}{Photocurrent measurement record}
\nomenclature[Fg2c]{$J(t)$}{Dyne detection measurement record}

\nomenclature[Fh]{$\eta$}{Quantum measurement efficiency}
\nomenclature[Fh6]{$T$}{Temperature}
\nomenclature[Fh7]{$n_\mathrm{th}$}{Thermal photon number}

\nomenclature[Hb1]{$\G$}{Ground state of the three-level atom}
\nomenclature[Hb2]{$\B$}{Bright state of the three-level atom}
\nomenclature[Hb3]{$\D$}{Dark state of the three-level atom}

\nomenclature[Hd1a]{$\omega$}{Resonance frequency}
\nomenclature[Hd1b]{$\omega_\mathrm{C}$}{Resonance frequency of readout cavity conditioned on the atom state $\G$}
\nomenclature[Hd1c]{$\omega_\mathrm{BG}$}{Bare resonance frequency of the $\G$ to $\B$ transition}
\nomenclature[Hd1d]{$\omega_\mathrm{DG}$}{Bare resonance frequency of the $\G$ to $\D$ transition}

\nomenclature[Hd2a]{$\Omega$}{Rabi drive amplitude}
\nomenclature[Hd2b]{$\Omega_{\mathrm{BG}}$}{Rabi drive (or drives $\Omega_{\mathrm{B0}}$ and $\Omega_{\mathrm{B1}}$) between $\B$ and $\G$}
\nomenclature[Hd2c]{$\Omega_{\mathrm{B0}}$}{First Rabi drive applied to the BG transition, at $\omega_{\mathrm{BG}}$}
\nomenclature[Hd2c]{$\Omega_{\mathrm{B1}}$}{Second Rabi drive applied to the BG transition, at  $\omega_{\mathrm{BG}} - \Delta_\mathrm{B1}$}
\nomenclature[Hd2c]{$\Omega_{\mathrm{DG}}$}{Rabi drive applied to the DG transition, at $\omega_{\mathrm{DG}} - \Delta_{\mathrm{DG}}$}

\nomenclature[Hd3]{$\Delta_{\mathrm{DG}}$}{Detuning of Rabi drive $\Omega_{\mathrm{DG}}$ from the bare DG transition frequency}
\nomenclature[Hd3]{$\Delta_{\mathrm{B1}}$}{Detuning of Rabi drive $\Omega_{\mathrm{B1}}$ from the bare BG transition frequency}

\nomenclature[Hd3]{$\deltatcatch$}{Catch signal duration, corresponding to the time of the complete absence of clicks, following the last click; The duration $\deltatcatch$ is divided in two phases, one lasting $\Delta t_{\mathrm{on}}$ and the other lasting $\Delta t_{\mathrm{off}}$ }
\nomenclature[Hd3b]{$\Delta t_{\mathrm{on}}$}{Duration of the first phase of catch protocol, during which all control drives are on}
\nomenclature[Hd3d]{$\Delta t_{\mathrm{off}}$}{Duration of the second phase of catch protocol, during which $\Omega_{\mathrm{DG}}$ is turned off}
\nomenclature[Hd3f]{$\Delta t_{\mathrm{mid}}$}{Time of mid-flight, in the complete presence of $\Omega_{\mathrm{DG}}$}
\nomenclature[Hd3g]{$\Delta t_{\mathrm{mid}}'$}{Time of mid-flight, in the absence of $\Omega_{\mathrm{DG}}$}

\nomenclature[Hd4b]{$\chi_\mathrm{B}, \chi_\mathrm{D}$}{Cross-Kerr coupling (dispersive shift) frequency between $\B$ or $\D$ and readout cavity, respectively}
\nomenclature[Hd4d]{$\chi_\mathrm{DB}$}{Cross-Kerr coupling frequency between the bright and dark transmon}

\nomenclature[Hd5]{$\alpha$}{Anharmonicity of a transmon qubit; also, occasionally used as arbitrary complex prefactor of coherent state}
\nomenclature[Hd5b]{$\alpha_\mathrm{B}, \alpha_\mathrm{D}$}{Anharmonicity of the bright and dark transmon, respectively }

\nomenclature[Hd7]{$\kappa$}{Energy decay rate of readout cavity}
\nomenclature[Hd8]{$\Gamma$}{Effective measurement rate of $\B$}
\nomenclature[Hd9]{$\bar{n}$}{Number of photons in readout cavity when driven resonantly}

\nomenclature[Hf1]{$Q$}{Quality factor}
\nomenclature[Hf2]{$S_{11}$}{Reflection scattering parameter}
\nomenclature[Hf3]{$S_{21}$}{Transmission scattering parameter}

\nomenclature[Hh3]{$I_\mathrm{rec},Q_\mathrm{rec}$}{Demodulated in-quadrature and out-of-quadrature measurement outcome of the heterodyne detection, respectively}

\nomenclature[Hf6]{$R_i(\theta)$}{Rotation around the $i$ axis by angle $\theta$}
\nomenclature[Hf7]{$\theta_{\mathrm{I}}$}{Rotation angle of intervention pulse in the reverse protocol}
\nomenclature[Hf7]{$\varphi_{\mathrm{I}}$}{Angle defining the axis, X', of the intervention pulse in the reverse protocol}
\nomenclature[Hf8]{$P_{\mathrm{G}},P_{\mathrm{D}}$}{Population in the ground and dark state, respectively}

\nomenclature[Hh2]{$T_\mathrm{int}$}{Integration time}
\nomenclature[Hh5]{$T_1$}{Energy relaxation time}
\nomenclature[Hh6]{$T_\mathrm{2R}$}{Ramsey coherence time}
\nomenclature[Hh7]{$T_\mathrm{2E}$}{Hahn echo coherence time}

\nomenclature[Hi1]{$X_\mathrm{DG},Y_\mathrm{DG},Z_\mathrm{DG}$}{Bloch vector components corresponding to the DG manifold}


\nomenclature[Ja]{$C$}{Capacitance}
\nomenclature[Ja]{$L$}{Inductance}
\nomenclature[Jb]{$L_j$}{Linear inductance of Josephson device $j$}
\nomenclature[Jb1]{$E_j$}{Energy scale of Josephson device $j$}
\nomenclature[Jb2]{$E_C$}{Transmon charging energy}

\nomenclature[Jcb1]{$p_{mj}$}{Energy participation ratio of Josephson device $j$ in LJC mode $m$}
\nomenclature[Jcb2]{$\phi_{mj}$}{Reduced zero-point fluctuation of Josephson device $j$ in LJC mode $m$}

\nomenclature[Jd1]{$\hat H$}{Hamiltonian of complete Josephson system ($\hat H = \hat  H_\mathrm{lin} + \hat H_\mathrm{nl}$)}
\nomenclature[Jd2]{$\hat H_\mathrm{lin}$}{Linearized $\hat H$ about operating point}
\nomenclature[Jd3]{$\hat H_\mathrm{nl}$}{Purely non-linear terms in $\hat H$}

\nomenclature[Jf]{$\hat c$,$\hat c^\dag$}{Readout cavity creation and annihilation operators}
\nomenclature[Jf]{$\hat b$,$\hat b^\dag$}{Bright transmon creation and annihilation operators}
\nomenclature[Jf]{$\hat d$,$\hat d^\dag$}{Dark transmon creation and annihilation operators}


\printnomenclature[2cm]

\phantomsection
\addcontentsline{toc}{chapter}{Acknowledgments}
\chapter*{Acknowledgments}
\doublespacing

\noindent\lettrine{I}{} am delighted to take this opportunity to acknowledge the people I learned from the most and those who supported me during my doctoral research at Yale. In this short Acknowledgements section, it is infeasible to properly thank everyone. I apologize in advance for any potential shortcomings. This is particularly relevant for those I worked with most closely during the initial years of my Ph.D., before I changed course by proposing and carrying out the quantum jump project in the final two years. The product of these two years constitutes the remainder of this Dissertation. 

To set the stage for the acknowledgements below, let me first briefly recount the origin and story of this work. In the summer of 2015, I traveled to Scotland to participate in the Scottish Universities Summer School in Physics (SUSSP71) by partly self-funding my participation. There I heard a cogent lecture by Howard J. Carmichael, which radically changed the direction of my doctoral scientific inquiry. Howard presented his gedanken experiment for catching and reversing a quantum jump mid-flight, which made the striking prediction that the nature of quantum jumps could be continuous and coherent. The discussion emphasized that a test of the conclusions remains infeasible, since the requisite experimental conditions remain far out of reach of atomic physics (see Chap.~1). Excited by the ideas, however, I brainstormed and ran simulations to find a possible realization of the gedanken experiment, but in a different domain --- superconducting quantum circuits. Although much was unclear in the leap from the quantum optics to the superconducting realm, I reached out to Howard and found him very receptive to my still developing ideas.  After an initial rebuff of my proposed experiment on my return to Yale, I spent the fall and early months of the next year developing the concepts and implementation in detail to prove their validity, and the feasibility of the project. Following many lessons and in-depth discussions with Michel, and the input of Howard and the people acknowledged below, the experiment was successfully realized just short of two years later. The predictions were not only experimentally confirmed but were further expanded to demonstrate the coherent, continuous, and deterministic evolution of the quantum jump even in the absence of a coherent drive on the jump transition. 

Thus, first and foremost, I would like to express my deep and heartfelt gratitude to my dissertation advisor, \textsc{Michel H. Devoret}.  Michel’s reputation as a great mentor and a brilliant physicist, agreed upon by all sources, notably preceded him as early as during my undergraduate days at Berkeley. Since I joined Michel’s group, I have only developed an ever-growing admiration for his breadth of knowledge and deep thinking. It has been a privilege and a pleasure to learn from and work closely with Michel. I am endlessly thankful for the countless hours and legions of lessons he so elegantly delivered on physics, writing, aesthetics in science, and innumerable other subjects. I am especially grateful to him for allowing me to take the unusual path of proposing and carrying out a complex, original experiment with little relation to any other project in the lab or to my previous work. I deeply appreciate this unique opportunity and I am especially thankful for his support, trust, and belief in the ideas of a young graduate student.  Michel also taught me how to communicate science clearly and to follow the highest standards, to pay attention to every detail, including font choices and color combinations. Michel continues to be my role model for a scientist of the highest caliber. I will dearly miss our inspired, in-depth conversations on science and beyond. I’ll also miss the collaborative and knowledge-rich environment of Michel’s \textsc{Quantronics Laboratory (Qlab)} on the fourth floor in Becton.

As related above, it was my good fortune to meet \textsc{Howard J. Carmichael} at SUSSP71, where I was inspired by his lecture on quantum jumps. It has been a pleasure and a privilege to work with Howard, who also inspires me with his record of pathbreaking advances in quantum optics theory and his seminal role in the foundation of quantum trajectory theory [the term was coined by him in \cite{Carmichael1993}]. His open reception of my ideas since the beginning, his generosity with his time, and his continued support has been beyond measure. I am endlessly grateful to Howard, as well as his student, \textsc{Ricardo Gutiérrez-Jáuregui}, whom I also had the pleasure of meeting at SUSSP71, for the indispensable theoretical modeling of the final experimental results. The project greatly benefitted from the discussions with and theoretical contributions of both Howard and Ricardo. I am particularly indebted to Howard for his thoughtful edits and input in the writing and revising of the paper we have submitted, and the enlightening lessons I picked up along the way. 

I deeply appreciate the theoretical discussions during the inception of the project with Professor \textsc{Mazyar Mirrahimi}, which were of significant help in navigating the landscape of cascaded non-linear parametric processes in circuit quantum electrodynamics (cQED), used in the readout of the atom.  I am grateful to Mazyar for patiently accommodating the many questions I had. I also want to express my deep gratitude to Professor \textsc{Steven M. Girvin} for the cQED and quantum trajectory discussions we had, for his detailed reading and edits of this dissertation manuscript, and for his kind manner of teaching and enlightening his students with countless deep insights, and warm encouragement. I feel indebted to Professor \textsc{Jack Harris}, who advised me early on at Yale on a project in optomechanics in his lab, and also provided careful and thoughtful comments on this dissertation manuscript. Jack has always been inspirational and supportive of my efforts. I also owe a debt to all the professors and senior scientists who taught me a great deal of what I know, \textsc{Rob Schoelkopf, Luigi Frunzio, Liang Jiang, Doug Stone}, and those who significantly broadened my horizons, and improved my understanding of physics: \textsc{Daniel Prober, Leonid Glazmann, Peter Rakich, David DeMille, Hui Cao, Yoram Alhassid, Paul Fleury, Sean E. Barrett}.

During the quantum jump project, I had the pleasure of working very closely with \textsc{Shantanu Mundhada}. Shantanu was instrumental to the success of the project by contributing a great deal with his aid in fabrication and the initial DiTransmon design of the device. \textsc{Philip Reinhold} had developed an outstanding Python platform for the control of the FPGA, and helped me a great deal in debugging the control code. \textsc{Shyam Shankar} aided in the fabrication of the device and provided general support to the lab in his kind and patient manner. I appreciate the many fruitful discussions with \textsc{Victor V. Albert, Matti P. Silveri,} and \textsc{Nissim Ofek}. Victor in particular addressed an aspect of the Lindblad theoretical modeling regarding the waiting-time distribution. It was Nissim and \textsc{Yehan Liu} who spearheaded the initial FPGA development.  In later discussions, I benefited from insightful conversations with \textsc{Howard M. Wiseman, Klaus  Mølmer, Birgitta Whaley, Juan P. Garrahan, Ananda Roy, Joachim Cohen,} and \textsc{Katarzyna Macieszczak.} 

In my earlier days at Yale, I learned much about low-temperature experimental physics from \textsc{Ioan Pop} and \textsc{Nick Masluk}. I had the good fortune to work with them and \textsc{Archana Kamal} (who inspired me with her dual mastery of experiment and theory) on the development of a superinductance with a Josephson junction array \citep{Masluk2012, Minev2012-APSMM}. Ioan and I, after spending three months repairing nearly every part of our dilution system, continued to work together for the next five years. We demonstrated the first superconducting whispering-gallery mode resonators (WGMR), which achieved the highest quality factors of planar or quasi-planar quantum structures at the time \citep{Minev2013}. These led us to demonstrate the first multi-layer (2.5D), flip-chip cQED architecture \citep{Minev2015-patent, Minev2016}, which demonstrated the successful unification of the advantages of the planar (2D) and three-dimension (3D) cQED architectures \citep{Minev2013-APSMM,Minev2014-APSMM,Minev2015-APSMM,Serniak2015-APSMM}. During the later phase of this project, I had the pleasure of working with \textsc{Kyle Serniak}, whom I thank for his many hours in the cleanroom. During these first years, I greatly benefited from Ioan’s mentorship, his energetic and cheerful character, and the pleasure of wonderful gatherings hosted by Cristina and him; additionally, our sailing lessons. While none of the work described in this paragraph is featured in this Dissertation --- as it could form an orthogonal, independent dissertation --- its results are detailed in the cited literature. 

During the development of the 2.5D architecture, I came up with an alternative idea for the quantization of black-box quantum circuits --- the energy-participation ratio (EPR) approach to the design of quantum Josephson circuits \citep{Minev2018-EPR}. I am grateful to Michel and to \textsc{Zaki Leghtas} for their support along the way for this unanticipated project. More generally, I had the extreme pleasure of working closely with Zaki and learning a great deal of physics from him in lab and over countless dinners. During the EPR project, I was privileged to coach a number of talented undergraduate students, whose enthusiasm and time I am thankful for: \textsc{Dominic Kwok, Samuel Haig, Chris Pang, Ike Swetlitz, Devin Cody, Antonio Martinez,} and \textsc{Lysander Christakis}.

Overall, many students and post-docs in \textsc{Qlab} and \textsc{RSL} contributed to the success of my time at Yale. I would like to thank them all. I have been fortunate to remain close friends and colleagues with my incoming class, \textsc{Kevin Chou, Eric Jin, Uri Vool, Theresa Brecht}, and \textsc{Jacob Blumoff}, and to learn dancing with Kevin. It has been a particular pleasure to work more closely with \textsc{Serge Rosenblum, Chan U Lei, Zhixin Wang, Vladimir Sivak, Steven Touzard}, and \textsc{Evan Zalys-Geller}. As part of \textsc{Qlab}, I had the privilege to work, although more indirectly, with a number of excellent postdoctoral researchers, including  \textsc{Michael Hatridge, Baleegh Abdo, Ioannis Tsioutsios, Philippe Campagne-Ibarcq, Gijs de Lange, Angela Kou}, and \textsc{Alexander Grimm}. I also had the pleasure to occasionally collaborate with a number of the other graduate students in \textsc{Qlab}, including \textsc{Anirudh Narla, Clarke Smith, Nick Frattini, Jaya Venkatraman, Max Hays, Xu Xiao, Alec Eickbusch, Spencer Diamond, Flavius Schackert, Katrina Sliwa,} and \textsc{Kurtis Geerlings}.

Our work was always mutually supported and very closely intertwined with that of Rob Scholekopf’s lab, and I thank \textsc{Hanhee Paik, Gerhard Kirchmair, Luyan Sun, Chen Wang, Reinier Heeres, Yiwen Chu, Brian Lester,} and \textsc{Vijay Jain}.  There are also many graduate students in Rob’s group I would like to acknowledge: \textsc{Andrei Petrenko, Matthew Reagor, Brian Vlastakis, Eric Holland, Matthew Reed,  Adam Sears, Christopher Axline, Luke Burkhart, Wolfgang Pfaff,   Yvonne Gao, Lev Krayzman, Christopher Wang, Taekwan Yoon, Jacob Curtis}, and \textsc{Sal Elder}. I benefited from a number of thoughtful theoretical discussions with \textsc{Linshu Li} and \textsc{William R Sweeney}.  

Our department would not run without the endless support and help provided to us by \textsc{Giselle M. DeVito, Maria P. Rao, Theresa Evangeliste}, and \textsc{Nuch Graves}, or that provided by \textsc{Florian Carle}and \textsc{Racquel Miller} for the \textsc{Yale Quantum Institute (YQI)}.

My time at Yale would not be what it was without \textsc{Open Labs}, a science outreach and  careers pathways not-for-profit I founded in 2012, and the innumerable, wonderful people who helped me develop it into a nation-wide organization that has reached over 3,000 young scholars and parents and coached more than several hundred graduate students. In 2015, I had the good fortune to meet two of my best friends and kindred spirits, \textsc{Darryl Seligman} and \textsc{Sharif Kronemer}. I am deeply grateful to Darryl for his immeasurable effort in spearheading the expansion of Open Labs from Yale to Princeton, Columbia, Penn, and Harvard, and to Sharif for further growing and shaping Open Labs into a long-lasting sustainable organization. I would like to thank \textsc{Maria Parente} and \textsc{Claudia Merson} for believing in my nascent idea of Open Labs and providing support through \textsc{Yale Pathways to Science}. There are far too many other key people to thank for their volunteer work with Open Labs, but I must acknowledge \textsc{Jordan Feyngold, Ian Weaver, Munazza Alam, Aida Behmard, Matt Grobis, Kirsten Blancato, Nicole Melso, Shannon Leslie, Diane Yu, Lina Kroehling, Christian Watkins, Arvin Kakekhani}, and \textsc{Charles Brown}. More recently, I wish to express my gratitude to Yale for recognizing me with the Yale-Jefferson Award for Public Service and to the American Physical Society (APS) and National Science Foundation (NSF) for supporting Open Labs with an outreach grant award. 

Beyond the world of physics, I was fortunate to meet some of my best friends whose support and good cheer I am deeply thankful for, including \textsc{Rick Yang, Brian Tennyson, Xiao Sun, Rasmus Kyng, Marius Constantin, Stafford Sheehan}, and \textsc{Luis J.P.~Lorenzo}.  I am also very grateful to \textsc{Olga Laur} for her unstinting support during the writing of this work. Finally, the people whose contributions are the greatest yet the least directly visible are my family members. There are no words to describe the incomparable debt I owe each of you, especially to my parents \textsc{Lora} and \textsc{Kris}, and to those painfully no longer among us --- my grandparents, \textsc{Angelina} and \textsc{Tzvetko}, to whom I dedicate my dissertation. Your richness of knowledge, creativity in science and art, and unconditional love remain a beacon of inspirational light. Thank you!

\begin{center}
	
	``\emph{If I have seen a little further, it is by standing on the shoulders of Giants.}''  \\ - Isaac Newton, "Letter from Sir Isaac Newton to Robert Hooke,'' \\Historical Society of Pennsylvania.
	
\end{center} 



\singlespacing




\doublespacing
\pagestyle{fancy} 

\newpage
\mainmatter

\graphicspath{{img/}} 

\chapter*{Overview\label{chap:Preamble}}

\addcontentsline{toc}{chapter}{Overview}

\emph{\begin{changemargin}{0.6cm}{0.6cm}  
Can there be, despite the indeterminism of quantum physics, a possibility
to know if a quantum jump is about to occur or not?\end{changemargin}
}

Chapter~\ref{chap:Introduction-and-overview} opens by introducing
the notion of a quantum jump between discrete energy levels of a quantum
system, a theoretical idea introduced by Bohr in 1913 \citep{Bohr1913}
--- yet, one whose existence was experimentally observed only seven
decades later \citep{Nagourney1986,Sauter1986,Bergquist1986}, in
a single atomic three-level system. Section~\ref{sec:Principle-of-the}
outlines our proposal to map out the dynamics of a quantum jump from
the ground, $\G$, to an excited, $\D$, state of a three-level superconducting
system. We propose a protocol to catch quantum jumps mid-flight and,
further, to reverse them prior to their completion. The proposal critically
hinges on achieving near unit-measurement efficiency, as discussed,
and experimentally demonstrated, in Sec.~\ref{sec:Unconditioned-monitoring-of}.
Building on this, the catch and reverse experimental protocols and
measurement results are presented in Sections~\ref{sec:Catching-the-quantum}
and~\ref{sec:Reversing-the-quantum}. These results directly demonstrate
that the answer to the above-posed question can indeed be in the affirmative.
Section~\ref{sec:Reversing-the-quantum} summarizes the experimental
results that demonstrate the deterministic prevention of the completion
of jumps; this experiment thereby precludes quantum jumps from occurring
altogether. A control experiment in which the feedback intervention
does not exploit the deterministic character of the completed jumps
is presented. Before proceeding to the remainder of the thesis, Section~\ref{sec:Discussion-of-main-results}
provides a brief discussion of the main results and their implications
for the hundred-year-long debate on the nature and reality of quantum
jumps. The section concludes by providing an outlook for the implications
of the results for future experiments. The remaining chapters, whose
individually aim is described in the following paragraphs, provide
further support to the main conclusions presented in Chapter~\ref{chap:Introduction-and-overview}
and devoted special attention to explicating the theory and experimental
methodology of the work.

Chapter~\ref{chap:Quantum-Trajectory-Theory} develops the essential
background needed to gain access to the core ideas and results of
quantum measurement theory and its formulation, which lead to the
catch and reverse theoretical prediction and modeling of the experiment.
The basic notions of the formalism are introduced in view of specific
examples. Building on this background, Chapter~\ref{chap:theoretical-description-jumps}
develops the quantum trajectory description of the quantum jumps observed
in the three-level atom. The basic ideas as well as the rigorous,
quantitative description of the continuous, coherent, and deterministic
evolution of a completed quantum jump is presented. Finally, the realistic
model of the experiment including known imperfections is developed.
 Chapter~\ref{chap:Experimental-Methods} details the experimental
methods, including our approach to the design of the superconducting
quantum devices developed in \citet{Minev2018-EPR}. Section~\ref{subsec:Energy-participation-ratio}
provides a nutshell introduction to this approach, referred to as
the energy-participation-ratio (EPR) approach and used to design and
optimize both the dissipative and Hamiltonian parameters of our circuit-quantum-electrodynamic
(cQED) systems.

Chapter~\ref{chap:Experimental-results} presents the results of
control experiments that further support the conclusions reached in
Chapter~\ref{chap:Introduction-and-overview}. The comparison between
the experimental results and the predictions of the quantum trajectory
theory developed in Chapter~\ref{chap:theoretical-description-jumps}
is provided in Sec.~\ref{subsec:Comparison-between-theory}. Chapter~\ref{chap:Conclusion-and-perspective}
summarizes the results of this dissertation  and discusses future
research directions.

\chapter{Introduction and main results\label{chap:Introduction-and-overview}}

\addcontentsline{lof}{chapter}{Introduction and overview\lofpost}

\singlespacing 
\epigraph{
If all this damned quantum jumping were really to stay,  I should be sorry I ever got involved with quantum theory.}
{Erwin Schr\"odinger\\ \textit{Brit. J. Philos. Sci. III}, 109 (1952)} 
\doublespacing  \noindent Bohr conceived of quantum jumps \citet{Bohr1913} in 1913, and while
Einstein elevated their hypothesis to the level of a quantitative
rule with his AB coefficient theory \citep{Einstein1916,Einstein1917},
Schrödinger strongly objected to their existence \citep{Schrodinger1952}.
The nature and existence of quantum jumps remained a subject of controversy
for seven decades until they were directly observed in a single system
\citep{Nagourney1986,Sauter1986,Bergquist1986}. Since then, quantum
jumps have been observed in a variety of atomic \citep{Basche1995,Peil1999,Gleyzes2007,Guerlin2007}
and solid-state \citep{Jelezko2002,Neumann2010,Robledo2011,Vijay2011,Hatridge2013}
systems. Recently, quantum jumps have been recognized as an essential
phenomenon in quantum feedback control \citep{Deleglise2008,Sayrin2001},
and in particular, for detecting and correcting decoherence-induced
errors in quantum information systems \citep{Sun2013,Ofek2016}.

Here, we focus on the canonical case of quantum jumps between two
levels indirectly monitored by a third --- the case that corresponds
to the original observation of quantum jumps in atomic physics \citep{Nagourney1986,Sauter1986,Bergquist1986},
see the level diagram of Fig.~\ref{fig:setup}a. A surprising prediction
emerges according to quantum trajectory theory (see \citet{Carmichael1993,Porrati1987,Ruskov2007}
and Chapter\,\ref{chap:Quantum-Trajectory-Theory}): not only does
the state of the system evolve continuously during the jump between
the ground $\ket{\mathrm{G}}$ and excited $\ket{\mathrm{D}}$ state,
but it is predicted that there is always a latency period prior to
the jump, during which it is possible to acquire a signal that warns
of the imminent occurrence of the jump (see Chapter~\ref{chap:theoretical-description-jumps}
for the theoretical analysis and mathematical treatment). This advance
warning signal consists of a rare, particular lull in the excitation
of the ancilla state $\ket{\mathrm{B}}$. The acquisition of this
signal requires the time-resolved detection of\textit{ every} de-excitation
of $\ket{\mathrm{B}}$. Instead, exploiting the specific advantages
of superconducting artificial atoms and their quantum-limited readout
chain, we designed an experiment that implements with maximum fidelity
and minimum latency the detection of the advance warning signal occurring
before the quantum jump (see rest of Fig.~\ref{fig:setup}).

\section{Principle of the experiment \label{sec:Principle-of-the}}

\begin{figure}
\centering{}\includegraphics[width=130mm]{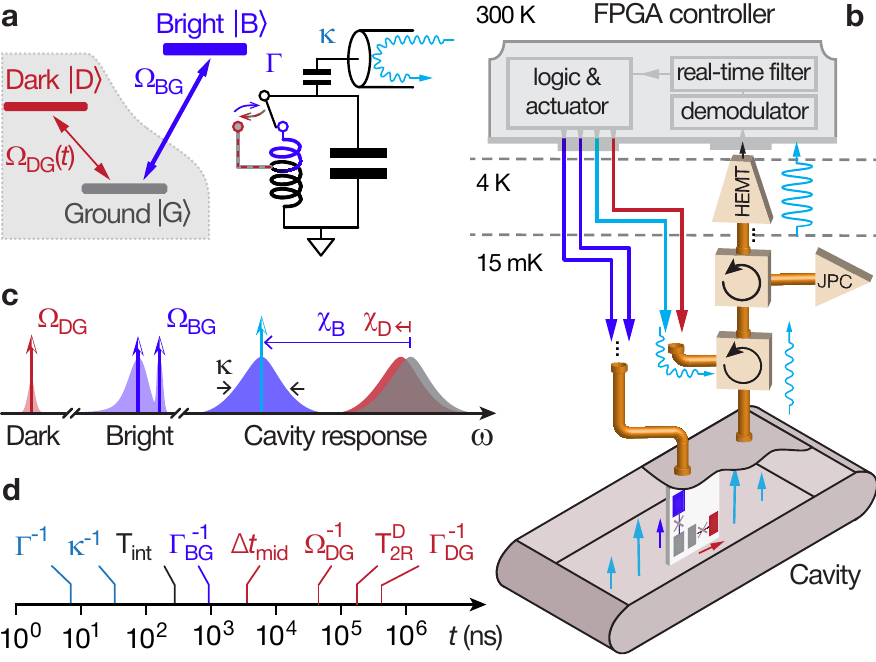} \caption[Principle of the experiment]{\label{fig:setup}\textbf{Principle of the experiment.} \textbf{a,}
Three-level atom possessing a hidden transition (shaded region) between
its ground $\ket{\mathrm{G}}$ and dark $\ket{\mathrm{D}}$ state,
driven by Rabi drive $\Omega_{\mathrm{DG}}(t)$. Quantum jumps between
$\ket{\mathrm{G}}$ and $\ket{\mathrm{D}}$ are indirectly monitored
by a stronger Rabi drive $\Omega_{\mathrm{BG}}$ between $\ket{\mathrm{G}}$
and the bright state $\ket{\mathrm{B}}$, whose occupancy is continuously
monitored at rate $\Gamma$ by an auxiliary oscillator (LC circuit
on right), itself measured in reflection by continuous-wave microwave
light (depicted in light blue). When the atom is in $\ket{\mathrm{B}}$,
the LC circuit resonance frequency shifts to a lower frequency than
when the atom is in $\ket{\mathrm{G}}$ or $\ket{\mathrm{D}}$ (effect
schematically represented by switch). Therefore, the probe tone performs
a $\ket{\mathrm{B}}$/not-$\ket{\mathrm{B}}$ measurement on the atom,
and is blind to any superposition of $\ket{\mathrm{G}}$ and $\ket{\mathrm{D}}$.
\textbf{b,} The actual atom and LC oscillator used in the experiment
is a superconducting circuit consisting of two strongly-hybridized
transmon qubits placed inside a readout resonator cavity at 15\,mK.
Control signals for the atom and cavity are supplied by a room-temperature
field-programmable gate array (FPGA) controller. This fast electronics
monitors the reflected signal from the cavity, and after demodulation
and filtering, actuates the control signals. The amplifier chain includes
circulators (curved arrows) and amplifiers (triangles and trapezoids).
\textbf{c,} Frequency landscape of atom and cavity responses, overlaid
with the control tones shown as vertical arrows. The cavity pull $\chi$
of the atom is nearly identical for $\ket{\mathrm{G}}$ and $\ket{\mathrm{D}}$,
but markedly distinct for $\ket{\mathrm{B}}$. The BG drive is bi-chromatic
in order to address the bright transition independently of the cavity
state. \textbf{d,} Hierarchy of timescales involved in the experiment,
which are required to span 5 orders of magnitude. Symbols explained
in text, and summarized in Table~\ref{tab:Summary-of-timescales.}.}
\end{figure}

First, we developed a superconducting artificial atom with the necessary
V-shape level structure (see Fig.~\ref{fig:setup}a and Section~\ref{sec:circuit-design}).
It consists, besides the ground level $\ket{\mathrm{G}}$, of one
protected, dark level $\ket{\mathrm{D}}$ --- engineered to not couple
to any dissipative environment or any measurement apparatus --- and
one ancilla level $\ket{\mathrm{B}}$, whose occupation is monitored
at rate $\Gamma$. Quantum jumps between $\left|\mathrm{G}\right>$
and $\left|\mathrm{D}\right>$ are induced by a weak Rabi drive $\Omega_{\mathrm{DG}}$
--- although this drive might eventually be turned off during the
jump, as explained later. Since a direct measurement of the dark level
is not possible, the jumps are monitored using the Dehmelt shelving
scheme \citep{Nagourney1986}. Thus, the occupation of $\left|\mathrm{G}\right>$
is linked to that of $\left|\mathrm{B}\right>$ by the strong Rabi
drive $\Omega_{\mathrm{BG}}$ ($\Omega_{\mathrm{DG}}\ll\Omega_{\mathrm{BG}}\ll\Gamma$).
In the atomic physics shelving scheme \citep{Nagourney1986,Sauter1986,Bergquist1986},
an excitation to $\ket{\mathrm{B}}$ is recorded with a photodetector
by detecting the emitted photons from $\ket{\mathrm{B}}$ as it cycles
back to $\G$. From the detection events --- referred to in the following
as ``clicks'' --- one infers the occupation of $\ket{\mathrm{G}}$.
On the other hand, from a prolonged absence of clicks (to be defined
precisely in Chapter~\ref{chap:theoretical-description-jumps}),
one infers that a quantum jump from $\ket{\mathrm{G}}$ to $\ket{\mathrm{D}}$
has occurred. Due to the poor collection efficiency and dead-time
of photon counters in atomic physics \citep{Volz2011}, it is exceedingly
difficult to detect every individual click required to faithfully
register the origin in time of  the advance warning signal. However,
superconducting systems present the advantage of high collection efficiencies
\citep{Vijay2012,Riste2013,Murch2013,Weber2014,Roch2014,deLange2014,Campagne2016-Fluorescence},
as their microwave photons are emitted into one-dimensional waveguides
and are detected with the same detection efficiencies as optical photons.
Furthermore, rather than monitoring the direct fluorescence of the
$\ket{\mathrm{B}}$ state, we monitor its occupation by dispersively
coupling it to an ancilla readout cavity. This further improves the
fidelity of the detection of the de-excitation from $\ket{\mathrm{B}}$
(effective collection efficiency of photons emitted from $\ket{\mathrm{B}}$).

The readout cavity, schematically depicted in Fig.~\ref{fig:setup}a
by an LC circuit, is resonant at $\omega_{\mathrm{C}}=8979.64\,\mathrm{MHz}$
and cooled to 15~mK. Its dispersive coupling to the atom results
in a conditional shift of its resonance frequency by $\chi_{\mathrm{B}}/2\pi=-5.08\pm0.2\,\mathrm{MHz}$
($\chi_{\mathrm{D}}/2\pi=-0.33\pm0.08\,\mathrm{MHz}$) when the atom
is in $\ket{\mathrm{B}}$ ($\ket{\mathrm{D}}$), see Fig.~\ref{fig:setup}c.
The engineered large asymmetry between $\chi_{\mathrm{B}}$ and $\chi_{\mathrm{D}}$
together with the cavity coupling rate to the output waveguide, $\kappa/2\pi=3.62\pm0.05\,\mathrm{MHz}$,
renders the cavity response markedly resolving for $\ket{\mathrm{B}}$
vs.~not-$\ket{\mathrm{B}}$, yet non-resolving \citep{Gambetta2011-Purcell,Riste2013,Roch2014}
for $\ket{\mathrm{G}}$ vs.~$\ket{\mathrm{D}}$, thus preventing
information about the dark transition from reaching the environment.
When probing the cavity response at $\omega_{\mathrm{C}}-\chi_{\mathrm{B}}$,
the cavity either remains empty, when the atom is in $\ket{\mathrm{G}}$
or $\ket{\mathrm{D}}$, or fills with $\bar{n}=5\pm0.2$ photons when
the atom is in $\ket{\mathrm{B}}$. This readout scheme yields a transduction
of the $\ket{\mathrm{B}}$-occupancy signal with five-fold amplification,
which is an important advantage to overcome the noise of the following
amplification stages. To summarize, in this readout scheme, the cavity
probe inquires: Is the atom in $\ket{\mathrm{B}}$ or not? The time
needed to arrive at an answer with a confidence level of 68\% (signal-to-noise
ratio of 1) is $\Gamma^{-1}\approx1/\left(\kappa\bar{n}\right)=8.8\,\mathrm{ns}$
for an ideal amplifier chain (see Chapter~\ref{chap:theoretical-description-jumps}).

Importantly, the engineered near-zero coupling between the cavity
and the $\ket{\mathrm{D}}$ state protects the $\ket{\mathrm{D}}$
state from harmful effects, including Purcell relaxation, photon shot-noise
dephasing, and the yet unexplained residual measurement-induced relaxation
in superconducting qubits \citep{Slichter2016-T1vsNbar}. We have
measured the following coherence times for the $\ket{\mathrm{B}}$
state: energy relaxation $T_{1}^{\mathrm{D}}=116\pm5\,\mathrm{\mu s}$,
Ramsey coherence $T_{\mathrm{2R}}^{\mathrm{D}}=120\pm5\,\mathrm{\mu s}$,
and Hahn echo $T_{\mathrm{2E}}^{\mathrm{D}}=162\pm6\,\mathrm{\mu s}$.
While protected, the $\ket{\mathrm{D}}$ state is indirectly quantum-non-demolition
(QND) read out by the combination of the V-structure, the drive between
$\ket{\mathrm{G}}$ and $\ket{\mathrm{B}}$, and the fast $\ket{\mathrm{B}}$-state
monitoring. In practice, we can access the population of $\ket{\mathrm{D}}$
using an 80~ns unitary pre-rotation among the levels followed by
a projective measurement of $\ket{\mathrm{B}}$ (see Chapter\,\ref{chap:Experimental-results}). 

Once the state of the readout cavity is imprinted with information
about the occupation of $\ket{\mathrm{B}}$, photons leak through
the cavity output port into a superconducting waveguide, which is
connected to the amplification chain, see Fig.~\ref{fig:setup}b,
where they are amplified by a factor of $10^{12}$. The first stage
of amplification is a quantum-limited Josephson parametric converter
(JPC) \citep{Bergeal2010}, followed by a high-electron-mobility transistor
(HEMT) amplifier at 4 K. The overall quantum efficiency of the amplification
chain is $\eta=0.33\pm0.03$. At room temperature, the heterodyne
signal is demodulated by a home-built field-programmable gate array
(FPGA) controller, with a 4\,ns clock period for logic operations.
The measurement record consists of a time series of two quadrature
outcomes, $I_{\mathrm{rec}}$ and $Q_{\mathrm{rec}}$, every 260 ns,
which is the integration time $T_{\mathrm{int}}$, from which the
FPGA controller estimates the state of the atom in real time. To reduce
the influence of noise, the controller applies a real-time, hysteretic
IQ filter (see Section\,\ref{subsec:IQ-filter}), and then, from
the estimated atom state, the control drives of the atom and readout
cavity are actuated, realizing feedback control. 

\section{Unconditioned monitoring of the quantum jumps\label{sec:Unconditioned-monitoring-of}}

\begin{figure*}
\centering{}\includegraphics[width=150mm]{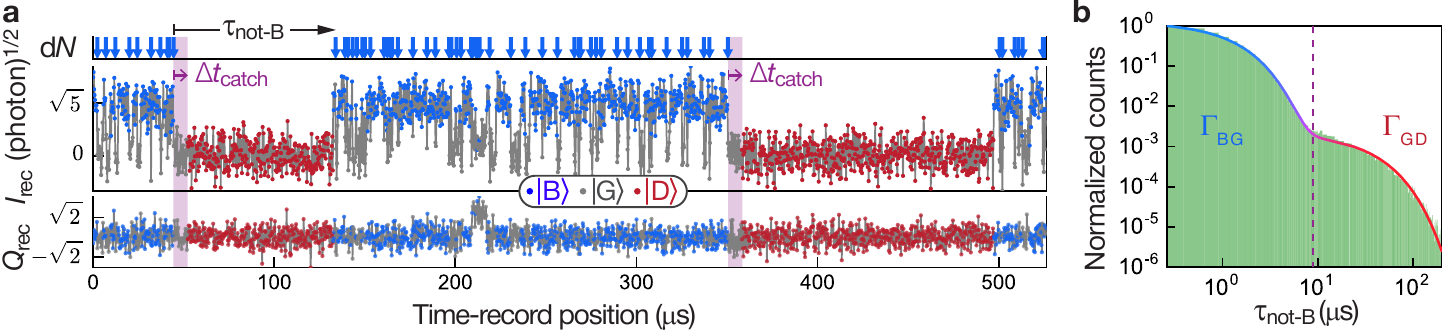} \caption[Unconditioned monitoring of quantum jumps in the 3-level system]{\label{fig:jumps} \textbf{Unconditioned monitoring of quantum jumps
in the 3-level system.} \textbf{a,} Typical measurement of integrated,
with duration $T_{\mathrm{int}},$ quadratures $I_{\mathrm{rec}}$
and $Q_{\mathrm{rec}}$ of signal reflected from readout cavity as
a function of time. The color of the dots (see legend) denotes the
state of the atom estimated by a real-time filter implemented with
the FPGAs (see Section\,\ref{subsec:IQ-filter}). On top, the vertical
arrows indicate ``click'' events ($\mathrm{d}N$) corresponding
to the inferred state changing from $\ket{\mathrm{B}}$ to not-$\ket{\mathrm{B}}$.
The symbol $\tau_{\operatorname{not-B}}$ corresponds to the time
spent in not-$\ket{\mathrm{B}}$, which is the time between two clicks
minus the last duration spent in $\ket{\mathrm{B}}$. An advance warning
that a jump to $\ket{\mathrm{D}}$ is occurring is triggered when
\textit{no} click has been observed for a duration $\Delta t_{\mathrm{catch}}$,
which is chosen between 1 and 12$\,\mathrm{\mu s}$ at the start of
the experiment. 
\textbf{b,} Log-log plot of the histogram of $\tau_{\operatorname{not-B}}$
(shaded green) for 3.2 s of continuous data of the type of panel (a).
Solid line is a bi-exponential fit defining jump rates $\Gamma_{\mathrm{BG}}=\left(0.99\pm0.06\,\mathrm{\mu s}\right)^{-1}$
and $\Gamma_{\mathrm{GD}}=\left(30.8\pm0.4\,\mathrm{\mu s}\right)^{-1}$.}
\end{figure*}

Having described the setup of the experiment, we proceed to report
its results. The field reflected out of the cavity is monitored in
a free-running protocol, for which the atom is subject to the continuous
Rabi drives $\Omega_{\mathrm{BG}}$ and $\Omega_{\mathrm{DG}}$, as
depicted in Fig.~\ref{fig:setup}. Figure~\ref{fig:jumps}a shows
a typical trace of the measurement record, displaying the quantum
jumps of our three-level artificial atom. For most of the duration
of the record, $I_{\mathrm{rec}}$ switches rapidly between a low
and high value, corresponding to approximately 0 ($\ket{\mathrm{G}}$
or $\ket{\mathrm{D}}$) and 5 ($\ket{\mathrm{B}}$) photons in the
cavity, respectively. The spike in $Q_{\mathrm{rec}}$ at $t=210\,\mathrm{\mu s}$
is recognized by the FPGA logic as a short-lived excursion of the
atom to a higher excited state (see Section \ref{subsec:IQ-filter}).
The corresponding state of the atom, estimated by the FPGA controller,
is depicted by the color of the dots. A change from $\ket{\mathrm{B}}$
to not-$\ket{\mathrm{B}}$ is equivalent to a ``click'' event, in
that it corresponds to the emission of a photon from $\ket{\mathrm{B}}$
to $\ket{\mathrm{G}}$, whose occurrence time is shown by the vertical
arrows in the inferred record $\mathrm{d}N\left(t\right)$ (top).
We could also indicate upward transitions from $\ket{\mathrm{G}}$
to $\ket{\mathrm{B}}$, corresponding to photon absorption events
(not emphasized here), which would not be detectable in the atomic
case. 

In the record, the detection of clicks stops completely at $t=45\,\mathrm{\mu s}$,
which reveals a quantum jump from $\ket{\mathrm{G}}$ to $\ket{\mathrm{D}}$.
The state $\ket{\mathrm{D}}$ survives for $90\,\mathrm{\mu s}$ before
the atom returns to $\ket{\mathrm{G}}$ at $t=135\,\mathrm{\mu s}$,
when the rapid switching between $\ket{\mathrm{G}}$ and $\ket{\mathrm{B}}$
resumes until a second quantum jump to the dark state occurs at $t=350\,\mathrm{\mu s}$.
Thus, the record presents jumps from $\ket{\mathrm{G}}$ to $\ket{\mathrm{D}}$
in the form of click interruptions.

In Fig.~\ref{fig:jumps}b, which is based on the continuous tracking
of the quantum jumps for 3.2\,s, a histogram of the time spent in
not-$\ket{\mathrm{B}}$, $\tau_{\operatorname{not-B}}$, is shown.
The panel also shows a fit of the histogram by a bi-exponential curve
that models two interleaved Poisson processes. This yields the average
time the atom rests in $\ket{\mathrm{G}}$ before an excitation to
$\ket{\mathrm{B}}$, $\Gamma_{\mathrm{BG}}^{-1}=0.99\pm0.06\,\mathrm{\mu s}$,
and the average time the atom stays up in $\ket{\mathrm{D}}$ before
returning to $\ket{\mathrm{G}}$ and being detected, $\Gamma_{\mathrm{GD}}^{-1}=30.8\pm0.4\,\mathrm{\mu s}$.
The average time between two consecutive $\ket{\mathrm{G}}$ to $\ket{\mathrm{D}}$
jumps is $\Gamma_{\mathrm{DG}}^{-1}=220\pm5\thinspace\mathrm{\mu s}$.
The corresponding rates depend on the atom drive amplitudes ($\Omega_{\mathrm{DG}}$
and $\Omega_{\mathrm{BG}}$) and the measurement rate $\Gamma$ (see
Chapter\,\ref{chap:theoretical-description-jumps}). Crucially, all
the rates in the system must be distributed over a minimum of 5 orders
of magnitude, as shown in Fig.\,\ref{fig:jumps}d. 

\section{Catching the quantum jump\label{sec:Catching-the-quantum}}

\begin{figure}
\centering{}\includegraphics[width=130mm]{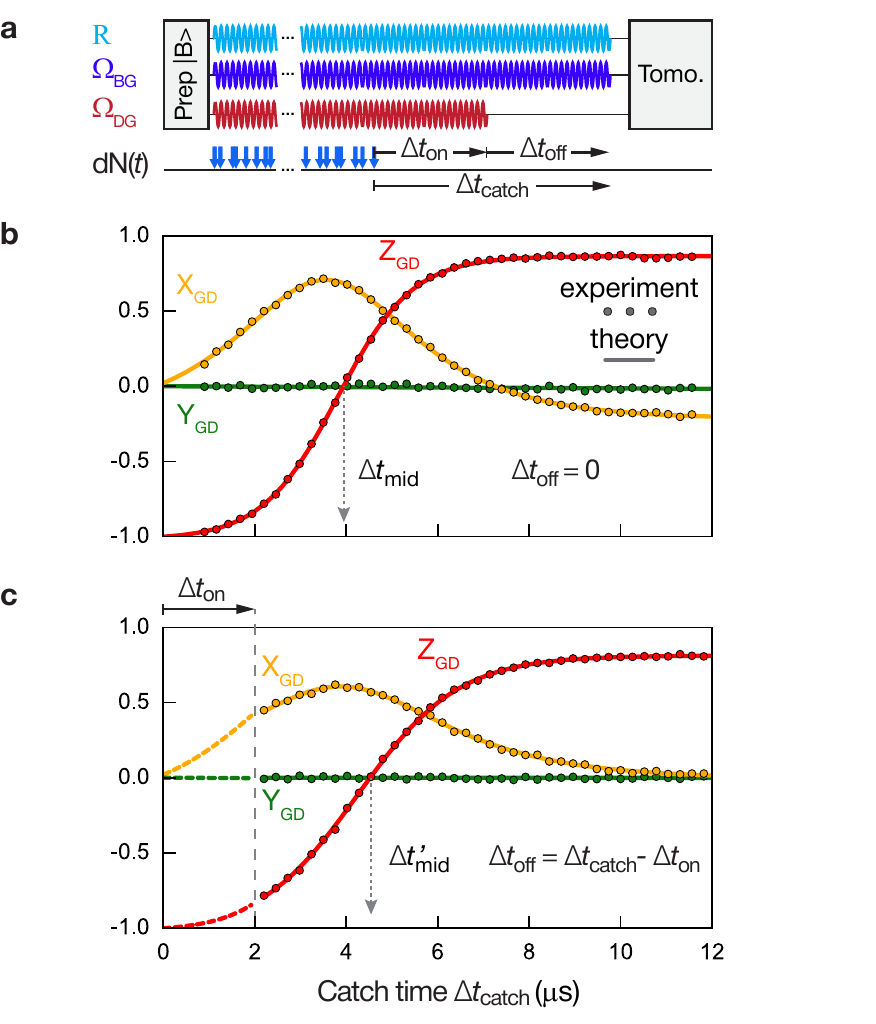} \caption[Catching the quantum jump mid-flight]{\label{fig:catch}\textbf{Catching the quantum jump mid-flight.}
\textbf{a,} The atom is initially prepared in $\ket{\mathrm{B}}$.
The readout tone ($\mathrm{R}$) and atom Rabi drive $\Omega_{\mathrm{BG}}$
are turned on until the catch condition is fulfilled, consisting of
the detection of a click followed by the absence of click detections
for a total time $\Delta t_{\mathrm{catch}}$. The Rabi drive $\Omega_{\mathrm{DG}}$
starts with $\Omega_{\mathrm{BG}}$, but can be shut off prematurely,
prior to the end of $\Delta t_{\mathrm{catch}}$. A tomography measurement
is performed after $\Delta t_{\mathrm{catch}}$. \textbf{b \& c,}
Conditional tomography revealing the continuous, coherent, and, surprisingly,
deterministic flight (when completed) of the quantum jump from $\ket{\mathrm{G}}$
to $\ket{\mathrm{D}}$. The error bars are smaller than the size of
the dots. The mid-flight time $\Delta t_{\mathrm{mid}}$ is defined
by $Z_{\mathrm{GD}}=0$. The jump proceeds even when $\Omega_{\mathrm{DG}}$
is turned off at the beginning of the flight (panel c), $\Delta t_{\mathrm{on}}=2\,\mathrm{\mu s}$.
Data obtained from $6.8\times10^{6}$ experimental realizations. Solid
lines: theoretical prediction (see Sec.~\ref{subsec:Comparison-between-theory}).
Dashed lines in panel c: theory curves for the $\Delta t_{\mathrm{on}}$
interval, reproduced from panel b. The data suggests that an advance-warning
signal of the jump can be provided by a no-click period for catch
time $\Delta t_{\mathrm{catch}}=\Delta t_{\mathrm{mid}}$, at which
half of the jumps will complete.}
\end{figure}

Having observed the quantum jumps in the free-running protocol, we
proceed to conditionally actuate the system control tones in order
to tomographically reconstruct the time dynamics of the quantum jump
from $\ket{\mathrm{G}}$ to $\ket{\mathrm{D}}$, see Fig.\,\ref{fig:catch}a.
Like previously, after initiating the atom in $\ket{\mathrm{B}}$,
the FPGA controller continuously subjects the system to the atom drives
($\Omega_{\mathrm{BG}}$ and $\Omega_{\mathrm{DG}}$) and to the readout
tone ($\mathrm{R}$). However, in the event that the controller detects
a single click followed by the complete absence of clicks for a total
time $\Delta t_{\operatorname{catch}}$, the controller suspends all
system drives, thus freezing the system evolution, and performs tomography,
as explained in Section \ref{subsec:Tomography-of-three-level}. Note
that in each realization, the tomography measurement yields a single
+1 or -1 outcome, one bit of information for a single density matrix
component. We also introduce a division of the duration $\Delta t_{\operatorname{catch}}$
into two phases, one lasting $\Delta t_{\mathrm{on}}$ during which
$\Omega_{\mathrm{DG}}$ is left on and one lasting $\Delta t_{\mathrm{off}}=\Delta t_{\operatorname{catch}}-\Delta t_{\mathrm{on}}$
during which $\Omega_{\mathrm{DG}}$ is turned off. As we explain
below, this has the purpose of demonstrating that the evolution of
the jump is not simply due to the Rabi drive between $\ket{\mathrm{G}}$
and $\ket{\mathrm{D}}$.

In Fig.\,\ref{fig:catch}b, we show the dynamics of the jump mapped
out in the full presence of the Rabi drive, $\Omega_{\mathrm{GD}}$,
by setting $\Delta t_{\mathrm{off}}=0$. From $3.4\times10^{6}$ experimental
realizations we reconstruct, as a function of $\Delta t_{\operatorname{catch}}$,
the quantum state, and present the evolution of the jump from $\ket{\mathrm{G}}$
to $\ket{\mathrm{D}}$ as the normalized, conditional GD tomogram
(see Section \ref{subsec:Tomography-of-three-level}). For $\Delta t_{\operatorname{catch}}<2\,\mathrm{\mu s}$,
the atom is predominantly detected in $\ket{\mathrm{G}}$ ($Z_{\mathrm{GD}}=-1$),
whereas for $\Delta t_{\operatorname{catch}}>10\,\mathrm{\mu s}$,
it is predominantly detected in $\ket{\mathrm{D}}$ ($Z_{\mathrm{GD}}=+1$).
Imperfections, due to excitations to higher levels, reduce the maximum
observed value to $Z_{\mathrm{GD}}=+0.9$ (see Section\,\ref{subsec:Error-analysis}). 

For intermediate no-click times, between $\Delta t_{\operatorname{catch}}=2\,\mathrm{\mu s}$
and $\Delta t_{\operatorname{catch}}=10\,\mathrm{\mu s}$, the state
of the atom evolves continuously and coherently from $\ket{\mathrm{G}}$
to $\ket{\mathrm{D}}$ --- the flight of the quantum jump. The time
of mid flight, $\Delta t_{\mathrm{mid}}\equiv3.95\,\mathrm{\mu s}$,
is markedly shorter than the Rabi period $2\pi/\Omega_{\mathrm{DG}}=50\,\mathrm{\mu s}$,
and is given by the function $\Delta t_{\text{mid}}=\left(\frac{\Omega_{\mathrm{BG}}^{2}}{2\Gamma}\right)^{-1}\ln\left(\frac{\Omega_{\mathrm{BG}}^{2}}{\Omega_{\mathrm{DG}}\Gamma}+1\right)$,
in which $\Omega_{\mathrm{DG}}$ enters logarithmically (see Section
\ref{sec:thry:3lvl-atom-simple-log}). The maximum coherence of the
superposition, corresponding to $\sqrt{X_{\mathrm{GD}}^{2}+Y_{\mathrm{GD}}^{2}}$,
during the flight is $0.71\pm0.005$, quantitatively understood to
be limited by several small imperfections (see Section\,\ref{subsec:Error-analysis}).

Motivated by the exact quantum trajectory theory, we fit the experimental
data with the analytic form of the jump evolution, $\mathrm{Z}_{\text{GD}}(\Delta t_{\operatorname{catch}})=a+b\tanh(\Delta t_{\operatorname{catch}}/\tau+c)$,
$\mathrm{X}_{\text{GD}}(\Delta t_{\operatorname{catch}})=a'+b'\operatorname{sech}(\Delta t_{\operatorname{catch}}/\tau'+c')$,
and $\mathrm{Y}_{\text{GD}}(\Delta t_{\operatorname{catch}})=0$.
We compare the fitted jump parameters ($a,a',b,b',c,c',\tau,\tau'$)
to those calculated from the theory and numerical simulations using
independently measured system characteristics (see Section \ref{subsec:Comparison-between-theory}).

By repeating the experiment with $\Delta t_{\text{on}}=2\,\mathrm{\mu s}$,
in Fig.~\ref{fig:catch}c, we show that the jump proceeds even if
the GD drive is shut off at the beginning of the no-click period.
The jump remains coherent and only differs from the previous case
in a minor renormalization of the overall amplitude and timescale.
The mid-flight time of the jump, $\Delta t_{\mathrm{mid}}'$, is given
by an updated formula (see Chapter \ref{chap:theoretical-description-jumps}).
The results demonstrate that the role of the Rabi drive $\Omega_{\mathrm{DG}}$
is to initiate the jump and provide a reference for the phase of its
evolution\footnote{A similar phase reference for a non-unitary, yet deterministic, evolution
induced by measurement was previously found in a different context
in: N. Katz, M. Ansmann, R. C. Bialczak, E. Lucero, R. McDermott,
M. Neeley, M. Steffen, E. M. Weig, A. N. Cleland, J. M. Martinis,
and A. N. Korotkov, Science (New York, N.Y.) 312, 1498 (2006).}. Note that the $\Delta t_{\mathrm{catch}}\gg\Delta t_{\mathrm{mid}}$
non-zero steady state value of $X_{\mathrm{GD}}$ in Fig.~\ref{fig:catch}b
is the result of the competition between the Rabi drive $\Omega_{\mathrm{DG}}$
and the effect of the measurement of $\ket{\mathrm{B}}$. This is
confirmed in Fig.~\ref{fig:catch}c, where $\Omega_{\mathrm{DG}}=0$,
and where there is no offset in the steady state value.

The results of Fig.~\ref{fig:catch} demonstrate that despite the
unpredictability of the jumps from $\ket{\mathrm{G}}$ to $\ket{\mathrm{D}}$,
they are preceded by an identical no-click record. While the jump
starts at a random time and can be prematurely interrupted by a click,
the deterministic nature of the flight comes as a surprise given the
quantum fluctuations in the heterodyne record $I_{\mathrm{rec}}$
during the jump --- an island of predictability in a sea of uncertainty. 

\section{Reversing the quantum jump\label{sec:Reversing-the-quantum}}

\begin{figure}
\centering{}\includegraphics[width=130mm]{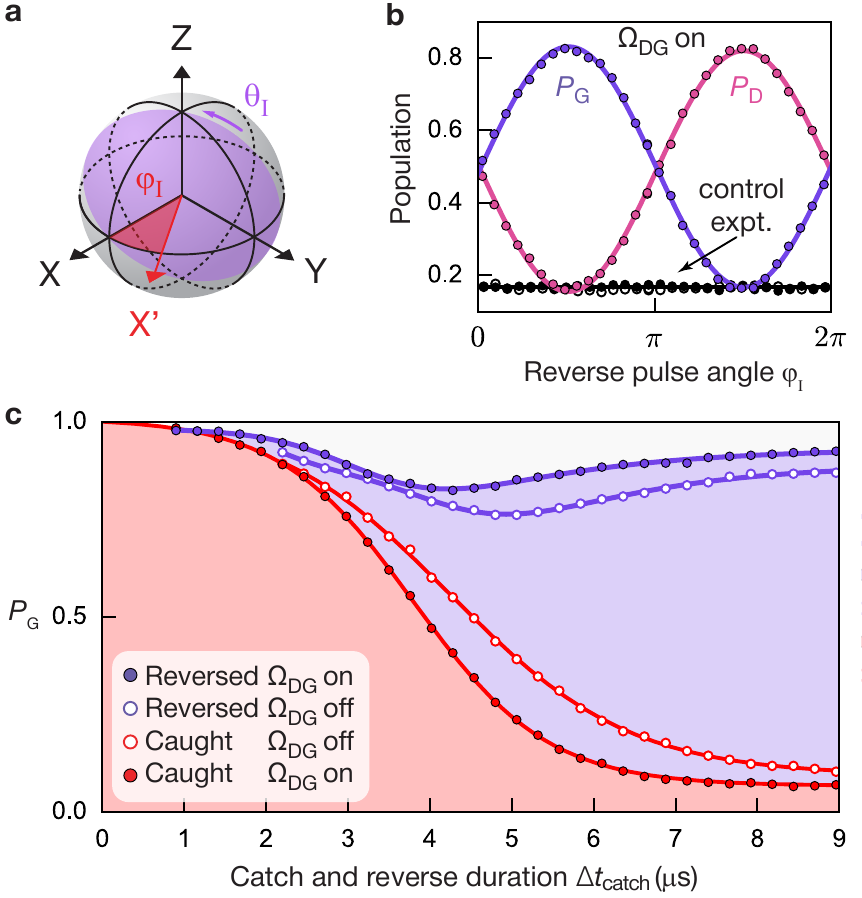}
\caption[Reversing the quantum jump mid-flight]{\label{fig:reverse}\textbf{Reversing the quantum jump mid-flight.}
\textbf{a,} Bloch sphere of the GD manifold, showing the axis X' for
the jump reversal, defined by the azimuthal angle $\varphi_{\mathrm{I}}$.
The angle of the intervention pulse is $\theta_{\mathrm{I}}$. \textbf{b,}
Success probabilities $P_{\mathrm{G}}$ (purple) and $P_{\mathrm{D}}$
(orange) to reverse to $\ket{\mathrm{G}}$ and complete to $\ket{\mathrm{D}}$
the quantum jump mid-flight at $\Delta t_{\mathrm{catch}}=\Delta t_{\mathrm{mid}}$,
with $\theta_{\mathrm{I}}=\pi/2$, in the presence of the Rabi drive
$\Omega_{\mathrm{DG}}$. The error bars are smaller than the size
of the dots. Black dots: success probability for $\ket{\mathrm{G}}$
(closed dots) and $\ket{\mathrm{D}}$ (open dots) in a control experiment
where intervention is applied at random times along the record, rather
than at $\Delta t_{\mathrm{catch}}$. \textbf{c,} Optimal success
of reverse protocol (purple) as a function of $\Delta t_{\mathrm{catch}}$.
The FPGA controller is programmed with the optimal $\left\{ \theta_{I}\left(\Delta t_{\mathrm{catch}}\right),\varphi_{I}\left(\Delta t_{\mathrm{catch}}\right)\right\} $.
Closed and open dots correspond to $\Delta t_{\mathrm{on}}=\Delta t_{\mathrm{catch}}$
and $\Delta t_{\mathrm{on}}=2\,\mathrm{\mu s}$, respectively. Red
points show the corresponding open-loop (no intervention) results
from Fig.~\ref{fig:catch}b and~c.}
\end{figure}

In Fig.~\ref{fig:reverse}b, we show that by choosing $\Delta t_{\operatorname{catch}}=\Delta t_{\mathrm{mid}}$
for the no-click period to serve as an advance warning signal, we
reverse the quantum jump\footnote{Reversal of quantum jumps have been theoretically considered in different
contexts, see H. Mabuchi and P. Zoller, Phys. Rev. Lett. 76, 3108
(1996) and R. Ruskov, A. Mizel, and A. N. Korotkov, Phys. Rev. B 75,
220501(R) (2007).} in the presence of $\Omega_{\mathrm{DG}}$; the same result is found
when $\Omega_{\mathrm{DG}}$ is off, see Section \ref{subsec:Incoherent-Bright-drive}.
The reverse pulse characteristics are defined in Fig.~\ref{fig:reverse}a.
For $\varphi_{\mathrm{I}}=\pi/2$, our feedback protocol succeeds
in reversing the jump to $\ket{\mathrm{G}}$ with $83.1\%\pm0.3\%$
fidelity, while for $\varphi_{\mathrm{I}}=3\pi/2$, the protocol completes
the jump to $\ket{\mathrm{D}}$, with $82.0\%\pm0.3\%$ fidelity.
In a control experiment, we repeat the protocol by applying the reverse
pulse at random times, rather than those determined by the advance
warning signal. Without the advance warning signal, the measured populations
only reflect those of the ensemble average.

In a final experiment, we programmed the controller with the optimal
reverse pulse parameters $\left\{ \theta_{I}\left(\Delta t_{\mathrm{catch}}\right),\varphi_{I}\left(\Delta t_{\mathrm{catch}}\right)\right\} $,
and as shown in Fig.~\ref{fig:reverse}c, we measured the success
of the reverse protocol as a function of the catch time, $\Delta t_{\mathrm{catch}}$.
The closed/open dots indicate the results for $\Omega_{\mathrm{DG}}$
on/off, while the solid curves are theory fits motivated by the exact
analytic expressions (see Chapter \ref{chap:theoretical-description-jumps}).
The complementary red dots and curves reproduce the open-loop results
of Fig.~\ref{fig:catch} for comparison.

\section{Discussion of main results\label{sec:Discussion-of-main-results}}

From the experimental results of Fig.~\ref{fig:jumps}a one can infer,
consistent with Bohr's initial intuition and the original ion experiments,
that quantum jumps are random and discrete. Yet, the results of Fig.~\ref{fig:catch}
support a contrary view, consistent with that of Schrödinger: the
evolution of the jump is coherent and continuous. Noting the difference
in time scales in the two figures, we interpret the coexistence of
these seemingly opposed point of views as a unification of the discreteness
of countable events like jumps with the continuity of the deterministic
Schrödinger’s equation. Furthermore, although all $6.8\times10^{6}$
recorded jumps (Fig.~\ref{fig:catch}) are entirely independent of
one another and stochastic in their initiation and termination, the
tomographic measurements as a function of $\Delta t_{\mathrm{catch}}$
explicitly show that all jump evolutions follow an essentially identical,
predetermined path in Hilbert space --- not a randomly chosen one
--- and, in this sense, they are deterministic. These results are
further corroborated by the reversal experiments shown in Fig.~\ref{fig:reverse},
which exploit the continuous, coherent, and deterministic nature of
the jump evolution and critically hinge on priori knowledge of the
Hilbert space path. With this knowledge ignored in the control experiment
of Fig.~\ref{fig:reverse}b, failure of the reversal is observed.

In conclusion (see Chapter~\ref{chap:Conclusion-and-perspective}
for an expanded discussion), these experiments revealing the coherence
of the jump, promote the view that a single quantum system under efficient,
continuous observation is characterized by a time-dependent state
vector inferred from the record of previous measurement outcomes,
and whose meaning is that of an objective, generalized degree of freedom.
The knowledge of the system on short timescales is not incompatible
with an unpredictable switching behavior on long time scales. The
excellent agreement between experiment and theory including known
experimental imperfections (see Sec.~\ref{subsec:Comparison-between-theory})
thus provides support to the modern quantum trajectory theory and
its reliability for predicting the performance of real-time intervention
techniques in the control of single quantum systems.

\chapter{Quantum measurement theory\label{chap:Quantum-Trajectory-Theory}}

\addcontentsline{lof}{chapter}{Quantum trajectory theory\lofpost}
\addcontentsline{lot}{chapter}{Quantum trajectory theory\lofpost} 

\singlespacing 
\epigraph{
In quantum physics, you don't see what you get, you get what you see.}
{A.N. Korotkov\\Private communication} 
\doublespacing\noindent\lettrine{T}{his} chapter provides a general background
to the central ideas and results of quantum measurement theory. It
begins with a prelude, Section~\ref{sec:Primer:-classical-vs.},
where the elementary notions of the measurement formalism are introduced.
These are developed, in Sections~\ref{subsec:Classical-measurement-theory-basic-concept}
and~\ref{subsec:Classical-model-of}, within the framework of probability
theory. For simplicity, the initial discussion is concerned with measurements
of classical systems. Section~\ref{subsec:Quantum-model-of} extends
the discussion to measurements of quantum systems, and it is seen
that many of the concepts developed in the classical setting directly
carry over. The tack of this approach makes it easy to discern the
classical from the quantum aspects of measurements. The ideas and
results of Sec.~\ref{sec:Primer:-classical-vs.} are extended to
time-continuous measurements in Sec.~\ref{sec:Introductory-example:-qubit}
by way of a specific example before generalizing to arbitrary systems.
Specifically, we construct a microscopic description of the homodyne
monitoring of a qubit, using only two-level ancillary systems. Although
the time-discrete model is simple and is readily solved, it contains
sufficient generality to illustrate the principal ideas of continuous
quantum measurements. The concept of a stochastic path taken by the
state of a monitored quantum system over time, known as its\emph{
quantum trajectory}, naturally emerges from the discussion. A higher
degree of mathematical rigor of the description follows in Section.~\ref{subsec:Continuous-limit:-Wiener},
which takes the continuous limit of our time-discrete model, thus
allowing the natural development of the basic notions of stochastic
calculus; in particular, the calculus of a Wiener process (Gaussian
white noise) is mathematically formulated. Finally, Section~\ref{sec:Quantum-trajectory-theory}
generalizes the results of the former section to formulate the general
theory of quantum measurements and quantum trajectories. This framework
sets the stage for the description of the quantum jumps experiment
presented in Chapter~\ref{chap:theoretical-description-jumps} (see
also \ref{chap:Experimental-results}). Suggestions for further reading
on the formulation of quantum trajectory theory are provided in Section~\ref{sec:Further-reading-traj-thry}.

\section{Prelude: from classical to quantum measurements\label{sec:Primer:-classical-vs.}}

This section provides an introduction to the basic concepts of measurement
theory. Before discussing a measurement of a quantum system, it is
helpful to develop and to understand the description of general, disturbing
classical measurements.\footnote{For further reading on classical measurement theory, we suggest Refs.~\citet{wiseman2010book}
and~\citet{jacobs2014book}. Our notation closely follows that of
Wiseman's book.} One finds that the probabilistic formulation of these greatly parallels
that of quantum measurements. In this way, it provides a closest approach
to the quantum one from the simpler, classical framework. Notably,
many key ideas carry over — but, with a few modifications that prove
profound and lead to the departure of the quantum measurements from
classical ones. For concreteness, throughout the discussion, we keep
the simplest possible example in view as we develop the theory, usually
based on a classical or quantum bit. While self-contained and thorough,
our discussion cannot hope to be exhaustive, and hence, for further
reading, we refer the reader to the references suggested in Section~\ref{sec:Further-reading-traj-thry}.

\subsection{Classical measurement theory: basic concepts\label{subsec:Classical-measurement-theory-basic-concept}}

Let us begin with the absolute minimum needed to discuss a measurement
of a classical system. Leading with the example of the simplest, smallest
classical system, a bit, we first establish the notions needed to
describe the system and then the measurement.

\paragraph{The simplest, smallest classical system — a bit.}

The simplest, smallest classical system is one that, at a given time,
can be described by only one of two possible configurations.\footnote{In some contexts, the term 'state' is sometimes employed instead of
the term 'configuration'. However, within the context of classical
measurement theory, the term 'state' is typically reserved for probability
distributions only, which will be introduced shortly. The motivation
for this choice of terminology is by analogy with quantum measurement
theory, where the state of the system describes, in effect, a probability
distribution.} A concrete, familiar example is that of a coin on a table, which
is either heads or tails. More generally, such a system with two possible
configurations (a bit worth of information) could represent any number
of physical situations; for instance, the bit could represent the
tilt of a mechanical seesaw on a playground (the seesaw is tilted
either to the left or to the right), or, for instance, in a classical
computer, it could represent the digital logic bit corresponding to
the thresholded voltage value at the output of a transistor (for example,
the two configuration could be that the voltage is less than five
volts or not).

\emph{Description of the system.} Continuing with the coin example,
the configuration of the coin is specified by a single property, corresponding
to the binary question: Is the coin tails, or not? Mathematically,
this property can be specified by a variable, which we will denote
$S$, and which takes only one of two \emph{values.}\footnote{Of course, we could use a representation where the binary values $S$
can take are ``H'' for heads and ``T'' for tails. We could then
endow these symbols, ``H'' and ``T,'' with an algebraic structure.
However, a more familiar and systematic approach is to use ordinary,
real numbers, as employed in the following.}\emph{ }Specifically, in anticipation of the discussion of a qubit,\footnote{We choose the values $+1$ and $-1$, rather than $0$ and $1$, in
order to parallel the later discussion of a quantum bit, and the outcome
of the Pauli $Z$ measurement. For completeness, the values $S=1$
and $S=-1$ are analogous to the ground ($\ket{+z}$) and excited
($\ket{-z}$) state of a qubit, respectively, which are introduced
in Sec.~\ref{subsec:Quantum-model-of}.} let us choose to assign the value $S=1$  to correspond to the coin
when it is tails and $S=-1$  for heads.\footnote{Note that at this stage, we assume no time dynamics of the system.
This will be introduced in the following.} Analogously, a general classical system is described by its \emph{configuration},
which is specified by a set of\emph{ variables, }each of which describes
an intrinsic property of the system, such as a degree of freedom.
These properties and variables are known to have \emph{objective,
definite values} for a classical system.

\emph{From perfect to probabilistic measurements. }In principle, a
perfect measurement of a classical system can be performed to unambiguously
obtain the values of the system variables, even without disturbing
the system. As such, an observer of the system can perform measurements
to determine the unambiguous configuration of the system. In this
case, the observer acquires complete information about the system,
and learns everything there is to know about it. If the system is
also deterministic, then the observer has thus additionally gained
complete knowledge of the result of all possible future measurements
on the system. Under these conditions, the description of the system
is exhaustive, and there isn't much more to say about measurements.
However, these ideal conditions are often not met in practice. Measurements
are often imperfect, ensembles of non-identical system have to be
considered, etc.  These situations require a description of the system
and measurements that is inherently probabilistic. This description
is the concern of classical measurement theory. In the following,
we first focus on the case of a probabilistic classical system, whose
description is somewhat analogous to that of an ensemble of quantum
systems.

\begin{figure}
\begin{centering}
\includegraphics[scale=1.6]{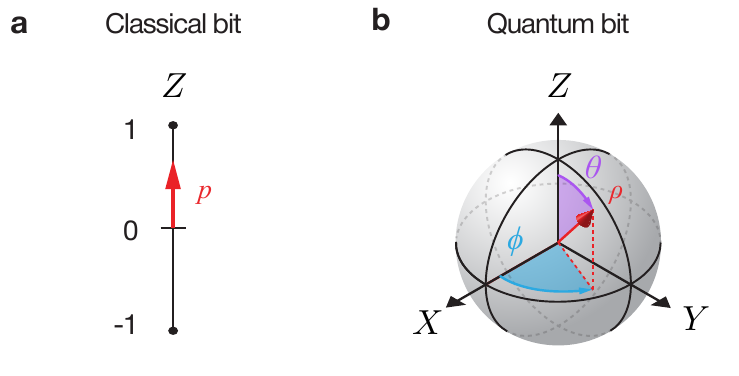}
\par\end{centering}
\caption[Geometric representation of the state of a classical and quantum bit]{\textbf{Geometric representation of the state of a classical and
quantum bit.} \textbf{\label{fig:Geometric-representation-of}a, }State
of a classical bit system represented as the one-dimensional probability
vector $p$ on the line segment Z between $-1$ and $1$ (see Eq.~(\ref{eq:ExpecL1})
for the definition of Z). \textbf{b, }State of quantum bit (qubit)
represented as the three-dimensional Bloch vector.  Unlike the classical
bit, the qubit has three observables (X, Y, and Z), which do not commute.
The quantum state of the qubit, $\rho$, is bounded by the unit sphere.
The surface of the sphere contains all pure states, which can be parametrized
by the angles $\phi$ and $\theta$. }
\end{figure}

\paragraph{\emph{Probabilistic bit system with perfect measurements.}}

For concreteness, consider a coin that is prepared probabilistically,
such as by a coin toss. Following the toss, an observer can perform
a measurement of the coin variable $S$, which yields a measurement
result. Formally, we should distinguish the measurement result obtained
by the observer from the actual value of the system property $S$.
For completeness, let's denote the variable of the \emph{measurement}
\emph{result} of $S$ as $m_{S}$. By analogy with $S$, we could
assign $m_{S}=-1$ and $m_{S}=1$ to results that corresponds to heads
and tails, respectively. The distinction between the measurement result
$m_{S}$ and system variable $S$ is crucial for imperfect and quantum
measurements. However, for simplicity, let us first proceed by assuming
perfect, classical measurements where there is no confusion between
$S$ and $m_{S}$, i.e, $m_{S}=S$. In this case, $m_{S}$ is redundant,
and for the following discussion there is no need to insist on the
distinction.

\emph{State of the system — a probability distribution.} To describe
the expected outcome of a measurement on the system, we introduce
the concept of the system \emph{state}.\footnote{For the following discussion, it suffices to adopt the point of view
that the state of the system represents \emph{subjective }knowledge
of the observer regarding the system. } The state describes the probability of a configuration to be the
system state. In other words, mathematically, the state is a probability
distribution over all possible system configurations, which form a
space known as the\emph{ configuration space}, denoted $\mathbb{S}$;
for the bit, $\mathbb{S}\equiv\left\{ S=1,S=-1\right\} $. The probability
for the coin be in the tails configuration, $S=1$, is then written
as $\Pr{S=1}$. More generally, the probability that the variable
$S$ of the system will have the value $s$ is $\Pr{S=s}$; for a
bit, $s\in\left\{ 1,-1\right\} $.\footnote{In this section, we employ the convention that capital letters denote
variables (typically, random ones) and lower case letters denote values.} This description of the classical system in terms of a probability
distribution, $\Pr{S=s}$, is analogous to the density matrix description
of a quantum system.\footnote{However, note that, as a probability, $\Pr{S=s}$ is a real and positive
number between 0 and 1.} Motivated by the analogy, we express the state of the coin bit as
a vector of probabilities, 
\begin{equation}
\vec{S}\equiv\begin{pmatrix}\Pr{S=1}\\
\Pr{S=-1}
\end{pmatrix}\,.\label{eq:SforBit}
\end{equation}
Keeping in mind the constraint that a measurement always yields a
result, one observes that the sum of the probabilities must be one.
Mathematically, the state vector $L^{1}$ norm is constrained, $\left|\vec{S}\right|_{\mathrm{L1}}=\sum_{s}\Pr{S=s}=1$,
where $\left|\cdot\right|_{\mathrm{L1}}$ denotes the $L^{1}$ norm.
This property is analogous to the unit-trace property of the density
matrix of a quantum state. Using this constrain, the state of the
coin, Eq.~(\ref{eq:SforBit}), can be simplified to a single information
parameter $p$, which denotes the bias of the coin,

\begin{equation}
\vec{S}=\begin{pmatrix}\frac{1+p}{2}\\
\frac{1-p}{2}
\end{pmatrix},\label{eq:BitBloch}
\end{equation}
The bias parameter $p$ is a number between $-1$ and 1,\footnote{Mathematically, $p$ is a number in the convex hull defined by $\mathbb{S}$.}
and, since it specifies the system state, is a quantity of central
importance. It can be viewed as the classical analog of the Bloch
vector of a quantum bit. In a sense, it represents a kind of one-dimensional
probability vector, which constitutes a geometric representation of
the system state; see Fig.~\ref{fig:Geometric-representation-of}a.

\paragraph{Operations on the system. }

An operation on the system results in a change of its configuration.
For the example of a coin, there are only two possible operations:
i) the coin is flipped (the logical negation operation, \emph{not})
or ii) the coin is left as is (the \emph{identity} operation). Working
within the framework of an ensemble of systems, an operation (one
that is applied to all systems in the ensemble) results in a change
of the state of the system that can be described by a linear map.
The state of the system ensemble after the operation, denoted $S'$,
can then be written as $\vec{S}'=U\vec{S}$, where the linear map
is represented by a configuration-transition matrix, denoted $U$.
For the coin, the two possible operations, the \emph{identity} and
\emph{not,} take the following forms
\begin{equation}
I\equiv\begin{pmatrix}1 & 0\\
0 & 1
\end{pmatrix}\text{\ and\ }\sigma_{x}\equiv\begin{pmatrix}0 & 1\\
1 & 0
\end{pmatrix}\,,\label{eq:IandSigmaX}
\end{equation}
respectively. The bit-flip Pauli matrix is denoted $\sigma_{x}$.

\paragraph{Perfect classical measurement of a system ensemble.}

Consider the long-run average value a series of repeated measurements
of the coin variable $S$, for the example of randomly prepared coins.
The expected mean value of $S$ is the weighted average of the results,
defined as
\begin{equation}
\E S\equiv\sum_{s}s\Pr{S=s}\,,\label{eq:EofS}
\end{equation}
where $\E{\cdot}$ represents ``expectation value of'' and the sum
is taken over all possible values $s$ of $S$. In matrix form, recalling
Eq.~(\ref{eq:BitBloch}), Eq.~(\ref{eq:EofS}) simplifies to

\begin{equation}
\E S=\left|\sigma_{Z}\vec{S}\right|_{\mathrm{L1}}=p\,,\label{eq:ExpecL1}
\end{equation}
where $p$ is non-negative and we have introduced the measurement
operator $\sigma_{Z}$, associated with the variable $S$ and given
by the Pauli matrix

\begin{equation}
\sigma_{Z}\equiv\begin{pmatrix}1 & 0\\
0 & -1
\end{pmatrix}\,.\label{eq:SigmaZ-CM}
\end{equation}
The matrix formulation given by Eq.~(\ref{eq:ExpecL1}) for the expectation
value of a classical measurement bears marked resemblance to that
employed with quantum systems. For a measurement on a quantum bit,
the expectation value of the $Z$ component of its spin is given by
$\Tr{\hat{\sigma}_{z}\rho}$, where $\rho$ is the qubit density matrix,
$\hat{\sigma}_{z}$ is the Pauli Z operator, represented by the matrix
given in Eq.~(\ref{eq:SigmaZ-CM}), and $\Tr{\cdot}$ denotes the
trace function. 

\begin{table}
\begin{centering}
\renewcommand*\arraystretch{1.5}
\begin{tabular}{l|c|>{\raggedright}p{0.5\columnwidth}}
\textbf{Concept} & \textbf{Symbols} & \textbf{Definition / Description}\tabularnewline
\hline 
\hline 
\multicolumn{3}{c}{\textbf{\rule{0pt}{5ex}Basic concepts}}\tabularnewline
\hline 
variable & $S$,$E$ & Describes intrinsic property of system, has definite value independent
of measurement apparatus\tabularnewline
\hline 
variable value & $s,e$ & Specific value that a variable can take\tabularnewline
\hline 
probability & $\Pr{S=s}$ & Probability that variable $S$ has value $s$\tabularnewline
\hline 
configuration & $\left\{ S\right\} $,$\left\{ E\right\} $,$\left\{ S,E\right\} $ & Set of all system variables\tabularnewline
\hline 
configuration space & $\mathbb{S}$,$\mathbb{E}$,$\mathbb{J}$ & Set of all possible system configurations\tabularnewline
\hline 
state  & $\vec{S},\vec{E},\vec{J}$ & Probability distribution on the configuration space, represented as
a vector\tabularnewline
\hline 
expectation value & $\E S$ & Expected (mean) value of repeated measurements of $S$, see Eq.~(\ref{eq:EofS})
\tabularnewline
\end{tabular}
\par\end{centering}
\caption[Basic concepts of classical measurement theory]{\textbf{Basic concepts of classical measurement theory.} }
\end{table}

\paragraph{Composite system. }

Extending the coin example, consider a composite system consisting
of two coins. The first coin is described by the variable $S$, or
in the ensemble situation, by the state $\vec{S}$, defined over the
configuration space $\mathbb{S}\equiv\left\{ S=1,S=-1\right\} $.
The second coin is similarly described by a single variable, $E$,
and a state $\vec{E}=\begin{pmatrix}\frac{1+p_{E}}{2}\\
\frac{1-p_{E}}{2}
\end{pmatrix}$, where $p_{E}$ is the coin bias. Its configuration space is $\mathbb{E}\equiv\left\{ E=1,E=-1\right\} $.
The configuration of the composite system consists of the simultaneous
specification of all variables, namely, $S$ and $E$. The set of
all possible configurations of the composite system is
\begin{align}
\mathbb{J} & =\mathbb{S}\otimes\mathbb{E}\label{eq:compsite-sys}\\
 & =\left\{ 1_{S},-1_{S}\right\} \otimes\left\{ 1_{E},-1_{E}\right\} \nonumber \\
 & =\left\{ 1_{S}1_{E},\,1_{S}-1_{E},\,-1_{S}1_{E},\,-1_{S}-1_{E}\right\} \,,\nonumber 
\end{align}
where $\otimes$ denotes the tensor product, and where, momentarily,
we have used the notation where $1_{S}$ stands for $S=1$.\footnote{So that no confusion arises, we note that the dimension of the composite
system space is not the sum but is the product of the subsystem dimensions,
i.e., $\dim\mathbb{J}=\dim\mathbb{S}\times\dim\mathbb{E}$, where
$\dim$ represents ``dimension of.''} The state of the composite system is a probability distribution over
$\mathbb{J}$, which can be represented by a 4-dimensional probability
vector, $\vec{J}$. When the two subsystems are uncorrelated, the
composite state is separable, and can be written as a simple product
of the states of the constituent subsystems, $\vec{J}=\vec{S}\otimes\vec{E}$.
However, when the subsystems are correlated, this is no longer possible.
For concreteness, consider the case where the two coins are prepared
randomly but always the same, the correlated randomness of the two
systems is described by the composite state  $\vec{J}=\left(\frac{1}{2},0,0,\frac{1}{2}\right)^{\intercal}$,
where $^{\intercal}$ denotes the transposition operation. More generally,
an operation that represents an interaction between the two coins
results in statistical correlations between them, and thus renders
the composite state inseparable. These features generally carry over
to the description of composite quantum systems, but standard statistical
correlations are replaced by entanglement. In the following subsection,
Sec.~\ref{subsec:Classical-model-of}, we explore the effect of an
interaction between the two coins.

\subsection{Classical toy model of system-environment interaction\label{subsec:Classical-model-of}}

For a more general discussion of measurements, it is necessary to
consider the interaction of the system with another, which probes
it and is often referred to as the \emph{environment}. In this subsection,
we consider the minimal limit of this model, where both the system
and environment are bits. Further, to introduce only the essentials
for now, we consider only the effect of a single interaction between
the classical system and environment, and discuss the effect of the
interaction on the system transfer of information. In the following
subsection, Sec.~\ref{subsec:Quantum-model-of}, we consider the
analogous quantum case, consisting of the interaction between a system
and environment, each of which is quantum bit. In Sec.~\ref{sec:Introductory-example:-qubit},
we generalize the toy model to the time-continuous homodyne monitoring
of a quantum bit. 

\begin{figure}
\begin{centering}
\includegraphics[scale=1.5]{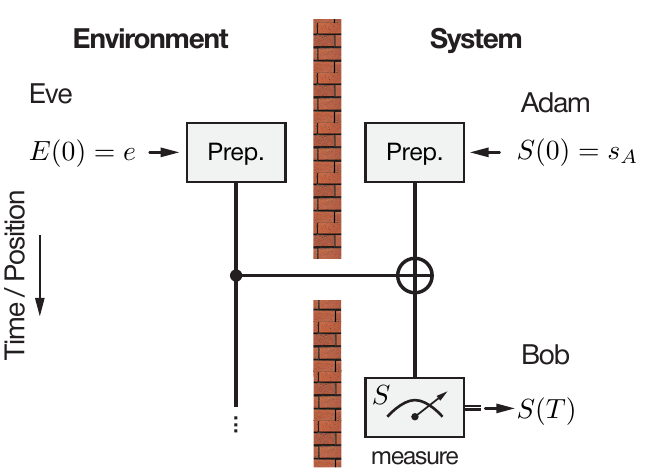}
\par\end{centering}
\caption[Classical toy model of the interaction between the system and environment]{\textbf{\label{fig:Reversibility-of-classical}Classical} \textbf{toy
model of the interaction between the system and environment.} Circuit
of the interaction between the system, with agents Adam and Bob, and
the environment, with agent Eve. Vertical lines depict the bits of
the system and environment, initially prepared by Adam and Eve in
the states $S\left(0\right)=s_{A}$ and $E\left(T\right)=e$, respectively.
The two bits interact via a controlled-NOT (cNOT) gate. Bob measures
the system at time $T$, obtaining the value $S\left(T\right)$. Brick
wall depicts the lack of communication between the agents of the system
and environment. }
\end{figure}

\paragraph{Classical toy model. }

Continuing with the example of two coins, we label one as the ``system''
and the other as the ``environment.'' For definitiveness, consider
the case where the system coin belongs to Adam, who aims to employ
it to communicate with Bob. To achieve this, at time $t=0$, Adam
prepares his coin in the configuration $S\left(0\right)=s_{A}$, where
$s_{A}$ is the bit value Adam hopes to communicate. He sends the
coin flying to Bob, who receives it at time $T$, and measures it
to obtain the value of $S\left(T\right)$. If the coin flies undisturbed,
$S\left(T\right)=S\left(0\right)$, and Bob faithfully receives Adam's
bit. 

However, during its flight, the coin unavoidably interacts with a
second flying coin, which belongs to an agent, Eve, who, at time $t=0$,
has prepared her coin in the configuration $E\left(0\right)=e$, where
$E$ is the variable describing her coin, and which is unknown to
Adam and Bob. For concreteness, suppose the interaction between the
two coins is described by the controlled-NOT (cNOT) gate, 
\begin{equation}
\mathrm{cNOT}\equiv I\otimes\begin{pmatrix}1 & 0\\
0 & 0
\end{pmatrix}+\sigma_{x}\otimes\begin{pmatrix}0 & 0\\
0 & 1
\end{pmatrix}=\begin{pmatrix}1 & 0 & 0 & 0\\
0 & 0 & 0 & 1\\
0 & 0 & 1 & 0\\
0 & 1 & 0 & 0
\end{pmatrix}\,,\label{eq:cNOT}
\end{equation}
where I and $\sigma_{x}$ are defined according to Eq.~(\ref{eq:IandSigmaX}).
Matrices associated with operations on the system (resp: environment)
are placed to the left (resp: right) of the tensor product. Given
that Adam and Bob lack knowledge of Eve's bit value, $e$, but are
aware of the interaction, to what degree can they communicate, i.e.,
what is the effect of the interaction on the value, $S\left(T\right)$,
measured by Bob?  More importantly, what action can Bob undertake
to undo the effect of the interaction, so as to obtain Adam's bit,
$S\left(T\right)=S\left(0\right)$? 

\paragraph{Evolution of the state, and Bob's information gain. }

Employing the formalism developed in Section~\ref{subsec:Classical-measurement-theory-basic-concept},
the initial state of the composite system, consisting of the two coins,
is described by the state vector $\vec{J}\left(0\right)=\vec{S}\left(0\right)\otimes\vec{E}\left(0\right),$
where the initial states of the system and environment are $\vec{S}\left(0\right)=\begin{pmatrix}\frac{1+s_{A}}{2}\\
\frac{1-s_{A}}{2}
\end{pmatrix}$ and $\vec{E}\left(0\right)=\begin{pmatrix}\frac{1+p_{E}}{2}\\
\frac{1-p_{E}}{2}
\end{pmatrix},$ respectively. The variable $p_{E}$ denotes the bias of Eve's coin,
see Eq.~(\ref{eq:BitBloch}). Following the interaction, the composite
system state is given by $\vec{J}\left(T\right)=\mathrm{cNOT\,}\vec{J}\left(0\right)$.
The expected mean value of Bob's measurement of $S\left(T\right)$,
represented by the matrix $I\otimes\sigma_{z}$, is given by, recalling
Eq.~(\ref{eq:ExpecL1}),
\begin{equation}
\E{S\left(T\right)}=\left|\left(I\otimes\sigma_{z}\right)\vec{J}\right|_{\mathrm{L1}}=p_{E}s_{A}.\label{eq:EsSamAdam}
\end{equation}
To understand Eq.~(\ref{eq:EsSamAdam}), consider three limiting
cases: i) Eve always prepares her coin facing up, $e=1$, corresponding
to a maximal coin bias, $p_{E}=1$. Since for $e=1$ the interaction
with her coin has no effect on Adam's coin, Bob faithfully receives
Adam's bit every time, $\E{S\left(T\right)}=s_{A}$. ii) Eve always
prepares her coin facing down, $e=-1$. Since her coin bias is now
$p_{E}=-1$, Bob always receives Adam's coin flipped $\E{S\left(T\right)}=-s_{A}$.
While inconvenient for Bob, by flipping each coin he receives (a deterministic
action), he could recover the bit. The effect of Eve's coin is to
change the encoding of the information, but has not resulted in its
loss. iii) Eve prepares her coin completely randomly, $p_{E}=0$.
On average, Bob receives no information from Adam, $\E{S\left(T\right)}=0$!
Eve has  randomly scrambled the encoding of the information for each
of the realizations, which, from Bob's point of view, results in the
complete loss of the initial system information encoded by Adam. More
generally, for an arbitrary coin bias $p_{E}$, the information shared
between Adam and Bob is characterized by the correlation between the
initial and final configurations of the system, which is given by
the bias of Eve's bit, $\E{S\left(T\right)S\left(0\right)}=p_{E}$,
which can be understood as the result of information transfer between
the system and environment, facilitated by the cNOT interaction, however,
where the ``information'' propagating to the system from the environment
is random noise. 

While the transfer of information between Adam and Bob is degraded
by the influence of the interaction with Eve's bit, in principle no
information has been erased, because the cNOT interaction is reversible.
For the case where $p_{E}=0$, Adams bit, $s_{A}$, is not transferred
to Bob at all; rather, it is encoded in the correlation between the
system and environment, $\E{S\left(T\right)E\left(T\right)}=\left|\left(\sigma_{z}\otimes\sigma_{z}\right)\vec{J}\right|_{\mathrm{L1}}=s_{A}$,
which is inaccessible to Adam and Bob, who only have control over
the system coin, and, hence, only access to $S$. To summarize, the
interaction between the system and a randomly prepared random environment
results in loss of information and injection of noise into the system,
as far as the system alone is concerned. Nevertheless, from the vantage
point of the composite system, no information is lost; rather, it
is transferred into correlations between the system and environment. 

\paragraph{Recovering the information.}

To recover Adam's bit, Bob requires access to Eve's physical coin
or knowledge of $e$, the specific value of her coin for \emph{each}
realization. First, consider the latter case, where Bob learns $e$.
Recalling that $\mathrm{cNOT}^{2}=I\otimes I$, before Bob performs
a measurement, he can undo the interaction effect by preparing an
ancillary, third, coin with the value $e$, by employing it to perform
a second cNOT operation on his coin, thus reversing the first. Applied
to each realization, this procedure results in faithful communication,
$\E{S\left(T\right)S\left(0\right)}=1$. Notably, Bob can also reverse
the interaction effect\emph{ after} performing his measurement by
essentially applying the second cNOT operation virtually, i.e., when
$e=1$, $s_{A}=s_{B}$, while otherwise, $s_{A}=-s_{B}$. We remark
that any operations performed by Eve on her coin after the system-environment
interaction have no consequences for Bob. To summarize, the examples
highlights three distinct aspects regarding the recovery of information
in the classical setting:
\begin{enumerate}
\item Eve's physical system was not required, only information about its
initial configuration, $E\left(0\right)=e$.
\item The effect of the interaction can be reversed \emph{before} or \emph{after}
Bob's measurement. 
\item Operations on Eve's coin subsequent to the interaction have no consequences.
\end{enumerate}
All three of these features break down in the quantum setting, as
discussed in the following subsection. 

\subsection{Quantum toy model\label{subsec:Quantum-model-of}}

Rather than communicating with classical bits (coins), consider the
situation where Adam and Bob communicate with quantum bits (qubits),
and Eve too employs a qubit. Before proceeding, we briefly review
the basic qubit concepts.

\paragraph{Quantum bit. }

While the fundamental concept of classical information is the bit,
which represents the minimal classical system, the fundamental concept
of quantum information is the \emph{quantum bit}, or \emph{qubit}
for short, which represents the minimal physical quantum system. A
qubit has two basis states, $\ket{+z}$ and $\ket{-z}$. A pure state
of the qubit  is described by the state $\ket{\psi}=\cos\left(\frac{\theta}{2}\right)\ket{+z}+e^{i\phi}\sin\left(\frac{\theta}{2}\right)\ket{+z}$,
where the angles $\theta$ and $\phi$, which fall in the range $0\leq\theta\leq\pi$
and $0\leq\phi\leq2\pi$, define a point on the unit sphere, known
as the \emph{Bloch sphere}, see Fig.~\ref{fig:Geometric-representation-of}b.
 More generally, a statistical ensemble of pure states, a \emph{mixed}
qubit state is described by the density matrix 
\begin{equation}
\rho=\frac{1}{2}\left(\hat{I}+X\hat{\sigma}_{x}+Y\hat{\sigma}_{y}+Z\hat{\sigma}_{z}\right)\,,
\end{equation}
where $X$, $Y$, $Z$ are real numbers parameterizing the state,
given by the averages of the Pauli operators, $X\equiv\Tr{\hat{\sigma}_{x}\rho}$,
\emph{et cetera}, where $\Tr{\cdot}$ denotes the trace operation.
The matrix representation of the identity, $\hat{I}$, and Pauli $\hat{\sigma}_{x}$
and $\hat{\sigma}_{z}$ operators is given in Eqs.~(\ref{eq:IandSigmaX})
and (\ref{eq:SigmaZ-CM}), while that of Pauli operator Y is $\hat{\sigma}_{y}=\begin{pmatrix}0 & -i\\
i & 0
\end{pmatrix}$, where $i$ is the unit imaginary. The Bloch vector, $\left(X,Y,Z\right)^{\intercal}$,
provides an important geometrical representation of the state of the
qubit, and as discussed in Sec.~\ref{subsec:Classical-measurement-theory-basic-concept}
is the analog of the coin bias $p$. For a pure state, the Bloch vector
extends to the surface of the Bloch sphere, while for mixed states,
it lies in the interior. Notably, it admits the spherical parameterization:
\begin{align}
X & =r\sin\left(\theta\right)\cos\left(\phi\right)\,,\nonumber \\
Y & =r\sin\left(\theta\right)\sin\left(\phi\right)\,,\nonumber \\
Z & =r\cos\left(\theta\right)\,,\label{eq:XYZBLoch}
\end{align}
where the angles $\theta$ and $\phi$ are defined as for pure states
and $r$ is the length of the Bloch vector, a number between 0 and
1. Notably, in the Bloch representation, mutually orthogonal state
vectors are \emph{not} represented by orthogonal Bloch vectors, but
rather, by opposite Bloch vectors, which specify antipodal points
on the sphere. 

\paragraph{Quantum toy model. }

Returning to the toy model example of the interaction between two
systems (recall Fig.~\ref{fig:Reversibility-of-classical}, which
depicts the analogous classical model), we consider the case where
at time $t=0$ Adam prepares his qubit in the pure state $\ket{\psi\left(0\right)}$,
with corresponding Bloch vector components $X\left(0\right)$, $Y\left(0\right)$,
and $Z\left(0\right)$, while Eve prepares her qubit in the pure state
$\ket{+x}$, where $\ket{+x}=\frac{1}{\sqrt{2}}\left(\ket{+z}+\ket{+z}\right)$.
Unlike in the classical toy model, Adam has a choice regarding the
encoding of his information — the orientation of the Bloch vector,
which has no classical analog. Both qubits are sent flying. A controlled-NOT
interaction occurs, described by the operator $\mathrm{cNOT}=\hat{I}\otimes\kb{+z}{+z}+\hat{\sigma}_{x}\otimes\kb{-z}{-z}$,
where operators on the left (resp., right) of the tensor product,
denoted $\otimes$, act on the system (resp., environment). Notably,
the matrix representation of the cNOT operator is the same that of
the classical cNOT gate, given in Eq.~(\ref{eq:cNOT}). After the
interaction Bob receives the system qubit, at time $T$. 

\paragraph{Effect of the interaction before the measurement. }

Before the measurement, the pure state of the composite system, $\ket{\Psi\left(T\right)}=\mathrm{cNOT}\left(\ket{\psi\left(0\right)}\ket{+x}\right)$,
is, in general, inseparable — it cannot be written as a simple product
of states of its component systems. On a mathematical level, this
result is the same as that for the classical model; however, the interpretation
and consequences are markedly distinct. Classically, the inseparability
represented statistical correlations between \emph{definite} configurations
of the system and environment. For the quantum model, the inseparability
represents entanglement between the system and environment — the system
cannot be fully described without considering the environment. Generally,
measurements of the entangled system are correlated with those of
the environment, and the system alone cannot be represented by a pure
state. The consequences of the system-environment entanglement are
at the heart of quantum measurement theory. 

Consider the reduced density matrix of the system qubit, found by
taking the partial trace over the environment, denoted $\mathrm{Tr}_{E}\left[\cdot\right]$,
\begin{equation}
\rho_{S}\left(T\right)=\mathrm{Tr}_{E}\left[\ketbra{\Psi\left(T\right)}{\Psi\left(T\right)}\right]=\frac{1}{2}\begin{pmatrix}1 & X\left(0\right)\\
X\left(0\right) & 1
\end{pmatrix}\,.\label{eq:rho_s-env-q}
\end{equation}
Evidently, entanglement in the composite state, the result of the
interaction between the system and environment results in the loss
of information from the point of view of the system. Specifically,
the $Y$ and $Z$ Bloch components prepared by Adam, $Y\left(0\right)$
and $Z\left(0\right)$, are absent in $\rho_{S}\left(T\right)$, despite
the deterministic preparation of the ancilla in a\emph{ pure} state
$\ket{+x}$. However, if Adam chose to encode his information along
the X component of the Bloch vector, it would propagate to Bob undisturbed
by the interaction with the environment, and Bob could receive it
by measuring $X$. It is the $X$ component that is preserved due
to the choice of the interaction and initial pure state of the environment.
Analogously to the classical case, no information is truly lost, but
rather, when viewed in the broader context of the composite system,
Adam's initial $Y$ and $Z$ qubit components are encoded in the YZ,
$\left\langle YZ\right\rangle \equiv\Tr{\left(\hat{\sigma}_{y}\otimes\hat{\sigma}_{z}\right)\rho}=Y\left(0\right)$,
and ZZ, $\left\langle ZZ\right\rangle \equiv\Tr{\left(\hat{\sigma}_{z}\otimes\hat{\sigma}_{z}\right)\rho}=Z\left(0\right)$,
correlations between the system and environment, respectively.

\paragraph{Recovering information before the measurement. }

In the classical case, by learning the initial configuration of the
environment, $E\left(0\right)=e$, Bob could undo the effect of the
system-environment interaction and could recover the state sent by
Adam before performing the measurement. In the quantum case, this
is not possible. Even though Bob can know the initial state, $\ket{+x}$,
of the environment and can clone it, by preparing a third ancilla
qubit in the state $\ket{+x}$, he cannot use this ancilla to perform
a second cNOT operation on the system so as to reverse (recall that
$\mathrm{cNOT}^{2}=\hat{I}$) the cNOT performed by the environment
qubit. This is a profound consequence of the entanglement between
the system and environment, and has no classical analog. The only
way to reverse the interaction is to use the physical qubit of the
environment to perform the second cNOT operation — \emph{no} clone
will suffice. 

\paragraph{Projective (von Neumann) measurement.\label{par:Projective-(von-Neumann)}}

For a classical system described by a state of maximal knowledge,
the result of any measurement can be determined with certainty. However,
for a quantum system described by a state of maximal knowledge, a
pure state, the result of a measurement is \emph{not}, in general,
determined. For definitiveness, consider the description of a perfect
projective (von Neumann) measurement performed by Bob on the $Z$
component of his qubit spin, with the associated  operator (\emph{observable})
$\hat{\sigma}_{z}$.  The measurement is described by the spectral
decomposition of the observable, $\hat{\sigma}_{z}=\sum_{r}r\hat{\pi}_{r}=\hat{\pi}_{1}-\hat{\pi}_{-1}$,
where $r$ is an eigenvalue, $r=1$ or $r=-1$, to which corresponds
a measurement result, and $\hat{\pi}_{r}$ is the projection operator
onto the eigenstate associated with $r$, $\hat{\pi}_{1}=\kb{+z}{+z}$
and $\hat{\pi}_{-1}=\kb{+z}{+z}$. The probability of obtaining an
outcome corresponding to the eigenvalue $r$ is
\begin{equation}
\wp_{r}=\Tr{\hat{\pi}_{r}\rho}\,.\label{eq:ProbOfprojMsr}
\end{equation}
According to the projection postulate of quantum mechanics,\footnote{Curiously, the modern formulation of the projection postulate is not
precisely that of von Neumann \citep{VonNeumann1932}, but contains
a correction due to Lüders \citep{Luders1951}. } the measurement leads to the projection (or ``collapse'')\footnote{W. Heisenberg introduced the idea of wavefunction collapse in 1927
\citep{Heisenberg1927}.} of the system state into an eigenstate of the measurement operator.
Immediately after the measurement, conditioned on the result $r$,
the state of the system is 
\begin{equation}
\rho_{r}=\frac{\hat{\pi}_{r}\rho\hat{\pi}_{r}}{\wp_{r}}\,.\label{eq:projectionPostulate}
\end{equation}
The evolution due to Eq.~(\ref{eq:projectionPostulate}) is markedly
non-linear in the state density, which appears in the denominator,
and represents a radical departure from the linear evolution encountered
with Schrödinger’s equation. Further, while a perfect measurement
of a classical system does \emph{not} alter its state, a perfect measurement
of a quantum system, in general, \emph{does} alter its state. This
non-linear disturbance has profound consequences.

Suppose, at time $T$, Bob performs a $Z$ measurement of his qubit
and obtains the result $r=1$, with probability, recalling Eqs.~(\ref{eq:rho_s-env-q})
and~(\ref{eq:ProbOfprojMsr}), $\wp_{1}=\Tr{\hat{\pi}_{1}\rho_{S}\left(T\right)}=\frac{1}{2}$.
Note that $\wp_{1}$ is independent of $X\left(0\right)$, $Y\left(0\right)$,
and $Z\left(0\right)$. The system state after the measurement is
$\rho_{1}=\hat{\pi}_{1}\rho_{S}\left(T\right)\hat{\pi}_{1}/\Tr{\rho_{S}\left(T\right)\hat{\pi}_{1}}=\begin{pmatrix}1 & 0\\
0 & 0
\end{pmatrix}$, corresponding to the pure state $\ket{+z}$. Notably, the potentially
recoverable information encoded by Adam, $X(0)$, is irreversibly
lost. From the point of view of the composite system, described by
the state $\ket{\Psi\left(T\right)}$, the measurement has projected
the state onto the measurement basis, according to the effect of the
projector $\hat{\pi}_{1}\otimes\hat{I}$. The state of the composite
system after the measurement, $\ket{+z}\ket{\psi\left(0\right)}$,
is pure and separable, i.e., the measurement has disentangled the
system and environment. In this toy model (and for this particular
measurement outcome), it just so happens that Adam's state is completely
teleported to Eve's qubit, a form of information transfer between
the two systems. To understand the situation a bit better, consider
the alternative, where $r=-1$, with the associated projector $\hat{\pi}_{-1}\otimes\hat{I}$.
The conditional state of the system after the measurement is again
obtained by employing Eq.~(\ref{eq:projectionPostulate}), yielding
$\ket{+z}\ket{\psi'}$ for the composite system, where the state $\ket{\psi'}$
has the same Bloch vector as Adam's initial state $\ket{\psi\left(0\right)}$
but with the $Y$ and $Z$ components flipped. This example illustrates
the more general feature that a measurement on either the system or
environment disentangles the two, resulting in a perfect correlation
between the measurement on one and the state of the other. Further,
it tends to lead to a transfer of information between the two subsystems.
We explore the profound consequences of these features in the following
section.

\section{Continuous quantum measurements: introduction to quantum trajectories
and stochastic calculus \label{sec:Introductory-example:-qubit} }

In this section, we consider a heuristic microscopic model of continuous
quantum measurements, which, although simple, contains sufficient
generality to introduce the principal ideas. Specifically, we model
the homodyne measurement of a qubit by a sequence of interactions
with a chain of identically prepared ancilla qubits.  A chain of
ancillary systems modeling the environment is known as a von Neumann
chain \citep{VonNeumann1932}. While the evolution due to the interaction
with each ancilla is unitary, ``deterministic,'' the addition of
a projective (von Neumann) measurement of each ancilla subsequent
to its interaction with the system results in the stochastic evolution
of the quantum state of the system — known as a\emph{ quantum trajectory}
\citep{Carmichael1993}. Due to the correlation between the state
of a measured ancilla and the resulting  state of the system, the
measurement results allow faithful tracking of the state trajectory
\citep{belavkin1987non,Carmichael1993,Gardiner1992-original-traj,Dalibard1992-original-traj,Korotkov1999-original-traj}.
After introducing the time-discrete version of the model, we take
its continuum limit, which allows us to introduce the fundamental
concepts of stochastic calculus. Specifically, we focus on introducing
the Wiener noise process and obtaining the stochastic differential
equations (SDEs) that describe the homodyne monitoring of the qubit.
Most of the results derived in this section carry over with little
modification to the following section, Sec.~\ref{sec:Quantum-trajectory-theory},
which establishes the general formulation of quantum measurement theory.
Time-discrete chain models have been discussed in Refs.~\citet{Caves1987-time-discrete,Attal2006,Tilloy2015,Korotkov2016-qm-bayes,Bardet2017}.

\subsection{Time-discrete model with flying spins}

\begin{figure}
\begin{centering}
\includegraphics[scale=1.4]{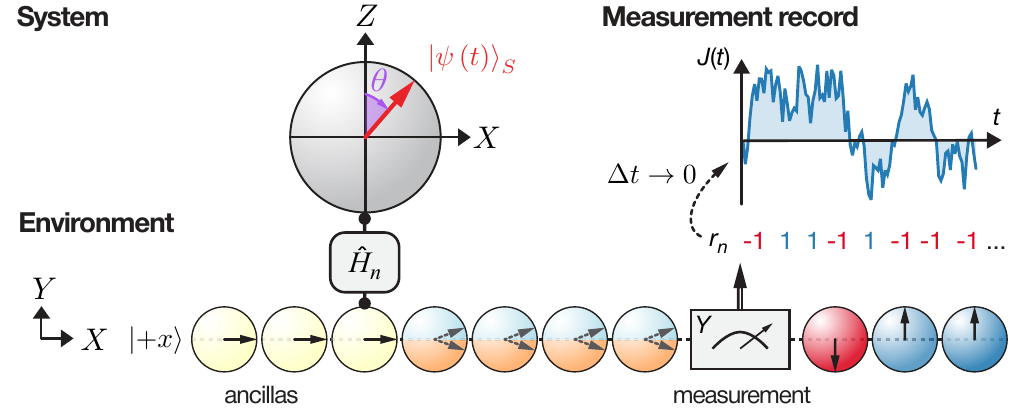}
\par\end{centering}
\caption[Homodyne monitoring of a quantum bit: time-discrete model]{\textbf{\label{fig:Ch2:SpinBathModel1}Homodyne monitoring of a quantum
bit: time-discrete model.} The qubit, whose Bloch vector lies in the
XZ plane, sequentially interacts with a chain of ancilla qubits, which
model the environment. At the beginning of each timestep, at time
$t$, the system is in a pure state, $\ket{\psi\left(t\right)}_{S}$.
During the $n$-th timestep, of length $\Delta t$, the qubit interacts,
subject to the Hamiltonian $\hat{H}_{n}$, with the $n$-th ancilla,
prepared in $\ket{+x}$, whereafter, the Y component of its spin is
projectively measured. The result of the measurement, $r_{n}$, which
is either -1 or 1, is recorded and accumulated; in the continuum limit,
$\Delta t\rightarrow0$, it leads to the homodyne signal $J\left(t\right)$,
a time-continuous stochastic (Weiner) process.}
\end{figure}

Time is discretized in small but finite bins of length $\Delta t$
labeled by the integer $n$, i.e.,~$t=n\Delta t$. During each timestep,
a single spin of the environment, referred to as the \emph{ancilla},
interacts with the system for time $\Delta t$, see Fig.~\ref{fig:Ch2:SpinBathModel1}.
For simplicity, assume each spin is identically prepared in the state
$\ket{+x}$. We employ the convention that the states $\ket{\pm x}$,
$\ket{\pm y}$, and $\ket{\pm z}$ denote eigenstates of the Pauli
X, Y, and Z operators, respectively. The interaction between the $n$-th
ancilla and the system is described by the Hamiltonian 
\begin{equation}
\hat{H}_{n}\equiv-\frac{\hbar\lambda}{2}\hat{\sigma}_{z}^{S}\otimes\hat{\sigma}_{z}^{\left(n\right)}\,,
\end{equation}
where $\lambda$ is the strength of the interaction, $\hbar$ is Plank's
constant, and $\hat{\sigma}_{z}^{S}$ and $\hat{\sigma}_{z}^{\left(n\right)}$
denote the Pauli Z operators of the system and ancilla, respectively.
For the time being, we assume that $\hat{H}_{n}$ is the only generator
of system evolution, and the system Hamiltonian is zero, $\hat{H}_{S}=0$.
Following the interaction, the ancilla is measured by a detector that
performs a projective measurement of the ancilla spin  Y component,
which yields the measurement result $r_{n}=-1$ or $r_{n}=1$. The
observer operating the measurement apparatus keeps track of the sum
total of the measurement results, the measurement signal: $J_{t}\equiv\sqrt{\Delta t}\sum_{n'=0}^{n}r_{n'}$. 

Note two assumptions  regarding the measurement: i) the ancilla qubits
are undisturbed during their flight from the system to the measurement
apparatus, and ii) the measurement apparatus performs a perfect measurement,
and does not add technical noise. These assumptions ensure no information
is lost in the measurement, nor spurious noise is added by it; i.e.,
the observer has perfect access to all information there is to know
in the environment, and is hence referred to as an \emph{omniscient}
\emph{observer}. 

\subparagraph{Evolution of the composite system. }

For simplicity, assume the state of the system at time $t$ is pure
and its Bloch vector lies in the XZ plane; i.e., it is described by
a single angle $\theta\left(t\right)$, 
\begin{align}
\ket{\psi\left(t\right)}_{S} & =\cos\left(\frac{\theta\left(t\right)}{2}\right)\ket{+z}_{S}+\sin\left(\frac{\theta\left(t\right)}{2}\right)\ket{-z}_{S}=\begin{pmatrix}\cos\left(\frac{\theta\left(t\right)}{2}\right)\\
\sin\left(\frac{\theta\left(t\right)}{2}\right)
\end{pmatrix}\,.\label{eq:SpinBath:sysState}
\end{align}
The state of the composite system at time $t$, consisting of the
$n$-th ancilla and the system qubit, is $\ket{\Psi\left(t\right)}=\ket{\psi\left(t\right)}_{S}\otimes\ket{+x}_{n}$,
and for duration $\Delta t$ evolves subject to the Hamiltonian $\hat{H}_{n}$.
The total evolution is given by the propagator $\hat{U}\left(t,t+\Delta t\right)=\exp\left(-i\hat{H}_{n}\Delta t/\hbar\right)$,
and the composite-system state after the interaction is 
\begin{equation}
\ket{\Psi\left(t+\Delta t\right)}=\hat{U}\left(t,t+\Delta t\right)\ket{\Psi\left(t\right)}\,,
\end{equation}
Anticipating the ancilla Y measurement, we express $\ket{\Psi\left(t+\Delta t\right)}$
in terms of the measurement operator eigenstates. The measurement
operator on the ancilla alone is the Pauli Y operator, $\hat{\sigma}_{y}^{\left(n\right)}$,
with eigenstates $\ket{-y}_{n}$ and $\ket{+y}_{n}$, in terms of
which, 
\begin{equation}
\ket{\Psi\left(t+\dt\right)}=\ket{\tilde{\psi}_{-1}\left(t+\Delta t\right)}_{S}\otimes\ket{-y}_{n}+\ket{\tilde{\psi}_{+1}\left(t+\Delta t\right)}_{S}\otimes\ket{+y}_{n}\,,\label{eq:PsiQubitEnvSpin}
\end{equation}
where the parameter $\epsilon\equiv\lambda\Delta t$ characterizes
the measurement strength and the un-normalized\footnote{By convention, a tilde will indicate an unnormalized state, with a
norm less than one.} system states $\ket{\tilde{\psi}_{\pm1}\left(t+\Delta t\right)}_{S}$
are
\begin{equation}
\ket{\tilde{\psi}_{\pm1}\left(t+\Delta t\right)}_{S}\equiv\begin{pmatrix}\cos\left(\frac{\theta\left(t\right)}{2}\right)\cos\left(\frac{\pi/2\pm\epsilon}{2}\right)\\
\sin\left(\frac{\theta\left(t\right)}{2}\right)\sin\left(\frac{\pi/2\pm\epsilon}{2}\right)
\end{pmatrix}_{S}\,.\label{eq:SpinBath:psiTilde}
\end{equation}
The state of the composite system following the interaction, Eq.~(\ref{eq:PsiQubitEnvSpin}),
is not separable. The interaction has entangled the system and environment,
as discussed of Sec.~\ref{subsec:Quantum-model-of}.

\paragraph{Projective (von Neumann) measurement of the ancilla. }

\begin{figure}
\begin{centering}
\includegraphics[scale=1.4]{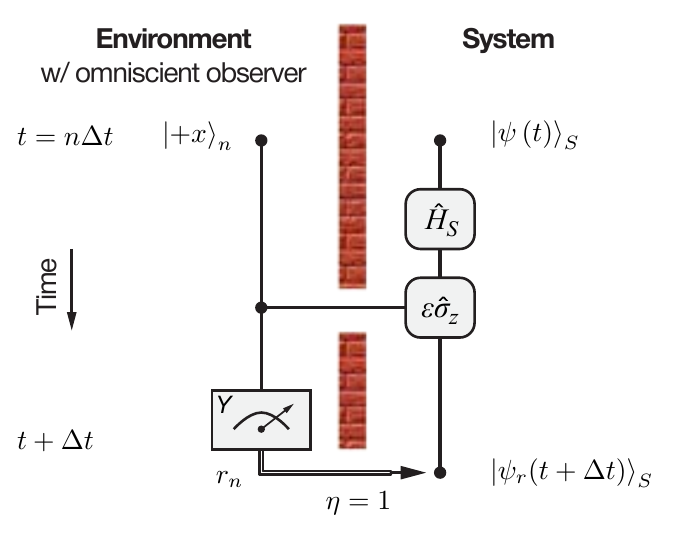}
\par\end{centering}
\caption[Circuit representation of the $n$-th timestep of the quantum trajectory]{\textbf{\label{fig:Ch2:SpinBath}Circuit representation of the $n$-th
timestep of the quantum trajectory. }At time $t=n\Delta t$, the system,
described by the state $\ket{\psi\left(t\right)}_{S}$, is subjected
to the system Hamiltonian $\hat{H}_{S}$ and the interaction with
the $n$-th ancilla, characterized by the parameter $\epsilon$. Every
ancilla is prepared in $\ket{+x}$. Following the interaction, the
detector projectively measures the Y component of the ancilla spin,
yielding the result $r_{n}$, which provides the information necessary
to update the state of the system. In the case of the omniscient observer,
characterized by unit quantum measurement efficiency, $\eta=1$, at
the end of the timestep, immediately after $t+\Delta$, the system
state, $\ket{\psi_{r}\left(t+\Delta t\right)}_{S}$, conditioned on
the measurement result is pure. Contrast with Fig.~\ref{fig:Reversibility-of-classical}. }
\end{figure}

The action of the measurement apparatus on the composite system is
described, recalling the discussion on Pg.~\pageref{par:Projective-(von-Neumann)},
by decomposing the measurement operator $\hat{Y}_{n}=\hat{I}\otimes\hat{\sigma}_{y}$
in terms of its  eigenstate projectors, $\hat{\pi}_{\pm}\equiv\hat{I}_{S}\otimes\left(\kb{\pm y}{\pm y}\right)_{n}$;
note, $\hat{Y}=\hat{\pi}_{+}-\hat{\pi}_{-}$. According to the von~Neumann
postulate, the projectors yield the probability for obtaining the
results $r_{n}=-1$ and $r_{n}=1$ from the measurement, 
\begin{align}
\wp_{r}\left(t\right) & =\bra{\Psi\left(t+\Delta t\right)}\hat{\pi}_{r}\ket{\Psi\left(t+\Delta t\right)}\nonumber \\
 & =\braket{\tilde{\psi}_{r}\left(t+\Delta t\right)}{\tilde{\psi}_{r}\left(t+\Delta t\right)}\nonumber \\
 & =\frac{1}{2}\left(1-r_{n}\sin\left(\epsilon\right)\cos\left(\theta\left(t\right)\right)\right)\,,\label{eq:PflyingSpinPM1}
\end{align}
as well as the state of the composite system immediately after the
measurement, conditioned on the result $r_{n}$, 
\begin{equation}
\ket{\Psi_{r}\left(t+\Delta t\right)}=\ket{\tilde{\psi}_{r}\left(t+\Delta t\right)}_{S}\ket{y:=r}_{n}/\sqrt{\wp_{r}\left(t\right)},\label{eq:Psir_afterMsr}
\end{equation}
where $\ket{y:=r}_{n}$ denotes ancilla state $\ket{+y}_{n}$ (resp.,
$\ket{-y}_{n}$) for $r=1$ (resp., $r=-1$). The measurement has
transformed the entanglement between the system and environment, evident
in the non-separable state $\ket{\Psi\left(t+\Delta t\right)}$, Eq.~(\ref{eq:PsiQubitEnvSpin}),
into a correlation between the pure state of the system and environment
after the measurement, evident in the separable, non-entangled conditional
state $\ket{\Psi_{r}\left(t+\Delta t\right)}$, Eq.~(\ref{eq:Psir_afterMsr}).
Assuming the ancilla never interacts with the system again, it is
unnecessary to retain it in the description of the measurement; removing
it from Eq.~(\ref{eq:Psir_afterMsr}), we obtain the pure state of
the system alone at time $t+\Delta t$:
\begin{align}
\ket{\psi_{r}\left(t+\Delta t\right)}_{S} & =\frac{1}{\sqrt{\wp_{r}\left(t\right)}}\ket{\tilde{\psi}_{r}\left(t+\Delta t\right)}_{S}=\begin{pmatrix}\cos\left(\frac{\theta_{r}\left(t+\dt\right)}{2}\right)\\
\sin\left(\frac{\theta_{r}\left(t+\dt\right)}{2}\right)
\end{pmatrix}\,.\label{eq:SpinModel:postState}
\end{align}
From the point of view of the observer, the entanglement is transformed
by the measurement into a classical correlation between the result
$r_{n}$ and the final conditional state of the system, $\ket{\psi_{r}\left(t+\Delta t\right)}_{S}$.
Figure~\ref{fig:Ch2:SpinBath} summarizes the steps of the model
and the conditional state update.

\paragraph{Solution for the conditional state update. }

To explicitly solve Eq.~(\ref{eq:SpinModel:postState}) for the updated
angle of the qubit system conditioned on the measurement result $r_{n}$,
$\theta_{r}\left(t+\Delta t\right)$, one can use Eqs.~(\ref{eq:PflyingSpinPM1})
and~(\ref{eq:SpinBath:psiTilde}), following trigonometric manipulation,
to obtain, without any approximations, an explicit relation (\citeauthor{MHD})
between the Bloch angle at the start and end of the timestep: 
\begin{equation}
\boxed{\tan\left(\frac{\theta_{r}\left(t+\Delta t\right)}{2}\right)=\tan\left(\frac{\theta\left(t\right)}{2}\right)\tan\left(\frac{\pi/2+r_{n}\epsilon}{2}\right)\,.}\label{eq:SpinBath:Ch2:tan}
\end{equation}
In the following section, Sec.~\ref{subsec:Random-walk-on}, this
seemingly non-linear equation is transformed into a linear equation
by a hyperbolic transformation of the circular angle $\theta$, and
is solved exactly. Nonetheless, for the continuum-limit discussion
in Sec.~\ref{subsec:Continuous-limit:-Wiener}, consider the solution
of Eq.~(\ref{eq:SpinBath:Ch2:tan}) in the limit of weak interactions,
$\epsilon\ll1$, to order $\epsilon$: 
\begin{equation}
\mathrm{d}\theta\left(t\right)\equiv\theta\left(t+\Delta t\right)-\theta\left(t\right)\approx\epsilon r_{n}X\left(t\right)\,,\label{eq:SPinBath:ThPM}
\end{equation}
where we have defined the Bloch angle increment, $\mathrm{d}\theta\left(t\right)$,
and $X\left(t\right)$ is the X component of the Bloch vector, $X\left(t\right)=\sin\left(\theta\left(t\right)\right)$. 

\paragraph{Interpretation and remarks. }

The system measurement dynamics are described in entirety by Eqs.~(\ref{eq:PflyingSpinPM1}),~(\ref{eq:SpinModel:postState}),
and~(\ref{eq:SPinBath:ThPM}). To make the discussion more concrete,
consider the particular case where the system and ancilla do not interact,
$\epsilon=0$. The measurement results are completely random, $\wp_{r}=\frac{1}{2}$,
uncorrelated with the system; similarly, the system state is independent
of the measurement results, $r_{n}$; in fact, there is no state evolution,
$\mathrm{d}\theta\left(t\right)=0$. Consider the more interesting
case of weak interactions, $\epsilon\ll1$. Measurement results are
correlated with the Z component of the system Bloch vector, $\wp_{r}=\frac{1}{2}\left(1-\epsilon r_{n}Z\left(t\right)\right)$,
where $Z\left(t\right)=\cos\left(\theta\left(t\right)\right)$. Nonetheless,
 due to $\epsilon\ll1$, the two measurement results still occur with
nearly equal probability, and the record consists of random noise,
but with a slight bias that correlates it with $Z$. Thus, the value
of $Z$ can be obtained from the instantaneous average of the measurement
results, $\E{r_{n}}=-\epsilon Z$. In time, from the point of view
of the observer, a long sequence of measurements gradually results
in the complete measurement of $Z$, obtained from the noisy measurement
record. A peculiar feature of the weak interaction regime, $\epsilon\ll1$,
is that amplitude of the noise is essentially constant for all measurement
strengths, its variance is $\mathrm{Var}\left[r_{n}\right]=1-\left(\epsilon Z\right)^{2}\approx1$.
This origin of the randomness can be interpreted to be quantum in
nature, since the system and environment are in pure states at all
times. Specifically, it is due to the incompatibility (orthogonality)
between the initial state of the ancilla, $\ket{+x}_{n}$, and the
eigenstates, $\ket{\pm y}_{n}$, of the measurement observable. 

The random measurement result, $r_{n}$, is correlated with a small
``kick'' on the state of the system, described by Eq.~(\ref{eq:SPinBath:ThPM}).
Conditioned on the result $r_{n}=1$ (resp., $r_{n}=-1$) the system
experiences a downward (resp., upward) kick corresponding to the circular
increment $\mathrm{d}\theta\left(t\right)=\epsilon r_{n}\mathrm{sgn}\left(X\left(t\right)\right)\sqrt{1-Z\left(t\right)}$,
whose magnitude is largest for $Z=0$, but vanishing in the limit
where $Z$ approaches $\pm1$; the sign function is denoted $\mathrm{sgn}$.
This state-dependent nature of the back-action kicks leads to the
eventual projection of the state onto one of the eigenstate of the
system observable, $\hat{\sigma}_{z}$, as discussed in Sec.~\ref{subsec:Random-walk-on}. 

The form of the backaction depends on the ancilla quantity being measured
by the apparatus; for example, a measurement of a quadrature other
than the ancilla $Y$ quadrature yields a different form of the measurement
backaction. More generally, we emphasize that no matter what ancilla
quantity is measured, so long as the measurement is projective and
complete knowledge about the ancilla is obtained, the ancilla is collapsed
onto a single unique state. From this, it follows that the system
cannot be entangled with the ancilla and for this reason the system
is left in a pure state. 

\paragraph{Generalized measurements. }

By introducing an ancilla that interacts unitarily with the system
and is subsequently measured, we obtained evolution equations for
the pure state of the quantum system conditioned on the measurement
result $r_{n}$, and could otherwise disregard the ancilla in the
measurement description. The ancilla scheme realizes an indirect measurement
of the system, which gradually obtains information about the system
and disturbs it in a manner that is indescribable with the von Neumann
formulation, summarized by Eqs.~(\ref{eq:ProbOfprojMsr}) and (\ref{eq:projectionPostulate}).
The example of this section belongs to a more general class of measurements,
referred to as \emph{generalized measurements}. A powerful theorem
by Neumark, see Sec.~9-6 of Ref.~\citet{PeresBook}, proved that
any generalized measurement can be formulated essentially according
to the scheme presented so far, where an auxiliary quantum system
is introduced, it interacts unitarily with the system, and is subsequently
projectively measured, in the traditional von~Neumann sense. The
effect of the generalized measurement on the system can be completely
described by system operators, denoted $\hat{M}_{r}$, that are not
in general Hermitian. For our example, the \emph{measurement operator,}\footnote{The measurement operator is sometimes referred to as a Kraus operator.}\emph{
$\hat{M}_{r}$,} follows directly from Eq.~(\ref{eq:Psir_afterMsr}),
\begin{equation}
\hat{M}_{r}\left(t\right)=\prescript{}{n}{\braOket{y:=r}{\hat{U}\left(t,t+\Delta t\right)}{+x}_{n}}\label{eq:MrSpinBath}
\end{equation}
Note that $\ket{+x}_{n}$ is the \emph{initial} ancilla state for
the $n$-th timestep, while $\ket{y:=r}_{n}$ is the \emph{final}
ancilla state, follwing the projective measurement, while $\hat{U}$
is the \emph{composite} system propagator. Since $\hat{M}_{r}$ in
Eq.~(\ref{eq:MrSpinBath}) is not\emph{ }Hermitian, it does not belong
to the traditional notion of an 'observable', and the outcomes $r_{n}$
are not the eigenvalues of $\hat{M}_{r}$, but serve merely as labels.
The measurement operators $\hat{M}_{1}$ and $\hat{M}_{-1}$, which
are non-orthogonal ($\hat{M}_{1}\hat{M}_{-1}\neq0$) link the system
state with the set of measurement probabilities $\wp_{r}$, and formally,
their operator set, $\left\{ \hat{M}_{r}^{\dagger}\hat{M}_{r}:r\right\} $,
constitutes a \emph{positive-operator-valued measure} (POVM) on the
space of results, see Sec.~2.2.6 of Ref.~\citet{NielsenChuangBook}.
In general, the measurement operators generalize von Neumann's postulate
in the following way:
\begin{align}
\wp_{r}\left(t\right)= & \Tr{\hat{M}_{r}\rho\hat{M}_{r}^{\dagger}}\,,\label{eq:GenMsrPosProb}\\
\rho_{r}\left(t+\Delta t\right)= & \hat{M}_{r}\rho\left(t\right)\hat{M}_{r}^{\dagger}/\wp_{r}\left(t\right)\,,\label{eq:GenMsrPosProj}
\end{align}
where $\wp_{r}\left(t\right)$ is the probability to obtain the measurement
outcome $r$ and $\rho_{r}\left(t+\Delta t\right)$ is the state
of the system immediately \emph{after} the measurement, conditioned
on the result $r$. Note that technically, the generalized projection
postulate does not introduce anything fundamentally new beyond von~Neumann's
postulate, since it follows from Neumark's theorem that considering
a larger quantum system with\emph{ }projective (von~Neumann) measurements
and unitary operations is completely equivalent.

\subsection{Geometric representation of a continuous measurement: random walk
on a hyperbola\label{subsec:Random-walk-on}}

\begin{figure}
\begin{centering}
\includegraphics[scale=1.4]{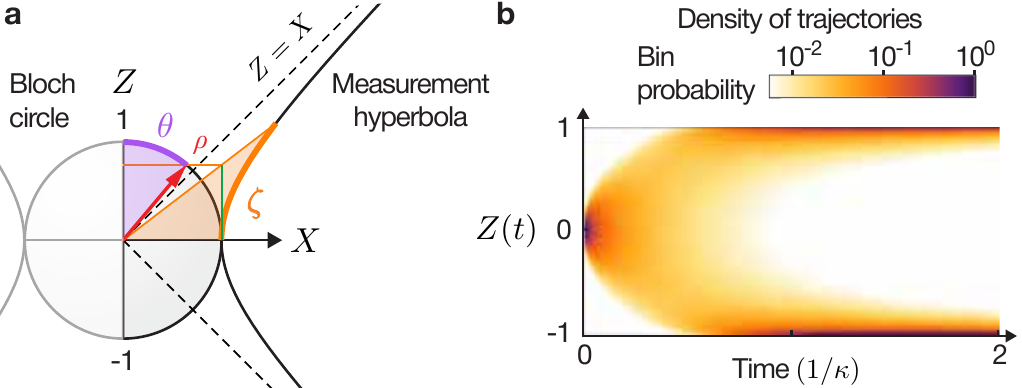}
\par\end{centering}
\caption[Random walk on the measurement hyperbola.]{\textbf{Random walk on the measurement hyperbola.\label{fig:Random-walk-on-hyperb}
a,} Circular and hyperbolic geometric representations of the pure
qubit state, $\rho$, parametrized by the circular and hyperbolic
angles $\theta$ and $\zeta$, respectively, obeying $\cos\left(\theta\right)=\tanh\zeta$.
The circle depicts a slice though the XZ plane of the Bloch sphere,
which is well-suited to represent unitary dynamics. The random walk
of $\zeta$ due to the measurement takes place on the unit hyperbola,
with asymptotes defined by the lines $Z=\pm X$. \textbf{b,} Histogram
of quantum trajectory densities obtained from simulations of the flying-spin
model. All trajectories (not shown here) begin with the initial state
defined by the Bloch coordinates $X\left(0\right)=1$ and $Z\left(0\right)=0$.
The time axis is scaled in units of the measurement rate, $\kappa$.
}
\end{figure}

In this subsection, we present a geometric representation of the measurement
dynamics. Section~\ref{subsec:Quantum-model-of} presented the geometric
representation of the qubit state as a point on the Bloch sphere,
with coordinates $X$, $Y$, and $Z$, or for a pure state, as a point
on the surface of the sphere, parameterized by the angles $\theta$
and $\phi$, Eq.~(\ref{eq:XYZBLoch}). For our qubit example, $Y=0$
by assumption,  hence its pure state, $\ket{\psi}_{S}$, can be represented
on the Bloch \emph{circle}, see Fig.~\ref{fig:Random-walk-on-hyperb}a,
parametrized by a single\footnote{For simplicity, we have assumed that $X>0$, which corresponds to
the angle $\phi=0$. Recall that since $0\leq\theta\leq\pi$, the
left half of the Bloch circle, $X<0$, corresponds to angle $\phi=\pi$.} circular angle $\theta$: $X=\sin\left(\theta\right)$ and $Z=\cos\left(\theta\right)$.
This geometric representation is particularly well suited to describing
unitary operations, which describe rotations in Hilbert space. For
concreteness, consider the state evolution subject to the Rabi Hamiltonian
$\hat{H}_{S}=\frac{1}{2}\hbar\omega\hat{\sigma}_{y}$, where, without
assumptions on the timestep, $\Delta t$, the effect of the propagator
$U\left(t,t+\Delta t\right)=\exp\left(-i\hat{H}_{S}\Delta t/\hbar\right)$
on the state is given by the simple linear equation: 
\begin{equation}
\theta\left(t+\Delta t\right)-\theta\left(t\right)=\omega\Delta t\,.\label{eq:SpinBathHRabi}
\end{equation}
The complexity of Eq.~(\ref{eq:SpinBath:Ch2:tan}) indicates that
the circular representation is \emph{not} well suited to describe
the evolution due to the measurement. Rather, we show that a natural
representation for measurement dynamics is a hyperbolic one. 

\subparagraph{Hyperbolic representation.}

We map the Bloch circle onto the standard hyperbola according to the
equation $Z=\cos\left(\theta\right)=\tanh\zeta$, where $\zeta$ is
the hyperbolic angle, the analogue of the circular angle $\theta$,
see Fig.~\ref{fig:Random-walk-on-hyperb}a. In terms of the hyperbolic
representation, without any approximations, Eq.~(\ref{eq:SpinBath:Ch2:tan})
is transformed into a simple linear equation, analogous to that of
Eq.~(\ref{eq:SpinBathHRabi}), 

\begin{equation}
\zeta_{r}\left(t+\Delta t\right)-\zeta\left(t\right)=-r_{n}\xi\,,\label{eq:HyperbolicRotationMsr}
\end{equation}
where $\zeta_{r}\left(t+\Delta t\right)$ is the hyperbolic angle
of the system after the measurement, conditioned on the result $r_{n}$,
$\zeta\left(t\right)$ is the hyperbolic angle before the interaction
with the ancilla, and $\xi$ is the hyperbolic increment, $\tanh\left(\xi\right)=\sin\left(\epsilon\right)$.
In view of the Eq.~(\ref{eq:HyperbolicRotationMsr}), the measurement
backaction kicks are understood as hyperbolic rotations of a definite
amplitude $\xi$, but random orientation $r_{n}$. By iterating the
calculation of Eq.~(\ref{eq:HyperbolicRotationMsr}) $N$ times,
one can obtain the stochastic path taken by the qubit state, its  quantum
trajectory, which, understood in terms of the stochastic difference
equation $\mathrm{d}\zeta_{r}\left(t\right)\equiv\zeta_{r}\left(t+\Delta t\right)-\zeta\left(t\right)=-r_{n}\xi$,
is a random walk on a hyperbola.

The circular and hyperbolic coordinate transformations together with
Eqs.~(\ref{eq:SpinBathHRabi}) and~(\ref{eq:HyperbolicRotationMsr})
can be employed to construct a finite-difference numerical scheme
to calculate the quantum trajectory of the qubit subject to homodyne
monitoring. Notably, since the difference equations were derived \emph{without}
approximations, especially with regard to the size of $\Delta t$,
they guarantee a physical system state for all parameters and at all
time, features which offer some practical advantages for numerical
simulations. In the following subsection, Sec.~\ref{subsec:Continuous-limit:-Wiener},
we construct the continuum limit of the model and formally derive
the differential equations for the quantum trajectory of the qubit.

\subsection{Continuum limit: Wiener noise and stochastic calculus\label{subsec:Continuous-limit:-Wiener}}

In this section, we take the appropriate limit in which the measurement
becomes continuous. In this limit, the interaction time, $\Delta t$,
becomes infinitely small while the measurement strength, $\lambda$,
becomes infinitely large. Since each short sequence of individual
measurements carries an infinitesimal amount of information in this
limit, we coarse-grain the measurement record in the following way.
The evolution up to time $t$ is divided in $m$ intervals of total
duration $\dt$, while each of these intervals is further subdivided
in $l$ yet smaller intervals, each of duration $\Delta t$; i.e.,
$t=n\Delta t=ml\Delta t$ and $\dt=l\Delta t$. The interaction amplitude,
$\lambda$, is chosen to be $\lambda=\sqrt{\kappa/\Delta t}$, where
$\kappa$ denotes the interaction rate. Subject to appropriate scaling,
by a factor $\sqrt{\Delta t}$, the sum of all measurement results
up to time $t$ is the measurement signal $J_{t}=\sqrt{\Delta t}\sum_{m'=0}^{m}\sum_{l'=0}^{l}r_{m'l'}$.
During a time interval $\dt$, beginning at $t=ml\Delta t$ and ending
at $t'=\left(m+1\right)l\Delta t$, the signal changes by 
\begin{equation}
\mathrm{d}J_{t}=J_{\left(m+1\right)l\Delta t}-J_{ml\Delta t}=\sqrt{\Delta t}\sum_{k=ml}^{\left(m+1\right)l}r_{k}\:.\label{eq:dJ-stoch-diff-eq}
\end{equation}
Equation~(\ref{eq:dJ-stoch-diff-eq}) is known as a \emph{stochastic
difference equation, }because the difference increment in the variable
$J_{t}$ is a random variable. Assuming the state of the system does
not appreciably change over the time interval $\dt$, the measurement
results, $r_{k}$, are independent and identically distributed binary
random variables described by the probability function $\wp\left[r_{k}\right]=\frac{1}{2}\left(1-\epsilon r_{k}Z\left(t\right)\right),$
recall Eq.~(\ref{eq:PflyingSpinPM1}), where $\epsilon=\sqrt{\kappa\Delta t}$.
It follows that the mean and variance of the measurement increment
are 
\begin{align}
\E{\mathrm{d}J_{t}} & =l\sqrt{\Delta t}\E{r_{k}}=-\sqrt{\kappa}Z\left(t\right)\dt\,,\label{eq:EdJ-qubithomo}\\
\mathrm{Var}\left[\mathrm{d}J_{t}\right]= & l\Delta t\mathrm{Var}\left[r_{k}\right]\approx\dt\,,\label{eq:Var-dJ-qubithomo}
\end{align}
respectively, while, according to the central limit theorem, all higher-order
cumulants of the probability distribution for $\mathrm{d}J_{t}$ vanish
in the limit of large $l$. Working in this limit, and employing Eqs.~(\ref{eq:EdJ-qubithomo})
and~(\ref{eq:Var-dJ-qubithomo}), the stochastic difference equation,
Eq.~(\ref{eq:dJ-stoch-diff-eq}), can be taken to  the continuum
limit,\footnote{For completeness, we note that Eq.~(\ref{eq:QubitHomo:dJ}) is a
special case of Eq.~(\ref{eq:Jhom}). }

\begin{equation}
\boxed{\mathrm{d}J\left(t\right)=-\sqrt{\kappa}Z\left(t\right)\dt+\dW\left(t\right)\,},\label{eq:QubitHomo:dJ}
\end{equation}
where the infinitesimal term $\dW\left(t\right)$ represents Gaussian
white noise, and is known as the \emph{Wiener increment}. It obeys
the following canonical relations and probability density: 
\begin{align}
\E{\dW\left(t\right)} & =0\,, & \dW\left(t\right)\dt & =0\,,\label{eq:dW-relation-1}\\
\E{\dW\left(t\right)^{2}} & =\dt\,, & \wp\left(\dW\right) & =\frac{\exp\left(-\dW^{2}/(2\dt)\right)}{\sqrt{2\pi\dt}}\,.\label{eq:dW-relation-2}
\end{align}

\paragraph{Stochastic calculus. }

Equation~(\ref{eq:QubitHomo:dJ}) is known as a\emph{ stochastic
differential equation} (SDE) because the infinitesimal differential
is not completely determined, but is a random variable. Notably, in
obtaining this equation, we took the value of the function $Z\left(t\right)$
at the \emph{beginning} of the timestep $\dt$. In general, because
the stochastic noise is not smooth and not differentiable if we had
taken the value of the function at the\emph{ end} of the time-interval
$\dt$ we would have obtained a different form of the SDE. By taking
the value of the function at the beginning of the timestep, the SDE
we obtained is said to be in \emph{Itô form}; otherwise, it would
have been in \emph{Stratonovich form}. The two forms are not equivalent
in a straightforward manner, but for our purposes, it will suffice
to only consider the Itô form; see Chapter~3 of Ref.~\citet{Jacobs2010Book}
for an in-depth discussion. From Eqs.~(\ref{eq:dW-relation-1}) and
(\ref{eq:dW-relation-2}) it follows that the SDE solution only depends
on terms proportional to $\dt,$ $\dW,$ and $\dW^{2},$ while all
other terms of the form $\dt^{p}\dW^{q}$ are vanishing. Importantly,
in an unusual departure from the rules of differential calculus, in
the continuum limit, one can set $\dW^{2}=\dt$, a result known as
\emph{Itô}'s \emph{rule}. We note that $\dW\left(t\right)$ is an
idealized Gaussian noise process in that it has a perfect delta function
correlation in time, which implies its Markovianity, but also that
it has a white noise power spectral density, non-zero for \emph{all}
frequencies.

\paragraph{Itô SDE for the Bloch vector. }

In our example, the Bloch vector of the qubit can be parameterized\footnote{For simplicity, we have assumed $X\left(t\right)>0$ for all $t$.}
by its Z component, $\vec{S}\left(t\right)=\left(\sqrt{1-Z\left(t\right)^{2}},0,Z\left(t\right)\right)^{\intercal}$.
Thus, it suffices to derive an SDE for $Z\left(t\right)$, which is
obtained from the system state, $\ket{\psi_{J}\left(t\right)}$, conditioned
on the measurement signal $J\left(t\right)$ by taking the expectation
value of $\hat{\sigma}_{z}$, $Z\left(t\right)=\braOket{\psi_{J}\left(t\right)}{\hat{\sigma}_{z}}{\psi_{J}\left(t\right)}$.
The SDE is derived by first Taylor expanding $Z\left(t\right)=\cos\left(\theta\left(t\right)\right)$
to first order in $\dt$, $\mathrm{d}Z\left(t\right)\approx-\sqrt{1-Z\left(t\right)^{2}}\mathrm{d}\theta\left(t\right)-\frac{1}{2}Z\left(t\right)\mathrm{d}\theta\left(t\right)^{2}$,
where we have retained terms to second order in $\mathrm{d}\theta\left(t\right)$.
This is necessary because $\mathrm{d}\theta\left(t\right)$, recalling
Eq.~(\ref{eq:SPinBath:ThPM}), is proportional to $\dW$, and $\dW^{2}=\dt$.
By summing $l$ times over the difference equation and performing
the same coarse graining employed to arrive at Eq.~(\ref{eq:QubitHomo:dJ}),
we arrive at the Itô form of the SDE for the Z component of the Bloch
vector of a qubit subject to heterodyne monitoring of $\hat{\sigma}_{z}$,
\begin{equation}
\boxed{\mathrm{d}Z\left(t\right)=-\sqrt{\kappa}\left(1-Z\left(t\right)^{2}\right)\dW\left(t\right)\,.}\label{eq:QubitHomoToyConti}
\end{equation}
Equation~(\ref{eq:QubitHomoToyConti}) is a non-linear diffusion
equation with a state-dependent diffusion coefficient, $D\left(Z\right)=\kappa\left(1-Z^{2}\right)^{2}$,
which is maximized for $Z\left(t\right)=0$, but approaches zero for
$Z=\pm1$. Mathematically, it is for this reason, that in the limit
$t\rightarrow\infty$, the system tends to one of the pointer states
($Z=\pm1$) of the measurement, where diffusion vanishes, and states
cluster.

\section{Quantum trajectory theory \label{sec:Quantum-trajectory-theory}}

The preceding sections introduced the general background required
to develop quantum trajectory theory. With the aid of specific examples,
important overriding themes were highlighted, which will carry us
forward in this section, and play a key role in understanding photon-counting,
homodyne, and heterodyne measurements.

\subsection{Photodetection \label{subsec:Photodetection}}

Photodetection is the minimal time-continuous measurement scheme —
at each moment in time, the detector records one of only two possible
results: $r=0$ (``no-click'') or $r=1$ (``click''). As discussed
in Sec.~\ref{sec:Introductory-example:-qubit}, the measurement result
communicates some knowledge about the system state and the unavoidable
disturbance caused to it by the measurement itself.\footnote{From an operational point of view, the state of the system is strictly
speaking only \emph{our} knowledge about the probabilities for outcomes
of \emph{future} measurements of the system.} The information gain as well as the action of the disturbance are
encoded in the measurement operators, $\hat{M}_{r}$, recall Eq.~(\ref{eq:MrSpinBath}).
These operators, also known as Kraus operators, generalize unitary
evolution due to a system Hamiltonian, $\hat{H}$, so as to include
the effect of the measurement process. Microscopically, the measurement
operators, $\hat{M}_{r}$, can be understood to describe the unitary
interaction between the system and another, auxiliary one, which is
subsequently measured by a von Neumann measurement apparatus, see
discussion on Neumark's theorem in Sec.~\ref{sec:Introductory-example:-qubit}.
In this section, we will only concern ourselves with the system evolution
subject to measurement, and will make no further reference to the
auxiliary system, other than to specify the system operator $\hat{c}$
that couples the two. The minimal set of infinitesimal measurement
operators, which corresponding to the no-click ($r=0$) and click
($r=1$) evolution, are
\begin{align}
\hat{M}_{0}\left(\dt\right) & =\hat{1}-\left(\half\hat{c}^{\dagger}\hat{c}+i\hat{H}\right)\dt\,,\label{eq:photonM0}\\
\hat{M}_{1}\left(\dt\right) & =\sqrt{dt}\hat{c}\,,\label{eq:photonM1}
\end{align}
in units where $\hbar=1$. One can verify that $\hat{M}_{0}\left(\dt\right)$
and $\hat{M}_{1}\left(\dt\right)$ form a positive-operator-valued
measure (POVM) on the space of results. Hence, they form a resolution
of the identity, $\hat{M}_{0}^{\dagger}\left(\dt\right)\hat{M}_{0}\left(\dt\right)+\hat{M}_{1}^{\dagger}\left(\dt\right)\hat{M}_{1}\left(\dt\right)=\hat{1}+\mathcal{O}\left(\dt^{2}\right)$,
guaranteeing that the law of total probability is satisfied, i.e.,
a measurement yields an outcome with probability 1. The probability
for a specific outcome, $r=0$ or $r=1$, is is given by the generalized
measurement postulate, Eq.~(\ref{eq:GenMsrPosProb}), $\wp_{r}\left(\dt\right)=\left\langle \hat{M}_{r}^{\dagger}\left(\dt\right)\hat{M}_{r}\left(\dt\right)\right\rangle $,
\begin{align}
\wp_{0}\left(\dt\right) & =1-\dt\left\langle \hat{c}^{\dagger}\hat{c}\right\rangle +\mathcal{O}\left(\dt^{2}\right)\,,\label{eq:PhotonP0}\\
\wp_{1}\left(\dt\right) & =\dt\left\langle \hat{c}^{\dagger}\hat{c}\right\rangle \,.\label{eq:PhotonP1}
\end{align}
We define the time-continuous photodetection measurement record to
be the number of photodetections up to time $t$, denoted $N\left(t\right)$.
It follows that the infinitesimal measurement increment, denoted $\dN\left(t\right)$,
is a \emph{point-process}, also known as the \emph{Poisson process},
defined by 
\begin{align}
\dN\left(t\right)^{2} & =\dN\left(t\right)\,,\label{eq:Photon-dN-2}\\
\E{\dN\left(t\right)} & =\wp_{1}\left(\dt\right)\,,\label{eq:Photon-dN-E}
\end{align}
where $\E{\cdot}$ denotes the expectation value in the classical
sense, see Sec.~\ref{subsec:Classical-measurement-theory-basic-concept}.
In the continuum limit, the detector photocurrent, $I\left(t\right)=\dN\left(t\right)/\dt$,
consists of a series of Dirac $\delta$-functions at the times of
the clicks. 

\paragraph{Stochastic master equation (SME) for ideal photodetection. }

Ideal photodetection is the limit where the photodetector collects
the entirety of the system output field and adds no technical noise,
i.e., the quantum measurement efficiency is one, $\eta=1$. According
to the generalized projection postulate, Eq.~(\ref{eq:GenMsrPosProj}),
the state of the system after a measurement at time $t$ conditioned
on the measurement result $r=0$ or $r=1$ is 
\begin{equation}
\rho_{r}\left(t+\dt\right)=\frac{\hat{M}_{r}\left(\dt\right)\rho\left(t\right)\hat{M}_{r}^{\dagger}\left(\dt\right)}{\wp_{r}\left(\dt\right)}\,.\label{eq:click:rhoPLUS}
\end{equation}
In the continuum limit, where the instantaneous measurement outcome
is the photocurrent $I\left(t\right),$ the two possible states, $\rho_{0}\left(t+\dt\right)$
and $\rho_{1}\left(t+\dt\right)$, can be combined in a single stochastic
differential equation (SDE) for the posterior system state $\rho_{I}\left(t+\dt\right)$
conditioned on $I\left(t\right)$, resulting in the state differential,
in Itô form,
\begin{align}
\mathrm{d}\rho_{I}\left(t\right) & =\rho_{I}\left(t+\dt\right)-\rho_{I}\left(t\right)\\
 & =\dN\left(t\right)\left(\rho_{1}\left(t+\dt\right)-\hat{1}\right)+\left(1-\dN\left(t\right)\right)\left(\rho_{0}\left(t+\dt\right)-\hat{1}\right)\,,\label{eq:SME-prior}
\end{align}
which can be simplified by Taylor expanding the denominator of $\rho_{0}\left(t+\dt\right)$,
retaining terms to order $\dt$, and employing the stochastic calculus
rule\footnote{Technically, $\dN\left(t\right)\dt$ is not strictly zero. However,
because the mean of $\dN$ is infinitesimal, $\dN\left(t\right)$
is negligible when compared with $\dt$, and so are all higher-order
products containing both $\dN$ and $\dt$. }
\begin{equation}
\dN\left(t\right)\dt=0\,
\end{equation}
in order to obtain the \emph{stochastic master equation} (SME) for
photodetection, in the Schrödinger picture and in Itô form,
\begin{equation}
\boxed{\mathrm{d}\rho_{I}\left(t\right)=\left(\dN\left(t\right)\mathcal{G}\left[\hat{c}\right]+\dt\mathcal{H}\left[\hat{H}_{\mathrm{eff}}\right]\right)\rho_{I}\left(t\right)\,,}\label{eq:photon:SME}
\end{equation}
where the superoperators $\mathcal{G}\left[\hat{c}\right]\rho$ and
$\mathcal{H}\left[\hat{c}\right]\rho$ are defined by
\begin{align}
\mathcal{G}\left[\hat{c}\right]\rho\equiv & \frac{\hat{c}^{\dagger}\rho\hat{c}}{\Tr{\hat{c}^{\dagger}\rho\hat{c}}}-\rho\,,\label{eq:PhotoG}\\
\mathcal{H}\left[\hat{c}\right]\rho\equiv & \hat{c}\rho+\rho\hat{c}^{\dagger}-\Tr{\hat{c}\rho+\rho\hat{c}^{\dagger}}\rho\label{eq:PhotoH}\\
= & \left(\hat{c}-\left\langle \hat{c}\right\rangle \right)\rho+\rho\left(\hat{c}^{\dagger}-\left\langle \hat{c}^{\dagger}\right\rangle \right)\,,
\end{align}
The superoperator $\mathcal{G}$ results in point-like discontinuous
state evolution, while $\mathcal{H}$ results in smooth, continuous,
but non-unitary evolution generated by the effective non-Hermitian
Hamiltonian 
\begin{equation}
\hat{H}_{\text{eff}}\equiv\hat{H}-i\frac{1}{2}\hat{c}^{\dagger}\hat{c}\,.\label{eq:H_eff_noclick}
\end{equation}
Notably, the trace terms in Eqs.~(\ref{eq:PhotoG}) and (\ref{eq:PhotoH})
make the photodetection SME, Eq.~(\ref{eq:photon:SME}), \emph{nonlinear}
in the state density, $\rho_{I}$. The origin of the trace term in
$\mathcal{H}$, namely $\Tr{\hat{c}\rho+\rho\hat{c}^{\dagger}}$,
is the Taylor expansion of the denominator of Eq.~(\ref{eq:click:rhoPLUS}),
which gives the no-click probability, $\wp_{0}\left(\dt\right)$.
As discussed in Sec.~\ref{sec:Introductory-example:-qubit}, the
role of this term is to preserve the state density trace for all time,
$\Tr{\rho_{I}\left(t\right)}=1$. The solution of the SME is a stochastic
path taken by the conditional state over time, known as a \emph{quantum
trajectory}, a term\emph{ }coined in Ref.\emph{~\citet{Carmichael1993}. }

\paragraph{Stochastic Schrödinger equation (SSE) for photodetection. }

For a system in a pure state, $\rho_{I}\left(t\right)=\kb{\psi_{I}\left(t\right)}{\psi_{I}\left(t\right)}$,
the SME, Eq.~(\ref{eq:photon:SME}), preserves the purity of the
state for all times. It follows that the state evolution is described
by a type of Schrödinger equation, known as the \emph{stochastic Schrödinger
equation }(SSE), 
\begin{equation}
\boxed{\mathrm{d}\ket{\psi_{I}\left(t\right)}=\left[\dt\left(\frac{\left\langle \hat{c}^{\dagger}\hat{c}\right\rangle \left(t\right)}{2}-\frac{\hat{c}^{\dagger}\hat{c}}{2}-i\hat{H}\right)+\dN\left(t\right)\left(\frac{\hat{c}}{\sqrt{\left\langle \hat{c}^{\dagger}\hat{c}\right\rangle \left(t\right)}}-\hat{1}\right)\right]\ket{\psi_{I}\left(t\right)}\,,}\label{eq:SSE-click}
\end{equation}
which is nonlinear in the state $\ket{\psi_{I}\left(t\right)}$. The
non-linear terms contain the expectation value $\left\langle \hat{c}^{\dagger}\hat{c}\right\rangle \left(t\right)=\frac{1}{2}\braOket{\psi_{I}\left(t\right)}{\hat{c}^{\dagger}\hat{c}}{\psi_{I}\left(t\right)}$,
which gives the click probability $\wp_{1}$, see Eq.~(\ref{eq:PhotonP1}).
Because these non-linear terms render analytic treatment of the equations
particularly difficult in general, a linear description of the state
evolution is desired, and can be accomplished as described next. The
term $\frac{1}{2}\dt\left\langle \hat{c}^{\dagger}\hat{c}\right\rangle \left(t\right)\ket{\psi_{I}\left(t\right)}$,
which updates the observer's state-of-knowledge in a non-linear way,
mathematically, ensures the proper normalization of $\ket{\psi_{I}\left(t\right)}$
for all times; however, normalization need not be enforced for each
infinitesimal timestep $\dt$. Instead, it can ``manually'' be enforced
for the no-click periods, $I\left(t\right)=0$, by first first solving
for the un-normalized system state, denoted with a tilde, $\ket{\tilde{\psi}_{I=0}\left(t\right)}$,
then normalizing it, $\ket{\psi_{I=0}\left(t\right)}=\ket{\tilde{\psi}_{I=0}\left(t\right)}/\braket{\tilde{\psi}_{I=0}\left(t\right)}{\tilde{\psi}_{I=0}\left(t\right)}$,
where the effective Schrödinger equation for $\ket{\tilde{\psi}_{I=0}\left(t\right)}$
is
\begin{equation}
\boxed{i\frac{\mathrm{d}}{\dt}\ket{\tilde{\psi}_{I=0}\left(t\right)}=\hat{H}_{\mathrm{eff}}\ket{\tilde{\psi}_{I=0}\left(t\right)}\,,}\label{eq:SSE-click-1}
\end{equation}
where $\hat{H}_{\mathrm{eff}}$ is the non-Hermitian Hamiltonian,
Eq.~(\ref{eq:H_eff_noclick}). Since Eq.~(\ref{eq:SSE-click-1})
is linear, it is generally easier to solve for the time-dynamics.
Calculation of system averages still requires normalizing $\ket{\tilde{\psi}_{I=0}\left(t\right)}$
by its the state norm, $\braket{\tilde{\psi}_{I}\left(t\right)}{\tilde{\psi}_{I}\left(t\right)}$,
which gives the probability of no-clicks occurring for duration $t$.
We remark that Eq.~(\ref{eq:SSE-click-1}) corresponds to tracking
the sub-ensemble of quantum trajectories that contain no clicks, or
mathematically, to the repetitive application of the measurement operator
$\hat{M}_{0}$; hence, in general, in the limit $t\rightarrow\infty$,
it leads to the decay of the norm to zero. 

\paragraph{Unconditioned evolution: master equation for photodetection. }

By averaging over all possible evolutions due to all measurement outcomes
at each instant, one can obtain the \emph{unconditioned} evolution
of the quantum state, denoted $\rho\left(t+\dt\right)$, from Eq.~(\ref{eq:photon:SME}).
Simplifying the average, $\rho\left(t+\dt\right)=\sum_{r}\wp_{r}\rho_{r}\left(t+\dt\right)$,
\begin{align}
\rho\left(t+\dt\right) & =\hat{M}_{0}\left(\dt\right)\rho\hat{M}_{0}^{\dagger}\left(\dt\right)+\hat{M}_{1}\left(\dt\right)\rho\hat{M}_{1}^{\dagger}\left(\dt\right)\\
 & =\rho\left(t\right)-i\left[\hat{H},\rho\left(t\right)\right]\dt+\mathcal{D}\left[\hat{c}\right]\rho\left(t\right)\dt\,,\label{eq:MEforPhootdetection}
\end{align}
where the superoperator $\mathcal{D}$ is defined to be 
\begin{equation}
\mathcal{D}\left[\hat{c}\right]\rho\equiv\hat{c}\rho\hat{c}^{\dagger}-\frac{1}{2}\left\{ \hat{c},\rho\right\} _{+}\,.\label{eq:LindbladSuperop}
\end{equation}
Eq.~(\ref{eq:MEforPhootdetection}) for the unconditioned state evolution
is known as the \emph{master equation, }in Lindblad\emph{ }form \citep{lindblad1976}.\footnote{For a comprehensive summary of the properties of the master equation
and Lindbladians, see Refs.~\citet{AlbertVV2014-Lindblad,Albert2016-Lindblad}.} Unlike the SME, it is linear in the state density, $\rho$, and yields
deterministic state evolution, since there are no stochastic increments,
$\dN$ or $\dW$. Notably, the master equation is very general and
makes no reference to photodetection, other than specifying the system
operator $\hat{c}$ subject to detection, although it does not specify
how. As will be evident from the following section, the same master
equation is obtained for heterodyne detection of $\hat{c}$, see Sec.~\ref{subsec:Homodyne-and-heterodyne}.
The two SMEs corresponding to the same master equation are known as
\emph{unravellings}\footnote{The\emph{ }term 'unraveling' was coined in Ref. \citet{Carmichael1993}.}\emph{
}of it. We note that the unravellings of the master equation are not
unique.

\subparagraph{Imperfect detection.}

Imperfect conditions limit the observer's access to information regarding
the system and generally result in excess noise. The effect of imperfections
can be modeled by considering an ideal photodetector that is, however,
sensitive to only a fraction $\eta$ of the system output field. This
fraction, known as the \emph{quantum measurement efficiency}, is a
real number between zero and one, $0\leq\eta\leq1$. Because of the
loss of information due to imperfect detection, over time, the system
state will, in general, become mixed, with purity less than one, $0\leq\Tr{\rho^{2}}<1$.
To account for the imperfections, SME, Eq.~(\ref{eq:photon:SME}),
is be modified in the following way, see Sec.~4.8.1 of Ref.~\citet{wiseman2010book}:

\begin{equation}
\boxed{\mathrm{d}\rho_{I}\left(t\right)=\left(\dN\left(t\right)\mathcal{G}\left[\sqrt{\eta}\hat{c}\right]+\dt\mathcal{H}\left[-i\hat{H}-\eta\frac{1}{2}\hat{c}^{\dagger}\hat{c}\right]+\dt\left(1-\eta\right)\mathcal{D}\left[\hat{c}\right]\right)\rho_{I}\left(t\right)\,,}\label{eq:photon:SME-loss}
\end{equation}
where the jump probability for each timestep is obtained by also replacing
$\hat{c}$ with $\sqrt{\eta}\hat{c}$ in Eq.~(\ref{eq:Photon-dN-E}),
$\E{\dN\left(t\right)}=\eta\Tr{\hat{c}^{\dagger}\hat{c}\rho}\dt$.
Considering the two limits $\eta\rightarrow0$ and $\eta\rightarrow1$,
one can associate the Lindblad superoperator $\mathcal{D}$ with information
loss, and $\mathcal{H}$ and $\mathcal{G}$ with information gain
due to the measurement.

\subsection{Homodyne and heterodyne detection\label{subsec:Homodyne-and-heterodyne}}

The measurements described so far are not sensitive to the phase of
the system output field, but only its amplitude. In the following,
we describe \emph{dyne} measurements\emph{, homodyne} or \emph{heterodyne,
}which provide information about the phase and a qualitatively different
(diffusive) trajectory unraveling.

\paragraph{Physical implementation. }

\begin{figure}
\begin{centering}
\includegraphics[scale=1.4]{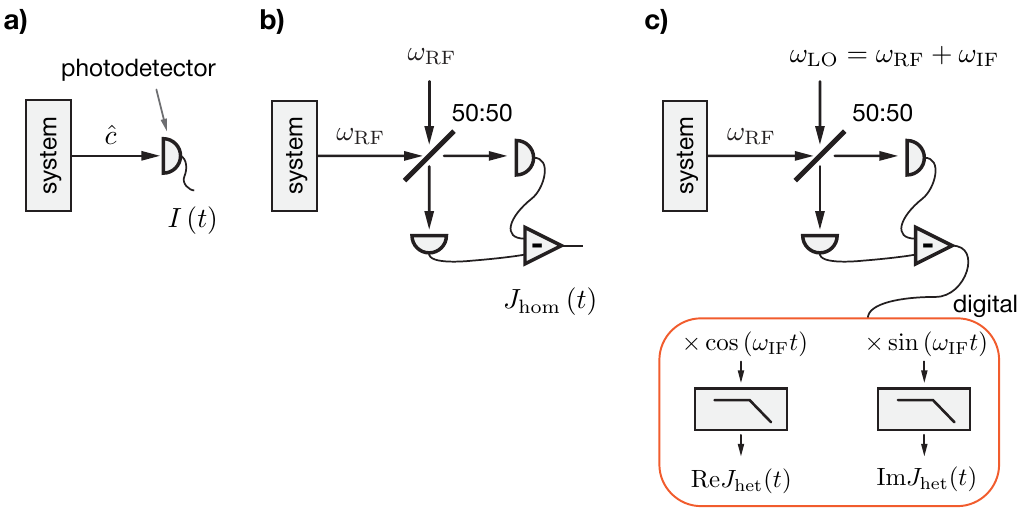}
\par\end{centering}
\caption[Schematic representations of a photo, homodyne, and heterodyne detection
schemes.]{\label{fig:Schematic-representation-of}Schematic representations
of a (a) photo, (b) homodyne, and (c) heterodyne detection schemes.
\textbf{a, }The system output field, proportional to the system coupling
operator $\hat{c}$, is directly monitored with a photodetector, whose
photocurrent $I\left(t\right)$ is the measurement record. \textbf{b,}
Optical balanced homodyne detection: system output field, assumed
with carrier frequency $\omega_{\mathrm{RF}}$, is interfered on a
50:50 beam splitter with a strong local oscillator (LO) tone at the
carrier frequency, $\omega_{\mathrm{LO}}=\omega_{\mathrm{RF}}$. The
measurement record, $J_{\mathrm{hom}}\left(t\right)$, is obtained
from the  difference of the photodetector currents on each output
arm of the beamsplitter. \textbf{c,} Balanced heterodyne detection
scheme (with digital demodulation): LO frequency is detuned by an
intermediate frequency value, $\omega_{\mathrm{IF}},$ where $\omega_{\mathrm{IF}}\ll\omega_{\mathrm{RF}},\omega_{\mathrm{LO}}$.
The difference of the photodetector currents of each arm, which oscillates
at $\omega_{\mathrm{IF}}$, is digitally demodulated to obtain the
in-phase {[}out-of-phase{]} quadrature $\mathrm{Re}J_{\mathrm{het}}\left(t\right)$
{[}$\mathrm{Im}J_{\mathrm{het}}\left(t\right)${]} by digitally mixing
the signal with a reference one, $\cos\left(\omega_{\mathrm{IF}}t\right)$
{[}$\sin\left(\omega_{\mathrm{IF}}t\right)${]}, and low-pass filtering
the output to reject tones above $\omega_{\mathrm{IF}}$. Digital
panel schematic inspired by Ref.~\citep{Campagne2016-Fluorescence}.
}
\end{figure}

Dyne detection is realized by mixing the system output signal with
a local-oscillator (LO) tone, see Fig.~\ref{fig:Schematic-representation-of}.
For a system carrier frequency, conventionally termed the radio frequency
(RF), $\omega_{\mathrm{RF}}$ and LO frequency $\omega_{\mathrm{LO}}$,
the lower sideband of the mixed-signal is at intermediate frequency
(IF), $\omega_{\mathrm{IF}}=\omega_{\mathrm{LO}}-\omega_{\mathrm{RF}}$.
In homodyne detection, the LO is tuned in resonance with the system
carrier, resulting in a direct-current (DC) IF signal, $\omega_{\mathrm{IF}}=0$,
proportional to a quadrate of the RF signal that depends on the LO
phase. The IF signal is typically sampled and digitally processed.
On the other hand, in heterodyne detection, the LO frequency is significantly
detuned from the RF, $\omega_{\mathrm{IF}}\gg0$, resulting in a an
oscillatory IF signal, which is demodulated to extract the information-bearing
\emph{in-quadrature} (I) and \emph{out-of-quadrature }(Q) components.
Note that the heterodyne measurement record consists of a series of
not one but \emph{ two} values, I and Q. Heterodyne detection is equivalent
to two concurrent homodyne detections with LO phases $90^{\circ}$
apart. 

\paragraph{Homodyne measurement record. }

The homodyne measurement signal is mathematically described by a function,
$J_{\mathrm{hom}}\left(t\right)$, that is real and continuous everywhere
but differentiable nowhere, see Sec.~\ref{subsec:Continuous-limit:-Wiener}.
The measurement signal gradually reveals information about a system
operator of the form $\hat{c}+\hat{c}^{\dagger}$, where $\hat{c}$
is the operator coupled to the measurement apparatus, which for the
example of Sec.~\ref{subsec:Continuous-limit:-Wiener} is $\hat{c}=-\frac{\sqrt{\kappa}}{2}\hat{\sigma}_{z}$.
In Itô form, the measurement increment is
\begin{equation}
\mathrm{d}J_{\mathrm{hom}}\left(t\right)=\left\langle \hat{c}+\hat{c}^{\dagger}\right\rangle \left(t\right)\dt+\dW\left(t\right)\,,\label{eq:Jhom}
\end{equation}
where $\dW\left(t\right)$ is the stochastic Wiener increment satisfying
the canonical relations given in Eqs.~(\ref{eq:dW-relation-1})
and~(\ref{eq:dW-relation-2}).

\paragraph{Heterodyne measurement record. }

The heterodyne measurement signal consists of two functions: the in-phase,
$J_{I}\left(t\right)$, and out-of-phase, $J_{Q}\left(t\right)$,
quadrature functions, which can be combined in a single complex function,
$J_{\mathrm{het}}\left(t\right)\equiv\frac{1}{2}\left(J_{I}\left(t\right)+iJ_{Q}\left(t\right)\right)$,
continuous everywhere but differentiable nowhere.  In heterodyne
detection, $J_{\mathrm{het}}\left(t\right)$ gradually reveals information
about a system operator $\hat{c}$, which need not be Hermitian but
which can be decomposed into the sum of two Hermitian operators, corresponding
to two observables, know as the quadrature operators, 
\begin{equation}
\hat{I}\equiv\hat{c}+\hat{c}^{\dagger}\ \text{and}\ \hat{Q}\equiv-i\left(\hat{c}-\hat{c}^{\dagger}\right)\,,
\end{equation}
so that $\hat{c}=\frac{1}{2}\left(\hat{I}+i\hat{Q}\right)$. The Itô
form of the measurement increment is 
\begin{equation}
\mathrm{d}J_{\mathrm{het}}\left(t\right)=\left\langle \hat{c}\right\rangle \left(t\right)\dt+\dZ\left(t\right)\,,\label{eq:Jhet-perfect}
\end{equation}
where $\dZ\equiv\frac{1}{\sqrt{2}}\left(\dW_{I}\left(t\right)+i\dW_{Q}\left(t\right)\right)$
is the complex Wiener increment, the sum of two independent Wiener
increments, $\dW_{I}\left(t\right)$ and $\dW_{Q}\left(t\right)$,
that satisfy $\E{\dW_{I}\left(t\right)\dW_{Q}\left(t\right)}=0$,
so that $\dZ\left(t\right)^{*}\dZ\left(t\right)=\dt$ and $\dZ\left(t\right)^{2}=0$.
We note that $\dZ$ is obtained by making the substitution $e^{i\omega_{{\rm RF}}t}\dW\rightarrow\dZ$
in the heterodyne derivation.

For concreteness, consider the example of a qubit coupled to the environment
where an observer performs heterodyne detection of the qubit fluoresce
\citep{Campagne-Ibarcq2014,Campagne2016-Fluorescence,Naghiloo2016}.
The system-environment coupling is given by the non-Hermitian operator
$\hat{c}=\hat{\sigma}_{-}\equiv\kb{+z}{-z}$, which is decomposed
into the two Hermitian quadrature operators $\hat{I}=\hat{\sigma}_{x}$
and $\hat{Q}=-\hat{\sigma}_{y}$. Note the minus sign in $\hat{Q}$.
The heterodyne detection of $\hat{\sigma}_{-}$ can be understood
as a homodyne detection of the observable $\hat{I}$ and a concurrent
homodyne detection of the observable $\hat{Q}$, each with efficiency
$\eta=1/2$, see below. Consider the example where the qubit is replaced
by a cavity, the coupling operator is $\hat{c}=\hat{a}$, where $\hat{a}$
is the annihilation operator, and the whose cavity output field is
subject to heterodyne monitoring, which reveals information about
$\hat{I}=\hat{a}+\hat{a}^{\dagger}$ and $\hat{Q}=-i\left(\hat{a}-\hat{a}^{\dagger}\right)$.
For a coherent state in the cavity, $\ket{\alpha\left(t\right)}$,
the measurement record gradually reveals its complex amplitude, $\E{\mathrm{d}J_{\mathrm{het}}\left(t\right)/\dt}=\braOket{\alpha\left(t\right)}{\frac{1}{2}\left(\hat{I}+\hat{i}\hat{Q}\right)}{\alpha\left(t\right)}=\alpha\left(t\right)$. 

\paragraph{Measurement operators and the SME for perfect dyne detection.}

At an instant in time, the noisy heterodyne record, $J_{\mathrm{het}}\left(t\right)$,
relates the measurement outcome to the quantum trajectory evolution
according to the action of the measurement operator (see discussion
on Pg.~\pageref{eq:MrSpinBath})
\begin{equation}
\hat{M}_{J}=\hat{1}-i\hat{H}\dt-\frac{1}{2}\hat{c}^{\dagger}\hat{c}\dt+J_{\mathrm{het}}^{*}\left(t\right)\hat{c}\dt\,.\label{eq:hetero:msrmap}
\end{equation}
The measurement operator for homodyne detection, also denoted $\hat{M}_{J}$,
is obtained by making the substitution $J_{\mathrm{het}}^{*}\left(t\right)\rightarrow J_{\mathrm{hom}}\left(t\right)$
in Eq.~(\ref{eq:hetero:msrmap}). Notably, the non-orthogonal set
of measurement operators for dyne detection, $\left\{ \hat{M}_{J}:J\right\} $,
is continuous, in contrast with that of photodetection, which consists
of two elements, $\left\{ \hat{M}_{0},\hat{M}_{1}\right\} ,$ since
there are only two possible measurement outcomes, click or no click.
The system state conditioned on the record at time $t$, denoted $\rho_{J}$,
is obtained by employing the generalized measurement postulate, Eq.~(\ref{eq:GenMsrPosProj}),

\begin{equation}
\rho_{J}\left(t+\dt\right)=\frac{\hat{M}_{J}\rho_{J}\left(t\right)\hat{M}_{J}^{\dagger}}{\Tr{\hat{M}_{J}\rho_{J}\left(t\right)\hat{M}_{J}^{\dagger}}}\,.\label{eq:click:rhoPLUS-1}
\end{equation}
Equation~(\ref{eq:click:rhoPLUS-1}) is simplified by Taylor expanding
the denominator to order $\dt$ and writing the infinitesimal state
change, in Itô form, $\mathrm{d}\rho_{J}\left(t\right)=\rho_{J}\left(t+\dt\right)-\rho_{J}\left(t\right)$,
thus obtaining the SME for perfect heterodyne detection, in the Schrödinger
picture,
\begin{equation}
\mathrm{d}\rho_{J}\left(t\right)=\left[-i\dt[\hat{H},\cdot]+\dt\mathcal{D}[\hat{c}]+\dZ^{*}\left(t\right)\mathcal{H}[\hat{c}]\right]\rho_{J}\left(t\right)\,,\label{eq:SME-Heterodyne}
\end{equation}
where the superoperators $\mathcal{D}$ and $\mathcal{H}$ are defined
in Eqs.~(\ref{eq:LindbladSuperop}) and~(\ref{eq:PhotoH}), respectively.
Equation~(\ref{eq:SME-Heterodyne}) has to be solved jointly with
Eq.~(\ref{eq:Jhet-perfect}). The homodyne SME is obtained by making
the substitution $\dZ^{*}\left(t\right)\rightarrow\dW\left(t\right)$
in Eq.~(\ref{eq:SME-Heterodyne}).  

\paragraph{SME for imperfect measurements.}

Measurement imperfections (see discussion on Pg.~\ref{eq:photon:SME-loss})
are primarily due to: i) losses associated with the propagation of
the system output field to the measurement apparatus, characterized
by a quantum efficiency $\eta_{\mathrm{prop}}$, and ii) finite detector
efficiency, $\eta_{\mathrm{det}}$. The measurement chain efficiency
is given by the product of those of it sub-components, $\eta=\eta_{\mathrm{prop}}\eta_{\mathrm{det}}$,
and is used to modify Eq.~(\ref{eq:Jhet-perfect}) to account for
imperfections by making the substitution $\hat{c}\rightarrow\sqrt{\eta}\hat{c}$,
\begin{equation}
\boxed{\mathrm{d}J_{\mathrm{het}}\left(t\right)=\sqrt{\eta}\left\langle \hat{c}\right\rangle \left(t\right)\dt+\dZ\left(t\right)\,.}\label{eq:dJimperf:het}
\end{equation}
Similarly, the homodyne measurement increment is $\mathrm{d}J_{\mathrm{hom}}\left(t\right)=\sqrt{\eta}\left\langle \hat{c}+\hat{c}^{\dagger}\right\rangle \left(t\right)\dt+\dW\left(t\right)$.
In Eq.~(\ref{eq:dJimperf:het}), the effect of an efficiency less
than one, $\eta<1$, is to reduce the measurement signal amplitude,
$\left\langle \hat{c}\right\rangle $, relative to the noise, $\dZ$,
resulting in a degraded signal-to-noise (SNR) ratio. In the extreme
limit $\eta\rightarrow0$, the measurement is entirely noise, and
the SNR is zero. Only in the limit $\eta\rightarrow1$, as discussed
in Sec.~\ref{sec:Introductory-example:-qubit}, can the noise be
interpreted as entirely due to quantum vacuum fluctuations.  The
trajectory evolution associated with the noisy signal, $J_{\mathrm{het}}\left(t\right)$,
is obtained by making the substitution $\hat{c}\rightarrow\sqrt{\eta}\hat{c}$
in the innovator, $\mathcal{H}$, term of the heterodyne SME, Eq.~(\ref{eq:SME-Heterodyne}),
which is responsible for the information gain due to the measurement,
thus obtaining the SME for finite-efficiency heterodyne detection,
\begin{equation}
\boxed{\mathrm{d}\rho_{J}\left(t\right)=\left[-i\dt[\hat{H},\cdot]+\dt\mathcal{D}[\hat{c}]+\dZ^{*}\left(t\right)\mathcal{H}[\sqrt{\eta}\hat{c}]\right]\rho_{J}\left(t\right)\,,}\label{eq:SME-heterodyne-imperfect}
\end{equation}
which upon the the substitution $\dZ^{*}\left(t\right)\rightarrow\dW\left(t\right)$
becomes the SME for finite-efficiency homodyne detection. Equations~(\ref{eq:SME-heterodyne-imperfect})
and~(\ref{eq:dJimperf:het}) have to be solved simultaneously.

\subparagraph{Qualitative comparison of dyne- vs. photo- detection trajectories. }

In Sec.~\ref{subsec:Photodetection}, we considered the stochastic
evolution of the conditional quantum state of a system subject to
photodetection, a measurement scheme that results in one of two possible
outcomes, $r=0$ and $r=1$, at each moment in time. The state evolution
was marked by two qualitatively distinct possibilities: i) smooth,
continuous, deterministic-like evolution due to the non-Hermitian
Hamiltonian $\hat{H}_{\mathrm{eff}}$, associated with $r=0$, or
ii) discontinuous, point-like, jumpy evolution due to the action of
the superoperator term $\mathcal{G}$, associated with the occasional
outcome $r=1$. Both the measurement record and the state evolution
of dyne detection are, in a sense, antithetic to those of dyne detection,
which is characterized by a (Gaussian-distributed) infinite continuum
of possible measurement outcomes and neither smooth nor jumpy state
evolution. Rather, the evolution is \emph{diffusive} \citep{Gisin1992},
a consequence of the Gaussian-distributed measurement outcomes. While
in photodetection, a click could result in a substantial amount of
information about the system being acquired at an instant in time,
such an event is not possible with dyne monitoring, where the noisy
signal, $J\left(t\right)$, only \emph{gradually} reveals information
about the state of the system. It is only in this gradual sense that
dyne measurements collapse the system state to an eigenstate of the
measurement operator, see discussion of Sec.~\ref{subsec:Continuous-limit:-Wiener}.

\section{Further reading \label{sec:Further-reading-traj-thry}}

For further reading, we suggest the following books, and, where applicable,
note sections closely related to some of the topics discussed in more
depth in this chapter:
\begin{itemize}
\item \citet{Carmichael1993} \& \citet{Carmichael2008-Book2} —  The formulation
of quantum trajectory theory is presented; key terms, such as 'unraveling',
are introduced. Section 17.2 of Ref.~\citet{Carmichael2008-Book2}
treats the example of a quantum bit subject to continuous photodetection,
comparing the state evolution in quantum measurement theory with that
of classical measurement theory. 
\item \citet{Gardiner2004} — Chapter~3 provides a useful derivation of
input-output theory by treating the one-dimensional transmission line.
Focus on the Heisenberg formulation of quantum measurements. 
\item \citet{wiseman2010book} — Classical measurement theory is introduced
in Chapter~2. 
\item \citet{Jacobs2010Book} — Introduction to \emph{classical} stochastic
differential equations (SDE).  
\item \citet{Girvin2014} — Chapter 3 discusses  quantum measurements
in the context of circuit quantum electrodynamics (cQED), which is
introduced in the remainder of the notes.
\item \citet{SteckQuantumNotes} — Lecture notes on quantum trajectories,
SDE numerical methods, and a number of related topics in quantum optics. 
\end{itemize}
We conclude this chapter with an amusing quote by H. Mabuchi:
\begin{center}
``\emph{The quantum measurement problem refers to a set of people}.''\\{[}the
set who have a problem with the theory of quantum measurements{]}
\citep{Fuchs2003-qminfo}.
\par\end{center}

\chapter{Theoretical description of quantum jumps\label{chap:theoretical-description-jumps} }

\addcontentsline{lof}{chapter}{Theoretical description of quantum jumps\lofpost}  

\singlespacing 
\setlength{\epigraphwidth}{.6\textwidth}
\epigraph{
Photons, the quanta of light, are countable and discrete, and one assumes they come and go in jumps. Einstein proposed it so --- though only as a pragmatic step  ... Yet the Schrödinger equation is deterministic and nothing within its jurisdiction jumps. What then to make of this unlikely marriage where the continuous is to somehow cavort with the discrete.}
{H.J. Carmichael\\New Zealand Science Review\\ Vol. 72 (2) 2015} 
\setlength{\epigraphwidth}{.4\textwidth} 
\doublespacing\noindent\lettrine{T}{his} chapter presents the quantum trajectory
description of the Dehmelt electron-shelving scheme and the catch-and-reverse
circuit quantum electrodynamics (cQED) experiment. Section~\ref{sec:Fluorescence-monitored-by}
discusses quantum jumps in the three-level atom subject to fluorescence
photodetection. The minimal idealized model with coherent Rabi drives
is considered in Section \ref{sec:thry:3lvl-atom-simple-log}. To
better conceptualize important aspects of the measurement dynamics,
Sec.~\ref{subsec:Knowledge-driven-force-in} considers the simpler
case of a three-level atom subject only to measurement and no competing
coherent dynamics; i.e., the Dark Rabi drive is zero, $\Odg=0$. The
character of the unavoidable state-disturbance due to the back-action
of the measurement is examined in depth, and the notion of \emph{measurement-backaction
effective force} and its geometrical representation are introduced.
Section~\ref{subsec:Incoherent-Bright-drive} continues the description
of quantum jumps and considers the flight of the jump in the presence
of an incoherent Bright drive and the conditional interruption of
$\Odg$ ($\Delta t_{\mathrm{off}}$). Section~\ref{sec:Heterodyne-monitoring-of}
presents the trajectory description of the cQED experiment including
all known imperfections. Section~\ref{subsec:Simulation-of-linear}
discusses the Monte Carlo simulation of the linear Stochastic Schrödinger
equation (SSE), employed in the comparison between theoretical predictions
and experimental results, see Sec.~\ref{subsec:Comparison-between-theory}. 

\section{Fluorescence monitored by photon counts \label{sec:Fluorescence-monitored-by}}

\subsection{Dehmelt electron-shelving scheme and quantum jumps \label{sec:thry:3lvl-atom-simple-log}}

As discussed in Chapter~\ref{chap:Introduction-and-overview}, \textit{\emph{t}}he
experiments with trapped ions \citep{Nagourney1986,Sauter1986,Bergquist1986}
monitor intermittent fluorescence from the bright state $\B$ to track
jumps between $\G$ and $\D$ \citep{Cook1985}. In the simplest three-level
scheme \citep{Bergquist1986} and using coherent radiation to excite
both the BG and DG transitions, the master equation, Eq.~(\ref{eq:MEforPhootdetection}),
for the reduced density operator $\rho$ of the three-level system,
written in the interaction picture, is
\begin{equation}
\frac{\mathrm{d}\rho}{\dt}\left(t\right)=(i\hbar)^{-1}[\hat{H}_{{\rm drive}},\rho\left(t\right)]+\gamma_{{\rm B}}{\cal D}\left[\kb{\mathrm{G}}{\mathrm{B}}\right]\rho\left(t\right)+\gamma_{{\rm D}}{\cal D}\left[\kb{\mathrm{G}}{\mathrm{G}}\right]\rho\left(t\right),\label{eq:PhotonLindblad}
\end{equation}
where ${\cal D}[\hat{c}]\cdot=\hat{c}\cdot\hat{c}^{\dagger}-\frac{1}{2}\{\hat{c}^{\dagger}\hat{c},\cdot\}$
denotes the Lindblad superoperator, defined in Eq.~(\ref{eq:LindbladSuperop}),
$\gamma_{{\rm B}}$ and $\gamma_{{\rm D}}$ are radiative decay rates
of the B and D level, respectively, and the drive Hamiltonian is
\begin{equation}
\hat{H}_{{\rm drive}}=i\hbar\frac{\Omega_{{\rm BG}}}{2}\big(\kb{\mathrm{B}}{\mathrm{G}}-\kb{\mathrm{G}}{\mathrm{B}}\big)+i\hbar\frac{\Omega_{{\rm DG}}}{2}\big(\kb{\mathrm{D}}{\mathrm{G}}-\kb{\mathrm{G}}{\mathrm{D}}\big),\label{eq:drive_coherent_fluorescence}
\end{equation}
with $\Omega_{{\rm BG}}$ and $\Omega_{{\rm DG}}$ the Rabi drives.

\paragraph{Quantum trajectory description.}

The quantum trajectory description \citep{Carmichael1993,Dalibard1992-original-traj,Dum1992}
unravels $\rho$ into an ensemble of pure states, see Sec.~\ref{subsec:Photodetection},
whose ket vectors evolve along stochastic paths conditioned on the
clicks of \emph{imaginary} photon detectors that monitor fluorescence
from $\B$ and, much less frequently, from $\D$. Working in the limit
of the omniscient observer,  corresponding to unit quantum measurement
efficiency, $\eta=1$, for both the B and D click records, denoted
$\dN_{\mathrm{B}}\left(t\right)$ and $\dN_{\mathrm{D}}\left(t\right)$,
respectively, see Eqs.~(\ref{eq:Photon-dN-2}) and~(\ref{eq:Photon-dN-E}),
and corresponding to the quantum jump operators $\hat{c}_{\mathrm{B}}=\sqrt{\gamma_{{\rm B}}}\kb{\mathrm{G}}{\mathrm{B}}$
and $\hat{c}_{\mathrm{D}}=\sqrt{\gamma_{{\rm D}}}\kb{\mathrm{G}}{\mathrm{D}}$,
respectively, the non-linear stochastic Schrödinger equation (SSE)
in Itô form, Eq.~(\ref{eq:SSE-click}), is
\begin{align}
\mathrm{d}\ket{\psi_{I}\left(t\right)}= & \dt\left(-i\hat{H}+\frac{\gamma_{\mathrm{B}}}{2}\left(\left\langle \kb{\mathrm{B}}{\mathrm{B}}\right\rangle \left(t\right)-\kb{\mathrm{B}}{\mathrm{B}}\right)+\frac{\gamma_{\mathrm{D}}}{2}\left(\left\langle \kb{\mathrm{D}}{\mathrm{D}}\right\rangle \left(t\right)-\kb{\mathrm{D}}{\mathrm{D}}\right)\right)\ket{\psi_{I}\left(t\right)}\nonumber \\
 & +\dN_{\mathrm{B}}\left(t\right)\left(\frac{\kb{\mathrm{G}}{\mathrm{B}}}{\sqrt{\left\langle \kb{\mathrm{B}}{\mathrm{B}}\right\rangle \left(t\right)}}-\hat{1}\right)+\dN_{\mathrm{D}}\left(t\right)\left(\frac{\kb{\mathrm{G}}{\mathrm{D}}}{\sqrt{\left\langle \kb{\mathrm{D}}{\mathrm{D}}\right\rangle \left(t\right)}}-\hat{1}\right)\,.\label{eq:PhotonPsiI}
\end{align}
The terms proportional to $\dN_{\mathrm{B}}\left(t\right)$ and $\dN_{\mathrm{D}}\left(t\right)$
reset the ket vector to $\G$ with instantaneous probability $\gamma_{\mathrm{B}}\left\langle \kb{\mathrm{B}}{\mathrm{B}}\right\rangle \left(t\right)\dt$
and $\gamma_{\mathrm{D}}\left\langle \kb{\mathrm{D}}{\mathrm{D}}\right\rangle \left(t\right)\dt$,
respectively, and correspond to the state-disturbance due to the detection
of a click on the B and D detectors, respectively. Otherwise, when
no click is observed on either detector, the state follows a deterministic
evolution as a coherent superposition, governed by the terms proportional
to $\dt$. 

\paragraph{Linear SSE and no-click evolution. }

To describe the conditional no-click evolution, as discussed on Pg.~\pageref{eq:SSE-click-1},
it is analytically favorable to work with the linear form of the SSE,
obtained by suppressing the expectation value terms in Eq.~(\ref{eq:PhotonPsiI}),
and defining the \emph{un-normalized} quantum state, 
\begin{equation}
\ket{\tilde{\psi}\left(\Delta t_{{\rm catch}}\right)}=\cg\left(\Delta t_{{\rm catch}}\right)\G+\cb\left(\Delta t_{{\rm catch}}\right)\B+\cd\left(\Delta t_{{\rm catch}}\right)\D\,,\label{eq:GBD-ket}
\end{equation}
where $\Delta t_{{\rm catch}}$ denotes the no-click duration. Immediately
after a click, marked by $\deltatcatch=0$, the state $\ket{\tilde{\psi}\left(\Delta t_{{\rm catch}}\right)}$
is reset with the coefficients $\cg\left(\Delta t_{{\rm catch}}\right)=1$
and $\cb\left(\Delta t_{{\rm catch}}\right)=\cd\left(\Delta t_{{\rm catch}}\right)=0.$
Conditioned on no clicks, the evolution of the un-normalized state
is governed by the effective non-Hermitian Hamiltonian, see Eq.~(\ref{eq:H_eff_noclick}),
\begin{equation}
\hat{H}_{\mathrm{eff}}=\hat{H}_{{\rm drive}}-i\hbar\frac{\gamma_{{\rm B}}}{2}\kb{\mathrm{B}}{\mathrm{B}}-i\hbar\frac{\gamma_{{\rm D}}}{2}\kb{\mathrm{D}}{\mathrm{D}}\,,\label{eq:ClickHeff}
\end{equation}
and the Schrödinger-type equation 
\begin{equation}
i\hbar\frac{\mathrm{d}\ket{\tilde{\psi}\left(\Delta t_{{\rm catch}}\right)}}{\mathrm{d}\Delta t_{{\rm catch}}}=\hat{H}_{\mathrm{eff}}\ket{\tilde{\psi}\left(\Delta t_{{\rm catch}}\right)}\,.\label{eq:NonHermitianShcr}
\end{equation}
Due to the purely imaginary-valued terms in $\hat{H}_{\mathrm{eff}}$,
the norm of the state $\ket{\tilde{\psi}\left(\Delta t_{{\rm catch}}\right)}$
decays as a function of $\Delta t_{{\rm catch}}$ and gives the probability
that the no-click evolution has continued without click interruptions
for duration $\Delta t_{{\rm catch}}$. In  the limit $\deltatcatch\rightarrow0$,
the norm of the ket approaches zero. The evolution of the state can
be described by a matrix equation for the state coefficients, using
Eqs.~(\ref{eq:GBD-ket}), (\ref{eq:ClickHeff}), and~(\ref{eq:NonHermitianShcr}),
\begin{equation}
\frac{\mathrm{d}}{\mathrm{d}\Delta t_{{\rm catch}}}\begin{pmatrix}\cg\\
\cb\\
\cd
\end{pmatrix}=\frac{1}{2}\left(\begin{matrix}0 & -\Omega_{{\rm BG}} & -\Omega_{{\rm DG}}\\
\Omega_{{\rm BG}} & -\gamma_{{\rm B}} & 0\\
\Omega_{{\rm DG}} & 0 & -\gamma_{{\rm D}}
\end{matrix}\right)\begin{pmatrix}\cg\\
\cb\\
\cd
\end{pmatrix}.\label{eqn:3x3_system}
\end{equation}
In general this $3\times3$ system does not have a closed solution
in simple form, although there is a particularly simple solution under
conditions that produce intermittent fluorescence, i.e., rare jumps
from $\G$ to $\D$ {[}``shelving'' in the dark state \citep{Nagourney1986}{]}
interspersed as intervals of fluorescence ``off'' in a background
of fluorescence ``on''. The conditions follow naturally if $\D$
is a metastable state \citep{Nagourney1986,Sauter1986,Bergquist1986}
whose lifetime $\gamma_{{\rm D}}^{-1}$ is extremely long on the scale
of the mean time, 
\begin{equation}
\tau_{\mathrm{BG}}=\frac{\Omega_{{\rm BG}}^{2}}{\gamma_{\mathrm{B}}}\,,\label{eq:GammaBG}
\end{equation}
between photon detector clicks for a weak $\Omega_{{\rm BG}}$ Rabi
drive. Subject to $(\Omega_{{\rm DG}},\gamma_{{\rm D}})\ll\Omega_{{\rm BG}}^{2}/\gamma_{{\rm B}}\ll\gamma_{{\rm B}}$,
one way to solve Eq.~(\ref{eqn:3x3_system}) is to first adiabatically
eliminate the fast time dynamics of the B level, by setting the time
derivative of the B coefficient to zero, 
\begin{eqnarray}
\frac{\mathrm{d}\cb}{\mathrm{d}\Delta t_{{\rm catch}}} & = & 0\,,\label{eq:Cbadiabatic}
\end{eqnarray}
Solving Eq.~(\ref{eq:Cbadiabatic}), $\cb=\frac{\Omega_{\mathrm{BG}}}{\gamma_{\mathrm{B}}}\cg$,
allows one to eliminate the B level from the description of the dynamics
and to extract the effective GD dynamics, to obtain the un-normalized
state conditioned on the detection of no clicks,
\begin{align}
\ket{\tilde{\psi}\left(\deltatcatch\right)}= & \exp\left(-\frac{\Omega_{{\rm BG}}^{2}}{2\gamma_{\mathrm{B}}}\deltatcatch\right)\left(\G+\frac{\Omega_{\mathrm{BG}}}{\gamma_{\mathrm{B}}}\B\right)\nonumber \\
 & +\left[\exp\left(-\frac{\gamma_{\mathrm{D}}}{2}\deltatcatch\right)-\exp\left(-\frac{\Omega_{{\rm BG}}^{2}}{2\gamma_{\mathrm{B}}}\deltatcatch\right)\right]\frac{\gamma_{\mathrm{B}}\Omega_{\mathrm{D}}}{\Omega_{{\rm BG}}^{2}}\D\,.\label{eq:psiPhotonNoClick}
\end{align}
Note that $\ket{\tilde{\psi}\left(\deltatcatch\right)}$ has purely
real coefficients, since $\hat{H}_{\mathrm{eff}}$ has purely imaginary
ones. The Bloch vector components of the normalized GD manifold evolution
conditioned on no-clicks are obtained by normalizing the state, Eq.~(\ref{eq:psiPhotonNoClick}),
\begin{eqnarray}
{\rm Z}_{{\rm GD}}(\Delta t_{{\rm catch}}) & = & \frac{W_{{\rm DG}}(\Delta t_{{\rm catch}})-W_{{\rm DG}}^{-1}(\Delta t_{{\rm catch}})}{W_{{\rm DG}}(\Delta t_{{\rm catch}})+W_{{\rm DG}}^{-1}(\Delta t_{{\rm catch}})},\label{eqn:Z}\\
{\rm X}_{{\rm GD}}(\Delta t_{{\rm catch}}) & = & \frac{2}{W_{{\rm DG}}(\Delta t_{{\rm catch}})+W_{{\rm DG}}^{-1}(\Delta t_{{\rm catch}})},\label{eqn:X}\\
{\rm Y}_{{\rm GD}}(\Delta t_{{\rm catch}}) & = & 0,\label{eqn:Y}
\end{eqnarray}
where we have defined the ratio \citep{Porrati1987}
\begin{equation}
W_{{\rm DG}}(\Delta t_{{\rm catch}})\equiv\frac{\cd(\Delta t_{{\rm catch}})}{\cg(\Delta t_{{\rm catch}})}.\label{eqn:W_DG_definition}
\end{equation}
Notably, as an alternative to the adiabatic method employed to solve
Eq.~(\ref{eqn:3x3_system}), one can instead directly write down
the equation of motion for $W_{{\rm DG}}$ within the same approximations,
\begin{equation}
\frac{\mathrm{d}W_{{\rm DG}}}{\mathrm{d}\Delta t_{{\rm catch}}}=\frac{\Omega_{{\rm BG}}^{2}}{2\gamma_{{\rm B}}}W_{{\rm DG}}+\frac{\Omega_{{\rm DG}}}{2},\label{eqn:W_DG_equation_coherent}
\end{equation}
which, with the initial condition $W_{{\rm DG}}(0)=0$, has the solution
\begin{equation}
W_{{\rm DG}}(\Delta t_{{\rm catch}})=\frac{\Omega_{{\rm DG}}}{\Omega_{{\rm BG}}^{2}/\gamma_{{\rm B}}}\left[\exp\left(\frac{\Omega_{{\rm BG}}^{2}}{2\gamma_{{\rm B}}}\Delta t_{{\rm catch}}\right)-1\right],\label{eq:WDG-coh}
\end{equation}
and also yields the Bloch components, Eqs.~(\ref{eqn:Z}), (\ref{eqn:X}),
and~(\ref{eqn:Y}). The timescale of the transition, the mid-flight
time of the quantum jump, $\Delta t_{{\rm mid}}$, is found by setting
the G and D coefficients equal to each other, $\cg\left(\Delta t_{{\rm mid}}\right)=\cd\left(\Delta t_{{\rm mid}}\right),$
\begin{equation}
\boxed{\Delta t_{{\rm mid}}=\left(\frac{\Omega_{{\rm BG}}^{2}}{2\gamma_{{\rm B}}}\right)^{-1}\ln\left(\frac{\Omega_{{\rm BG}}^{2}/\gamma_{{\rm B}}}{\Omega_{{\rm DG}}}+1\right).}\label{eq:t_mid_coherent}
\end{equation}
For strong monitoring, $\Omega_{{\rm BG}}^{2}\gamma_{{\rm B}}^{-1}\gg\Omega_{{\rm DG}}$,
the +1 in Eq.~(\ref{eq:t_mid_coherent}) can be dropped. Working
in this limit, Eqs.~(\ref{eqn:Z})–(\ref{eqn:Y}) provide simple
formulas for the continuous, deterministic, and coherent evolution
of the completed quantum jump:
\begin{eqnarray}
{\rm Z}_{{\rm GD}}(\Delta t_{{\rm catch}}) & = & {\rm tanh}\left[\frac{\Omega_{{\rm BG}}^{2}}{2\gamma_{B}}(\Delta t_{{\rm catch}}-\Delta t_{{\rm mid}})\right],\label{eqn:Z_approx}\\
{\rm X}_{{\rm GD}}(\Delta t_{{\rm catch}}) & = & {\rm sech}\left[\frac{\Omega_{{\rm BG}}^{2}}{2\gamma_{B}}(\Delta t_{{\rm catch}}-\Delta t_{{\rm mid}})\right],\label{eqn:X_approx}\\
{\rm Y}_{{\rm GD}}(\Delta t_{{\rm catch}}) & = & 0.\label{eqn:Y_approx}
\end{eqnarray}
These formulas, derived in the strong-monitoring limit, execute a
perfect jump, ${\rm Z}_{{\rm GD}}(\infty)=1$, ${\rm X}_{{\rm GD}}(\infty)={\rm Y}_{{\rm GD}}(\infty)=0$.
Departures from the ideal limit can be transparently analyzed by adopting
an incoherent Bright drive, see Sec.~\ref{subsec:Incoherent-Bright-drive}.
An elegant analysis of the no-click evolution for arbitrary amplitude
of the Dark Rabi drive can be found in Refs.~\citet{Ruskov2007}
and~\citet{Ruskov2009-unpublished}. For an interesting connection
to of the three-level intermittent dynamics to dynamical phase transitions,
see Refs.~\citet{Lesanovsky2013} and~\citet{Garrahan2018}.
\begin{figure}
\begin{centering}
\includegraphics[scale=1.4]{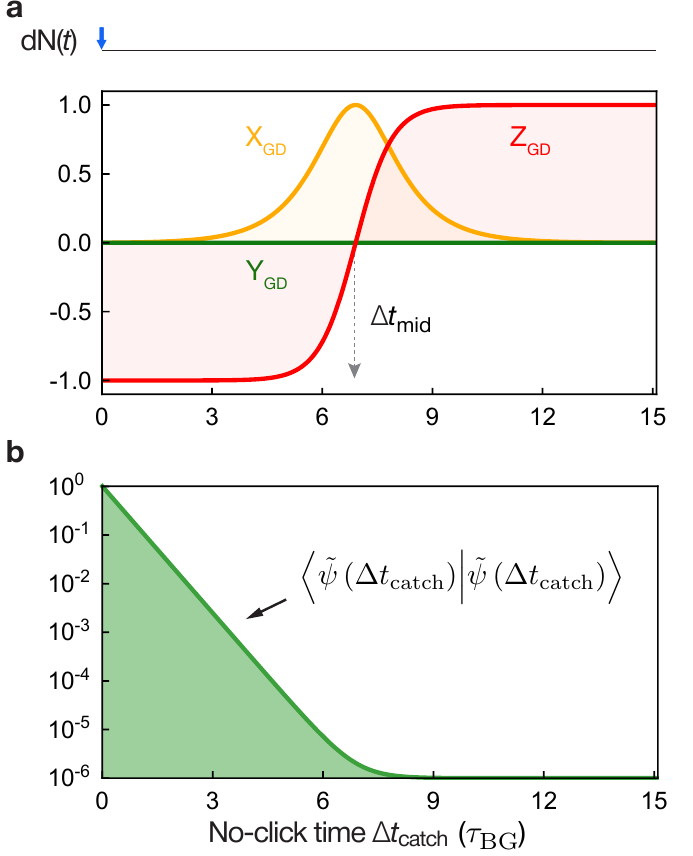}
\par\end{centering}
\caption[Conditional no-click evolution of the jump from $\G$ to $\D$: ideal
photodetection theory.]{\textbf{\label{fig:Conditional-no-click-evolution}Conditional no-click
evolution of the jump from $\G$ to $\D$: ideal photodetection theory.
a, }A typical quantum trajectory for a jump from $\G$ to $\D$ represented
as the GD Bloch vector ($X_{\mathrm{GD}}$, $Y_{\mathrm{GD}}$, $Z_{\mathrm{GD}}$),
conditioned on \emph{no }clicks, $\dN\left(t\right)=0$, for duration
$\deltatcatch$. The Rabi drives are $\Omega_{\mathrm{DG}}=10^{-5}$
and $\Omega_{\mathrm{BG}}=0.1$ in units of the decay rate $\gamma_{\mathrm{B}}$.
Time axis is scaled in units of the mean time between detector clicks,
$\tau_{\mathrm{BG}}=(\Omega_{{\rm BG}}^{2}/\gamma_{\mathrm{B}})^{-1}$.
Time scale of the jump flight is the mid-flight time $\Delta t_{\mathrm{mid}}$,
defined by $Z_{\mathrm{GD}}=0$. \textbf{b,} Log plot of the norm
of the un-normalized no-click state, $\braket{\tilde{\psi}\left(\Delta t_{\mathrm{catch}}\right)}{\tilde{\psi}\left(\Delta t_{\mathrm{catch}}\right)}$,
as a function of $\deltatcatch$, in units of $\tau_{\mathrm{BG}}$.
Parameters of the plot correspond to those of panel a. }
\end{figure}

\paragraph{Remarks on the state evolution. }

The evolution of the GD manifold Bloch vector ($X_{\mathrm{GD}}\left(\deltatcatch\right),$
$Y_{\mathrm{GD}}\left(\deltatcatch\right),$ $Z_{\mathrm{GD}}\left(\deltatcatch\right)$)
conditioned on \emph{no }clicks, $\dN\left(\deltatcatch\right)=0$,
for duration $\deltatcatch$, is plotted in Fig.~\ref{fig:Conditional-no-click-evolution}a.
The partial tomogram visually shows that the predicted evolution of
the quantum jump from $\ket{\mathrm{G}}$ to $\ket{\mathrm{D}}$ is
continuous and coherent, $X_{\mathrm{GD}}^{2}+Y_{\mathrm{GD}}^{2}+Z_{\mathrm{GD}}^{2}=1$
for all no-click times, $\deltatcatch$. The measurement record, $\dN$,
and the predicted trajectory is identical for any two jumps from $\G$
to $\D$. The time axis has been scaled in units of the mean time,
$\tau_{\mathrm{BG}}=(\Omega_{{\rm BG}}^{2}/\gamma_{\mathrm{B}})^{-1}$,
between photon detector clicks. This time can also be understood as
the inverse of the information-gain rate of the measurement about
the G level. To expand on this, for definitiveness, consider the situation
where the atom is initialized in $\G$, but this information is not
shared with the observer operating the photon detector. By measurement,
how long does it take the observer to statistically deduce that the
atom is in $\G$ or not?  The measurement drive $\Omega_{\mathrm{BG}}$
is actuated and the observer  monitors the detector for clicks. If
the atom is in $\G,$ on average, the detector records a click after
time $\tau_{\mathrm{BG}}$, and informs the observer that the atom
is definitively in $\G$. Alternatively, if the atom was initialized
in $\D,$ no clicks would be recorded. As the detector does not record
a click for durations longer than $\tau_{{\rm BG}}$, the observer
becomes increasingly confident that the atom could not be in $\G$
since it becomes exponentially unlikely that a click has not yet been
observed, see Fig.~\ref{fig:Conditional-no-click-evolution}b, and
the alternative conclusion, that the atom is in $\D$, becomes increasingly
likely. Although this information-gain consideration is carried out
from the point of view of the observer, and with classical measurements
would bear no consequence for the objective state of the system, with
quantum measurements the gain of information about the system by virtue
of a measurement is necessarily accompanied by a result-correlated
state-disturbance (back-action). In Hilbert space, the disturbance
can be viewed as a \emph{measurement-backaction effective force},
as discussed in the following subsection, Sec.~\ref{subsec:Knowledge-driven-force-in}.

\paragraph{Probability of no-click record.}

Fig.~\ref{fig:Conditional-no-click-evolution}b shows a plot of the
conditional no-click state norm, $\braket{\psi\left(\deltatcatch\right)}{\psi\left(\deltatcatch\right)}$,
as a function of the no-click duration, $\deltatcatch$. The norm
initially decays exponentially with a time-constant $\tau_{{\rm BG}}$,
during which time, the atom remains essentially in $\G$, as indicated
by the $Z_{{\rm GD}}$ Bloch component in panel a, which is roughly
equal to $-1$. However, as the no-click duration approaches  mid-flight
time of the jump, $\deltatcatch\approx\Delta t_{{\rm mid}}$, the
decay of the norm slows down dramatically, since the atom transitions
from $\G$ to $\D$, in which state one can stop expecting the rapid
occurrence of clicks. The quantum jump from $\G$ to $\D$ can be
observed in the tomogram shown in panel (a). For no-click duration
$\deltatcatch\gg\Delta t_{{\rm mid}}$, the decay of the norm initially
appears flat, however, on longer time-scales (not shown) it is seen
that it also follows an exponential decay law with a much longer time
constant, $\tau_{{\rm DG}}\gg\tau_{{\rm BG}}$, corresponding to the
waiting-time for the jump back down, from $\D$ to $\G$.

\paragraph{\textit{\emph{Application of the photon counting model to the experiment.}}}

The photon-counting theory presented in this section provides the
background to the experiment along with a link to the original ion
experiments. It captures a core set of the ideas, even though the
monitoring of $|{\rm B\rangle}$ implemented in the experiment is
diffusive — the opposite limit of the point-process description presented
here, see Sec.~\ref{sec:Heterodyne-monitoring-of}. Nevertheless,
the photon-counting theory even provides a quantitative first approximation
of the experimental results. For definitiveness, consider the flight
of the quantum jump shown in Fig.~\ref{fig:catch}b. The measured
mid-flight time, $\Delta t_{\mathrm{mid}}=3.95~\mathrm{\mu s}$, is
predicted, in a first approximation, by Eq.~(\ref{eq:t_mid_coherent}).
Using the (independently measured) values of the experimental parameters,
summarized in Table~\ref{tab:system-params} (setting $\Omega_{\mathrm{BG}}$
equal to $\Omega_{\mathrm{B0}}=2\pi\times1.2$~MHz, the BG drive
when the atom is not in $\ket{\rm B}$) and extracting the effective
measurement rate of $|{\rm B\rangle}$, $\gamma_{\mathrm{B}}=2\pi\times9.0$~MHz
(which follows from Eq.~(\ref{eq:GammaBG}) where $\Gamma_{\mathrm{BG}}=2\pi\times1.01$~MHz,
the average click rate on the BG transition), Eq.~(\ref{eq:t_mid_coherent})
predicts $\Delta t_{{\rm mid}}\approx4.3~\mathrm{\mu s}$ — in fair
agreement with the observed value $\Delta t_{{\rm mid}}=3.95~\mathrm{\mu s}$.
The photodetection theory presented in in Sec.~\ref{subsec:Incoherent-Bright-drive}
further improves the agreement. These calculations serve to generally
illustrate the theory and ideas of the experiment; the quantitative
comparison between theory and experiment is only presented in Sec.~\ref{subsec:Comparison-between-theory}.

\subsection{Measurement-backaction effective force in the absence of the Dark
Rabi drive\label{subsec:Knowledge-driven-force-in}}

While in Sec.~\ref{sec:thry:3lvl-atom-simple-log} we considered
the coherent dynamics of the three-level atom in the presence of  both
unitary evolution, due to the Rabi drive $\Odg$, and the competing
non-unitary state collapse, due to the measurement, in this subsection,
we examine the simpler case where only measurement dynamics are at
play, i.e., $\Odg=0$. In this simpler situation, some important features
of the measurement, consisting of the Rabi drive $\Obg$ and the monitoring
of the B level at the rate $\gamma_{{\rm B}}$, are more easily discussed.
In particular, we pay attention to the notion of a\emph{ measurement-backaction
effective force}, the special force that unavoidably disturbs of the
quantum state due to the measurement. 

For definitiveness, consider the situation where  the three-level
atom is prepared in an initial superposition involving the G and D
levels, $\ket{\psi\left(0\right)}=\mathcal{N}\left(\G+\epsilon\D\right)$,
where $\epsilon\ll1$ and $\mathcal{N}$ is the ket normalization
factor; for simplicity, assume $\D$ is completely decoupled form
the environment, $\gamma_{{\rm D}}=0$. To measure the atom, only
the Rabi drive $\Obg$ is turned on. One of two qualitatively distinct
measurement records is observed: either clicks are recorded indefinitely
or no clicks are ever recorded, which can qualitatively be understood
in view of the following considerations. When the BG drive is first
turned on, some of the initial population from the G level is transferred
to the B level due to the steering force of $\Obg$. However, even
as a tiny amount of population is deposited in $\B$, the strong coupling
with the environment and the photodetector dampens the transfer and
quickly yields the detection of a click, with probability $\wp_{1}\left(\dt\right)=\left\langle \hat{c}_{{\rm B}}^{\dagger}\hat{c}_{{\rm B}}\right\rangle \left(t\right)\dt$,
where $\hat{c}_{\mathrm{B}}=\sqrt{\gamma_{{\rm B}}}\kb{\mathrm{G}}{\mathrm{B}}$
is the  jump operator. The click resets the atom to the ground state,
$\G$. Once the atom is completely in $\G$, the amplitude of $\D$
is zero, and since $\Odg=0$, the atom can never transition to $\D$
subsequently. The remainder of the history proceeds as described above,
the atom remains predominantly in $\G$ and continues to fluoresce
though the partial excitation and subsequent relaxation of $\B$,
by means of $\Obg$ and the detection, $\gamma_{{\rm B}}$, respectively.
In this way, the Dehmelt electron scheme implements a measurement
with result $\G$, occurring with an approximate probability $1-\epsilon^{2}$.
The alternative set of trajectories, where no clicks are observed
is quantitatively analyzed in the following, and occur with approximate
probability $\epsilon^{2}$.

\paragraph{No-click trajectory. }

The \emph{normalized} state of the three-level atom conditioned on
no-clicks, \footnote{Notationally, we employ capital letters for the coefficients of the
normalized sate, and lower-case letters for those of the un-normalized
state.} 
\begin{equation}
\ket{\psi_{I=0}\left(t\right)}=\cgn\left(t\right)\G+\cbn\left(t\right)\B+\cdn\left(t\right)\D\,,\label{eq:psiPhotonNL}
\end{equation}
evolves according to the non-linear Schrödinger equation, see Eq.~(\ref{eq:psiPhotonNoClick}),
\begin{equation}
\frac{\mathrm{d}}{\mathrm{d}t}\ket{\psi_{I=0}\left(t\right)}=\left(-i\hat{H}-\half\hat{c}_{{\rm B}}^{\dagger}\hat{c}_{{\rm B}}+\half\left\langle \hat{c}_{{\rm B}}^{\dagger}\hat{c}_{{\rm B}}\right\rangle \left(t\right)\right)\ket{\psi_{I=0}\left(t\right)}\:,\label{eq:NL-SSe-Cond-3}
\end{equation}
which in terms of the normalized state coefficients ($\cgn,\cbn$,
and $\cdn$) yields the set of coupled non-linear equations,
\begin{align}
\ddt\cbn\left(t\right)= & \frac{1}{2}\gamma_{{\rm B}}\cbn\left(t\right)^{3}+\frac{1}{2}\Omega_{{\rm BG}}\cgn\left(t\right)-\half\gamma_{{\rm B}}\cbn\left(t\right),\label{eq:Cbn}\\
\ddt\cgn\left(t\right)= & \frac{1}{2}\gamma_{{\rm B}}\cbn\left(t\right)^{2}\cgn\left(t\right)-\frac{1}{2}\Omega_{{\rm BG}}\cbn\left(t\right),\label{eq:Cgn}\\
\ddt\cdn\left(t\right)= & \frac{1}{2}\gamma_{{\rm B}}\cbn\left(t\right)^{2}\cdn\left(t\right),\label{eq:Cdn}
\end{align}
The measurement terms in Eqs.~(\ref{eq:Cbn})-(\ref{eq:Cdn}) associated
with information gain are the non-linear ones. As discussed in Sec.~\ref{sec:Quantum-trajectory-theory},
they originate from the normalization term in the conditional state
update and give rise to non-unitary state dynamics, resulting in a
drastic departure from the usual Schrödinger equation. To analyze
their effect and gain some physical intuition, we introduce a graphical
representation of the Hilbert space of the three-level atom.

\paragraph{$\mathbb{R}$-qutrit sphere representation. }

\begin{figure}
\begin{centering}
\includegraphics[scale=1.5]{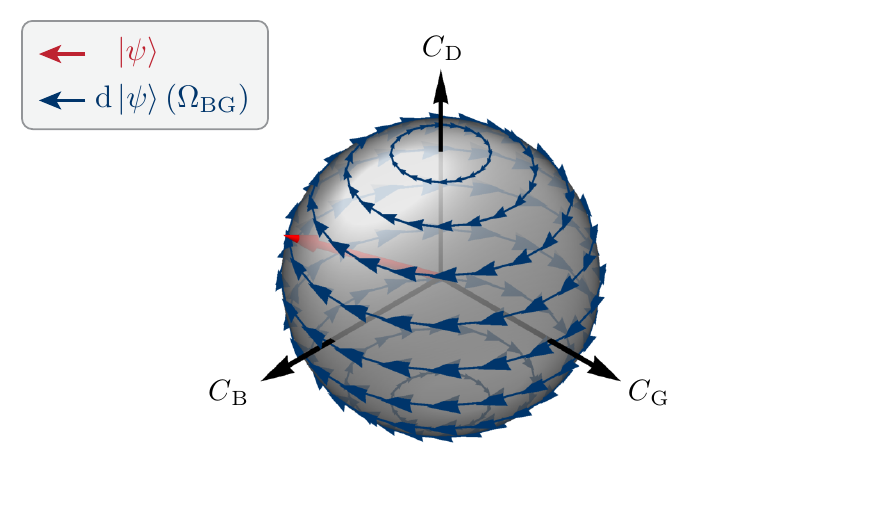}
\par\end{centering}
\caption[Geometrical representation of a qutrit state with real coefficients:\textbf{
$\mathbb{\mathbb{R}}$}-qutrit sphere]{\label{fig:R-qutrit-sphere-representation.}\textbf{Geometrical representation
of a qutrit state with real coefficients:} $\mathbb{\mathbb{R}}$\textbf{-qutrit
sphere. }Geometric representation of the Hilbert space of pure states
of a qutrit, $\ket{\psi}=\cbn\B+\cgn\G+\cdn\D$, with real-valued
coefficients, notably isomorphic to the special orthogonal group $SO\left(3\right)$.
Overlaid vector field represents the infinitesimal state change, $\mathrm{d}\ket{\psi}$,
due to the Rabi drive $\Omega_{{\rm BG}}.$}
\end{figure}

It follows from the real coefficients of Eqs.~(\ref{eq:Cbn})-(\ref{eq:Cdn})
and the initial conditions that $\cgn,\cbn,$ and $\cdn$ are constrained
to be real. Hence a pure state of the atom admits a geometrical representation
as a point on the unit sphere defined by the state norm condition
$\cbn^{2}+\cgn^{2}+\cdn^{2}=1$, see Fig.~\ref{fig:R-qutrit-sphere-representation.}.
We nickname this representation the '$\mathbb{R}$-qutrit sphere'.
Unlike the Bloch sphere representation where orthogonal state vectors
are represented by \emph{antiparallel} vectors, in the $\mathbb{R}$-qutrit
sphere representation, orthogonal state vectors are actually represented
by orthogonal vectors, extending from the origin to the surface of
the sphere. Notably, the sphere contains states of the two-level sub-manifolds
that are \emph{not} represented on the Bloch sphere, those with a
``global phase.''\footnote{The special unitary Lie group $SU(2)$ is not isomorphic to the special
orthogonal Lie group $SO(3)$, but is a double cover of it.} The addition of the third level allows, in general, the observation
of the global phase because it can be measured relative to the phase
of the third level. Consequently, a Rabi rotation between two levels
is no longer $2\pi$ periodic but $4\pi$ periodic.  Unitary evolution
is represented by rotations on the sphere; for instance, the evolution
due to the Bright Rabi drive, $\Obg$, is a rotation about the D axis,
and the corresponding infinitesimal-state-change vector field, $\mathrm{d}\ket{\psi}\left(\Obg\right)=\frac{1}{2}\left(\cgn\B-\cbn\G\right)\Obg\dt$,
is plotted on the surface of the sphere in Fig.~\ref{fig:R-qutrit-sphere-representation.}.
Note that the length of the vectors is largest at the GB equator and
approaches zero toward the D poles. The vector field representation
is useful in the analysis of the non-linear measurement-backaction
effective force  due to the renormalization terms and we hope can
provide a more intuitive understanding of the interplay between the
coherent Rabi and the stochastic measurement dynamics.  Before elaborating
on the geometrical representation of the measurement dynamics, it
is useful to first algebraically solve Eqs.~(\ref{eq:Cbn})-(\ref{eq:Cdn}).

\paragraph{Measurement-backaction  force steers atom towards $\D$. }

Although there is no Rabi drive or measurement directly applied to
the Dark level, conditioned on not detecting a click, according to
Eq.~(\ref{eq:Cdn}), a force is nonetheless exerted by the B-level
monitoring that steers the atom toward the Dark level. Specifically,
the rate of change of the D level amplitude, $\ddt\cdn$, is given
by an anti-damping term with a state-dependent rate proportional to
the B level population, $\cbn\left(t\right)^{2}$, and measurement
rate, $\gamma_{{\rm B}}$. Solving Eq.~(\ref{eq:Cdn}), one finds
\begin{equation}
\cdn\left(t\right)=\cdn\left(0\right)\exp\left(\int_{0}^{t}\mathrm{d}t'\frac{1}{2}\gamma_{{\rm B}}\cbn\left(t\right)^{2}\right)\,.\label{eq:Cdsol}
\end{equation}
In this sense, the renormalization of the conditional state amounts
to a (non-unitary) measurement-backaction force on $\D$, which is
linked to the population of $\B.$ To explicitly solve Eq.~(\ref{eq:Cdsol}),
we need to solve the remaining equations, Eqs.~(\ref{eq:Cbn}) and
(\ref{eq:Cgn}). 

\paragraph{Adiabatic elimination of the Bright state dynamics.}

Because Eq.~(\ref{eq:Cbn}) is non-linear, we consider the B level
dynamics and their adiabatic elimination with greater care. Eq.~(\ref{eq:Cbn}),
contains both a damping, $-\half\gamma_{{\rm B}}\cbn$, and an anti-damping,
$\frac{1}{2}\gamma_{{\rm B}}\cbn{}^{3}$, term. These cancel out perfectly
only if the atom is either entirely in $\B$, $\cbn=\pm1$, or not
at all in $\B$, $\cbn=0$; otherwise, $\left|\cbn\right|<1$, the
damping dominates, steering $\cbn$ in the direction of zero. In the
extreme case, where $\Omega_{{\rm BG}}=0$, one can explicitly solve
the B level dynamics, $\cbn\left(t\right)^{2}=\left[1+\left(\cbn\left(0\right)^{-2}-1\right)\exp\left(\gamma_{{\rm B}}t\right)\right]^{-1},$
which for small initial populations, $\cbn\left(0\right)^{2}\ll1$
rapidly decays to a stable zero equilibrium at a rate $\frac{1}{2}\gamma_{{\rm B}}$,
$\cbn\left(t\right)\approx\cbn\left(0\right)\exp\left(-\frac{1}{2}\gamma_{{\rm B}}t\right)$.
Since $\gamma_{{\rm B}}$ is the fastest timescale in the problem
and the B dynamics are convergent, we can adiabatically eliminate
$\cbn$ by setting $\ddt\cbn\left(t\right)=0$; solving the cubic
equation, one finds three solution branches,
\begin{equation}
\cbnb\left(t\right)=\begin{cases}
-1-\frac{\Obg}{2\gamma_{{\rm B}}}\cgn\left(t\right)+\mathcal{O}\left(\left(\Obg/\gamma_{{\rm B}}\right)^{2}\right)\\
1-\frac{\Obg}{2\gamma_{{\rm B}}}\cgn\left(t\right)+\mathcal{O}\left(\left(\Obg/\gamma_{{\rm B}}\right)^{2}\right)\\
\frac{\Obg}{\gamma_{{\rm B}}}\cgn\left(t\right)+\mathcal{O}\left(\left(\Obg/\gamma_{{\rm B}}\right)^{3}\right)\,.
\end{cases}\label{eq:Cb-bar}
\end{equation}
Operating the three-level atom in the limit where the $\B$ level
is never appreciably populated, we employ the third solution branch,
$\cbnb\left(t\right)=\frac{\Obg}{\gamma_{{\rm B}}}\cgn\left(t\right)$,
in Eq.~(\ref{eq:Cgn}) to find the effective equation of motion for
the G level dynamics, 
\begin{equation}
\ddt\cgn\left(t\right)=-\tau_{\mathrm{BG}}^{-1}\left[1-\cgn\left(t\right)^{2}\right]\cgn\left(t\right),\label{eq:ddtCGn}
\end{equation}
which are now completely decoupled from the other levels. In Eq.~(\ref{eq:ddtCGn}),
we identify a damping and an anti-damping term with a constant and
G-population dependent, $\cgn^{2}$, rate, respectively. The scale
of both terms is given by the parameter $\tau_{\mathrm{BG}}^{-1}=\Omega_{{\rm BG}}^{2}/2\gamma_{\mathrm{B}}$,
which is the expected rate of clicks when the atom is in $\G$. By
eliminating the B level, we have obtained an explicit relation for
the effective monitoring of the G level, which occurs at a rate $\tau_{{\rm BG}}^{-1}$,
which can also be interpreted as the quantum Zeno rate \citep{Misra1977,Gambetta2008-qm-traj,Matsuzaki2010,Vijay2011,Slichter2016-T1vsNbar,Harrington2017,Hacohen-Gourgy2018}.
The numerator $\Obg^{2}$ is proportional to the population transfer
rate from $\G$ to $\B$ while the denominator $\gamma_{{\rm B}}$
gives the rate of projection from $\B$ to $\G$. Solving Eq.~(\ref{eq:ddtCGn})
and substituting its solution in Eq.~(\ref{eq:Cdsol}), one finds
\begin{figure}
\begin{centering}
\includegraphics[scale=1.5]{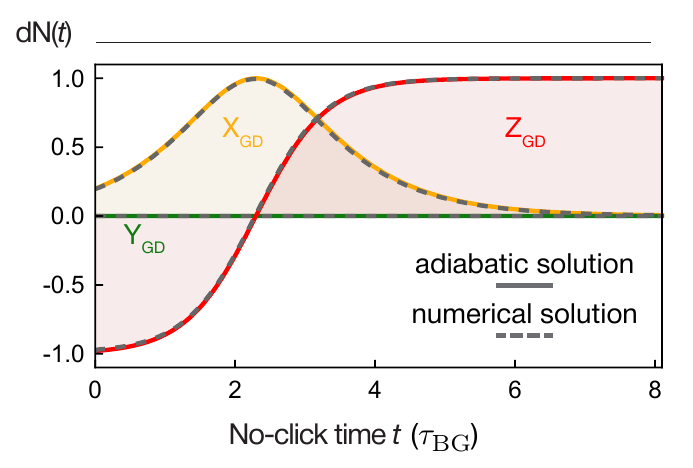}
\par\end{centering}
\caption[Adiabatic solution for the no-click GD manifold trajectory of a superposition
state measured with $\Odg=0$]{\textbf{\label{fig:3:adiabatic-sol}Adiabatic solution for the no-click
GD manifold trajectory of a superposition state measured with }$\Odg=0$\textbf{.
}Adiabatic-approximation (solid lines) and numerical (dashed lines)
solution for the partial tomogram of the GD manifold of the no-click
quantum trajectory of an initial superposition state $\ket{\psi\left(0\right)}=\mathcal{N}\left(\G+\epsilon\D\right)$,
where $\epsilon=0.1$ and $\mathcal{N}=\left(1+\epsilon^{2}\right)^{-1/2}$.
The Bright Rabi drive is $\Obg=0.1$, in units of the decay rate $\gamma_{{\rm B}}$.
Time axis scaled in units of $\tau_{\mathrm{BG}}=\left(\Obg^{2}/\gamma_{{\rm BG}}\right)^{-1}.$}
\end{figure}
\begin{align}
\cgn\left(t\right)^{2}= & \frac{p_{{\rm G}}}{p_{{\rm G}}+\left(1-p_{{\rm G}}\right)e^{2t/\tau_{{\rm BG}}}}\,,\label{eq:Cgnoclicksol}\\
\cdn\left(t\right)^{2}= & \frac{p_{{\rm D}}e^{t/\tau_{{\rm BG}}}}{p_{{\rm D}}+\left(1-p_{{\rm D}}\right)e^{2t/\tau_{{\rm BG}}}}\,,\label{eq:CdNoClickSol}
\end{align}
where $p_{{\rm G}}\equiv\cgn\left(0\right)^{2}$ and $p_{{\rm D}}\equiv\cdn\left(0\right)^{2}$
are the initial conditions. Note, for $p_{{\rm D}}=0$, the above
solution for $\cdn$ is always zero. The evolution of the GD Bloch
vector conditioned on no clicks for the initial state $\ket{\psi\left(0\right)}=\mathcal{N}\left(\G+\epsilon\D\right)$
is obtained by substituting Eq.~(\ref{eq:Cgnoclicksol}) and (\ref{eq:CdNoClickSol})
in Eqs.~(\ref{eqn:Z})-(\ref{eqn:Y}), 
\begin{align}
{\rm Z}_{{\rm GD}}(t) & ={\rm tanh}\left[t/\tau_{{\rm BG}}+\mathrm{arctanh}\left[{\rm Z}_{{\rm GD}}(0)\right]\right],\label{eq:Zgd-Adi}\\
{\rm X}_{{\rm GD}}(t) & ={\rm sech}\left[t/\tau_{{\rm BG}}+\mathrm{arctanh}\left[{\rm Z}_{{\rm GD}}(0)\right]\right],\\
{\rm Y}_{{\rm GD}}(t) & =0.\label{eq:Ygd-Adi}
\end{align}
We note that a few results of this subsection, especially Eqs.~(\ref{eq:Zgd-Adi})-(\ref{eq:Ygd-Adi}),
bear resemblance to results from Sec.~\ref{sec:thry:3lvl-atom-simple-log},
yet we stress that  the two situations are fundamentally distinct,
and the resemblance must be considered with care. For instance, we
note that the mid-flight time $\Delta t_{\mathrm{mid}}$ cannot be
recovered from the simpler situation considered here, where no quantum
jumps occur and there is no competition between unitary dynamics due
to $\Omega_{{\rm DG}}$ and the measurement. 

In Fig.~\ref{fig:3:adiabatic-sol}a, we plot the adiabatic-approximation
solution to the non-linear Schrödinger evolution, Eq.~(\ref{eq:NL-SSe-Cond-3}),
for the GD manifold Bloch vector conditioned on no clicks, Eqs.~(\ref{eq:Zgd-Adi})-(\ref{eq:Ygd-Adi}),
obtained in the limit $\Obg\ll\gamma_{B}$. Overlaid (dashed lines)
is the corresponding numerically calculated solution to Eq.~(\ref{eq:NL-SSe-Cond-3}).
Even for modest separation of timescales, $\Obg/\gamma_{{\rm B}}=0.1$
in the plot, the two solutions appear nearly indistinguishable. The
initial atom state, $\epsilon=0.1$, is gradually projected to $\D$
on a timescale given by $\tau_{{\rm BG}}$ and evolves in a characteristically
non-unitary manner. Notably, the state remains pure at all times,
and in the limit $t\gg\tau_{{\rm BG}}$ remains essentially in $\D$,
indefinitely. Importantly, for times $t$ on the other of $\tau_{{\rm BG}}$,
the projection can (but need not) be interrupted by the detection
of a click, which would project the state to $\G$, and occurs with
total probability $\approx1-\epsilon^{2}$. 
\begin{figure}
\begin{centering}
\includegraphics[scale=1.4]{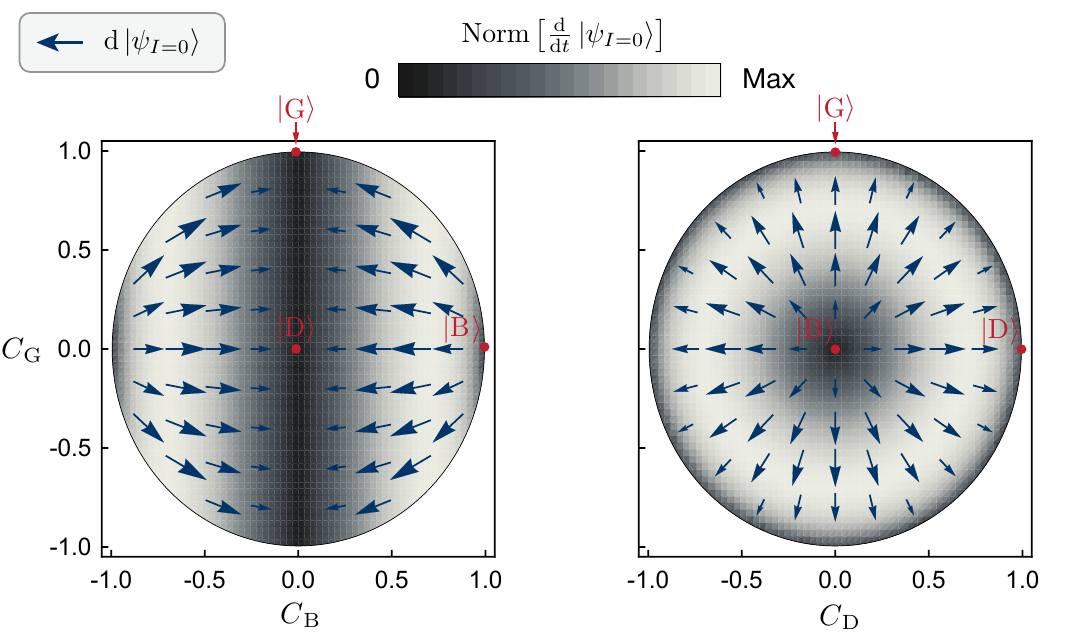}
\par\end{centering}
\caption[Geometrical representation of the no-click measurement-backaction
force for $\Obg=0$]{\textbf{\label{fig:Knowledge-driven-force-in-1}Geometrical representation
of the no-click measurement-backaction  force for $\Obg=0$.} Shown
are two projections of $\mathbb{\mathbb{\mathbb{R}}}$-qutrit sphere
overlaid with the measurement-force vector field, $\frac{\mathrm{d}}{\mathrm{d}t}\ket{\psi_{I=0}}$,
due to the monitoring of the B level with $\Obg=0$. Color of density
plot indicates the relative magnitude of the change, $\mathrm{Norm}\left[\frac{\mathrm{d}}{\mathrm{d}t}\ket{\psi_{I=0}}\right]$.}
\end{figure}

\paragraph{Hilbert space representation of the measurement dynamics.}

It is useful to consider a geometric representation of the measurement
dynamics and in particular of the non-linear measurement-backaction
force. For simplicity, first consider the measurement force due only
to the monitoring of the B level, in the absence of the Bright Rabi
drive, $\Obg=0$. This  force can be represented as a vector field
on the surface of the $\mathbb{R}$-qutrit sphere, see Fig.~\ref{fig:Knowledge-driven-force-in-1}.
The vector field is calculated from Eqs.~(\ref{eq:Zgd-Adi})-(\ref{eq:Ygd-Adi})
for the change in the state conditioned on detecting no clicks, 
\begin{equation}
\frac{\mathrm{d}}{\mathrm{d}t}\ket{\psi_{I=0}}=\frac{1}{2}\gamma_{{\rm B}}\cbn{}^{2}\begin{pmatrix}\cbn-1\\
\cgn\\
\cdn
\end{pmatrix}\,.\label{eq:ddtPsiI-vec}
\end{equation}
The colormap in Fig.~\ref{fig:Knowledge-driven-force-in-1} depicts
the relative magnitude of the change, $\mathrm{Norm}\left[\frac{\mathrm{d}}{\mathrm{d}t}\ket{\psi_{I=0}}\right],$
which we note is only zero in two special cases: i) when the atom
is $\pm\B$, corresponding to the points $\left(\pm1,0,0\right),$
and ii) when the atom is in a\emph{ }state involving exclusively $\G$
and $\D$ but not $\B$. The latter is special in that it corresponds
to an entire manifold of states, the GD equatorial circle, which can
be visually recognized in Fig.~\ref{fig:Knowledge-driven-force-in-1}
as the dark vertical stripe at the center of the left panel and the
dark circular perimeter of the disk in the right panel. All other
states, not covered under the latter two cases, are superpositions
involving $\B$. From the vector field plot, it is evident that these
states are guided by the force away from the $\B$ poles and toward
the GD equator. It is precisely this feature of the measurement force
that results in the gradual projection of the state conditioned on
no clicks — it is the unavoidable disturbance of the atom due to the
information-gain of the no-click measurement outcomes, which lead
the observer to gradually learn that the atom is not in $\D$, thus
resulting in the increased likelihood that it is in $\G$ or $\D$.
This dynamics embody the message of the Chapter~\ref{chap:Quantum-Trajectory-Theory}
epigraph, ``In quantum physics you don't see what you get, you get
what you see.''

\begin{figure}
\begin{centering}
\includegraphics[scale=1.4]{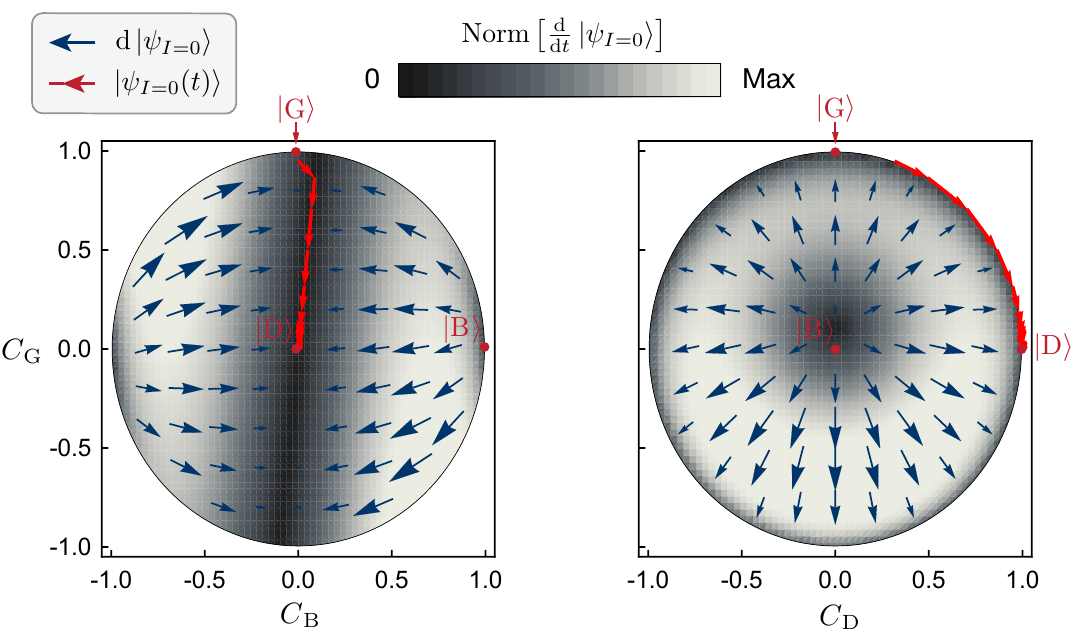}
\par\end{centering}
\caption[Geometrical representation of the measurement-backaction force and
a no-click trajectory with $\Obg=0.1\gamma_{{\rm B}}$]{\textbf{\label{fig:Knowledge-driven-force-in-2}Geometrical representation
of the measurement-backaction force and a no-click trajectory with
$\Obg=0.1\gamma_{{\rm B}}$.}  Two projections of the $\mathbb{\mathbb{\mathbb{R}}}$-qutrit
sphere overlaid with the measurement-force vector field $\frac{\mathrm{d}}{\mathrm{d}t}\ket{\psi_{I=0}}$
(blue arrows) and the path of the quantum trajectory from Fig.~(\ref{fig:3:adiabatic-sol})
(red arrows), depicting the gradual projection of a superposition
state to $\D$ conditioned on no clicks. Density plot indicates relative
field magnitude, $\mathrm{Norm}\left[\frac{\mathrm{d}}{\mathrm{d}t}\ket{\psi_{I=0}}\right]$.}
\end{figure}

In Fig.~\ref{fig:Knowledge-driven-force-in-2}, we plot the measurement
vector field in the presence of the Bright Rabi drive $\Omega_{\mathrm{BG}}$,
\begin{equation}
\frac{\mathrm{d}}{\mathrm{d}t}\ket{\psi_{I=0}}=\frac{1}{2}\gamma_{{\rm B}}\cbn{}^{2}\begin{pmatrix}\cbn-1\\
\cgn\\
\cdn
\end{pmatrix}+\frac{1}{2}\Obg\begin{pmatrix}\cgn\\
-\cbn\\
0
\end{pmatrix}\,,\label{eq:ddtPsiI-vec-1}
\end{equation}
with $\Obg=0.1\gamma_{{\rm B}}$. The Bright Rabi drive, visually
represented on the $\mathbb{R}$-qutrit sphere in Fig.~\ref{fig:R-qutrit-sphere-representation.},
perturbs the measurement field, shown in Fig.~\ref{fig:Knowledge-driven-force-in-1},
by linking the B and G levels and lifting the degeneracy of the measurement,
represented in GD equator. Visually, this is evident in the tilt of
the vertical dark stripe in the center of the left-panel colormap.
In the right panel, it is also evident that $\B$ is no longer an
equilibrium point; the equilibrium has been shifted in the direction
of $\G$ by an amount proportional $\Obg/\gamma_{{\rm B}},$ see Eq.~(\ref{eq:Cb-bar}),
and made metastable. The red arrows depict the path of the quantum
trajectory in Hilbert of the gradual projection of an initial superposition
state of $\G$ and $\D$, for the same parameters as employed in Fig.~(\ref{fig:3:adiabatic-sol}),
where $\epsilon=0.1$. Initially, the state is quickly steered in
the direction of $\B$ by the force of $\Obg$. However, as the state
moves in the direction of $\B$, the motion is quickly opposed by
the no-click measurement-backaction force, which grows larger in amplitude
in this direction. The two forces do not precisely cancel each other
out, because of the slight mismatch in angles. The net force, albeit
small, steers the atom towards the GD equator and with a slight tilt
toward $\D$. The opposition of the $\Obg$ drive and the measurement
back-action ``trap'' the state in the ridge where the two forces
nearly cancel each other out, the nearly vertical dark stripe in the
left panel, and the small angular mismatch slowly carries the state
in the direction of $\D$, an equilibirum point, where all forces
are zero.

\subsection{Incoherent Bright drive and Dark drive off\label{subsec:Incoherent-Bright-drive}}

In this section, we consider the case of quantum jumps in the three-level
atom subject to photodetection and an incoherent Bright drive, rather
than a Rabi one, see Sec.~\ref{sec:thry:3lvl-atom-simple-log}. The
situation analyzed in this section is somewhat more analogous to the
cQED experiment where the Bright Rabi drive consists of a bi-chromatic
tone that unconditionally addresses the BG transition, independent
of the population of the readout cavity, necessitated by the  dispersive
pull of the readout cavity on the BG frequency. The bi-chromatic drive
effectively acts as an incoherent drive. The incoherent Bright drive
photodetection theory presented here sheds some further light on the
dynamics of the quantum jump from $\G$ to $\D$. Features such as
the non-zero coherence, $X_{{\rm GD}}$, and in the limit $\deltatcatch\gg\Delta t_{{\rm mid}}$
are discussed.\footnote{The following derivation is due to H.J. Carmichael and R. Gutierrez-Jauregui.}

Replacing the coherent Rabi drive $\Omega_{{\rm BG}}$ by an incoherent
drive $\Gamma_{{\rm BG}}$ in  the master equation of the three-level
atom in the interaction picture, Eq.~(\ref{eq:PhotonLindblad}) becomes
\begin{equation}
\frac{\mathrm{d}\rho}{\mathrm{d}t}=(i\hbar)^{-1}[\hat{H}_{{\rm drive}},\rho]+\Gamma_{{\rm BG}}{\cal D}\left[\kb{\rm B}{\rm G}\right]\rho+(\gamma_{{\rm B}}+\Gamma_{{\rm BG}}){\cal D}\left[\kb{\rm G}{\rm B}\right]\rho+\gamma_{{\rm D}}{\cal D}\left[\kb{\rm G}{\rm D}\right]\rho,
\end{equation}
and Eq.~(\ref{eq:ClickHeff}) becomes
\begin{equation}
\hat{H}_{{\rm drive}}=i\hbar\frac{\Omega_{{\rm DG}}}{2}\big(\kb{\rm D}{\rm G}-\kb{\rm G}{\rm D}\big).
\end{equation}
The strong-monitoring assumption, $\tau_{{\rm BG}}^{-1}\ll\gamma_{{\rm B}}$,
is also carried over from Sec.~\ref{sec:thry:3lvl-atom-simple-log}
by assuming $\Gamma_{{\rm BG}}\ll\gamma_{{\rm B}}$, i.e., the time
between clicks in fluorescence is essentially the same as the time
separating photon absorptions from the incoherent drive, as absorption
is rapidly followed by fluorescence ($\gamma_{{\rm B}}+\Gamma_{{\rm BG}}\gg\Gamma_{{\rm BG}}$).
This brings a useful simplification, since, following each reset to
$\G$, the unnormalized state evolves in the GD-subspace, 
\begin{equation}
i\hbar\frac{\mathrm{d}\ket{\tilde{\psi}}}{\mathrm{d}\Delta t_{{\rm catch}}}=\left(\hat{H}_{{\rm drive}}-i\hbar\frac{\Gamma_{{\rm BG}}}{2}\kb{\rm G}{\rm G}-i\hbar\frac{\gamma_{{\rm D}}}{2}\kb{\rm D}{\rm D}\right)\ket{\tilde{\psi}},
\end{equation}
thus replacing Eqs.~(\ref{eqn:3x3_system}) and (\ref{eqn:W_DG_equation_coherent})
by the simpler $2\times2$ system 
\begin{equation}
\frac{\mathrm{d}}{\mathrm{d}\Delta t_{{\rm catch}}}\begin{pmatrix}\cg\\
\cd
\end{pmatrix}=\frac{1}{2}\left(\begin{matrix}-\Gamma_{{\rm BG}} & -\Omega_{{\rm DG}}\\
\Omega_{{\rm DG}} & -\gamma_{{\rm D}}
\end{matrix}\right)\begin{pmatrix}\cg\\
\cd
\end{pmatrix},
\end{equation}
and, in the limit $\gamma_{{\rm D}}\ll\Gamma_{{\rm BG}}$, the equation
of motion for $W_{{\rm DG}}$, defined in Eq.~(\ref{eqn:W_DG_definition}),
is
\begin{equation}
\frac{\mathrm{d}W_{{\rm DG}}}{\mathrm{d}\Delta t_{{\rm catch}}}=\frac{\Gamma_{{\rm BG}}}{2}W_{{\rm DG}}+\frac{\Omega_{{\rm DG}}}{2}(1+W_{{\rm DG}}^{2}),\label{eqn:W_DG_equation_incoherent}
\end{equation}
with solution, for $W_{{\rm DG}}(0)=0$, 
\begin{equation}
W_{{\rm DG}}(\Delta t_{{\rm catch}})=\frac{\exp\left[(V-V^{-1})\Omega_{{\rm DG}}\Delta t_{{\rm catch}}/2\right]-1}{V-V^{-1}\exp\left[(V-V^{-1})\Omega_{{\rm DG}}\Delta t_{{\rm catch}}/2\right]},\label{eqn:W_DG_incoherent}
\end{equation}
where 
\begin{equation}
V=\frac{1}{2}\frac{\Gamma_{{\rm BG}}}{\Omega_{{\rm DG}}}+\sqrt{\frac{1}{4}\left(\frac{\Gamma_{{\rm BG}}}{\Omega_{{\rm DG}}}\right)^{2}-1}.
\end{equation}
In Ref.~\citet{Ruskov2007}, a general form of the Bloch vector equations
for arbitrary amplitude of the Rabi drive was found. Inversion of
the condition $W_{{\rm DG}}(\Delta t_{{\rm mid}})=1$ gives the characteristic
time scale 
\begin{equation}
\Delta t_{{\rm mid}}=2\left[(V-V^{-1})\Omega_{{\rm DG}}\right]^{-1}\ln\left(\frac{V+1}{V^{-1}+1}\right).\label{eqn:t_mid_incoherent}
\end{equation}
Although Eqs.~(\ref{eqn:W_DG_incoherent}) and (\ref{eqn:t_mid_incoherent})
replace Eqs.~(\ref{eq:WDG-coh}) and (\ref{eq:t_mid_coherent}),
under strong monitoring, $\Gamma_{{\rm BG}}\gg\Omega_{{\rm DG}}$,
they revert to the latter with the substitution $\Omega_{{\rm BG}}^{2}/2\gamma_{{\rm B}}\to\Gamma_{{\rm BG}}/2$,;
in this way, Eqs.~(\ref{eqn:Z})–(\ref{eqn:Y}) are recovered with
the same substitution. More generally, $W_{{\rm DG}}(\Delta t_{{\rm catch}})$
goes to infinity at finite $\Delta t_{{\rm catch}}$, changes sign,
and returns from infinity to settle on the steady value $W_{{\rm DG}}(\infty)=-V$.
The singular behavior marks a trajectory passing through the north
pole of Bloch sphere. It yields the long-time solution 
\begin{equation}
{\rm Z}_{{\rm GD}}(\infty)=\sqrt{1-4\left(\frac{\Omega_{{\rm DG}}}{\Gamma_{{\rm BG}}}\right)^{2}},\qquad{\rm X}_{{\rm GD}}(\infty)=-2\frac{\Omega_{{\rm DG}}}{\Gamma_{{\rm BG}}},\qquad{\rm Y}_{{\rm GD}}(\infty)=0,\label{eq:ZGDXGD-long-term}
\end{equation}
in contrast to the perfect jump of Eqs.~(\ref{eqn:Z_approx})–(\ref{eqn:Y_approx}).
The long-term coherence and lower-than-one value of Z were observed
in the experiments, see Fig.~\ref{fig:catch}b. They can be understood
as the equilibrium point of the coherent Rabi drive $\Omega_{{\rm DG}}$
attempting to rotate the state from $\D$ back to $\G$ perfectly
balanced against the measurement-backaction force of the no-click
measurement steering the atom toward $\D$, recall discussion of the
measurement vector field, see Figs.~\ref{fig:Knowledge-driven-force-in-1}
and~\ref{fig:Knowledge-driven-force-in-2}. 

\paragraph{\textit{\emph{Dark drive off.}}}

Turing the Dark drive off shortly after a click demonstrates the connection
between the flight of a quantum jump and a projective measurement.
From the point of view of the trajectory equations, the only change
is the setting of $\Omega_{{\rm DG}}$ to zero at time $\Delta t_{{\rm on}}$
on the right-hand side of Eqs.~(\ref{eqn:W_DG_equation_coherent})
and (\ref{eqn:W_DG_equation_incoherent}). Subsequently, $W_{{\rm DG}}(\Delta t_{{\rm catch}})$
continues its exponential growth at rate $\Omega_{{\rm BG}}^{2}/2\gamma_{B}$
{[}Eq.~(\ref{eqn:W_DG_equation_coherent}){]} or $\Gamma_{{\rm BG}}/2$
{[}Eq.~(\ref{eqn:W_DG_equation_incoherent}){]}. Equations (\ref{eqn:Z})–(\ref{eqn:Y})
for the GD Bloch components still hold, but now with 
\begin{equation}
\Delta t_{{\rm mid}}=\left(\frac{\Omega_{{\rm BG}}^{2}}{2\gamma_{B}},\frac{\Gamma_{{\rm BG}}}{2}\right)^{-1}\ln\big[W_{{\rm DG}}^{-2}(\Delta t_{{\rm on}})\big].
\end{equation}
which can provide an estimate of $\Delta t'_{\mathrm{mid}}$, specifying
the time at which $Z_{\mathrm{GD}}=0$.

The evolution during $\Delta t_{\mathrm{off}}$, in the absence of
$\Omega_{\mathrm{DG}}$, in effect realizes a projective measurement
of whether the state of the atom is $\G$ or $|{\rm D}\rangle$, similar
to the one analyzed in Sec.~\ref{subsec:Knowledge-driven-force-in},
where the normalized state at $\Delta t_{{\rm on}}$ is 
\begin{equation}
\frac{|\psi(\Delta t_{{\rm on}})\rangle}{\sqrt{\mathcal{N}(\Delta t_{{\rm on}})}}=\frac{\cg(\Delta t_{{\rm on}})\G+\cd(\Delta t_{{\rm on}})\D}{\sqrt{\mathcal{N}(\Delta t_{{\rm on}})}},\label{eqn:initial_state}
\end{equation}
with $\mathcal{N}(\Delta t_{{\rm on}})=\cg^{2}(\Delta t_{{\rm on}})+\cd^{2}(\Delta t_{{\rm on}})$
the probability for the jump to reach $\Delta t_{{\rm catch}}=\Delta t_{{\rm on}}$
after a click reset to $\G$ at $\Delta t_{{\rm catch}}=0$. The probability
for the jump to continue to $\Delta t_{{\rm catch}}>\Delta t_{{\rm on}}$
(given $\Delta t_{{\rm on}}$ is reached) is then 
\begin{equation}
\frac{\mathcal{N}(\Delta t_{{\rm catch}})}{\mathcal{N}(\Delta t_{{\rm on}})}=\frac{C_{{\rm D}}^{2}(\Delta t_{{\rm on}})}{\mathcal{N}(\Delta t_{{\rm on}})}+\frac{C_{{\rm G}}^{2}(\Delta t_{{\rm on}})}{\mathcal{N}(\Delta t_{{\rm on}})}\exp\left[-\left(\frac{\Omega_{{\rm BG}}^{2}}{\gamma_{{\rm B}}},\Gamma_{{\rm BG}}\right)\Delta t_{{\rm catch}}\right].\label{eq:completion-prob}
\end{equation}

\paragraph{Completed and aborted evolutions of the jump transition.}

In this simple model, the probability for the trajectory to complete
— for the measurement to yield the result $|{\rm D}\rangle$ — is
obtained in the limit $\Delta t_{\mathrm{catch}}\rightarrow\infty$,
and, as expected, is equal to the probability to occupy the state
$|{\rm D}\rangle$ at time $\Delta t_{{\rm on}}$; i.e., the completion
probability is $P_{\mathrm{D}}(\Delta t_{{\rm on}})=C_{{\rm D}}^{2}(\Delta t_{{\rm on}})/{{\rm Norm}(\Delta t_{{\rm on}})}$.
It is helpful to illustrate this idea with an example. Consider the
catch experiment of Fig.~\ref{fig:catch}b in the absence of the
Dark Rabi drive, $\Omega_{\mathrm{DG}}$. From $Z_{\mathrm{GD}}$,
we can estimate that out of all the trajectories that pass though
the $\Delta t_{{\rm on}}$ mark approximately $P_{\mathrm{D}}(\Delta t_{{\rm on}})=(1+Z_{\mathrm{GD}}(\Delta t_{{\rm on}}))/2\approx8\%$
fully complete without an interruption. On the other hand, for those
that pass the $\Delta t'_{\mathrm{mid}}$ mark, approximately 50\%
complete. It follows from Eq.~(\ref{eq:completion-prob}), that the
probability of the evolution to complete increases the further along
the trajectory is. Although some of the jump evolutions abort at random,
importantly, every single jump evolution that completes, and is thus
recorded as a jump, follows \textit{not} a random but an identical
path in Hilbert space, i.e., a deterministic one. This path (of \textit{any}
jump) is determined by Eq.~(\ref{eqn:W_DG_incoherent}), or, in the
simpler model, by the Eqs.~(\ref{eqn:Z_approx})-(\ref{eqn:Y_approx})
for the components of the GD Bloch vector.

\section{Heterodyne monitoring of readout cavity coupled to three-level atom\label{sec:Heterodyne-monitoring-of}}

\subsection{Description of cQED experiment \label{subsec:Description-of-cQED}}

In Chapter 1, we described the cQED experiment  involving a superconducting
atom with the necessary V-shape level structure (see Fig.~\ref{fig:setup}a
or Sec.~\ref{sec:circuit-design}) subject to heterodyne monitoring
of $\B$ by means a dispersively coupled readout cavity. The three-level
atom is formed form two heavily hybridized transmon modes, which are
coupled by means of a cross-Kerr interaction to the readout cavity
in an asymmetric way, $\chi_{\mathrm{BC}}\gg\chi_{\mathrm{DC}}$.
In the following, we present the quantum trajectory description of
the heterodyne monitoring, including imperfections.

\paragraph{System Hamiltonian. }

In the lab frame, the Hamiltonian of the system is, see also Sec.~\ref{sec:circuit-design},
\begin{equation}
\hat{H}_{{\rm lab}}=\hat{H}_{0}+\hat{H}_{I}+\hat{H}_{d}\left(t\right)\,,
\end{equation}
where the Hamiltonian of the uncoupled three-level atom and cavity,
is 
\begin{equation}
\hat{H}_{0}=\hbar\omega_{{\rm DG}}\kb{\rm D}{\rm D}+\hbar\omega_{{\rm BG}}\kb{\rm B}{\rm B}+\hbar\omega_{C}\hat{c}^{\dagger}\hat{c}\,,
\end{equation}
where $\omega_{{\rm DG}}$, $\omega_{{\rm BG}}$, $\omega_{{\rm C}}$
are the Dark, Bright, and cavity mode frequency, respectively, $\hat{c}$
is the cavity amplitude operator, the atom-cavity interaction Hamiltonian
is 
\begin{equation}
\hat{H}_{I}=\hat{c}^{\dagger}\hat{c}\left[\hbar\chi_{{\rm B}}\kb{\rm B}{\rm B}+\hbar\chi_{{\rm D}}\kb{\rm D}{\rm D}\right]\:,
\end{equation}
where the shift of the cavity frequency conditioned on $\B$ ($\D$)
is $\chi_{{\rm B}}$ ($\chi_{{\rm D}}$), and the Hamiltonian of the
atom Rabi drives and readout probe tone is
\begin{align}
\hat{H}_{d}\left(t\right)= & -\frac{i\hbar}{2}\left[\kappa\sqrt{\bar{n}}\hat{c}e^{i\left(\omega_{{\rm C}}+\Delta_{{\rm R}}\right)t}+\Omega_{{\rm DG}}^{*}e^{i\left(\omega_{{\rm DG}}+\Delta_{{\rm DG}}\right)t}\kb{\rm G}{\rm D}\right.\nonumber \\
 & \left.+\Omega_{{\rm B0}}^{*}e^{i\omega_{{\rm BG}}t}\kb{\rm G}{\rm B}+\Omega_{{\rm B1}}^{*}e^{i\left(\omega_{{\rm BG}}+\Delta_{{\rm B1}}\right)t}\kb{\rm G}{\rm B}+\mathrm{H.c.}\right]\,,
\end{align}
where $\bar{n}$ is the steady state number of photons in the cavity
when driven resonantly, $\Delta_{{\rm R}}$, $\Delta_{{\rm DG}},$
and $\Delta_{{\rm B1}}$ are the drive detunings from the bare mode
frequencies. The first Bright Rabi tone, $\Omega_{{\rm B0}}$, addresses
the Bright transition when the cavity is unpopulated, while the second
tone, $\Omega_{{\rm B0}}$, addresses the BG transition  when the
cavity is populated, see frequency spectrum in Fig.~\ref{fig:setup}c.
Moving to the rotating frame at the drive frequencies, defined by
the ket transformation $\ket{\psi(t)}=U(t)\ket{\psi_{\mathrm{lab}}(t)}$,
where $U(t)=\exp\left(u\left(t\right)/i\hbar\right)$ and $u(t)=\hbar t\left[\left(\omega_{{\rm C}}+\Delta_{{\rm R}}\right)a^{\dagger}a+\omega_{{\rm BG}}\ket{\rm B}\bra{\rm B}+\left(\omega_{{\rm DG}}+\Delta_{{\rm DG}}\right)\ket{\rm D}\bra{\rm D}\right]$,
the Hamiltonian in the rotating frame is
\begin{equation}
\hat{H}\left(t\right)=\hat{H}_{{\rm R}}+\hat{H}_{{\rm drive}}\left(t\right)\,,
\end{equation}
where $\hat{H}_{\mathrm{R}}$ is a time-independent Hamiltonian,
\begin{equation}
\hat{H}_{{\rm R}}=-\hbar\Delta_{{\rm R}}\hat{c}^{\dagger}\hat{c}+i\hbar\frac{\kappa}{2}\sqrt{\bar{n}}(\hat{c}^{\dagger}-\hat{c})+\hbar\big(\chi_{{\rm B}}\kb{\rm B}{\rm B}+\chi_{{\rm D}}\kb{\rm D}{\rm D}\big)\hat{c}^{\dagger}\hat{c},
\end{equation}
and $\hat{H}_{{\rm drive}}$ is the time-dependent Hamiltonian of
the atom Rabi drives, 
\begin{equation}
\hat{H}_{{\rm drive}}\left(t\right)=i\hbar\left[\frac{\Omega_{{\rm BG}}(t)}{2}\kb{\rm B}{\rm G}-\frac{\Omega_{{\rm BG}}^{*}(t)}{2}\kb{\rm G}{\rm B}\right]+i\hbar\frac{\Omega_{{\rm DG}}}{2}\left(\kb{\rm D}{\rm G}-\kb{\rm G}{\rm D}\right).
\end{equation}
The bi-chromatic drive, which unselectively addresses the BG transition,
$\Omega_{{\rm BG}}(t)=\Omega_{{\rm B}0}+\Omega_{{\rm B}1}\exp(-i\Delta_{{\rm B}1}t)$
replaces the Rabi drive $\Omega_{{\rm BG}}$ of Eq.~(\ref{eq:drive_coherent_fluorescence}).

\paragraph{Measurement record.}

The readout cavity input-output coupling is given by the jump operator
$\sqrt{\kappa}\hat{c}$. It follows from Eq.~(\ref{eq:dJimperf:het})
that the heterodyne measurement-record increment is
\begin{equation}
\mathrm{d}J_{\mathrm{het}}\left(t\right)=\sqrt{\eta\kappa}\left\langle \hat{c}\right\rangle \left(t\right)\dt+\dZ\left(t\right)\,,\label{eq:Record}
\end{equation}
where $\dZ$ is a complex Wiener increment, see discussion below Eq.~(\ref{eq:Jhet-perfect}),
and $\eta$ is the quantum efficiency of the readout and amplification
chain. The record, $\mathrm{d}J_{\mathrm{het}}\left(t\right)$, is
scaled — to units of (readout cavity photon number)$^{1/2}$ — and
filtered to generate the simulated quadratures $I_{{\rm rec}}$ and
$Q_{{\rm rec}}$ of the measurement record: 
\begin{eqnarray}
dI_{{\rm rec}} & = & -\frac{\kappa_{{\rm filter}}}{2}\left[I_{{\rm rec}}dt-\left(\eta\frac{\kappa}{2}\right)^{-1/2}{\rm Re}(\mathrm{d}J_{\mathrm{het}})\right],\label{eqn:simulated_I_int}\\
dQ_{{\rm rec}} & = & -\frac{\kappa_{{\rm filter}}}{2}\left[Q_{{\rm rec}}dt-\left(\eta\frac{\kappa}{2}\right)^{-1/2}{\rm Im}(\mathrm{d}J_{\mathrm{het}})\right],\label{eqn:simulated_Q_int}
\end{eqnarray}
where $\kappa_{{\rm filter}}$ is the bandwidth of the amplifier chain.
In practice, it is assured that $\kappa_{{\rm filter}}$ is the fastest
rate in the problem, $\kappa_{{\rm filter}}\gg\kappa$, so that its
effect is largely negligible.  

\subsection{Simulation of Stochastic Schrödinger equation (SSE) \label{subsec:Simulation-of-linear}}

The quantum trajectory unraveling monitors the reflected probe with
efficiency $\eta$ and accounts for residual photon loss through random
jumps. It follows that the linear stochastic Schrödinger equation
combines a continuous evolution (heterodyne readout channel), 
\begin{equation}
{\rm d}\ket{\psi\left(t\right)}=\left[\frac{1}{i\hbar}\left(\hat{H}_{{\rm drive}}+\hat{H}_{{\rm R}}-i\hbar\frac{\kappa}{2}\hat{c}^{\dagger}\hat{c}\right)\dt+\sqrt{\eta\kappa}\mathrm{d}J_{\mathrm{het}}^{*}\left(t\right)\hat{c}\right]\ket{\psi\left(t\right)},\label{eqn:SSE_continuous}
\end{equation}
with the point-like process (photon loss), 
\begin{equation}
\ket{\psi}\to\hat{c}\ket{\psi}\qquad\text{at rate}\qquad(1-\eta)\kappa\frac{\braOket{\psi}{\hat{c}^{\dagger}\hat{c}}{\psi}}{\braket{\psi}{\psi}}.
\end{equation}
Note that for perfect quantum efficiency, $\eta=1$, the rate of the
photon loss channel goes to zero. We emphasize that expectation values
are performed over the normalized state; importantly, when calculating
the measurement record increment $\mathrm{d}J_{\mathrm{het}}^{*}\left(t\right)$,
see Eq.~(\ref{eq:Record}), $\left\langle \hat{c}\right\rangle \left(t\right)=\braOket{\psi\left(t\right)}{\hat{c}}{\psi\left(t\right)}/\braket{\psi\left(t\right)}{\psi\left(t\right)}.$

\paragraph{\textit{\emph{Independently measured imperfections.}}\emph{ }}

To more realistically model the cQED experiment, we need to account
for the small experimental non-idealities associated with the device
performance; namely, the finite energy relaxation lifetime of the
levels ($T_{1}$), the finite dephasing time of the levels ($T_{2}^{*}$),
which is generally smaller than the bound imposed by the lifetime,
$T_{2}^{*}<T_{1}$, and the finite temperature of the device ($n_{{\rm th}}$).
Specifically, we supplement the stochastic Schrödinger equation by
spontaneous and thermal jumps on both the $\G$ to $\B$ and $\G$
to $\D$ transitions ($n_{{\rm th}}^{{\rm B}}$ and $n_{{\rm th}}^{{\rm D}}$)
and by pure dephasing of the ${\rm GB}$ and ${\rm GD}$ coherences
($\gamma_{{\rm B}}^{\phi}$ and $\gamma_{{\rm D}}^{\phi}$). With
these processes included, the term 
\[
-i\hbar\left\{ \left[\frac{\gamma_{{\rm B}}}{2}(n_{{\rm th}}^{{\rm B}}+1)+\gamma_{{\rm B}}^{\phi}\right]\kb{\rm B}{\rm B}+\left[\frac{\gamma_{{\rm D}}}{2}(n_{{\rm th}}^{{\rm D}}+1)+\gamma_{{\rm D}}^{\phi}\right]\kb{\rm D}{\rm D}+\frac{\gamma_{{\rm B}}n_{{\rm th}}^{{\rm B}}+\gamma_{{\rm D}}n_{{\rm th}}^{{\rm D}}}{2}|\kb{\rm G}{\rm G}\right\} 
\]
is added to the non-Hermitian Hamiltonian, $\hat{H}_{{\rm drive}}+\hat{H}_{{\rm R}}-i\hbar\frac{\kappa}{2}\hat{c}^{\dagger}\hat{c}$,
on the right-hand side of Eq.~(\ref{eqn:SSE_continuous}), with the
additional three point-processes: 
\begin{align}
|\psi\rangle\to|{\rm G}\rangle & \quad\text{at rate}\quad\gamma_{{\rm B}}(n_{{\rm th}}^{{\rm B}}+1)\frac{\langle\psi|{\rm B}\rangle\langle{\rm B}|\psi\rangle}{\langle\psi|\psi\rangle}+\gamma_{{\rm D}}(n_{{\rm th}}^{{\rm D}}+1)\frac{\langle\psi|{\rm D}\rangle\langle{\rm D}|\psi\rangle}{\langle\psi|\psi\rangle},\\
\noalign{\vskip4pt}|\psi\rangle\to|{\rm B}\rangle & \quad\text{at rate}\quad\gamma_{{\rm B}}n_{{\rm th}}^{{\rm B}}\frac{\langle\psi|{\rm G}\rangle\langle{\rm G}|\psi\rangle}{\langle\psi|\psi\rangle}+2\gamma_{{\rm B}}^{\phi}\frac{\langle\psi|{\rm B}\rangle\langle{\rm B}|\psi\rangle}{\langle\psi|\psi\rangle},\\
\noalign{\vskip4pt}|\psi\rangle\to|{\rm D}\rangle & \quad\text{at rate}\quad\gamma_{{\rm D}}n_{{\rm th}}^{{\rm D}}\frac{\langle\psi|{\rm G}\rangle\langle{\rm G}|\psi\rangle}{\langle\psi|\psi\rangle}+2\gamma_{{\rm D}}^{\phi}\frac{\langle\psi|{\rm D}\rangle\langle{\rm D}|\psi\rangle}{\langle\psi|\psi\rangle}.
\end{align}
In the simulation, the parameters $\gamma_{{\rm B,D}}$, $n_{{\rm th}}^{{\rm B,D}}$,
and $\gamma_{{\rm B,D}}^{\phi}$ are mapped to the independently measured
parameters $T_{{\rm B,D}}^{1}$, $n_{{\rm th}}^{{\rm G,D}}$, and
$T_{2{\rm R}}^{{\rm B,D}}$ listed in Table \ref{fig:T1-vs-nbar}.

\paragraph{\textit{\emph{Leakage from the }}\emph{GBD}\textit{\emph{-manifold.}}\emph{ }}

Because the three-level atom is realized from two transmon qubits,
the three-state manifold, $\left\{ \G,\B,\D\right\} $, is not strictly
closed, and transitions to higher excited states are sometimes observed.
This imperfection is modeled in the SSE simulation with the addition
of the further term 
\[
-i\hbar\left\{ \frac{\gamma_{{\rm FG}}}{2}|{\rm G}\rangle\langle{\rm G}|+\frac{\gamma_{{\rm FD}}}{2}|{\rm D}\rangle\langle{\rm D}|+\frac{\gamma_{{\rm GF}}+\gamma_{{\rm DF}}}{2}|{\rm F}\rangle\langle{\rm F}|\right\} 
\]
to the non-Hermitian Hamiltonian, and the associated additional random
jumps, 
\begin{eqnarray}
|\psi\rangle\to|{\rm F}\rangle &  & \qquad\hbox{at\ rate}\qquad\gamma_{{\rm FG}}\frac{\langle\psi|{\rm G}\rangle\langle{\rm G}|\psi\rangle}{\langle\psi|\psi\rangle}+\gamma_{{\rm FD}}\frac{\langle\psi|{\rm D}\rangle\langle{\rm D}|\psi\rangle}{\langle\psi|\psi\rangle},\\
\noalign{\vskip4pt}|\psi\rangle\to|{\rm G}\rangle &  & \qquad\hbox{at\ rate}\qquad\gamma_{{\rm GF}}\frac{\langle\psi|{\rm F}\rangle\langle{\rm F}|\psi\rangle}{\langle\psi|\psi\rangle},\\
\noalign{\vskip4pt}|\psi\rangle\to|{\rm D}\rangle &  & \qquad\hbox{at\ rate}\qquad\gamma_{{\rm DF}}\frac{\langle\psi|{\rm F}\rangle\langle{\rm F}|\psi\rangle}{\langle\psi|\psi\rangle},\label{eqn:jumps_to_F}
\end{eqnarray}
where $|{\rm F}\rangle$ models the all higher level by a single catch-all
higher excited state. The results of the simulation are presented
in Sec.~\ref{subsec:Comparison-between-theory}.

\chapter{Experimental methods\label{chap:Experimental-Methods} }

\addcontentsline{lof}{chapter}{Experimental methods\lofpost}  
  
\singlespacing
\epigraph{ 
If I knew what I was doing, it wouldn't be called research.}    
{Albert Einstein\\
See \citet{hawken2010natural}}   
\doublespacing\noindent  \noindent\lettrine{T}{he} design of the superconducting three-level
atom and readout cavity is presented in Sec.~\ref{sec:circuit-design}.
It was optimized subject to the constrains of the experiment (Hamiltonian
and dissipative) with the energy-participation ratio (EPR) approach,
presented in Sec.~\ref{subsec:Energy-participation-ratio}. The methodology
of the finite-element numerical simulations employed to engineer the
electromagnetic (EM) properties of the distributed circuit is presented
in Secs.~\ref{subsec:Calculation-of-EPR} and~\ref{subsec:Calculation-of-Hamiltonian}.
Sample fabrication is discussed in Sec.~\ref{sec:Fabrication-of-sample},
while design and assembly of the sample holder are discussed in Sec.~\ref{sec:Sample-holder-materials}.
Particular attention is paid to material selection, a care continued
in Sec.~\ref{sec:Cryogenic-setup}, where aspects of the cryogenic
setup of the experiment are discussed, including sample thermalization,
surface preparation, light-tightness, and magnetic shielding. The
microwave setup of the experiment is discussed in Sec.~\ref{sec:Microwave-setup}.
For further information on experimental methods employed in circuit
quantum electrodynamics (cQED) experiments see Refs.~\citet{Geerlings2013,Reed2013,Reagor2016,Brecht2017-Thesis}.

\section{Sample design \label{sec:circuit-design}}

\paragraph{Overview. }

\begin{figure}
\centering{}\includegraphics[scale=1.4]{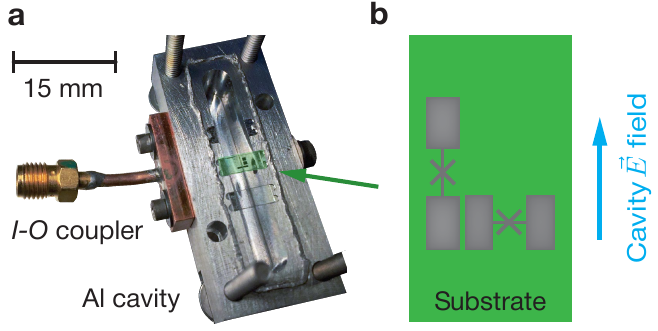}
\caption[Sample and chip layout]{\textbf{Sample and chip layout.\label{fig:Sample-design-assembly}}
\textbf{a,} Photograph of Darkmon chip ($2.9\times7$~mm, sapphire)
in the aluminum (Al) cavity, which serves as the sample holder, shown
with upper half removed. Green arrow points to location of the chip.
Also visible: input-output (\emph{I-O}\textbf{) }SMA coupler and frequency
tuning screw (right side). \textbf{ b, }Not-to-scale schematic representation
of the Bright (vertical) and Dark (horizontal) transmon qubits. Vertical
blue arrow indicates the orientation of the electric field of the
fundamental ($\mathrm{TE}_{\mathrm{101}}$) cavity mode. }
\end{figure}

The superconducting artificial atom presented in Sec.~\ref{sec:Principle-of-the},
see Fig.~\ref{fig:Sample-design-assembly}a, consists of two coupled
transmon qubits \citep{Koch2007,Schreier2008-transmon,Paik2011} fabricated
on a 2.9~mm-by-7~mm double-side-polished c-plane sapphire wafer
with the Al/Al$\text{O}_{\text{x}}$/Al bridge-free electron-beam
lithography technique \citep{Lecocq2011-bridge-free,Rigetti2009};
for fabrication methodology, see Sec.~\ref{sec:Fabrication-of-sample}.
The first transmon (B) is aligned with the electric field of the fundamental
$\mathrm{TE}_{\mathrm{101}}$ mode of the aluminum rectangular cavity
(alloy 6061; dimensions: 5.08~mm by 35.5~mm by 17.8~mm), while
the second transmon (D) is oriented perpendicular to the first and
positioned $170\,\mathrm{\mu m}$ adjacent to it, see Fig.~\ref{fig:Sample-design-assembly}b.
The inductance of the Josephson junction of each transmon (nominally,
9~nH for both B and D), the placement and dimensions of each transmon,
and the geometry of the cavity were designed and optimized using finite-element
electromagnetic analysis and the energy-participation-ratio (EPR)
method\footnote{Z.K. Minev \emph{et al.}, in preparation.}, as discussed
in Sec.~\ref{subsec:Energy-participation-ratio}. 

\paragraph{Hamiltonian and level diagram.}

\begin{figure}
\centering{}\includegraphics[scale=1.7]{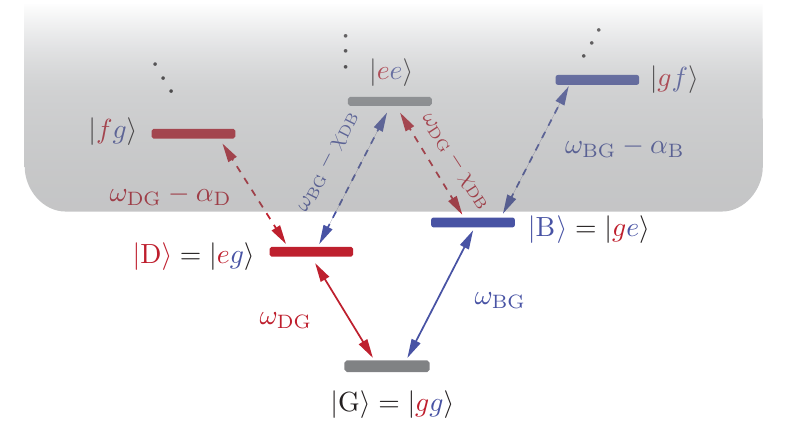}
\caption[Darkmon energy-level diagram]{\textbf{Darkmon energy-level diagram.\label{fig:Darkmon-energy-level-diagram}}
Energy level diagram of the hybridized Dark and Bright transmon qubits.
Red (Blue) color denotes association with the Dark (Bright) transmon,
while grey denotes strong association with both transmons. The strong
non-linear, dispersive interactions in the circuit, self-Kerr ($\alpha_{{\rm B,D}}$)
and cross-Kerr ($\chi_{{\rm DB}}$), allow the lowest-lying three
levels to be isolated, and for the two qubit system to be employed
as a three-level one with a V-shape structure. }
\end{figure}

Under the rotating-wave approximation and in the low-excitation limit,
see Sec.~\ref{subsec:Calculation-of-Hamiltonian}, the effective
Hamiltonian of the device, consisting of the Dark, Bright, and cavity
modes, is energy conserving, 
\begin{align}
\hat{\overline{H}}/\hbar= & \omega_{{\rm D}}\hat{n}_{\emph{{\rm D}}}+\hbar\omega_{{\rm B}}\hat{n}_{{\rm B}}+\hbar\omega_{{\rm C}}\hat{n}_{{\rm C}}\nonumber \\
 & -\frac{1}{2}\alpha_{{\rm D}}\hat{n}_{\emph{{\rm D}}}\left(\hat{n}_{\emph{{\rm D}}}-\hat{1}\right)-\frac{1}{2}\alpha_{{\rm B}}\hat{n}_{{\rm B}}\left(\hat{n}_{{\rm B}}-\hat{1}\right)\nonumber \\
 & +\chi_{{\rm DB}}\hat{n}_{\emph{{\rm D}}}\hat{n}_{{\rm B}}+\chi_{{\rm DC}}\hat{n}_{\emph{{\rm D}}}\hat{n}_{{\rm C}}+\chi_{{\rm BC}}\hat{n}_{{\rm B}}\hat{n}_{{\rm C}}\,,\label{eq:HDarkmon}
\end{align}
where $\hat{n}_{{\rm D,B,C}}$ are the Dark, Bright, and cavity photon-number
operators, $\alpha_{{\rm D,B}}$ are the Dark and Bright qubit anharmonicities,
also referred to as self-Kerr frequencies, $\chi_{{\rm DB}}$ is the
dispersive cross-Kerr frequency shift between the Dark and Bright
modes, while $\chi_{{\rm DC,BC}}$ are the dispersive shifts between
the cavity and the two transmons. The energy level structure of the
two-transmon composite system is schematically represented in Fig.~\ref{fig:Darkmon-energy-level-diagram}.
When the anharmonicities, $\alpha_{{\rm B}}$ and $\alpha_{{\rm D}},$
are relatively large, typically in the range 100 to 300 MHz, see Table~\ref{tab:system-params}
for device parameters, the level structure becomes sufficiently anharmonic
and we can restrict our attention to the manifold of the four lowest
energy states, $\left\{ \ket{gg},\ket{eg},\ket{ge},\ket{ee}\right\} $,
where the first (second) letter refers to the Dark (Bright) transmon.
When the two qubits are uncoupled, $\chi_{{\rm DB}}=0$, the transitions
among the levels contain degeneracies, and $\ket{ge}$ and $\ket{eg}$
cannot be addressed individually. In the limit where the coupling,
$\chi_{{\rm DB}}$, is large, in practice, on the order of 100 MHz,
the degeneracy is lifted, and the $\ket{ge}$ and $\ket{eg}$ states
become independent, allowing us to further restrict our attention
to the three lowest-lying states. We label $\ket{gg}$ simply as $\G$,
$\ket{eg}$ as $\D$, and $\ket{ge}$ as $\B.$ In reference to the
protected Dark state, $\D$, which is engineered to be decoupled from
the environment and readout cavity, we  nickname the device ``Darkmon.''

\paragraph{Unique design constraints. }

In addition to the required large transmon anharmonicity, $\alpha_{{\rm B}}$,
$\alpha_{{\rm D}}$, and cross-Kerr, $\chi_{{\rm DB}},$ frequencies,
a few somewhat unique, decoherence related, constraints were required,
notably, at odds with the large couplings. First, catching and reversing
the quantum jump from $\G$ to $\D$ coherently and with high fidelity
required five orders of magnitude in timescales, see Table~\ref{tab:Summary-of-timescales.},
thus imposing the constraint that the $\D$ level coherences be minimally
at the $100\ \mathrm{\mu s}$ level, both the energy-relaxation and
dephasing times, $T_{1}^{{\rm {\rm D}}},T_{{\rm 2R}}^{{\rm D}}\geq100\,\mathrm{\mu s}$.
The regime of long energy relaxation, $T_{1}^{{\rm {\rm D}}},$ is
accessible with the state-of-the-art Purcell-filtered three-dimensional
(3D) transmon qubits \citep{Paik2011,Wang2015,Dial2016}, but long
quantum coherences, $T_{{\rm 2R}}^{{\rm D}}$, are far more difficult
to achieve, and are generally obtained by decoupling the transmon
qubit from the readout cavity \citep{Gambetta2006-dephasing,Rigetti2012},
thus making a tradeoff between quantum coherence and the ability to
perform a fast readout. In the quantum jumps experiment, this tradeoff
is not permissible, a fast readout of the $\D$ is required simultaneously
with long coherences. 

To maximize the coherence properties of $\D$, we designed the Dark
transmon to be decoupled from all dissipative environments, including
the readout cavity and input-output (I-O) coupler. Removing the coupling
between $\D$ and the cavity is advantageous in three important ways:
it protects $\D$ from i) dephasing due to cavity photon shot noise
\citep{Gambetta2006-dephasing,Wang2019-cav-atten,Wang2018-cav-atten-aps},
ii) energy relaxation through the cavity by means of the Purcell effect
\citep{Gambetta2011-Purcell,Srinivasan2011,Diniz2013,Dumur2015,Novikov2015,Zhang2017,Roy2017-3qubits},
and iii) measurement-induced energy relaxation \citep{Boissonneault2009-Photon-induced-relax,Slichter2016-T1vsNbar},
see Fig.~(\ref{fig:T1-vs-nbar}). Decoupling $\D$ might seem like
a tradeoff at first, since $\D$ can no longer be directly measured
through the cavity. However, the strong coupling between the Dark
transmon and the Bright transmon, together with the special nature
of the V-shape level structure and the $\B$/not-$\B$ dispersive
readout, can be employed to nonetheless achieve a fast and faithful
readout, see Sec.~\ref{subsec:Tomography-of-three-level}. The associated
challenge is that when two transmon qubits are strongly coupled, the
D level needs to remain otherwise isolated and coherent at the same
time that the B level is strongly coupled to the low-quality (low-Q)
cavity. The coupling between $\B$ and the cavity is necessitated
to yield a large dispersive shift, $\chi_{{\rm BC}},$ used to realize
the $\B$/not-$\B$ readout; however, this coupling is accompanied
by a degree of energy relaxation by means of the Purcell effect inherited
by $\B$ \citep{Gambetta2011-Purcell}. The dissipation in $\B$ can
in turn be inherited by the coupled state $\D$, due to the hybridization
between the two transmons, $\chi_{{\rm DB}}$, if the design is not
carefully optimized. 

\begin{figure}
\centering{}\includegraphics[scale=1.3]{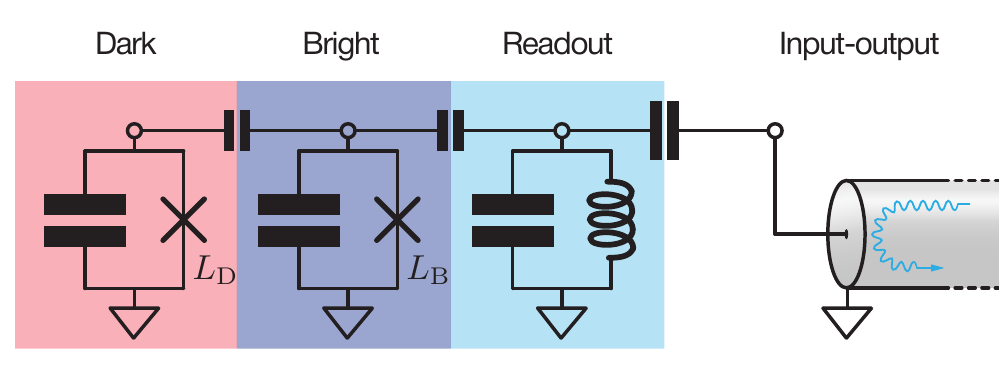}
\caption[Effective circuit model of Darkmon system]{\textbf{Effective circuit model of Darkmon system.\label{fig:Darkmon-circuit}}
Dark and Bright transmons represented as lumped-element junction-capacitor
circuits, junction denoted by cross, coupled an LC circuit representing
the readout cavity. $L_{\mathrm{D,B}}$ denote the Dark and Bright
Josephson inductances, corresponding to the horizontal and vertical
junctions in Fig.~\ref{fig:Sample-design-assembly}, respectively.}
\end{figure}

On a conceptual level, the Darkmon device and the couplings among
the levels can be understood in terms of an effective circuit model,
see Fig.~\ref{fig:Darkmon-circuit}, where each mode is represented
by a single LC oscillator, with the two qubits having the inductors
replaced by non-linear Josephson tunnel junctions. The Dark resonator
is capacitively coupled to the Bright one, which is capacitively coupled
to the readout resonator, which is capacitively coupled to the input-output
transmission line. The Bright and readout resonators can be seen to
act as a two-pole filter shielding the Dark resonator from the dissipative
effect of the transmission line. While the model is conceptually useful
to analyze the qualitative behavior of the circuit, it cannot produce
reliable quantitative results. Instead, engineering the highly asymmetric
set of couplings while isolating the $\D$ level was achieved by means
of an iterative search over the design geometry  with the energy participation
ratio (EPR) approach. In the following, we briefly summarize the methodology. 

\subsection{Energy-participation-ratio (EPR) approach\label{subsec:Energy-participation-ratio}}

The design of distributed circuits with the aim of obtaining a desired
Hamiltonian and set of environmental couplings has attracted a lot
of interest \citep{Nigg2012,Bourassa2012,Solgun2014,Solgun2015,Smith2016},
but a general solution to this inverse problem appears to be out of
reach. Instead, one applies a search algorithm over the direct problem
— a circuit is chosen, the non-linear mixing and dissipation parameters
are calculated, the circuit is modified, and the process is repeated
in search of the target parameters. A broadly-applicable approach
based on the concept of the energy-participation ratio (EPR) of the
nonlinear elements (Josephson devices) in the circuit allows the efficient
calculation of the Hamiltonian. In the following, we briefly outline
the EPR procedure and finite-element methodology employed in the design
of the sample.

\subsubsection{Josephson tunnel junction}

\paragraph{Non-linear, flux-controlled inductor.}

From the point of view of circuit theory \citep{Yurke1984,Devoret1997,Girvin2014,Vool2017},
a Josephson tunnel junction \citep{Josephson1962} is a two-terminal,
non-linear, flux-controlled, lumped-element inductor, whose constitutive
current-flux relationship is 
\begin{equation}
I\left(t\right)=I_{c}\sin\left(\Phi\left(t\right)/\phi_{0}\right)\,,\label{eq:IcJJ}
\end{equation}
where $I_{c}$ is the \emph{critical current} of the junction, a phenomenological
parameter, $\phi_{0}\equiv\hbar/2e$ is the \emph{reduced flux quantum},
and $\Phi\left(t\right)$ is the\emph{ generalized flux} across the
junction, which has the same dimension as magnetic flux,
\begin{equation}
\Phi\left(t\right)\equiv\int_{-\infty}^{t}\mathrm{d}\tau V\left(\tau\right)\,,
\end{equation}
where $V$ is the \emph{instantaneous voltage} \emph{drop} across
the junction terminals. The differential inductance presented by the
junction is $L_{J}/\cos\left(\Phi\right),$ where $L_{J}\equiv\phi_{0}/I_{c}$
is known as the \emph{Josephson inductance}. The quantity $E_{J}\equiv\phi_{0}I_{c}$
is known as the\emph{ Josephson energy}, see Eq.~(\ref{eq:JosEnergy}).
Since there are two Josephson junctions in the Darkmon device, we
label the junction variables with a subscript $j\in\left\{ \mathrm{V,H}\right\} $,
where ${\rm V}$ and ${\rm H}$ denote the vertical and horizontal
junctions, respectively. From Eq.~(\ref{eq:IcJJ}) it follows that
the potential energy function of the $j$-th junction is a function
of flux, 
\begin{equation}
\mathcal{E}_{j}\left(\Phi_{j}\right)=E_{j}\left(1-\cos\left(\Phi_{j}/\phi_{0}\right)\right)\,,\label{eq:JosEnergy}
\end{equation}
where $E_{j}$ and $\Phi_{j}$ are the Josephson energy and generalized
flux of the $j$-th junction. 

\paragraph{Linear and non-linear contributions. }

Dropping constant terms, the potential energy of the Josephson junctions,
Eq.~(\ref{eq:JosEnergy}), can conceptually be separated in two,
corresponding to terms associated with the linear-response and non-linear
response of the junction, $\mathcal{E}_{j\mathrm{,lin}}$ and $\mathcal{E}_{j\mathrm{,nl}}$,
respectively, 
\begin{equation}
\mathcal{E}_{j}\left(\Phi_{j}\right)\equiv\mathcal{E}_{j\mathrm{,lin}}\left(\Phi_{j}\right)+\mathcal{E}_{j\mathrm{,nl}}\left(\Phi_{j}\right)\,,\label{eq:app:circuit: defn of junc energy}
\end{equation}
where \begin{subequations} \label{eq:app:circuit: defn of junc}
\begin{eqnarray}
\mathcal{E}_{j\mathrm{,lin}}\left(\Phi_{j}\right) & = & \frac{1}{2}E_{j}\left(\frac{\Phi_{j}}{\phi_{0}}\right)^{2}\,,\label{eq:app:circuit: defn oof Uj lin}\\
\mathcal{E}_{j\mathrm{,nl}}\left(\Phi_{j}\right) & = & E_{j}\sum_{p=3}^{\infty}c_{jp}\left(\frac{\Phi_{j}}{\phi_{0}}\right)^{p}\,,\label{eq:app:circuit: defn oof Uj nl}
\end{eqnarray}
\end{subequations} where $c_{jp}$ are the \textit{dimensionless}
coefficients of the Taylor series of $\mathcal{E}_{j}$,
\begin{equation}
c_{jp}=\begin{cases}
\frac{\left(-1\right)^{p/2+1}}{p!} & \text{for even }p\\
0 & \text{for odd }p
\end{cases}\,.\label{eq:app:Jos potentual Ujn}
\end{equation}

\subsubsection{Distributed circuit with non-linear lumped elements }

Although electromagnetic (EM) structures are often classified as planar
\citep{Blais2004,Wallraff2004,Barends2013,FYan2016} (2D), quasi-planar
\citep{Minev2013,Minev2016,Brecht2016,Rosenberg2017} (2.5D), or three-dimensional
\citep{Paik2011,Rigetti2012,Reagor2016-cavity,Axline2016} (3D), it
is possible to treat these classes on equal footing within the EPR
framework. Aside from the junctions, the distributed EM circuit of
the readout cavity with the Darkmon chip can be described by a quadratic
Hamiltonian function, $\mathcal{H}_{{\rm EM}}$, that depends on the
device geometry and the material properties. Analytic treatment of
this function is impractical, but finite-element (FE) numerical simulations
are adept at handling systems described by quadratic energy functions
and finding their eigenmodes \citep{Louisell1973,Jin2014}. The Hamiltonian
of the Josephson circuit, consisting of the EM and Josephson elements,
is
\begin{equation}
\mathcal{H}=\mathcal{H}_{\mathrm{lin}}+\mathcal{H}_{\mathrm{nl}}\,,
\end{equation}
where its quadratic part is 
\begin{equation}
\mathcal{H}_{\mathrm{lin}}\equiv\mathcal{H}_{{\rm EM}}+\sum_{j\in J}\frac{1}{2}E_{j}\left(\Phi_{j}/\phi_{0}\right)^{2}\,,\label{eq:HLin}
\end{equation}
while its non-linear part, originating from the non-linearity of the
Josephson junctions, is 
\begin{equation}
\mathcal{H}_{\mathrm{nl}}\equiv\sum_{j\in J}\sum_{p=3}^{\infty}E_{j}c_{jp}\left(\Phi_{j}/\phi_{0}\right)^{p}\,,
\end{equation}
where, for notational convenience, $J\equiv\left\{ \mathrm{V,H}\right\} .$
The quadratic Hamiltonian, $\mathcal{H}_{\mathrm{lin}}$, corresponds
to the \emph{linearized Josephson circuit} (LJC), which can be numerically
simulated with FE EM methods to find its eigenfrequencies, $\omega_{m}$,
and modal field distributions, consisting of the electric field, $\vec{E}_{m}\left(\vec{r}\right)$,
and magnetic field, $\vec{H}_{m}\left(\vec{r}\right)$, eigenvectors
over the simulation domain, where $\vec{r}$ is the spatial coordinate.
For our device, we restrict our attention to the lowest three eigenmodes,
the Dark, Bright, and readout cavity modes, labeled ${\rm D},$ $\mathrm{B}$,
and $\mathrm{C}$, respectively; i.e., $m\in M\equiv\left\{ \mathrm{D,B,C}\right\} $.
Quantizing $\mathcal{H}_{\mathrm{lin}}$, the quantum Hamiltonian
of the LJC can thus be expressed 
\begin{align}
\hat{H}_{\mathrm{lin}} & =\sum_{m\in M}\hbar\omega_{m}\hat{a}_{m}^{\dagger}\hat{a}_{m}\;,\label{eq:Hlin-multi}
\end{align}
where $\hat{a}_{m}$ is the $m$-th mode amplitude (annihilation operator).
Importantly, the frequencies $\omega_{m}$ should be seen as an intermediate
parameter entering in the calculation of the rest of the quantum Josephson
Hamiltonian, 
\begin{align}
\hat{H}_{\mathrm{nl}} & =\sum_{j\in J}\sum_{p=3}^{\infty}E_{j}c_{jp}\hat{\phi}_{j}^{p}\,.\label{ex:Hnl-multi}
\end{align}
While $E_{j}$ and $c_{jp}$ are known from the fabrication of the
circuit devices, the quantum operators $\hat{\phi}_{j}\equiv\hat{\Phi}_{j}/\phi_{0}$
remain to be expressed in terms of the mode amplitudes. It can be
shown that $\hat{\phi}_{j}$ is a linear combination of the latter,
\begin{equation}
\hat{\phi}_{j}=\sum_{m\in M}\phi_{mj}\left(\hat{a}_{m}^{\dagger}+\hat{a}_{m}\right)\;,\label{eq:multijj:ZPF defn}
\end{equation}
where $\phi_{mj}$ are the dimensionless, \textit{real}-valued zero-point
fluctuations (ZPF) of mode $m$ at the position of the junction $j$.
Calculation of $\hat{H}$ is now reduced to computing $\phi_{mj}$.
We achieve this by employing the energy participation ratio.

\subsubsection{Energy participation ratio }

We define the EPR $p_{mj}$ of junction $j$ in eigenmode $m$ to
be the fraction of the total inductive energy that is stored in the
junction, \begin{subequations} \label{eq:Pmj} 
\begin{align}
p_{mj} & \equiv\frac{\text{Inductive energy stored in junction }j}{\text{Inductive energy stored in mode }m}\label{eq:EPR-defn-verbal}\\
 & =\frac{\langle n_{m}|\colon\frac{1}{2}E_{j}\hat{\phi}_{j}^{2}\colon|n_{m}\rangle}{\langle n_{m}|\frac{1}{2}\hat{H}_{\mathrm{lin}}|n_{m}\rangle}\;,
\end{align}
\end{subequations}where we have taken the normal-ordered \citep{Gerry2005}
expectation values over the state $\ket{n_{m}}$, a Fock excitation
in mode $m$. The normal-ordering, denoted by $\colon\colon$, nulls
the parasitic effect of vacuum-energy contributions. 

The EPR $p_{mj}$ is computed from the FE eigenfield solutions $\vec{E}_{m}(\vec{r})$
and $\vec{H}_{m}(\vec{r})$ as explained in Sec.~\ref{subsec:Calculation-of-EPR}.
It is a bounded real number, $0\leq p_{mj}\leq1$. For example, a
participation of 0 means that junction $j$ is not excited by mode
$m$, while a participation of $1$ means that it is the only inductive
element excited by the mode. It can be shown that the values $\phi_{mj}^{2}$
and $p_{mj}$ are directly proportional to each other, 
\begin{equation}
\boxed{\phi_{mj}^{2}=p_{mj}\frac{\hbar\omega_{m}}{2E_{j}}\;.}\label{eq:pmj_multi_zpf}
\end{equation}
Equation~(\ref{eq:pmj_multi_zpf}) constitutes the bridge between
the classical solution of the LJC and the quantum Hamiltonian $\hat{H}$
of the full Josephson system, and, as detailed below, is very useful
for practical applications.

\paragraph{Fundamental design constraints. }

The ZPF $\phi_{mj}$ are not independent of each other, since the
EPRs are submitted to three types of constraints. These are of practical
importance, as they are useful guides in evaluating the performance
of possible designs and assessing their limitations. It is possible
to show the EPRs obey one sum rule per junction $j$ and one set of
inequalities per mode $m$, \begin{subequations} 
\begin{align}
\sum_{m\in M}p_{mj}=1 & \,,\label{eq:sum_pj=00003D1}\\
0\leq\sum_{j\in J}p_{mj}\leq1 & \,.\label{eq:sum_pj<1}
\end{align}
\end{subequations} In practice, Eq.~(\ref{eq:sum_pj=00003D1}) can
be exploited only if $M$ contains the total number of relevant modes
of the system, otherwise the sum of the EPR is bounded by one, rather
than equal to one. The final fundamental EPR relation concerns the
orthogonality of the EPRs. Solving Eq.~(\ref{eq:pmj_multi_zpf})
explicitly for the ZPF, 
\begin{equation}
\phi_{mj}=S_{mj}\sqrt{p_{mj}\hbar\omega/2E_{j}}\;,\label{eq:zpf-general-Smj-Pmj}
\end{equation}
where $S_{mj}=+1$ or $S_{mj}=-1$ is the \textit{EPR sign bit} of
Josephson device $j$ in mode $m$. The EPR sign bit encodes the relative
current direction across the Josephson device. Specifically, only
the \textit{relative} value between $S_{mj}$ and $S_{mj'}$ for $j\neq j'$
has physical significance. The EPR sign bit $S_{mj}$ is calculated
during the process of calculating $p_{mj}$, from the field solution
$\vec{H}(\vec{r})$, see Eq.~(\ref{eq:app:Smj calc}). The EPRs obey
the orthogonality relationship 
\begin{align}
\sum_{m\in M}S_{mj}S_{mj'}\sqrt{p_{mj}p_{mj'}}\, & =0\,,\label{eq:pmjpmj' orthogonality}
\end{align}
valid when all relevant modes are considered.

\subsection{Calculation of the EPR \label{subsec:Calculation-of-EPR}}

\paragraph{Modeling the Josephson junction. }

\begin{figure}[th]
\centering{}\includegraphics[scale=1.1]{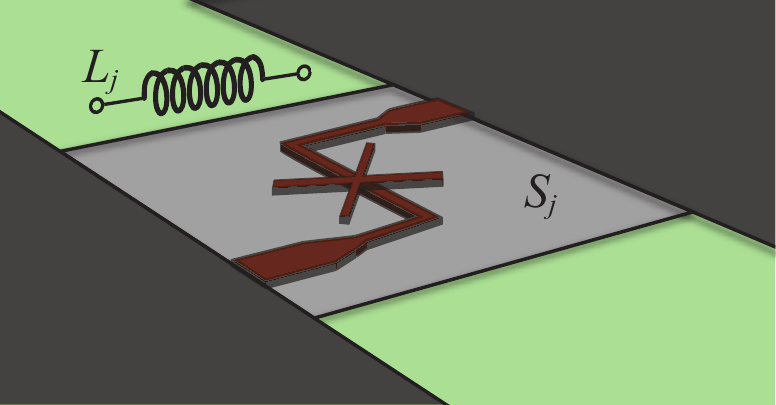} \caption[Finite-element model of linearized Josephson junction]{\textbf{Finite-element model of linearized Josephson junction.}\label{fig:Appdx:JJ-FE-model}
Not-to-scale schematic representation of the finite-element model
of the linearized Josephson junction (location marked by cross) connected
by wire leads (elevated brown trace) to two large metal pads (dark
rectangles). Since the geometry of the junction and leads is in deep-sub-wavelength
regime, it can typically be ignored, and the inductance presented
by the junction, $L_{j}$, graphically represented by black inductor
symbol with two open terminals, can be modeled by a single lumped-element
inductive sheet, $S_{j}$, in the FE simulation, depicted by light
grey rectangle. Green background represents the substrate.}
\end{figure}

In our device, as in most cQED experiments, the physical dimensions
of the Josephson junction ($\approx10^{-7}$~m) are approximately
5 orders of magnitude smaller than the wavelength of the modes of
interest ($\approx10^{-2}$~m), making the junction geometry unimportant,
other than its role in establishing the value of the Josephson inductance
$L_{j}$. Similarly, the lead wires leading up to the junction from
the transmon pads are deep-sub-wavelength features, and it follows
that, typically, their geometry is also unimportant, and can be ignored
altogether, aside from any kinetic inductance contribution. In view
of this, in the FE simulation, we model the $j$-th junction as a
single, two-dimensional rectangular sheet, $S_{j}$, see Fig.~\ref{fig:Appdx:JJ-FE-model},
acting as lumped-element inductor with linear inductance $L_{j}$,
Eq.~(\ref{eq:app:circuit: defn oof Uj lin}). The sheet is assigned
a surface-impedance boundary condition that links the tangental electric
field, $\vec{E}_{\parallel},$ to the tangental magnetic field, $\vec{H}_{\parallel}$,
on the surface of the sheet, $\vec{E}_{\parallel}=Z_{S}(\hat{n}\times\vec{H}_{\parallel}),$
where $\hat{n}$ is the surface normal vector and $Z_{S}$ is the
complex surface impedance, which is calculated so that the total sheet
inductance is $L_{j}$. Note that in the EM context, a hat symbol
denotes a unit vector, not a quantum operator.

\begin{figure}[ht]
\centering{}\includegraphics[width=4.3in]{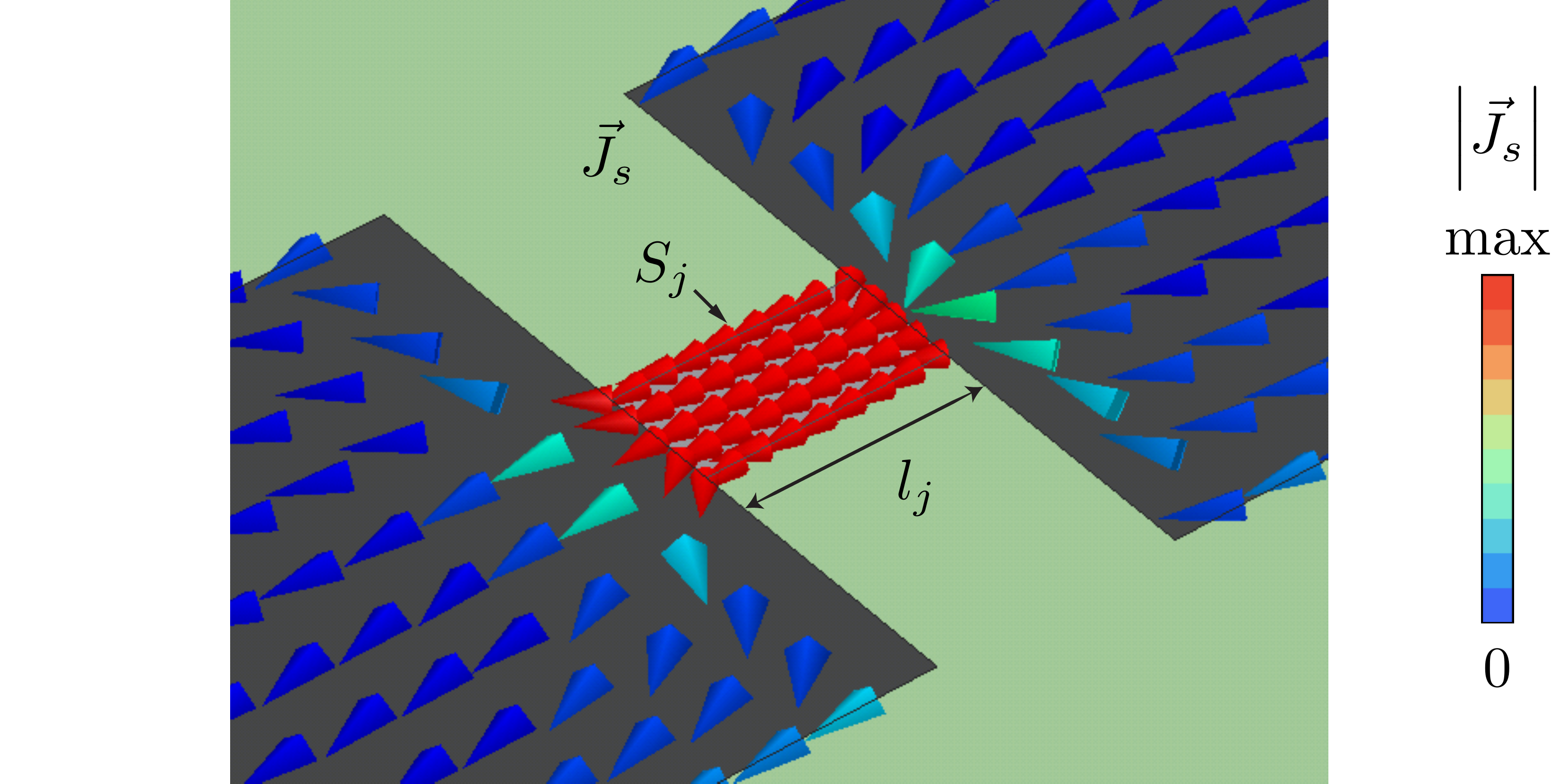}
\caption[Finite-element simulation of a transmon device]{\label{fig:appx:FE JJsurf}\textbf{Finite-element simulation of a
transmon device.} Plot of the surface-current density, $\vec{J}_{S}$,
of a transmon qubit mode, obtained with finite-element electromagnetic
eigenmode simulation; red (blue) indicates maximum (minimum) current
magnitude. Josephson junction (center rectangle) is modeled by a single
inductive sheet ($S_{j}$) with length $l_{j}$, spanning the distance
between the two transmon pads (dark rectangles). Green background
represents the transmon chip. }
\end{figure}

After the design geometry and boundary conditions are established,
 additional fine-mesh operations on crucial features of interest,
such as the junction rectangles, are applied to speed up the solver
convergence, which can be diagnosed by examining the parameters $\omega_{m}$
and $p_{mj}$ as a function of adaptive pass number. At each pass,
the FE solver provides the modal frequencies, $\omega_{m},$ and the
electric, $\vec{E}_{\mathrm{max}}(\vec{r})$, and magnetic, $\vec{H}_{\mathrm{max}}(\vec{r})$,
phasors. The electric field at a point $\vec{r}$ in the volume of
the device, $V$, at time $t$ is 
\[
\vec{E}\left(\vec{r},t\right)=\mathrm{Re}\,\vec{E}_{\mathrm{max}}\left(x,y,z\right)e^{j\omega_{m}t}\,.
\]
The total magnetic and electric field energies of a mode can be computed
from the eigenfields \citep{Pozar}:
\begin{eqnarray}
\mathcal{E}_{\mathrm{elec}} & = & \frac{1}{4}\mathrm{Re}\int_{V}\mathrm{d}v\vec{E}_{\text{max}}^{*}\overleftrightarrow{\epsilon}\vec{E}_{\text{max}}\;,\label{eq:app:FE:  W_E  W_H  integrals}\\
\mathcal{E}_{\mathrm{mag}} & = & \frac{1}{4}\mathrm{Re}\int_{V}\mathrm{d}v\vec{H}_{\text{max}}^{*}\overleftrightarrow{\mu}\vec{H}_{\text{max}}\;,\label{eq:app:FE:  W_H  integral}
\end{eqnarray}
where the spatial integral is performed over total volume, $V$, of
the device, and $\overleftrightarrow{\epsilon}$ ($\overleftrightarrow{\mu}$)
is the electric-permittivity (magnetic-permeability) tensor. While
the magnetic and electric energies are typically equal on resonance
\citep{Pozar}, when lumped elements are included in the model, the
more general equality is between the capacitive, $\mathcal{E}_{\mathrm{cap}},$
and inductive, $\mathcal{E}_{\mathrm{ind}},$ energies, $\mathcal{E}_{\mathrm{cap}}=\mathcal{E}_{\mathrm{ind}}\,.$
For our design, the capacitive energy is stored entirely in the electric
fields, $\mathcal{E}_{\mathrm{cap}}=\mathcal{E}_{\mathrm{elec}}$,
but the inductive energy is stored both in the magnetic fields and
in the kinetic inductance of the Josephson junctions, $\mathcal{E}_{\mathrm{mag}}=\mathcal{E}_{\mathrm{ind}}+\mathcal{E}_{\mathrm{kin}}$,
where $\mathcal{E}_{\mathrm{kin}}$ is the total energy stored in
the kinetic inductors, $\mathcal{H}-\mathcal{H}_{\mathrm{EM}}$, see
Eq.~(\ref{eq:HLin}); it follows, 
\begin{equation}
\mathcal{E}_{\mathrm{cap}}=\mathcal{E}_{\mathrm{ind}}+\mathcal{E}_{\mathrm{kin}}\,,\label{eq:CapIndKinEnegry}
\end{equation}
which, for a single-junction device, implies that the EPR of the junction
in mode $m$ is 
\begin{equation}
p_{m}=\frac{\mathcal{E}_{\mathrm{elec}}-\mathcal{E}_{\mathrm{mag}}}{\mathcal{E}_{\mathrm{elec}}}\:.
\end{equation}
For a device with multiple junctions, such as the Darkmon, it follows
from Eq.~(\ref{eq:EPR-defn-verbal}), that the EPR of junction $j$
in mode $m$ is 
\begin{equation}
p_{mj}=\frac{1}{2}L_{j}I_{mj}{}^{2}/\mathcal{E}_{\mathrm{ind}}\;,\label{eq:appx:FE:pmj}
\end{equation}
where $I_{mj}$ is the junction peak current, which can be calculated
from the surface-current density, $\vec{J_{s}^{m}}$, of the junction
sheet $S_{j}$, 
\begin{equation}
\left|I_{mj}\right|=l_{j}^{-1}\int_{S_{j}}\mathrm{d}s\left|\vec{J_{s}^{m}}\right|\;,\label{eq:appx:FE:ImjAbs}
\end{equation}
where $l_{j}$ is the length of the sheet, see Fig.~\ref{fig:appx:FE JJsurf}.

The calculation of the EPR sign bits, $S_{mj}\in\left\{ -1,1\right\} $,
requires the definition of a convention for the junction orientation,
which is accomplished by supplementing the FE model with a directed
line, $\mathrm{DL}_{j}$, along the length of the junction sheet $S_{j}$.
The actual orientation of the line is irrelevant, so long as it spans
the two terminals of the junction. The sign of the current along the
line can be used as the sign bit 
\begin{equation}
S_{mj}=\sign\int_{\mathrm{DL}_{j}}\mathrm{d}\vec{l}\cdot\vec{J_{s}^{m}}\;.\label{eq:app:Smj calc}
\end{equation}

\paragraph{Remarks.}

The convergence of the EPR extracted from local field quantities,
$I_{mj},$ can be enhanced by renormalizing the EPRs so as to enforce
the global condition given by Eq.~(\ref{eq:CapIndKinEnegry}), $\sum_{j\in J}p_{mj}=\mathcal{E}_{\mathrm{kin}}/\mathcal{E}_{\mathrm{ind}}$.
The eigenmode simulation approach affords several distinct advantages.
No prior knowledge of the mode frequencies is required to execute
the simulation. The solver can be queried to solve for the $N$-lowest
eigenmodes. If only information on modes above a particular frequency
is desired, this cutoff frequency can also be supplied to the solver.
From a single mesh and simulation, the FE solver returns \emph{complete}
information for all modes of interest — the parameters $\omega_{m}$,
$p_{mj},$ and $S_{mj}$ of the Hamiltonian and, as shown in Sec.~\ref{subsec:Dissipation-budget},
the dissipation budget. These features play nicely into the iterative
nature of the design optimization, and make the eigenmode design-optimization
process easy to automate, provided freely to the community in our
software package pyEPR.\footnote{http://github.com/zlatko-minev/pyEPR}
In the optimization of the Drakmon device, the finite-element software
of choice was \emph{Ansys high-frequency electromagnetic-field simulator}
\emph{(HFSS)}.

\subsection{Calculation of Hamiltonian parameters with the EPR\label{subsec:Calculation-of-Hamiltonian}}

The quantities $\omega_{m}$, $p_{mj},$ and $S_{mj}$ obtained from
the FE eigenmode solution together with Eqs.~(\ref{eq:multijj:ZPF defn}),
and~(\ref{eq:pmj_multi_zpf}) completely specify $\hat{H}_{\mathrm{nl}}$,
Eq.~(\ref{ex:Hnl-multi}). The multitude of non-linear interactions
contained in $\hat{H}_{\mathrm{nl}}$ mix the LJC modes. However,
operating the Darkmon in the dispersive regime \citep{Blais2004,Koch2007},
defined by $\omega_{k}-\omega_{m}\gg E_{j}c_{jp}\left<\hat{\phi}_{j}^{p}\right>$
for all $p\geq3$, we can restrict our attention to the leading order
correction of $\hat{H}_{\mathrm{nl}}$ to $\hat{H}_{\mathrm{lin}}$
to account for the device spectrum, see level diagram of Fig.~\ref{fig:Darkmon-energy-level-diagram}.
The leading-order correction is given by the $p=4$ terms that survive
the rotating-wave approximation (RWA) \citep{Carmichael2008-Book2,Gardiner2004},
representing energy-conserving interactions. To leading order, in
the RWA, after normal-ordering, $\hat{H}_{\mathrm{nl}}$ reduces to
the effective Hamiltonian 
\begin{equation}
\hat{\overline{H}}_{4}/\hbar=-\sum_{m\in M}\Delta_{m}\hat{a}_{m}^{\dagger}\hat{a}_{m}+\frac{\alpha_{m}}{2}\hat{a}_{m}^{\dagger2}\hat{a}_{m}^{2}+\frac{1}{2}\sum_{m\neq n}\chi_{mn}\hat{a}_{m}^{\dagger}\hat{a}_{m}\hat{a}_{n}^{\dagger}\hat{a}_{n}\,,\label{eq:H4 - RWA main}
\end{equation}
which when combined with $\hat{H}_{\mathrm{lin}}$, Eq.~(\ref{eq:Hlin-multi}),
yields Eq.~(\ref{eq:HDarkmon}). The Lamb shift, $\Delta_{m}=\frac{1}{2}\sum_{n\in M}\chi_{mn}\,,$
represents the dressing of the linear mode $m$ by the zero-point
vacuum energy of all $M$ modes. From Eq.~(\ref{eq:H4 - RWA main}),
it follows that the measured transition frequency between $\G$ and
$\B$ is $\omega_{\mathrm{BG}}=\omega_{\mathrm{B}}-\Delta_{\mathrm{B}},$
where $\omega_{\mathrm{B}}$ is the LJC Bright eigenmode frequency
and $\Delta_{\mathrm{B}}$ is the Bright mode Lamb shift; a similar
conclusion holds for the GD transition. The Kerr frequencies are found
from the EPR,
\begin{equation}
\chi_{mn}=-\sum_{j\in J}\frac{\hbar\omega_{m}\omega_{n}}{4E_{j}}p_{mj}p_{nj}\label{eq:chi_mj=00003D}
\end{equation}
and $\alpha_{m}=\chi_{mm}/2$. Remarkably, from Eq.~(\ref{eq:chi_mj=00003D})
it becomes evident that the EPRs are essentially the only free parameters
subject to optimization and design in engineering the non-linear couplings,
since the frequencies, $\omega_{m}$, and Josephson energies, $E_{j}$,
are constrained to a narrow range due to practical considerations.
Notably, Eq.~(\ref{eq:chi_mj=00003D}) embodies the structure of
a spatial-mode overlap in the EPRs.

\subsection{Calculation of dissipation budget with the EPR \label{subsec:Dissipation-budget}}

In this subsection, we summarize the methodology employed in minimizing
dissipation in the Darkmon device. This is achieved by optimizing
the geometry of the design (in parallel with the Hamiltonian parameter
optimization) with the aim of minimizing the susceptibility of each
mode to the various unavoidable material and input–output losses.
For each design variation, we compute the bound on the modal quality
factors by constructing a \emph{dissipation budget}, which consists
of the loss expected due to each lossy element in the design. By manipulating
the geometry, the budget can be favorably altered, to a degree. The
calculation of the dissipation parameters is detailed in the following. 

Losses can be classified as either \emph{capacitive, }proportional
to the electric field intensity, $\left|\vec{E}\right|^{2}$, or \emph{inductive},
proportional to the magnetic field intensity, $\left|\vec{H}\right|^{2}$.
The total loss due to a material is proportional to its energy participation
in the mode, $p^{l}$, a geometric quantity related to the field distribution,
and its intrinsic quality, $Q$, a material property. The intrinsic
material quality, $Q$, can typically only be bounded, while $p^{l}$
can be calculated from the eigenfields. The total capacitive and inductive
losses, characterized by $Q_{\text{cap}}$ and $Q_{\text{ind}},$
respectively, sum together with the loss due to input-output coupling,
$Q_{\mathrm{rad}}$, and give the upper bound on the quality factor
of an EM mode, $Q_{\text{total}}$, \citep{Zmuidzinas2012,Geerlings2013,Reagor2016}
\begin{equation}
\frac{1}{Q_{\text{total}}}=\frac{1}{Q_{\text{cap}}}+\frac{1}{Q_{\text{ind}}}+\frac{1}{Q_{\text{rad}}}\,.
\end{equation}
In the following, we explicate the calculation of each quantity. We
note that the EPR treats dissipation and Hamiltonian parameters on
equal footing, and all quantities, Hamiltonian and dissipative, are
extracted from a single eigensolution. 

\paragraph{Capacitive losses.}

Capacitive losses, proportional to the intensity of the electric field,
$\left|\vec{E}\right|^{2}$, can originate from bulk or surface of
materials. Dielectrics, such as the substrate of the Darkmon device,
constitute the primary source of bulk loss \citep{Martinis2014,Dial2016,Kamal2016-anneal},
and, unfortunately, every surface in a device possesses a near-unavoidable,
lossy, surface dielectric layer, possibly due to chemical residues,
condensation, dust, etc. \citep{Martinis2014,Wang2015}. Regardless
of the microscopic origin of the dielectric losses, the loss properties
of the $l$-th dielectric are characterized by a catch-all quality
factor $Q_{\text{cap}}^{l}$ (or equivalently the inverse of the loss
tangent) and the EPR of the dielectric in the mode, $p_{\mathrm{cap}}^{l}$
— the fraction of capacitive energy stored in dielectric element $l$.
For a bulk dielectric, the dissipative EPR is given by 
\begin{equation}
p_{\text{cap,bulk}}^{l}=\frac{1}{\mathcal{E}_{\mathrm{elec}}}\frac{1}{4}\mathrm{Re}\int_{V_{l}}\mathrm{d}v\vec{E}_{\text{max}}^{*}\overleftrightarrow{\epsilon}\vec{E}_{\text{max}}\,,
\end{equation}
where the integral is carried over the volume of the $l$-th dielectric
element, namely $V_{l}$. The dissipative EPR of a surface dielectric,
$p_{\mathrm{cap,surf}}^{l}$, can be approximated by 
\begin{equation}
p_{\text{cap,surf}}^{l}=\frac{1}{\mathcal{E}_{\mathrm{elec}}}\frac{t_{l}\epsilon_{l}}{4}\mathrm{Re}\int_{\text{surf}_{l}}\mathrm{d}s\left|\vec{E}_{\text{max}}\right|^{2}\,,
\end{equation}
where the surface layer thickness is $t_{l}$, and its dielectric
permittivity is $\epsilon_{l}$. The total capacitive loss in the
mode is the EPR-weighted sum of the individual contributions \citep{Zmuidzinas2012,Geerlings2013},
\begin{equation}
\frac{1}{Q_{\text{cap}}}=\sum_{l}{\frac{p_{\text{cap}}^{l}}{Q_{\text{cap}}^{l}}}\,.\label{eq:cap-loss}
\end{equation}

\paragraph{Inductive losses.}

Electric currents flowing in metals or metal-metal seams can result
in inductive losses. The bound on the mode inductive-loss quality
factor $Q_{\text{ind}}$ is a weighted sum of the intrinsic material
quality $Q_{\text{ind}}^{l}$ of each lossy inductive element $l$,
analogous to Eq.~(\ref{eq:cap-loss}), 
\begin{equation}
\frac{1}{Q_{\text{ind}}}=\sum_{l}{\frac{p_{\text{ind}}^{l}}{Q_{\text{ind}}^{l}}}\;,
\end{equation}
where $p_{\text{ind}}^{l}$ is the inductive-loss EPR of element $l$.
For a metal surface, this can be calculated from the eigenfield solutions,
\begin{equation}
p_{\text{ind,surf}}^{l}=\frac{1}{\mathcal{E}_{\mathrm{mag}}}\frac{\lambda_{0}\mu_{l}}{4}\mathrm{Re}\int_{\text{surf}_{l}}\mathrm{d}s\left|\vec{H}_{\text{max},\parallel}\right|^{2}\;,\label{eq:appx:dissip:p ind surf}
\end{equation}
where $\lambda_{0}$ is the metal skin depth at $\omega_{m}$, and
$\mu_{l}$ is the magnetic permeability of the surface (typically,
$\mu_{l}=\mu_{0}$). In the case of superconductors $p_{\text{ind,surf}}^{l}$
is the kinetic inductance fraction \citep{Gao2008,Zmuidzinas2012},
commonly denoted $\alpha$. Normal metals typically have an inductive
quality factor $Q_{\text{ind,surf}}^{l}$ of approximately one \citep{Pozar}.
Bulk superconducting aluminum has been measured to have an inductive
quality factor $Q_{\text{ind,surf}}$ bounded to be better than a
few thousand \citep{Reagor2013}. Meanwhile, the bound on the quality
of thin-film Al has been measured to be better than $10^{5}$ \citep{Minev2013}. 

\subparagraph{Seam losses. }

A distinct loss mechanisms occurs at the seam of two metals \citep{Brecht2015}.
For instance, a common source of such loss is the seam used in superconducting
cavities. In the FE model, the seam can be modeled by a line, denoted
$\text{seam}_{l}$, between the two mating metallic surfaces. The
seam inductive participation is 
\begin{equation}
p_{\text{ind,seam}}^{l}=\frac{1}{\mathcal{E}_{\mathrm{mag}}}\frac{\lambda_{0}t_{l}\mu_{l}}{4}\mathrm{Re}\int_{\text{seam}_{l}}\mathrm{d}l\left|\vec{H}_{\text{max},\perp}\right|^{2}\;,\label{eq:Pseam}
\end{equation}
where the seam thickness is denoted $t_{l}$, its magnetic permeability
$\mu_{l}$, and its the penetration depth $\lambda_{0}$. It is convenient
to recast the seam loss contribution 
\begin{equation}
\frac{p_{\text{ind,seam}}^{l}}{Q_{\text{seam}}}=\frac{1}{g_{\text{seam}}}\frac{\int_{\text{seam}}\left|\vec{J}_{s}\times\vec{l}\right|^{2}dl}{\omega\mu_{0}\int_{\text{vol}}\left|H_{\text{max}}\right|^{2}dV}\;,
\end{equation}
in terms of a seam admittance $g_{\text{seam}}$, which is defined
in Ref.~\citet{Brecht2015}.

\section{Sample fabrication\label{sec:Fabrication-of-sample}}

Samples were fabricated on 430~$\text{\ensuremath{\mu}}$m thick,
double-side-polished, c-plane sapphire wafers, grown with the edge-defined
film-fed growth (EFG) technique, with the bridge-free junction fabrication
method, see Refs. \citet{Lecocq2011-bridge-free,Pop2012-Junction,Pop2011-Thesis,Reagor2016}.
We defined the sample pattern, both large and fine structures, with
a 100~kV electron-beam pattern generator (\emph{Raith EBPG 5000+})
in a single step on a PMAA/MAA resist bilayer. In the following,
we describe each step of the fabrication process in detail, and we
hope that by adding some additional information about each step and
motivation behind it, a reader who is new to the subject will benefit. 

\paragraph{Cleaning the wafer. }

First, the sapphire wafer is solvent cleaned under a chemical hood
in a two-step $N$-Methyl-2-pyrrolidone (NMP) process. The solvent
removes dust, organic residues, and oils on the wafer surface. For
our samples, we heated the wafer to $90\ensuremath{~\ensuremath{^{\circ}}\text{C}}$
for 10 minutes in an NMP bath, then sonicated it in the bath for another
10~minutes. After removing the wafer from the bath, if it is left
out to dry on its own, the NMP would evaporate quickly and leave undesirable
residue behind. Instead, we rinsed the wafer in an acetone bath, followed
by a methanol one, before finally blow drying it with filtered nitrogen
gas. Methanol has low evaporation pressure and under the blow drying
tends to take away the residues, rather than simply evaporating and
leaving residues behind. An acid should not be used to clean the sapphire
wafer, since the wafer is costly and already polished.

\paragraph{Spinning the positive resist bi-layer. }

The copolymer resist (\emph{Microchem} \emph{EL-13}) is spun onto
the cleaned wafer at 2,000~revolutions per minute (r.p.m.) for 100
seconds,  then, it is baked for 5~minutes at $200~\ensuremath{^{\circ}}\text{C}$.
The PMMA resist (\emph{Microchem A-4}) is spun on top of the first
at 2000~r.p.m. for 100~seconds. The wafer is baked at $200~\ensuremath{^{\circ}}\text{C}$
for 15~minutes, yielding a thickness of about 200~nm. It is worth
noting why PMMA is the resist of choice: it offers high, nm-sized
resolution, simplicity and ease of handling, no sensitivity to white
light, nor shelf or film life issues, and is easily dissolved, qualities
that make it the ideal resist for this type of nanofabrication.

Before proceeding to patterning, fabrication on sapphire requires
an extra step: the anti-charging layer. Whereas most silicon substrates
are conducive enough to prevent electron beam deflection during the
e-beam writing, sapphire substrates are not. The buildup of charge
is mitigated by depositing  a thin (10~nm) anti-charging layer of
gold on the wafer. In terms of metals, gold is an excellent choice,
as it is inert, does not have an oxide, and has notably high electrical
conductivity. Alternatively, we have also used aluminum for the anti-charging
layer (13~nm thick). 

\paragraph{Writing and developing the pattern. }

Both large and fine structures, including the Josephson tunnel junction
are patterned in a single step with the 100~kV EBPG, following which,
the gold layer is removed by submerging the wafer in a potassium-iodide/iodine
etch solution for 10~seconds. Next, the wafer is rinsed in water
and the resist is developed in a 3:1~IPA:water mixture at $6~\ensuremath{^{\circ}}\text{C}$
for 2~minutes. After development, the pattern is inspected under
an optical microscope.

\paragraph{Plasma cleaning, deposition, and oxidation. }

The wafer is loaded in the electron-beam evaporation system, a multi-chamber\emph{
Plassys UMS300 UHV}. To prepare the surfaces for deposition and reduce
the amount of aging of the Josephson junction, the exposed sample
surfaces are subjected to a 1~minute of oxygen-argon plasma cleaning,
under a pressure of $3\times10^{-3}$~mbar. In this procedure, the
etch removed 30~nm from the upper resist layer; however, ideally,
one would use a shorter duration and larger pressure \citep{Pop2012-Junction},
which was not available. Next, the sample is transferred from the
load-lock to the deposition chamber, where an automated titanium sweep
is performed to absorb residual gases in the deposition chamber. 
At an angle of $19$~degrees, 20~nm of Aluminum is deposited onto
the sample, following which, the sample is transferred to the oxidation
chamber, where it is exposed to a 3:17 oxygen:argon mixture for 10~minutes
at 100~Torr. This forms an approximately 1~nm thick aluminum oxide
layer, the insulating barrier of the $\text{Al/Al\ensuremath{\mathrm{O}_{\mathrm{x}}}/Al}$
Josephson tunnel junctions. The sample is returned to the deposition
chamber, where the second and final layer of aluminum (30~nm) is
deposited at an angle of $-19$~degrees. Next, rather than directly
removing the sample from the evaporation system and allowing the exposed
aluminum surfaces to uncontrollably oxidize in air, the surfaces are
passivated with a final oxidation step at 50~Torr for 10~minutes.
The aluminum forms a self-limiting oxide capping layer.

\paragraph{Liftoff.}

The sample is placed in a heated bath of NMP solvent at $70~\ensuremath{^{\circ}}\text{C}$
for two hours. It is then sonicated for 1~minute, while still in
the NMP, following which, the NMP is cleaned with acetone, methanol,
IPA, and, finally, a dry nitrogen blow gun. The solvent ``lifts off''
the unwanted metal from the wafer by dissolving the resist underneath
it, thus leaving the bath full of aluminum flakes. Note that NMP
can't be heated much above $100~\ensuremath{^{\circ}}\text{C}$, since
that will likely damage the $\text{Al/Al\ensuremath{\mathrm{O}_{\mathrm{x}}}/Al}$
junctions.  

\paragraph{Dicing.}

A protective coating of optical resist (\emph{SC1827}) is spun at
1,500~r.p.m. for 120~seconds and baked at $90~\ensuremath{^{\circ}}\text{C}$
for 5~minutes. The sample is loaded in the dicer (\emph{ADT ProVecturs
7100}), which then is calibrated, aligned, and which then performs
the dicing. The diced chips are cleaned with acetone, methanol, and
dry nitrogen, and are stored until they ready to be mounted in the
sample holder. 

\paragraph*{Sample selection.}

The diced chips are cleaned (NMP, Acetone, methanol, nitrogen air)
and visually examined under an optical microscope where the normal-state
resistance, $R_{N}$, of their Josephson junctions is measured. This
is performed under an optical microscope (\emph{Copra Optical Inc.
SMZ800}) with probe needles (\emph{Quater-Research H-20242}) lowered
to contact the transmon pads on either side of the junction, taking
care to properly ground all object in contact with the sample and
to minimize unavoidable scratching of the pad during the probing.
The measurement of $R_{N}$ provides a good estimate of the junction
Josephson energy, $E_{J}$, by an extrapolation from room temperature
to the operating sample temperature, at approximately $15$~mK, using
the Ambegaokar-Baratoff relation \citep{Ambegaokar1963},
\begin{equation}
E_{J}=\frac{1}{2}\frac{h\Delta}{\left(2e\right)^{2}}R_{N}^{-1},
\end{equation}
where $\Delta$ is the superconducting gap of aluminum. The chip
closest matching the Josephson energies, $E_{J}$, of the EPR-designed
vertical and horizontal junctions is selected for mounting in the
sample holder. 

\section{Sample holder \label{sec:Sample-holder-materials}}

In this section, we describe the methodology used in the design of
the sample holder and readout cavity, while paying special attention
to the motivation underlying the design choices. The boundary conditions
of the readout cavity, depicted in Fig.~\ref{fig:setup}, are formed
from the superconducting inner walls of the chip sample holder, composed
of two main halves, see Fig.~\ref{fig:Sample-holder-and-pins}c,
and based on the ideas presented in Ref.~\citet{Paik2011}. Before
a discussion on the design geometry, we focus on the selection of
its materials.

\subsection{Material losses and selection \label{subsec:sample-holder-material-selection}}

\paragraph{Readout cavity considerations.}

The inner walls of the sample holder establish the boundary conditions
of the readout cavity mode, and hence have a large inductive, $p_{\text{ind,surf}}^{l}$,
and dielectric, $p_{\text{cap,surf}}^{l},$ surface-loss participation
ratio, see Sec.~\ref{subsec:Dissipation-budget}. It follows that
the material quality of the cavity inner walls is important in determining
the readout quality factor, $Q_{\mathrm{R}}$. However, since this
mode is purposefully made low-Q, by coupling it strongly to the input-output
(\emph{I-O}) couplers, the importance of the sample-wall material
is greatly reduced, and could in principle be rather lossy. For instance,
in some designs, copper, which has an inductive quality factor of
unity,  $Q=1$, has been used, thus limiting $Q_{\mathrm{R}}$ to
several thousand. Under these conditions, a fraction of the readout
cavity signal is lost to the walls of the cavity, rather than to the
\emph{I-O} couplers. Nonetheless, the \emph{I-O} coupling is engineered
to be larger still, so that most of the signal in the readout cavity
makes it to the amplifier chain, and a high quantum measurement efficiency,
$\eta$, can still be obtained. 

\paragraph{Qubit mode considerations. }

However, the Bright and Dark qubit modes, while predominantly spatially
localized to the sapphire substrate region, have a small fraction
of their fields extending to the inner walls of the readout cavity.
Although the lossy energy participation ratios, $p_{\text{ind,surf}}^{l}$
and $p_{\text{cap,surf}}^{l}$, are exponentially small ($\lesssim10^{-5}$),
so that the cavity walls participate on the part-per-million level,
a normal-metal wall ($Q_{\mathrm{ind}}^{l}\approx1$) could limit
the qubit quality factors significantly, making lifetimes on the order
of $T_{\mathrm{1}}^{\mathrm{D}}\approx100\ \mathrm{\mu s}$ out of
reach. For this reason, we employ a low-loss superconducting material
for the sample holder, and clean its surfaces with care, see discussion
on Pg.~\pageref{par: Surface-preparation}. Specifically, we machined
the readout cavity from 6061 aluminum alloy, which is typically found
in the construction of aircraft structures. Notably, it is a good
superconductor, and due to its hardened structure offers a machining
advantage over regular aluminum, which is too soft. 

We remark that in other experiments, involving high-Q storage mode
cavities, the cavities are often machined from high-purity 4N (99.99\%
pure) aluminum (or sometimes, 5N), which is very soft, and thus difficult
to machine. Further, the machining forms deep cracks ($\approx100\ \mathrm{\mu m}$),
where machining oils and dirt seep in, and hence, the surfaces require
a more involved chemical etch process to remove about 150$\ \mathrm{\mu}$m
of the surface; see the dissertation of M. Reagor \citep{Reagor2016}.

\begin{figure}
\begin{centering}
\includegraphics[width=1\columnwidth]{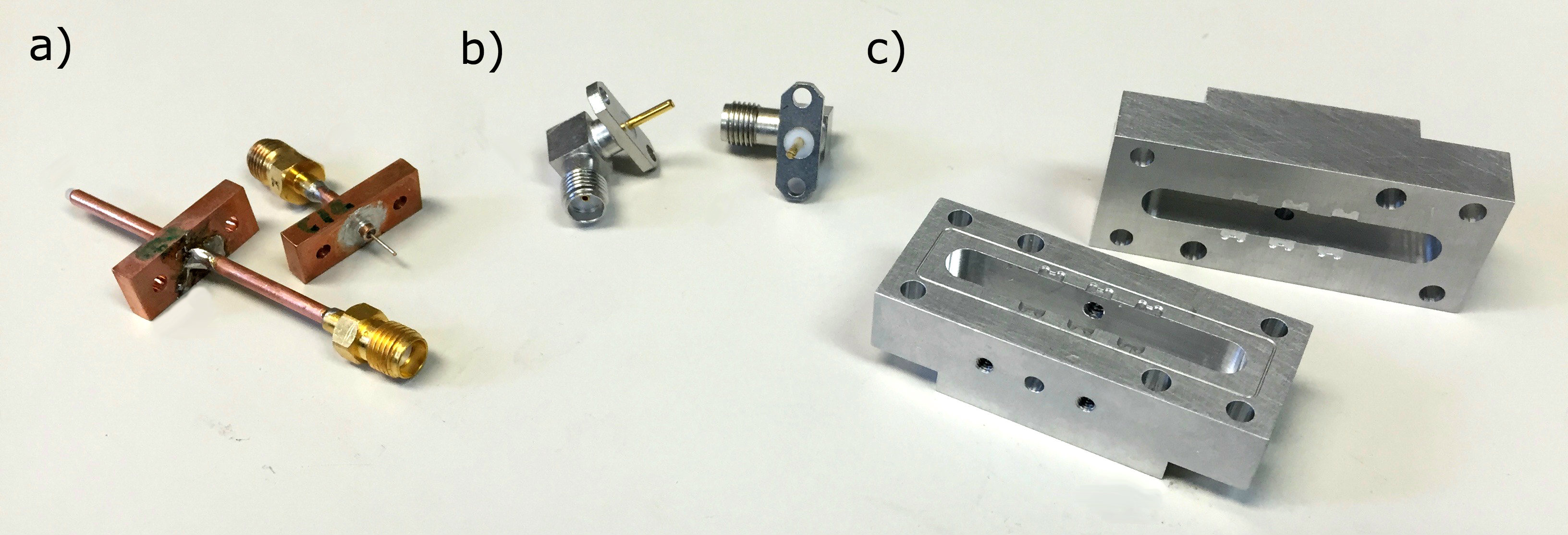}
\par\end{centering}
\caption[Non-magnetic couplers and sampler holder ]{\label{fig:Sample-holder-and-pins}\textbf{Non-magnetic couplers
and sampler holder.} \textbf{a/b, }Photograph of two generations of
custom-made, non-magnetic, \emph{SubMiniature version A} (SMA), input-output
(\emph{I-O}) pin couplers. \textbf{c,} Photograph of disassembled
sample holder, inside walls forms boundary condition for the readout
cavity mode. Three machined grooves on the mating surfaces of the
two halves provide placement slots for samples. Mating surface of
lower-half has groove encircling the inside cavity, employed with
an indium wire to seal the two halves. Large through holes visible
on the mating surfaces provide means to fasten sample holder and attach
it to a cold finger in the dilution refrigerator. Small hole visible
on the interior back wall of the lower half allows for screw-tuning
of the readout cavity frequency.}
\end{figure}

\paragraph*{Non-magnetic input-output couplers.}

It has been recognized that commercial flange-based\emph{ I-O} couplers
contain magnetic ferrite impurities with fields at the ten milligauss
level, which although small, due to the close spatial proximity of
the coupler to the thin-film superconducting pads of the transmon,
as well as the Josephson junction, could introduce vortices in the
films, and it is suspected, generally degrade the superconductor performance.
To achieve better control of the electromagnetic environment and to
reduce potential losses due to magnetic impurities, we employed custom-made
pins from non-magnetic materials, such as copper and brass. 

Panels (a) and (b) of Fig.~\ref{fig:Sample-holder-and-pins} show
two generations of non-magnetic, \emph{SubMiniature version A} (SMA)
pin-couplers. The first generation, see panel (a), was made in-house
from a standard, SMA copper cable with two female connectors. After
testing the quality of the cable, by checking its insertion loss with
a vector network analyzer (VNA), the cable was cut in two, partially
stripped (external shielding and teflon) to expose the center conductor,
which serves as the pin inside the readout cavity, and soldered (non-magnetic
solder) to a custom copper flange. The flange can then be mounted
to the outside surface of the readout cavity, see panel (c), by non-magnetic,
brass or aluminum, screws. Panel (b) shows a second generation of
non-magnetic couplers, made from beryllium-copper, and custom-ordered,
courtesy of Christopher Axline. In general, components used with the
sample holder, placed inside the enclosing magnetic shielding, were
tested for magnetic compatibility with a magnetometer inside a magnetically
shielded box at room-temperature. 

\paragraph*{Sample-holder seam.}

To enclose the Darkmon chip in the sample holder, the sample holder
is designed as two separate halves, see Fig.~\ref{fig:Sample-holder-and-pins}c.
As discussed in Sec.~\ref{subsec:Dissipation-budget}, the placement
of the seam in the design is important, as it determines the seam-loss
participation ratio, $p_{\text{ind,seam}}^{l}$. The seam in placed
at the minimum of the current field profile of the readout cavity
mode. No perfect symmetry exists in the design; it is broken by the\emph{
I-O} couplers and the sample chip, so even at this location, the participation
is not strictly zero for either the readout cavity or qubit modes.
For this reason, it is important that the seam quality is as high
as possible. In the following, we remark on seam properties at the
microscopic level, and the use of an indium seal for improved electrical
contact.

\emph{Seam quality at the microscopic level.\label{disc:Seam-quality}
}Even under high pressure, applied by the fastening action of the
sample holder screws, see Fig.~\ref{fig:Sample-holder-and-pins}c,
the mating faces of the two halves of the sample holder do not join
well at the atomic level. Three interface regions can be identified:
i) \emph{metal-to-metal} \emph{regions}, the rarest, where aluminum
atoms from both halves are in physical contact, allowing superconducting
current to flow undisturbed, ii)\emph{ semi-conducting regions}, more
common, where contaminants, typically dielectric, result in resistive
conductance, and iii) \emph{non-conducting regions, }typically, the
most common, where electrical flow is altogether prohibited, for either
the region is a vacuum gap or is dominated by a thick non-conductive
film of oxides, sulphides, etc. The physical contact area is typically
less than a tenth of the area of the mating surfaces.

\emph{Higher-quality seam. }To create a seam with higher electrical
conductivity, one can first increase the force applied to form the
bond to fracture the native oxide layer of the mating surfaces and
to yield a greater physical contact area, region (i). However, this
method is rather limited in applicability with our design, because
the constraint of non-magnetic screws excludes nearly all hard-material
screws, including stainless steel ones, due to magnetic impurities.
The soft screws we use, aluminum and brass, lack the impurities needed
to make them withstand larger forces, and tend to break and strip
easily. Mostly, we relied on a soft-metal seal, a thin wire gasket
placed in a small groove in one of the mating faces. A number of materials
metals are conventionally used as soft-metal seals, such as copper,
aluminum, indium, etc. The seal of choice in the cQED community is
indium, typically used in cryogenic hermetic seals and low-temperature
solder with melting point of 47~°C, since it remains soft and malleable
even at cryogenic temperatures and is a superconductor. The indium
gasket has been observed to increase the internal quality factor of
a superconducting cavity by several orders of magnitude and its estimated
seam conductance is $g_{\mathrm{seam}}\gtrsim10^{6}/\mathrm{\Omega m}$
\citep{Brecht2017-Thesis}. In our experiment, we used an un-greased
99.99\% indium wire of 0.020~in. diameter to form the seam gasket.

\subsection{Assembly}

After the surfaces of the machined sample holder are cleaned, see
discussion on Pg.~\pageref{par: Surface-preparation}, the Darkmon
chip selected from the diced wafer, Sec.~\ref{sec:Fabrication-of-sample},
is cleaned (NMP, Acetone, methanol, nitrogen air) and, under an optical
microscope, is immediately placed in the central groove of the sample
holder, see Fig.~\ref{fig:Sample-design-assembly}a. The groove is
designed to be larger (at minimum 5\%) than the dimensions of the
chip to account for the differential thermal contraction between aluminum
and sapphire (mostly, the aluminum contracts) and machining tolerances.
The precision of the dicing saw (\emph{ADT ProVecturs 7100}) is high,
several microns when recently calibrated, and the chip will not exceed
the diced margin by more than a few microns, but it can fall quite
short of that, because, unlike silicon, when sapphire is diced, it
shatters around the edges, much like glass, and forms a jagged edge.
For this reason, when placed in the groove, the chip can rattle about,
and requires anchoring, accomplished by placing four small bits of
indium on its four corners and pressing them down to fill the corner
circular pockets machined in the groove, see Fig.~\ref{fig:Sample-holder-and-pins}c.
To minimize contamination during the mounting, the chip is placed
face down in the groove, although, we note that even dust on the back
side of the chip can contribute to loss in the qubit modes, though,
its participation ratio will be far smaller than if it were on the
front face. 

When the sample is well anchored, the indium gasket is laid down in
the gasket groove, as visible in Fig.~\ref{fig:Sample-design-assembly}a,
and the top half of the sample holder is mounted on top, fastened
tightly with even-pressure to allow the indium to distribute evenly.
For the screws, we used aircraft-alloy 7075 (\emph{McMaster}/\emph{Fastener
Express}), with less than 1\% iron impurities. These screws, as discussed
earlier, are rather soft, and to achieve a higher compression between
the two halves at cryogenic temperatures, we used the screws with
\label{par:molybdenum}molybdenum washers, which provide differential
contraction — the linear thermal contraction for molybdenum (aluminum)
between room temperature and 4~K is 0.095\% (0.415\%). Molybdenum
is compatible with the non-magnetic requirement. After 10 minutes,
the indium seal relaxes, and the screws can be further tightened,
with even pressure.

\section{Cryogenic setup\label{sec:Cryogenic-setup} }

The embedding environment of the sample is nearly as important as
the properties of the sample itself in achieving long-coherences and
desired performance. For this reason, in this section, we focus on
a few notable aspects of the cryogenic setup, and pay particular attention
to motivations. For the most part, our cryogenic setup is rather standard
in the field of cQED. For a more general discussion of low-temperature
cryogenics, see Refs.~\citet{ventura2010book} and \citet{pobell2013book}.

\subsection{Material selection\label{subsec:Material-selection}}

As already emphasized, the material selection of components used in
the setup is of prime importance. For this reason, we feel it worthwhile
to note a few overriding principles, corroborated by experience, employed
when selecting materials for the cryogenic setup of a cQED experiment,
which have to be compatible with operation at milikelvin temperatures
and high vacuum ($<10^{-1}$ \,Pa).

\paragraph{Tested and well-understood cryogenic materials. }

In the cryogenic setup of our experiment, we employed only materials
that have been exhaustively studied, characterized, and established
in regular laboratory use. There are only a handful suitable for cQED
experiments, which, in the solid-state, can classified as: i) ambient-pressure
superconductors: aluminum, niobium, indium, titanium, tin, molybdenum,
and niobium-titanium (NbTi), ii) normal metals: copper, brass, gold,
beryllium, stainless steel, and iii) dielectrics: silicon, sapphire,
quartz, nylon, Teflon, Stycast, poly(methyl methacrylate) (PMMA).
The listed materials are the most common ones; for material properties,
see Refs.~\citet{ventura2010book} and \citet{pobell2013book}. 

\paragraph*{Simplicity and homogeneity.}

The simplest and smallest number of materials were employed in the
cryogenic setup. The Darkmon sample (including \emph{I-O} pins, seam
gaskets, screws, \emph{et cetera})  consisted of essentially three
types of atoms — aluminum, oxygen, and indium. Beyond the materials
employed in the sample holder, the properties of commercial components
were found to vary among manufacturers, for instance, the residual
magnetic impurity levels measured in screws and SMA connectors, barrels,
adapters, \emph{et cetera} varied across manufacturers. For this reason,
prior to use in the setup, all components were screened with a magnetometer,
especially when employed inside magnetically shielded compartments.

\paragraph{Aluminum. }

Chief among the materials employed was aluminum (Al), and hence, we
pay special attention to its properties. From a structural standpoint,
Al is ``light-weight,'' having one-third the density and stiffness
of steel and copper. However, unlike steel, it is free of magnetic
impurities and has 59\% of the thermal and electrical conductivity
of copper at room temperature. Importantly, when aluminum oxidizes,
it forms a protective coating of amorphous aluminum oxide ($\mathrm{AlO_{x}}$)
that is thermodynamically favored to self limit growth to a thickness
of merely one nanometer. The $\mathrm{AlO_{x}}$ layer is special
in that it has one of the highest hardness coefficients of all oxides,
even greater than glass, making it an excellent (unavoidable) encapsulation
layer, rendering Al highly resistant to corrosion, but also making
the formation of a very-conductive Al-Al seam difficult, as discussed
on Pg.~\pageref{disc:Seam-quality}, and resulting in a surface dielectric
layer with a loss tangent, see Sec.~\ref{subsec:Dissipation-budget}.

\subsection{Thermalization}

The design and implementation of a high-thermal-conductivity link
between the Darkmon sample and the main source of cooling power in
the dilution refrigerator, the mixing chamber pot, is crucial to take
undesired heat away from the sample. At low temperatures, the task
is complicated since the rate limiting factor in the heat transfer
becomes the \emph{contact} \emph{thermal resistance, $R_{\mathrm{C}},$}
found at the interface of two mating surfaces, intricately dependent
upon on the interface properties and difficult to control. The temperature
discontinuity, $\Delta T$, across two mating surfaces is 
\begin{equation}
\Delta T=\frac{R_{\mathrm{C}}}{A}\dot{Q}\,,
\end{equation}
where $A$ is the surface area and $\dot{Q}$ is the power flowing
through the surface. It is seen that to minimize $\Delta T$ once
can increase the contact area, $A$, or decrease the geometry-independent
resistance, $R_{\mathrm{C}}.$ In the following, we describe the thermal
link setup of the sample and briefly outline the strategies employed
to minimize $R_{C}$ across the various interfaces.

\textbf{Gold plating and welding. }The sample was mounted on a cold-finger
attached to the mezzanine mixing-chamber plate. The plate is \emph{gold
plated }($\approx5\mu m$ by electroplating) to achieve higher thermal
conductivity. As a soft metal, gold allows for a larger 'real' surface-area
contact, and due to its chemical inertness, also provides protection
from oxidation of the surface, which keeps $R_{\mathrm{C}}$ low across
multiple uses and over time. The cold finger is not gold-plated, due
to the cost and long-lead time of the process. It is machined from
two oxygen-free high-thermal-conductivity (OFHC) copper blocks, which
are \emph{welded together }to minimize the number of contact joints,
essentially eliminating $R_{\mathrm{C}}$ altogether. Of course, the
bulk thermal resistance of the OFHC copper block remains, but it is
rather small and the cross-sectional area of the cold-finger block
is rather large, providing a good thermal link.

\textbf{Pressure and differential contraction. }The sample is fixed
to the cold finger with aluminum screws. The cold finger is mounted
on the mezzanine mixing-chamber plate with stainless steel screws,
which allow greater pressure. To maximize the pressure across all
joints, \emph{molybdenum washers }were used everywhere to provide
further differential-contraction pressure, see discussion on Pg.~\pageref{par:molybdenum}. 

\textbf{Thermal straps. }The cold finger thermalization link contains
several unavoidable joins, an issue that can be circumvented to a
degree with the use of a flexible heat strap (also known as \emph{thermal
braid}). Directly mounted on the sample was a small copper block welded
to a thick OFHC copper heat strap (models \emph{P6-501} and \emph{P7-501}
from \emph{TAI}) that extended to the mixing chamber plate without
interruption. The strap has the advantage of being durable, flexible,
and reusable.

\textbf{\label{par: Surface-preparation}Surface preparation. }The
physical and chemical condition of the surfaces forming the contact
determine $R_{\mathrm{C}}$. When a component is machined, the stresses
applied to the surface create dislocations and riddle the surface
with extremely narrow (order of a few nanometers) but deep (hundreds
of nanometers) cracks, into which machining oils seep. The cracks
and oil residues degrade the surface quality, visually, electrically,
and thermally. For this reason, all surfaces involved in forming a
thermal link were prepared in the following way. First, they were
cleaned abrasively with scotch bright, buffing, and  fine sandpaper,
removing the top surface layer and resulting in a shiny mirror finish.
Second, the mating component was cleaned chemically. Typically, by
sonication in an anionic detergent solution (\emph{Alconox 1\%}),
followed by acetone, then IPA, and finally blow dried with dry nitrogen
air. For more aggressive cleaning, we used a powerful oxidizing agent,
nitric acid ($\mathrm{HNO_{3}}$), to etch the surface. The acid and
oxidized impurities were then washed away with deionized water, followed
by an acetone and isopropanol (IPA) bath, and finally a nitrogen blow
dry. The components were then mounted immediately, before a substantial
oxide layer could form. For previously treated components that required
remounting, the mating surface was cleaned with blue solder flux,
which has a high concentration of nitric acid, immediately prior to
mounting.

\subsection{Light and magnetic shielding}

The quality factor of superconducting microwave aluminum resonators
(and qubits) is known to strongly depend on the quality of the infrared
and magnetic shielding of the embedding environment \citep{Barends2011-qp,Wang2014,Kreikebaum2016}.
Stray infrared light that is absorbed by a superconductor creates
quasiparticles, which reduces the overall quality factor of the superconducting
surface. This effect is especially pronounced in aluminum, whose superconducting
gap is rather low, $\approx88$~GHz \citep{Barends2011-qp,deVisser2011-qp}.
The effect of stray light can be largely mitigated with multistage
infrared shielding. Magnetic fields at the surface of the superconductor
can suppress the superconducting gap and introduce vortices, which
generally reduce the inductive surface quality, although, under certain
conditions, the vortices can act as quasiparticle traps, and can result
in overall higher quality \citep{Wang2014,Vool2014-qp}. 

\paragraph{Light shielding.}

Black-body radiation from warmer stages in the dilution refrigerator
was blocked by reflective thermal shields enclosing the mixing chamber
space and special care with taken to prevent line-of-sight leaks through
screw holes, \emph{et cetera}. However, \emph{low frequency} photons
are more difficult to shield against, and several further strategies
were employed to  make a light-tight sample space for the Darkmon
device. In addition to the indium seal between the two metal halves
of the sample holder, the sample holder was wrapped in three layers
of aluminized mylar foil, secured with copper tape. In some cooldowns,
the inside of the magnetic shield housing the sample (see below) was
lined with infrared absorbing epoxy \citep{Barends2011-qp,Rigetti2012}.
Coaxial thermalization and infrared filters (teflon replaced by\emph{
Eccosorb CR-110} as the dielectric) were used on the input and output
lines of the sample. 

\paragraph{Magnetic shielding. }

A high-magnetic-permeability, $\mu$-metal (\emph{Amumetal A4K}) can
enclosed an aluminum superconducting cylinder which housed the sample.
In this configuration, the $\mu$-metal shield allows the superconducting
shield to cool through its critical temperature in a lower magnetic
field, lowering the possibility of vortex trapping. Both shields were
thermally anchored to the mixing-chamber base plate by thermal straps,
while the $\mu$-metal shield was also anchored to the cold-finger
by direct contact. Components employed within the can were tested
for magnetic impurities with a magnetometer. Special care was taken
to avoid markings and paint that could be magnetic; paint on components,
such as directional couplers, was stripped with a solvent bath (typically,
Acetone).

\section{Microwave setup\label{sec:Microwave-setup} }

\paragraph{Room temperature.}

The control tones depicted in Fig.~\ref{fig:setup} were each generated
from individual microwave generators ($\Omega_{\mathrm{D}}$ and $\Omega_{\mathrm{B0}}$:
\textit{Agilent N5183A}; readout cavity tone R and $\Omega_{\mathrm{B1}}$:
\textit{Vaunix LabBrick LMS-103-13} and LMS-802-13, respectively).
To achieve IQ control, the generated tones were mixed (\textit{Marki
Microwave Mixers IQ-0618LXP} for the cavity and \textit{IQ-0307LXP}
for $\Omega_{\mathrm{B0}},\Omega_{\mathrm{B1}}$, and $\Omega_{\mathrm{D}}$)
with intermediate-frequency (IF) signals synthesized by the 16 bit
digital-to-analog converters (DACs) of the integrated FPGA controller
system (\textit{Innovative Integration VPXI-ePC}). Prior to mixing,
each analog output was filtered by a $50\,\Omega$ low pass filter
(\textit{Mini-Circuits BLP-300+}) and attenuated by a minimum of 10\,dB.
The radio-frequency (RF) output was amplified at room temperature
(\textit{MiniCircuits ZVA-183-S+}) and filtered by \textit{Mini-Circuits}
coaxial bandpass filters. The output signal was further pulse modulated
by the FPGA with high-isolation SPST switches (\textit{Analog Device
HMC-C019}), which provided additional 80\,dB isolation when the control
drives were turned off. The signals were subsequently routed to the
input lines of the refrigerator. 

At room temperature, following the cryogenic high-electron mobility
amplifier (HEMT; \textit{Low Noise Factory LNF-LNC7\_10A}), the signal
were amplified by 28\,dB (\textit{Miteq AFS3-00101200-35-ULN}) before
being mixed down (\textit{Marki image reject double-balanced mixer
IRW-0618}) to an intermediate frequency (IF) of 50\,MHz, where they
were band-pass filtered (\textit{Mini-Circuits SIF-50+}) and further
amplified by a cascaded preamplifier (\textit{Stanford Research Systems
SR445A}), before finally digitization by the FPGA analog-to-digital
converters (ADC).

\paragraph{Cryogenic.}

The experiments were carried out in a cryogen-free dilution refrigerator
(\textit{Oxford Triton 200}). Our input-output cryogenic setup is
nearly identical to that described in~\citet{Ofek2016} and~\citet{Minev2016},
aside from the differences evident in the schematic of our setup (see
Figs.~\ref{fig:setup}b) Notably, for the output lines between the
sample and the HEMT, we employed low-loss superconducting cables (\emph{CoaxCo
Ltd. SC-086/50-NbTi-NbTi PTFE}). The input line had a 12~GHz low-pass
filter (\emph{K\&L 6L250-12000/T26000-OP/O}) and the output line had
two broadband isolators (\emph{Quinstar CWJ1019-K414}), providing
a total of 36~dB of reverse isolation between the HEMT and the JPC.
Since the experiment spanned more than a dozen cool-downs, we note
that regular retightening of all cryogenic SMA connectors and screws
was observed to yield overall better performance.

\chapter{Additional experimental results\label{chap:Experimental-results} }

\addcontentsline{lof}{chapter}{Experimental results\lofpost}  
\addcontentsline{lot}{chapter}{Experimental results\lofpost}

\singlespacing 
\epigraph{
A strong claim of violation [of Bell's inequality] should be supported by at least a 5 sigma deviation.}
{Alain Aspect\\Rosenthal Lecture, 2018} 
\doublespacing\noindent \noindent\lettrine{T}{his} chapter presents experimental results
and control experiments that support the main experimental results
and conclusions presented in Chapter~\ref{chap:Introduction-and-overview}.
The characterization of the Hamiltonian parameters, coherence properties,
and other non-idealities of the two-transmon, one-readout-cavity device
employed in the experiment is discussed in Sec.~\ref{sec:Characterization-of-the}.
The calibration  of the tomography and control pulses and relevant
control experiments are discussed in Sec.~\ref{sec:Control-of-the}.
A summary of the  drive amplitudes and frequencies can be found in
Sec.~\ref{subsec:Atom-and-cavity}. Details of the experimental flow
of the catch and reverse protocol with regard to the FPGA controller
are discussed in Sec.~\ref{sec:Catching-and-reversing}. A comparison
between the predictions of the quantum trajectory description of the
experiment developed in Chapter~\ref{chap:theoretical-description-jumps}
and the main experimental results is presented in Sec.~\ref{subsec:Comparison-between-theory}.

\section{Characterization of the system\label{sec:Characterization-of-the}}

In this section, we describe the characterization of the Hamiltonian
and coherence parameters of the two-transmon, one-readout-cavity device
employed in the experiment. In reference to the protected Dark level,
which is engineered to be decoupled from the environment and readout
cavity, we  nickname the device ``Darkmon.''  The low-excitation
manifold of the Darkmon device is well described by the approximate
dispersive Hamiltonian, see Sec.~\ref{sec:circuit-design}, 
\begin{align}
\hat{H}/\hbar= & \omega_{\mathrm{B}}\hat{b}^{\dagger}\hat{b}-\frac{1}{2}\alpha_{\mathrm{B}}\hat{b}^{\dagger2}\hat{b}^{2}+\omega_{\mathrm{D}}\hat{d}^{\dagger}\hat{d}-\frac{1}{2}\alpha_{\mathrm{D}}\hat{d}^{\dagger2}\hat{d}^{2}-\chi_{\mathrm{DB}}\hat{b}^{\dagger}\hat{b}\hat{d}^{\dagger}\hat{d}\label{eq:Hamiltonian-of-sys}\\
 & \left(\omega_{\mathrm{C}}+\chi_{\mathrm{B}}\hat{b}^{\dagger}\hat{b}+\chi_{\mathrm{D}}\hat{d}^{\dagger}\hat{d}\right)\hat{c}^{\dagger}\hat{c}\,,\nonumber 
\end{align}
where $\omega_{\mathrm{D,B,C}}$ are the Dark, Bright, and cavity
mode frequencies, $\hat{d}$, $\hat{b},$ $\hat{c}$ are the respective
mode amplitude (annihilation) operators, $\alpha_{{\rm D}}$ ($\alpha_{{\rm B}}$)
is the Dark (Bright) transmon anharmonicity, $\chi_{\mathrm{D}}$
($\chi_{\mathrm{B}}$) is the dispersive shift between the Dark (Bright)
transmon and the readout cavity, and $\chi_{\mathrm{DB}}$ is the
dispersive shift between the two qubits. The Dark, $\ket{\mathrm{D}}$,
and Bright, $\B,$ states correspond to a single excitation in the
Dark and Bright transmon modes, $\hat{d}^{\dagger}\ket0$ and $\hat{b}^{\dagger}\ket0$,
respectively; see Fig.~\ref{fig:Darkmon-energy-level-diagram} for
a level diagram of the low-energy manifold. 

The readout cavity frequency was spectroscopically measured in reflection
\citep{Geerlings2013}, $\omega_{\mathrm{C}}/2\pi=8979.640\text{\,MHz}$,
and the extracted cavity linewidth agreed well with an independent
measurement of the energy-relaxation rate of the cavity extracted
from a time-domain ring-down measurement, $\kappa/2\pi=3.62\pm0.05\,\mathrm{MHz}.$
The cavity was observed to be well over-coupled; i.e., the coupling
quality factor, $Q_{c}$, dominated the internal quality factor, $Q_{i}$;
making it difficult to precisely extract $Q_{i}.$ The frequency and
anharmonicity of the B transmon were $\omega_{\mathrm{B}}/2\pi=5570.349\,\mathrm{MHz}$
and $\alpha_{\mathrm{B}}/2\pi=195\,\mathrm{MHz}$, respectively, measured
with two-tone pulsed spectroscopy \citep{Geerlings2013,Reagor2016}.
The frequency and anharmonicity of the D transmon, $\omega_{\mathrm{D}}/2\pi=4845.255\,\mathrm{MHz}$
and $\alpha_{\mathrm{D}}/2\pi=152\,\mathrm{MHz}$, respectively, were
measured in a modified two-tone spectroscopy sequence, where the $\G$
level was mapped to the $\B$ level at the end of the spectroscopy
sequence, before the readout, with $\pi$-pulse on the BG transition.
In a similar two-tone spectroscopy experiment, which included a pre-rotation
on either the BG or DG transition, and a measurement rotation after
the probe tone is turned off but before the readout tone is actuated,
the cross-Kerr coupling between the two qubits was measured to be
$\chi_{\mathrm{DB}}/2\pi=61\pm2\,\mathrm{MHz}$. In a standard energy-relaxation
experiment \citep{Geerlings2013}, the $\ket{\mathrm{B}}$ lifetime
was measured to be $T_{\mathrm{1}}^{\mathrm{B}}=28\pm2\,\mathrm{\mu s}$,
which we believe is limited by the Purcell effect with the readout
cavity mode, based on a finite-element calculation, see Sec.~\ref{subsec:Calculation-of-EPR}.
The Ramsey coherence time of $\B$ was $T_{\mathrm{2}}^{\mathrm{R,B}}=18\pm1\,\mathrm{\mu s}$,
possibly limited by photon shot noise \citep{Gambetta2006-dephasing,Rigetti2012}.
The measured Hamiltonian and coherence parameters of the device are
summarized in Table~\ref{tab:system-params}, where the drive parameters
employed in the experiment can also be found. 

\begin{table}
\begin{centering}
\renewcommand*\arraystretch{1.5}
\hspace*{-1.2cm} 
\begin{tabular}{rl|rl|rl}
\multicolumn{2}{c|}{\textbf{Readout cavity}} & \multicolumn{2}{c|}{\textbf{BG transition}} & \multicolumn{2}{c}{\textbf{DG transition}}\tabularnewline
\hline 
\hline 
\multicolumn{6}{c}{\textbf{\rule{0pt}{5ex}Mode frequencies and non-linear parameters}}\tabularnewline
\hline 
\textbf{\rule{0pt}{4ex}}$\omega_{\mathrm{C}}/2\pi=$ & $8979.640\text{\,MHz}$ & ~$\omega_{\mathrm{BG}}/2\pi=$ & $5570.349\text{\,MHz}$ & ~$\omega_{\mathrm{DG}}/2\pi=$ & $4845.255\text{\,MHz}$\tabularnewline
 &  & $\chi_{\mathrm{B}}/2\pi=$ & $-5.08\pm0.2\,\mathrm{MHz}$~ & $\chi_{\mathrm{D}}/2\pi=$ & $-0.33\pm0.08\,\mathrm{MHz}$\tabularnewline
 &  & $\alpha_{\mathrm{B}}/2\pi=$ & $195\pm2\,\mathrm{MHz}$ & $\alpha_{\mathrm{D}}/2\pi=$ & $152\pm2\,\mathrm{MHz}$\tabularnewline
 &  & \multicolumn{4}{c}{$\chi_{\mathrm{DB}}/2\pi=61\pm2\,\mathrm{MHz}$}\tabularnewline
\multicolumn{6}{c}{\textbf{\rule{0pt}{5ex}Coherence related parameters}}\tabularnewline
\hline 
\textbf{\rule{0pt}{5ex}}$\kappa/2\pi$= & $3.62\pm0.05\,\mathrm{MHz}$ & $T_{\mathrm{1}}^{\mathrm{B}}=$ & $28\pm2\,\mathrm{\mu s}$ & $T_{\mathrm{1}}^{\mathrm{D}}=$ & $116\pm5\,\mathrm{\mu s}$\tabularnewline
$\eta=$ & $0.33\pm0.03$ & $T_{\mathrm{2\mathrm{R}}}^{\mathrm{B}}=$ & $18\pm1\,\mathrm{\mu s}$ & $T_{\mathrm{2\mathrm{R}}}^{\mathrm{D}}=$ & $120\pm5\,\mathrm{\mu s}$\tabularnewline
$T_{\mathrm{int}}=$ & $260.0\,\mathrm{ns}$ & $T_{\mathrm{2\mathrm{E}}}^{\mathrm{B}}=$ & $25\pm2\,\mathrm{\mu s}$ & $T_{\mathrm{2\mathrm{E}}}^{\mathrm{D}}=$ & $162\pm6\,\mathrm{\mu s}$\tabularnewline
$n_{\mathrm{th}}^{\mathrm{C}}\le$ & $0.0017\pm0.0002$ & $n_{\mathrm{th}}^{\mathrm{B}}\le$ & $0.01\pm0.005$ & $n_{\mathrm{th}}^{\mathrm{D}}\leq$ & $0.05\pm0.01$\tabularnewline
\multicolumn{6}{c}{\textbf{\rule{0pt}{5ex}Drive amplitude and detuning parameters}}\tabularnewline
\hline 
\textbf{\rule{0pt}{5ex}}$\bar{n}=$ & $5.0\pm0.2$ & ~$\Omega_{\mathrm{B}0}/2\pi=$ & $1.20\pm0.01\,\mathrm{MHz}$ & ~$\Omega_{\mathrm{DG}}/2\pi=$ & $20\pm2\,\mathrm{kHz}$\tabularnewline
 &  & ~$\Omega_{\mathrm{B}1}/2\pi=$ & $0.60\pm0.01\,\mathrm{MHz}$ &  & \tabularnewline
$\Delta_{\mathrm{R}}=$ & $\chi_{\mathrm{B}}$ & ~$\Delta_{\mathrm{B}1}/2\pi=$ & $-30.0\,\text{MHz}$ & ~$\Delta_{\mathrm{DG}}/2\pi=$ & $-275.0\,\text{kHz}$\tabularnewline
\end{tabular}
\par\end{centering}
\caption[Compilation of experimental parameters]{\textbf{\label{tab:system-params}Compilation of  experimental parameters.}}
\end{table}

\subsection{Measurement-induced relaxation\textit{ $T_{1}(\bar{n})$\label{subsec:Measurement-induced-relaxation}}}

\begin{figure}
\begin{centering}
\includegraphics[scale=1.5]{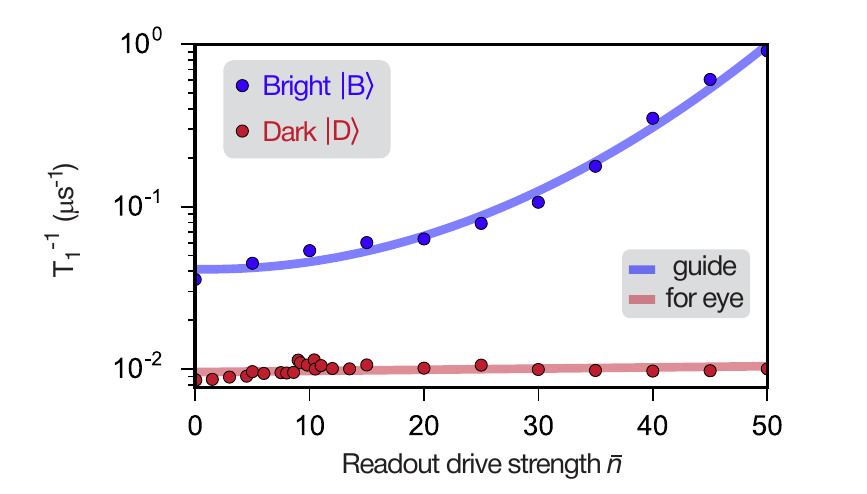}
\par\end{centering}
\caption[Measurement-induced energy relaxation $T_{1}(\bar{n})$]{\label{fig:T1-vs-nbar}\textbf{Measurement-induced energy relaxation
$T_{1}(\bar{n})$.} Energy relaxation rate ($T_{1}^{-1}$) of $\ket{\mathrm{B}}$
(blue dots) and $\ket{\mathrm{D}}$ (red dots) as a function of $\bar{n}$,
measured with the following protocol: after the atom is prepared in
either $\ket{\mathrm{B}}$ or $\ket{\mathrm{D}}$, the readout tone
($\mathrm{R}$) is turned on for duration $t_{\mathrm{read}}$ with
amplitude $\bar{n}$ (corresponding to the number of steady-state
photons in the readout cavity when excited on resonance), whereafter,
the population of the initial state is measured. As in all other experiments,
the readout drive is applied at the $\ket{\mathrm{B}}$ cavity frequency
($\omega_{\mathrm{C}}-\chi_{\mathrm{B}}$). The relaxation rates are
extracted from exponential fits of the population decay as a function
of $t_{\mathrm{read}}$, from $1.3\times10^{7}$ experimental realizations.
The solids lines are guides to the eye: blue line indicates the rapid
degradation of $T_{1}^{\mathrm{B}}$ as a function of the readout
strength, while the red line indicates the nearly constants $T_{1}^{\mathrm{D}}$
of the protected dark level.}
\end{figure}

It has been established in the superconducting qubit community \citep{Boissonneault2009-Photon-induced-relax,Slichter2012,Sank2016-T1vsNbar,Slichter2016-T1vsNbar}
that as a function of the number of photons circulating in the readout
cavity, $\bar{n},$ the energy-relaxation time, $T_{1}$, of a dispersively
coupled qubit is degraded. In Fig.~\ref{fig:T1-vs-nbar}, we show
a measurement of the $T_{1}$ lifetime of the $\B$ and $\D$ states
as a function of the readout drive strength, in units of the number
of circulating photons, $\bar{n}$, when the drive is resonant; the
measurement protocol is explained in the figure caption. As typically
observed in cQED experiments, the Bright level, which is directly
coupled to the readout cavity, exhibits a large parasitic measurement-induced
energy relaxation, $T_{1}^{\mathrm{B}}\left(\bar{n}\right)$ -- its
lifetime suffers more than an order of magnitude degradation. On the
other hand, perhaps surprisingly, the lifetime, $T_{1}^{\mathrm{D}},$
of the Dark state, $\ket{\mathrm{D}}$, remains essentially unaffected,
even at very large drive strengths, $\bar{n}\approx50$. In this sense,
the Dark level is protected from the $T_{1}\left(\bar{n}\right)$
parasitic effect.

\section{Control of the three-level atom\label{sec:Control-of-the}}

\subsection{Qubit pulses\label{subsec:Qubit-pulses-calib}}

The implementation of precise and coherent manipulation of the three-level
atom is important for the tomographic reconstruction of the flight
of the quantum jump as well the ability to faithfully reverse it.
One of the main sources of pulse infidelity is typically decoherence,
but the rather long coherence time of the Darkmon device relative
to the duration of the pulses employed in the experiment make it largely
unimportant, and instead, place emphasis on the technical details
of pulse generation and Hamiltonian non-idealities, such as leakage
to higher excited states. 

Mitigation of main technical non-idealities. The effect of the zero-order
hold of the FPGA digital-to-analog converter (DAC) was mitigated by
installing a 270~MHz low pass filter (\textit{Mini-Circuits BLP-300+})
on each of the analog output channels, see Sec.~\ref{sec:Microwave-setup}.
All microwave tones were generated with single-sideband IQ-controlled
modulation at a base intermediate frequency (IF) of 50~MHz, and the
lower radio-frequency (RF) sideband was used for the control tones
(detuned 50~MHz below the local oscillator (LO) frequency). The IQ
mixers were calibrated with a four stage iterative routine to minimize
carrier leakage, by tuning the DC offsets of the I and Q channels,
and to suppress the RF image, by minimizing the quadrature skew and
IQ gain imbalance. The LO leakage could typically be suppressed to
$\approx-70$~dB relative to the RF tone. Spurious intermodulation
tones generated by higher-order non-linear terms present in the mixers
{[}i.e., third-order intercept-point (IP3) products{]} were generally
negligible as the mixers were not typically driven near saturation,
but bandpass filters were installed on the RF outputs of all mixers
to nonetheless suppress any spurious tones. Excess noise from the
following RF amplifier (\textit{MiniCircuits ZVA-183-S+}) was suppressed
by 80\,dB when the control drives were turned off by use of a high-isolation
SPST switch (\textit{Analog Device HMC-C019}).

The pulses applied to the Dark and Bright transition were calibrated
with a combination of Rabi, derivative removal via adiabatic gate
(DRAG) \citep{JChow2010-DRAG}, \emph{All-XY} \citep{Reed2013}, and
amplitude pulse train sequences \citep{Bylander2011}. Pulse timings
and delays, especially between the analog channels and the SPST switch
digital markers, were calibrated with a wide-bandwidth oscilloscope
with ultra-low jitter (\emph{Keysight 86100D Infiniium DCA-X}). The
alignment was verified by performing a Gaussian qubit $\pi$ pulse
on the GB transition and varying the delay between the rise of the
SPST digital marker and the signal on the analog IQ pair playing the
pulse.

\subsection{Tomography of the three-level atom\label{subsec:Tomography-of-three-level}}

At the end of each experimental realization, we performed one of 15
rotation sequences on the atom that transferred information about
one component of the density matrix, $\hat{\rho}_{a}$, to the population
of $\ket{\mathrm{B}}$, which was measured with a 600~ns square pulse
on the readout cavity. Pulses were calibrated as discussed in Sec.~\ref{subsec:Qubit-pulses-calib}.
The readout signal was demodulated with the appropriate digital filter
function required to realize temporal mode matching \citep{Eichler2012-itinerant-entanglement}.
To remove the effect of potential systematic offset errors in the
readout signal, we subtracted the measurement results of operator
components of $\hat{\rho}_{a}$ and their opposites. From the measurement
results of this protocol, we reconstructed the density matrix $\hat{\rho}_{a}$,
and subsequently parametrized it the useful form 
\begin{equation}
\hat{\rho}_{a}=\begin{pmatrix}\frac{N}{2}\left(1-Z_{\mathrm{GD}}\right) & \frac{N}{2}\left(X_{\mathrm{GD}}+iY_{\mathrm{GD}}\right) & R_{\mathrm{BG}}+iI_{\mathrm{BG}}\\
\frac{N}{2}\left(X_{\mathrm{GD}}-iY_{\mathrm{GD}}\right) & \frac{N}{2}\left(1+Z_{\mathrm{GD}}\right) & R_{\mathrm{BD}}+iI_{\mathrm{BD}}\\
R_{\mathrm{BG}}-iI_{\mathrm{BG}} & R_{\mathrm{BD}}-iI_{\mathrm{BD}} & 1-N
\end{pmatrix},\label{eq:rhoa}
\end{equation}
where $X_{\mathrm{GD}},Y_{\mathrm{GD}},$ and $Z_{\mathrm{GD}}$ are
the Bloch vector components of the GD manifold, $N$ is the total
population of the $\ket{\mathrm{G}}$ and $\ket{\mathrm{D}}$ states,
while $R_{\mathrm{BG}},R_{\mathrm{BD}},I_{\mathrm{BG}}$ and $I_{\mathrm{BD}}$
are the coherences associated with $\ket{\mathrm{B}}$, relative to
the GD manifold. The measured population in $\ket{\mathrm{B}}$, $1-N$,
remains below 0.03 during the quantum jump, see Fig.~\ref{fig:tomo_tmid}.
Tomographic reconstruction was calibrated and verified by preparing
Clifford states, accounting for the readout fidelity of 97\%.

\paragraph{Control experiment.}

\begin{figure}
\begin{centering}
\includegraphics[scale=1.2]{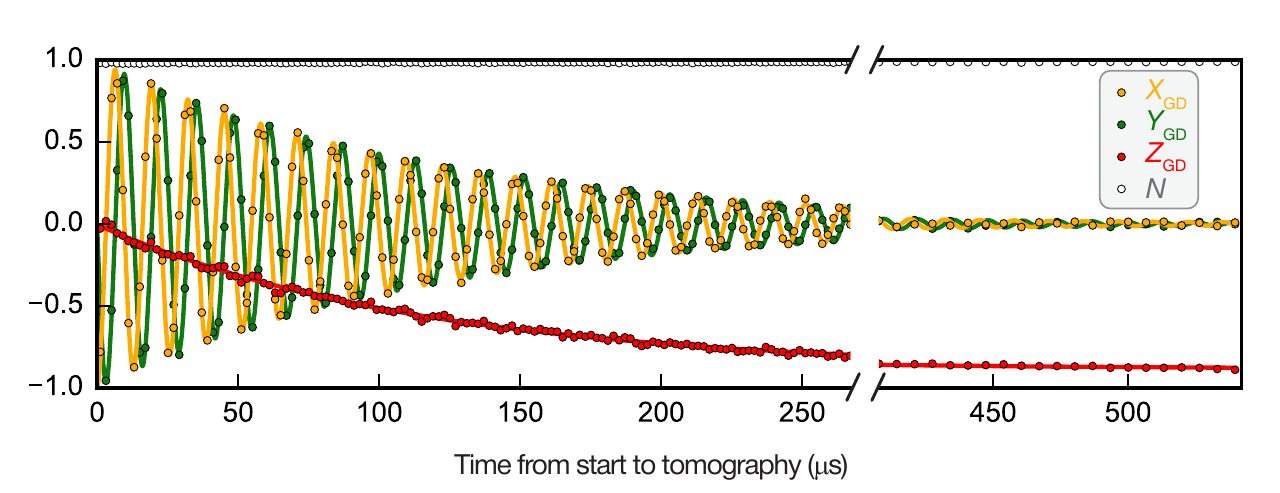}
\par\end{centering}
\caption[Control experiment: time-resolved tomogram of the free evolution of
a DG superposition]{\label{fig:time-resolved-tomo-D}\textbf{Control experiment: time-resolved
tomogram of the free evolution of a DG superposition. }The atom is
prepared in $\frac{1}{\sqrt{2}}\left(\D-\G\right)$ and tomography
is performed after a varied delay. Dots: reconstructed conditional
GD tomogram ($X_{\mathrm{DG}},Y_{\mathrm{DG}}$, and $Z_{\mathrm{DG}}$)
and population in DG manifold, $N$, see Eq.~(\ref{eq:rhoa}). Solid
lines: theoretical fits. }
\end{figure}

\begin{figure}
\centering{}\includegraphics[width=120mm]{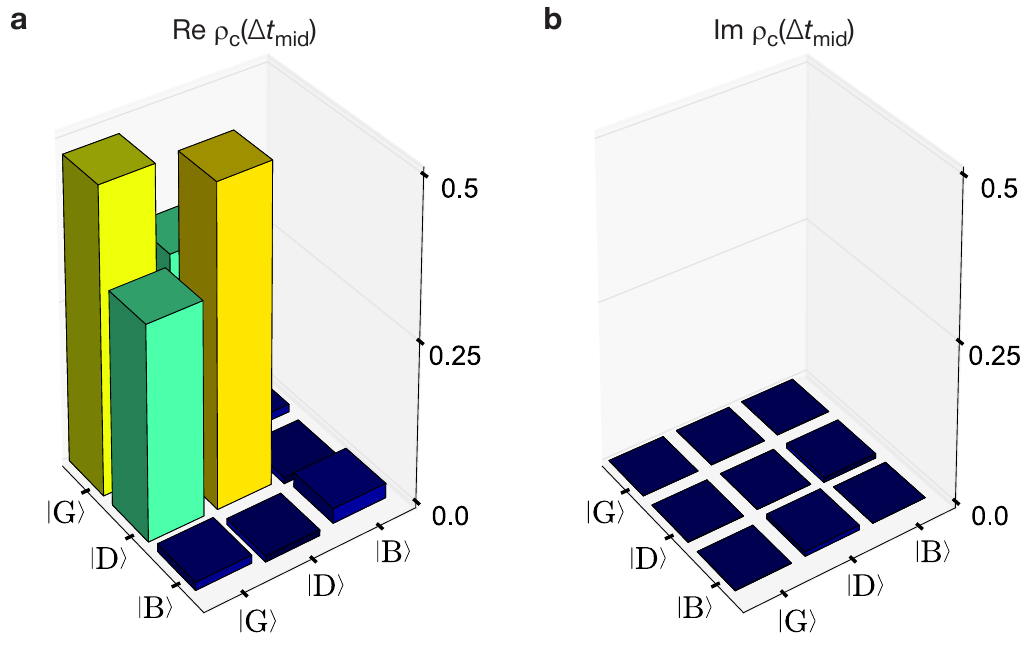}
\caption[Mid-flight tomogram]{\label{fig:tomo_tmid}\textbf{Mid-flight tomogram. }The plots show
the real (a) and imaginary (b) parts of the conditional density matrix,
$\rho_{\mathrm{c}}$, at the mid flight of the quantum jump ($\Delta t_{\mathrm{catch}}=\Delta t_{\mathrm{mid}}$),
in the presence of the Rabi drive from $\ket{\mathrm{G}}$ to $\ket{\mathrm{D}}$
($\Delta t_{\mathrm{off}}=0$). The population of the $\ket{\mathrm{B}}$
state is 0.023, and the magnitude of all imaginary components is less
than 0.007. }
\end{figure}

In Fig.~\ref{fig:time-resolved-tomo-D}, we show the results of a
control experiment where we verified the Ramsey coherence ($T_{\mathrm{2R}}^{\mathrm{D}}$)
and energy relaxation ($T_{\mathrm{1}}^{\mathrm{D}}$) times of the
DG transition with our tomography method. Solid lines are fitted theoretical
curves for the free evolution of the prepared initial state $\frac{1}{\sqrt{2}}\left(\D-\G\right)$.
The $T_{\mathrm{2R}}^{\mathrm{D}}=119.2\,\mathrm{\mu s}$ value gained
from the simultaneous fit of $X_{\mathrm{DG}}(t)$ and $Y_{\mathrm{DG}}(t)$
matches the lifetime independently obtained from a standard $T_{\mathrm{2R}}$
measurement. Similarly, the value of $T_{\mathrm{1}}^{\mathrm{D}}=115.4\,\mathrm{\mu s}$
extracted from an exponential fit of $Z_{\mathrm{DG}}(t)$ matches
the value obtained from a standard $T_{\mathrm{1}}$ measurement.
We note that our tomography procedure is well calibrated and skew-free,
as evident in the zero steady-state values of $X_{\mathrm{DG}}$ and
$Y_{\mathrm{DG}}$. The steady state $Z_{\mathrm{DG}}$ corresponds
to the thermal population of the dark state $n_{\mathrm{th}}^{\mathrm{D}}$.
It has recently been shown that residual thermal populations in cQED
systems can be significantly reduced by properly thermalizing the
input-output lines \citep{Yeh2017Atten,Wang2019-cav-atten}.

\paragraph{Mid-flight tomogram.}

In the presence of the coherent Rabi drive $\Odg$ (corresponding
to catch parameter $\Delta t_{\mathrm{off}}=0$), the complete tomogram
of the three-level atom was reconstructed, and a slice at the mid-flight
time, $\Delta t_{\mathrm{mid}}$, is shown in Fig.~\ref{fig:tomo_tmid}.
All imaginary components of the reconstructed conditional density
matrix, $\rho_{\mathrm{c}}$, are negligibly small, less than 0.007,
as expected, see Sec.~\ref{subsec:Comparison-between-theory}, for
well-calibrated tomographic phase control. The population of the $\B$
state, 0.023, is nearly negligible as well, as it is conditioned away
by the IQ filter, see Sec.~\ref{subsec:IQ-filter}.

\subsection{Atom and cavity drives \label{subsec:Atom-and-cavity}}

In all experiments, unless noted otherwise, the following drive parameters
were used: The DG Rabi drive, $\Omega_{\mathrm{DG}}$, was applied
275~kHz below $\omega_{\mathrm{D}}$ to account for the Stark shift
of the cavity. The BG drive, $\Omega_{\mathrm{BG}}$, was realized
as a bi-chromatic tone in order to unselectively address the BG transition,
which was broadened and Stark shifted due to the coupling between
$\ket{\mathrm{B}}$ and the readout cavity. Specifically, we addressed
transitions from $\ket{\mathrm{G}}$ to $\ket{\mathrm{B}}$ with a
Rabi drive $\Omega_{\mathrm{B0}}/2\pi=1.20\pm0.01\,$~MHz at frequency
$\omega_{\mathrm{BG}}$, whereas transitions from $\ket{\mathrm{B}}$
to $\ket{\mathrm{G}}$ were addressed with a Rabi drive $\Omega_{\mathrm{B1}}/2\pi=0.60\pm0.01\,$~MHz
tuned 30~MHz below $\omega_{\mathrm{BG}}$. This bi-chromatic scheme
provided the ability to tune the up-click and down-click rates independently,
but otherwise essentially functioned as an incoherent broad-band source.
In Table~\ref{tab:Summary-of-timescales.}, we summarize the hierarchy
of timescales established by the drive amplitudes and frequencies
as well as the relevant decoherence properties of the atom.

\begin{table}[!ht]
\begin{centering}
\addtolength{\tabcolsep}{2pt} 
\renewcommand*{\arraystretch}{1.5}
\begin{tabular}{cc>{\raggedright}p{0.65\columnwidth}}
\hline 
\textbf{Symbol}  & \textbf{Value}  & \textbf{Description}\tabularnewline
\hline 
$\Gamma^{-1}$  & $\approx8.8$\,ns  & Effective measurement time of $\ket{\mathrm{B}}$, approximately given
by $1/\kappa\bar{n}$, where $\bar{n}=5\pm0.2$ in the main experiment\tabularnewline
$\kappa^{-1}$  & $44.0\pm0.06$\,ns  & Readout cavity lifetime\tabularnewline
$T_{\mathrm{int}}$  & 260.0\,ns  & Integration time of the measurement record, set in the controller
at the beginning of the experiment \tabularnewline
$\Gamma_{\mathrm{BG}}^{-1}$  & $0.99\pm0.06\,\mathrm{\mu s}$  & Average time the atom rests in $\ket{\mathrm{G}}$ before an excitation
to $\ket{\mathrm{B}}$, see Fig.~\ref{fig:jumps}b\tabularnewline
$\Delta t_{\mathrm{mid}}$  & $3.95\,\mathrm{\mu s}$  & No-click duration for reaching $Z_{\mathrm{GD}}=0$ in the flight
of the quantum jump from $\ket{\mathrm{G}}$ to $\ket{\mathrm{D}}$,
in the full presence of $\Omega_{\mathrm{DG}}$, see Fig.~\ref{fig:catch}b\tabularnewline
$\Gamma_{\mathrm{GD}}^{-1}$  & $30.8\pm0.4\,\mathrm{\mu s}$  & Average time the atom stays in $\ket{\mathrm{D}}$ before returning
to $\ket{\mathrm{G}}$ and being detected, see Fig.~\ref{fig:jumps}b\tabularnewline
$T_{1}^{\mathrm{D}}$  & $116\pm5\,\mathrm{\mu s}$  & Energy relaxation time of $\ket{\mathrm{D}}$\tabularnewline
$T_{2\mathrm{R}}^{\mathrm{D}}$  & $120\pm5\,\mathrm{\mu s}$  & Ramsey coherence time of $\ket{\mathrm{D}}$\tabularnewline
$T_{2\mathrm{E}}^{\mathrm{D}}$  & $162\pm6\,\mathrm{\mu s}$  & Echo coherence time of $\ket{\mathrm{D}}$\tabularnewline
$\Gamma_{\mathrm{DG}}^{-1}$  & $220\pm5\,\mathrm{\mu s}$  & Average time between two consecutive $\ket{\mathrm{G}}$ to $\ket{\mathrm{D}}$
jumps\tabularnewline
\end{tabular}
\par\end{centering}
\caption[Summary of timescales]{\textbf{Summary of timescales. }List of the characteristic timescales
involved in the catch and reverse experiment. The Hamiltonian parameters
of the system are summarized in Sec.~\ref{sec:Characterization-of-the}.
\textbf{\label{tab:Summary-of-timescales.}}}
\end{table}

\section{Monitoring quantum jumps in real time }

\subsection{IQ filter \label{subsec:IQ-filter}}

To mitigate the effects of imperfections in the atom readout scheme
in extracting a $\ket{\mathrm{B}}$/not-$\ket{\mathrm{B}}$ result,
we applied a two-point, hysteretic IQ filter, implemented on the FPGA
controller in real time. The filter is realized by comparing the present
quadrature record values $\left\{ I_{\mathrm{rec}},Q_{\mathrm{rec}}\right\} $,
with three thresholds ($I_{\mathrm{B}},I_{\bar{\mathrm{B}}},$ and
$Q_{\mathrm{B}}$) in the following way: 
\begin{center}
\begin{tabular}{>{\raggedleft}m{20mm}|>{\centering}p{30mm}|>{\centering}p{30mm}|>{\centering}p{30mm}}
\textbf{Input: } & $Q_{\mathrm{rec}}\geq Q_{\mathrm{B}}$ or

$I_{\mathrm{rec}}>I_{\mathrm{B}}$  & $Q_{\mathrm{rec}}<Q_{\mathrm{B}}$ and

$I_{\mathrm{rec}}<I_{\bar{\mathrm{B}}}$  & $Q_{\mathrm{rec}}<Q_{\mathrm{B}}$ and

$I_{\bar{\mathrm{B}}}\leq I_{\mathrm{rec}}\leq I_{\mathrm{B}}$\tabularnewline
\hline 
\textbf{Output: } & $\ket{\mathrm{B}}$  & not-$\ket{\mathrm{B}}$  & previous\tabularnewline
\end{tabular}
\par\end{center}

\noindent The filter and thresholds were selected to provide a best
estimate of the time of a click, operationally understood as a change
in the filter output from $\ket{\mathrm{B}}$ to not-$\ket{\mathrm{B}}$.
The $I_{{\mathrm{B}}}$ and $I_{\bar{\mathrm{B}}}$ thresholds were
chosen 1.5 standard deviations away from the I-quadrature mean of
the $\ket{\mathrm{B}}$ and not-$\ket{\mathrm{B}}$ distributions,
respectively. The $Q_{{\mathrm{B}}}$ threshold was chosen 3 standard
deviations away from the Q-quadrature mean. Higher excited states
of the atom were selected out by $Q_{\mathrm{rec}}$ values exceeding
the $Q_{\mathrm{B}}$ threshold. 

\subsection{Unconditioned monitoring}

\begin{figure}
\centering{}\includegraphics[width=145mm]{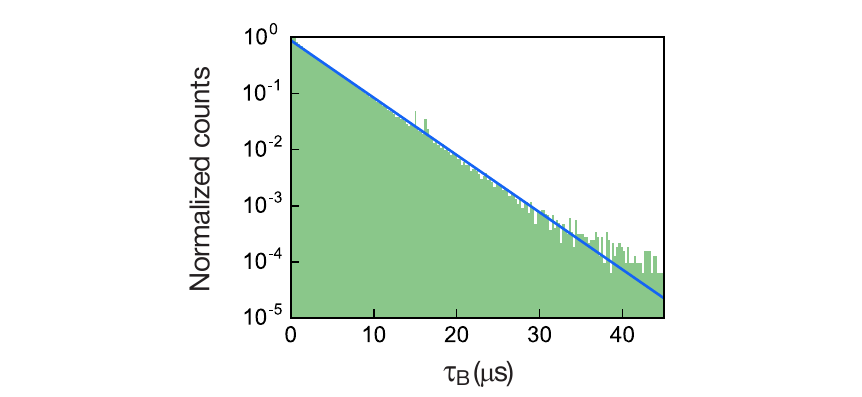}
\caption[Waiting time to switch from a $\B$ to not-$\B$ state assignment
result]{\label{fig:B-wait-time}\textbf{Waiting time to switch from a $\B$
to not-$\B$ state assignment result.} Semi-log plot of the histogram
(shaded green) of the duration of times corresponding to $\B$-measurement
results, $\tau_{\operatorname{B}}$, for 3.2~s of continuous data
of the type shown in Fig.~\ref{fig:jumps}a. Solid line is an exponential
fit, which yields a $4.2\pm0.03\,\mathrm{\mu s}$ time constant. }
\end{figure}

In Sec.~\ref{sec:Unconditioned-monitoring-of}, we described a protocol
for the unconditioned monitoring of the quantum jumps where the atom
is subject to the continuous Rabi drives $\Omega_{\mathrm{BG}}$ and
$\Omega_{\mathrm{DG}}$, as depicted in Fig.~\ref{fig:setup}. From
the continuous tracking of the quantum jumps, over 3.2~s. of data,
we histogrammed the times, $\tau_{\operatorname{not-B}}$, spent in
not-$\ket{\mathrm{B}}$, Fig.~\ref{fig:jumps}b. In Fig.~\ref{fig:B-wait-time},
we show the complimentary histogram for the times, $\tau_{{\rm B}},$
spent in $\B$, which is unlike the latter, in that it follows a single
exponential decay law. This single Poisson process character follows
from the fact that the $\B$ measurement result collapses the atom
to a single state, $\B$, unlike the not-$\B$ result. The average
time spent in $\B$, extracted from the fit, is $\bar{\tau}_{{\rm B}}=4.2\pm0.03\,\mathrm{\mu s}$.

\pagebreak{}

\section{Catching and reversing the jump \label{sec:Catching-and-reversing}}

\subsection{Experiment flow}

\begin{figure}
\begin{centering}
\includegraphics[angle=-90,scale=1.1]{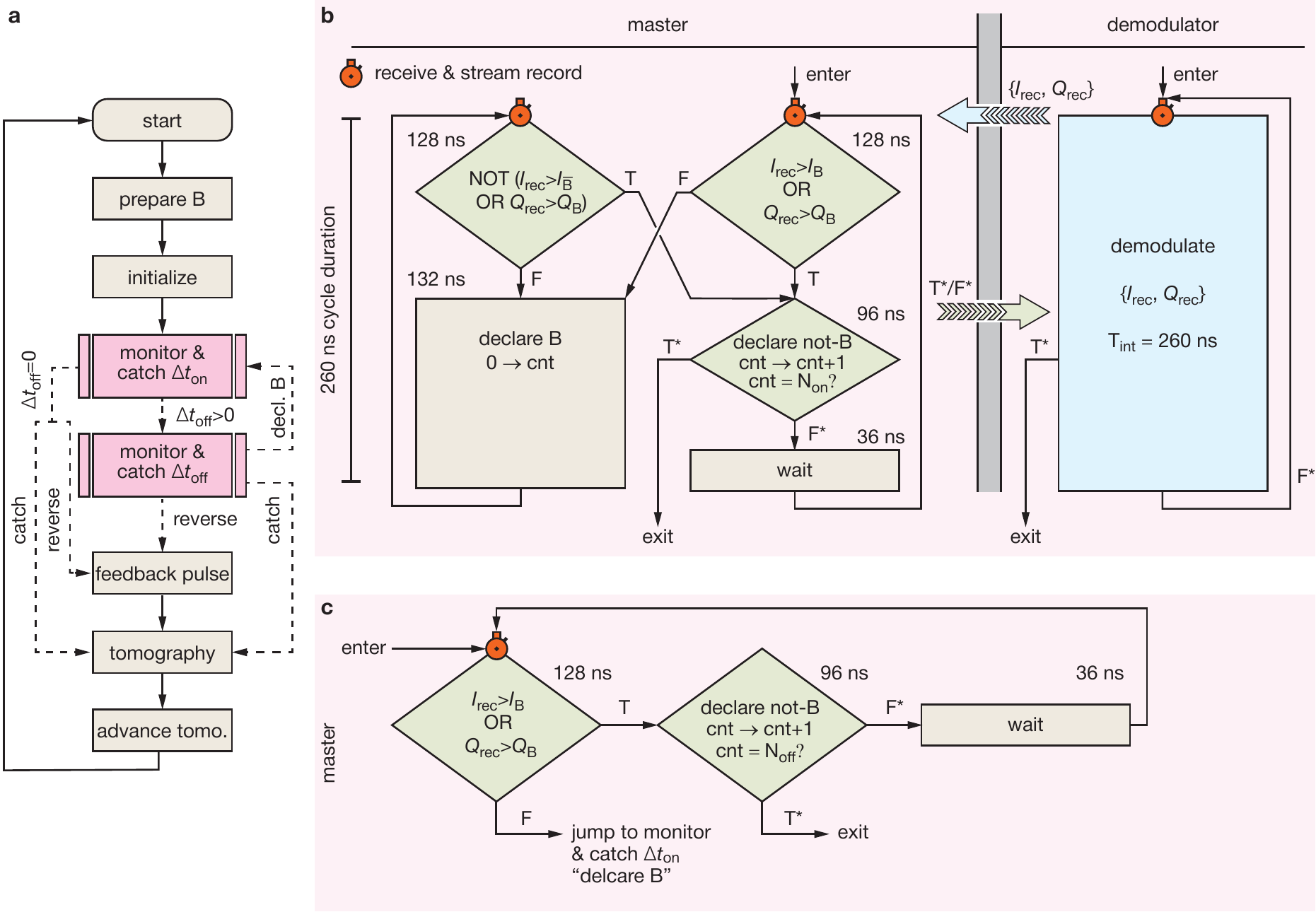}
\par\end{centering}
\caption[Experiment flow]{\label{fig:experiment-flow}\textbf{Experiment flow. }See text for
detailed description.}
\end{figure}
Figure \ref{fig:experiment-flow}a shows a flowchart representation
of steps involved in the catch and reverse protocol. In the following,
we describe each block in the diagram in the order in which it would
be executed.

\paragraph{Start:}

internal memory registers are set to zero \citep{Ofek2016,Liu2016Thesis},
including the no-click counter ``cnt,'' defined below. 

\paragraph{Prepare B:}

controller deterministically prepares the atom in $\ket{\mathrm{B}}$,
a maximally conservative initial state, with measurement-based feedback
\citep{Riste2012-qubit-measure-reset}. 

\paragraph{Initialize:}

controller turns on the atom ($\Omega_{\mathrm{BG}}$ and $\Omega_{\mathrm{DG}}$)
and cavity drives ($\mathrm{R}$) and begins demodulation. 

\paragraph{Monitor and catch $\Delta t_{\mathrm{on}}$: }

with all drives on ($\Omega_{\mathrm{BG}},\Omega_{\mathrm{DG}}$,
and $\mathrm{R}$), the controller actively monitors the cavity output
signal until it detects no-clicks for duration $\Delta t_{\mathrm{on}}$,
as described in panel (b), whereafter, the controller proceeds to
``monitor and catch $\Delta t_{\mathrm{off}}$'' in the case that
$\Delta t_{\mathrm{off}}>0$; otherwise, for $\Delta t_{\mathrm{off}}=0$,
the controller proceeds to ``tomography'' (``feedback pulse'')
for the catch (reverse) protocol.

\paragraph{Monitor and catch $\Delta t_{\mathrm{off}}$: }

with the Rabi drive $\Omega_{\mathrm{DG}}$ off, while keeping the
drives $\Omega_{\mathrm{BG}}$ and R on, the controller continues
to monitor the output signal. The controller exits the routine only
if it detects a click, proceeding to the ``declare B'' step of the
``monitor and catch $\Delta t_{\mathrm{on}}$'' routine, or if no
further clicks are detected for the pre-defined duration $\Delta t_{\mathrm{off}}$,
proceeding to ``tomography'' (``feedback pulse'') for the catch
(reverse) protocol.

\paragraph{Feedback pulse: }

with all the continuous drives turned off, the controller performs
a pulse on the DG transition of the atom, defined by the two angles
$\left\{ \theta_{I}\left(\Delta t_{\mathrm{catch}}\right),\varphi_{I}\left(\Delta t_{\mathrm{catch}}\right)\right\} $.

\paragraph{Tomography: }

controller performs next-in-order tomography sequence (see Sec.~\ref{subsec:Tomography-of-three-level})
while the demodulator finishes processing the final data in its pipeline.

\paragraph{Advance tomo.: }

tomography sequence counter is incremented, and after a $50~\mathrm{\mu s}$
delay, the next realization of the experiment is started.

\subsubsection{Logic and timing of catch subroutines}

\paragraph{Monitor and catch $\Delta t_{\mathrm{on}}$.}

Figure \ref{fig:experiment-flow}b shows a concurrent-programming
flowchart representation of the ``monitor and catch $\Delta t_{\mathrm{on}}$''
routine. Displayed are the master and demodulator modules of the controller.
The demodulator outputs a pair of 16 bit signed integers, $\left\{ I_{\mathrm{rec}},Q_{\mathrm{rec}}\right\} $,
every $T_{\mathrm{int}}=260$~ns, which is routed to the master module,
as depicted by the large left-pointing arrow. The master module implements
the IQ filter (see Sec.~\ref{subsec:IQ-filter}) and tracks the number
of consecutive not-$\ket{\mathrm{B}}$ measurement results with the
counter cnt. The counter thus keeps track of the no-click time elapsed
since the last click, which is understood as a change in the measurement
result from $\ket{\mathrm{B}}$ to not-$\ket{\mathrm{B}}$. When the
counter reaches the critical value $N_{\mathrm{on}}$, corresponding
to $\Delta t_{\mathrm{on}}$, the master and demodulator modules synchronously
exit the current routine, see the T{*} branch of the ``declare not-B''
decision block. Until this condition is fulfilled (F{*}), the two
modules proceed within the current routine as depicted by the black
flowlines. 

To minimize latency and maximize computation throughput, the master
and demodulator were designed to be independent sequential processes
running concurrently on the FPGA controller, communicating strictly
through synchronous message passing, which imposed stringent synchronization
and execution time constraints. All master inter-module logic was
constrained to run at a 260~ns cycle, the start of which necessarily
was imposed to coincide with a ``receive \& stream record'' operation,
here, denoted by the stopwatch. In other words, this imposed the algorithmic
constraint that all flowchart paths staring at a stopwatch and ending
in a stopwatch, itself or other, were constrained to a 260\,ns execution
timing. A second key timing constraint was imposed by the time required
to propagate signals between the different FPGA cards, which corresponded
to a minimum branching-instruction duration of 76~ns.

\paragraph{Monitor and catch $\Delta t_{\mathrm{off}}$.}

Figure \ref{fig:experiment-flow}c shows a concurrent-programming
flowchart representation of the master module of the ``monitor and
catch $\Delta t_{\mathrm{off}}$'' routine. The corresponding demodulation-module
flowchart is identical to that shown of panel (b); hence, it is not
shown. This routine functions in following manner: If a $\ket{\mathrm{B}}$
outcome is detected, the controller jumps to the ``declare B'' block
of the monitor \& catch $\Delta t_{\mathrm{on}}$ routine; otherwise,
when only not-$\ket{\mathrm{B}}$ outcomes are observed, and the counter
reaches the critical value $N_{\mathrm{off}}$, corresponding to $\Delta t_{\mathrm{catch}}=\Delta t_{\mathrm{on}}+\Delta t_{\mathrm{off}}$,
the controller exits the routine.

\section{Comparison between theory and experiment \label{subsec:Comparison-between-theory}\label{subsec:Error-analysis}}

In this section, we present the comparison between the results of
the quantum jumps experiment and the predictions of the quantum trajectory
theory of the experiment developed in Chapter~\ref{chap:theoretical-description-jumps}.
The results agree with the theoretical predictions, accounting for
known imperfections, essentially without adjustable parameters. Simulation
plots courtesy of H.J. Carmichael. 

\subsection{Simulated data sets}

\paragraph{\textit{\emph{Independently measured parameters.}}}

The parameters used in the Monte Carlo simulation described in Sec.~\ref{subsec:Simulation-of-linear}
are listed in Table~\ref{table:table2}. Nearly all are set to the
value at the center of the range quoted in Table~\ref{tab:system-params},
with three exceptions: i) $T_{1}^{{\rm B}}$ and $T_{1}^{{\rm D}}$
are set to lower values in response to the photon number dependence
of the readout displayed in Fig.~\ref{fig:T1-vs-nbar}, ii) $\Omega_{{\rm DG}}/2\pi$
is set higher, but still falls inside the experimental error bars,
and iii) $n_{{\rm th}}^{{\rm C}}=0$. Notably, of the three exceptions,
only $\Omega_{{\rm DG}}/2\pi$ has a noticeable effect on the comparison
between simulated and experimental data sets.

\paragraph{\textit{\emph{Leakage from the }}\emph{GBD}\textit{\emph{-manifold.}}\emph{ }}

As discussed in Sec.~\ref{sec:circuit-design}, see Fig.~\ref{fig:Darkmon-energy-level-diagram},
the Darkmon system has higher excited states, which are generally
unimportant, but do contribute a small imperfection that needs to
be considered to qualitatively account for the results. As discussed
in Sec.~\ref{subsec:Simulation-of-linear}, we model the effect of
leakage from the GBD manifold by adding a single additional higher-excited
state level, denoted $\ket{\mathrm{F}}.$ The additional random jumps
to state $|{\rm F}\rangle$ are governed by four parameters that are
not independently measured; they serve as fitting parameters, required
to bring the simulation into agreement with the asymptotic behavior
of ${\rm Z}(\Delta t_{{\rm catch}})$, which, without leakage to $|{\rm F}\rangle$,
settles to a value higher than is measured in the experiment. The
evolution of the ${\rm X}(\Delta t_{{\rm catch}})$ is largely unaffected
by the assignment of these parameters, where any change that does
occur can be offset by adjusting $\Omega_{{\rm DG}}/2\pi$ while staying
within the experimental error bars.

\paragraph{\textit{\emph{Ensemble average.}}\emph{ }}

\begin{figure}[t]
\centering{}\includegraphics[width=145mm]{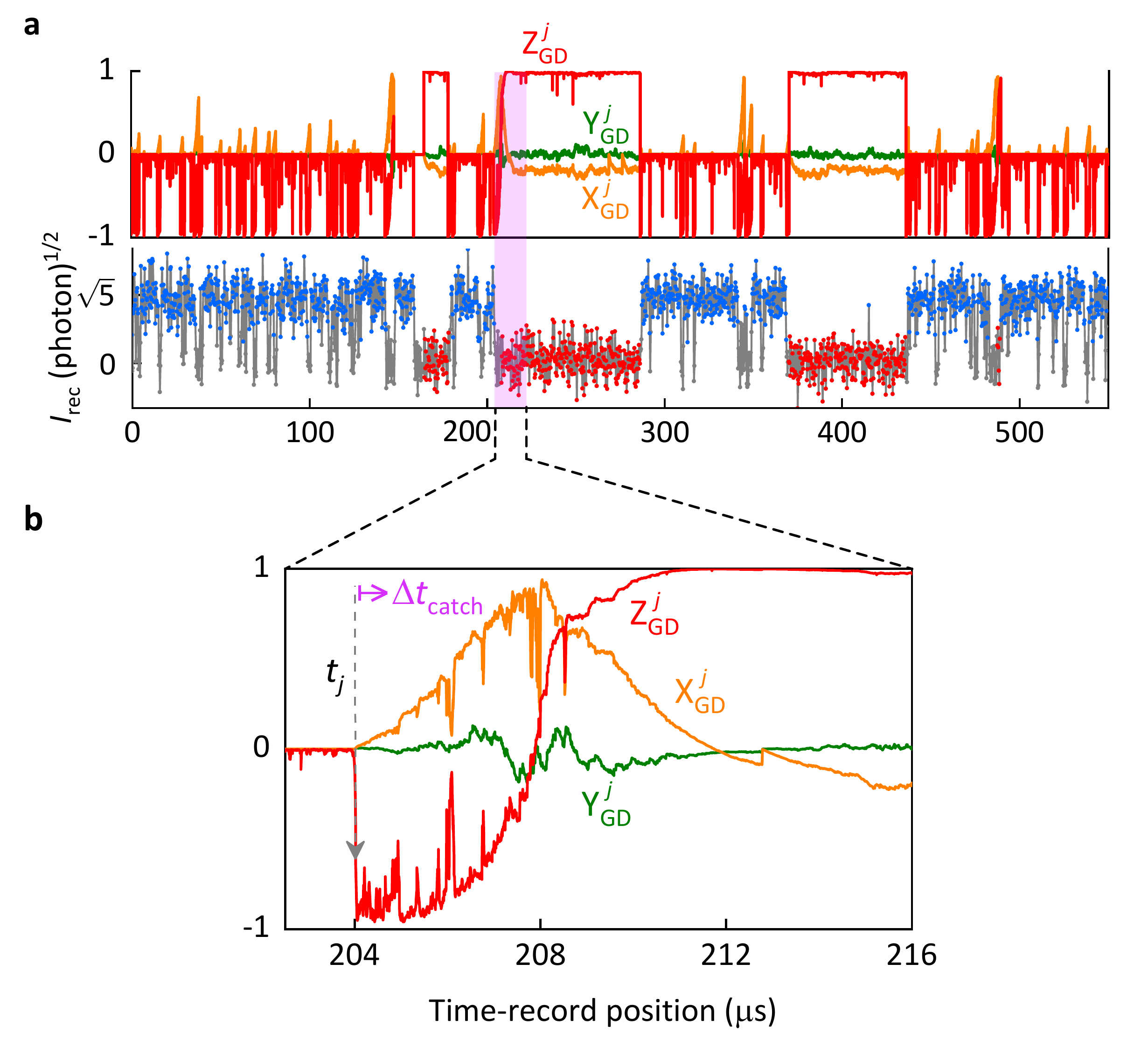}\caption[Sampling of the Monte-Carlo simulation (courtesy of H.J. Carmichael)]{\label{fig:monte-carlo} \textbf{Sampling of the Monte-Carlo simulation.}
\textbf{a,} Simulated measurement quadrature $I_{{\rm rec}}$ and
correlated trajectory computed from Eqs.~(\ref{eqn:correlated_Z_GD})
and (\ref{eqn:correlated_X_GD=000026Y_GD}). Three sample intervals
are shown. The earliest corresponds to leakage from the GBD-manifold,
where a jump from $|{\rm G}\rangle$ to $|{\rm F}\rangle$ is followed
by a jump from $|{\rm F}\rangle$ to $|{\rm D}\rangle$. The second
and third sample intervals correspond to direct transitions from $|{\rm G}\rangle$
to $|{\rm D}\rangle$, which are continuously monitored and the object
of the experiment. \textbf{b,} Expanded view of the shaded region
of the second sample interval in panel (a). The evolution is continuous
but not smooth, due to backaction noise from the continuously monitored
readout. This feature is in sharp contrast to the perfect ``no-click''
readout upon which the simple theory of Sec.~\ref{sec:Fluorescence-monitored-by}
is based. Figure courtesy of H.J. Carmichael.}
\end{figure}

\begin{figure}
\centering{}\includegraphics[width=120mm]{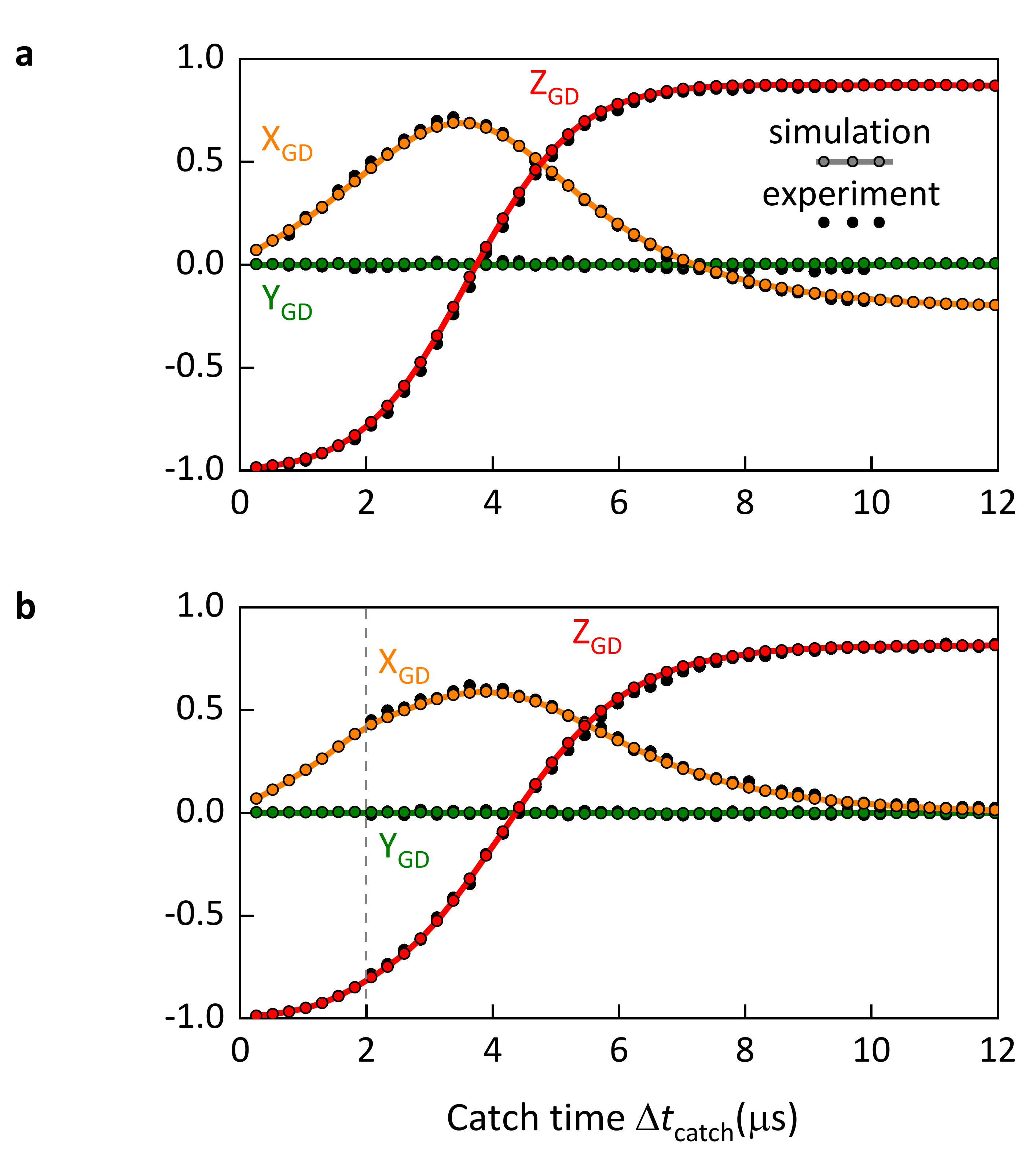}\caption[Comparison between simulation and experiment (courtesy of H.J. Carmichael)]{\label{fig:simulation_vs_experiment} \textbf{Comparison between
simulation and experiment.} \textbf{a,} Simulated data set obtained
with Rabi drive $\Omega_{{\rm DG}}$ turned on for the entire $\Delta t_{{\rm catch}}$;
parameters taken from Table \ref{table:table2} and leakage from the
GBD-manifold included with $(\gamma_{{\rm FG}},\gamma_{{\rm FD}})/2\pi=0.38\mkern2mu {\rm kHz}$
and $(\gamma_{{\rm GF}},\gamma_{{\rm DF}})/2\pi=11.24\mkern2mu {\rm kHz}$.
\textbf{b,} Simulated data set obtained with Rabi drive $\Omega_{{\rm DG}}$
turned off at time $\Delta t_{{\rm on}}=2\mkern2mu \mu{\rm s}$; parameters
taken from Table \ref{table:table2} and leakage from the GBD-manifold
included with $\gamma_{{\rm FG}}/2\pi=0.217\mkern2mu {\rm kHz}$,
$\gamma_{{\rm FD}}/2\pi=4.34\mkern2mu {\rm kHz}$, $\gamma_{{\rm GF}}/2\pi=11.08\mkern2mu {\rm kHz}$,
and $\gamma_{{\rm DF}}/2\pi=15.88\mkern2mu {\rm kHz}$. When leakage
from the GBD-manifold is omitted, the ${\rm Z}_{{\rm GD}}$ curve
rises more sharply and settles to a value that is 10\% (20\%) higher
in panel (a) (panel (b)). Figure courtesy of H.J. Carmichael.}
\end{figure}

Simulated data sets are computed as an ensemble average by sampling
an ongoing Monte Carlo simulation, numerically implementing the model
outlined in Eqs.~(\ref{eqn:SSE_continuous})--(\ref{eqn:jumps_to_F}).
Quadratures $I_{{\rm rec}}$ and $Q_{{\rm rec}}$ are computed from
Eqs.~(\ref{eqn:simulated_I_int}) and (\ref{eqn:simulated_Q_int}),
digitized with integration time $T_{{\rm int}}=260\mkern2mu {\rm ns}$,
and then, as in the experiment, a hysteric filter is used to locate
``click'' events ($\Delta t_{{\rm catch}}=0$) corresponding to
an inferred change of state from $|{\rm B}\rangle$ to not-$|{\rm B}\rangle$.
During the subsequent sampling interval ($\Delta t_{{\rm catch}}\geq0$),
the four quantities 
\begin{equation}
\big({\rm Z}_{{\rm GD}}^{j},{\rm X}_{{\rm GD}}^{j},{\rm Y}_{{\rm GD}}^{j},{\rm P}_{{\rm BB}}^{j}\big)(\Delta t_{{\rm catch}})=\big({\rm Z}_{{\rm GD}}^{{\rm rec}},{\rm X}_{{\rm GD}}^{{\rm rec}},{\rm Y}_{{\rm GD}}^{{\rm rec}},{\rm P}_{{\rm BB}}^{{\rm rec}}\big)(t_{j}+\Delta t_{{\rm catch}}),
\end{equation}
with $t_{j}$ is the click time and 
\begin{eqnarray}
{\rm Z}_{{\rm GD}}^{{\rm rec}}(t) & = & \frac{\langle{\rm D}|\psi(t)\rangle\langle\psi(t)|{\rm D}\rangle-\langle{\rm G}|\psi(t)\rangle\langle\psi(t)|{\rm G}\rangle}{\langle\psi(t)|\psi(t)\rangle},\label{eqn:correlated_Z_GD}\\
\noalign{\vskip4pt}{\rm X}_{{\rm GD}}^{{\rm rec}}(t)+i{\rm Y}_{{\rm GD}}^{{\rm rec}}(t) & = & 2\frac{\langle{\rm D}|\psi(t)\rangle\langle\psi(t)|{\rm G}\rangle}{\langle\psi(t)|\psi(t)\rangle},\label{eqn:correlated_X_GD=000026Y_GD}\\
\noalign{\vskip4pt}{\rm P}_{{\rm BB}}^{{\rm rec}}(t) & = & \frac{\langle{\rm B}|\psi(t)\rangle\langle\psi(t)|{\rm B}\rangle}{\langle\psi(t)|\psi(t)\rangle},
\end{eqnarray}
are computed, and running sums of each are updated. The sample terminates
when the measurement record indicates a change of state from not-$|{\rm B}\rangle$
back to $|{\rm B}\rangle$. Finally, for comparison with the experiment,
Bloch vector components are recovered from the average over sample
intervals via the formula 
\begin{equation}
\big({\rm Z}_{{\rm GD}},{\rm X}_{{\rm GD}},{\rm Y}_{{\rm GD}}\big)(\Delta t_{{\rm catch}})=\frac{\sum_{j}^{N(\Delta t_{{\rm catch}})}\big({\rm Z}_{{\rm GD}}^{j},{\rm X}_{{\rm GD}}^{j},{\rm Y}_{{\rm GD}}^{j}\big)(\Delta t_{{\rm catch}})}{N(\Delta t_{{\rm catch}})-\sum_{j}^{N(\Delta t_{{\rm catch}})}P_{{\rm BB}}^{j}(\Delta t_{{\rm catch}})},\label{eqn:ensemble_average}
\end{equation}
where $N(\Delta t_{{\rm catch}})$ is the number of sample intervals
that extend up to, or beyond, the time $\Delta t_{{\rm catch}}$.
The simulation and sampling procedure is illustrated in Fig.~\ref{fig:monte-carlo},
and a comparison between the experiment and the simulation is provided
in Fig.~\ref{fig:simulation_vs_experiment}.

The simulated and measured Bloch vector components are fit with expressions
motivated by Eqs.~(\ref{eqn:Z_approx})-(\ref{eqn:Y_approx}) and~(\ref{eq:ZGDXGD-long-term}),
modified to account for the effect of non-idealities in the experiment,
\begin{align}
\mathrm{Z}_{\text{GD}}(\Delta t_{\operatorname{catch}}) & =\ensuremath{a+b\tanh(\Delta t_{\operatorname{catch}}/\tau+c)},\\
\ensuremath{\mathrm{X}_{\text{GD}}(\Delta t_{\operatorname{catch}})} & =a'+b'\operatorname{sech}(\Delta t_{\operatorname{catch}}/\tau'+c')\,,\\
\mathrm{Y}_{\text{GD}}(\Delta t_{\operatorname{catch}}) & =0\,.
\end{align}
The fit parameters ($a,a',b,b',c,c',\tau,\tau'$) for the simulated
and experimental data shown in Fig.~\ref{fig:simulation_vs_experiment}
are compared in Table~\ref{tab:Comparison-of-parameters}. As imposed
by Eq.~(\ref{eq:ZGDXGD-long-term}), in the absence of $\Omega_{{\rm DG}}$
(turned off at time $\Delta t_{{\rm on}}=2\mkern2mu \mu{\rm s}$)
$a'$, the offset of $\mathrm{X}_{\text{GD}}$, is strictly enforced
to be zero. The extracted simulation and experiment parameters are
found to agree at the percent level.

\begin{table}[t]
\begin{centering}
\renewcommand*\arraystretch{1.5}
\begin{tabular}{rl|rl|rl}
\multicolumn{2}{c|}{\textbf{Readout cavity}} & \multicolumn{2}{c|}{\textbf{BG transition}} & \multicolumn{2}{c}{\textbf{DG transition}}\tabularnewline
\hline 
\hline 
\multicolumn{6}{c}{\textbf{\rule{0pt}{5ex}Non-linear parameters}}\tabularnewline
\hline 
 &  & $\chi_{\mathrm{B}}/2\pi=$ & $-5.08\,\mathrm{MHz}$~ & $\chi_{\mathrm{D}}/2\pi=$ & $-0.33\,\mathrm{MHz}$\tabularnewline
\multicolumn{6}{c}{\textbf{\rule{0pt}{5ex}Coherence related parameters}}\tabularnewline
\hline 
\textbf{\rule{0pt}{5ex}}$\kappa/2\pi$= & $3.62\mathrm{MHz}$ & $T_{\mathrm{1}}^{\mathrm{B}}=$ & $15\,\mathrm{\mu s}$ & $T_{\mathrm{1}}^{\mathrm{D}}=$ & $105\,\mathrm{\mu s}$\tabularnewline
$\eta=$ & $0.33$ & $T_{\mathrm{2\mathrm{R}}}^{\mathrm{B}}=$ & $18\,\mathrm{\mu s}$ & $T_{\mathrm{2\mathrm{R}}}^{\mathrm{D}}=$ & $120\,\mathrm{\mu s}$\tabularnewline
$T_{\mathrm{int}}=$ & $260.0\,\mathrm{ns}$ &  &  &  & \tabularnewline
$n_{\mathrm{th}}^{\mathrm{C}}=$ & $0$ & $n_{\mathrm{th}}^{\mathrm{B}}=$ & $0.01$ & $n_{\mathrm{th}}^{\mathrm{D}}=$ & $0.05$\tabularnewline
\multicolumn{6}{c}{\textbf{\rule{0pt}{5ex}Drive amplitude and detuning parameters}}\tabularnewline
\hline 
\textbf{\rule{0pt}{5ex}}$\bar{n}=$ & $5.0$ & ~$\Omega_{\mathrm{B}0}/2\pi=$ & $1.20\,\mathrm{MHz}$ & ~$\Omega_{\mathrm{DG}}/2\pi=$ & $21.6\,\mathrm{kHz}$\tabularnewline
 &  & ~$\Omega_{\mathrm{B}1}/2\pi=$ & $0.60\,\mathrm{MHz}$ &  & \tabularnewline
$\Delta_{\mathrm{R}}=$ & $\chi_{\mathrm{B}}$ & ~$\Delta_{\mathrm{B}1}/2\pi=$ & $-30.0\,\text{MHz}$ & ~$\Delta_{\mathrm{DG}}/2\pi=$ & $-274.5\,\text{kHz}$\tabularnewline
\end{tabular}
\par\end{centering}
\caption[Compilation of the simulation parameters]{\textbf{Compilation of the simulation parameters.}}
\label{table:table2} 
\end{table}

\clearpage{}

\begin{table}
\begin{centering}
\begin{tabular}{c}
\textbf{(a)} In presence of $\Omega_{\mathrm{DG}}$\tabularnewline
\tabularnewline
\centering{}\renewcommand*{\arraystretch}{1.2} %
\begin{tabular}{cc|crcllrrlcr}
Parameter  & \multicolumn{1}{c|}{} &  & \multicolumn{3}{c}{Experiment} &  & \multicolumn{3}{c}{Simulation} &  & Error\tabularnewline
\hline 
\rule{0pt}{3ex} $a$  &  &  & -0.07  & $\pm$  & 0.005  &  & -0.07  & $\pm$  & 0.005  &  & 0.5\%\tabularnewline
$a'$  &  &  & -0.21  & $\pm$  & 0.005  &  & -0.22  & $\pm$  & 0.005  &  & 2\%\tabularnewline
$b$  &  &  & 0.94  & $\pm$  & 0.005  &  & 0.95  & $\pm$  & 0.005  &  & 1\%\tabularnewline
$b'$  &  &  & 0.93  & $\pm$  & 0.005  &  & 0.91  & $\pm$  & 0.005  &  & 2\%\tabularnewline
$c$  &  &  & -2.32  & $\pm$  & 0.03  &  & -2.27  & $\pm$  & 0.03  &  & 2\%\tabularnewline
$c'$  &  &  & -2.04  & $\pm$  & 0.03  &  & -2.05  & $\pm$  & 0.03  &  & 0.5\%\tabularnewline
$\tau$  &  &  & 1.64  & $\pm$  & 0.01  &  & 1.65  & $\pm$  & 0.01  &  & 0.5\%\tabularnewline
$\tau'$  &  &  & 1.74  & $\pm$  & 0.01  &  & 1.76  & $\pm$  & 0.01  &  & 1\%\tabularnewline
\end{tabular}\tabularnewline
\tabularnewline
\textbf{(b)} In absence of $\Omega_{\mathrm{DG}}$\tabularnewline
\tabularnewline
\centering{}\renewcommand*{\arraystretch}{1.2} %
\begin{tabular}{cc|crcllrrlcr}
Parameter & \multicolumn{1}{c|}{} &  & \multicolumn{3}{c}{Experiment} &  & \multicolumn{3}{c}{Simulation} &  & Error\tabularnewline
\hline 
\rule{0pt}{3ex} $a$ &  &  & -0.11 & $\pm$ & 0.005 &  & -0.10 & $\pm$ & 0.005 &  & 8\%\tabularnewline
$a'$ &  &  & 0 & $\pm$ & 0 &  & 0 & $\pm$ & 0 &  & 0\%\tabularnewline
$b$ &  &  & 0.92 & $\pm$ & 0.008 &  & 0.91 & $\pm$ & 0.008 &  & 1\%\tabularnewline
$b'$ &  &  & 0.61 & $\pm$ & 0.005 &  & 0.60 & $\pm$ & 0.005 &  & 2\%\tabularnewline
$c$ &  &  & -1.96 & $\pm$ & 0.05 &  & -2.10 & $\pm$ & 0.05 &  & 7\%\tabularnewline
$c'$ &  &  & -1.97 & $\pm$ & 0.05 &  & -2.05 & $\pm$ & 0.05 &  & 4\%\tabularnewline
$\tau$ &  &  & 2.17 & $\pm$ & 0.05 &  & 2.03 & $\pm$ & 0.05 &  & 6\%\tabularnewline
$\tau'$ &  &  & 1.98 & $\pm$ & 0.05 &  & 1.92 & $\pm$ & 0.05 &  & 3\%\tabularnewline
\end{tabular}\tabularnewline
\tabularnewline
\end{tabular}
\par\end{centering}
\caption[Comparison between parameters extracted from the simulation and those
from the experiment. ]{\label{tab:Comparison-of-parameters} \textbf{Comparison between
parameters extracted from the simulation and those from the experiment}.
\textbf{a,} Parameters obtained from fits of the simulated and measured
data for the catch protocol in the presence of the Rabi drive $\Omega_{{\rm DG}}$
throughout the entire duration of the quantum jump, data shown in
Fig.~\ref{fig:simulation_vs_experiment}a. \textbf{b,} Parameters
obtained from fits of the simulated and measured data for the catch
protocol in the absence of the $\Omega_{{\rm DG}}$ during the flight
of the quantum jump for $\Delta t_{\mathrm{on}}=2\mathrm{\ \mu s}$,
data shown in Fig.~\ref{fig:simulation_vs_experiment}b. }
\end{table}

\clearpage{}

\subsection{Error budget\label{sec:Budget-coherence}}

In this section, we examine the effect of the various imperfections
and dissipation channels on the fidelity of the catch protocol. 

\paragraph{\textit{\emph{Imperfections.}}}

The various imperfections are expected to reduce the maximum coherence
recovered in the measurement of ${\rm X}_{{\rm GD}}(\Delta t_{{\rm catch}})$.
They include: 
\begin{enumerate}
\item Readout errors when inferring $|{\rm B}\rangle$ to not-$|{\rm B}\rangle$
transitions and the reverse. Such errors affect the assignment of
$\Delta t_{{\rm catch}}$, which can be either too short or too long
to correlate correctly with the true state of the system. 
\item Leaks from the GBD-manifold to higher excited states. Importantly,
these errors mimic a $|{\rm B}\rangle$ to not-$|{\rm B}\rangle$
transition, as in the first sample interval of Fig.~\ref{fig:monte-carlo},
but the anticipated coherent evolution within the GBD-manifold does
not occur. In this manner, the excitations to higher states lead to
false detections.
\item Thermal jumps from $|{\rm G}\rangle$ to $|{\rm D}\rangle$. Such
incoherent transitions contribute in a similar way to ${\rm Z}_{{\rm GD}}(\Delta t_{{\rm catch}})$,
while making no contribution to the measured coherence. 
\item Direct dephasing of the DG-coherence, $T_{\mathrm{2R}}^{\mathrm{D}}$. 
\item Partial distinguishability of $|{\rm G}\rangle$ and $|{\rm D}\rangle$.
The readout cavity is not entirely empty of photons when the state
is not-$|{\rm B}\rangle$, in which case the cross-Kerr interaction
$\chi_{{\rm D}}|{\rm D}\rangle\langle{\rm D}|\hat{c}^{\dagger}\hat{c}$
shifts the $\Omega_{{\rm DG}}$ Rabi drive from resonance; hence,
backaction noise is transferred from the photon number to ${\rm X_{{\rm GD}}}(\Delta t_{{\rm catch}})$. 
\end{enumerate}

\paragraph{\textit{\emph{Budget for lost coherence.}}\emph{ }}

\begin{figure}[h]
\begin{centering}
\includegraphics[width=130mm]{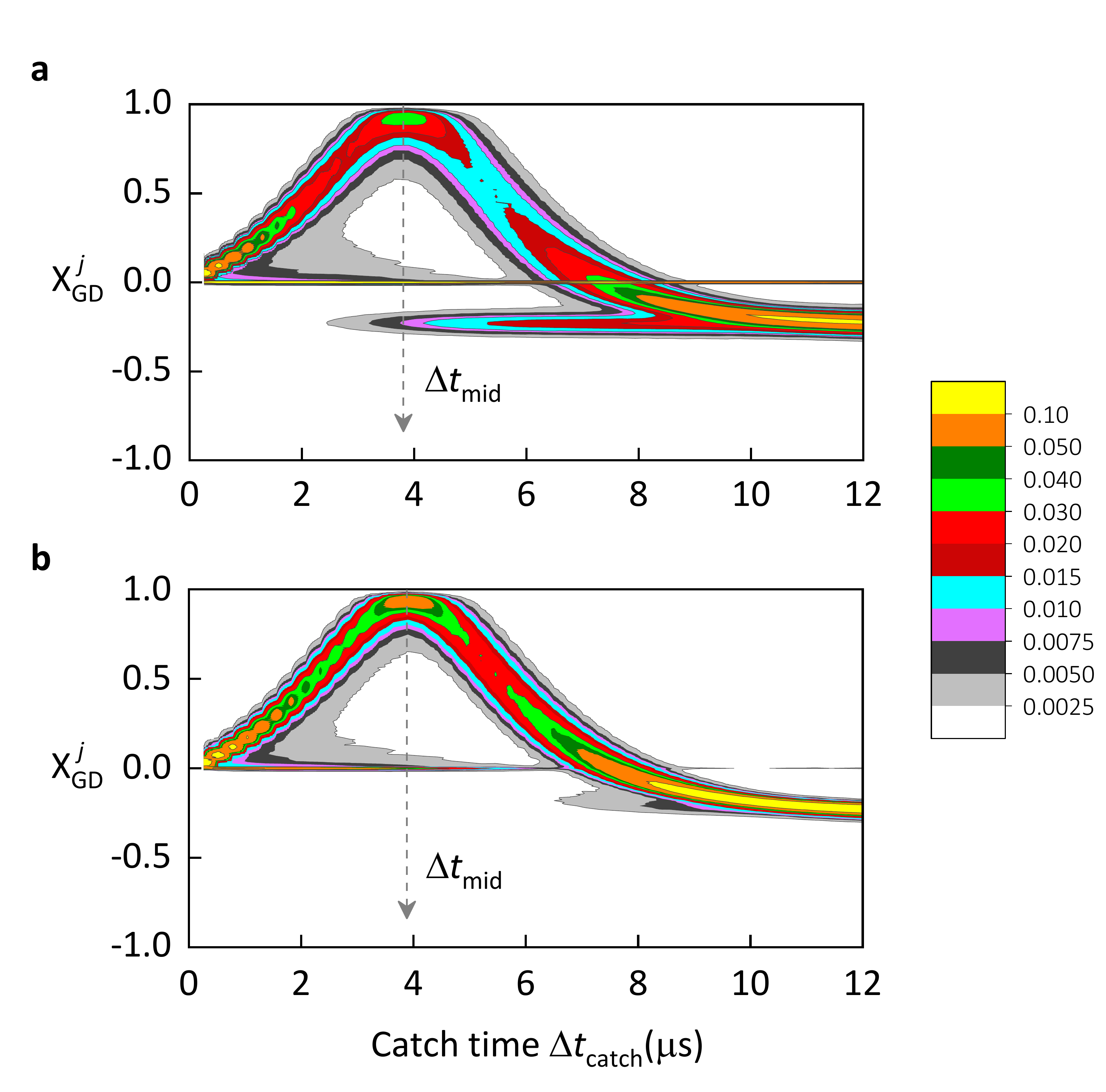}
\par\end{centering}
\centering{}\caption[Coherence loss through sample to sample fluctuations (courtesy of
H.J. Carmichael)]{\label{fig:coherence_loss}\textbf{Coherence loss through sample
to sample fluctuations.} \textbf{a,} Contour plot of the distribution
of ${\rm X}_{{\rm GD}}^{j}(\Delta t_{{\rm catch}})$ samples corresponding
to the simulated data set displayed in panel (a) of Fig.~\ref{fig:simulation_vs_experiment}.
\textbf{b,} Same as panel (a) but with transitions between $|{\rm B}\rangle$
and not-$|{\rm B}\rangle$ identified in the ket $|\psi\rangle$ rather
than from the simulated measurement record, and with changed parameters:
$(\gamma_{{\rm FG}},\gamma_{{\rm FD}},\gamma_{{\rm GF}},\gamma_{{\rm DF}})/2\pi=0$,
$n_{{\rm th}}^{{\rm B}}=n_{{\rm th}}^{{\rm D}}=0$, $T_{2}^{{\rm D}}=2T_{1}^{{\rm D}}$,
and $\chi_{{\rm D}}/2\pi=0$. Figure courtesy of H.J. Carmichael.}
\end{figure}

The maximum coherence reported in the experiment is $0.71\pm0.005$.
In the simulation it is a little lower at 0.69. By removing the imperfections
from the simulation, one by one, we can assign a fraction of the total
coherence loss to each. Readout errors are eliminated by identifying
transitions between $|{\rm B}\rangle$ and not-$|{\rm B}\rangle$
in the ket $|\psi\rangle$ rather than from the simulated measurement
record; all other imperfections are turned off by setting some parameter
to zero. The largest coherence loss comes from readout errors, whose
elimination raises the ${\rm X}_{{\rm GD}}(\Delta t_{{\rm catch}})$
maximum by 0.09. The next largest comes from leakage to higher excited
states, which raises the maximum by a further 0.06. Setting $\chi_{{\rm D}}$
to zero adds a further 0.04, and thermal transitions and pure dephasing
together add 0.02. Figure~\ref{fig:coherence_loss} illustrates the
change in the distribution of ${\rm X}_{{\rm GD}}^{j}(\Delta t_{{\rm catch}})$
samples underlying the recovery of coherence. The removal of the finger
pointing to the left in panel (a) is mainly brought about by the elimination
of readout errors, while the reduced line of zero coherence marks
the elimination of leakage to higher excited states. Aside from these
two largest changes, there is also a sharpening of the distribution,
at a given $\Delta t_{{\rm catch}}$, when moving from panel (a) to
panel (b). Having addressed the five listed imperfections, a further
10\% loss remains unaccounted for, i.e., the distribution of panel
(b) is not a line passing through ${\rm X}_{{\rm GD}}^{j}(\Delta t_{{\rm mid}})=1$.
The final 10\% is explained by the heterodyne detection backaction
noise, a function of the drive and measurement parameters, displayed
in panel (b) of Fig.~\ref{fig:monte-carlo}. 

\section{Signal-to-noise ratio (SNR) and de-excitation measurement efficiency\label{subsec:Signal-to-noise-ratio-(SNR)}}

The catch protocol hinges on the efficient detection of de-excitations
from $|{\rm B}\rangle$ to $|{\rm G}\rangle$, as discussed in more
detail in Chapter~\ref{chap:theoretical-description-jumps}. In atomic
physics, de-excitations are typically monitored by a \emph{direct}
detection method, employing a photodetector. Alternatively, de-excitations
can be monitored by an\emph{ indirect} method, as done in our experiment.
In this section, we discuss the efficiency of both methods. For the
indirect method, using simple analytics, we estimate the \emph{total}
efficiency of time-continuous, uninterrupted monitoring of de-excitations
from $|{\rm B}\rangle$ to $|{\rm G}\rangle$ to be $\eta_{\mathrm{eff,clk}}=0.90\pm0.01$
for the parameters of our experiment, with integration time $T_{\mathrm{int}}=0.26\,\mathrm{\mu s}$.
The simple analysis of this section complements the numerical one
of the previous section, Sec.~\ref{sec:Budget-coherence}.

\emph{Direct monitoring method in atomic physics.} The direct method
monitors for a $|{\rm B}\rangle$ de-excitation by collecting and
absorbing the photon radiated in the de-excitation. The \emph{total}
measurement efficiency of this method is limited by i) collection
efficiency --- the fraction of emitted photons collected by the detector
in its own input spatial modes (for instance, as determined by the
solid angle) --- typically falls in the range 0.1 - 50\%, \citep{Volz2011}
ii) the efficiency of detecting the absorption of a single photon,
which falls in the range 1 - 90\%, \citep{Eisaman2011} and iii) non-idealities
of the photodetector apparatus, including its dead time, dark counts,
jitter, etc. \citep{Eisaman2011} The combination of these inefficiencies
presents an almost insurmountable challenge in experimental atomic
physics for realizing continuous, time-resolved detection of nearly
every single photon emitted by the three-level atom, required to faithfully
catch the jump. 

\emph{Direct monitoring method with superconducting circuits.} While
technologically very different, the direct monitoring method with
superconducting circuits is conceptually similar to atomic method
but can readily achieve high collection efficiencies \citep{Katz2008,Vijay2011,Riste2012-qubit-measure-reset,Vijay2012,Hatridge2013,Murch2013a,deLange2014,Roch2014,Weber2014,Campagne-Ibarcq2014,Macklin2015,Campagne2016-Fluorescence,Campagne-Ibarcq2016,Hacohen-Gourgy2016-non-comm,Naghiloo2016,White2016,Ficheux2017,Naghiloo2017-thermo,Tan2017,Hacohen-Gourgy2018,Heinsoo2018,Bultink2018}.
 However, the energy of the emitted microwave photon is exceedingly
small --- $23\text{ }\mathrm{\mu eV}$, about a part per 100,000
of the energy of a single optical photon --- which essentially forbids
the direct detection of the photon with near-unit efficiency. This
is because the propagating photon is unavoidably subjected to significant
loss, added spurious noise, amplifier non-idealities, etc. In our
experiment, these imperfections reduce the full measurement/amplification
chain efficiency from its ideal value \citep{Hatridge2013,Macklin2015,Bultink2018}
of 1 to a modest $\eta=0.33\pm0.03$, corresponding to the direct
detection of approximately only one out of every three single photons
--- insufficient for the catch protocol.

\subsubsection{Indirect monitoring method with superconducting circuits}

Alternatively, the indirect monitoring method couples the atom to
an ancillary degree of freedom, which is itself monitored in place
of the atom. In our experiment, the atom is strongly, dispersively
coupled to the ancillary readout cavity. The cavity scatters a probe
tone, whose phase shift constitutes the readout signal, as discussed
in Chapter~\ref{chap:theoretical-description-jumps}. Since the probe
tone can carry itself many photons, this scheme increases the signal-to-noise
ratio ($\mathrm{SNR}$) and, hence, the total efficiency ($\eta_{\mathrm{eff,clk}}$)
of detecting a $|{\rm B}\rangle$ de-excitation. Note that the efficiency
$\eta_{\mathrm{eff,clk}}$ should not be confused with the efficiency
of a photodetector or the efficiency $\eta$ of the measurement/amplification
chain, since $\eta_{\mathrm{eff,clk}}$ includes the effect of all
readout imperfections and non-idealities, state discrimination and
assignment errors, etc. see below. In the remainder of this section,
we estimate the SNR and efficiency $\eta_{\mathrm{eff,clk}}$ of the
experiment.

\emph{SNR of the indirect (dispersive) method.}  The output of the
measurement and amplification chain monitoring the readout cavity
is proportional to the complex heterodyne measurement record $\zeta\left(t\right)$,
which obeys the It\^{o} stochastic differential equation, see Eq.~(\ref{eq:Record}),\footnote{Since the bandwidth of the measurement chain, $\kappa_{\mathrm{filter}}$,
is significantly larger than that, $\kappa$, of the readout cavity,
$\kappa_{\mathrm{filter}}\gg\kappa$, we can neglect the effect of
$\kappa_{\mathrm{filter}}$ for simplicity of discussion, see Eqs.~(\ref{eqn:simulated_I_int})
and~(\ref{eqn:simulated_Q_int}).} 
\begin{equation}
\mathrm{d}\zeta\left(t\right)=\sqrt{\eta\kappa}\frac{\langle\psi\left(t\right)|\hat{a}|\psi\left(t\right)\rangle}{\langle\psi\left(t\right)|\psi\left(t\right)\rangle}\mathrm{d}t+\mathrm{d}Z\left(t\right),\label{eq:heterodyne-current2}
\end{equation}
where $\hat{a}$ is the cavity amplitude operator in the Schr\"{o}dinger
picture, $\eta$ is the total measurement efficiency of the amplification
chain --- again, not to be confused with the de-excitation measurement
efficiency, $\eta_{\mathrm{eff,clk}}$ --- and $\mathrm{d}Z$ is
the complex Wiener process increment, defined below Eq.~(\ref{eq:heterodyne-current2}).
A somewhat counterintuitive property of Eq.~(\ref{eq:heterodyne-current2})
is that  the heterodyne record increment $\mathrm{d}\zeta\left(t\right)$
is stochastic and noisy even when $\eta=1$, the case of ideal measurement
in which no signal is lost --- the stochastic term, $\mathrm{d}Z$,
represents pure quantum vacuum fluctuations, which are inherent in
the case of heterodyne detection \citep{Carmichael1993,Plenio1998,wiseman2010book}.
Due to the unavoidable presence of these fluctuations, only an infinitesimal
amount of information about the system can be extracted from $\mathrm{d}\zeta$
at an instant of time. Finite amount of information is extracted by
integrating $\mathrm{d}\zeta$ for a finite duration $T_{\mathrm{int}}$,
\begin{equation}
s\equiv I_{\mathrm{rec}}+iQ_{\mathrm{rec}}\equiv\int_{0}^{T_{\mathrm{int}}}\mathrm{d}\zeta\left(t\right)\,,\label{eq:s=00003D}
\end{equation}
where $I_{\mathrm{rec}}$ and $Q_{\mathrm{rec}}$ are the in- and
out-of-phase quadrature components of one segment of the record. What
does $s$ correspond to? Its value depends on $\mathrm{d}\zeta$,
which depends on the state of the cavity, $|\psi\rangle$, which itself
depends on the occupation of $\ket{\mathrm{B}}$ --- and therefore
$s$ contains the occupation of $\ket{\mathrm{B}}$. A de-excitation
of $\ket{\mathrm{B}}$ to $\ket{\mathrm{G}}$ can thus be detected
by monitoring $s$, whose value is different for the two states, since
the cavity is generally in the coherent state $\ket{\alpha_{\mathrm{B}}}$
or $\ket{\alpha_{\mathrm{G}}}$ when the atom is in $\ket{\mathrm{B}}$
or $\ket{\mathrm{G}}$, respectively. For the moment, assuming the
atom and cavity do not change states during the course of the measurement
duration $T_{\mathrm{int}}$, the stochastic integral in Eq.~(\ref{eq:s=00003D})
explicitly evaluates to 
\begin{equation}
s_{\mathrm{B,G}}=\left\{ \sqrt{\eta\kappa}\mathrm{Re}\left[\alpha_{\mathrm{B,G}}\right]T_{\mathrm{int}}+\frac{1}{\sqrt{2}}W_{I}\left(T_{\mathrm{int}}\right)\right\} +i\left\{ -\sqrt{\eta\kappa}\mathrm{Im}\left[\alpha_{\mathrm{B,G}}\right]T_{\mathrm{int}}+\frac{1}{\sqrt{2}}W_{Q}\left(T_{\mathrm{int}}\right)\right\} \,,\label{eq:s=00003DExplicit}
\end{equation}
where $W_{I,Q}$ denote independent Wiener processes, obeying the
conventional rules, $\mathrm{E}\left[W\left(t\right)\right]=0$ and
$\mathrm{Var}\left[W\left(t\right)\right]=t^{2}$. Equation~(\ref{eq:s=00003DExplicit})
shows that the distribution of the stochastic variable $s$ is a Gaussian
blob in the IQ plane centered at $\bar{s}_{\mathrm{B,G}}\equiv\operatorname{E}\left[s_{\mathrm{B,G}}\right]=\sqrt{\eta\gamma}T_{\mathrm{int}}\alpha_{\mathrm{B,G}}$
with width determined by the variance $\sigma_{\mathrm{B,G}}^{2}\equiv\operatorname{Var}\left[s_{\mathrm{B,G}}\right]=\frac{1}{2}T_{\mathrm{int}}$.
We can thus define the SNR of the experiment by comparing the distance
between the two pointer distributions to their width, 
\begin{equation}
\mathrm{SNR}\equiv\left|\frac{\bar{s}_{\mathrm{B}}-\bar{s}_{\mathrm{G}}}{\sigma_{\mathrm{B}}+\sigma_{\mathrm{G}}}\right|^{2}\,,\label{eq:SNR-defn}
\end{equation}
where the B (resp., G) subscript denotes signals conditioned on the
atom being in $\ket{\mathrm{B}}$ (resp., $\ket{\mathrm{G}}$). In
terms of $\ket{\alpha_{\mathrm{B}}}$ and $\ket{\alpha_{\mathrm{G}}}$,
\begin{equation}
\mathrm{SNR}=\frac{1}{2}\eta\kappa T_{\mathrm{int}}\left|\alpha_{\mathrm{B}}-\alpha_{\mathrm{G}}\right|^{2}\,,
\end{equation}
which can be expressed in terms of the parameters of the experiment,
summarized in Table~\ref{tab:system-params},
\begin{equation}
\mathrm{SNR}=\frac{1}{2}\eta\kappa T_{\mathrm{int}}\left[\cos\left(\arctan\left(\frac{\kappa}{2\chi_{\mathrm{BG}}}\right)\right)\right]^{2}\bar{n}\,,\label{eq:SNR-expression}
\end{equation}
Holding other parameters fixed, according to Eq.~(\ref{eq:SNR-expression}),
the SNR can be increased arbitrarily by increasing $\bar{n}$, which
can be readily done by increasing the amplitude of the cavity probe
tone. A higher SNR for $s$ corresponds to a higher SNR for measuring
an atom de-excitation, since $s$ is a proxy of the $\ket{\mathrm{B}}$
population. Thus, the indirect cavity monitoring can overcome the
typical degradation in SNR imposed by the inefficiencies and non-idealities
of the measurement chain, $\eta$. In practice, the SNR increase
with $\bar{n}$ is bounded from above, since with sufficiently high
$\bar{n}$ spurious non-linear effects become significant \citep{Boissonneault2008,Boissonneault2009-Photon-induced-relax,Minev2013,Sank2016-T1vsNbar,Khezri2016,Bultink2016,Khezri2017,Walter2017,Lescanne2018,Verney2018,Serniak2018}.
The cavity and non-linear coupling to the atom serve in effect as
a rudimentary embedded pre-amplifier at the site of the atom, which
transduces with amplification the de-excitation signal before its
SNR is degraded during propagation and further processing.

\emph{Discrimination efficiency of the indirect method.} While the
SNR provides a basic characterization of the measurement, it is useful
to convert it to a number between 0 and 1, which is called the discrimination
efficiency, $\eta_{\mathrm{disc}}$. It quantifies the degree to which
the two Gaussian distributions of $s$ are distinguishable \citep{Gambetta2007-ProtocolsMsr},
\begin{equation}
\eta_{\mathrm{disc}}=\frac{1}{2}\operatorname{erfc}\left[-\sqrt{\frac{\mathrm{SNR}}{2}}\right]\,,\label{eq:eta-msr}
\end{equation}
where $\operatorname{erfc}$ denotes the complementary error function.
Equation~(\ref{eq:eta-msr}) shows that increasing the SNR by separating
the $s_{\mathrm{B}}$ and $s_{\mathrm{G}}$ distributions far beyond
their spread, $\sigma_{\mathrm{B/G}}$, provides only marginal gain
as $\eta_{\mathrm{disc}}$ saturates to 1. Next, we calculate the
SNR and $\eta_{\mathrm{disc}}$ for the parameters of the experiment
and discuss corrections due to readout non-idealities. 

\emph{A first comparison to the experiment. }A first estimate of the
SNR and $\eta_{\mathrm{disc}}$ of the experiment are provided by
Eqs.~(\ref{eq:SNR-expression}) and~(\ref{eq:eta-msr}). Using the
parameters of the experiment, summarized in Table~\ref{tab:system-params},
from these two equations, we find $\mathrm{SNR}=4.3\pm0.6$ and $\eta_{\mathrm{disc}}=0.98\pm0.007$.
Using data from the experiment, in particular, a second long IQ record
trace, represented by a short segment in Fig.~2a, we find the SNR
of the jumps experiment, by fitting the histogram of the trace with
a bi-Gaussian distribution, to be $\mathrm{SNR}=3.8\pm0.4$, corresponding
to $\eta_{\mathrm{\mathrm{disc}}}=0.96\pm0.01$. The measured values
are slightly lower than the analytics predict due to readout imperfections
not included in the calculation so far, such as state transitions
during $T_{\mathrm{int}}$, cavity transient dynamics, additional
pointer-state distributions, etc.

\emph{Effective click detection efficiency. }The dominant next-order
error is due to atom state transitions during the measurement window,
$T_{\mathrm{int}}$, which contributes an assignment error of approximately
$1-\eta_{\mathrm{asg}}=1-\exp\left(T_{\mathrm{int}}/\tau_{\mathrm{B}}\right)=0.06\pm0.001$
to the detection of a $|\mathrm{B}\rangle$ de-excitation. Combining
$\eta_{\mathrm{disc}}$ with $\eta_{\mathrm{asg}}$, we obtain the
total efficiency for detecting $\ket{\mathrm{B}}$ de-excitations
$\eta_{\mathrm{eff,clk}}=\eta_{\mathrm{disc}}\eta_{\mathrm{asg}}=0.90\pm0.01$,
consistent with the total readout efficiency of $0.91$ that is independently
estimated using the trajectory numerics, see Sec.~\ref{sec:Budget-coherence}.

\chapter{Conclusions and perspectives\label{chap:Conclusion-and-perspective}}

\addcontentsline{lof}{chapter}{Conclusions and perspectives \lofpost}

\singlespacing 
\epigraph{
Technological forecasting is even harder than weather forecasting.}
{Rolf Landauer} 
\doublespacing  \noindent

\section{Conclusions}

In conclusion, we experimentally demonstrated that, despite the fundamental
indeterminism of quantum physics in the context of the monitoring
of the evolution of a system, it is possible to detect an advance
warning that signals the occurrence of an event, the quantum jump
from the ground state ($\G$) to the excited state $(\D)$ of a three-level
superconducting atom, prior to its complete occurrence (Sec.~\ref{sec:Catching-the-quantum}).
While the quantum jump begins at a random time and can be prematurely
interrupted by a click, a quantum jumps that completes follows a \emph{continuous},
\emph{deterministic}, and \emph{coherent} ``flight,'' which comes
as a surprise in view of standard textbooks on quantum mechanics.
The special nature of the transition was exploited to catch the jump
and reverse it to the ground state, $\G$. Additionally, the state
of the atom was tomographically reconstructed, $\rho_{c}$, as a function
of the duration of the catch signal, $\Delta t_{\mathrm{catch}}$,
from $6.8\times10^{6}$ individual experimental realizations, each
catching a single jump occurring at a random time. At the mid-flight
time of the quantum jump, $\Delta t_{{\rm mid}}$, the atom was observed
to be in coherent superposition of $\G$ (equivalent to no jump) and
$\D$ (equivalent to a jump), with state purity $\Tr{\rho_{c}^{2}}=0.75\pm0.004$.
Even when conditionally turning off the Rabi drive between $\G$ and
$\D$, $\Odg$, at the beginning of the jump, $\Delta t_{\mathrm{on}}=2\ \mathrm{\mu s}$,
the flight of the quantum jump was observed to nonetheless proceed
in coherent, deterministic, and essentially identical manner, despite
the absence of the coherent Rabi drive. This demonstrated  that the
role of $\Omega_{\mathrm{DG}}$ is to initiate the jump and set its
phase but is otherwise unimportant, and that the dynamics of the flight
are (essentially) entirely governed by the measurement-backaction
force due to the measurement, discussed in Chapter~\ref{chap:theoretical-description-jumps}. 

The jump coherence and deterministic-like character (any two jumps
take the same gradual flight) provide a small island of predictability
in a sea of uncertainty that was exploited, see Sec.~\ref{sec:Reversing-the-quantum},
to reverse the quantum jump to the ground state, thus precluding its
occurrence. When applied at the mid-flight time, $\Delta t_{\mathrm{mid}}$,
the protocol succeeded in reversing the jump to $\G$ with $82.0\%\pm0.3\%$
fidelity. Remarkably, under ideal conditions, every jump that would
complete is detected by the warning signal and reversed, thus eliminating
\emph{all} quantum jumps from $\G$ to $\D$, and preventing the atom
from ever reaching $\D$. Jumps that would not complete and are reversed
by the protocol meet their fate faster by the warning-based intervention.

In Sec.~\ref{subsec:Comparison-between-theory}, we showed that the
experimental results agree essentially without adjustable parameters
with the theory predictions, accounting for known experimental imperfections,
such as finite quantum measurement efficiency, $\eta$, temperature,
$n_{\mathrm{th}},$ dephasing mechanisms, $T_{1}$ and $T_{2}$, etc.
The agreement testifies for the validity, reliability, and predictive
power of quantum trajectory theory and suggests its critical role
in the practical development of real-time feedback techniques for
quantum system control. 

On a technological level, we developed a three-level superconducting
atom with distinct features of interest. By decoupling one of the
states, $\D$, from both the readout cavity and the environment, we
demonstrated a protected qubit design with notable quantum coherence,
$T_{\mathrm{2R}}^{\mathrm{D}}=120\pm5\,\mathrm{\mu s}$, importantly,
\emph{without }sacrificing measurement efficiency or speed, as typically
necessitated when decoupling a level, see Sec.~\ref{sec:circuit-design}.
Integral to the implementation was the design optimization with the
energy participation ratio (EPR) approach, as described in Sec.~\ref{sec:circuit-design}.
In Sec.~\ref{subsec:Measurement-induced-relaxation}, we demonstrated
the ability to populate the readout cavity with a large number of
photons without degrading the coherence properties of $\D$ due to
measurement-induced relaxation, $T_{1}\left(\bar{n}\right)$ .

\section{Perspectives}

In the following, we discuss a few possible research directions that
build on the catch and reverse experiment and the development of the
Darkmon system, listed in ascending order of difficulty. 

\paragraph{Fundamental tests. }

The Darkmon three-level atom is a particularly versatile platform
for fundamental tests in quantum physics. Two unique aspects of the
$\B$/not-$\B$ measurement are important for fundamental tests: i)
it is information-asymmetric, and ii) its is degenerate. For definitiveness,
consider the situation where a measurement is performed on the atom
prepared in an unknown initial state. At first, the observer has no
knowledge of the system, i..e, zero bits of information. Performing
a measurement and obtaining a B outcome, the observer learns that
the measurement has projected the atom in the definite, pure state
$\B$, and now posses complete knowledge of the system, and has thus
gained $\mathcal{I}=\ln_{2}3\approx1.6$~bits of information, although
the initial state  remains unknown. In contrast, in obtaining a not-B
outcome, only $\mathcal{I}=\ln_{2}\left(3/2\right)\approx0.6$~bits
of information are gained, since the atom could still be in $\G$
or $\D$. The measurement has left behind 1~bit of the initial-state
information. Importantly, the $\B$/not-$\B$ measurement does not
disturb this bit, and preserves its quantum coherence. Since it leaves
behind a manifold of states untouched, it is known as \emph{degenerate
measurement.}

\emph{Contextuality.} — Degenerate measurements  are required to
perform tests of Kochen-Specker contextuality \citep{KochenSpecker1967},
which reveals an essential aspect of the nonclassical nature of quantum
measurements and constrains hidden variable theories; it can be viewed
as a complement to Bell's theorem. It follows from the degenerate
measurement requirement that a qutrit is the simplest system in which
contextuality can be observed, and the Darkmon system with its notable
control and coherence properties could prove a well-suited testbed
for rigorous tests \citep{Mermin1993-rev,Klyachko2008,Yu2012,Szangolies2015-book}.

\emph{Wavefunction collapse and the arrow of time. }— There is a growing
interest in the community to experimentally investigate the dynamics
of the wavefunction ``collapse'' \citep{Katz2006,Katz2008,Murch2013a,Hatridge2013,Weber2014,Campagne-Ibarcq2014,Campagne2016-Fluorescence,Jordan2016-fluorescence,Naghiloo2016,Tan2017,Harrington2017}
and associated fundamental questions. An interesting research direction
is to investigate the emergence of the apparent irreversibility of
the collapse, which, it is argued, yields the arrow of time in quantum
physics. Recently, theoretical work has emerged that suggests ways
in which quantum trajectory experiments can begin to probe this outstanding
question regarding the origin of the arrow of time with the tools
of quantum trajectory theory \citep{Dressel2017-arrow-of-time,Jordan2017-Janus}.
Focus so far has been almost exclusively on two level systems, but
important fundamental features in quantum physics, such as Kochen-Specker
contextuality, only emerge in systems of larger than two dimension.
The arrow of time is an especially interesting direction in view of
the results presented here, where we reverse quantum jumps prior to
their complete occurrence. We believe the Darkmon system and its degenerate
measurement could offer a unique vantage point on the problem. 

\emph{Thermodynamics.} — In a related research direction, the Darkmon
system could be employed to probe the emergence of thermodynamics
in continuously monitored systems, a question of active research in
the community. Specifically, understanding (and formulating) the fundamental
laws of thermodynamics in the quantum domain, as well as notions such
as work, heat, and Maxwell's demon, with applications to heat engines,
are under investigation and can be explored with quantum trajectories
in the multi-level system \citep{Alonso2016-thermo,Naghiloo2017-thermo,Elouard2017,Cottet2017}.

\paragraph{Protected qubit with faithful readout. }

Technologically, pursuing further the ideas demonstrated with the
Darkmon system could lead to improved qubit coherences and measurement
capabilities within the cQED architecture with the aim of addressing
the third and fourth DiVincenzo criteria for practical quantum computation
\citep{DiVincenzo2000-criteria} . 

In the development of fast and high-fidelity superconducting qubit
readout, a number of non-linear process have been employed \citep{cooper2004,Astafiev2004-readout,Siddiqi2006,Lupascu2006-readout,Mallet2009-readout,Reed2010-readout},
but the linear dispersive readout, by means of a low-Q cavity \citep{Wallraff2005-readout,Blais2004,Johnson2012-readout,Riste2012-readout},
is adopted most widely. While the cavity inhibits the spontaneous
relaxation of the qubit, it introduces three additional loss mechanisms:
i) energy relaxation ($T_{1}$) due to the Purcell effect \citep{Esteve1986,Koch2007,Neeley2008-Purcell},
ii) qubit dephasing ($T_{\phi}$) due to the photon shot noise of
the readout cavity \citep{Blais2004,Schuster2005-ACStark,Gambetta2006-dephasing,Schuster2007,Gambetta2008-qm-traj,Sears2012-PhotonShot,Rigetti2012},
often dominated by residual thermal population, $n_{\mathrm{th}}$,
and iii) measurement-induced qubit energy relaxation ($T_{1}\left(\bar{n}\right)$)
\citep{Boissonneault2009-Photon-induced-relax,Slichter2012,Sank2016-T1vsNbar,Slichter2016-T1vsNbar}.
In contrast, the GD qubit in the Darkmon device is decoupled from
all three dissipation channels, while still benefiting from the cavity
properties, and not sacrificing the ability to perform a fast readout
or to monitor the atom continuously. Practically advantageous is that
the design is hardware-efficient (simple) in the sense that it does
not require additional control gates, such as fast-flux lines or DC
gates, or other high-power pump tones. 

An interesting idea to pursue further stems from the use of the readout
cavity in the catch and reverse experiment to provide amplification
(and transduction) of the $\B$ signal \emph{prior to} its transmission
to the following quantum-limited amplifier. Reducing losses in the
transmission, $\eta$, is an outstanding challenge in the field. However,
a strategy to overcome this problem is indicated by the design: the
addition of a built-in gain element at the site of the sample that
provides sufficient amplification to overcome transmission losses
and whose coupling to the readout signal can be tuned independently
of the gain (in the experiment, by means of $\Obg$). First, we note
that direct monitoring of the $\B$ signal, by means of fluorescence
detection, would have prohibited the faithful execution of the catch
and reverse protocol, since a large number of the click signals would
have been lost in transmission, due to $\eta$. In contrast, in the
experiment, the $\B$/not-$\B$ signal was effectively amplified fivefold
(with frequency transduction) by the readout scheme by use of the
large disperse shift, $\chi_{{\rm BC}}\gg\kappa_{C}$, and the cavity
probe tone, $\bar{n}$. In contrast to the usual dispersive readout
scheme, where the use of a large probe signal, $\bar{n},$ results
in degradation of the signal-to-noise ratio and qubit coherence due
to the $T_{1}\left(\bar{n}\right)$ effect, the $\D$ level was shown
to be essentially immune to this, see Sec.~\ref{subsec:Measurement-induced-relaxation}.
This could provide the ability to use strong pump tones to activate
interesting non-linear interactions \citep{Mundhada2017,Mundhada2018}
without harming $\D$, and to implement a gain element or to couple
$\B$ to the readout cavity in a dissipation engineered manner. A
specific form of the latter would realize a coupling term proportional
to $\kb{\rm G}{\rm B}\left(\hat{a}+\hat{a}^{\dagger}\right)$, where
$\hat{a}$ is the cavity annihilation operator that would operate
as follows: if the atom is not-in-$\B$, the cavity is empty, otherwise,
as $\Obg$ steers the atom from $\G$ to $\B$, the cavity fills with
$\bar{n}\gg1$ photons and activates a strong dissipation channel
of $\B$ that repopulates $\G$ before $\B$ is ever appreciably populated.
Similar-in-spirit dissipation channels have been realized, e.g., with
the double-drive reset-of-population (DDROP) protocol \citep{Geerlings2013-reset}.
If sufficient gain is achieved, no quantum-limited amplifier is required,
and the scheme would simplify the setup and  transmission losses,
$\eta$.

\paragraph{Distillation and single-photon detector. }

The degenerate measurement of the Darkmon atom, perhaps employed
with the lowest four levels, could make it an interesting candidate
for magic-state and entanglement distillation protocols \citep{Bennett1996,Bravyi2005}.
Interestingly, the detection of a quantum jump from $\G$ to $\D$
can be viewed as the absorption and detection of a photon from the
input-output transmission line. In this sense, the three-level monitoring
scheme implements a photodetection apparatus for single flying microwave
photons in cQED. The device could be optimized with this goal in mind
to address the outstanding challenge of detecting itinerant microwave
photons with high efficiency \citep{Chen2011-FlyingPhoton,Fan2014-photonFly,Inomata2016-FlyingPhoton,Narla2016}.
In contrast to previous work on this subject, which focused on operating
detectors in a time-gated mode, the Darkmon scheme affords the advantage
of time-resolved, time-continuous photodetection with gain. It is
possible these advantages can be exploited for catching and releasing
flying Fock states \citep{Kalb2017,Campagne2017-entangle2}.

\paragraph{Stochastic drive, $\Omega_{{\rm DG}}$, and reversal. }

Realizable with the current device, one could catch and reverse the
quantum jump from $\G$ to $\D$ in the presence of a stochastic drive
from $\G$ to $\D$, $\Omega_{\mathrm{DG}}\left(t\right)$. The stochastic
drive more realistically models the effect of the environment and
breaks the feature of identical jumps; i.e., any two jumps no longer
look identical. The phase of the mid-flight superposition between
$\G$ and $\D$ is determined by the details of the stochastic $\Odg$
phase during the initial period of the jump, $\deltatcatch\ll\Delta t_{\mathrm{mid}}.$
Nonetheless, our prediction is that if the phase fluctuations of $\Odg$
at the beginning of the jump are known, one could still successfully
reverse the jump mid-flight with the appropriate coherent intervention.
This could be implemented by generating $\Odg$ with the FPGA controller,
on the fly, and having the controller calculate the correct intervention
angles, $\left\{ \theta_{I}\left(\deltatcatch\right),\varphi_{I}\left(\deltatcatch\right)\right\} $.
The reverse could be studied as a function of the bandwidth of the
noisy signal, $\Omega_{\mathrm{DG}}$ (practically, this could be
made as large as 25~MHz), thus exploring the crossover from jump
dynamics due to deterministic forces and those of the environment,
perhaps shedding further light on decoherence and measurement irreversibility.

\paragraph{Phase-agnostic reversal and quantum error correction.}

Calculation of the angles, $\left\{ \theta_{I}\left(\deltatcatch\right),\varphi_{I}\left(\deltatcatch\right)\right\} $,
becomes increasingly difficult for larger noise bandwidths. An alternative
strategy is to implement a phase-agnostic reversal. This could be
achieved with dissipation engineering \citep{Poyatos1996-qm-res-eng}
to conditionally dynamically cool to atom the ground state. Practical
cooling protocols have been experimentally demonstrated in cQED \citep{Valenzuela2006,Grajcar2008,Murch2012-bath-eng,Geerlings2013-reset,LiuYehan2016}. 

\begin{figure}
\begin{centering}
\includegraphics[scale=1.5]{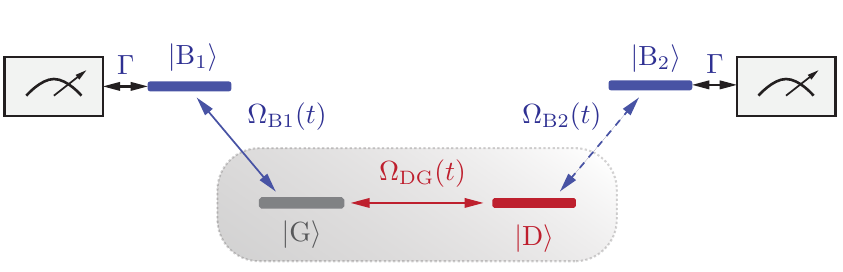}
\par\end{centering}
\caption[Four-level atom with two counter-steering measurements]{\label{fig:4lvlatom}\textbf{Four-level atom with two counter-steering
measurements.} Sketch of a modified Darkmon atom consisting of four
levels: ground, $\G$, Dark, $\D$, a first ``Bright,'' $\ket{\mathrm{B}_{1}}$,
and a second ``Bright,'' $\ket{\mathrm{B}_{2}}$. Both Bright levels
are monitored at rate $\Gamma$, while controlled-actuated Rabi drives
$\Omega_{\mathrm{B1}}\left(t\right)$ and $\Omega_{\mathrm{B1}}\left(t\right)$
turn on the effective monitoring of $\G$ and $\D,$ respectively.
A potentially stochastic Rabi drive $\Omega_{\mathrm{DG}}\left(t\right)$
links $\G$ and $\D$.}
\end{figure}

Instead of dissipation engineering, the jump could be reversed by
means of a measurement-backaction force due to another measurement.
This will likely have to be probabilistic, unless adaptive measurements
\citep{Wiseman1995-adaptive,Jacobs2003-adaptive-msr} or measured-based
quantum steering is employed \citep{schrodinger1935gegenwartige,Murch2013a,Wiseman2007-steering}.
Specifically, we propose to investigate a four-level scheme that builds
on the Darkmon, see Fig.~\ref{fig:4lvlatom}. The ground state, $\G$,
is monitored though a Bright state, $\ket{\mathrm{B}_{1}}$, by means
of Rabi drive, $\Omega_{\mathrm{B1}}\left(t\right)$, and the photodetection
of $\ket{\mathrm{B}_{1}}$ at rate $\Gamma.$ Similarly, the Dark
level, $\D$, is monitored by coupling it with a Rabi drive $\Omega_{\mathrm{B2}}\left(t\right)$
to a second Bright state, $\ket{\mathrm{B}_{2}}$, monitored at rate
$\Gamma.$ Conditioned on no clicks, the $\G$ measurement steers
the atom toward $\D,$ see Chapter~\ref{chap:theoretical-description-jumps}.
In contrast, conditioned on no clicks, the $\D$ measurement steers
the atom toward $\G$. Both forces are phase agnostic. Since they
oppose each other, one can be used to undo the effect of the other
with proper conditioning and control of the Rabi drives. If $\G$
is measured subject to the Dark Rabi drive $\Odg$ from $\G$ to $\D$,
which could be deterministic or stochastic, while $\Omega_{\mathrm{B2}}=0$,
the protocol demonstrated in Chapter~1 is implemented. When the signal
warning of the occurrence of the quantum jump from $\G$ to $\D$
is detected, $\Omega_{\mathrm{B1}}$ is shut off. If the catch time
is set to $\Delta t_{\mathrm{mid}}$, the state of the GD superposition
is known to be on the GD equator, but its phase may be unknown. If
$\Omega_{\mathrm{B2}}$ is turned on and the record is conditioned
on no clicks, the jump should be reversed, no matter what the superposition
phase is. More generally, the opposition of the two counter-steering
no-click measurements offers a unique testbed for studying non-commuting
simultaneous measurements, a topic of rising interest in the field
\citep{Jordan2005-non-coom,Ruskov2010,Hacohen-Gourgy2016-non-comm,Perarnau-Llobet2017,Atalaya2017,Lewalle2017,Patti2017,Ficheux2017,Chantasri2018}. 

Quantum jumps are intimately involved in the detection and correction
of errors in quantum information systems \citep{Sun2013,Ofek2016}.
A controller continuously monitors an error syndrome, often parity,
such as that of a cavity state \citep{Ofek2016,Cohen2017-cw-parity-msr}
or multi-qubit stabilizer \citep{Huembeli2017}, and detects jumps
in the measurement record, which signal the occurrence of an error.
The error needs to corrected. Catch and reversing the quantum jump
of an error syndrome prior to its occurrence could prevent the error
from manifesting fully. A research direction that could be explored
is to couple the GD transition to the parity operator of a long-lived
quantum-memory cavity \citep{Kirchmair2013} that encodes a logical
quantum state \citep{Cochrane1999-catCode,Mirrahimi2014,Leghtas2015,Michael2016,LiLinshu2017,Touzard2017}.
The parity bit of the cavity state would be continuously mapped onto
the GD manifold. The state $\G$ would indicate \emph{no} error, while
$\D$ would indicate that an error has occurred. If the noise process
driving the parity bit flips has sufficiently narrow bandwidth, it
may be possible to extend the catch and reverse protocol to intervene
in the occurrence of the parity-bit error. By monitoring the $\B$/not-$\B$
measurement record as discussed in Chapter~1, the controller would
detect the advance-warning signal and immediately perform a phase-agnostic
reversal of the jump to prevent the error. Applications of the jump
reversal protocol to quantum error correction schemes present an interesting
and open direction for theoretical and experimental research. 

\paragraph{Closing statement. }

We hope the catch and reverse experiment offers a new vantage point
on the state-disturbing nature of measurements and the interplay between
deterministic forces and the necessarily-stochastic ones due to quantum
measurements. More generally, we hope it could provide a conceptually
simple but striking illustration to help write operationally-based,
rather than postulate-based, textbooks for quantum mechanics.

\medskip

\addcontentsline{toc}{chapter}{Bibliography}
\singlespacing
\nocite{apsrev41Control} 
\bibliographystyle{apsrmp4-1}
\bibliography{library,bib2,revtex_mod} 

\begin{thebibliography}{236}%
\makeatletter
\providecommand \@ifxundefined [1]{%
 \@ifx{#1\undefined}
}%
\providecommand \@ifnum [1]{%
 \ifnum #1\expandafter \@firstoftwo
 \else \expandafter \@secondoftwo
 \fi
}%
\providecommand \@ifx [1]{%
 \ifx #1\expandafter \@firstoftwo
 \else \expandafter \@secondoftwo
 \fi
}%
\providecommand \natexlab [1]{#1}%
\providecommand \enquote  [1]{``#1''}%
\providecommand \bibnamefont  [1]{#1}%
\providecommand \bibfnamefont [1]{#1}%
\providecommand \citenamefont [1]{#1}%
\providecommand \href@noop [0]{\@secondoftwo}%
\providecommand \href [0]{\begingroup \@sanitize@url \@href}%
\providecommand \@href[1]{\@@startlink{#1}\@@href}%
\providecommand \@@href[1]{\endgroup#1\@@endlink}%
\providecommand \@sanitize@url [0]{\catcode `\\12\catcode `\$12\catcode
  `\&12\catcode `\#12\catcode `\^12\catcode `\_12\catcode `\%12\relax}%
\providecommand \@@startlink[1]{}%
\providecommand \@@endlink[0]{}%
\providecommand \url  [0]{\begingroup\@sanitize@url \@url }%
\providecommand \@url [1]{\endgroup\@href {#1}{\urlprefix }}%
\providecommand \urlprefix  [0]{URL }%
\providecommand \Eprint [0]{\href }%
\providecommand \doibase [0]{http://dx.doi.org/}%
\providecommand \selectlanguage [0]{\@gobble}%
\providecommand \bibinfo  [0]{\@secondoftwo}%
\providecommand \bibfield  [0]{\@secondoftwo}%
\providecommand \translation [1]{[#1]}%
\providecommand \BibitemOpen [0]{}%
\providecommand \bibitemStop [0]{}%
\providecommand \bibitemNoStop [0]{.\EOS\space}%
\providecommand \EOS [0]{\spacefactor3000\relax}%
\providecommand \BibitemShut  [1]{\csname bibitem#1\endcsname}%
\let\auto@bib@innerbib\@empty
\bibitem [{\citenamefont {Albert}\ \emph {et~al.}(2016)\citenamefont {Albert},
  \citenamefont {Bradlyn}, \citenamefont {Fraas},\ and\ \citenamefont
  {Jiang}}]{Albert2016-Lindblad}%
  \BibitemOpen
  \bibfield  {author} {\bibinfo {author} {\bibfnamefont {V.~V.}\ \bibnamefont
  {Albert}}, \bibinfo {author} {\bibfnamefont {B.}~\bibnamefont {Bradlyn}},
  \bibinfo {author} {\bibfnamefont {M.}~\bibnamefont {Fraas}}, \ and\ \bibinfo
  {author} {\bibfnamefont {L.}~\bibnamefont {Jiang}}} (\bibinfo {year}
  {2016}),\ \bibfield  {title} {\enquote {\bibinfo {title} {{Geometry and
  Response of Lindbladians}},}\ }\href {\doibase 10.1103/PhysRevX.6.041031}
  {\bibfield  {journal} {\bibinfo  {journal} {Physical Review X}\ }\textbf
  {\bibinfo {volume} {6}}~(\bibinfo {number} {4}),\ \bibinfo {pages}
  {041031}}\BibitemShut {NoStop}%
\bibitem [{\citenamefont {Albert}\ and\ \citenamefont
  {Jiang}(2014)}]{AlbertVV2014-Lindblad}%
  \BibitemOpen
  \bibfield  {author} {\bibinfo {author} {\bibfnamefont {V.~V.}\ \bibnamefont
  {Albert}}, \ and\ \bibinfo {author} {\bibfnamefont {L.}~\bibnamefont
  {Jiang}}} (\bibinfo {year} {2014}),\ \bibfield  {title} {\enquote {\bibinfo
  {title} {{Symmetries and conserved quantities in Lindblad master
  equations}},}\ }\href {\doibase 10.1103/PhysRevA.89.022118} {\bibfield
  {journal} {\bibinfo  {journal} {Physical Review A}\ }\textbf {\bibinfo
  {volume} {89}}~(\bibinfo {number} {2}),\ \bibinfo {pages}
  {022118}}\BibitemShut {NoStop}%
\bibitem [{\citenamefont {Alonso}\ \emph {et~al.}(2016)\citenamefont {Alonso},
  \citenamefont {Lutz},\ and\ \citenamefont {Romito}}]{Alonso2016-thermo}%
  \BibitemOpen
  \bibfield  {author} {\bibinfo {author} {\bibfnamefont {J.~J.}\ \bibnamefont
  {Alonso}}, \bibinfo {author} {\bibfnamefont {E.}~\bibnamefont {Lutz}}, \ and\
  \bibinfo {author} {\bibfnamefont {A.}~\bibnamefont {Romito}}} (\bibinfo
  {year} {2016}),\ \bibfield  {title} {\enquote {\bibinfo {title}
  {{Thermodynamics of Weakly Measured Quantum Systems}},}\ }\href {\doibase
  10.1103/PhysRevLett.116.080403} {\bibfield  {journal} {\bibinfo  {journal}
  {Physical Review Letters}\ }\textbf {\bibinfo {volume} {116}}~(\bibinfo
  {number} {8}),\ \bibinfo {pages} {080403}}\BibitemShut {NoStop}%
\bibitem [{\citenamefont {Ambegaokar}\ and\ \citenamefont
  {Baratoff}(1963)}]{Ambegaokar1963}%
  \BibitemOpen
  \bibfield  {author} {\bibinfo {author} {\bibfnamefont {V.}~\bibnamefont
  {Ambegaokar}}, \ and\ \bibinfo {author} {\bibfnamefont {A.}~\bibnamefont
  {Baratoff}}} (\bibinfo {year} {1963}),\ \bibfield  {title} {\enquote
  {\bibinfo {title} {{Tunneling Between Superconductors}},}\ }\href {\doibase
  10.1103/PhysRevLett.11.104} {\bibfield  {journal} {\bibinfo  {journal}
  {Physical Review Letters}\ }\textbf {\bibinfo {volume} {11}}~(\bibinfo
  {number} {2}),\ \bibinfo {pages} {104--104}}\BibitemShut {NoStop}%
\bibitem [{\citenamefont {Astafiev}\ \emph {et~al.}(2004)\citenamefont
  {Astafiev}, \citenamefont {Pashkin}, \citenamefont {Yamamoto}, \citenamefont
  {Nakamura},\ and\ \citenamefont {Tsai}}]{Astafiev2004-readout}%
  \BibitemOpen
  \bibfield  {author} {\bibinfo {author} {\bibfnamefont {O.}~\bibnamefont
  {Astafiev}}, \bibinfo {author} {\bibfnamefont {Y.~A.}\ \bibnamefont
  {Pashkin}}, \bibinfo {author} {\bibfnamefont {T.}~\bibnamefont {Yamamoto}},
  \bibinfo {author} {\bibfnamefont {Y.}~\bibnamefont {Nakamura}}, \ and\
  \bibinfo {author} {\bibfnamefont {J.~S.}\ \bibnamefont {Tsai}}} (\bibinfo
  {year} {2004}),\ \bibfield  {title} {\enquote {\bibinfo {title} {{Single-shot
  measurement of the Josephson charge qubit}},}\ }\href {\doibase
  10.1103/PhysRevB.69.180507} {\bibfield  {journal} {\bibinfo  {journal}
  {Physical Review B}\ }\textbf {\bibinfo {volume} {69}}~(\bibinfo {number}
  {18}),\ \bibinfo {pages} {180507}}\BibitemShut {NoStop}%
\bibitem [{\citenamefont {Atalaya}\ \emph {et~al.}(2017)\citenamefont
  {Atalaya}, \citenamefont {Hacohen-Gourgy}, \citenamefont {Martin},
  \citenamefont {Siddiqi}, \citenamefont {Korotkov},\ and\ \citenamefont {{N.
  Korotkov}}}]{Atalaya2017}%
  \BibitemOpen
  \bibfield  {author} {\bibinfo {author} {\bibfnamefont {J.}~\bibnamefont
  {Atalaya}}, \bibinfo {author} {\bibfnamefont {S.}~\bibnamefont
  {Hacohen-Gourgy}}, \bibinfo {author} {\bibfnamefont {L.~S.}\ \bibnamefont
  {Martin}}, \bibinfo {author} {\bibfnamefont {I.}~\bibnamefont {Siddiqi}},
  \bibinfo {author} {\bibfnamefont {A.~N.}\ \bibnamefont {Korotkov}}, \ and\
  \bibinfo {author} {\bibfnamefont {A.}~\bibnamefont {{N. Korotkov}}}}
  (\bibinfo {year} {2017}),\ \href {http://arxiv.org/abs/1702.08077} {\emph
  {\bibinfo {title} {{Correlators in simultaneous measurement of non-commuting
  qubit observables}}}},\ \Eprint {http://arxiv.org/abs/1702.08077}
  {arXiv:1702.08077} \BibitemShut {NoStop}%
\bibitem [{\citenamefont {Attal}\ and\ \citenamefont
  {Pautrat}(2006)}]{Attal2006}%
  \BibitemOpen
  \bibfield  {author} {\bibinfo {author} {\bibfnamefont {S.}~\bibnamefont
  {Attal}}, \ and\ \bibinfo {author} {\bibfnamefont {Y.}~\bibnamefont
  {Pautrat}}} (\bibinfo {year} {2006}),\ \bibfield  {title} {\enquote {\bibinfo
  {title} {{From Repeated to Continuous Quantum Interactions}},}\ }\href
  {\doibase 10.1007/s00023-005-0242-8} {\bibfield  {journal} {\bibinfo
  {journal} {Annales Henri Poincar{\'{e}}}\ }\textbf {\bibinfo {volume}
  {7}}~(\bibinfo {number} {1}),\ \bibinfo {pages} {59--104}}\BibitemShut
  {NoStop}%
\bibitem [{\citenamefont {Axline}\ \emph {et~al.}(2016)\citenamefont {Axline},
  \citenamefont {Reagor}, \citenamefont {Heeres}, \citenamefont {Reinhold},
  \citenamefont {Wang}, \citenamefont {Shain}, \citenamefont {Pfaff},
  \citenamefont {Chu}, \citenamefont {Frunzio},\ and\ \citenamefont
  {Schoelkopf}}]{Axline2016}%
  \BibitemOpen
  \bibfield  {author} {\bibinfo {author} {\bibfnamefont {C.}~\bibnamefont
  {Axline}}, \bibinfo {author} {\bibfnamefont {M.}~\bibnamefont {Reagor}},
  \bibinfo {author} {\bibfnamefont {R.}~\bibnamefont {Heeres}}, \bibinfo
  {author} {\bibfnamefont {P.}~\bibnamefont {Reinhold}}, \bibinfo {author}
  {\bibfnamefont {C.}~\bibnamefont {Wang}}, \bibinfo {author} {\bibfnamefont
  {K.}~\bibnamefont {Shain}}, \bibinfo {author} {\bibfnamefont
  {W.}~\bibnamefont {Pfaff}}, \bibinfo {author} {\bibfnamefont
  {Y.}~\bibnamefont {Chu}}, \bibinfo {author} {\bibfnamefont {L.}~\bibnamefont
  {Frunzio}}, \ and\ \bibinfo {author} {\bibfnamefont {R.~J.}\ \bibnamefont
  {Schoelkopf}}} (\bibinfo {year} {2016}),\ \bibfield  {title} {\enquote
  {\bibinfo {title} {{An architecture for integrating planar and 3D cQED
  devices}},}\ }\href {\doibase 10.1063/1.4959241} {\bibfield  {journal}
  {\bibinfo  {journal} {Applied Physics Letters}\ }\textbf {\bibinfo {volume}
  {109}}~(\bibinfo {number} {4}),\ \bibinfo {pages} {42601}}\BibitemShut
  {NoStop}%
\bibitem [{\citenamefont {Bardet}(2017)}]{Bardet2017}%
  \BibitemOpen
  \bibfield  {author} {\bibinfo {author} {\bibfnamefont {I.}~\bibnamefont
  {Bardet}}} (\bibinfo {year} {2017}),\ \bibfield  {title} {\enquote {\bibinfo
  {title} {{Classical and Quantum Parts of the Quantum Dynamics: The
  Discrete-Time Case}},}\ }\href {\doibase 10.1007/s00023-016-0517-2}
  {\bibfield  {journal} {\bibinfo  {journal} {Annales Henri Poincar{\'{e}}}\
  }\textbf {\bibinfo {volume} {18}}~(\bibinfo {number} {3}),\ \bibinfo {pages}
  {955--981}}\BibitemShut {NoStop}%
\bibitem [{\citenamefont {Barends}\ \emph {et~al.}(2013)\citenamefont
  {Barends}, \citenamefont {Kelly}, \citenamefont {Megrant}, \citenamefont
  {Sank}, \citenamefont {Jeffrey}, \citenamefont {Chen}, \citenamefont {Yin},
  \citenamefont {Chiaro}, \citenamefont {Mutus}, \citenamefont {Neill},
  \citenamefont {O'Malley}, \citenamefont {Roushan}, \citenamefont {Wenner},
  \citenamefont {White}, \citenamefont {Cleland},\ and\ \citenamefont
  {Martinis}}]{Barends2013}%
  \BibitemOpen
  \bibfield  {author} {\bibinfo {author} {\bibfnamefont {R.}~\bibnamefont
  {Barends}}, \bibinfo {author} {\bibfnamefont {J.}~\bibnamefont {Kelly}},
  \bibinfo {author} {\bibfnamefont {A.}~\bibnamefont {Megrant}}, \bibinfo
  {author} {\bibfnamefont {D.}~\bibnamefont {Sank}}, \bibinfo {author}
  {\bibfnamefont {E.}~\bibnamefont {Jeffrey}}, \bibinfo {author} {\bibfnamefont
  {Y.}~\bibnamefont {Chen}}, \bibinfo {author} {\bibfnamefont {Y.}~\bibnamefont
  {Yin}}, \bibinfo {author} {\bibfnamefont {B.}~\bibnamefont {Chiaro}},
  \bibinfo {author} {\bibfnamefont {J.}~\bibnamefont {Mutus}}, \bibinfo
  {author} {\bibfnamefont {C.}~\bibnamefont {Neill}}, \bibinfo {author}
  {\bibfnamefont {P.}~\bibnamefont {O'Malley}}, \bibinfo {author}
  {\bibfnamefont {P.}~\bibnamefont {Roushan}}, \bibinfo {author} {\bibfnamefont
  {J.}~\bibnamefont {Wenner}}, \bibinfo {author} {\bibfnamefont {T.~C.}\
  \bibnamefont {White}}, \bibinfo {author} {\bibfnamefont {A.~N.}\ \bibnamefont
  {Cleland}}, \ and\ \bibinfo {author} {\bibfnamefont {J.~M.}\ \bibnamefont
  {Martinis}}} (\bibinfo {year} {2013}),\ \bibfield  {title} {\enquote
  {\bibinfo {title} {{Coherent josephson qubit suitable for scalable quantum
  integrated circuits}},}\ }\href {\doibase 10.1103/PhysRevLett.111.080502}
  {\bibfield  {journal} {\bibinfo  {journal} {Physical Review Letters}\
  }\textbf {\bibinfo {volume} {111}}~(\bibinfo {number} {8}),\ \bibinfo {pages}
  {080502}},\ \Eprint {http://arxiv.org/abs/1304.2322} {arXiv:1304.2322}
  \BibitemShut {NoStop}%
\bibitem [{\citenamefont {Barends}\ \emph {et~al.}(2011)\citenamefont
  {Barends}, \citenamefont {Wenner}, \citenamefont {Lenander}, \citenamefont
  {Chen}, \citenamefont {Bialczak}, \citenamefont {Kelly}, \citenamefont
  {Lucero}, \citenamefont {O'Malley}, \citenamefont {Mariantoni}, \citenamefont
  {Sank}, \citenamefont {Wang}, \citenamefont {White}, \citenamefont {Yin},
  \citenamefont {Zhao}, \citenamefont {Cleland}, \citenamefont {Martinis},\
  and\ \citenamefont {Baselmans}}]{Barends2011-qp}%
  \BibitemOpen
  \bibfield  {author} {\bibinfo {author} {\bibfnamefont {R.}~\bibnamefont
  {Barends}}, \bibinfo {author} {\bibfnamefont {J.}~\bibnamefont {Wenner}},
  \bibinfo {author} {\bibfnamefont {M.}~\bibnamefont {Lenander}}, \bibinfo
  {author} {\bibfnamefont {Y.}~\bibnamefont {Chen}}, \bibinfo {author}
  {\bibfnamefont {R.~C.}\ \bibnamefont {Bialczak}}, \bibinfo {author}
  {\bibfnamefont {J.}~\bibnamefont {Kelly}}, \bibinfo {author} {\bibfnamefont
  {E.}~\bibnamefont {Lucero}}, \bibinfo {author} {\bibfnamefont
  {P.}~\bibnamefont {O'Malley}}, \bibinfo {author} {\bibfnamefont
  {M.}~\bibnamefont {Mariantoni}}, \bibinfo {author} {\bibfnamefont
  {D.}~\bibnamefont {Sank}}, \bibinfo {author} {\bibfnamefont {H.}~\bibnamefont
  {Wang}}, \bibinfo {author} {\bibfnamefont {T.~C.}\ \bibnamefont {White}},
  \bibinfo {author} {\bibfnamefont {Y.}~\bibnamefont {Yin}}, \bibinfo {author}
  {\bibfnamefont {J.}~\bibnamefont {Zhao}}, \bibinfo {author} {\bibfnamefont
  {A.~N.}\ \bibnamefont {Cleland}}, \bibinfo {author} {\bibfnamefont {J.~M.}\
  \bibnamefont {Martinis}}, \ and\ \bibinfo {author} {\bibfnamefont {J.~J.~A.}\
  \bibnamefont {Baselmans}}} (\bibinfo {year} {2011}),\ \bibfield  {title}
  {\enquote {\bibinfo {title} {{Minimizing quasiparticle generation from stray
  infrared light in superconducting quantum circuits}},}\ }\href {\doibase
  10.1063/1.3638063} {\bibfield  {journal} {\bibinfo  {journal} {Applied
  Physics Letters}\ }\textbf {\bibinfo {volume} {99}}~(\bibinfo {number}
  {11}),\ \bibinfo {pages} {113507}}\BibitemShut {NoStop}%
\bibitem [{\citenamefont {Basche}\ \emph {et~al.}(1995)\citenamefont {Basche},
  \citenamefont {Kummer},\ and\ \citenamefont {Brauchle}}]{Basche1995}%
  \BibitemOpen
  \bibfield  {author} {\bibinfo {author} {\bibfnamefont {T.}~\bibnamefont
  {Basche}}, \bibinfo {author} {\bibfnamefont {S.}~\bibnamefont {Kummer}}, \
  and\ \bibinfo {author} {\bibfnamefont {C.}~\bibnamefont {Brauchle}}}
  (\bibinfo {year} {1995}),\ \bibfield  {title} {\enquote {\bibinfo {title}
  {{Direct spectroscopic observation of quantum jumps of a single molecule}},}\
  }\href {\doibase doi:10.1038/373132a0} {\bibfield  {journal} {\bibinfo
  {journal} {Nature}\ }\textbf {\bibinfo {volume} {373}}~(\bibinfo {number}
  {6510}),\ \bibinfo {pages} {132--134}}\BibitemShut {NoStop}%
\bibitem [{\citenamefont {Belavkin}(1987)}]{belavkin1987non}%
  \BibitemOpen
  \bibfield  {author} {\bibinfo {author} {\bibfnamefont {V.}~\bibnamefont
  {Belavkin}}} (\bibinfo {year} {1987}),\ \bibfield  {title} {\enquote
  {\bibinfo {title} {{Non-Demolition Measurement and Control in Quantum
  Dynamical Systems}},}\ }in\ \href {\doibase 10.1007/978-3-7091-2971-5_19}
  {\emph {\bibinfo {booktitle} {Information Complexity and Control in Quantum
  Physics}}}\ (\bibinfo  {publisher} {Springer Vienna},\ \bibinfo {address}
  {Vienna})\ pp.\ \bibinfo {pages} {311--329}\BibitemShut {NoStop}%
\bibitem [{\citenamefont {Bennett}\ \emph {et~al.}(1996)\citenamefont
  {Bennett}, \citenamefont {DiVincenzo}, \citenamefont {Smolin},\ and\
  \citenamefont {Wootters}}]{Bennett1996}%
  \BibitemOpen
  \bibfield  {author} {\bibinfo {author} {\bibfnamefont {C.~H.}\ \bibnamefont
  {Bennett}}, \bibinfo {author} {\bibfnamefont {D.~P.}\ \bibnamefont
  {DiVincenzo}}, \bibinfo {author} {\bibfnamefont {J.~A.}\ \bibnamefont
  {Smolin}}, \ and\ \bibinfo {author} {\bibfnamefont {W.~K.}\ \bibnamefont
  {Wootters}}} (\bibinfo {year} {1996}),\ \bibfield  {title} {\enquote
  {\bibinfo {title} {{Mixed-state entanglement and quantum error
  correction}},}\ }\href {\doibase 10.1103/PhysRevA.54.3824} {\bibfield
  {journal} {\bibinfo  {journal} {Physical Review A}\ }\textbf {\bibinfo
  {volume} {54}}~(\bibinfo {number} {5}),\ \bibinfo {pages}
  {3824--3851}}\BibitemShut {NoStop}%
\bibitem [{\citenamefont {Bergeal}\ \emph {et~al.}(2010)\citenamefont
  {Bergeal}, \citenamefont {Schackert}, \citenamefont {Metcalfe}, \citenamefont
  {Vijay}, \citenamefont {Manucharyan}, \citenamefont {Frunzio}, \citenamefont
  {Prober}, \citenamefont {Schoelkopf}, \citenamefont {Girvin},\ and\
  \citenamefont {Devoret}}]{Bergeal2010}%
  \BibitemOpen
  \bibfield  {author} {\bibinfo {author} {\bibfnamefont {N.}~\bibnamefont
  {Bergeal}}, \bibinfo {author} {\bibfnamefont {F.}~\bibnamefont {Schackert}},
  \bibinfo {author} {\bibfnamefont {M.}~\bibnamefont {Metcalfe}}, \bibinfo
  {author} {\bibfnamefont {R.}~\bibnamefont {Vijay}}, \bibinfo {author}
  {\bibfnamefont {V.~E.}\ \bibnamefont {Manucharyan}}, \bibinfo {author}
  {\bibfnamefont {L.}~\bibnamefont {Frunzio}}, \bibinfo {author} {\bibfnamefont
  {D.~E.}\ \bibnamefont {Prober}}, \bibinfo {author} {\bibfnamefont {R.~J.}\
  \bibnamefont {Schoelkopf}}, \bibinfo {author} {\bibfnamefont {S.~M.}\
  \bibnamefont {Girvin}}, \ and\ \bibinfo {author} {\bibfnamefont {M.~H.}\
  \bibnamefont {Devoret}}} (\bibinfo {year} {2010}),\ \bibfield  {title}
  {\enquote {\bibinfo {title} {{Phase-preserving amplification near the quantum
  limit with a Josephson ring modulator}},}\ }\href {\doibase
  10.1038/nature09035} {\bibfield  {journal} {\bibinfo  {journal} {Nature}\
  }\textbf {\bibinfo {volume} {465}}~(\bibinfo {number} {7294}),\ \bibinfo
  {pages} {64--68}},\ \Eprint {http://arxiv.org/abs/0912.3407}
  {arXiv:0912.3407} \BibitemShut {NoStop}%
\bibitem [{\citenamefont {Bergquist}\ \emph {et~al.}(1986)\citenamefont
  {Bergquist}, \citenamefont {Hulet}, \citenamefont {Itano},\ and\
  \citenamefont {Wineland}}]{Bergquist1986}%
  \BibitemOpen
  \bibfield  {author} {\bibinfo {author} {\bibfnamefont {J.~C.}\ \bibnamefont
  {Bergquist}}, \bibinfo {author} {\bibfnamefont {R.~G.}\ \bibnamefont
  {Hulet}}, \bibinfo {author} {\bibfnamefont {W.~M.}\ \bibnamefont {Itano}}, \
  and\ \bibinfo {author} {\bibfnamefont {D.~J.}\ \bibnamefont {Wineland}}}
  (\bibinfo {year} {1986}),\ \bibfield  {title} {\enquote {\bibinfo {title}
  {{Observation of Quantum Jumps in a Single Atom}},}\ }\href {\doibase
  10.1103/PhysRevLett.57.1699} {\bibfield  {journal} {\bibinfo  {journal}
  {Physical Review Letters}\ }\textbf {\bibinfo {volume} {57}}~(\bibinfo
  {number} {14}),\ \bibinfo {pages} {1699--1702}}\BibitemShut {NoStop}%
\bibitem [{\citenamefont {Blais}\ \emph {et~al.}(2004)\citenamefont {Blais},
  \citenamefont {Huang}, \citenamefont {Wallraff}, \citenamefont {Girvin},\
  and\ \citenamefont {Schoelkopf}}]{Blais2004}%
  \BibitemOpen
  \bibfield  {author} {\bibinfo {author} {\bibfnamefont {A.}~\bibnamefont
  {Blais}}, \bibinfo {author} {\bibfnamefont {R.-S.}\ \bibnamefont {Huang}},
  \bibinfo {author} {\bibfnamefont {A.}~\bibnamefont {Wallraff}}, \bibinfo
  {author} {\bibfnamefont {S.~M.}\ \bibnamefont {Girvin}}, \ and\ \bibinfo
  {author} {\bibfnamefont {R.~J.}\ \bibnamefont {Schoelkopf}}} (\bibinfo {year}
  {2004}),\ \bibfield  {title} {\enquote {\bibinfo {title} {{Cavity quantum
  electrodynamics for superconducting electrical circuits: An architecture for
  quantum computation}},}\ }\href {\doibase 10.1103/PhysRevA.69.062320}
  {\bibfield  {journal} {\bibinfo  {journal} {Physical Review A}\ }\textbf
  {\bibinfo {volume} {69}}~(\bibinfo {number} {6}),\ \bibinfo {pages}
  {062320}},\ \Eprint {http://arxiv.org/abs/0402216} {arXiv:0402216 [cond-mat]}
  \BibitemShut {NoStop}%
\bibitem [{\citenamefont {Bohr}(1913)}]{Bohr1913}%
  \BibitemOpen
  \bibfield  {author} {\bibinfo {author} {\bibfnamefont {N.}~\bibnamefont
  {Bohr}}} (\bibinfo {year} {1913}),\ \bibfield  {title} {\enquote {\bibinfo
  {title} {{On the Constitution of Atoms and Molecules}},}\ }\href@noop {}
  {\bibfield  {journal} {\bibinfo  {journal} {Philosophical Magazine}\ }\textbf
  {\bibinfo {volume} {26}},\ \bibinfo {pages} {476--502}}\BibitemShut {NoStop}%
\bibitem [{\citenamefont {Boissonneault}\ \emph {et~al.}(2008)\citenamefont
  {Boissonneault}, \citenamefont {Gambetta},\ and\ \citenamefont
  {Blais}}]{Boissonneault2008}%
  \BibitemOpen
  \bibfield  {author} {\bibinfo {author} {\bibfnamefont {M.}~\bibnamefont
  {Boissonneault}}, \bibinfo {author} {\bibfnamefont {J.~M.}\ \bibnamefont
  {Gambetta}}, \ and\ \bibinfo {author} {\bibfnamefont {A.}~\bibnamefont
  {Blais}}} (\bibinfo {year} {2008}),\ \bibfield  {title} {\enquote {\bibinfo
  {title} {{Nonlinear dispersive regime of cavity QED: The dressed dephasing
  model}},}\ }\href {\doibase 10.1103/PhysRevA.77.060305} {\bibfield  {journal}
  {\bibinfo  {journal} {Physical Review A}\ }\textbf {\bibinfo {volume}
  {77}}~(\bibinfo {number} {6}),\ \bibinfo {pages} {060305}}\BibitemShut
  {NoStop}%
\bibitem [{\citenamefont {Boissonneault}\ \emph {et~al.}(2009)\citenamefont
  {Boissonneault}, \citenamefont {Gambetta},\ and\ \citenamefont
  {Blais}}]{Boissonneault2009-Photon-induced-relax}%
  \BibitemOpen
  \bibfield  {author} {\bibinfo {author} {\bibfnamefont {M.}~\bibnamefont
  {Boissonneault}}, \bibinfo {author} {\bibfnamefont {J.~M.}\ \bibnamefont
  {Gambetta}}, \ and\ \bibinfo {author} {\bibfnamefont {A.}~\bibnamefont
  {Blais}}} (\bibinfo {year} {2009}),\ \bibfield  {title} {\enquote {\bibinfo
  {title} {{Dispersive regime of circuit QED: Photon-dependent qubit dephasing
  and relaxation rates}},}\ }\href {\doibase 10.1103/PhysRevA.79.013819}
  {\bibfield  {journal} {\bibinfo  {journal} {Physical Review A}\ }\textbf
  {\bibinfo {volume} {79}}~(\bibinfo {number} {1}),\ \bibinfo {pages}
  {013819}}\BibitemShut {NoStop}%
\bibitem [{\citenamefont {Bourassa}\ \emph {et~al.}(2012)\citenamefont
  {Bourassa}, \citenamefont {Beaudoin}, \citenamefont {Gambetta},\ and\
  \citenamefont {Blais}}]{Bourassa2012}%
  \BibitemOpen
  \bibfield  {author} {\bibinfo {author} {\bibfnamefont {J.}~\bibnamefont
  {Bourassa}}, \bibinfo {author} {\bibfnamefont {F.}~\bibnamefont {Beaudoin}},
  \bibinfo {author} {\bibfnamefont {J.~M.}\ \bibnamefont {Gambetta}}, \ and\
  \bibinfo {author} {\bibfnamefont {A.}~\bibnamefont {Blais}}} (\bibinfo {year}
  {2012}),\ \bibfield  {title} {\enquote {\bibinfo {title}
  {{Josephson-junction-embedded transmission-line resonators: From Kerr medium
  to in-line transmon}},}\ }\href {\doibase 10.1103/PhysRevA.86.013814}
  {\bibfield  {journal} {\bibinfo  {journal} {Physical Review A}\ }\textbf
  {\bibinfo {volume} {86}}~(\bibinfo {number} {1}),\ \bibinfo {pages}
  {013814}},\ \Eprint {http://arxiv.org/abs/arXiv:1204.2237v2}
  {arXiv:arXiv:1204.2237v2} \BibitemShut {NoStop}%
\bibitem [{\citenamefont {Bravyi}\ and\ \citenamefont
  {Kitaev}(2005)}]{Bravyi2005}%
  \BibitemOpen
  \bibfield  {author} {\bibinfo {author} {\bibfnamefont {S.}~\bibnamefont
  {Bravyi}}, \ and\ \bibinfo {author} {\bibfnamefont {A.}~\bibnamefont
  {Kitaev}}} (\bibinfo {year} {2005}),\ \bibfield  {title} {\enquote {\bibinfo
  {title} {{Universal quantum computation with ideal Clifford gates and noisy
  ancillas}},}\ }\href {\doibase 10.1103/PhysRevA.71.022316} {\bibfield
  {journal} {\bibinfo  {journal} {Physical Review A}\ }\textbf {\bibinfo
  {volume} {71}}~(\bibinfo {number} {2}),\ \bibinfo {pages}
  {022316}}\BibitemShut {NoStop}%
\bibitem [{\citenamefont {Brecht}(2017)}]{Brecht2017-Thesis}%
  \BibitemOpen
  \bibfield  {author} {\bibinfo {author} {\bibfnamefont {T.}~\bibnamefont
  {Brecht}}} (\bibinfo {year} {2017}),\ \emph {\bibinfo {title} {{Micromachined
  Quantum Circuits}}},\ \href
  {http://rsl.yale.edu/sites/default/files/files/RSL{\_}Theses/Brecht{\_}Thesis{\_}Final{\_}ScreenVersion.pdf}
  {Ph.D. thesis}\ (\bibinfo  {school} {Yale Univeristy})\BibitemShut {NoStop}%
\bibitem [{\citenamefont {Brecht}\ \emph {et~al.}(2016)\citenamefont {Brecht},
  \citenamefont {Pfaff}, \citenamefont {Wang}, \citenamefont {Chu},
  \citenamefont {Frunzio}, \citenamefont {Devoret},\ and\ \citenamefont
  {Schoelkopf}}]{Brecht2016}%
  \BibitemOpen
  \bibfield  {author} {\bibinfo {author} {\bibfnamefont {T.}~\bibnamefont
  {Brecht}}, \bibinfo {author} {\bibfnamefont {W.}~\bibnamefont {Pfaff}},
  \bibinfo {author} {\bibfnamefont {C.}~\bibnamefont {Wang}}, \bibinfo {author}
  {\bibfnamefont {Y.}~\bibnamefont {Chu}}, \bibinfo {author} {\bibfnamefont
  {L.}~\bibnamefont {Frunzio}}, \bibinfo {author} {\bibfnamefont {M.~H.}\
  \bibnamefont {Devoret}}, \ and\ \bibinfo {author} {\bibfnamefont {R.~J.}\
  \bibnamefont {Schoelkopf}}} (\bibinfo {year} {2016}),\ \bibfield  {title}
  {\enquote {\bibinfo {title} {{Multilayer microwave integrated quantum
  circuits for scalable quantum computing}},}\ }\href {\doibase
  10.1038/npjqi.2016.2} {\bibfield  {journal} {\bibinfo  {journal} {npj Quantum
  Information}\ }\textbf {\bibinfo {volume} {2}}~(\bibinfo {number} {1}),\
  \bibinfo {pages} {16002}},\ \Eprint {http://arxiv.org/abs/1509.01127}
  {arXiv:1509.01127} \BibitemShut {NoStop}%
\bibitem [{\citenamefont {Brecht}\ \emph {et~al.}(2015)\citenamefont {Brecht},
  \citenamefont {Reagor}, \citenamefont {Chu}, \citenamefont {Pfaff},
  \citenamefont {Wang}, \citenamefont {Frunzio}, \citenamefont {Devoret},\ and\
  \citenamefont {Schoelkopf}}]{Brecht2015}%
  \BibitemOpen
  \bibfield  {author} {\bibinfo {author} {\bibfnamefont {T.}~\bibnamefont
  {Brecht}}, \bibinfo {author} {\bibfnamefont {M.}~\bibnamefont {Reagor}},
  \bibinfo {author} {\bibfnamefont {Y.}~\bibnamefont {Chu}}, \bibinfo {author}
  {\bibfnamefont {W.}~\bibnamefont {Pfaff}}, \bibinfo {author} {\bibfnamefont
  {C.}~\bibnamefont {Wang}}, \bibinfo {author} {\bibfnamefont {L.}~\bibnamefont
  {Frunzio}}, \bibinfo {author} {\bibfnamefont {M.~H.}\ \bibnamefont
  {Devoret}}, \ and\ \bibinfo {author} {\bibfnamefont {R.~J.}\ \bibnamefont
  {Schoelkopf}}} (\bibinfo {year} {2015}),\ \bibfield  {title} {\enquote
  {\bibinfo {title} {{Demonstration of superconducting micromachined
  cavities}},}\ }\href {\doibase 10.1063/1.4935541} {\bibfield  {journal}
  {\bibinfo  {journal} {Applied Physics Letters}\ }\textbf {\bibinfo {volume}
  {107}}~(\bibinfo {number} {19}),\ \bibinfo {pages} {192603}},\ \Eprint
  {http://arxiv.org/abs/1509.01119} {arXiv:1509.01119} \BibitemShut {NoStop}%
\bibitem [{\citenamefont {Bultink}\ \emph {et~al.}(2016)\citenamefont
  {Bultink}, \citenamefont {Rol}, \citenamefont {O'Brien}, \citenamefont {Fu},
  \citenamefont {Dikken}, \citenamefont {Dickel}, \citenamefont {Vermeulen},
  \citenamefont {de~Sterke}, \citenamefont {Bruno}, \citenamefont {Schouten},\
  and\ \citenamefont {DiCarlo}}]{Bultink2016}%
  \BibitemOpen
  \bibfield  {author} {\bibinfo {author} {\bibfnamefont {C.~C.}\ \bibnamefont
  {Bultink}}, \bibinfo {author} {\bibfnamefont {M.~A.}\ \bibnamefont {Rol}},
  \bibinfo {author} {\bibfnamefont {T.~E.}\ \bibnamefont {O'Brien}}, \bibinfo
  {author} {\bibfnamefont {X.}~\bibnamefont {Fu}}, \bibinfo {author}
  {\bibfnamefont {B.~C.~S.}\ \bibnamefont {Dikken}}, \bibinfo {author}
  {\bibfnamefont {C.}~\bibnamefont {Dickel}}, \bibinfo {author} {\bibfnamefont
  {R.~F.~L.}\ \bibnamefont {Vermeulen}}, \bibinfo {author} {\bibfnamefont
  {J.~C.}\ \bibnamefont {de~Sterke}}, \bibinfo {author} {\bibfnamefont
  {A.}~\bibnamefont {Bruno}}, \bibinfo {author} {\bibfnamefont {R.~N.}\
  \bibnamefont {Schouten}}, \ and\ \bibinfo {author} {\bibfnamefont
  {L.}~\bibnamefont {DiCarlo}}} (\bibinfo {year} {2016}),\ \bibfield  {title}
  {\enquote {\bibinfo {title} {{Active Resonator Reset in the Nonlinear
  Dispersive Regime of Circuit QED}},}\ }\href {\doibase
  10.1103/PhysRevApplied.6.034008} {\bibfield  {journal} {\bibinfo  {journal}
  {Physical Review Applied}\ }\textbf {\bibinfo {volume} {6}}~(\bibinfo
  {number} {3}),\ \bibinfo {pages} {034008}}\BibitemShut {NoStop}%
\bibitem [{\citenamefont {Bultink}\ \emph {et~al.}(2018)\citenamefont
  {Bultink}, \citenamefont {Tarasinski}, \citenamefont {Haandb{\ae}k},
  \citenamefont {Poletto}, \citenamefont {Haider}, \citenamefont {Michalak},
  \citenamefont {Bruno},\ and\ \citenamefont {DiCarlo}}]{Bultink2018}%
  \BibitemOpen
  \bibfield  {author} {\bibinfo {author} {\bibfnamefont {C.~C.}\ \bibnamefont
  {Bultink}}, \bibinfo {author} {\bibfnamefont {B.}~\bibnamefont {Tarasinski}},
  \bibinfo {author} {\bibfnamefont {N.}~\bibnamefont {Haandb{\ae}k}}, \bibinfo
  {author} {\bibfnamefont {S.}~\bibnamefont {Poletto}}, \bibinfo {author}
  {\bibfnamefont {N.}~\bibnamefont {Haider}}, \bibinfo {author} {\bibfnamefont
  {D.~J.}\ \bibnamefont {Michalak}}, \bibinfo {author} {\bibfnamefont
  {A.}~\bibnamefont {Bruno}}, \ and\ \bibinfo {author} {\bibfnamefont
  {L.}~\bibnamefont {DiCarlo}}} (\bibinfo {year} {2018}),\ \bibfield  {title}
  {\enquote {\bibinfo {title} {{General method for extracting the quantum
  efficiency of dispersive qubit readout in circuit QED}},}\ }\href {\doibase
  10.1063/1.5015954} {\bibfield  {journal} {\bibinfo  {journal} {Applied
  Physics Letters}\ }\textbf {\bibinfo {volume} {112}}~(\bibinfo {number}
  {9}),\ \bibinfo {pages} {092601}}\BibitemShut {NoStop}%
\bibitem [{\citenamefont {Bylander}\ \emph {et~al.}(2011)\citenamefont
  {Bylander}, \citenamefont {Gustavsson}, \citenamefont {Yan}, \citenamefont
  {Yoshihara}, \citenamefont {Harrabi}, \citenamefont {Fitch}, \citenamefont
  {Cory}, \citenamefont {Nakamura}, \citenamefont {Tsai},\ and\ \citenamefont
  {Oliver}}]{Bylander2011}%
  \BibitemOpen
  \bibfield  {author} {\bibinfo {author} {\bibfnamefont {J.}~\bibnamefont
  {Bylander}}, \bibinfo {author} {\bibfnamefont {S.}~\bibnamefont
  {Gustavsson}}, \bibinfo {author} {\bibfnamefont {F.}~\bibnamefont {Yan}},
  \bibinfo {author} {\bibfnamefont {F.}~\bibnamefont {Yoshihara}}, \bibinfo
  {author} {\bibfnamefont {K.}~\bibnamefont {Harrabi}}, \bibinfo {author}
  {\bibfnamefont {G.}~\bibnamefont {Fitch}}, \bibinfo {author} {\bibfnamefont
  {D.~G.}\ \bibnamefont {Cory}}, \bibinfo {author} {\bibfnamefont
  {Y.}~\bibnamefont {Nakamura}}, \bibinfo {author} {\bibfnamefont {J.-S.}\
  \bibnamefont {Tsai}}, \ and\ \bibinfo {author} {\bibfnamefont {W.~D.}\
  \bibnamefont {Oliver}}} (\bibinfo {year} {2011}),\ \bibfield  {title}
  {\enquote {\bibinfo {title} {{Noise spectroscopy through dynamical decoupling
  with a superconducting flux qubit}},}\ }\href {\doibase 10.1038/nphys1994}
  {\bibfield  {journal} {\bibinfo  {journal} {Nature Physics}\ }\textbf
  {\bibinfo {volume} {7}}~(\bibinfo {number} {7}),\ \bibinfo {pages}
  {565--570}}\BibitemShut {NoStop}%
\bibitem [{\citenamefont {Campagne-Ibarcq}\ \emph {et~al.}(2014)\citenamefont
  {Campagne-Ibarcq}, \citenamefont {Bretheau}, \citenamefont {Flurin},
  \citenamefont {Auff{\`{e}}ves}, \citenamefont {Mallet},\ and\ \citenamefont
  {Huard}}]{Campagne-Ibarcq2014}%
  \BibitemOpen
  \bibfield  {author} {\bibinfo {author} {\bibfnamefont {P.}~\bibnamefont
  {Campagne-Ibarcq}}, \bibinfo {author} {\bibfnamefont {L.}~\bibnamefont
  {Bretheau}}, \bibinfo {author} {\bibfnamefont {E.}~\bibnamefont {Flurin}},
  \bibinfo {author} {\bibfnamefont {A.}~\bibnamefont {Auff{\`{e}}ves}},
  \bibinfo {author} {\bibfnamefont {F.}~\bibnamefont {Mallet}}, \ and\ \bibinfo
  {author} {\bibfnamefont {B.}~\bibnamefont {Huard}}} (\bibinfo {year}
  {2014}),\ \bibfield  {title} {\enquote {\bibinfo {title} {{Observing
  interferences between past and future quantum states in resonance
  fluorescence}},}\ }\href {\doibase 10.1103/PhysRevLett.112.180402} {\bibfield
   {journal} {\bibinfo  {journal} {Physical Review Letters}\ }\textbf {\bibinfo
  {volume} {112}}~(\bibinfo {number} {18}),\ \bibinfo {pages} {180402}},\
  \Eprint {http://arxiv.org/abs/arXiv:1311.5605v1} {arXiv:arXiv:1311.5605v1}
  \BibitemShut {NoStop}%
\bibitem [{\citenamefont {Campagne-Ibarcq}\ \emph
  {et~al.}(2016{\natexlab{a}})\citenamefont {Campagne-Ibarcq}, \citenamefont
  {Jezouin}, \citenamefont {Cottet}, \citenamefont {Six}, \citenamefont
  {Bretheau}, \citenamefont {Mallet}, \citenamefont {Sarlette}, \citenamefont
  {Rouchon},\ and\ \citenamefont {Huard}}]{Campagne-Ibarcq2016}%
  \BibitemOpen
  \bibfield  {author} {\bibinfo {author} {\bibfnamefont {P.}~\bibnamefont
  {Campagne-Ibarcq}}, \bibinfo {author} {\bibfnamefont {S.}~\bibnamefont
  {Jezouin}}, \bibinfo {author} {\bibfnamefont {N.}~\bibnamefont {Cottet}},
  \bibinfo {author} {\bibfnamefont {P.}~\bibnamefont {Six}}, \bibinfo {author}
  {\bibfnamefont {L.}~\bibnamefont {Bretheau}}, \bibinfo {author}
  {\bibfnamefont {F.}~\bibnamefont {Mallet}}, \bibinfo {author} {\bibfnamefont
  {A.}~\bibnamefont {Sarlette}}, \bibinfo {author} {\bibfnamefont
  {P.}~\bibnamefont {Rouchon}}, \ and\ \bibinfo {author} {\bibfnamefont
  {B.}~\bibnamefont {Huard}}} (\bibinfo {year} {2016}{\natexlab{a}}),\
  \bibfield  {title} {\enquote {\bibinfo {title} {{Using Spontaneous Emission
  of a Qubit as a Resource for Feedback Control}},}\ }\href {\doibase
  10.1103/PhysRevLett.117.060502} {\bibfield  {journal} {\bibinfo  {journal}
  {Physical Review Letters}\ }\textbf {\bibinfo {volume} {117}}~(\bibinfo
  {number} {6}),\ \bibinfo {pages} {060502}},\ \Eprint
  {http://arxiv.org/abs/1602.05479} {arXiv:1602.05479} \BibitemShut {NoStop}%
\bibitem [{\citenamefont {Campagne-Ibarcq}\ \emph
  {et~al.}(2016{\natexlab{b}})\citenamefont {Campagne-Ibarcq}, \citenamefont
  {Six}, \citenamefont {Bretheau}, \citenamefont {Sarlette}, \citenamefont
  {Mirrahimi}, \citenamefont {Rouchon},\ and\ \citenamefont
  {Huard}}]{Campagne2016-Fluorescence}%
  \BibitemOpen
  \bibfield  {author} {\bibinfo {author} {\bibfnamefont {P.}~\bibnamefont
  {Campagne-Ibarcq}}, \bibinfo {author} {\bibfnamefont {P.}~\bibnamefont
  {Six}}, \bibinfo {author} {\bibfnamefont {L.}~\bibnamefont {Bretheau}},
  \bibinfo {author} {\bibfnamefont {A.}~\bibnamefont {Sarlette}}, \bibinfo
  {author} {\bibfnamefont {M.}~\bibnamefont {Mirrahimi}}, \bibinfo {author}
  {\bibfnamefont {P.}~\bibnamefont {Rouchon}}, \ and\ \bibinfo {author}
  {\bibfnamefont {B.}~\bibnamefont {Huard}}} (\bibinfo {year}
  {2016}{\natexlab{b}}),\ \bibfield  {title} {\enquote {\bibinfo {title}
  {{Observing Quantum State Diffusion by Heterodyne Detection of
  Fluorescence}},}\ }\href {\doibase 10.1103/PhysRevX.6.011002} {\bibfield
  {journal} {\bibinfo  {journal} {Physical Review X}\ }\textbf {\bibinfo
  {volume} {6}}~(\bibinfo {number} {1}),\ \bibinfo {pages} {011002}},\ \Eprint
  {http://arxiv.org/abs/1511.01415} {arXiv:1511.01415} \BibitemShut {NoStop}%
\bibitem [{\citenamefont {Campagne-Ibarcq}\ \emph {et~al.}(2017)\citenamefont
  {Campagne-Ibarcq}, \citenamefont {Zalys-Geller}, \citenamefont {Narla},
  \citenamefont {Shankar}, \citenamefont {Reinhold}, \citenamefont {Burkhart},
  \citenamefont {Axline}, \citenamefont {Pfaff}, \citenamefont {Frunzio},
  \citenamefont {Schoelkopf},\ and\ \citenamefont
  {Devoret}}]{Campagne2017-entangle2}%
  \BibitemOpen
  \bibfield  {author} {\bibinfo {author} {\bibfnamefont {P.}~\bibnamefont
  {Campagne-Ibarcq}}, \bibinfo {author} {\bibfnamefont {E.}~\bibnamefont
  {Zalys-Geller}}, \bibinfo {author} {\bibfnamefont {A.}~\bibnamefont {Narla}},
  \bibinfo {author} {\bibfnamefont {S.}~\bibnamefont {Shankar}}, \bibinfo
  {author} {\bibfnamefont {P.}~\bibnamefont {Reinhold}}, \bibinfo {author}
  {\bibfnamefont {L.~D.}\ \bibnamefont {Burkhart}}, \bibinfo {author}
  {\bibfnamefont {C.~J.}\ \bibnamefont {Axline}}, \bibinfo {author}
  {\bibfnamefont {W.}~\bibnamefont {Pfaff}}, \bibinfo {author} {\bibfnamefont
  {L.}~\bibnamefont {Frunzio}}, \bibinfo {author} {\bibfnamefont {R.~J.}\
  \bibnamefont {Schoelkopf}}, \ and\ \bibinfo {author} {\bibfnamefont {M.~H.}\
  \bibnamefont {Devoret}}} (\bibinfo {year} {2017}),\ \bibfield  {title}
  {\enquote {\bibinfo {title} {{Deterministic remote entanglement of
  superconducting circuits through microwave two-photon transitions}},}\ }\href
  {http://arxiv.org/abs/1712.05854} {\ }\Eprint
  {http://arxiv.org/abs/1712.05854} {arXiv:1712.05854} \BibitemShut {NoStop}%
\bibitem [{\citenamefont {Carmichael}(1993)}]{Carmichael1993}%
  \BibitemOpen
  \bibfield  {author} {\bibinfo {author} {\bibfnamefont {H.~J.}\ \bibnamefont
  {Carmichael}}} (\bibinfo {year} {1993}),\ \href {\doibase
  10.1007/978-3-540-47620-7} {\emph {\bibinfo {title} {An Open Systems Approach
  to Quantum Optics}}}\ (\bibinfo  {publisher} {Springer, Berlin,
  Heidelberg})\BibitemShut {NoStop}%
\bibitem [{\citenamefont {Carmichael}(1999)}]{Carmichael2008-Book2}%
  \BibitemOpen
  \bibfield  {author} {\bibinfo {author} {\bibfnamefont {H.~J.}\ \bibnamefont
  {Carmichael}}} (\bibinfo {year} {1999}),\ \href {\doibase
  10.1007/978-3-540-71320-3} {\emph {\bibinfo {title} {{Statistical Methods in
  Quantum Optics 1}}}},\ Theoretical and Mathematical Physics\ (\bibinfo
  {publisher} {Springer Berlin Heidelberg},\ \bibinfo {address} {Berlin,
  Heidelberg})\BibitemShut {NoStop}%
\bibitem [{\citenamefont {Caves}\ and\ \citenamefont
  {Milburn}(1987)}]{Caves1987-time-discrete}%
  \BibitemOpen
  \bibfield  {author} {\bibinfo {author} {\bibfnamefont {C.~M.}\ \bibnamefont
  {Caves}}, \ and\ \bibinfo {author} {\bibfnamefont {G.~J.}\ \bibnamefont
  {Milburn}}} (\bibinfo {year} {1987}),\ \bibfield  {title} {\enquote {\bibinfo
  {title} {{Quantum-mechanical model for continuous position measurements}},}\
  }\href {\doibase 10.1103/PhysRevA.36.5543} {\bibfield  {journal} {\bibinfo
  {journal} {Physical Review A}\ }\textbf {\bibinfo {volume} {36}}~(\bibinfo
  {number} {12}),\ \bibinfo {pages} {5543--5555}}\BibitemShut {NoStop}%
\bibitem [{\citenamefont {Chantasri}\ \emph {et~al.}(2018)\citenamefont
  {Chantasri}, \citenamefont {Atalaya}, \citenamefont {Hacohen-Gourgy},
  \citenamefont {Martin}, \citenamefont {Siddiqi},\ and\ \citenamefont
  {Jordan}}]{Chantasri2018}%
  \BibitemOpen
  \bibfield  {author} {\bibinfo {author} {\bibfnamefont {A.}~\bibnamefont
  {Chantasri}}, \bibinfo {author} {\bibfnamefont {J.}~\bibnamefont {Atalaya}},
  \bibinfo {author} {\bibfnamefont {S.}~\bibnamefont {Hacohen-Gourgy}},
  \bibinfo {author} {\bibfnamefont {L.~S.}\ \bibnamefont {Martin}}, \bibinfo
  {author} {\bibfnamefont {I.}~\bibnamefont {Siddiqi}}, \ and\ \bibinfo
  {author} {\bibfnamefont {A.~N.}\ \bibnamefont {Jordan}}} (\bibinfo {year}
  {2018}),\ \bibfield  {title} {\enquote {\bibinfo {title} {{Simultaneous
  continuous measurement of noncommuting observables: Quantum state
  correlations}},}\ }\href {\doibase 10.1103/PhysRevA.97.012118} {\bibfield
  {journal} {\bibinfo  {journal} {Physical Review A}\ }\textbf {\bibinfo
  {volume} {97}}~(\bibinfo {number} {1}),\ \bibinfo {pages}
  {012118}}\BibitemShut {NoStop}%
\bibitem [{\citenamefont {Chen}\ \emph {et~al.}(2011)\citenamefont {Chen},
  \citenamefont {Hover}, \citenamefont {Sendelbach}, \citenamefont {Maurer},
  \citenamefont {Merkel}, \citenamefont {Pritchett}, \citenamefont {Wilhelm},\
  and\ \citenamefont {McDermott}}]{Chen2011-FlyingPhoton}%
  \BibitemOpen
  \bibfield  {author} {\bibinfo {author} {\bibfnamefont {Y.-F.}\ \bibnamefont
  {Chen}}, \bibinfo {author} {\bibfnamefont {D.}~\bibnamefont {Hover}},
  \bibinfo {author} {\bibfnamefont {S.}~\bibnamefont {Sendelbach}}, \bibinfo
  {author} {\bibfnamefont {L.}~\bibnamefont {Maurer}}, \bibinfo {author}
  {\bibfnamefont {S.~T.}\ \bibnamefont {Merkel}}, \bibinfo {author}
  {\bibfnamefont {E.~J.}\ \bibnamefont {Pritchett}}, \bibinfo {author}
  {\bibfnamefont {F.~K.}\ \bibnamefont {Wilhelm}}, \ and\ \bibinfo {author}
  {\bibfnamefont {R.}~\bibnamefont {McDermott}}} (\bibinfo {year} {2011}),\
  \bibfield  {title} {\enquote {\bibinfo {title} {{Microwave Photon Counter
  Based on Josephson Junctions}},}\ }\href {\doibase
  10.1103/PhysRevLett.107.217401} {\bibfield  {journal} {\bibinfo  {journal}
  {Physical Review Letters}\ }\textbf {\bibinfo {volume} {107}}~(\bibinfo
  {number} {21}),\ \bibinfo {pages} {217401}}\BibitemShut {NoStop}%
\bibitem [{\citenamefont {Chow}\ \emph {et~al.}(2010)\citenamefont {Chow},
  \citenamefont {DiCarlo}, \citenamefont {Gambetta}, \citenamefont {Motzoi},
  \citenamefont {Frunzio}, \citenamefont {Girvin},\ and\ \citenamefont
  {Schoelkopf}}]{JChow2010-DRAG}%
  \BibitemOpen
  \bibfield  {author} {\bibinfo {author} {\bibfnamefont {J.~M.}\ \bibnamefont
  {Chow}}, \bibinfo {author} {\bibfnamefont {L.}~\bibnamefont {DiCarlo}},
  \bibinfo {author} {\bibfnamefont {J.~M.}\ \bibnamefont {Gambetta}}, \bibinfo
  {author} {\bibfnamefont {F.}~\bibnamefont {Motzoi}}, \bibinfo {author}
  {\bibfnamefont {L.}~\bibnamefont {Frunzio}}, \bibinfo {author} {\bibfnamefont
  {S.~M.}\ \bibnamefont {Girvin}}, \ and\ \bibinfo {author} {\bibfnamefont
  {R.~J.}\ \bibnamefont {Schoelkopf}}} (\bibinfo {year} {2010}),\ \bibfield
  {title} {\enquote {\bibinfo {title} {{Optimized driving of superconducting
  artificial atoms for improved single-qubit gates}},}\ }\href {\doibase
  10.1103/PhysRevA.82.040305} {\bibfield  {journal} {\bibinfo  {journal}
  {Physical Review A}\ }\textbf {\bibinfo {volume} {82}}~(\bibinfo {number}
  {4}),\ \bibinfo {pages} {040305}}\BibitemShut {NoStop}%
\bibitem [{\citenamefont {Cochrane}\ \emph {et~al.}(1999)\citenamefont
  {Cochrane}, \citenamefont {Milburn},\ and\ \citenamefont
  {Munro}}]{Cochrane1999-catCode}%
  \BibitemOpen
  \bibfield  {author} {\bibinfo {author} {\bibfnamefont {P.~T.}\ \bibnamefont
  {Cochrane}}, \bibinfo {author} {\bibfnamefont {G.~J.}\ \bibnamefont
  {Milburn}}, \ and\ \bibinfo {author} {\bibfnamefont {W.~J.}\ \bibnamefont
  {Munro}}} (\bibinfo {year} {1999}),\ \bibfield  {title} {\enquote {\bibinfo
  {title} {{Macroscopically distinct quantum-superposition states as a bosonic
  code for amplitude damping}},}\ }\href {\doibase 10.1103/PhysRevA.59.2631}
  {\bibfield  {journal} {\bibinfo  {journal} {Physical Review A}\ }\textbf
  {\bibinfo {volume} {59}}~(\bibinfo {number} {4}),\ \bibinfo {pages}
  {2631--2634}}\BibitemShut {NoStop}%
\bibitem [{\citenamefont {Cohen}\ \emph {et~al.}(2017)\citenamefont {Cohen},
  \citenamefont {Smith}, \citenamefont {Devoret},\ and\ \citenamefont
  {Mirrahimi}}]{Cohen2017-cw-parity-msr}%
  \BibitemOpen
  \bibfield  {author} {\bibinfo {author} {\bibfnamefont {J.}~\bibnamefont
  {Cohen}}, \bibinfo {author} {\bibfnamefont {W.~C.}\ \bibnamefont {Smith}},
  \bibinfo {author} {\bibfnamefont {M.~H.}\ \bibnamefont {Devoret}}, \ and\
  \bibinfo {author} {\bibfnamefont {M.}~\bibnamefont {Mirrahimi}}} (\bibinfo
  {year} {2017}),\ \bibfield  {title} {\enquote {\bibinfo {title}
  {{Degeneracy-Preserving Quantum Nondemolition Measurement of Parity-Type
  Observables for Cat Qubits}},}\ }\href {\doibase
  10.1103/PhysRevLett.119.060503} {\bibfield  {journal} {\bibinfo  {journal}
  {Physical Review Letters}\ }\textbf {\bibinfo {volume} {119}}~(\bibinfo
  {number} {6}),\ \bibinfo {pages} {060503}}\BibitemShut {NoStop}%
\bibitem [{\citenamefont {Cook}\ and\ \citenamefont {Kimble}(1985)}]{Cook1985}%
  \BibitemOpen
  \bibfield  {author} {\bibinfo {author} {\bibfnamefont {R.~J.}\ \bibnamefont
  {Cook}}, \ and\ \bibinfo {author} {\bibfnamefont {H.~J.}\ \bibnamefont
  {Kimble}}} (\bibinfo {year} {1985}),\ \bibfield  {title} {\enquote {\bibinfo
  {title} {{Possibility of Direct Observation of Quantum Jumps}},}\ }\href
  {\doibase 10.1103/PhysRevLett.54.1023} {\bibfield  {journal} {\bibinfo
  {journal} {Physical Review Letters}\ }\textbf {\bibinfo {volume}
  {54}}~(\bibinfo {number} {10}),\ \bibinfo {pages} {1023--1026}}\BibitemShut
  {NoStop}%
\bibitem [{\citenamefont {Cooper}\ \emph {et~al.}(2004)\citenamefont {Cooper},
  \citenamefont {Steffen}, \citenamefont {McDermott}, \citenamefont {Simmonds},
  \citenamefont {Oh}, \citenamefont {Hite}, \citenamefont {Pappas},\ and\
  \citenamefont {Martinis}}]{cooper2004}%
  \BibitemOpen
  \bibfield  {author} {\bibinfo {author} {\bibfnamefont {K.~B.}\ \bibnamefont
  {Cooper}}, \bibinfo {author} {\bibfnamefont {M.}~\bibnamefont {Steffen}},
  \bibinfo {author} {\bibfnamefont {R.}~\bibnamefont {McDermott}}, \bibinfo
  {author} {\bibfnamefont {R.~W.}\ \bibnamefont {Simmonds}}, \bibinfo {author}
  {\bibfnamefont {S.}~\bibnamefont {Oh}}, \bibinfo {author} {\bibfnamefont
  {D.~A.}\ \bibnamefont {Hite}}, \bibinfo {author} {\bibfnamefont {D.~P.}\
  \bibnamefont {Pappas}}, \ and\ \bibinfo {author} {\bibfnamefont {J.~M.}\
  \bibnamefont {Martinis}}} (\bibinfo {year} {2004}),\ \bibfield  {title}
  {\enquote {\bibinfo {title} {{Observation of Quantum Oscillations between a
  Josephson Phase Qubit and a Microscopic Resonator Using Fast Readout}},}\
  }\href {\doibase 10.1103/PhysRevLett.93.180401} {\bibfield  {journal}
  {\bibinfo  {journal} {Physical Review Letters}\ }\textbf {\bibinfo {volume}
  {93}}~(\bibinfo {number} {18}),\ \bibinfo {pages} {180401}}\BibitemShut
  {NoStop}%
\bibitem [{\citenamefont {Cottet}\ \emph {et~al.}(2017)\citenamefont {Cottet},
  \citenamefont {Jezouin}, \citenamefont {Bretheau}, \citenamefont
  {Campagne-Ibarcq}, \citenamefont {Ficheux}, \citenamefont {Anders},
  \citenamefont {Auff{\`{e}}ves}, \citenamefont {Azouit}, \citenamefont
  {Rouchon},\ and\ \citenamefont {Huard}}]{Cottet2017}%
  \BibitemOpen
  \bibfield  {author} {\bibinfo {author} {\bibfnamefont {N.}~\bibnamefont
  {Cottet}}, \bibinfo {author} {\bibfnamefont {S.}~\bibnamefont {Jezouin}},
  \bibinfo {author} {\bibfnamefont {L.}~\bibnamefont {Bretheau}}, \bibinfo
  {author} {\bibfnamefont {P.}~\bibnamefont {Campagne-Ibarcq}}, \bibinfo
  {author} {\bibfnamefont {Q.}~\bibnamefont {Ficheux}}, \bibinfo {author}
  {\bibfnamefont {J.}~\bibnamefont {Anders}}, \bibinfo {author} {\bibfnamefont
  {A.}~\bibnamefont {Auff{\`{e}}ves}}, \bibinfo {author} {\bibfnamefont
  {R.}~\bibnamefont {Azouit}}, \bibinfo {author} {\bibfnamefont
  {P.}~\bibnamefont {Rouchon}}, \ and\ \bibinfo {author} {\bibfnamefont
  {B.}~\bibnamefont {Huard}}} (\bibinfo {year} {2017}),\ \bibfield  {title}
  {\enquote {\bibinfo {title} {{Observing a quantum Maxwell demon at work}},}\
  }\href {\doibase 10.1073/pnas.1704827114} {\bibfield  {journal} {\bibinfo
  {journal} {Proceedings of the National Academy of Sciences}\ }\textbf
  {\bibinfo {volume} {114}}~(\bibinfo {number} {29}),\ \bibinfo {pages}
  {7561--7564}}\BibitemShut {NoStop}%
\bibitem [{\citenamefont {Dalibard}\ \emph {et~al.}(1992)\citenamefont
  {Dalibard}, \citenamefont {Castin},\ and\ \citenamefont
  {M{\o}lmer}}]{Dalibard1992-original-traj}%
  \BibitemOpen
  \bibfield  {author} {\bibinfo {author} {\bibfnamefont {J.}~\bibnamefont
  {Dalibard}}, \bibinfo {author} {\bibfnamefont {Y.}~\bibnamefont {Castin}}, \
  and\ \bibinfo {author} {\bibfnamefont {K.}~\bibnamefont {M{\o}lmer}}}
  (\bibinfo {year} {1992}),\ \bibfield  {title} {\enquote {\bibinfo {title}
  {{Wave-function approach to dissipative processes in quantum optics}},}\
  }\href {\doibase 10.1103/PhysRevLett.68.580} {\bibfield  {journal} {\bibinfo
  {journal} {Physical Review Letters}\ }\textbf {\bibinfo {volume}
  {68}}~(\bibinfo {number} {5}),\ \bibinfo {pages} {580--583}}\BibitemShut
  {NoStop}%
\bibitem [{\citenamefont {{De Lange}}\ \emph {et~al.}(2014)\citenamefont {{De
  Lange}}, \citenamefont {Riste}, \citenamefont {Tiggelman}, \citenamefont
  {Eichler}, \citenamefont {Tornberg}, \citenamefont {Johansson}, \citenamefont
  {Wallraff}, \citenamefont {Schouten},\ and\ \citenamefont
  {Dicarlo}}]{deLange2014}%
  \BibitemOpen
  \bibfield  {author} {\bibinfo {author} {\bibfnamefont {G.}~\bibnamefont {{De
  Lange}}}, \bibinfo {author} {\bibfnamefont {D.}~\bibnamefont {Riste}},
  \bibinfo {author} {\bibfnamefont {M.~J.}\ \bibnamefont {Tiggelman}}, \bibinfo
  {author} {\bibfnamefont {C.}~\bibnamefont {Eichler}}, \bibinfo {author}
  {\bibfnamefont {L.}~\bibnamefont {Tornberg}}, \bibinfo {author}
  {\bibfnamefont {G.}~\bibnamefont {Johansson}}, \bibinfo {author}
  {\bibfnamefont {A.}~\bibnamefont {Wallraff}}, \bibinfo {author}
  {\bibfnamefont {R.~N.}\ \bibnamefont {Schouten}}, \ and\ \bibinfo {author}
  {\bibfnamefont {L.}~\bibnamefont {Dicarlo}}} (\bibinfo {year} {2014}),\
  \bibfield  {title} {\enquote {\bibinfo {title} {{Reversing quantum
  trajectories with analog feedback}},}\ }\href {\doibase
  10.1103/PhysRevLett.112.080501} {\bibfield  {journal} {\bibinfo  {journal}
  {Physical Review Letters}\ }\textbf {\bibinfo {volume} {112}}~(\bibinfo
  {number} {8}),\ \bibinfo {pages} {080501}},\ \Eprint
  {http://arxiv.org/abs/1311.5472} {arXiv:1311.5472} \BibitemShut {NoStop}%
\bibitem [{\citenamefont {Del{\'{e}}glise}\ \emph {et~al.}(2008)\citenamefont
  {Del{\'{e}}glise}, \citenamefont {Dotsenko}, \citenamefont {Sayrin},
  \citenamefont {Bernu}, \citenamefont {Brune}, \citenamefont {Raimond},\ and\
  \citenamefont {Haroche}}]{Deleglise2008}%
  \BibitemOpen
  \bibfield  {author} {\bibinfo {author} {\bibfnamefont {S.}~\bibnamefont
  {Del{\'{e}}glise}}, \bibinfo {author} {\bibfnamefont {I.}~\bibnamefont
  {Dotsenko}}, \bibinfo {author} {\bibfnamefont {C.}~\bibnamefont {Sayrin}},
  \bibinfo {author} {\bibfnamefont {J.}~\bibnamefont {Bernu}}, \bibinfo
  {author} {\bibfnamefont {M.}~\bibnamefont {Brune}}, \bibinfo {author}
  {\bibfnamefont {J.-M.}\ \bibnamefont {Raimond}}, \ and\ \bibinfo {author}
  {\bibfnamefont {S.}~\bibnamefont {Haroche}}} (\bibinfo {year} {2008}),\
  \bibfield  {title} {\enquote {\bibinfo {title} {{Reconstruction of
  non-classical cavity field states with snapshots of their decoherence}},}\
  }\href {\doibase 10.1038/nature07288} {\bibfield  {journal} {\bibinfo
  {journal} {Nature}\ }\textbf {\bibinfo {volume} {455}}~(\bibinfo {number}
  {7212}),\ \bibinfo {pages} {510--514}},\ \Eprint
  {http://arxiv.org/abs/0809.1064} {arXiv:0809.1064} \BibitemShut {NoStop}%
\bibitem [{\citenamefont {Devoret}(1997)}]{Devoret1997}%
  \BibitemOpen
  \bibfield  {author} {\bibinfo {author} {\bibfnamefont {M.}~\bibnamefont
  {Devoret}}} (\bibinfo {year} {1997}),\ \bibfield  {title} {\enquote {\bibinfo
  {title} {{Quantum Fluctuations in Electrical Circuits}},}\ }in\ \href@noop {}
  {\emph {\bibinfo {booktitle} {Fluctuations Quantiques/Quantum
  Fluctuations}}},\ \bibinfo {editor} {edited by\ \bibinfo {editor}
  {\bibfnamefont {S.}~\bibnamefont {Reynaud}}, \bibinfo {editor} {\bibfnamefont
  {E.}~\bibnamefont {Giacobino}}, \ and\ \bibinfo {editor} {\bibfnamefont
  {J.}~\bibnamefont {Zinn-Justin}}},\ p.\ \bibinfo {pages} {351}\BibitemShut
  {NoStop}%
\bibitem [{\citenamefont {{Devoret, M.H.}}(2017)}]{MHD}%
  \BibitemOpen
  \bibfield  {author} {\bibinfo {author} {\bibnamefont {{Devoret, M.H.}}}}
  (\bibinfo {year} {2017}),\ \href@noop {} {}\bibinfo {note} {Private
  communication}\BibitemShut {NoStop}%
\bibitem [{\citenamefont {Dial}\ \emph {et~al.}(2016)\citenamefont {Dial},
  \citenamefont {McClure}, \citenamefont {Poletto}, \citenamefont {Keefe},
  \citenamefont {Rothwell}, \citenamefont {Gambetta}, \citenamefont {Abraham},
  \citenamefont {Chow},\ and\ \citenamefont {Steffen}}]{Dial2016}%
  \BibitemOpen
  \bibfield  {author} {\bibinfo {author} {\bibfnamefont {O.}~\bibnamefont
  {Dial}}, \bibinfo {author} {\bibfnamefont {D.~T.}\ \bibnamefont {McClure}},
  \bibinfo {author} {\bibfnamefont {S.}~\bibnamefont {Poletto}}, \bibinfo
  {author} {\bibfnamefont {G.~A.}\ \bibnamefont {Keefe}}, \bibinfo {author}
  {\bibfnamefont {M.~B.}\ \bibnamefont {Rothwell}}, \bibinfo {author}
  {\bibfnamefont {J.~M.}\ \bibnamefont {Gambetta}}, \bibinfo {author}
  {\bibfnamefont {D.~W.}\ \bibnamefont {Abraham}}, \bibinfo {author}
  {\bibfnamefont {J.~M.}\ \bibnamefont {Chow}}, \ and\ \bibinfo {author}
  {\bibfnamefont {M.}~\bibnamefont {Steffen}}} (\bibinfo {year} {2016}),\
  \bibfield  {title} {\enquote {\bibinfo {title} {{Bulk and surface loss in
  superconducting transmon qubits}},}\ }\href {\doibase
  10.1088/0953-2048/29/4/044001} {\bibfield  {journal} {\bibinfo  {journal}
  {Superconductor Science and Technology}\ }\textbf {\bibinfo {volume}
  {29}}~(\bibinfo {number} {4}),\ \bibinfo {pages} {044001}},\ \Eprint
  {http://arxiv.org/abs/1509.03859} {arXiv:1509.03859} \BibitemShut {NoStop}%
\bibitem [{\citenamefont {Diniz}\ \emph {et~al.}(2013)\citenamefont {Diniz},
  \citenamefont {Dumur}, \citenamefont {Buisson},\ and\ \citenamefont
  {Auff{\`{e}}ves}}]{Diniz2013}%
  \BibitemOpen
  \bibfield  {author} {\bibinfo {author} {\bibfnamefont {I.}~\bibnamefont
  {Diniz}}, \bibinfo {author} {\bibfnamefont {E.}~\bibnamefont {Dumur}},
  \bibinfo {author} {\bibfnamefont {O.}~\bibnamefont {Buisson}}, \ and\
  \bibinfo {author} {\bibfnamefont {A.}~\bibnamefont {Auff{\`{e}}ves}}}
  (\bibinfo {year} {2013}),\ \bibfield  {title} {\enquote {\bibinfo {title}
  {{Ultrafast quantum nondemolition measurements based on a diamond-shaped
  artificial atom}},}\ }\href {\doibase 10.1103/PhysRevA.87.033837} {\bibfield
  {journal} {\bibinfo  {journal} {Physical Review A}\ }\textbf {\bibinfo
  {volume} {87}}~(\bibinfo {number} {3}),\ \bibinfo {pages} {033837}},\ \Eprint
  {http://arxiv.org/abs/1302.3847} {arXiv:1302.3847} \BibitemShut {NoStop}%
\bibitem [{\citenamefont {DiVincenzo}(2000)}]{DiVincenzo2000-criteria}%
  \BibitemOpen
  \bibfield  {author} {\bibinfo {author} {\bibfnamefont {D.~P.}\ \bibnamefont
  {DiVincenzo}}} (\bibinfo {year} {2000}),\ \bibfield  {title} {\enquote
  {\bibinfo {title} {{The Physical Implementation of Quantum Computation}},}\
  }\href {\doibase
  10.1002/1521-3978(200009)48:9/11<771::AID-PROP771>3.0.CO;2-E} {\bibfield
  {journal} {\bibinfo  {journal} {Fortschritte der Physik}\ }\textbf {\bibinfo
  {volume} {48}}~(\bibinfo {number} {9-11}),\ \bibinfo {pages}
  {771--783}}\BibitemShut {NoStop}%
\bibitem [{\citenamefont {Dressel}\ \emph {et~al.}(2017)\citenamefont
  {Dressel}, \citenamefont {Chantasri}, \citenamefont {Jordan},\ and\
  \citenamefont {Korotkov}}]{Dressel2017-arrow-of-time}%
  \BibitemOpen
  \bibfield  {author} {\bibinfo {author} {\bibfnamefont {J.}~\bibnamefont
  {Dressel}}, \bibinfo {author} {\bibfnamefont {A.}~\bibnamefont {Chantasri}},
  \bibinfo {author} {\bibfnamefont {A.~N.}\ \bibnamefont {Jordan}}, \ and\
  \bibinfo {author} {\bibfnamefont {A.~N.}\ \bibnamefont {Korotkov}}} (\bibinfo
  {year} {2017}),\ \bibfield  {title} {\enquote {\bibinfo {title} {{Arrow of
  Time for Continuous Quantum Measurement}},}\ }\href {\doibase
  10.1103/PhysRevLett.119.220507} {\bibfield  {journal} {\bibinfo  {journal}
  {Physical Review Letters}\ }\textbf {\bibinfo {volume} {119}}~(\bibinfo
  {number} {22}),\ \bibinfo {pages} {220507}}\BibitemShut {NoStop}%
\bibitem [{\citenamefont {Dum}\ \emph {et~al.}(1992)\citenamefont {Dum},
  \citenamefont {Zoller},\ and\ \citenamefont {Ritsch}}]{Dum1992}%
  \BibitemOpen
  \bibfield  {author} {\bibinfo {author} {\bibfnamefont {R.}~\bibnamefont
  {Dum}}, \bibinfo {author} {\bibfnamefont {P.}~\bibnamefont {Zoller}}, \ and\
  \bibinfo {author} {\bibfnamefont {H.}~\bibnamefont {Ritsch}}} (\bibinfo
  {year} {1992}),\ \bibfield  {title} {\enquote {\bibinfo {title} {{Monte Carlo
  simulation of the atomic master equation for spontaneous emission}},}\ }\href
  {\doibase 10.1103/PhysRevA.45.4879} {\bibfield  {journal} {\bibinfo
  {journal} {Physical Review A}\ }\textbf {\bibinfo {volume} {45}}~(\bibinfo
  {number} {7}),\ \bibinfo {pages} {4879--4887}}\BibitemShut {NoStop}%
\bibitem [{\citenamefont {Dumur}\ \emph {et~al.}(2015)\citenamefont {Dumur},
  \citenamefont {K{\"{u}}ng}, \citenamefont {Feofanov}, \citenamefont {Weissl},
  \citenamefont {Roch}, \citenamefont {Naud}, \citenamefont {Guichard},\ and\
  \citenamefont {Buisson}}]{Dumur2015}%
  \BibitemOpen
  \bibfield  {author} {\bibinfo {author} {\bibfnamefont {{\'{E}}.}~\bibnamefont
  {Dumur}}, \bibinfo {author} {\bibfnamefont {B.}~\bibnamefont {K{\"{u}}ng}},
  \bibinfo {author} {\bibfnamefont {A.~K.}\ \bibnamefont {Feofanov}}, \bibinfo
  {author} {\bibfnamefont {T.}~\bibnamefont {Weissl}}, \bibinfo {author}
  {\bibfnamefont {N.}~\bibnamefont {Roch}}, \bibinfo {author} {\bibfnamefont
  {C.}~\bibnamefont {Naud}}, \bibinfo {author} {\bibfnamefont {W.}~\bibnamefont
  {Guichard}}, \ and\ \bibinfo {author} {\bibfnamefont {O.}~\bibnamefont
  {Buisson}}} (\bibinfo {year} {2015}),\ \bibfield  {title} {\enquote {\bibinfo
  {title} {{V-shaped superconducting artificial atom based on two inductively
  coupled transmons}},}\ }\href {\doibase 10.1103/PhysRevB.92.020515}
  {\bibfield  {journal} {\bibinfo  {journal} {Physical Review B}\ }\textbf
  {\bibinfo {volume} {92}}~(\bibinfo {number} {2}),\ \bibinfo {pages}
  {020515}},\ \Eprint {http://arxiv.org/abs/1501.04892} {arXiv:1501.04892}
  \BibitemShut {NoStop}%
\bibitem [{\citenamefont {Eichler}\ \emph {et~al.}(2012)\citenamefont
  {Eichler}, \citenamefont {Lang}, \citenamefont {Fink}, \citenamefont
  {Govenius}, \citenamefont {Filipp},\ and\ \citenamefont
  {Wallraff}}]{Eichler2012-itinerant-entanglement}%
  \BibitemOpen
  \bibfield  {author} {\bibinfo {author} {\bibfnamefont {C.}~\bibnamefont
  {Eichler}}, \bibinfo {author} {\bibfnamefont {C.}~\bibnamefont {Lang}},
  \bibinfo {author} {\bibfnamefont {J.~M.}\ \bibnamefont {Fink}}, \bibinfo
  {author} {\bibfnamefont {J.}~\bibnamefont {Govenius}}, \bibinfo {author}
  {\bibfnamefont {S.}~\bibnamefont {Filipp}}, \ and\ \bibinfo {author}
  {\bibfnamefont {A.}~\bibnamefont {Wallraff}}} (\bibinfo {year} {2012}),\
  \bibfield  {title} {\enquote {\bibinfo {title} {{Observation of Entanglement
  between Itinerant Microwave Photons and a Superconducting Qubit}},}\ }\href
  {\doibase 10.1103/PhysRevLett.109.240501} {\bibfield  {journal} {\bibinfo
  {journal} {Physical Review Letters}\ }\textbf {\bibinfo {volume}
  {109}}~(\bibinfo {number} {24}),\ \bibinfo {pages} {240501}}\BibitemShut
  {NoStop}%
\bibitem [{\citenamefont {Einstein}(1916)}]{Einstein1916}%
  \BibitemOpen
  \bibfield  {author} {\bibinfo {author} {\bibfnamefont {A.}~\bibnamefont
  {Einstein}}} (\bibinfo {year} {1916}),\ \bibfield  {title} {\enquote
  {\bibinfo {title} {{Strahlungs-emission und -absorption nach der
  Quantentheorie (Emission and absorption of radiation in quantum theory)}},}\
  }\href
  {http://www.informationphilosopher.com/solutions/scientists/einstein/1916{\_}A-B.html
  http://adsabs.harvard.edu/abs/1916DPhyG..18..318E} {\bibfield  {journal}
  {\bibinfo  {journal} {Verhandlungen der Deutschen Physikalischen
  Zeitschrift}\ }\textbf {\bibinfo {volume} {18}},\ \bibinfo {pages}
  {318--323}}\BibitemShut {NoStop}%
\bibitem [{\citenamefont {Einstein}(1917)}]{Einstein1917}%
  \BibitemOpen
  \bibfield  {author} {\bibinfo {author} {\bibfnamefont {A.}~\bibnamefont
  {Einstein}}} (\bibinfo {year} {1917}),\ \bibfield  {title} {\enquote
  {\bibinfo {title} {{Quantentheorie der Strahlung (On the quantum theory of
  radiation)}},}\ }\href {http://adsabs.harvard.edu/abs/1917PhyZ...18..121E}
  {\bibfield  {journal} {\bibinfo  {journal} {Physikalische Zeitschrift}\
  }\textbf {\bibinfo {volume} {18}},\ \bibinfo {pages} {121}}\BibitemShut
  {NoStop}%
\bibitem [{\citenamefont {Eisaman}\ \emph {et~al.}(2011)\citenamefont
  {Eisaman}, \citenamefont {Fan}, \citenamefont {Migdall},\ and\ \citenamefont
  {Polyakov}}]{Eisaman2011}%
  \BibitemOpen
  \bibfield  {author} {\bibinfo {author} {\bibfnamefont {M.~D.}\ \bibnamefont
  {Eisaman}}, \bibinfo {author} {\bibfnamefont {J.}~\bibnamefont {Fan}},
  \bibinfo {author} {\bibfnamefont {A.}~\bibnamefont {Migdall}}, \ and\
  \bibinfo {author} {\bibfnamefont {S.~V.}\ \bibnamefont {Polyakov}}} (\bibinfo
  {year} {2011}),\ \bibfield  {title} {\enquote {\bibinfo {title} {{Invited
  Review Article: Single-photon sources and detectors}},}\ }\href {\doibase
  10.1063/1.3610677} {\bibfield  {journal} {\bibinfo  {journal} {Review of
  Scientific Instruments}\ }\textbf {\bibinfo {volume} {82}}~(\bibinfo {number}
  {7}),\ \bibinfo {pages} {071101}}\BibitemShut {NoStop}%
\bibitem [{\citenamefont {Elouard}\ \emph {et~al.}(2017)\citenamefont
  {Elouard}, \citenamefont {Herrera-Mart{\'{i}}}, \citenamefont {Huard},\ and\
  \citenamefont {Auff{\`{e}}ves}}]{Elouard2017}%
  \BibitemOpen
  \bibfield  {author} {\bibinfo {author} {\bibfnamefont {C.}~\bibnamefont
  {Elouard}}, \bibinfo {author} {\bibfnamefont {D.}~\bibnamefont
  {Herrera-Mart{\'{i}}}}, \bibinfo {author} {\bibfnamefont {B.}~\bibnamefont
  {Huard}}, \ and\ \bibinfo {author} {\bibfnamefont {A.}~\bibnamefont
  {Auff{\`{e}}ves}}} (\bibinfo {year} {2017}),\ \bibfield  {title} {\enquote
  {\bibinfo {title} {{Extracting Work from Quantum Measurement in Maxwell's
  Demon Engines}},}\ }\href {\doibase 10.1103/PhysRevLett.118.260603}
  {\bibfield  {journal} {\bibinfo  {journal} {Physical Review Letters}\
  }\textbf {\bibinfo {volume} {118}}~(\bibinfo {number} {26}),\ \bibinfo
  {pages} {260603}}\BibitemShut {NoStop}%
\bibitem [{\citenamefont {Esteve}\ \emph {et~al.}(1986)\citenamefont {Esteve},
  \citenamefont {Devoret},\ and\ \citenamefont {Martinis}}]{Esteve1986}%
  \BibitemOpen
  \bibfield  {author} {\bibinfo {author} {\bibfnamefont {D.}~\bibnamefont
  {Esteve}}, \bibinfo {author} {\bibfnamefont {M.~H.}\ \bibnamefont {Devoret}},
  \ and\ \bibinfo {author} {\bibfnamefont {J.~M.}\ \bibnamefont {Martinis}}}
  (\bibinfo {year} {1986}),\ \bibfield  {title} {\enquote {\bibinfo {title}
  {{Effect of an arbitrary dissipative circuit on the quantum energy levels and
  tunneling of a Josephson junction}},}\ }\href {\doibase
  10.1103/PhysRevB.34.158} {\bibfield  {journal} {\bibinfo  {journal} {Physical
  Review B}\ }\textbf {\bibinfo {volume} {34}}~(\bibinfo {number} {1}),\
  \bibinfo {pages} {158--163}}\BibitemShut {NoStop}%
\bibitem [{\citenamefont {Fan}\ \emph {et~al.}(2014)\citenamefont {Fan},
  \citenamefont {Johansson}, \citenamefont {Combes}, \citenamefont {Milburn},\
  and\ \citenamefont {Stace}}]{Fan2014-photonFly}%
  \BibitemOpen
  \bibfield  {author} {\bibinfo {author} {\bibfnamefont {B.}~\bibnamefont
  {Fan}}, \bibinfo {author} {\bibfnamefont {G.}~\bibnamefont {Johansson}},
  \bibinfo {author} {\bibfnamefont {J.}~\bibnamefont {Combes}}, \bibinfo
  {author} {\bibfnamefont {G.~J.}\ \bibnamefont {Milburn}}, \ and\ \bibinfo
  {author} {\bibfnamefont {T.~M.}\ \bibnamefont {Stace}}} (\bibinfo {year}
  {2014}),\ \bibfield  {title} {\enquote {\bibinfo {title} {{Nonabsorbing
  high-efficiency counter for itinerant microwave photons}},}\ }\href {\doibase
  10.1103/PhysRevB.90.035132} {\bibfield  {journal} {\bibinfo  {journal}
  {Physical Review B}\ }\textbf {\bibinfo {volume} {90}}~(\bibinfo {number}
  {3}),\ \bibinfo {pages} {035132}}\BibitemShut {NoStop}%
\bibitem [{\citenamefont {Ficheux}\ \emph {et~al.}(2018)\citenamefont
  {Ficheux}, \citenamefont {Jezouin}, \citenamefont {Leghtas},\ and\
  \citenamefont {Huard}}]{Ficheux2017}%
  \BibitemOpen
  \bibfield  {author} {\bibinfo {author} {\bibfnamefont {Q.}~\bibnamefont
  {Ficheux}}, \bibinfo {author} {\bibfnamefont {S.}~\bibnamefont {Jezouin}},
  \bibinfo {author} {\bibfnamefont {Z.}~\bibnamefont {Leghtas}}, \ and\
  \bibinfo {author} {\bibfnamefont {B.}~\bibnamefont {Huard}}} (\bibinfo {year}
  {2018}),\ \bibfield  {title} {\enquote {\bibinfo {title} {{Dynamics of a
  qubit while simultaneously monitoring its relaxation and dephasing}},}\
  }\href {\doibase 10.1038/s41467-018-04372-9} {\bibfield  {journal} {\bibinfo
  {journal} {Nature Communications}\ }\textbf {\bibinfo {volume} {9}}~(\bibinfo
  {number} {1}),\ \bibinfo {pages} {1926}},\ \Eprint
  {http://arxiv.org/abs/1711.01208} {arXiv:1711.01208} \BibitemShut {NoStop}%
\bibitem [{\citenamefont {Fuchs}(2003)}]{Fuchs2003-qminfo}%
  \BibitemOpen
  \bibfield  {author} {\bibinfo {author} {\bibfnamefont {C.~A.}\ \bibnamefont
  {Fuchs}}} (\bibinfo {year} {2003}),\ \bibfield  {title} {\enquote {\bibinfo
  {title} {{Quantum mechanics as quantum information, mostly}},}\ }\href
  {\doibase 10.1080/09500340308234548} {\bibfield  {journal} {\bibinfo
  {journal} {Journal of Modern Optics}\ }\textbf {\bibinfo {volume}
  {50}}~(\bibinfo {number} {6-7}),\ \bibinfo {pages} {987--1023}}\BibitemShut
  {NoStop}%
\bibitem [{\citenamefont {Gambetta}\ \emph {et~al.}(2008)\citenamefont
  {Gambetta}, \citenamefont {Blais}, \citenamefont {Boissonneault},
  \citenamefont {Houck}, \citenamefont {Schuster},\ and\ \citenamefont
  {Girvin}}]{Gambetta2008-qm-traj}%
  \BibitemOpen
  \bibfield  {author} {\bibinfo {author} {\bibfnamefont {J.}~\bibnamefont
  {Gambetta}}, \bibinfo {author} {\bibfnamefont {A.}~\bibnamefont {Blais}},
  \bibinfo {author} {\bibfnamefont {M.}~\bibnamefont {Boissonneault}}, \bibinfo
  {author} {\bibfnamefont {A.~A.}\ \bibnamefont {Houck}}, \bibinfo {author}
  {\bibfnamefont {D.~I.}\ \bibnamefont {Schuster}}, \ and\ \bibinfo {author}
  {\bibfnamefont {S.~M.}\ \bibnamefont {Girvin}}} (\bibinfo {year} {2008}),\
  \bibfield  {title} {\enquote {\bibinfo {title} {{Quantum trajectory approach
  to circuit QED: Quantum jumps and the Zeno effect}},}\ }\href {\doibase
  10.1103/PhysRevA.77.012112} {\bibfield  {journal} {\bibinfo  {journal}
  {Physical Review A}\ }\textbf {\bibinfo {volume} {77}}~(\bibinfo {number}
  {1}),\ \bibinfo {pages} {012112}}\BibitemShut {NoStop}%
\bibitem [{\citenamefont {Gambetta}\ \emph {et~al.}(2006)\citenamefont
  {Gambetta}, \citenamefont {Blais}, \citenamefont {Schuster}, \citenamefont
  {Wallraff}, \citenamefont {Frunzio}, \citenamefont {Majer}, \citenamefont
  {Devoret}, \citenamefont {Girvin},\ and\ \citenamefont
  {Schoelkopf}}]{Gambetta2006-dephasing}%
  \BibitemOpen
  \bibfield  {author} {\bibinfo {author} {\bibfnamefont {J.}~\bibnamefont
  {Gambetta}}, \bibinfo {author} {\bibfnamefont {A.}~\bibnamefont {Blais}},
  \bibinfo {author} {\bibfnamefont {D.~I.}\ \bibnamefont {Schuster}}, \bibinfo
  {author} {\bibfnamefont {A.}~\bibnamefont {Wallraff}}, \bibinfo {author}
  {\bibfnamefont {L.}~\bibnamefont {Frunzio}}, \bibinfo {author} {\bibfnamefont
  {J.}~\bibnamefont {Majer}}, \bibinfo {author} {\bibfnamefont {M.~H.}\
  \bibnamefont {Devoret}}, \bibinfo {author} {\bibfnamefont {S.~M.}\
  \bibnamefont {Girvin}}, \ and\ \bibinfo {author} {\bibfnamefont {R.~J.}\
  \bibnamefont {Schoelkopf}}} (\bibinfo {year} {2006}),\ \bibfield  {title}
  {\enquote {\bibinfo {title} {{Qubit-photon interactions in a cavity:
  Measurement-induced dephasing and number splitting}},}\ }\href {\doibase
  10.1103/PhysRevA.74.042318} {\bibfield  {journal} {\bibinfo  {journal}
  {Physical Review A}\ }\textbf {\bibinfo {volume} {74}}~(\bibinfo {number}
  {4}),\ \bibinfo {pages} {042318}}\BibitemShut {NoStop}%
\bibitem [{\citenamefont {Gambetta}\ \emph {et~al.}(2007)\citenamefont
  {Gambetta}, \citenamefont {Braff}, \citenamefont {Wallraff}, \citenamefont
  {Girvin},\ and\ \citenamefont {Schoelkopf}}]{Gambetta2007-ProtocolsMsr}%
  \BibitemOpen
  \bibfield  {author} {\bibinfo {author} {\bibfnamefont {J.}~\bibnamefont
  {Gambetta}}, \bibinfo {author} {\bibfnamefont {W.~A.}\ \bibnamefont {Braff}},
  \bibinfo {author} {\bibfnamefont {A.}~\bibnamefont {Wallraff}}, \bibinfo
  {author} {\bibfnamefont {S.~M.}\ \bibnamefont {Girvin}}, \ and\ \bibinfo
  {author} {\bibfnamefont {R.~J.}\ \bibnamefont {Schoelkopf}}} (\bibinfo {year}
  {2007}),\ \bibfield  {title} {\enquote {\bibinfo {title} {{Protocols for
  optimal readout of qubits using a continuous quantum nondemolition
  measurement}},}\ }\href {\doibase 10.1103/PhysRevA.76.012325} {\bibfield
  {journal} {\bibinfo  {journal} {Physical Review A}\ }\textbf {\bibinfo
  {volume} {76}}~(\bibinfo {number} {1}),\ \bibinfo {pages}
  {012325}}\BibitemShut {NoStop}%
\bibitem [{\citenamefont {Gambetta}\ \emph {et~al.}(2011)\citenamefont
  {Gambetta}, \citenamefont {Houck},\ and\ \citenamefont
  {Blais}}]{Gambetta2011-Purcell}%
  \BibitemOpen
  \bibfield  {author} {\bibinfo {author} {\bibfnamefont {J.~M.}\ \bibnamefont
  {Gambetta}}, \bibinfo {author} {\bibfnamefont {A.~A.}\ \bibnamefont {Houck}},
  \ and\ \bibinfo {author} {\bibfnamefont {A.}~\bibnamefont {Blais}}} (\bibinfo
  {year} {2011}),\ \bibfield  {title} {\enquote {\bibinfo {title}
  {{Superconducting Qubit with Purcell Protection and Tunable Coupling}},}\
  }\href {\doibase 10.1103/PhysRevLett.106.030502} {\bibfield  {journal}
  {\bibinfo  {journal} {Physical Review Letters}\ }\textbf {\bibinfo {volume}
  {106}}~(\bibinfo {number} {3}),\ \bibinfo {pages} {030502}},\ \Eprint
  {http://arxiv.org/abs/1009.4470} {arXiv:1009.4470} \BibitemShut {NoStop}%
\bibitem [{\citenamefont {Gao}(2008)}]{Gao2008}%
  \BibitemOpen
  \bibfield  {author} {\bibinfo {author} {\bibfnamefont {J.}~\bibnamefont
  {Gao}}} (\bibinfo {year} {2008}),\ \emph {\bibinfo {title} {{The physics of
  superconducting microwave resonators}}},\ \href
  {http://thesis.library.caltech.edu/2530/1/thesismain{\%}7B{\_}{\%}7D0610.pdf}
  {Ph.D. thesis}\ (\bibinfo  {school} {California Institute of
  Technology})\BibitemShut {NoStop}%
\bibitem [{\citenamefont {Gardiner}\ and\ \citenamefont
  {Zoller}(2004)}]{Gardiner2004}%
  \BibitemOpen
  \bibfield  {author} {\bibinfo {author} {\bibfnamefont {C.}~\bibnamefont
  {Gardiner}}, \ and\ \bibinfo {author} {\bibfnamefont {P.}~\bibnamefont
  {Zoller}}} (\bibinfo {year} {2004}),\ \href@noop {} {\emph {\bibinfo {title}
  {{Quantum Noise: A Handbook of Markovian and Non-Markovian Quantum Stochastic
  Methods with Applications to Quantum Optics}}}}\ (\bibinfo  {publisher}
  {Springer-Verlag Berlin Heidelberg})\BibitemShut {NoStop}%
\bibitem [{\citenamefont {Gardiner}\ \emph {et~al.}(1992)\citenamefont
  {Gardiner}, \citenamefont {Parkins},\ and\ \citenamefont
  {Zoller}}]{Gardiner1992-original-traj}%
  \BibitemOpen
  \bibfield  {author} {\bibinfo {author} {\bibfnamefont {C.~W.}\ \bibnamefont
  {Gardiner}}, \bibinfo {author} {\bibfnamefont {A.~S.}\ \bibnamefont
  {Parkins}}, \ and\ \bibinfo {author} {\bibfnamefont {P.}~\bibnamefont
  {Zoller}}} (\bibinfo {year} {1992}),\ \bibfield  {title} {\enquote {\bibinfo
  {title} {{Wave-function quantum stochastic differential equations and
  quantum-jump simulation methods}},}\ }\href {\doibase
  10.1103/PhysRevA.46.4363} {\bibfield  {journal} {\bibinfo  {journal}
  {Physical Review A}\ }\textbf {\bibinfo {volume} {46}}~(\bibinfo {number}
  {7}),\ \bibinfo {pages} {4363--4381}}\BibitemShut {NoStop}%
\bibitem [{\citenamefont {Garrahan}\ and\ \citenamefont
  {Guţă}(2018)}]{Garrahan2018}%
  \BibitemOpen
  \bibfield  {author} {\bibinfo {author} {\bibfnamefont {J.~P.}\ \bibnamefont
  {Garrahan}}, \ and\ \bibinfo {author} {\bibfnamefont {M.}~\bibnamefont
  {Guţă}}} (\bibinfo {year} {2018}),\ \bibfield  {title} {\enquote {\bibinfo
  {title} {{Catching and reversing quantum jumps and thermodynamics of quantum
  trajectories}},}\ }\href {\doibase 10.1103/PhysRevA.98.052137} {\bibfield
  {journal} {\bibinfo  {journal} {Physical Review A}\ }\textbf {\bibinfo
  {volume} {98}}~(\bibinfo {number} {5}),\ \bibinfo {pages} {052137}},\ \Eprint
  {http://arxiv.org/abs/1808.00726} {arXiv:1808.00726} \BibitemShut {NoStop}%
\bibitem [{\citenamefont {Geerlings}\ \emph {et~al.}(2013)\citenamefont
  {Geerlings}, \citenamefont {Leghtas}, \citenamefont {Pop}, \citenamefont
  {Shankar}, \citenamefont {Frunzio}, \citenamefont {Schoelkopf}, \citenamefont
  {Mirrahimi},\ and\ \citenamefont {Devoret}}]{Geerlings2013-reset}%
  \BibitemOpen
  \bibfield  {author} {\bibinfo {author} {\bibfnamefont {K.}~\bibnamefont
  {Geerlings}}, \bibinfo {author} {\bibfnamefont {Z.}~\bibnamefont {Leghtas}},
  \bibinfo {author} {\bibfnamefont {I.~M.}\ \bibnamefont {Pop}}, \bibinfo
  {author} {\bibfnamefont {S.}~\bibnamefont {Shankar}}, \bibinfo {author}
  {\bibfnamefont {L.}~\bibnamefont {Frunzio}}, \bibinfo {author} {\bibfnamefont
  {R.~J.}\ \bibnamefont {Schoelkopf}}, \bibinfo {author} {\bibfnamefont
  {M.}~\bibnamefont {Mirrahimi}}, \ and\ \bibinfo {author} {\bibfnamefont
  {M.~H.}\ \bibnamefont {Devoret}}} (\bibinfo {year} {2013}),\ \bibfield
  {title} {\enquote {\bibinfo {title} {{Demonstrating a Driven Reset Protocol
  for a Superconducting Qubit}},}\ }\href {\doibase
  10.1103/PhysRevLett.110.120501} {\bibfield  {journal} {\bibinfo  {journal}
  {Physical Review Letters}\ }\textbf {\bibinfo {volume} {110}}~(\bibinfo
  {number} {12}),\ \bibinfo {pages} {120501}}\BibitemShut {NoStop}%
\bibitem [{\citenamefont {Geerlings}(2013)}]{Geerlings2013}%
  \BibitemOpen
  \bibfield  {author} {\bibinfo {author} {\bibfnamefont {K.~L.}\ \bibnamefont
  {Geerlings}}} (\bibinfo {year} {2013}),\ \emph {\bibinfo {title} {{Improving
  Coherence of Superconducting Qubits and Resonators}}},\ \href
  {http://qulab.eng.yale.edu/documents/theses/Kurtis{\%}7B{\_}{\%}7DImprovingCoherenceSuperconductingQubits.pdf
  http://qulab.eng.yale.edu/documents/theses/Kurtis{\_}ImprovingCoherenceSuperconductingQubits.pdf}
  {Ph.D. thesis}\ (\bibinfo  {school} {Yale University})\BibitemShut {NoStop}%
\bibitem [{\citenamefont {Gerry}\ and\ \citenamefont
  {Knight}(2005)}]{Gerry2005}%
  \BibitemOpen
  \bibfield  {author} {\bibinfo {author} {\bibfnamefont {C.}~\bibnamefont
  {Gerry}}, \ and\ \bibinfo {author} {\bibfnamefont {P.}~\bibnamefont
  {Knight}}} (\bibinfo {year} {2005}),\ \href@noop {} {\emph {\bibinfo {title}
  {{Introductory quantum optics}}}}\ (\bibinfo  {publisher} {Cambridge
  University Press})\BibitemShut {NoStop}%
\bibitem [{\citenamefont {Girvin}(2014)}]{Girvin2014}%
  \BibitemOpen
  \bibfield  {author} {\bibinfo {author} {\bibfnamefont {S.~M.}\ \bibnamefont
  {Girvin}}} (\bibinfo {year} {2014}),\ \bibfield  {title} {\enquote {\bibinfo
  {title} {{Circuit QED: superconducting qubits coupled to microwave
  photons}},}\ }\href {\doibase 10.1093/acprof:oso/9780199681181.003.0003}
  {\bibinfo  {journal} {Quantum Machines: Measurement and Control of Engineered
  Quantum Systems}\ ,\ \bibinfo {pages} {113--256}}\BibitemShut {NoStop}%
\bibitem [{\citenamefont {Gisin}\ and\ \citenamefont
  {Percival}(1992)}]{Gisin1992}%
  \BibitemOpen
\bibfield  {journal} {  }\bibfield  {author} {\bibinfo {author} {\bibfnamefont
  {N.}~\bibnamefont {Gisin}}, \ and\ \bibinfo {author} {\bibfnamefont {I.~C.}\
  \bibnamefont {Percival}}} (\bibinfo {year} {1992}),\ \bibfield  {title}
  {\enquote {\bibinfo {title} {{Wave-function approach to dissipative
  processes: are there quantum jumps?}}}\ }\href {\doibase
  10.1016/0375-9601(92)90264-M} {\bibfield  {journal} {\bibinfo  {journal}
  {Physics Letters A}\ }\textbf {\bibinfo {volume} {167}}~(\bibinfo {number}
  {4}),\ \bibinfo {pages} {315--318}}\BibitemShut {NoStop}%
\bibitem [{\citenamefont {Gleyzes}\ \emph {et~al.}(2007)\citenamefont
  {Gleyzes}, \citenamefont {Kuhr}, \citenamefont {Guerlin}, \citenamefont
  {Bernu}, \citenamefont {Del{\'{e}}glise}, \citenamefont {Busk}, \citenamefont
  {Brune}, \citenamefont {Raimond}, \citenamefont {Haroche}, \citenamefont
  {Deleglise}, \citenamefont {{Busk Hoff}}, \citenamefont {Brune},
  \citenamefont {Raimond},\ and\ \citenamefont {Haroche}}]{Gleyzes2007}%
  \BibitemOpen
  \bibfield  {author} {\bibinfo {author} {\bibfnamefont {S.~S.}\ \bibnamefont
  {Gleyzes}}, \bibinfo {author} {\bibfnamefont {S.}~\bibnamefont {Kuhr}},
  \bibinfo {author} {\bibfnamefont {C.}~\bibnamefont {Guerlin}}, \bibinfo
  {author} {\bibfnamefont {J.}~\bibnamefont {Bernu}}, \bibinfo {author}
  {\bibfnamefont {S.}~\bibnamefont {Del{\'{e}}glise}}, \bibinfo {author}
  {\bibfnamefont {U.}~\bibnamefont {Busk}}, \bibinfo {author} {\bibfnamefont
  {M.}~\bibnamefont {Brune}}, \bibinfo {author} {\bibfnamefont {J.-M.}\
  \bibnamefont {Raimond}}, \bibinfo {author} {\bibfnamefont {S.}~\bibnamefont
  {Haroche}}, \bibinfo {author} {\bibfnamefont {S.}~\bibnamefont {Deleglise}},
  \bibinfo {author} {\bibfnamefont {U.}~\bibnamefont {{Busk Hoff}}}, \bibinfo
  {author} {\bibfnamefont {M.}~\bibnamefont {Brune}}, \bibinfo {author}
  {\bibfnamefont {J.-M.}\ \bibnamefont {Raimond}}, \ and\ \bibinfo {author}
  {\bibfnamefont {S.}~\bibnamefont {Haroche}}} (\bibinfo {year} {2007}),\
  \bibfield  {title} {\enquote {\bibinfo {title} {{Observing the quantum jumps
  of light : birth and death of a photon in a cavity}},}\ }\href
  {http://dx.doi.org/10.1038/nature05589} {\bibfield  {journal} {\bibinfo
  {journal} {Nature}\ }\textbf {\bibinfo {volume} {446}}~(\bibinfo {number}
  {15}),\ \bibinfo {pages} {297}}\BibitemShut {NoStop}%
\bibitem [{\citenamefont {Grajcar}\ \emph {et~al.}(2008)\citenamefont
  {Grajcar}, \citenamefont {van~der Ploeg}, \citenamefont {Izmalkov},
  \citenamefont {Il'ichev}, \citenamefont {Meyer}, \citenamefont {Fedorov},
  \citenamefont {Shnirman},\ and\ \citenamefont {Sch{\"{o}}n}}]{Grajcar2008}%
  \BibitemOpen
  \bibfield  {author} {\bibinfo {author} {\bibfnamefont {M.}~\bibnamefont
  {Grajcar}}, \bibinfo {author} {\bibfnamefont {S.~H.~W.}\ \bibnamefont
  {van~der Ploeg}}, \bibinfo {author} {\bibfnamefont {A.}~\bibnamefont
  {Izmalkov}}, \bibinfo {author} {\bibfnamefont {E.}~\bibnamefont {Il'ichev}},
  \bibinfo {author} {\bibfnamefont {H.-G.}\ \bibnamefont {Meyer}}, \bibinfo
  {author} {\bibfnamefont {A.}~\bibnamefont {Fedorov}}, \bibinfo {author}
  {\bibfnamefont {A.}~\bibnamefont {Shnirman}}, \ and\ \bibinfo {author}
  {\bibfnamefont {G.}~\bibnamefont {Sch{\"{o}}n}}} (\bibinfo {year} {2008}),\
  \bibfield  {title} {\enquote {\bibinfo {title} {{Sisyphus cooling and
  amplification by a superconducting qubit}},}\ }\href {\doibase
  10.1038/nphys1019} {\bibfield  {journal} {\bibinfo  {journal} {Nature
  Physics}\ }\textbf {\bibinfo {volume} {4}}~(\bibinfo {number} {8}),\ \bibinfo
  {pages} {612--616}}\BibitemShut {NoStop}%
\bibitem [{\citenamefont {Guerlin}\ \emph {et~al.}(2007)\citenamefont
  {Guerlin}, \citenamefont {Bernu}, \citenamefont {Del{\'{e}}glise},
  \citenamefont {Sayrin}, \citenamefont {Gleyzes}, \citenamefont {Kuhr},
  \citenamefont {Brune}, \citenamefont {Raimond},\ and\ \citenamefont
  {Haroche}}]{Guerlin2007}%
  \BibitemOpen
  \bibfield  {author} {\bibinfo {author} {\bibfnamefont {C.}~\bibnamefont
  {Guerlin}}, \bibinfo {author} {\bibfnamefont {J.}~\bibnamefont {Bernu}},
  \bibinfo {author} {\bibfnamefont {S.}~\bibnamefont {Del{\'{e}}glise}},
  \bibinfo {author} {\bibfnamefont {C.}~\bibnamefont {Sayrin}}, \bibinfo
  {author} {\bibfnamefont {S.}~\bibnamefont {Gleyzes}}, \bibinfo {author}
  {\bibfnamefont {S.}~\bibnamefont {Kuhr}}, \bibinfo {author} {\bibfnamefont
  {M.}~\bibnamefont {Brune}}, \bibinfo {author} {\bibfnamefont {J.-M.}\
  \bibnamefont {Raimond}}, \ and\ \bibinfo {author} {\bibfnamefont
  {S.}~\bibnamefont {Haroche}}} (\bibinfo {year} {2007}),\ \bibfield  {title}
  {\enquote {\bibinfo {title} {{Progressive field-state collapse and quantum
  non-demolition photon counting.}}}\ }\href {\doibase 10.1038/nature06057}
  {\bibfield  {journal} {\bibinfo  {journal} {Nature}\ }\textbf {\bibinfo
  {volume} {448}}~(\bibinfo {number} {7156}),\ \bibinfo {pages} {889--93}},\
  \Eprint {http://arxiv.org/abs/0707.3880} {arXiv:0707.3880} \BibitemShut
  {NoStop}%
\bibitem [{\citenamefont {Hacohen-Gourgy}\ \emph {et~al.}(2018)\citenamefont
  {Hacohen-Gourgy}, \citenamefont {Garc{\'{i}}a-Pintos}, \citenamefont
  {Martin}, \citenamefont {Dressel},\ and\ \citenamefont
  {Siddiqi}}]{Hacohen-Gourgy2018}%
  \BibitemOpen
  \bibfield  {author} {\bibinfo {author} {\bibfnamefont {S.}~\bibnamefont
  {Hacohen-Gourgy}}, \bibinfo {author} {\bibfnamefont {L.~P.}\ \bibnamefont
  {Garc{\'{i}}a-Pintos}}, \bibinfo {author} {\bibfnamefont {L.~S.}\
  \bibnamefont {Martin}}, \bibinfo {author} {\bibfnamefont {J.}~\bibnamefont
  {Dressel}}, \ and\ \bibinfo {author} {\bibfnamefont {I.}~\bibnamefont
  {Siddiqi}}} (\bibinfo {year} {2018}),\ \bibfield  {title} {\enquote {\bibinfo
  {title} {{Incoherent Qubit Control Using the Quantum Zeno Effect}},}\ }\href
  {\doibase 10.1103/PhysRevLett.120.020505} {\bibfield  {journal} {\bibinfo
  {journal} {Physical Review Letters}\ }\textbf {\bibinfo {volume}
  {120}}~(\bibinfo {number} {2}),\ \bibinfo {pages} {020505}},\ \Eprint
  {http://arxiv.org/abs/1706.08577} {arXiv:1706.08577} \BibitemShut {NoStop}%
\bibitem [{\citenamefont {Hacohen-Gourgy}\ \emph {et~al.}(2016)\citenamefont
  {Hacohen-Gourgy}, \citenamefont {Martin}, \citenamefont {Flurin},
  \citenamefont {Ramasesh}, \citenamefont {Whaley},\ and\ \citenamefont
  {Siddiqi}}]{Hacohen-Gourgy2016-non-comm}%
  \BibitemOpen
  \bibfield  {author} {\bibinfo {author} {\bibfnamefont {S.}~\bibnamefont
  {Hacohen-Gourgy}}, \bibinfo {author} {\bibfnamefont {L.~S.}\ \bibnamefont
  {Martin}}, \bibinfo {author} {\bibfnamefont {E.}~\bibnamefont {Flurin}},
  \bibinfo {author} {\bibfnamefont {V.~V.}\ \bibnamefont {Ramasesh}}, \bibinfo
  {author} {\bibfnamefont {K.~B.}\ \bibnamefont {Whaley}}, \ and\ \bibinfo
  {author} {\bibfnamefont {I.}~\bibnamefont {Siddiqi}}} (\bibinfo {year}
  {2016}),\ \bibfield  {title} {\enquote {\bibinfo {title} {{Quantum dynamics
  of simultaneously measured non-commuting observables}},}\ }\href {\doibase
  10.1038/nature19762} {\bibfield  {journal} {\bibinfo  {journal} {Nature}\
  }\textbf {\bibinfo {volume} {538}}~(\bibinfo {number} {7626}),\ \bibinfo
  {pages} {491--494}}\BibitemShut {NoStop}%
\bibitem [{\citenamefont {Harrington}\ \emph {et~al.}(2017)\citenamefont
  {Harrington}, \citenamefont {Monroe},\ and\ \citenamefont
  {Murch}}]{Harrington2017}%
  \BibitemOpen
  \bibfield  {author} {\bibinfo {author} {\bibfnamefont {P.~M.}\ \bibnamefont
  {Harrington}}, \bibinfo {author} {\bibfnamefont {J.~T.}\ \bibnamefont
  {Monroe}}, \ and\ \bibinfo {author} {\bibfnamefont {K.~W.}\ \bibnamefont
  {Murch}}} (\bibinfo {year} {2017}),\ \bibfield  {title} {\enquote {\bibinfo
  {title} {{Quantum Zeno Effects from Measurement Controlled Qubit-Bath
  Interactions}},}\ }\href {\doibase 10.1103/PhysRevLett.118.240401} {\bibfield
   {journal} {\bibinfo  {journal} {Physical Review Letters}\ }\textbf {\bibinfo
  {volume} {118}}~(\bibinfo {number} {24}),\ \bibinfo {pages}
  {240401}}\BibitemShut {NoStop}%
\bibitem [{\citenamefont {Hatridge}\ \emph {et~al.}(2013)\citenamefont
  {Hatridge}, \citenamefont {Shankar}, \citenamefont {Mirrahimi}, \citenamefont
  {Schackert}, \citenamefont {Geerlings}, \citenamefont {Brecht}, \citenamefont
  {Sliwa}, \citenamefont {Abdo}, \citenamefont {Frunzio}, \citenamefont
  {Girvin}, \citenamefont {Schoelkopf},\ and\ \citenamefont
  {Devoret}}]{Hatridge2013}%
  \BibitemOpen
  \bibfield  {author} {\bibinfo {author} {\bibfnamefont {M.}~\bibnamefont
  {Hatridge}}, \bibinfo {author} {\bibfnamefont {S.}~\bibnamefont {Shankar}},
  \bibinfo {author} {\bibfnamefont {M.}~\bibnamefont {Mirrahimi}}, \bibinfo
  {author} {\bibfnamefont {F.}~\bibnamefont {Schackert}}, \bibinfo {author}
  {\bibfnamefont {K.}~\bibnamefont {Geerlings}}, \bibinfo {author}
  {\bibfnamefont {T.}~\bibnamefont {Brecht}}, \bibinfo {author} {\bibfnamefont
  {K.~M.}\ \bibnamefont {Sliwa}}, \bibinfo {author} {\bibfnamefont
  {B.}~\bibnamefont {Abdo}}, \bibinfo {author} {\bibfnamefont {L.}~\bibnamefont
  {Frunzio}}, \bibinfo {author} {\bibfnamefont {S.~M.}\ \bibnamefont {Girvin}},
  \bibinfo {author} {\bibfnamefont {R.~J.}\ \bibnamefont {Schoelkopf}}, \ and\
  \bibinfo {author} {\bibfnamefont {M.~H.}\ \bibnamefont {Devoret}}} (\bibinfo
  {year} {2013}),\ \bibfield  {title} {\enquote {\bibinfo {title} {{Quantum
  Back-Action of an Individual Variable-Strength Measurement}},}\ }\href
  {\doibase 10.1126/science.1226897} {\bibfield  {journal} {\bibinfo  {journal}
  {Science}\ }\textbf {\bibinfo {volume} {339}}~(\bibinfo {number} {6116}),\
  \bibinfo {pages} {178--181}}\BibitemShut {NoStop}%
\bibitem [{\citenamefont {Hawken}\ \emph {et~al.}(2010)\citenamefont {Hawken},
  \citenamefont {Lovins},\ and\ \citenamefont {Lovins}}]{hawken2010natural}%
  \BibitemOpen
  \bibfield  {author} {\bibinfo {author} {\bibfnamefont {P.}~\bibnamefont
  {Hawken}}, \bibinfo {author} {\bibfnamefont {A.~B.}\ \bibnamefont {Lovins}},
  \ and\ \bibinfo {author} {\bibfnamefont {L.~H.}\ \bibnamefont {Lovins}}}
  (\bibinfo {year} {2010}),\ \href
  {https://books.google.com/books?id=KiepOn7khp0C} {\emph {\bibinfo {title}
  {{Natural Capitalism: The Next Industrial Revolution}}}}\ (\bibinfo
  {publisher} {Earthscan})\BibitemShut {NoStop}%
\bibitem [{\citenamefont {Heinsoo}\ \emph {et~al.}(2018)\citenamefont
  {Heinsoo}, \citenamefont {Andersen}, \citenamefont {Remm}, \citenamefont
  {Krinner}, \citenamefont {Walter}, \citenamefont {Salath{\'{e}}},
  \citenamefont {Gasparinetti}, \citenamefont {Besse}, \citenamefont
  {Poto{\v{c}}nik}, \citenamefont {Wallraff},\ and\ \citenamefont
  {Eichler}}]{Heinsoo2018}%
  \BibitemOpen
  \bibfield  {author} {\bibinfo {author} {\bibfnamefont {J.}~\bibnamefont
  {Heinsoo}}, \bibinfo {author} {\bibfnamefont {C.~K.}\ \bibnamefont
  {Andersen}}, \bibinfo {author} {\bibfnamefont {A.}~\bibnamefont {Remm}},
  \bibinfo {author} {\bibfnamefont {S.}~\bibnamefont {Krinner}}, \bibinfo
  {author} {\bibfnamefont {T.}~\bibnamefont {Walter}}, \bibinfo {author}
  {\bibfnamefont {Y.}~\bibnamefont {Salath{\'{e}}}}, \bibinfo {author}
  {\bibfnamefont {S.}~\bibnamefont {Gasparinetti}}, \bibinfo {author}
  {\bibfnamefont {J.-C.}\ \bibnamefont {Besse}}, \bibinfo {author}
  {\bibfnamefont {A.}~\bibnamefont {Poto{\v{c}}nik}}, \bibinfo {author}
  {\bibfnamefont {A.}~\bibnamefont {Wallraff}}, \ and\ \bibinfo {author}
  {\bibfnamefont {C.}~\bibnamefont {Eichler}}} (\bibinfo {year} {2018}),\
  \bibfield  {title} {\enquote {\bibinfo {title} {{Rapid High-fidelity
  Multiplexed Readout of Superconducting Qubits}},}\ }\href {\doibase
  10.1103/PhysRevApplied.10.034040} {\bibfield  {journal} {\bibinfo  {journal}
  {Physical Review Applied}\ }\textbf {\bibinfo {volume} {10}}~(\bibinfo
  {number} {3}),\ \bibinfo {pages} {034040}},\ \Eprint
  {http://arxiv.org/abs/1801.07904} {arXiv:1801.07904} \BibitemShut {NoStop}%
\bibitem [{\citenamefont {Heisenberg}(1927)}]{Heisenberg1927}%
  \BibitemOpen
  \bibfield  {author} {\bibinfo {author} {\bibfnamefont {W.}~\bibnamefont
  {Heisenberg}}} (\bibinfo {year} {1927}),\ \bibfield  {title} {\enquote
  {\bibinfo {title} {{{\"{U}}ber den anschaulichen Inhalt der
  quantentheoretischen Kinematik und Mechanik (The actual content of quantum
  theoretical kinematics and mechanics)}},}\ }\href {\doibase
  10.1007/BF01397280} {\bibfield  {journal} {\bibinfo  {journal} {Zeitschrift
  f{\"{u}}r Physik}\ }\textbf {\bibinfo {volume} {43}}~(\bibinfo {number}
  {3-4}),\ 10.1007/BF01397280}\BibitemShut {NoStop}%
\bibitem [{\citenamefont {Huembeli}\ and\ \citenamefont
  {Nigg}(2017)}]{Huembeli2017}%
  \BibitemOpen
  \bibfield  {author} {\bibinfo {author} {\bibfnamefont {P.}~\bibnamefont
  {Huembeli}}, \ and\ \bibinfo {author} {\bibfnamefont {S.~E.}\ \bibnamefont
  {Nigg}}} (\bibinfo {year} {2017}),\ \bibfield  {title} {\enquote {\bibinfo
  {title} {{Towards a heralded eigenstate-preserving measurement of multi-qubit
  parity in circuit QED}},}\ }\href {\doibase 10.1103/PhysRevA.96.012313}
  {\bibfield  {journal} {\bibinfo  {journal} {Physical Review A}\ }\textbf
  {\bibinfo {volume} {96}}~(\bibinfo {number} {1}),\ \bibinfo {pages}
  {012313}}\BibitemShut {NoStop}%
\bibitem [{\citenamefont {Inomata}\ \emph {et~al.}(2016)\citenamefont
  {Inomata}, \citenamefont {Lin}, \citenamefont {Koshino}, \citenamefont
  {Oliver}, \citenamefont {Tsai}, \citenamefont {Yamamoto},\ and\ \citenamefont
  {Nakamura}}]{Inomata2016-FlyingPhoton}%
  \BibitemOpen
  \bibfield  {author} {\bibinfo {author} {\bibfnamefont {K.}~\bibnamefont
  {Inomata}}, \bibinfo {author} {\bibfnamefont {Z.}~\bibnamefont {Lin}},
  \bibinfo {author} {\bibfnamefont {K.}~\bibnamefont {Koshino}}, \bibinfo
  {author} {\bibfnamefont {W.~D.}\ \bibnamefont {Oliver}}, \bibinfo {author}
  {\bibfnamefont {J.-S.}\ \bibnamefont {Tsai}}, \bibinfo {author}
  {\bibfnamefont {T.}~\bibnamefont {Yamamoto}}, \ and\ \bibinfo {author}
  {\bibfnamefont {Y.}~\bibnamefont {Nakamura}}} (\bibinfo {year} {2016}),\
  \bibfield  {title} {\enquote {\bibinfo {title} {{Single microwave-photon
  detector using an artificial $\Lambda$-type three-level system}},}\ }\href
  {\doibase 10.1038/ncomms12303} {\bibfield  {journal} {\bibinfo  {journal}
  {Nature Communications}\ }\textbf {\bibinfo {volume} {7}},\ \bibinfo {pages}
  {12303}}\BibitemShut {NoStop}%
\bibitem [{\citenamefont {Jacobs}(2003)}]{Jacobs2003-adaptive-msr}%
  \BibitemOpen
  \bibfield  {author} {\bibinfo {author} {\bibfnamefont {K.}~\bibnamefont
  {Jacobs}}} (\bibinfo {year} {2003}),\ \bibfield  {title} {\enquote {\bibinfo
  {title} {{How to project qubits faster using quantum feedback}},}\ }\href
  {\doibase 10.1103/PhysRevA.67.030301} {\bibfield  {journal} {\bibinfo
  {journal} {Physical Review A}\ }\textbf {\bibinfo {volume} {67}}~(\bibinfo
  {number} {3}),\ \bibinfo {pages} {030301}}\BibitemShut {NoStop}%
\bibitem [{\citenamefont {Jacobs}(2010)}]{Jacobs2010Book}%
  \BibitemOpen
  \bibfield  {author} {\bibinfo {author} {\bibfnamefont {K.}~\bibnamefont
  {Jacobs}}} (\bibinfo {year} {2010}),\ \href {\doibase
  10.1017/CBO9780511815980} {\emph {\bibinfo {title} {{Stochastic Processes for
  Physicists}}}}\ (\bibinfo  {publisher} {Cambridge University Press},\
  \bibinfo {address} {Cambridge})\BibitemShut {NoStop}%
\bibitem [{\citenamefont {Jacobs}(2014)}]{jacobs2014book}%
  \BibitemOpen
  \bibfield  {author} {\bibinfo {author} {\bibfnamefont {K.}~\bibnamefont
  {Jacobs}}} (\bibinfo {year} {2014}),\ \href
  {https://books.google.com/books?id=dAhQBAAAQBAJ} {\emph {\bibinfo {title}
  {{Quantum Measurement Theory and its Applications}}}}\ (\bibinfo  {publisher}
  {Cambridge University Press})\BibitemShut {NoStop}%
\bibitem [{\citenamefont {Jelezko}\ \emph {et~al.}(2002)\citenamefont
  {Jelezko}, \citenamefont {Popa}, \citenamefont {Gruber}, \citenamefont
  {Tietz}, \citenamefont {Wrachtrup}, \citenamefont {Nizovtsev},\ and\
  \citenamefont {Kilin}}]{Jelezko2002}%
  \BibitemOpen
  \bibfield  {author} {\bibinfo {author} {\bibfnamefont {F.}~\bibnamefont
  {Jelezko}}, \bibinfo {author} {\bibfnamefont {I.}~\bibnamefont {Popa}},
  \bibinfo {author} {\bibfnamefont {A.}~\bibnamefont {Gruber}}, \bibinfo
  {author} {\bibfnamefont {C.}~\bibnamefont {Tietz}}, \bibinfo {author}
  {\bibfnamefont {J.}~\bibnamefont {Wrachtrup}}, \bibinfo {author}
  {\bibfnamefont {A.}~\bibnamefont {Nizovtsev}}, \ and\ \bibinfo {author}
  {\bibfnamefont {S.}~\bibnamefont {Kilin}}} (\bibinfo {year} {2002}),\
  \bibfield  {title} {\enquote {\bibinfo {title} {{Single spin states in a
  defect center resolved by optical spectroscopy}},}\ }\href {\doibase
  10.1063/1.1507838} {\bibfield  {journal} {\bibinfo  {journal} {Applied
  Physics Letters}\ }\textbf {\bibinfo {volume} {81}}~(\bibinfo {number}
  {12}),\ \bibinfo {pages} {2160--2162}}\BibitemShut {NoStop}%
\bibitem [{\citenamefont {Jin}(2014)}]{Jin2014}%
  \BibitemOpen
  \bibfield  {author} {\bibinfo {author} {\bibfnamefont {J.-M.}\ \bibnamefont
  {Jin}}} (\bibinfo {year} {2014}),\ \href@noop {} {\emph {\bibinfo {title}
  {{The finite element method in electromagnetics}}}}\ (\bibinfo  {publisher}
  {Wiley-IEEE Press})\BibitemShut {NoStop}%
\bibitem [{\citenamefont {Johnson}\ \emph {et~al.}(2012)\citenamefont
  {Johnson}, \citenamefont {Macklin}, \citenamefont {Slichter}, \citenamefont
  {Vijay}, \citenamefont {Weingarten}, \citenamefont {Clarke},\ and\
  \citenamefont {Siddiqi}}]{Johnson2012-readout}%
  \BibitemOpen
  \bibfield  {author} {\bibinfo {author} {\bibfnamefont {J.~E.}\ \bibnamefont
  {Johnson}}, \bibinfo {author} {\bibfnamefont {C.}~\bibnamefont {Macklin}},
  \bibinfo {author} {\bibfnamefont {D.~H.}\ \bibnamefont {Slichter}}, \bibinfo
  {author} {\bibfnamefont {R.}~\bibnamefont {Vijay}}, \bibinfo {author}
  {\bibfnamefont {E.~B.}\ \bibnamefont {Weingarten}}, \bibinfo {author}
  {\bibfnamefont {J.}~\bibnamefont {Clarke}}, \ and\ \bibinfo {author}
  {\bibfnamefont {I.}~\bibnamefont {Siddiqi}}} (\bibinfo {year} {2012}),\
  \bibfield  {title} {\enquote {\bibinfo {title} {{Heralded State Preparation
  in a Superconducting Qubit}},}\ }\href {\doibase
  10.1103/PhysRevLett.109.050506} {\bibfield  {journal} {\bibinfo  {journal}
  {Physical Review Letters}\ }\textbf {\bibinfo {volume} {109}}~(\bibinfo
  {number} {5}),\ \bibinfo {pages} {050506}}\BibitemShut {NoStop}%
\bibitem [{\citenamefont {Jordan}\ and\ \citenamefont
  {B{\"{u}}ttiker}(2005)}]{Jordan2005-non-coom}%
  \BibitemOpen
  \bibfield  {author} {\bibinfo {author} {\bibfnamefont {A.~N.}\ \bibnamefont
  {Jordan}}, \ and\ \bibinfo {author} {\bibfnamefont {M.}~\bibnamefont
  {B{\"{u}}ttiker}}} (\bibinfo {year} {2005}),\ \bibfield  {title} {\enquote
  {\bibinfo {title} {{Continuous Quantum Measurement with Independent Detector
  Cross Correlations}},}\ }\href {\doibase 10.1103/PhysRevLett.95.220401}
  {\bibfield  {journal} {\bibinfo  {journal} {Physical Review Letters}\
  }\textbf {\bibinfo {volume} {95}}~(\bibinfo {number} {22}),\ \bibinfo {pages}
  {220401}}\BibitemShut {NoStop}%
\bibitem [{\citenamefont {Jordan}\ \emph {et~al.}(2017)\citenamefont {Jordan},
  \citenamefont {Chantasri}, \citenamefont {Murch}, \citenamefont {Dressel},\
  and\ \citenamefont {Korotkov}}]{Jordan2017-Janus}%
  \BibitemOpen
  \bibfield  {author} {\bibinfo {author} {\bibfnamefont {A.~N.}\ \bibnamefont
  {Jordan}}, \bibinfo {author} {\bibfnamefont {A.}~\bibnamefont {Chantasri}},
  \bibinfo {author} {\bibfnamefont {K.}~\bibnamefont {Murch}}, \bibinfo
  {author} {\bibfnamefont {J.}~\bibnamefont {Dressel}}, \ and\ \bibinfo
  {author} {\bibfnamefont {A.~N.}\ \bibnamefont {Korotkov}}} (\bibinfo {year}
  {2017}),\ \bibfield  {title} {\enquote {\bibinfo {title} {{Janus sequences of
  quantum measurements and the arrow of time}},}\ }in\ \href {\doibase
  10.1063/1.4982767} {\emph {\bibinfo {booktitle} {AIP Conference
  Proceedings}}},\ Vol.\ \bibinfo {volume} {1841}\ (\bibinfo  {publisher}
  {American Institute of Physics})\ p.\ \bibinfo {pages} {020003}\BibitemShut
  {NoStop}%
\bibitem [{\citenamefont {Jordan}\ \emph {et~al.}(2016)\citenamefont {Jordan},
  \citenamefont {Chantasri}, \citenamefont {Rouchon},\ and\ \citenamefont
  {Huard}}]{Jordan2016-fluorescence}%
  \BibitemOpen
  \bibfield  {author} {\bibinfo {author} {\bibfnamefont {A.~N.}\ \bibnamefont
  {Jordan}}, \bibinfo {author} {\bibfnamefont {A.}~\bibnamefont {Chantasri}},
  \bibinfo {author} {\bibfnamefont {P.}~\bibnamefont {Rouchon}}, \ and\
  \bibinfo {author} {\bibfnamefont {B.}~\bibnamefont {Huard}}} (\bibinfo {year}
  {2016}),\ \bibfield  {title} {\enquote {\bibinfo {title} {{Anatomy of
  fluorescence: quantum trajectory statistics from continuously measuring
  spontaneous emission}},}\ }\href {\doibase 10.1007/s40509-016-0075-9}
  {\bibfield  {journal} {\bibinfo  {journal} {Quantum Studies: Mathematics and
  Foundations}\ }\textbf {\bibinfo {volume} {3}}~(\bibinfo {number} {3}),\
  \bibinfo {pages} {237--263}}\BibitemShut {NoStop}%
\bibitem [{\citenamefont {Josephson}(1962)}]{Josephson1962}%
  \BibitemOpen
  \bibfield  {author} {\bibinfo {author} {\bibfnamefont {B.}~\bibnamefont
  {Josephson}}} (\bibinfo {year} {1962}),\ \bibfield  {title} {\enquote
  {\bibinfo {title} {{Possible new effects in superconductive tunnelling}},}\
  }\href {\doibase 10.1016/0031-9163(62)91369-0} {\bibfield  {journal}
  {\bibinfo  {journal} {Physics Letters}\ }\textbf {\bibinfo {volume}
  {1}}~(\bibinfo {number} {7}),\ \bibinfo {pages} {251--253}}\BibitemShut
  {NoStop}%
\bibitem [{\citenamefont {Kalb}\ \emph {et~al.}(2017)\citenamefont {Kalb},
  \citenamefont {Reiserer}, \citenamefont {Humphreys}, \citenamefont
  {Bakermans}, \citenamefont {Kamerling}, \citenamefont {Nickerson},
  \citenamefont {Benjamin}, \citenamefont {Twitchen}, \citenamefont {Markham},\
  and\ \citenamefont {Hanson}}]{Kalb2017}%
  \BibitemOpen
  \bibfield  {author} {\bibinfo {author} {\bibfnamefont {N.}~\bibnamefont
  {Kalb}}, \bibinfo {author} {\bibfnamefont {A.~A.}\ \bibnamefont {Reiserer}},
  \bibinfo {author} {\bibfnamefont {P.~C.}\ \bibnamefont {Humphreys}}, \bibinfo
  {author} {\bibfnamefont {J.~J.~W.}\ \bibnamefont {Bakermans}}, \bibinfo
  {author} {\bibfnamefont {S.~J.}\ \bibnamefont {Kamerling}}, \bibinfo {author}
  {\bibfnamefont {N.~H.}\ \bibnamefont {Nickerson}}, \bibinfo {author}
  {\bibfnamefont {S.~C.}\ \bibnamefont {Benjamin}}, \bibinfo {author}
  {\bibfnamefont {D.~J.}\ \bibnamefont {Twitchen}}, \bibinfo {author}
  {\bibfnamefont {M.}~\bibnamefont {Markham}}, \ and\ \bibinfo {author}
  {\bibfnamefont {R.}~\bibnamefont {Hanson}}} (\bibinfo {year} {2017}),\
  \bibfield  {title} {\enquote {\bibinfo {title} {{Entanglement distillation
  between solid-state quantum network nodes}},}\ }\href {\doibase
  10.1126/science.aan0070} {\bibfield  {journal} {\bibinfo  {journal}
  {Science}\ }\textbf {\bibinfo {volume} {356}}~(\bibinfo {number} {6341}),\
  \bibinfo {pages} {928--932}}\BibitemShut {NoStop}%
\bibitem [{\citenamefont {Kamal}\ \emph {et~al.}(2016)\citenamefont {Kamal},
  \citenamefont {Yoder}, \citenamefont {Yan}, \citenamefont {Gudmundsen},
  \citenamefont {Hover}, \citenamefont {Sears}, \citenamefont {Welander},
  \citenamefont {Orlando}, \citenamefont {Gustavsson},\ and\ \citenamefont
  {Oliver}}]{Kamal2016-anneal}%
  \BibitemOpen
  \bibfield  {author} {\bibinfo {author} {\bibfnamefont {A.}~\bibnamefont
  {Kamal}}, \bibinfo {author} {\bibfnamefont {J.~L.}\ \bibnamefont {Yoder}},
  \bibinfo {author} {\bibfnamefont {F.}~\bibnamefont {Yan}}, \bibinfo {author}
  {\bibfnamefont {T.~J.}\ \bibnamefont {Gudmundsen}}, \bibinfo {author}
  {\bibfnamefont {D.}~\bibnamefont {Hover}}, \bibinfo {author} {\bibfnamefont
  {A.~P.}\ \bibnamefont {Sears}}, \bibinfo {author} {\bibfnamefont
  {P.}~\bibnamefont {Welander}}, \bibinfo {author} {\bibfnamefont {T.~P.}\
  \bibnamefont {Orlando}}, \bibinfo {author} {\bibfnamefont {S.}~\bibnamefont
  {Gustavsson}}, \ and\ \bibinfo {author} {\bibfnamefont {W.~D.}\ \bibnamefont
  {Oliver}}} (\bibinfo {year} {2016}),\ \bibfield  {title} {\enquote {\bibinfo
  {title} {{Improved superconducting qubit coherence with high-temperature
  substrate annealing}},}\ }\href {http://arxiv.org/abs/1606.09262} {\ }\Eprint
  {http://arxiv.org/abs/1606.09262} {arXiv:1606.09262} \BibitemShut {NoStop}%
\bibitem [{\citenamefont {Katz}\ \emph {et~al.}(2006)\citenamefont {Katz},
  \citenamefont {Ansmann}, \citenamefont {Bialczak}, \citenamefont {Lucero},
  \citenamefont {McDermott}, \citenamefont {Neeley}, \citenamefont {Steffen},
  \citenamefont {Weig}, \citenamefont {Cleland}, \citenamefont {Martinis},\
  and\ \citenamefont {Korotkov}}]{Katz2006}%
  \BibitemOpen
  \bibfield  {author} {\bibinfo {author} {\bibfnamefont {N.}~\bibnamefont
  {Katz}}, \bibinfo {author} {\bibfnamefont {M.}~\bibnamefont {Ansmann}},
  \bibinfo {author} {\bibfnamefont {R.~C.}\ \bibnamefont {Bialczak}}, \bibinfo
  {author} {\bibfnamefont {E.}~\bibnamefont {Lucero}}, \bibinfo {author}
  {\bibfnamefont {R.}~\bibnamefont {McDermott}}, \bibinfo {author}
  {\bibfnamefont {M.}~\bibnamefont {Neeley}}, \bibinfo {author} {\bibfnamefont
  {M.}~\bibnamefont {Steffen}}, \bibinfo {author} {\bibfnamefont {E.~M.}\
  \bibnamefont {Weig}}, \bibinfo {author} {\bibfnamefont {A.~N.}\ \bibnamefont
  {Cleland}}, \bibinfo {author} {\bibfnamefont {J.~M.}\ \bibnamefont
  {Martinis}}, \ and\ \bibinfo {author} {\bibfnamefont {A.~N.}\ \bibnamefont
  {Korotkov}}} (\bibinfo {year} {2006}),\ \bibfield  {title} {\enquote
  {\bibinfo {title} {{Coherent state evolution in a superconducting qubit from
  partial-collapse measurement.}}}\ }\href {\doibase 10.1126/science.1126475}
  {\bibfield  {journal} {\bibinfo  {journal} {Science}\ }\textbf {\bibinfo
  {volume} {312}}~(\bibinfo {number} {5779}),\ \bibinfo {pages}
  {1498--500}}\BibitemShut {NoStop}%
\bibitem [{\citenamefont {Katz}\ \emph {et~al.}(2008)\citenamefont {Katz},
  \citenamefont {Neeley}, \citenamefont {Ansmann}, \citenamefont {Bialczak},
  \citenamefont {Hofheinz}, \citenamefont {Lucero}, \citenamefont {O'Connell},
  \citenamefont {Wang}, \citenamefont {Cleland}, \citenamefont {Martinis},\
  and\ \citenamefont {Korotkov}}]{Katz2008}%
  \BibitemOpen
  \bibfield  {author} {\bibinfo {author} {\bibfnamefont {N.}~\bibnamefont
  {Katz}}, \bibinfo {author} {\bibfnamefont {M.}~\bibnamefont {Neeley}},
  \bibinfo {author} {\bibfnamefont {M.}~\bibnamefont {Ansmann}}, \bibinfo
  {author} {\bibfnamefont {R.~C.}\ \bibnamefont {Bialczak}}, \bibinfo {author}
  {\bibfnamefont {M.}~\bibnamefont {Hofheinz}}, \bibinfo {author}
  {\bibfnamefont {E.}~\bibnamefont {Lucero}}, \bibinfo {author} {\bibfnamefont
  {A.}~\bibnamefont {O'Connell}}, \bibinfo {author} {\bibfnamefont
  {H.}~\bibnamefont {Wang}}, \bibinfo {author} {\bibfnamefont {A.~N.}\
  \bibnamefont {Cleland}}, \bibinfo {author} {\bibfnamefont {J.~M.}\
  \bibnamefont {Martinis}}, \ and\ \bibinfo {author} {\bibfnamefont {A.~N.}\
  \bibnamefont {Korotkov}}} (\bibinfo {year} {2008}),\ \bibfield  {title}
  {\enquote {\bibinfo {title} {{Reversal of the Weak Measurement of a Quantum
  State in a Superconducting Phase Qubit}},}\ }\href {\doibase
  10.1103/PhysRevLett.101.200401} {\bibfield  {journal} {\bibinfo  {journal}
  {Physical Review Letters}\ }\textbf {\bibinfo {volume} {101}}~(\bibinfo
  {number} {20}),\ \bibinfo {pages} {200401}}\BibitemShut {NoStop}%
\bibitem [{\citenamefont {Khezri}\ and\ \citenamefont
  {Korotkov}(2017)}]{Khezri2017}%
  \BibitemOpen
  \bibfield  {author} {\bibinfo {author} {\bibfnamefont {M.}~\bibnamefont
  {Khezri}}, \ and\ \bibinfo {author} {\bibfnamefont {A.~N.}\ \bibnamefont
  {Korotkov}}} (\bibinfo {year} {2017}),\ \bibfield  {title} {\enquote
  {\bibinfo {title} {{Hybrid phase-space–Fock-space approach to evolution of
  a driven nonlinear resonator}},}\ }\href {\doibase
  10.1103/PhysRevA.96.043839} {\bibfield  {journal} {\bibinfo  {journal}
  {Physical Review A}\ }\textbf {\bibinfo {volume} {96}}~(\bibinfo {number}
  {4}),\ \bibinfo {pages} {043839}}\BibitemShut {NoStop}%
\bibitem [{\citenamefont {Khezri}\ \emph {et~al.}(2016)\citenamefont {Khezri},
  \citenamefont {Mlinar}, \citenamefont {Dressel},\ and\ \citenamefont
  {Korotkov}}]{Khezri2016}%
  \BibitemOpen
  \bibfield  {author} {\bibinfo {author} {\bibfnamefont {M.}~\bibnamefont
  {Khezri}}, \bibinfo {author} {\bibfnamefont {E.}~\bibnamefont {Mlinar}},
  \bibinfo {author} {\bibfnamefont {J.}~\bibnamefont {Dressel}}, \ and\
  \bibinfo {author} {\bibfnamefont {A.~N.}\ \bibnamefont {Korotkov}}} (\bibinfo
  {year} {2016}),\ \bibfield  {title} {\enquote {\bibinfo {title} {{Measuring a
  transmon qubit in circuit QED: Dressed squeezed states}},}\ }\href {\doibase
  10.1103/PhysRevA.94.012347} {\bibfield  {journal} {\bibinfo  {journal}
  {Physical Review A}\ }\textbf {\bibinfo {volume} {94}}~(\bibinfo {number}
  {1}),\ \bibinfo {pages} {012347}}\BibitemShut {NoStop}%
\bibitem [{\citenamefont {Kirchmair}\ \emph {et~al.}(2013)\citenamefont
  {Kirchmair}, \citenamefont {Vlastakis}, \citenamefont {Leghtas},
  \citenamefont {Nigg}, \citenamefont {Paik}, \citenamefont {Ginossar},
  \citenamefont {Mirrahimi}, \citenamefont {Frunzio}, \citenamefont {Girvin},\
  and\ \citenamefont {Schoelkopf}}]{Kirchmair2013}%
  \BibitemOpen
  \bibfield  {author} {\bibinfo {author} {\bibfnamefont {G.}~\bibnamefont
  {Kirchmair}}, \bibinfo {author} {\bibfnamefont {B.}~\bibnamefont
  {Vlastakis}}, \bibinfo {author} {\bibfnamefont {Z.}~\bibnamefont {Leghtas}},
  \bibinfo {author} {\bibfnamefont {S.~E.}\ \bibnamefont {Nigg}}, \bibinfo
  {author} {\bibfnamefont {H.}~\bibnamefont {Paik}}, \bibinfo {author}
  {\bibfnamefont {E.}~\bibnamefont {Ginossar}}, \bibinfo {author}
  {\bibfnamefont {M.}~\bibnamefont {Mirrahimi}}, \bibinfo {author}
  {\bibfnamefont {L.}~\bibnamefont {Frunzio}}, \bibinfo {author} {\bibfnamefont
  {S.~M.}\ \bibnamefont {Girvin}}, \ and\ \bibinfo {author} {\bibfnamefont
  {R.~J.}\ \bibnamefont {Schoelkopf}}} (\bibinfo {year} {2013}),\ \bibfield
  {title} {\enquote {\bibinfo {title} {{Observation of quantum state collapse
  and revival due to the single-photon Kerr effect.}}}\ }\href {\doibase
  10.1038/nature11902} {\bibfield  {journal} {\bibinfo  {journal} {Nature}\
  }\textbf {\bibinfo {volume} {495}}~(\bibinfo {number} {7440}),\ \bibinfo
  {pages} {205--209}},\ \Eprint {http://arxiv.org/abs/1211.2228}
  {arXiv:1211.2228} \BibitemShut {NoStop}%
\bibitem [{\citenamefont {Klyachko}\ \emph {et~al.}(2008)\citenamefont
  {Klyachko}, \citenamefont {Can}, \citenamefont {Binicioğlu},\ and\
  \citenamefont {Shumovsky}}]{Klyachko2008}%
  \BibitemOpen
  \bibfield  {author} {\bibinfo {author} {\bibfnamefont {A.~A.}\ \bibnamefont
  {Klyachko}}, \bibinfo {author} {\bibfnamefont {M.~A.}\ \bibnamefont {Can}},
  \bibinfo {author} {\bibfnamefont {S.}~\bibnamefont {Binicioğlu}}, \ and\
  \bibinfo {author} {\bibfnamefont {A.~S.}\ \bibnamefont {Shumovsky}}}
  (\bibinfo {year} {2008}),\ \bibfield  {title} {\enquote {\bibinfo {title}
  {{Simple Test for Hidden Variables in Spin-1 Systems}},}\ }\href {\doibase
  10.1103/PhysRevLett.101.020403} {\bibfield  {journal} {\bibinfo  {journal}
  {Physical Review Letters}\ }\textbf {\bibinfo {volume} {101}}~(\bibinfo
  {number} {2}),\ \bibinfo {pages} {020403}},\ \Eprint
  {http://arxiv.org/abs/0706.0126} {arXiv:0706.0126} \BibitemShut {NoStop}%
\bibitem [{\citenamefont {Koch}\ \emph {et~al.}(2007)\citenamefont {Koch},
  \citenamefont {Yu}, \citenamefont {Gambetta}, \citenamefont {Houck},
  \citenamefont {Schuster}, \citenamefont {Majer}, \citenamefont {Blais},
  \citenamefont {Devoret}, \citenamefont {Girvin},\ and\ \citenamefont
  {Schoelkopf}}]{Koch2007}%
  \BibitemOpen
  \bibfield  {author} {\bibinfo {author} {\bibfnamefont {J.}~\bibnamefont
  {Koch}}, \bibinfo {author} {\bibfnamefont {T.~M.}\ \bibnamefont {Yu}},
  \bibinfo {author} {\bibfnamefont {J.}~\bibnamefont {Gambetta}}, \bibinfo
  {author} {\bibfnamefont {A.~A.}\ \bibnamefont {Houck}}, \bibinfo {author}
  {\bibfnamefont {D.~I.}\ \bibnamefont {Schuster}}, \bibinfo {author}
  {\bibfnamefont {J.}~\bibnamefont {Majer}}, \bibinfo {author} {\bibfnamefont
  {A.}~\bibnamefont {Blais}}, \bibinfo {author} {\bibfnamefont {M.~H.}\
  \bibnamefont {Devoret}}, \bibinfo {author} {\bibfnamefont {S.~M.}\
  \bibnamefont {Girvin}}, \ and\ \bibinfo {author} {\bibfnamefont {R.~J.}\
  \bibnamefont {Schoelkopf}}} (\bibinfo {year} {2007}),\ \bibfield  {title}
  {\enquote {\bibinfo {title} {{Charge-insensitive qubit design derived from
  the Cooper pair box}},}\ }\href {\doibase 10.1103/PhysRevA.76.042319}
  {\bibfield  {journal} {\bibinfo  {journal} {Physical Review A}\ }\textbf
  {\bibinfo {volume} {76}}~(\bibinfo {number} {4}),\ \bibinfo {pages}
  {42319}}\BibitemShut {NoStop}%
\bibitem [{\citenamefont {Kochen}\ and\ \citenamefont
  {Specker}(1967)}]{KochenSpecker1967}%
  \BibitemOpen
  \bibfield  {author} {\bibinfo {author} {\bibfnamefont {S.}~\bibnamefont
  {Kochen}}, \ and\ \bibinfo {author} {\bibfnamefont {E.}~\bibnamefont
  {Specker}}} (\bibinfo {year} {1967}),\ \bibfield  {title} {\enquote {\bibinfo
  {title} {{The Problem of Hidden Variables in Quantum Mechanics}},}\ }\href
  {\doibase 10.1512/iumj.1968.17.17004} {\bibfield  {journal} {\bibinfo
  {journal} {Indiana University Mathematics Journal}\ }\textbf {\bibinfo
  {volume} {17}}~(\bibinfo {number} {1}),\ \bibinfo {pages}
  {59--87}}\BibitemShut {NoStop}%
\bibitem [{\citenamefont {Korotkov}(1999)}]{Korotkov1999-original-traj}%
  \BibitemOpen
  \bibfield  {author} {\bibinfo {author} {\bibfnamefont {A.~N.}\ \bibnamefont
  {Korotkov}}} (\bibinfo {year} {1999}),\ \bibfield  {title} {\enquote
  {\bibinfo {title} {{Continuous quantum measurement of a double dot}},}\
  }\href {\doibase 10.1103/PhysRevB.60.5737} {\bibfield  {journal} {\bibinfo
  {journal} {Physical Review B}\ }\textbf {\bibinfo {volume} {60}}~(\bibinfo
  {number} {8}),\ \bibinfo {pages} {5737--5742}}\BibitemShut {NoStop}%
\bibitem [{\citenamefont {Korotkov}(2016)}]{Korotkov2016-qm-bayes}%
  \BibitemOpen
  \bibfield  {author} {\bibinfo {author} {\bibfnamefont {A.~N.}\ \bibnamefont
  {Korotkov}}} (\bibinfo {year} {2016}),\ \bibfield  {title} {\enquote
  {\bibinfo {title} {{Quantum Bayesian approach to circuit QED measurement with
  moderate bandwidth}},}\ }\href {\doibase 10.1103/PhysRevA.94.042326}
  {\bibfield  {journal} {\bibinfo  {journal} {Physical Review A}\ }\textbf
  {\bibinfo {volume} {94}}~(\bibinfo {number} {4}),\ \bibinfo {pages}
  {042326}},\ \Eprint {http://arxiv.org/abs/1111.4016} {arXiv:1111.4016}
  \BibitemShut {NoStop}%
\bibitem [{\citenamefont {Kreikebaum}\ \emph {et~al.}(2016)\citenamefont
  {Kreikebaum}, \citenamefont {Dove}, \citenamefont {Livingston}, \citenamefont
  {Kim},\ and\ \citenamefont {Siddiqi}}]{Kreikebaum2016}%
  \BibitemOpen
  \bibfield  {author} {\bibinfo {author} {\bibfnamefont {J.~M.}\ \bibnamefont
  {Kreikebaum}}, \bibinfo {author} {\bibfnamefont {A.}~\bibnamefont {Dove}},
  \bibinfo {author} {\bibfnamefont {W.}~\bibnamefont {Livingston}}, \bibinfo
  {author} {\bibfnamefont {E.}~\bibnamefont {Kim}}, \ and\ \bibinfo {author}
  {\bibfnamefont {I.}~\bibnamefont {Siddiqi}}} (\bibinfo {year} {2016}),\
  \bibfield  {title} {\enquote {\bibinfo {title} {{Optimization of infrared and
  magnetic shielding of superconducting TiN and Al coplanar microwave
  resonators}},}\ }\href {\doibase 10.1088/0953-2048/29/10/104002} {\bibfield
  {journal} {\bibinfo  {journal} {Superconductor Science and Technology}\
  }\textbf {\bibinfo {volume} {29}}~(\bibinfo {number} {10}),\ \bibinfo {pages}
  {104002}}\BibitemShut {NoStop}%
\bibitem [{\citenamefont {Lecocq}\ \emph {et~al.}(2011)\citenamefont {Lecocq},
  \citenamefont {Pop}, \citenamefont {Peng}, \citenamefont {Matei},
  \citenamefont {Crozes}, \citenamefont {Fournier}, \citenamefont {Naud},
  \citenamefont {Guichard},\ and\ \citenamefont
  {Buisson}}]{Lecocq2011-bridge-free}%
  \BibitemOpen
  \bibfield  {author} {\bibinfo {author} {\bibfnamefont {F.}~\bibnamefont
  {Lecocq}}, \bibinfo {author} {\bibfnamefont {I.~M.}\ \bibnamefont {Pop}},
  \bibinfo {author} {\bibfnamefont {Z.}~\bibnamefont {Peng}}, \bibinfo {author}
  {\bibfnamefont {I.}~\bibnamefont {Matei}}, \bibinfo {author} {\bibfnamefont
  {T.}~\bibnamefont {Crozes}}, \bibinfo {author} {\bibfnamefont
  {T.}~\bibnamefont {Fournier}}, \bibinfo {author} {\bibfnamefont
  {C.}~\bibnamefont {Naud}}, \bibinfo {author} {\bibfnamefont {W.}~\bibnamefont
  {Guichard}}, \ and\ \bibinfo {author} {\bibfnamefont {O.}~\bibnamefont
  {Buisson}}} (\bibinfo {year} {2011}),\ \bibfield  {title} {\enquote {\bibinfo
  {title} {{Junction fabrication by shadow evaporation without a suspended
  bridge}},}\ }\href {\doibase 10.1088/0957-4484/22/31/315302} {\bibfield
  {journal} {\bibinfo  {journal} {Nanotechnology}\ }\textbf {\bibinfo {volume}
  {22}}~(\bibinfo {number} {31}),\ \bibinfo {pages} {315302}},\ \Eprint
  {http://arxiv.org/abs/1101.4576} {arXiv:1101.4576} \BibitemShut {NoStop}%
\bibitem [{\citenamefont {Leghtas}\ \emph {et~al.}(2015)\citenamefont
  {Leghtas}, \citenamefont {Touzard}, \citenamefont {Pop}, \citenamefont {Kou},
  \citenamefont {Vlastakis}, \citenamefont {Petrenko}, \citenamefont {Sliwa},
  \citenamefont {Narla}, \citenamefont {Shankar}, \citenamefont {Hatridge},
  \citenamefont {Reagor}, \citenamefont {Frunzio}, \citenamefont {Schoelkopf},
  \citenamefont {Mirrahimi},\ and\ \citenamefont {Devoret}}]{Leghtas2015}%
  \BibitemOpen
  \bibfield  {author} {\bibinfo {author} {\bibfnamefont {Z.}~\bibnamefont
  {Leghtas}}, \bibinfo {author} {\bibfnamefont {S.}~\bibnamefont {Touzard}},
  \bibinfo {author} {\bibfnamefont {I.~M.}\ \bibnamefont {Pop}}, \bibinfo
  {author} {\bibfnamefont {A.}~\bibnamefont {Kou}}, \bibinfo {author}
  {\bibfnamefont {B.}~\bibnamefont {Vlastakis}}, \bibinfo {author}
  {\bibfnamefont {A.}~\bibnamefont {Petrenko}}, \bibinfo {author}
  {\bibfnamefont {K.~M.}\ \bibnamefont {Sliwa}}, \bibinfo {author}
  {\bibfnamefont {A.}~\bibnamefont {Narla}}, \bibinfo {author} {\bibfnamefont
  {S.}~\bibnamefont {Shankar}}, \bibinfo {author} {\bibfnamefont {M.~J.}\
  \bibnamefont {Hatridge}}, \bibinfo {author} {\bibfnamefont {M.}~\bibnamefont
  {Reagor}}, \bibinfo {author} {\bibfnamefont {L.}~\bibnamefont {Frunzio}},
  \bibinfo {author} {\bibfnamefont {R.~J.}\ \bibnamefont {Schoelkopf}},
  \bibinfo {author} {\bibfnamefont {M.}~\bibnamefont {Mirrahimi}}, \ and\
  \bibinfo {author} {\bibfnamefont {M.~H.}\ \bibnamefont {Devoret}}} (\bibinfo
  {year} {2015}),\ \bibfield  {title} {\enquote {\bibinfo {title} {{Confining
  the state of light to a quantum manifold by engineered two-photon loss}},}\
  }\href {\doibase 10.1126/science.aaa2085} {\bibfield  {journal} {\bibinfo
  {journal} {Science}\ }\textbf {\bibinfo {volume} {347}}~(\bibinfo {number}
  {6224}),\ \bibinfo {pages} {853--857}}\BibitemShut {NoStop}%
\bibitem [{\citenamefont {Lesanovsky}\ \emph {et~al.}(2013)\citenamefont
  {Lesanovsky}, \citenamefont {van Horssen}, \citenamefont {Guţă},\ and\
  \citenamefont {Garrahan}}]{Lesanovsky2013}%
  \BibitemOpen
  \bibfield  {author} {\bibinfo {author} {\bibfnamefont {I.}~\bibnamefont
  {Lesanovsky}}, \bibinfo {author} {\bibfnamefont {M.}~\bibnamefont {van
  Horssen}}, \bibinfo {author} {\bibfnamefont {M.}~\bibnamefont {Guţă}}, \
  and\ \bibinfo {author} {\bibfnamefont {J.~P.}\ \bibnamefont {Garrahan}}}
  (\bibinfo {year} {2013}),\ \bibfield  {title} {\enquote {\bibinfo {title}
  {{Characterization of Dynamical Phase Transitions in Quantum Jump
  Trajectories Beyond the Properties of the Stationary State}},}\ }\href
  {\doibase 10.1103/PhysRevLett.110.150401} {\bibfield  {journal} {\bibinfo
  {journal} {Physical Review Letters}\ }\textbf {\bibinfo {volume}
  {110}}~(\bibinfo {number} {15}),\ \bibinfo {pages} {150401}}\BibitemShut
  {NoStop}%
\bibitem [{\citenamefont {Lescanne}\ \emph {et~al.}(2019)\citenamefont
  {Lescanne}, \citenamefont {Verney}, \citenamefont {Ficheux}, \citenamefont
  {Devoret}, \citenamefont {Huard}, \citenamefont {Mirrahimi},\ and\
  \citenamefont {Leghtas}}]{Lescanne2018}%
  \BibitemOpen
  \bibfield  {author} {\bibinfo {author} {\bibfnamefont {R.}~\bibnamefont
  {Lescanne}}, \bibinfo {author} {\bibfnamefont {L.}~\bibnamefont {Verney}},
  \bibinfo {author} {\bibfnamefont {Q.}~\bibnamefont {Ficheux}}, \bibinfo
  {author} {\bibfnamefont {M.~H.}\ \bibnamefont {Devoret}}, \bibinfo {author}
  {\bibfnamefont {B.}~\bibnamefont {Huard}}, \bibinfo {author} {\bibfnamefont
  {M.}~\bibnamefont {Mirrahimi}}, \ and\ \bibinfo {author} {\bibfnamefont
  {Z.}~\bibnamefont {Leghtas}}} (\bibinfo {year} {2019}),\ \bibfield  {title}
  {\enquote {\bibinfo {title} {{Escape of a Driven Quantum Josephson Circuit
  into Unconfined States}},}\ }\href {\doibase
  10.1103/PhysRevApplied.11.014030} {\bibfield  {journal} {\bibinfo  {journal}
  {Physical Review Applied}\ }\textbf {\bibinfo {volume} {11}}~(\bibinfo
  {number} {1}),\ \bibinfo {pages} {014030}},\ \Eprint
  {http://arxiv.org/abs/1805.05198} {arXiv:1805.05198} \BibitemShut {NoStop}%
\bibitem [{\citenamefont {Lewalle}\ \emph {et~al.}(2017)\citenamefont
  {Lewalle}, \citenamefont {Chantasri},\ and\ \citenamefont
  {Jordan}}]{Lewalle2017}%
  \BibitemOpen
  \bibfield  {author} {\bibinfo {author} {\bibfnamefont {P.}~\bibnamefont
  {Lewalle}}, \bibinfo {author} {\bibfnamefont {A.}~\bibnamefont {Chantasri}},
  \ and\ \bibinfo {author} {\bibfnamefont {A.~N.}\ \bibnamefont {Jordan}}}
  (\bibinfo {year} {2017}),\ \bibfield  {title} {\enquote {\bibinfo {title}
  {{Prediction and characterization of multiple extremal paths in continuously
  monitored qubits}},}\ }\href {\doibase 10.1103/PhysRevA.95.042126} {\bibfield
   {journal} {\bibinfo  {journal} {Physical Review A}\ }\textbf {\bibinfo
  {volume} {95}}~(\bibinfo {number} {4}),\ \bibinfo {pages}
  {042126}}\BibitemShut {NoStop}%
\bibitem [{\citenamefont {Li}\ \emph {et~al.}(2017)\citenamefont {Li},
  \citenamefont {Zou}, \citenamefont {Albert}, \citenamefont {Muralidharan},
  \citenamefont {Girvin},\ and\ \citenamefont {Jiang}}]{LiLinshu2017}%
  \BibitemOpen
  \bibfield  {author} {\bibinfo {author} {\bibfnamefont {L.}~\bibnamefont
  {Li}}, \bibinfo {author} {\bibfnamefont {C.-L.}\ \bibnamefont {Zou}},
  \bibinfo {author} {\bibfnamefont {V.~V.}\ \bibnamefont {Albert}}, \bibinfo
  {author} {\bibfnamefont {S.}~\bibnamefont {Muralidharan}}, \bibinfo {author}
  {\bibfnamefont {S.~M.}\ \bibnamefont {Girvin}}, \ and\ \bibinfo {author}
  {\bibfnamefont {L.}~\bibnamefont {Jiang}}} (\bibinfo {year} {2017}),\
  \bibfield  {title} {\enquote {\bibinfo {title} {{Cat Codes with Optimal
  Decoherence Suppression for a Lossy Bosonic Channel}},}\ }\href {\doibase
  10.1103/PhysRevLett.119.030502} {\bibfield  {journal} {\bibinfo  {journal}
  {Physical Review Letters}\ }\textbf {\bibinfo {volume} {119}}~(\bibinfo
  {number} {3}),\ \bibinfo {pages} {030502}}\BibitemShut {NoStop}%
\bibitem [{\citenamefont {Lindblad}(1976)}]{lindblad1976}%
  \BibitemOpen
  \bibfield  {author} {\bibinfo {author} {\bibfnamefont {G.}~\bibnamefont
  {Lindblad}}} (\bibinfo {year} {1976}),\ \bibfield  {title} {\enquote
  {\bibinfo {title} {{On the generators of quantum dynamical semigroups}},}\
  }\href {\doibase 10.1007/BF01608499} {\bibfield  {journal} {\bibinfo
  {journal} {Communications in Mathematical Physics}\ }\textbf {\bibinfo
  {volume} {48}}~(\bibinfo {number} {2}),\ \bibinfo {pages}
  {119--130}}\BibitemShut {NoStop}%
\bibitem [{\citenamefont {Liu}(2016)}]{Liu2016Thesis}%
  \BibitemOpen
  \bibfield  {author} {\bibinfo {author} {\bibfnamefont {Y.}~\bibnamefont
  {Liu}}} (\bibinfo {year} {2016}),\ \emph {\bibinfo {title} {{Quantum Feedback
  Control of Multiple Superconducting Qubits}}},\ \href@noop {} {Ph.D. thesis}\
  (\bibinfo  {school} {Yale University})\BibitemShut {NoStop}%
\bibitem [{\citenamefont {Liu}\ \emph {et~al.}(2016)\citenamefont {Liu},
  \citenamefont {Shankar}, \citenamefont {Ofek}, \citenamefont {Hatridge},
  \citenamefont {Narla}, \citenamefont {Sliwa}, \citenamefont {Frunzio},
  \citenamefont {Schoelkopf},\ and\ \citenamefont {Devoret}}]{LiuYehan2016}%
  \BibitemOpen
  \bibfield  {author} {\bibinfo {author} {\bibfnamefont {Y.}~\bibnamefont
  {Liu}}, \bibinfo {author} {\bibfnamefont {S.}~\bibnamefont {Shankar}},
  \bibinfo {author} {\bibfnamefont {N.}~\bibnamefont {Ofek}}, \bibinfo {author}
  {\bibfnamefont {M.}~\bibnamefont {Hatridge}}, \bibinfo {author}
  {\bibfnamefont {A.}~\bibnamefont {Narla}}, \bibinfo {author} {\bibfnamefont
  {K.~M.}\ \bibnamefont {Sliwa}}, \bibinfo {author} {\bibfnamefont
  {L.}~\bibnamefont {Frunzio}}, \bibinfo {author} {\bibfnamefont {R.~J.}\
  \bibnamefont {Schoelkopf}}, \ and\ \bibinfo {author} {\bibfnamefont {M.~H.}\
  \bibnamefont {Devoret}}} (\bibinfo {year} {2016}),\ \bibfield  {title}
  {\enquote {\bibinfo {title} {{Comparing and Combining Measurement-Based and
  Driven-Dissipative Entanglement Stabilization}},}\ }\href {\doibase
  10.1103/PhysRevX.6.011022} {\bibfield  {journal} {\bibinfo  {journal}
  {Physical Review X}\ }\textbf {\bibinfo {volume} {6}}~(\bibinfo {number}
  {1}),\ \bibinfo {pages} {011022}}\BibitemShut {NoStop}%
\bibitem [{\citenamefont {Louisell}(1973)}]{Louisell1973}%
  \BibitemOpen
  \bibfield  {author} {\bibinfo {author} {\bibfnamefont {W.~H. W.~H.}\
  \bibnamefont {Louisell}}} (\bibinfo {year} {1973}),\ \href@noop {} {\emph
  {\bibinfo {title} {{Quantum statistical properties of radiation}}}}\
  (\bibinfo  {publisher} {Wiley})\BibitemShut {NoStop}%
\bibitem [{\citenamefont {L{\"{u}}ders}(1951)}]{Luders1951}%
  \BibitemOpen
  \bibfield  {author} {\bibinfo {author} {\bibfnamefont {G.}~\bibnamefont
  {L{\"{u}}ders}}} (\bibinfo {year} {1951}),\ \bibfield  {title} {\enquote
  {\bibinfo {title} {{{\"{U}}ber die Zustands{\"{a}}nderung durch den
  Me{\ss}proze{\ss} (Concerning the state-change due to the measurement
  process)}},}\ }\href {\doibase 10.1002/andp.200610207} {\bibfield  {journal}
  {\bibinfo  {journal} {Annalen der Physik}\ }\textbf {\bibinfo {volume}
  {15}}~(\bibinfo {number} {9}),\ \bibinfo {pages} {663--670}}\BibitemShut
  {NoStop}%
\bibitem [{\citenamefont {Lupaşcu}\ \emph {et~al.}(2006)\citenamefont
  {Lupaşcu}, \citenamefont {Driessen}, \citenamefont {Roschier}, \citenamefont
  {Harmans},\ and\ \citenamefont {Mooij}}]{Lupascu2006-readout}%
  \BibitemOpen
  \bibfield  {author} {\bibinfo {author} {\bibfnamefont {A.}~\bibnamefont
  {Lupaşcu}}, \bibinfo {author} {\bibfnamefont {E.~F.~C.}\ \bibnamefont
  {Driessen}}, \bibinfo {author} {\bibfnamefont {L.}~\bibnamefont {Roschier}},
  \bibinfo {author} {\bibfnamefont {C.~J. P.~M.}\ \bibnamefont {Harmans}}, \
  and\ \bibinfo {author} {\bibfnamefont {J.~E.}\ \bibnamefont {Mooij}}}
  (\bibinfo {year} {2006}),\ \bibfield  {title} {\enquote {\bibinfo {title}
  {{High-Contrast Dispersive Readout of a Superconducting Flux Qubit Using a
  Nonlinear Resonator}},}\ }\href {\doibase 10.1103/PhysRevLett.96.127003}
  {\bibfield  {journal} {\bibinfo  {journal} {Physical Review Letters}\
  }\textbf {\bibinfo {volume} {96}}~(\bibinfo {number} {12}),\ \bibinfo {pages}
  {127003}}\BibitemShut {NoStop}%
\bibitem [{\citenamefont {Macklin}\ \emph {et~al.}(2015)\citenamefont
  {Macklin}, \citenamefont {O'Brien}, \citenamefont {Hover}, \citenamefont
  {Schwartz}, \citenamefont {Bolkhovsky}, \citenamefont {Zhang}, \citenamefont
  {Oliver},\ and\ \citenamefont {Siddiqi}}]{Macklin2015}%
  \BibitemOpen
  \bibfield  {author} {\bibinfo {author} {\bibfnamefont {C.}~\bibnamefont
  {Macklin}}, \bibinfo {author} {\bibfnamefont {K.}~\bibnamefont {O'Brien}},
  \bibinfo {author} {\bibfnamefont {D.}~\bibnamefont {Hover}}, \bibinfo
  {author} {\bibfnamefont {M.~E.}\ \bibnamefont {Schwartz}}, \bibinfo {author}
  {\bibfnamefont {V.}~\bibnamefont {Bolkhovsky}}, \bibinfo {author}
  {\bibfnamefont {X.}~\bibnamefont {Zhang}}, \bibinfo {author} {\bibfnamefont
  {W.~D.}\ \bibnamefont {Oliver}}, \ and\ \bibinfo {author} {\bibfnamefont
  {I.}~\bibnamefont {Siddiqi}}} (\bibinfo {year} {2015}),\ \bibfield  {title}
  {\enquote {\bibinfo {title} {{A near-quantum-limited Josephson traveling-wave
  parametric amplifier}},}\ }\href {\doibase 10.1126/science.aaa8525}
  {\bibfield  {journal} {\bibinfo  {journal} {Science}\ }\textbf {\bibinfo
  {volume} {350}}~(\bibinfo {number} {6258}),\ \bibinfo {pages}
  {307--310}}\BibitemShut {NoStop}%
\bibitem [{\citenamefont {Mallet}\ \emph {et~al.}(2009)\citenamefont {Mallet},
  \citenamefont {Ong}, \citenamefont {Palacios-Laloy}, \citenamefont {Nguyen},
  \citenamefont {Bertet}, \citenamefont {Vion},\ and\ \citenamefont
  {Esteve}}]{Mallet2009-readout}%
  \BibitemOpen
  \bibfield  {author} {\bibinfo {author} {\bibfnamefont {F.}~\bibnamefont
  {Mallet}}, \bibinfo {author} {\bibfnamefont {F.~R.}\ \bibnamefont {Ong}},
  \bibinfo {author} {\bibfnamefont {A.}~\bibnamefont {Palacios-Laloy}},
  \bibinfo {author} {\bibfnamefont {F.}~\bibnamefont {Nguyen}}, \bibinfo
  {author} {\bibfnamefont {P.}~\bibnamefont {Bertet}}, \bibinfo {author}
  {\bibfnamefont {D.}~\bibnamefont {Vion}}, \ and\ \bibinfo {author}
  {\bibfnamefont {D.}~\bibnamefont {Esteve}}} (\bibinfo {year} {2009}),\
  \bibfield  {title} {\enquote {\bibinfo {title} {{Single-shot qubit readout in
  circuit quantum electrodynamics}},}\ }\href {\doibase 10.1038/nphys1400}
  {\bibfield  {journal} {\bibinfo  {journal} {Nature Physics}\ }\textbf
  {\bibinfo {volume} {5}}~(\bibinfo {number} {11}),\ \bibinfo {pages}
  {791--795}}\BibitemShut {NoStop}%
\bibitem [{\citenamefont {Martinis}\ and\ \citenamefont
  {Megrant}(2014)}]{Martinis2014}%
  \BibitemOpen
  \bibfield  {author} {\bibinfo {author} {\bibfnamefont {J.~M.}\ \bibnamefont
  {Martinis}}, \ and\ \bibinfo {author} {\bibfnamefont {A.}~\bibnamefont
  {Megrant}}} (\bibinfo {year} {2014}),\ \bibfield  {title} {\enquote {\bibinfo
  {title} {{UCSB final report for the CSQ program: Review of decoherence and
  materials physics for superconducting qubits}},}\ }\href
  {http://arxiv.org/abs/1410.5793} {\ }\Eprint {http://arxiv.org/abs/1410.5793}
  {arXiv:1410.5793} \BibitemShut {NoStop}%
\bibitem [{\citenamefont {Masluk}\ \emph {et~al.}(2012)\citenamefont {Masluk},
  \citenamefont {Pop}, \citenamefont {Kamal}, \citenamefont {Minev},\ and\
  \citenamefont {Devoret}}]{Masluk2012}%
  \BibitemOpen
  \bibfield  {author} {\bibinfo {author} {\bibfnamefont {N.}~\bibnamefont
  {Masluk}}, \bibinfo {author} {\bibfnamefont {I.}~\bibnamefont {Pop}},
  \bibinfo {author} {\bibfnamefont {A.}~\bibnamefont {Kamal}}, \bibinfo
  {author} {\bibfnamefont {Z.}~\bibnamefont {Minev}}, \ and\ \bibinfo {author}
  {\bibfnamefont {M.}~\bibnamefont {Devoret}}} (\bibinfo {year} {2012}),\
  \bibfield  {title} {\enquote {\bibinfo {title} {{Microwave characterization
  of josephson junction arrays: Implementing a low loss superinductance}},}\
  }\href {\doibase 10.1103/PhysRevLett.109.137002} {\bibfield  {journal}
  {\bibinfo  {journal} {Physical Review Letters}\ }\textbf {\bibinfo {volume}
  {109}}~(\bibinfo {number} {13}),\ 10.1103/PhysRevLett.109.137002}\BibitemShut
  {NoStop}%
\bibitem [{\citenamefont {Matsuzaki}\ \emph {et~al.}(2010)\citenamefont
  {Matsuzaki}, \citenamefont {Saito}, \citenamefont {Kakuyanagi},\ and\
  \citenamefont {Semba}}]{Matsuzaki2010}%
  \BibitemOpen
  \bibfield  {author} {\bibinfo {author} {\bibfnamefont {Y.}~\bibnamefont
  {Matsuzaki}}, \bibinfo {author} {\bibfnamefont {S.}~\bibnamefont {Saito}},
  \bibinfo {author} {\bibfnamefont {K.}~\bibnamefont {Kakuyanagi}}, \ and\
  \bibinfo {author} {\bibfnamefont {K.}~\bibnamefont {Semba}}} (\bibinfo {year}
  {2010}),\ \bibfield  {title} {\enquote {\bibinfo {title} {{Quantum Zeno
  effect with a superconducting qubit}},}\ }\href {\doibase
  10.1103/PhysRevB.82.180518} {\bibfield  {journal} {\bibinfo  {journal}
  {Physical Review B}\ }\textbf {\bibinfo {volume} {82}}~(\bibinfo {number}
  {18}),\ \bibinfo {pages} {180518}}\BibitemShut {NoStop}%
\bibitem [{\citenamefont {Mermin}(1993)}]{Mermin1993-rev}%
  \BibitemOpen
  \bibfield  {author} {\bibinfo {author} {\bibfnamefont {N.~D.}\ \bibnamefont
  {Mermin}}} (\bibinfo {year} {1993}),\ \bibfield  {title} {\enquote {\bibinfo
  {title} {{Hidden variables and the two theorems of John Bell}},}\ }\href
  {\doibase 10.1103/RevModPhys.65.803} {\bibfield  {journal} {\bibinfo
  {journal} {Reviews of Modern Physics}\ }\textbf {\bibinfo {volume}
  {65}}~(\bibinfo {number} {3}),\ \bibinfo {pages} {803--815}}\BibitemShut
  {NoStop}%
\bibitem [{\citenamefont {Michael}\ \emph {et~al.}(2016)\citenamefont
  {Michael}, \citenamefont {Silveri}, \citenamefont {Brierley}, \citenamefont
  {Albert}, \citenamefont {Salmilehto}, \citenamefont {Jiang},\ and\
  \citenamefont {Girvin}}]{Michael2016}%
  \BibitemOpen
  \bibfield  {author} {\bibinfo {author} {\bibfnamefont {M.~H.}\ \bibnamefont
  {Michael}}, \bibinfo {author} {\bibfnamefont {M.}~\bibnamefont {Silveri}},
  \bibinfo {author} {\bibfnamefont {R.~T.}\ \bibnamefont {Brierley}}, \bibinfo
  {author} {\bibfnamefont {V.~V.}\ \bibnamefont {Albert}}, \bibinfo {author}
  {\bibfnamefont {J.}~\bibnamefont {Salmilehto}}, \bibinfo {author}
  {\bibfnamefont {L.}~\bibnamefont {Jiang}}, \ and\ \bibinfo {author}
  {\bibfnamefont {S.~M.}\ \bibnamefont {Girvin}}} (\bibinfo {year} {2016}),\
  \bibfield  {title} {\enquote {\bibinfo {title} {{New Class of Quantum
  Error-Correcting Codes for a Bosonic Mode}},}\ }\href {\doibase
  10.1103/PhysRevX.6.031006} {\bibfield  {journal} {\bibinfo  {journal}
  {Physical Review X}\ }\textbf {\bibinfo {volume} {6}}~(\bibinfo {number}
  {3}),\ \bibinfo {pages} {031006}}\BibitemShut {NoStop}%
\bibitem [{\citenamefont {Minev}\ \emph {et~al.}(2018)\citenamefont {Minev},
  \citenamefont {Leghtas}, \citenamefont {Mudhada}, \citenamefont {Pop},
  \citenamefont {Christakis}, \citenamefont {Schoelkopf},\ and\ \citenamefont
  {Devoret}}]{Minev2018-EPR}%
  \BibitemOpen
  \bibfield  {author} {\bibinfo {author} {\bibfnamefont {Z.}~\bibnamefont
  {Minev}}, \bibinfo {author} {\bibfnamefont {Z.}~\bibnamefont {Leghtas}},
  \bibinfo {author} {\bibfnamefont {S.}~\bibnamefont {Mudhada}}, \bibinfo
  {author} {\bibfnamefont {I.}~\bibnamefont {Pop}}, \bibinfo {author}
  {\bibfnamefont {L.}~\bibnamefont {Christakis}}, \bibinfo {author}
  {\bibfnamefont {R.}~\bibnamefont {Schoelkopf}}, \ and\ \bibinfo {author}
  {\bibfnamefont {M.}~\bibnamefont {Devoret}}} (\bibinfo {year} {2018}),\
  \bibfield  {title} {\enquote {\bibinfo {title} {{Energy Participation
  Approach to the Design of Quantum Josephson Circuits}},}\ }\href@noop {}
  {\bibinfo  {journal} {(in preparation)}\ }\BibitemShut {NoStop}%
\bibitem [{\citenamefont {Minev}\ \emph {et~al.}(2014)\citenamefont {Minev},
  \citenamefont {Pop}, \citenamefont {Serniak},\ and\ \citenamefont
  {Devoret}}]{Minev2014-APSMM}%
  \BibitemOpen
\bibfield  {journal} {  }\bibfield  {author} {\bibinfo {author} {\bibfnamefont
  {Z.}~\bibnamefont {Minev}}, \bibinfo {author} {\bibfnamefont
  {I.}~\bibnamefont {Pop}}, \bibinfo {author} {\bibfnamefont {K.}~\bibnamefont
  {Serniak}}, \ and\ \bibinfo {author} {\bibfnamefont {M.}~\bibnamefont
  {Devoret}}} (\bibinfo {year} {2014}),\ \bibfield  {title} {\enquote {\bibinfo
  {title} {{Qubit coupling to superconducting whispering gallery mode
  resonator}},}\ }in\ \href@noop {} {\emph {\bibinfo {booktitle} {APS March
  Meeting Abstracts}}},\ Vol.\ \bibinfo {volume} {2014},\ p.\ \bibinfo {pages}
  {M28.013}\BibitemShut {NoStop}%
\bibitem [{\citenamefont {Minev}\ \emph
  {et~al.}(2013{\natexlab{a}})\citenamefont {Minev}, \citenamefont {Pop},\ and\
  \citenamefont {Devoret}}]{Minev2013}%
  \BibitemOpen
  \bibfield  {author} {\bibinfo {author} {\bibfnamefont {Z.}~\bibnamefont
  {Minev}}, \bibinfo {author} {\bibfnamefont {I.~M.}\ \bibnamefont {Pop}}, \
  and\ \bibinfo {author} {\bibfnamefont {M.~H.}\ \bibnamefont {Devoret}}}
  (\bibinfo {year} {2013}{\natexlab{a}}),\ \bibfield  {title} {\enquote
  {\bibinfo {title} {{Planar superconducting whispering gallery mode
  resonators}},}\ }\href {\doibase 10.1063/1.4824201} {\bibfield  {journal}
  {\bibinfo  {journal} {Applied Physics Letters}\ }\textbf {\bibinfo {volume}
  {103}}~(\bibinfo {number} {14}),\ \bibinfo {pages} {142604}},\ \Eprint
  {http://arxiv.org/abs/arXiv:1308.1743v1} {arXiv:arXiv:1308.1743v1}
  \BibitemShut {NoStop}%
\bibitem [{\citenamefont {Minev}\ \emph
  {et~al.}(2015{\natexlab{a}})\citenamefont {Minev}, \citenamefont {Serniak},
  \citenamefont {Pop}, \citenamefont {Leghtas}, \citenamefont {Sliwa},
  \citenamefont {Frunzio}, \citenamefont {Schoelkopf},\ and\ \citenamefont
  {Devoret}}]{Minev2015-APSMM}%
  \BibitemOpen
  \bibfield  {author} {\bibinfo {author} {\bibfnamefont {Z.}~\bibnamefont
  {Minev}}, \bibinfo {author} {\bibfnamefont {K.}~\bibnamefont {Serniak}},
  \bibinfo {author} {\bibfnamefont {I.}~\bibnamefont {Pop}}, \bibinfo {author}
  {\bibfnamefont {Z.}~\bibnamefont {Leghtas}}, \bibinfo {author} {\bibfnamefont
  {K.}~\bibnamefont {Sliwa}}, \bibinfo {author} {\bibfnamefont
  {L.}~\bibnamefont {Frunzio}}, \bibinfo {author} {\bibfnamefont
  {R.}~\bibnamefont {Schoelkopf}}, \ and\ \bibinfo {author} {\bibfnamefont
  {M.}~\bibnamefont {Devoret}}} (\bibinfo {year} {2015}{\natexlab{a}}),\
  \bibfield  {title} {\enquote {\bibinfo {title} {{Coherences of transmon
  qubits embedded in superconducting whispering gallery mode resonators}},}\
  }in\ \href@noop {} {\emph {\bibinfo {booktitle} {APS March Meeting
  Abstracts}}},\ Vol.\ \bibinfo {volume} {2015},\ p.\ \bibinfo {pages}
  {Y39.009}\BibitemShut {NoStop}%
\bibitem [{\citenamefont {Minev}\ \emph {et~al.}(2016)\citenamefont {Minev},
  \citenamefont {Serniak}, \citenamefont {Pop}, \citenamefont {Leghtas},
  \citenamefont {Sliwa}, \citenamefont {Hatridge}, \citenamefont {Frunzio},
  \citenamefont {Schoelkopf},\ and\ \citenamefont {Devoret}}]{Minev2016}%
  \BibitemOpen
  \bibfield  {author} {\bibinfo {author} {\bibfnamefont {Z.}~\bibnamefont
  {Minev}}, \bibinfo {author} {\bibfnamefont {K.}~\bibnamefont {Serniak}},
  \bibinfo {author} {\bibfnamefont {I.~M.}\ \bibnamefont {Pop}}, \bibinfo
  {author} {\bibfnamefont {Z.}~\bibnamefont {Leghtas}}, \bibinfo {author}
  {\bibfnamefont {K.}~\bibnamefont {Sliwa}}, \bibinfo {author} {\bibfnamefont
  {M.}~\bibnamefont {Hatridge}}, \bibinfo {author} {\bibfnamefont
  {L.}~\bibnamefont {Frunzio}}, \bibinfo {author} {\bibfnamefont {R.~J.}\
  \bibnamefont {Schoelkopf}}, \ and\ \bibinfo {author} {\bibfnamefont {M.~H.}\
  \bibnamefont {Devoret}}} (\bibinfo {year} {2016}),\ \bibfield  {title}
  {\enquote {\bibinfo {title} {{Planar Multilayer Circuit Quantum
  Electrodynamics}},}\ }\href {\doibase 10.1103/PhysRevApplied.5.044021}
  {\bibfield  {journal} {\bibinfo  {journal} {Physical Review Applied}\
  }\textbf {\bibinfo {volume} {5}}~(\bibinfo {number} {4}),\ \bibinfo {pages}
  {044021}},\ \Eprint {http://arxiv.org/abs/1509.01619} {arXiv:1509.01619}
  \BibitemShut {NoStop}%
\bibitem [{\citenamefont {Minev}\ \emph
  {et~al.}(2013{\natexlab{b}})\citenamefont {Minev}, \citenamefont {Pop},
  \citenamefont {Kwok},\ and\ \citenamefont {Devoret}}]{Minev2013-APSMM}%
  \BibitemOpen
  \bibfield  {author} {\bibinfo {author} {\bibfnamefont {Z.~K.}\ \bibnamefont
  {Minev}}, \bibinfo {author} {\bibfnamefont {I.}~\bibnamefont {Pop}}, \bibinfo
  {author} {\bibfnamefont {D.}~\bibnamefont {Kwok}}, \ and\ \bibinfo {author}
  {\bibfnamefont {M.}~\bibnamefont {Devoret}}} (\bibinfo {year}
  {2013}{\natexlab{b}}),\ \bibfield  {title} {\enquote {\bibinfo {title}
  {{Coherence of Superconducting Whispering Gallery Resonators}},}\ }in\
  \href@noop {} {\emph {\bibinfo {booktitle} {APS March Meeting Abstracts}}},\
  Vol.\ \bibinfo {volume} {2013},\ p.\ \bibinfo {pages} {C25.003}\BibitemShut
  {NoStop}%
\bibitem [{\citenamefont {Minev}\ \emph {et~al.}(2012)\citenamefont {Minev},
  \citenamefont {Pop}, \citenamefont {Masluk}, \citenamefont {Kamal},\ and\
  \citenamefont {Devoret}}]{Minev2012-APSMM}%
  \BibitemOpen
  \bibfield  {author} {\bibinfo {author} {\bibfnamefont {Z.~K.}\ \bibnamefont
  {Minev}}, \bibinfo {author} {\bibfnamefont {I.}~\bibnamefont {Pop}}, \bibinfo
  {author} {\bibfnamefont {N.}~\bibnamefont {Masluk}}, \bibinfo {author}
  {\bibfnamefont {A.}~\bibnamefont {Kamal}}, \ and\ \bibinfo {author}
  {\bibfnamefont {M.}~\bibnamefont {Devoret}}} (\bibinfo {year} {2012}),\
  \bibfield  {title} {\enquote {\bibinfo {title} {{Ground Capacitance in
  Josephson Junction Arrays}},}\ }in\ \href@noop {} {\emph {\bibinfo
  {booktitle} {APS March Meeting Abstracts}}},\ Vol.\ \bibinfo {volume}
  {2012},\ p.\ \bibinfo {pages} {K1.097}\BibitemShut {NoStop}%
\bibitem [{\citenamefont {Minev}\ \emph
  {et~al.}(2015{\natexlab{b}})\citenamefont {Minev}, \citenamefont {Serniak},
  \citenamefont {Pop}, \citenamefont {Chu}, \citenamefont {Brecht},
  \citenamefont {Frunzio}, \citenamefont {Devoret},\ and\ \citenamefont
  {Schoelkopf}}]{Minev2015-patent}%
  \BibitemOpen
  \bibfield  {author} {\bibinfo {author} {\bibfnamefont {Z.~K.}\ \bibnamefont
  {Minev}}, \bibinfo {author} {\bibfnamefont {K.}~\bibnamefont {Serniak}},
  \bibinfo {author} {\bibfnamefont {I.~M.}\ \bibnamefont {Pop}}, \bibinfo
  {author} {\bibfnamefont {Y.}~\bibnamefont {Chu}}, \bibinfo {author}
  {\bibfnamefont {T.}~\bibnamefont {Brecht}}, \bibinfo {author} {\bibfnamefont
  {L.}~\bibnamefont {Frunzio}}, \bibinfo {author} {\bibfnamefont {M.~H.}\
  \bibnamefont {Devoret}}, \ and\ \bibinfo {author} {\bibfnamefont {R.~J.}\
  \bibnamefont {Schoelkopf}}} (\bibinfo {year} {2015}{\natexlab{b}}),\
  \bibfield  {title} {\enquote {\bibinfo {title} {{Techniques for coupling
  planar qubits to non-planar resonators and related systems and methods}},}\
  }\href {https://patents.google.com/patent/EP3262573A1/en} {\bibinfo
  {journal} {International Patent Publication No. WO/2016/138395; Priority
  Date: Feb. 27, 2015}\ }\BibitemShut {NoStop}%
\bibitem [{\citenamefont {Mirrahimi}\ \emph {et~al.}(2014)\citenamefont
  {Mirrahimi}, \citenamefont {Leghtas}, \citenamefont {Albert}, \citenamefont
  {Touzard}, \citenamefont {Schoelkopf}, \citenamefont {Jiang},\ and\
  \citenamefont {Devoret}}]{Mirrahimi2014}%
  \BibitemOpen
\bibfield  {journal} {  }\bibfield  {author} {\bibinfo {author} {\bibfnamefont
  {M.}~\bibnamefont {Mirrahimi}}, \bibinfo {author} {\bibfnamefont
  {Z.}~\bibnamefont {Leghtas}}, \bibinfo {author} {\bibfnamefont {V.~V.}\
  \bibnamefont {Albert}}, \bibinfo {author} {\bibfnamefont {S.}~\bibnamefont
  {Touzard}}, \bibinfo {author} {\bibfnamefont {R.~J.}\ \bibnamefont
  {Schoelkopf}}, \bibinfo {author} {\bibfnamefont {L.}~\bibnamefont {Jiang}}, \
  and\ \bibinfo {author} {\bibfnamefont {M.~H.}\ \bibnamefont {Devoret}}}
  (\bibinfo {year} {2014}),\ \bibfield  {title} {\enquote {\bibinfo {title}
  {{Dynamically protected cat-qubits: a new paradigm for universal quantum
  computation}},}\ }\href {\doibase 10.1088/1367-2630/16/4/045014} {\bibfield
  {journal} {\bibinfo  {journal} {New Journal of Physics}\ }\textbf {\bibinfo
  {volume} {16}}~(\bibinfo {number} {4}),\ \bibinfo {pages}
  {045014}}\BibitemShut {NoStop}%
\bibitem [{\citenamefont {Misra}\ and\ \citenamefont
  {Sudarshan}(1977)}]{Misra1977}%
  \BibitemOpen
  \bibfield  {author} {\bibinfo {author} {\bibfnamefont {B.}~\bibnamefont
  {Misra}}, \ and\ \bibinfo {author} {\bibfnamefont {E.~C.~G.}\ \bibnamefont
  {Sudarshan}}} (\bibinfo {year} {1977}),\ \bibfield  {title} {\enquote
  {\bibinfo {title} {{The Zeno's paradox in quantum theory}},}\ }\href
  {\doibase 10.1063/1.523304} {\bibfield  {journal} {\bibinfo  {journal}
  {Journal of Mathematical Physics}\ }\textbf {\bibinfo {volume}
  {18}}~(\bibinfo {number} {4}),\ \bibinfo {pages} {756--763}}\BibitemShut
  {NoStop}%
\bibitem [{\citenamefont {Mundhada}\ \emph {et~al.}(2018)\citenamefont
  {Mundhada}, \citenamefont {Grimm}, \citenamefont {Venkatraman}, \citenamefont
  {Minev}, \citenamefont {Touzard}, \citenamefont {Frattini}, \citenamefont
  {Sivak}, \citenamefont {Sliwa}, \citenamefont {Reinhold}, \citenamefont
  {Shankar}, \citenamefont {Mirrahimi},\ and\ \citenamefont
  {Devoret}}]{Mundhada2018}%
  \BibitemOpen
  \bibfield  {author} {\bibinfo {author} {\bibfnamefont {S.}~\bibnamefont
  {Mundhada}}, \bibinfo {author} {\bibfnamefont {A.}~\bibnamefont {Grimm}},
  \bibinfo {author} {\bibfnamefont {J.}~\bibnamefont {Venkatraman}}, \bibinfo
  {author} {\bibfnamefont {Z.}~\bibnamefont {Minev}}, \bibinfo {author}
  {\bibfnamefont {S.}~\bibnamefont {Touzard}}, \bibinfo {author} {\bibfnamefont
  {N.}~\bibnamefont {Frattini}}, \bibinfo {author} {\bibfnamefont
  {V.}~\bibnamefont {Sivak}}, \bibinfo {author} {\bibfnamefont
  {K.}~\bibnamefont {Sliwa}}, \bibinfo {author} {\bibfnamefont
  {P.}~\bibnamefont {Reinhold}}, \bibinfo {author} {\bibfnamefont
  {S.}~\bibnamefont {Shankar}}, \bibinfo {author} {\bibfnamefont
  {M.}~\bibnamefont {Mirrahimi}}, \ and\ \bibinfo {author} {\bibfnamefont
  {M.}~\bibnamefont {Devoret}}} (\bibinfo {year} {2018}),\ \bibfield  {title}
  {\enquote {\bibinfo {title} {{Experimental implementation of a Raman-assisted
  six-quanta process}},}\ }\href@noop {} {\bibfield  {journal} {\bibinfo
  {journal} {arXiv e-prints}\ ,\ \bibinfo {pages} {arXiv:1811.06589}}}\Eprint
  {http://arxiv.org/abs/1811.06589} {arXiv:1811.06589 [quant-ph]} \BibitemShut
  {NoStop}%
\bibitem [{\citenamefont {Mundhada}\ \emph {et~al.}(2017)\citenamefont
  {Mundhada}, \citenamefont {Grimm}, \citenamefont {Touzard}, \citenamefont
  {Vool}, \citenamefont {Shankar}, \citenamefont {Devoret},\ and\ \citenamefont
  {Mirrahimi}}]{Mundhada2017}%
  \BibitemOpen
  \bibfield  {author} {\bibinfo {author} {\bibfnamefont {S.~O.}\ \bibnamefont
  {Mundhada}}, \bibinfo {author} {\bibfnamefont {A.}~\bibnamefont {Grimm}},
  \bibinfo {author} {\bibfnamefont {S.}~\bibnamefont {Touzard}}, \bibinfo
  {author} {\bibfnamefont {U.}~\bibnamefont {Vool}}, \bibinfo {author}
  {\bibfnamefont {S.}~\bibnamefont {Shankar}}, \bibinfo {author} {\bibfnamefont
  {M.~H.}\ \bibnamefont {Devoret}}, \ and\ \bibinfo {author} {\bibfnamefont
  {M.}~\bibnamefont {Mirrahimi}}} (\bibinfo {year} {2017}),\ \bibfield  {title}
  {\enquote {\bibinfo {title} {{Generating higher-order quantum dissipation
  from lower-order parametric processes}},}\ }\href {\doibase
  10.1088/2058-9565/aa6e9d} {\bibfield  {journal} {\bibinfo  {journal} {Quantum
  Science and Technology}\ }\textbf {\bibinfo {volume} {2}}~(\bibinfo {number}
  {2}),\ \bibinfo {pages} {024005}}\BibitemShut {NoStop}%
\bibitem [{\citenamefont {Murch}\ \emph {et~al.}(2012)\citenamefont {Murch},
  \citenamefont {Vool}, \citenamefont {Zhou}, \citenamefont {Weber},
  \citenamefont {Girvin},\ and\ \citenamefont {Siddiqi}}]{Murch2012-bath-eng}%
  \BibitemOpen
  \bibfield  {author} {\bibinfo {author} {\bibfnamefont {K.~W.}\ \bibnamefont
  {Murch}}, \bibinfo {author} {\bibfnamefont {U.}~\bibnamefont {Vool}},
  \bibinfo {author} {\bibfnamefont {D.}~\bibnamefont {Zhou}}, \bibinfo {author}
  {\bibfnamefont {S.~J.}\ \bibnamefont {Weber}}, \bibinfo {author}
  {\bibfnamefont {S.~M.}\ \bibnamefont {Girvin}}, \ and\ \bibinfo {author}
  {\bibfnamefont {I.}~\bibnamefont {Siddiqi}}} (\bibinfo {year} {2012}),\
  \bibfield  {title} {\enquote {\bibinfo {title} {{Cavity-Assisted Quantum Bath
  Engineering}},}\ }\href {\doibase 10.1103/PhysRevLett.109.183602} {\bibfield
  {journal} {\bibinfo  {journal} {Physical Review Letters}\ }\textbf {\bibinfo
  {volume} {109}}~(\bibinfo {number} {18}),\ \bibinfo {pages}
  {183602}}\BibitemShut {NoStop}%
\bibitem [{\citenamefont {Murch}\ \emph
  {et~al.}(2013{\natexlab{a}})\citenamefont {Murch}, \citenamefont {Weber},
  \citenamefont {Beck}, \citenamefont {Ginossar},\ and\ \citenamefont
  {Siddiqi}}]{Murch2013}%
  \BibitemOpen
  \bibfield  {author} {\bibinfo {author} {\bibfnamefont {K.~W.}\ \bibnamefont
  {Murch}}, \bibinfo {author} {\bibfnamefont {S.~J.}\ \bibnamefont {Weber}},
  \bibinfo {author} {\bibfnamefont {K.~M.}\ \bibnamefont {Beck}}, \bibinfo
  {author} {\bibfnamefont {E.}~\bibnamefont {Ginossar}}, \ and\ \bibinfo
  {author} {\bibfnamefont {I.}~\bibnamefont {Siddiqi}}} (\bibinfo {year}
  {2013}{\natexlab{a}}),\ \bibfield  {title} {\enquote {\bibinfo {title}
  {{Reduction of the radiative decay of atomic coherence in squeezed
  vacuum}},}\ }\href {\doibase 10.1038/nature12264} {\bibfield  {journal}
  {\bibinfo  {journal} {Nature}\ }\textbf {\bibinfo {volume} {499}}~(\bibinfo
  {number} {7456}),\ \bibinfo {pages} {62--65}},\ \Eprint
  {http://arxiv.org/abs/1301.6276} {arXiv:1301.6276} \BibitemShut {NoStop}%
\bibitem [{\citenamefont {Murch}\ \emph
  {et~al.}(2013{\natexlab{b}})\citenamefont {Murch}, \citenamefont {Weber},
  \citenamefont {Macklin},\ and\ \citenamefont {Siddiqi}}]{Murch2013a}%
  \BibitemOpen
  \bibfield  {author} {\bibinfo {author} {\bibfnamefont {K.~W.}\ \bibnamefont
  {Murch}}, \bibinfo {author} {\bibfnamefont {S.~J.}\ \bibnamefont {Weber}},
  \bibinfo {author} {\bibfnamefont {C.}~\bibnamefont {Macklin}}, \ and\
  \bibinfo {author} {\bibfnamefont {I.}~\bibnamefont {Siddiqi}}} (\bibinfo
  {year} {2013}{\natexlab{b}}),\ \bibfield  {title} {\enquote {\bibinfo {title}
  {{Observing single quantum trajectories of a superconducting quantum bit}},}\
  }\href {\doibase 10.1038/nature12539} {\bibfield  {journal} {\bibinfo
  {journal} {Nature}\ }\textbf {\bibinfo {volume} {502}}~(\bibinfo {number}
  {7470}),\ \bibinfo {pages} {211--214}},\ \Eprint
  {http://arxiv.org/abs/1305.7270} {arXiv:1305.7270} \BibitemShut {NoStop}%
\bibitem [{\citenamefont {Naghiloo}\ \emph {et~al.}(2016)\citenamefont
  {Naghiloo}, \citenamefont {Foroozani}, \citenamefont {Tan}, \citenamefont
  {Jadbabaie},\ and\ \citenamefont {Murch}}]{Naghiloo2016}%
  \BibitemOpen
  \bibfield  {author} {\bibinfo {author} {\bibfnamefont {M.}~\bibnamefont
  {Naghiloo}}, \bibinfo {author} {\bibfnamefont {N.}~\bibnamefont {Foroozani}},
  \bibinfo {author} {\bibfnamefont {D.}~\bibnamefont {Tan}}, \bibinfo {author}
  {\bibfnamefont {A.}~\bibnamefont {Jadbabaie}}, \ and\ \bibinfo {author}
  {\bibfnamefont {K.~W.}\ \bibnamefont {Murch}}} (\bibinfo {year} {2016}),\
  \bibfield  {title} {\enquote {\bibinfo {title} {{Mapping quantum state
  dynamics in spontaneous emission}},}\ }\href {\doibase 10.1038/ncomms11527}
  {\bibfield  {journal} {\bibinfo  {journal} {Nature Communications}\ }\textbf
  {\bibinfo {volume} {7}},\ \bibinfo {pages} {11527}}\BibitemShut {NoStop}%
\bibitem [{\citenamefont {Naghiloo}\ \emph {et~al.}(2017)\citenamefont
  {Naghiloo}, \citenamefont {Tan}, \citenamefont {Harrington}, \citenamefont
  {Alonso}, \citenamefont {Lutz}, \citenamefont {Romito},\ and\ \citenamefont
  {Murch}}]{Naghiloo2017-thermo}%
  \BibitemOpen
  \bibfield  {author} {\bibinfo {author} {\bibfnamefont {M.}~\bibnamefont
  {Naghiloo}}, \bibinfo {author} {\bibfnamefont {D.}~\bibnamefont {Tan}},
  \bibinfo {author} {\bibfnamefont {P.~M.}\ \bibnamefont {Harrington}},
  \bibinfo {author} {\bibfnamefont {J.~J.}\ \bibnamefont {Alonso}}, \bibinfo
  {author} {\bibfnamefont {E.}~\bibnamefont {Lutz}}, \bibinfo {author}
  {\bibfnamefont {A.}~\bibnamefont {Romito}}, \ and\ \bibinfo {author}
  {\bibfnamefont {K.~W.}\ \bibnamefont {Murch}}} (\bibinfo {year} {2017}),\
  \bibfield  {title} {\enquote {\bibinfo {title} {{Thermodynamics along
  individual trajectories of a quantum bit}},}\ }\href
  {http://arxiv.org/abs/1703.05885} {\ }\Eprint
  {http://arxiv.org/abs/1703.05885} {arXiv:1703.05885} \BibitemShut {NoStop}%
\bibitem [{\citenamefont {Nagourney}\ \emph {et~al.}(1986)\citenamefont
  {Nagourney}, \citenamefont {Sandberg},\ and\ \citenamefont
  {Dehmelt}}]{Nagourney1986}%
  \BibitemOpen
  \bibfield  {author} {\bibinfo {author} {\bibfnamefont {W.}~\bibnamefont
  {Nagourney}}, \bibinfo {author} {\bibfnamefont {J.}~\bibnamefont {Sandberg}},
  \ and\ \bibinfo {author} {\bibfnamefont {H.}~\bibnamefont {Dehmelt}}}
  (\bibinfo {year} {1986}),\ \bibfield  {title} {\enquote {\bibinfo {title}
  {{Shelved optical electron amplifier: Observation of quantum jumps}},}\
  }\href {\doibase 10.1103/PhysRevLett.56.2797} {\bibfield  {journal} {\bibinfo
   {journal} {Physical Review Letters}\ }\textbf {\bibinfo {volume}
  {56}}~(\bibinfo {number} {26}),\ \bibinfo {pages} {2797--2799}}\BibitemShut
  {NoStop}%
\bibitem [{\citenamefont {Narla}\ \emph {et~al.}(2016)\citenamefont {Narla},
  \citenamefont {Shankar}, \citenamefont {Hatridge}, \citenamefont {Leghtas},
  \citenamefont {Sliwa}, \citenamefont {Zalys-Geller}, \citenamefont
  {Mundhada}, \citenamefont {Pfaff}, \citenamefont {Frunzio}, \citenamefont
  {Schoelkopf},\ and\ \citenamefont {Devoret}}]{Narla2016}%
  \BibitemOpen
  \bibfield  {author} {\bibinfo {author} {\bibfnamefont {A.}~\bibnamefont
  {Narla}}, \bibinfo {author} {\bibfnamefont {S.}~\bibnamefont {Shankar}},
  \bibinfo {author} {\bibfnamefont {M.}~\bibnamefont {Hatridge}}, \bibinfo
  {author} {\bibfnamefont {Z.}~\bibnamefont {Leghtas}}, \bibinfo {author}
  {\bibfnamefont {K.~M.}\ \bibnamefont {Sliwa}}, \bibinfo {author}
  {\bibfnamefont {E.}~\bibnamefont {Zalys-Geller}}, \bibinfo {author}
  {\bibfnamefont {S.~O.}\ \bibnamefont {Mundhada}}, \bibinfo {author}
  {\bibfnamefont {W.}~\bibnamefont {Pfaff}}, \bibinfo {author} {\bibfnamefont
  {L.}~\bibnamefont {Frunzio}}, \bibinfo {author} {\bibfnamefont {R.~J.}\
  \bibnamefont {Schoelkopf}}, \ and\ \bibinfo {author} {\bibfnamefont {M.~H.}\
  \bibnamefont {Devoret}}} (\bibinfo {year} {2016}),\ \bibfield  {title}
  {\enquote {\bibinfo {title} {{Robust Concurrent Remote Entanglement Between
  Two Superconducting Qubits}},}\ }\href {\doibase 10.1103/PhysRevX.6.031036}
  {\bibfield  {journal} {\bibinfo  {journal} {Physical Review X}\ }\textbf
  {\bibinfo {volume} {6}}~(\bibinfo {number} {3}),\ \bibinfo {pages}
  {031036}}\BibitemShut {NoStop}%
\bibitem [{\citenamefont {Neeley}\ \emph {et~al.}(2008)\citenamefont {Neeley},
  \citenamefont {Ansmann}, \citenamefont {Bialczak}, \citenamefont {Hofheinz},
  \citenamefont {Katz}, \citenamefont {Lucero}, \citenamefont {O'Connell},
  \citenamefont {Wang}, \citenamefont {Cleland},\ and\ \citenamefont
  {Martinis}}]{Neeley2008-Purcell}%
  \BibitemOpen
  \bibfield  {author} {\bibinfo {author} {\bibfnamefont {M.}~\bibnamefont
  {Neeley}}, \bibinfo {author} {\bibfnamefont {M.}~\bibnamefont {Ansmann}},
  \bibinfo {author} {\bibfnamefont {R.~C.}\ \bibnamefont {Bialczak}}, \bibinfo
  {author} {\bibfnamefont {M.}~\bibnamefont {Hofheinz}}, \bibinfo {author}
  {\bibfnamefont {N.}~\bibnamefont {Katz}}, \bibinfo {author} {\bibfnamefont
  {E.}~\bibnamefont {Lucero}}, \bibinfo {author} {\bibfnamefont
  {A.}~\bibnamefont {O'Connell}}, \bibinfo {author} {\bibfnamefont
  {H.}~\bibnamefont {Wang}}, \bibinfo {author} {\bibfnamefont {A.~N.}\
  \bibnamefont {Cleland}}, \ and\ \bibinfo {author} {\bibfnamefont {J.~M.}\
  \bibnamefont {Martinis}}} (\bibinfo {year} {2008}),\ \bibfield  {title}
  {\enquote {\bibinfo {title} {{Transformed dissipation in superconducting
  quantum circuits}},}\ }\href {\doibase 10.1103/PhysRevB.77.180508} {\bibfield
   {journal} {\bibinfo  {journal} {Physical Review B}\ }\textbf {\bibinfo
  {volume} {77}}~(\bibinfo {number} {18}),\ \bibinfo {pages} {180508}},\
  \Eprint {http://arxiv.org/abs/0801.2994} {arXiv:0801.2994} \BibitemShut
  {NoStop}%
\bibitem [{\citenamefont {von Neumann}(1932)}]{VonNeumann1932}%
  \BibitemOpen
  \bibfield  {author} {\bibinfo {author} {\bibfnamefont {J.}~\bibnamefont {von
  Neumann}}} (\bibinfo {year} {1932}),\ \href {\doibase
  10.1007/978-3-642-61409-5} {\emph {\bibinfo {title} {{Mathematische
  Grundlagen der Quantenmechanik (Mathematical Foundations of Quantum
  Mechanics)}}}}\ (\bibinfo  {publisher} {Springer},\ \bibinfo {address}
  {Berlin})\BibitemShut {NoStop}%
\bibitem [{\citenamefont {Neumann}\ \emph {et~al.}(2010)\citenamefont
  {Neumann}, \citenamefont {Beck}, \citenamefont {Steiner}, \citenamefont
  {Rempp}, \citenamefont {Fedder}, \citenamefont {Hemmer}, \citenamefont
  {Wrachtrup},\ and\ \citenamefont {Jelezko}}]{Neumann2010}%
  \BibitemOpen
  \bibfield  {author} {\bibinfo {author} {\bibfnamefont {P.}~\bibnamefont
  {Neumann}}, \bibinfo {author} {\bibfnamefont {J.}~\bibnamefont {Beck}},
  \bibinfo {author} {\bibfnamefont {M.}~\bibnamefont {Steiner}}, \bibinfo
  {author} {\bibfnamefont {F.}~\bibnamefont {Rempp}}, \bibinfo {author}
  {\bibfnamefont {H.}~\bibnamefont {Fedder}}, \bibinfo {author} {\bibfnamefont
  {P.~R.}\ \bibnamefont {Hemmer}}, \bibinfo {author} {\bibfnamefont
  {J.}~\bibnamefont {Wrachtrup}}, \ and\ \bibinfo {author} {\bibfnamefont
  {F.}~\bibnamefont {Jelezko}}} (\bibinfo {year} {2010}),\ \bibfield  {title}
  {\enquote {\bibinfo {title} {{Single-Shot Readout of a Single Nuclear
  Spin}},}\ }\href {\doibase 10.1126/science.1189075} {\bibfield  {journal}
  {\bibinfo  {journal} {Science}\ }\textbf {\bibinfo {volume} {329}}~(\bibinfo
  {number} {5991}),\ \bibinfo {pages} {542--544}}\BibitemShut {NoStop}%
\bibitem [{\citenamefont {Nielsen}\ and\ \citenamefont
  {Chuang}(2010)}]{NielsenChuangBook}%
  \BibitemOpen
  \bibfield  {author} {\bibinfo {author} {\bibfnamefont {M.~A.}\ \bibnamefont
  {Nielsen}}, \ and\ \bibinfo {author} {\bibfnamefont {I.~L.}\ \bibnamefont
  {Chuang}}} (\bibinfo {year} {2010}),\ \href {\doibase
  10.1017/CBO9780511976667} {\emph {\bibinfo {title} {{Quantum Computation and
  Quantum Information}}}}\ (\bibinfo  {publisher} {Cambridge University
  Press},\ \bibinfo {address} {Cambridge})\BibitemShut {NoStop}%
\bibitem [{\citenamefont {Nigg}\ \emph {et~al.}(2012)\citenamefont {Nigg},
  \citenamefont {Paik}, \citenamefont {Vlastakis}, \citenamefont {Kirchmair},
  \citenamefont {Shankar}, \citenamefont {Frunzio}, \citenamefont {Devoret},
  \citenamefont {Schoelkopf},\ and\ \citenamefont {Girvin}}]{Nigg2012}%
  \BibitemOpen
  \bibfield  {author} {\bibinfo {author} {\bibfnamefont {S.~E.}\ \bibnamefont
  {Nigg}}, \bibinfo {author} {\bibfnamefont {H.}~\bibnamefont {Paik}}, \bibinfo
  {author} {\bibfnamefont {B.}~\bibnamefont {Vlastakis}}, \bibinfo {author}
  {\bibfnamefont {G.}~\bibnamefont {Kirchmair}}, \bibinfo {author}
  {\bibfnamefont {S.}~\bibnamefont {Shankar}}, \bibinfo {author} {\bibfnamefont
  {L.}~\bibnamefont {Frunzio}}, \bibinfo {author} {\bibfnamefont {M.~H.}\
  \bibnamefont {Devoret}}, \bibinfo {author} {\bibfnamefont {R.~J.}\
  \bibnamefont {Schoelkopf}}, \ and\ \bibinfo {author} {\bibfnamefont {S.~M.}\
  \bibnamefont {Girvin}}} (\bibinfo {year} {2012}),\ \bibfield  {title}
  {\enquote {\bibinfo {title} {{Black-Box Superconducting Circuit
  Quantization}},}\ }\href {\doibase 10.1103/PhysRevLett.108.240502} {\bibfield
   {journal} {\bibinfo  {journal} {Physical Review Letters}\ }\textbf {\bibinfo
  {volume} {108}}~(\bibinfo {number} {24}),\ \bibinfo {pages}
  {240502}}\BibitemShut {NoStop}%
\bibitem [{\citenamefont {Novikov}\ \emph {et~al.}(2015)\citenamefont
  {Novikov}, \citenamefont {Sweeney}, \citenamefont {Robinson}, \citenamefont
  {Premaratne}, \citenamefont {Suri}, \citenamefont {Wellstood},\ and\
  \citenamefont {Palmer}}]{Novikov2015}%
  \BibitemOpen
  \bibfield  {author} {\bibinfo {author} {\bibfnamefont {S.}~\bibnamefont
  {Novikov}}, \bibinfo {author} {\bibfnamefont {T.}~\bibnamefont {Sweeney}},
  \bibinfo {author} {\bibfnamefont {J.~E.}\ \bibnamefont {Robinson}}, \bibinfo
  {author} {\bibfnamefont {S.~P.}\ \bibnamefont {Premaratne}}, \bibinfo
  {author} {\bibfnamefont {B.}~\bibnamefont {Suri}}, \bibinfo {author}
  {\bibfnamefont {F.~C.}\ \bibnamefont {Wellstood}}, \ and\ \bibinfo {author}
  {\bibfnamefont {B.~S.}\ \bibnamefont {Palmer}}} (\bibinfo {year} {2015}),\
  \bibfield  {title} {\enquote {\bibinfo {title} {{Raman coherence in a circuit
  quantum electrodynamics lambda system}},}\ }\href {\doibase
  10.1038/nphys3537} {\bibfield  {journal} {\bibinfo  {journal} {Nature
  Physics}\ }\textbf {\bibinfo {volume} {12}}~(\bibinfo {number} {1}),\
  \bibinfo {pages} {75--79}}\BibitemShut {NoStop}%
\bibitem [{\citenamefont {Ofek}\ \emph {et~al.}(2016)\citenamefont {Ofek},
  \citenamefont {Petrenko}, \citenamefont {Heeres}, \citenamefont {Reinhold},
  \citenamefont {Leghtas}, \citenamefont {Vlastakis}, \citenamefont {Liu},
  \citenamefont {Frunzio}, \citenamefont {Girvin}, \citenamefont {Jiang},
  \citenamefont {Mirrahimi}, \citenamefont {Devoret},\ and\ \citenamefont
  {Schoelkopf}}]{Ofek2016}%
  \BibitemOpen
  \bibfield  {author} {\bibinfo {author} {\bibfnamefont {N.}~\bibnamefont
  {Ofek}}, \bibinfo {author} {\bibfnamefont {A.}~\bibnamefont {Petrenko}},
  \bibinfo {author} {\bibfnamefont {R.}~\bibnamefont {Heeres}}, \bibinfo
  {author} {\bibfnamefont {P.}~\bibnamefont {Reinhold}}, \bibinfo {author}
  {\bibfnamefont {Z.}~\bibnamefont {Leghtas}}, \bibinfo {author} {\bibfnamefont
  {B.}~\bibnamefont {Vlastakis}}, \bibinfo {author} {\bibfnamefont
  {Y.}~\bibnamefont {Liu}}, \bibinfo {author} {\bibfnamefont {L.}~\bibnamefont
  {Frunzio}}, \bibinfo {author} {\bibfnamefont {S.~M.}\ \bibnamefont {Girvin}},
  \bibinfo {author} {\bibfnamefont {L.}~\bibnamefont {Jiang}}, \bibinfo
  {author} {\bibfnamefont {M.}~\bibnamefont {Mirrahimi}}, \bibinfo {author}
  {\bibfnamefont {M.~H.}\ \bibnamefont {Devoret}}, \ and\ \bibinfo {author}
  {\bibfnamefont {R.~J.}\ \bibnamefont {Schoelkopf}}} (\bibinfo {year}
  {2016}),\ \bibfield  {title} {\enquote {\bibinfo {title} {{Demonstrating
  Quantum Error Correction that Extends the Lifetime of Quantum
  Information}},}\ }\href {\doibase 10.1038/nature18949} {\bibfield  {journal}
  {\bibinfo  {journal} {Nature}\ }\textbf {\bibinfo {volume} {536}}~(\bibinfo
  {number} {7617}),\ \bibinfo {pages} {441--445}},\ \Eprint
  {http://arxiv.org/abs/1602.04768} {arXiv:1602.04768} \BibitemShut {NoStop}%
\bibitem [{\citenamefont {Paik}\ \emph {et~al.}(2011)\citenamefont {Paik},
  \citenamefont {Schuster}, \citenamefont {Bishop}, \citenamefont {Kirchmair},
  \citenamefont {Catelani}, \citenamefont {Sears}, \citenamefont {Johnson},
  \citenamefont {Reagor}, \citenamefont {Frunzio}, \citenamefont {Glazman},
  \citenamefont {Girvin}, \citenamefont {Devoret},\ and\ \citenamefont
  {Schoelkopf}}]{Paik2011}%
  \BibitemOpen
  \bibfield  {author} {\bibinfo {author} {\bibfnamefont {H.}~\bibnamefont
  {Paik}}, \bibinfo {author} {\bibfnamefont {D.~I.}\ \bibnamefont {Schuster}},
  \bibinfo {author} {\bibfnamefont {L.~S.}\ \bibnamefont {Bishop}}, \bibinfo
  {author} {\bibfnamefont {G.}~\bibnamefont {Kirchmair}}, \bibinfo {author}
  {\bibfnamefont {G.}~\bibnamefont {Catelani}}, \bibinfo {author}
  {\bibfnamefont {A.~P.}\ \bibnamefont {Sears}}, \bibinfo {author}
  {\bibfnamefont {B.~R.}\ \bibnamefont {Johnson}}, \bibinfo {author}
  {\bibfnamefont {M.~J.}\ \bibnamefont {Reagor}}, \bibinfo {author}
  {\bibfnamefont {L.}~\bibnamefont {Frunzio}}, \bibinfo {author} {\bibfnamefont
  {L.~I.}\ \bibnamefont {Glazman}}, \bibinfo {author} {\bibfnamefont {S.~M.}\
  \bibnamefont {Girvin}}, \bibinfo {author} {\bibfnamefont {M.~H.}\
  \bibnamefont {Devoret}}, \ and\ \bibinfo {author} {\bibfnamefont {R.~J.}\
  \bibnamefont {Schoelkopf}}} (\bibinfo {year} {2011}),\ \bibfield  {title}
  {\enquote {\bibinfo {title} {{Observation of high coherence in Josephson
  junction qubits measured in a three-dimensional circuit QED architecture}},}\
  }\href {\doibase 10.1103/PhysRevLett.107.240501} {\bibfield  {journal}
  {\bibinfo  {journal} {Physical Review Letters}\ }\textbf {\bibinfo {volume}
  {107}}~(\bibinfo {number} {24}),\ \bibinfo {pages} {240501}},\ \Eprint
  {http://arxiv.org/abs/1105.4652} {arXiv:1105.4652} \BibitemShut {NoStop}%
\bibitem [{\citenamefont {Patti}\ \emph {et~al.}(2017)\citenamefont {Patti},
  \citenamefont {Chantasri}, \citenamefont {Garc{\'{i}}a-Pintos}, \citenamefont
  {Jordan},\ and\ \citenamefont {Dressel}}]{Patti2017}%
  \BibitemOpen
  \bibfield  {author} {\bibinfo {author} {\bibfnamefont {T.~L.}\ \bibnamefont
  {Patti}}, \bibinfo {author} {\bibfnamefont {A.}~\bibnamefont {Chantasri}},
  \bibinfo {author} {\bibfnamefont {L.~P.}\ \bibnamefont
  {Garc{\'{i}}a-Pintos}}, \bibinfo {author} {\bibfnamefont {A.~N.}\
  \bibnamefont {Jordan}}, \ and\ \bibinfo {author} {\bibfnamefont
  {J.}~\bibnamefont {Dressel}}} (\bibinfo {year} {2017}),\ \bibfield  {title}
  {\enquote {\bibinfo {title} {{Linear feedback stabilization of a dispersively
  monitored qubit}},}\ }\href {\doibase 10.1103/PhysRevA.96.022311} {\bibfield
  {journal} {\bibinfo  {journal} {Physical Review A}\ }\textbf {\bibinfo
  {volume} {96}}~(\bibinfo {number} {2}),\ \bibinfo {pages} {022311}},\ \Eprint
  {http://arxiv.org/abs/1705.03878} {arXiv:1705.03878} \BibitemShut {NoStop}%
\bibitem [{\citenamefont {Peil}\ and\ \citenamefont
  {Gabrielse}(1999)}]{Peil1999}%
  \BibitemOpen
  \bibfield  {author} {\bibinfo {author} {\bibfnamefont {S.}~\bibnamefont
  {Peil}}, \ and\ \bibinfo {author} {\bibfnamefont {G.}~\bibnamefont
  {Gabrielse}}} (\bibinfo {year} {1999}),\ \bibfield  {title} {\enquote
  {\bibinfo {title} {{Observing the Quantum Limit of an Electron Cyclotron: QND
  Measurements of Quantum Jumps between Fock States}},}\ }\href {\doibase
  10.1103/PhysRevLett.83.1287} {\bibfield  {journal} {\bibinfo  {journal}
  {Physical Review Letters}\ }\textbf {\bibinfo {volume} {83}}~(\bibinfo
  {number} {7}),\ \bibinfo {pages} {1287--1290}}\BibitemShut {NoStop}%
\bibitem [{\citenamefont {Perarnau-Llobet}\ and\ \citenamefont
  {Nieuwenhuizen}(2017)}]{Perarnau-Llobet2017}%
  \BibitemOpen
  \bibfield  {author} {\bibinfo {author} {\bibfnamefont {M.}~\bibnamefont
  {Perarnau-Llobet}}, \ and\ \bibinfo {author} {\bibfnamefont {T.~M.}\
  \bibnamefont {Nieuwenhuizen}}} (\bibinfo {year} {2017}),\ \bibfield  {title}
  {\enquote {\bibinfo {title} {{Simultaneous measurement of two noncommuting
  quantum variables: Solution of a dynamical model}},}\ }\href {\doibase
  10.1103/PhysRevA.95.052129} {\bibfield  {journal} {\bibinfo  {journal}
  {Physical Review A}\ }\textbf {\bibinfo {volume} {95}}~(\bibinfo {number}
  {5}),\ \bibinfo {pages} {052129}}\BibitemShut {NoStop}%
\bibitem [{\citenamefont {Peres}(2002)}]{PeresBook}%
  \BibitemOpen
  \bibfield  {author} {\bibinfo {author} {\bibfnamefont {A.}~\bibnamefont
  {Peres}}} (\bibinfo {year} {2002}),\ \href {\doibase 10.1007/0-306-47120-5}
  {\emph {\bibinfo {title} {{Quantum Theory: Concepts and Methods}}}}\
  (\bibinfo  {publisher} {Springer Netherlands},\ \bibinfo {address}
  {Dordrecht})\BibitemShut {NoStop}%
\bibitem [{\citenamefont {Plenio}\ and\ \citenamefont
  {Knight}(1998)}]{Plenio1998}%
  \BibitemOpen
  \bibfield  {author} {\bibinfo {author} {\bibfnamefont {M.~B.}\ \bibnamefont
  {Plenio}}, \ and\ \bibinfo {author} {\bibfnamefont {P.~L.}\ \bibnamefont
  {Knight}}} (\bibinfo {year} {1998}),\ \bibfield  {title} {\enquote {\bibinfo
  {title} {{The quantum-jump approach to dissipative dynamics in quantum
  optics}},}\ }\href {\doibase 10.1103/RevModPhys.70.101} {\bibfield  {journal}
  {\bibinfo  {journal} {Reviews of Modern Physics}\ }\textbf {\bibinfo {volume}
  {70}}~(\bibinfo {number} {1}),\ \bibinfo {pages} {101--144}},\ \Eprint
  {http://arxiv.org/abs/9702007} {arXiv:9702007 [quant-ph]} \BibitemShut
  {NoStop}%
\bibitem [{\citenamefont {Pobell}(2013)}]{pobell2013book}%
  \BibitemOpen
  \bibfield  {author} {\bibinfo {author} {\bibfnamefont {F.}~\bibnamefont
  {Pobell}}} (\bibinfo {year} {2013}),\ \href
  {https://books.google.com/books?id=Gj{\_}8CAAAQBAJ} {\emph {\bibinfo {title}
  {{Matter and Methods at Low Temperatures}}}}\ (\bibinfo  {publisher}
  {Springer Berlin Heidelberg})\BibitemShut {NoStop}%
\bibitem [{\citenamefont {Pop}(2011)}]{Pop2011-Thesis}%
  \BibitemOpen
  \bibfield  {author} {\bibinfo {author} {\bibfnamefont {I.~M.}\ \bibnamefont
  {Pop}}} (\bibinfo {year} {2011}),\ \emph {\bibinfo {title} {{Quantum
  phase-slips in Josephson junction chains}}},\ \href
  {https://tel.archives-ouvertes.fr/tel-00586075} {\bibinfo {type} {Theses}}\
  (\bibinfo  {school} {Universit{\'{e}} de Grenoble})\BibitemShut {NoStop}%
\bibitem [{\citenamefont {Pop}\ \emph {et~al.}(2012)\citenamefont {Pop},
  \citenamefont {Fournier}, \citenamefont {Crozes}, \citenamefont {Lecocq},
  \citenamefont {Matei}, \citenamefont {Pannetier}, \citenamefont {Buisson},\
  and\ \citenamefont {Guichard}}]{Pop2012-Junction}%
  \BibitemOpen
  \bibfield  {author} {\bibinfo {author} {\bibfnamefont {I.~M.}\ \bibnamefont
  {Pop}}, \bibinfo {author} {\bibfnamefont {T.}~\bibnamefont {Fournier}},
  \bibinfo {author} {\bibfnamefont {T.}~\bibnamefont {Crozes}}, \bibinfo
  {author} {\bibfnamefont {F.}~\bibnamefont {Lecocq}}, \bibinfo {author}
  {\bibfnamefont {I.}~\bibnamefont {Matei}}, \bibinfo {author} {\bibfnamefont
  {B.}~\bibnamefont {Pannetier}}, \bibinfo {author} {\bibfnamefont
  {O.}~\bibnamefont {Buisson}}, \ and\ \bibinfo {author} {\bibfnamefont
  {W.}~\bibnamefont {Guichard}}} (\bibinfo {year} {2012}),\ \bibfield  {title}
  {\enquote {\bibinfo {title} {{Fabrication of stable and reproducible
  submicron tunnel junctions}},}\ }\href {\doibase 10.1116/1.3673790}
  {\bibfield  {journal} {\bibinfo  {journal} {Journal of Vacuum Science {\&}
  Technology B, Nanotechnology and Microelectronics: Materials, Processing,
  Measurement, and Phenomena}\ }\textbf {\bibinfo {volume} {30}}~(\bibinfo
  {number} {1}),\ \bibinfo {pages} {010607}}\BibitemShut {NoStop}%
\bibitem [{\citenamefont {Porrati}\ and\ \citenamefont
  {Putterman}(1987)}]{Porrati1987}%
  \BibitemOpen
  \bibfield  {author} {\bibinfo {author} {\bibfnamefont {M.}~\bibnamefont
  {Porrati}}, \ and\ \bibinfo {author} {\bibfnamefont {S.}~\bibnamefont
  {Putterman}}} (\bibinfo {year} {1987}),\ \bibfield  {title} {\enquote
  {\bibinfo {title} {{Wave-function collapse due to null measurements: The
  origin of intermittent atomic fluorescence}},}\ }\href {\doibase
  10.1103/PhysRevA.36.929} {\bibfield  {journal} {\bibinfo  {journal} {Physical
  Review A}\ }\textbf {\bibinfo {volume} {36}}~(\bibinfo {number} {2}),\
  \bibinfo {pages} {929--932}}\BibitemShut {NoStop}%
\bibitem [{\citenamefont {Poyatos}\ \emph {et~al.}(1996)\citenamefont
  {Poyatos}, \citenamefont {Cirac},\ and\ \citenamefont
  {Zoller}}]{Poyatos1996-qm-res-eng}%
  \BibitemOpen
  \bibfield  {author} {\bibinfo {author} {\bibfnamefont {J.~F.}\ \bibnamefont
  {Poyatos}}, \bibinfo {author} {\bibfnamefont {J.~I.}\ \bibnamefont {Cirac}},
  \ and\ \bibinfo {author} {\bibfnamefont {P.}~\bibnamefont {Zoller}}}
  (\bibinfo {year} {1996}),\ \bibfield  {title} {\enquote {\bibinfo {title}
  {{Quantum Reservoir Engineering with Laser Cooled Trapped Ions}},}\ }\href
  {\doibase 10.1103/PhysRevLett.77.4728} {\bibfield  {journal} {\bibinfo
  {journal} {Physical Review Letters}\ }\textbf {\bibinfo {volume}
  {77}}~(\bibinfo {number} {23}),\ \bibinfo {pages} {4728--4731}}\BibitemShut
  {NoStop}%
\bibitem [{\citenamefont {Pozar}(2011)}]{Pozar}%
  \BibitemOpen
  \bibfield  {author} {\bibinfo {author} {\bibfnamefont {D.~M.}\ \bibnamefont
  {Pozar}}} (\bibinfo {year} {2011}),\ \href
  {http://books.google.com/books/about/Microwave{\%}7B{\_}{\%}7DEngineering.html?id=Zys5YgEACAAJ{\%}7B{\&}{\%}7Dpgis=1}
  {\emph {\bibinfo {title} {{Microwave Engineering}}}},\ \bibinfo {edition}
  {4th}\ ed.,\ Vol.\ \bibinfo {volume} {2011}\ (\bibinfo  {publisher} {John
  Wiley and Sons},\ \bibinfo {address} {Hoboken, NJ})\BibitemShut {NoStop}%
\bibitem [{\citenamefont {Reagor}\ \emph {et~al.}(2013)\citenamefont {Reagor},
  \citenamefont {Paik}, \citenamefont {Catelani}, \citenamefont {Sun},
  \citenamefont {Axline}, \citenamefont {Holland}, \citenamefont {Pop},
  \citenamefont {Masluk}, \citenamefont {Brecht}, \citenamefont {Frunzio},
  \citenamefont {Devoret}, \citenamefont {Glazman},\ and\ \citenamefont
  {Schoelkopf}}]{Reagor2013}%
  \BibitemOpen
  \bibfield  {author} {\bibinfo {author} {\bibfnamefont {M.}~\bibnamefont
  {Reagor}}, \bibinfo {author} {\bibfnamefont {H.}~\bibnamefont {Paik}},
  \bibinfo {author} {\bibfnamefont {G.}~\bibnamefont {Catelani}}, \bibinfo
  {author} {\bibfnamefont {L.}~\bibnamefont {Sun}}, \bibinfo {author}
  {\bibfnamefont {C.}~\bibnamefont {Axline}}, \bibinfo {author} {\bibfnamefont
  {E.}~\bibnamefont {Holland}}, \bibinfo {author} {\bibfnamefont {I.~M.}\
  \bibnamefont {Pop}}, \bibinfo {author} {\bibfnamefont {N.~A.}\ \bibnamefont
  {Masluk}}, \bibinfo {author} {\bibfnamefont {T.}~\bibnamefont {Brecht}},
  \bibinfo {author} {\bibfnamefont {L.}~\bibnamefont {Frunzio}}, \bibinfo
  {author} {\bibfnamefont {M.~H.}\ \bibnamefont {Devoret}}, \bibinfo {author}
  {\bibfnamefont {L.}~\bibnamefont {Glazman}}, \ and\ \bibinfo {author}
  {\bibfnamefont {R.~J.}\ \bibnamefont {Schoelkopf}}} (\bibinfo {year}
  {2013}),\ \bibfield  {title} {\enquote {\bibinfo {title} {{Reaching 10 ms
  single photon lifetimes for superconducting aluminum cavities}},}\ }\href
  {\doibase 10.1063/1.4807015} {\bibfield  {journal} {\bibinfo  {journal}
  {Applied Physics Letters}\ }\textbf {\bibinfo {volume} {102}}~(\bibinfo
  {number} {19}),\ \bibinfo {pages} {192604}}\BibitemShut {NoStop}%
\bibitem [{\citenamefont {Reagor}\ \emph {et~al.}(2016)\citenamefont {Reagor},
  \citenamefont {Pfaff}, \citenamefont {Axline}, \citenamefont {Heeres},
  \citenamefont {Ofek}, \citenamefont {Sliwa}, \citenamefont {Holland},
  \citenamefont {Wang}, \citenamefont {Blumoff}, \citenamefont {Chou},
  \citenamefont {Hatridge}, \citenamefont {Frunzio}, \citenamefont {Devoret},
  \citenamefont {Jiang},\ and\ \citenamefont {Schoelkopf}}]{Reagor2016-cavity}%
  \BibitemOpen
  \bibfield  {author} {\bibinfo {author} {\bibfnamefont {M.}~\bibnamefont
  {Reagor}}, \bibinfo {author} {\bibfnamefont {W.}~\bibnamefont {Pfaff}},
  \bibinfo {author} {\bibfnamefont {C.}~\bibnamefont {Axline}}, \bibinfo
  {author} {\bibfnamefont {R.~W.}\ \bibnamefont {Heeres}}, \bibinfo {author}
  {\bibfnamefont {N.}~\bibnamefont {Ofek}}, \bibinfo {author} {\bibfnamefont
  {K.}~\bibnamefont {Sliwa}}, \bibinfo {author} {\bibfnamefont
  {E.}~\bibnamefont {Holland}}, \bibinfo {author} {\bibfnamefont
  {C.}~\bibnamefont {Wang}}, \bibinfo {author} {\bibfnamefont {J.}~\bibnamefont
  {Blumoff}}, \bibinfo {author} {\bibfnamefont {K.}~\bibnamefont {Chou}},
  \bibinfo {author} {\bibfnamefont {M.~J.}\ \bibnamefont {Hatridge}}, \bibinfo
  {author} {\bibfnamefont {L.}~\bibnamefont {Frunzio}}, \bibinfo {author}
  {\bibfnamefont {M.~H.}\ \bibnamefont {Devoret}}, \bibinfo {author}
  {\bibfnamefont {L.}~\bibnamefont {Jiang}}, \ and\ \bibinfo {author}
  {\bibfnamefont {R.~J.}\ \bibnamefont {Schoelkopf}}} (\bibinfo {year}
  {2016}),\ \bibfield  {title} {\enquote {\bibinfo {title} {{Quantum memory
  with millisecond coherence in circuit QED}},}\ }\href {\doibase
  10.1103/PhysRevB.94.014506} {\bibfield  {journal} {\bibinfo  {journal}
  {Physical Review B}\ }\textbf {\bibinfo {volume} {94}}~(\bibinfo {number}
  {1}),\ \bibinfo {pages} {014506}},\ \Eprint {http://arxiv.org/abs/1508.05882}
  {arXiv:1508.05882} \BibitemShut {NoStop}%
\bibitem [{\citenamefont {Reagor}(2016)}]{Reagor2016}%
  \BibitemOpen
  \bibfield  {author} {\bibinfo {author} {\bibfnamefont {M.~J.}\ \bibnamefont
  {Reagor}}} (\bibinfo {year} {2016}),\ \emph {\bibinfo {title}
  {{Superconducting Cavities for Circuit Quantum Electrodynamics}}},\
  \href@noop {} {Ph.D. thesis}\ (\bibinfo  {school} {Yale
  University})\BibitemShut {NoStop}%
\bibitem [{\citenamefont {Reed}(2013)}]{Reed2013}%
  \BibitemOpen
  \bibfield  {author} {\bibinfo {author} {\bibfnamefont {M.~D.}\ \bibnamefont
  {Reed}}} (\bibinfo {year} {2013}),\ \emph {\bibinfo {title} {{Entanglement
  and Quantum Error Correction with Superconducting Qubits}}},\ \href
  {https://books.google.com/books?id=7mORBQAAQBAJ{\&}dq=allxy+reed{\&}source=gbs{\_}navlinks{\_}s}
  {Ph.D. thesis}\ (\bibinfo  {school} {Yale University})\BibitemShut {NoStop}%
\bibitem [{\citenamefont {Reed}\ \emph {et~al.}(2010)\citenamefont {Reed},
  \citenamefont {DiCarlo}, \citenamefont {Johnson}, \citenamefont {Sun},
  \citenamefont {Schuster}, \citenamefont {Frunzio},\ and\ \citenamefont
  {Schoelkopf}}]{Reed2010-readout}%
  \BibitemOpen
  \bibfield  {author} {\bibinfo {author} {\bibfnamefont {M.~D.}\ \bibnamefont
  {Reed}}, \bibinfo {author} {\bibfnamefont {L.}~\bibnamefont {DiCarlo}},
  \bibinfo {author} {\bibfnamefont {B.~R.}\ \bibnamefont {Johnson}}, \bibinfo
  {author} {\bibfnamefont {L.}~\bibnamefont {Sun}}, \bibinfo {author}
  {\bibfnamefont {D.~I.}\ \bibnamefont {Schuster}}, \bibinfo {author}
  {\bibfnamefont {L.}~\bibnamefont {Frunzio}}, \ and\ \bibinfo {author}
  {\bibfnamefont {R.~J.}\ \bibnamefont {Schoelkopf}}} (\bibinfo {year}
  {2010}),\ \bibfield  {title} {\enquote {\bibinfo {title} {{High-Fidelity
  Readout in Circuit Quantum Electrodynamics Using the Jaynes-Cummings
  Nonlinearity}},}\ }\href {\doibase 10.1103/PhysRevLett.105.173601} {\bibfield
   {journal} {\bibinfo  {journal} {Physical Review Letters}\ }\textbf {\bibinfo
  {volume} {105}}~(\bibinfo {number} {17}),\ \bibinfo {pages}
  {173601}}\BibitemShut {NoStop}%
\bibitem [{\citenamefont {Rigetti}\ \emph {et~al.}(2012)\citenamefont
  {Rigetti}, \citenamefont {Gambetta}, \citenamefont {Poletto}, \citenamefont
  {Plourde}, \citenamefont {Chow}, \citenamefont {C{\'{o}}rcoles},
  \citenamefont {Smolin}, \citenamefont {Merkel}, \citenamefont {Rozen},
  \citenamefont {Keefe}, \citenamefont {Rothwell}, \citenamefont {Ketchen},\
  and\ \citenamefont {Steffen}}]{Rigetti2012}%
  \BibitemOpen
  \bibfield  {author} {\bibinfo {author} {\bibfnamefont {C.}~\bibnamefont
  {Rigetti}}, \bibinfo {author} {\bibfnamefont {J.~M.}\ \bibnamefont
  {Gambetta}}, \bibinfo {author} {\bibfnamefont {S.}~\bibnamefont {Poletto}},
  \bibinfo {author} {\bibfnamefont {B.~L.~T.}\ \bibnamefont {Plourde}},
  \bibinfo {author} {\bibfnamefont {J.~M.}\ \bibnamefont {Chow}}, \bibinfo
  {author} {\bibfnamefont {A.~D.}\ \bibnamefont {C{\'{o}}rcoles}}, \bibinfo
  {author} {\bibfnamefont {J.~A.}\ \bibnamefont {Smolin}}, \bibinfo {author}
  {\bibfnamefont {S.~T.}\ \bibnamefont {Merkel}}, \bibinfo {author}
  {\bibfnamefont {J.~R.}\ \bibnamefont {Rozen}}, \bibinfo {author}
  {\bibfnamefont {G.~A.}\ \bibnamefont {Keefe}}, \bibinfo {author}
  {\bibfnamefont {M.~B.}\ \bibnamefont {Rothwell}}, \bibinfo {author}
  {\bibfnamefont {M.~B.}\ \bibnamefont {Ketchen}}, \ and\ \bibinfo {author}
  {\bibfnamefont {M.}~\bibnamefont {Steffen}}} (\bibinfo {year} {2012}),\
  \bibfield  {title} {\enquote {\bibinfo {title} {{Superconducting qubit in a
  waveguide cavity with a coherence time approaching 0.1 ms}},}\ }\href
  {\doibase 10.1103/PhysRevB.86.100506} {\bibfield  {journal} {\bibinfo
  {journal} {Physical Review B}\ }\textbf {\bibinfo {volume} {86}}~(\bibinfo
  {number} {10}),\ \bibinfo {pages} {100506}}\BibitemShut {NoStop}%
\bibitem [{\citenamefont {Rigetti}(2009)}]{Rigetti2009}%
  \BibitemOpen
  \bibfield  {author} {\bibinfo {author} {\bibfnamefont {C.~T.}\ \bibnamefont
  {Rigetti}}} (\bibinfo {year} {2009}),\ \emph {\bibinfo {title} {{Quantum
  Gates for Superconducting Qubits}}},\ \href@noop {} {Ph.D. thesis}\ (\bibinfo
   {school} {Yale University})\BibitemShut {NoStop}%
\bibitem [{\citenamefont {Rist{\`{e}}}\ \emph
  {et~al.}(2012{\natexlab{a}})\citenamefont {Rist{\`{e}}}, \citenamefont
  {Bultink}, \citenamefont {Lehnert},\ and\ \citenamefont
  {DiCarlo}}]{Riste2012-qubit-measure-reset}%
  \BibitemOpen
  \bibfield  {author} {\bibinfo {author} {\bibfnamefont {D.}~\bibnamefont
  {Rist{\`{e}}}}, \bibinfo {author} {\bibfnamefont {C.~C.}\ \bibnamefont
  {Bultink}}, \bibinfo {author} {\bibfnamefont {K.~W.}\ \bibnamefont
  {Lehnert}}, \ and\ \bibinfo {author} {\bibfnamefont {L.}~\bibnamefont
  {DiCarlo}}} (\bibinfo {year} {2012}{\natexlab{a}}),\ \bibfield  {title}
  {\enquote {\bibinfo {title} {{Feedback Control of a Solid-State Qubit Using
  High-Fidelity Projective Measurement}},}\ }\href {\doibase
  10.1103/PhysRevLett.109.240502} {\bibfield  {journal} {\bibinfo  {journal}
  {Physical Review Letters}\ }\textbf {\bibinfo {volume} {109}}~(\bibinfo
  {number} {24}),\ \bibinfo {pages} {240502}}\BibitemShut {NoStop}%
\bibitem [{\citenamefont {Rist{\`{e}}}\ \emph {et~al.}(2013)\citenamefont
  {Rist{\`{e}}}, \citenamefont {Dukalski}, \citenamefont {Watson},
  \citenamefont {{De Lange}}, \citenamefont {Tiggelman}, \citenamefont
  {Blanter}, \citenamefont {Lehnert}, \citenamefont {Schouten},\ and\
  \citenamefont {Dicarlo}}]{Riste2013}%
  \BibitemOpen
  \bibfield  {author} {\bibinfo {author} {\bibfnamefont {D.}~\bibnamefont
  {Rist{\`{e}}}}, \bibinfo {author} {\bibfnamefont {M.}~\bibnamefont
  {Dukalski}}, \bibinfo {author} {\bibfnamefont {C.~A.}\ \bibnamefont
  {Watson}}, \bibinfo {author} {\bibfnamefont {G.}~\bibnamefont {{De Lange}}},
  \bibinfo {author} {\bibfnamefont {M.~J.}\ \bibnamefont {Tiggelman}}, \bibinfo
  {author} {\bibfnamefont {Y.~M.}\ \bibnamefont {Blanter}}, \bibinfo {author}
  {\bibfnamefont {K.~W.}\ \bibnamefont {Lehnert}}, \bibinfo {author}
  {\bibfnamefont {R.~N.}\ \bibnamefont {Schouten}}, \ and\ \bibinfo {author}
  {\bibfnamefont {L.}~\bibnamefont {Dicarlo}}} (\bibinfo {year} {2013}),\
  \bibfield  {title} {\enquote {\bibinfo {title} {{Deterministic entanglement
  of superconducting qubits by parity measurement and feedback}},}\ }\href
  {\doibase 10.1038/nature12513} {\bibfield  {journal} {\bibinfo  {journal}
  {Nature}\ }\textbf {\bibinfo {volume} {502}}~(\bibinfo {number} {7471}),\
  \bibinfo {pages} {350--354}},\ \Eprint {http://arxiv.org/abs/1306.4002}
  {arXiv:1306.4002} \BibitemShut {NoStop}%
\bibitem [{\citenamefont {Rist{\`{e}}}\ \emph
  {et~al.}(2012{\natexlab{b}})\citenamefont {Rist{\`{e}}}, \citenamefont {van
  Leeuwen}, \citenamefont {Ku}, \citenamefont {Lehnert},\ and\ \citenamefont
  {DiCarlo}}]{Riste2012-readout}%
  \BibitemOpen
  \bibfield  {author} {\bibinfo {author} {\bibfnamefont {D.}~\bibnamefont
  {Rist{\`{e}}}}, \bibinfo {author} {\bibfnamefont {J.~G.}\ \bibnamefont {van
  Leeuwen}}, \bibinfo {author} {\bibfnamefont {H.-S.}\ \bibnamefont {Ku}},
  \bibinfo {author} {\bibfnamefont {K.~W.}\ \bibnamefont {Lehnert}}, \ and\
  \bibinfo {author} {\bibfnamefont {L.}~\bibnamefont {DiCarlo}}} (\bibinfo
  {year} {2012}{\natexlab{b}}),\ \bibfield  {title} {\enquote {\bibinfo {title}
  {{Initialization by Measurement of a Superconducting Quantum Bit Circuit}},}\
  }\href {\doibase 10.1103/PhysRevLett.109.050507} {\bibfield  {journal}
  {\bibinfo  {journal} {Physical Review Letters}\ }\textbf {\bibinfo {volume}
  {109}}~(\bibinfo {number} {5}),\ \bibinfo {pages} {050507}}\BibitemShut
  {NoStop}%
\bibitem [{\citenamefont {Robledo}\ \emph {et~al.}(2011)\citenamefont
  {Robledo}, \citenamefont {Childress}, \citenamefont {Bernien}, \citenamefont
  {Hensen}, \citenamefont {Alkemade},\ and\ \citenamefont
  {Hanson}}]{Robledo2011}%
  \BibitemOpen
  \bibfield  {author} {\bibinfo {author} {\bibfnamefont {L.}~\bibnamefont
  {Robledo}}, \bibinfo {author} {\bibfnamefont {L.}~\bibnamefont {Childress}},
  \bibinfo {author} {\bibfnamefont {H.}~\bibnamefont {Bernien}}, \bibinfo
  {author} {\bibfnamefont {B.}~\bibnamefont {Hensen}}, \bibinfo {author}
  {\bibfnamefont {P.~F.~A.}\ \bibnamefont {Alkemade}}, \ and\ \bibinfo {author}
  {\bibfnamefont {R.}~\bibnamefont {Hanson}}} (\bibinfo {year} {2011}),\
  \bibfield  {title} {\enquote {\bibinfo {title} {{High-fidelity projective
  read-out of a solid-state spin quantum register}},}\ }\href {\doibase
  10.1038/nature10401} {\bibfield  {journal} {\bibinfo  {journal} {Nature}\
  }\textbf {\bibinfo {volume} {477}}~(\bibinfo {number} {7366}),\ \bibinfo
  {pages} {574--578}},\ \Eprint {http://arxiv.org/abs/1301.0392v1}
  {arXiv:1301.0392v1} \BibitemShut {NoStop}%
\bibitem [{\citenamefont {Roch}\ \emph {et~al.}(2014)\citenamefont {Roch},
  \citenamefont {Schwartz}, \citenamefont {Motzoi}, \citenamefont {Macklin},
  \citenamefont {Vijay}, \citenamefont {Eddins}, \citenamefont {Korotkov},
  \citenamefont {Whaley}, \citenamefont {Sarovar},\ and\ \citenamefont
  {Siddiqi}}]{Roch2014}%
  \BibitemOpen
  \bibfield  {author} {\bibinfo {author} {\bibfnamefont {N.}~\bibnamefont
  {Roch}}, \bibinfo {author} {\bibfnamefont {M.~E.}\ \bibnamefont {Schwartz}},
  \bibinfo {author} {\bibfnamefont {F.}~\bibnamefont {Motzoi}}, \bibinfo
  {author} {\bibfnamefont {C.}~\bibnamefont {Macklin}}, \bibinfo {author}
  {\bibfnamefont {R.}~\bibnamefont {Vijay}}, \bibinfo {author} {\bibfnamefont
  {A.~W.}\ \bibnamefont {Eddins}}, \bibinfo {author} {\bibfnamefont {A.~N.}\
  \bibnamefont {Korotkov}}, \bibinfo {author} {\bibfnamefont {K.~B.}\
  \bibnamefont {Whaley}}, \bibinfo {author} {\bibfnamefont {M.}~\bibnamefont
  {Sarovar}}, \ and\ \bibinfo {author} {\bibfnamefont {I.}~\bibnamefont
  {Siddiqi}}} (\bibinfo {year} {2014}),\ \bibfield  {title} {\enquote {\bibinfo
  {title} {{Observation of measurement-induced entanglement and quantum
  trajectories of remote superconducting qubits}},}\ }\href {\doibase
  10.1103/PhysRevLett.112.170501} {\bibfield  {journal} {\bibinfo  {journal}
  {Physical Review Letters}\ }\textbf {\bibinfo {volume} {112}}~(\bibinfo
  {number} {17}),\ \bibinfo {pages} {170501}},\ \Eprint
  {http://arxiv.org/abs/1402.1868} {arXiv:1402.1868} \BibitemShut {NoStop}%
\bibitem [{\citenamefont {Rosenberg}\ \emph {et~al.}(2017)\citenamefont
  {Rosenberg}, \citenamefont {Kim}, \citenamefont {Das}, \citenamefont {Yost},
  \citenamefont {Gustavsson}, \citenamefont {Hover}, \citenamefont {Krantz},
  \citenamefont {Melville}, \citenamefont {Racz}, \citenamefont {Samach},
  \citenamefont {Weber}, \citenamefont {Yan}, \citenamefont {Yoder},
  \citenamefont {Kerman},\ and\ \citenamefont {Oliver}}]{Rosenberg2017}%
  \BibitemOpen
  \bibfield  {author} {\bibinfo {author} {\bibfnamefont {D.}~\bibnamefont
  {Rosenberg}}, \bibinfo {author} {\bibfnamefont {D.}~\bibnamefont {Kim}},
  \bibinfo {author} {\bibfnamefont {R.}~\bibnamefont {Das}}, \bibinfo {author}
  {\bibfnamefont {D.}~\bibnamefont {Yost}}, \bibinfo {author} {\bibfnamefont
  {S.}~\bibnamefont {Gustavsson}}, \bibinfo {author} {\bibfnamefont
  {D.}~\bibnamefont {Hover}}, \bibinfo {author} {\bibfnamefont
  {P.}~\bibnamefont {Krantz}}, \bibinfo {author} {\bibfnamefont
  {A.}~\bibnamefont {Melville}}, \bibinfo {author} {\bibfnamefont
  {L.}~\bibnamefont {Racz}}, \bibinfo {author} {\bibfnamefont {G.~O.}\
  \bibnamefont {Samach}}, \bibinfo {author} {\bibfnamefont {S.~J.}\
  \bibnamefont {Weber}}, \bibinfo {author} {\bibfnamefont {F.}~\bibnamefont
  {Yan}}, \bibinfo {author} {\bibfnamefont {J.}~\bibnamefont {Yoder}}, \bibinfo
  {author} {\bibfnamefont {A.~J.}\ \bibnamefont {Kerman}}, \ and\ \bibinfo
  {author} {\bibfnamefont {W.~D.}\ \bibnamefont {Oliver}}} (\bibinfo {year}
  {2017}),\ \bibfield  {title} {\enquote {\bibinfo {title} {{3D integrated
  superconducting qubits}},}\ }\href {http://arxiv.org/abs/1706.04116} {\ ,\
  \bibinfo {pages} {1--6}}\Eprint {http://arxiv.org/abs/1706.04116}
  {arXiv:1706.04116} \BibitemShut {NoStop}%
\bibitem [{\citenamefont {Roy}\ \emph {et~al.}(2017)\citenamefont {Roy},
  \citenamefont {Kundu}, \citenamefont {Chand}, \citenamefont {Hazra},
  \citenamefont {Nehra}, \citenamefont {Cosmic}, \citenamefont {Ranadive},
  \citenamefont {Patankar}, \citenamefont {Damle},\ and\ \citenamefont
  {Vijay}}]{Roy2017-3qubits}%
  \BibitemOpen
  \bibfield  {author} {\bibinfo {author} {\bibfnamefont {T.}~\bibnamefont
  {Roy}}, \bibinfo {author} {\bibfnamefont {S.}~\bibnamefont {Kundu}}, \bibinfo
  {author} {\bibfnamefont {M.}~\bibnamefont {Chand}}, \bibinfo {author}
  {\bibfnamefont {S.}~\bibnamefont {Hazra}}, \bibinfo {author} {\bibfnamefont
  {N.}~\bibnamefont {Nehra}}, \bibinfo {author} {\bibfnamefont
  {R.}~\bibnamefont {Cosmic}}, \bibinfo {author} {\bibfnamefont
  {A.}~\bibnamefont {Ranadive}}, \bibinfo {author} {\bibfnamefont {M.~P.}\
  \bibnamefont {Patankar}}, \bibinfo {author} {\bibfnamefont {K.}~\bibnamefont
  {Damle}}, \ and\ \bibinfo {author} {\bibfnamefont {R.}~\bibnamefont {Vijay}}}
  (\bibinfo {year} {2017}),\ \bibfield  {title} {\enquote {\bibinfo {title}
  {{Implementation of Pairwise Longitudinal Coupling in a Three-Qubit
  Superconducting Circuit}},}\ }\href {\doibase
  10.1103/PhysRevApplied.7.054025} {\bibfield  {journal} {\bibinfo  {journal}
  {Physical Review Applied}\ }\textbf {\bibinfo {volume} {7}}~(\bibinfo
  {number} {5}),\ \bibinfo {pages} {054025}},\ \Eprint
  {http://arxiv.org/abs/1610.07915} {arXiv:1610.07915} \BibitemShut {NoStop}%
\bibitem [{\citenamefont {Ruskov}\ \emph {et~al.}(2009)\citenamefont {Ruskov},
  \citenamefont {Dobrovitski},\ and\ \citenamefont
  {Harmon}}]{Ruskov2009-unpublished}%
  \BibitemOpen
  \bibfield  {author} {\bibinfo {author} {\bibfnamefont {R.}~\bibnamefont
  {Ruskov}}, \bibinfo {author} {\bibfnamefont {V.~V.}\ \bibnamefont
  {Dobrovitski}}, \ and\ \bibinfo {author} {\bibfnamefont {B.~N.}\ \bibnamefont
  {Harmon}}} (\bibinfo {year} {2009}),\ \bibfield  {title} {\enquote {\bibinfo
  {title} {{Manipulation of double-dot spin qubit by continuous noisy
  measurement}},}\ }\href {http://arxiv.org/abs/0906.0425} {\ }\Eprint
  {http://arxiv.org/abs/0906.0425} {arXiv:0906.0425} \BibitemShut {NoStop}%
\bibitem [{\citenamefont {Ruskov}\ \emph {et~al.}(2010)\citenamefont {Ruskov},
  \citenamefont {Korotkov},\ and\ \citenamefont {M{\o}lmer}}]{Ruskov2010}%
  \BibitemOpen
  \bibfield  {author} {\bibinfo {author} {\bibfnamefont {R.}~\bibnamefont
  {Ruskov}}, \bibinfo {author} {\bibfnamefont {A.~N.}\ \bibnamefont
  {Korotkov}}, \ and\ \bibinfo {author} {\bibfnamefont {K.}~\bibnamefont
  {M{\o}lmer}}} (\bibinfo {year} {2010}),\ \bibfield  {title} {\enquote
  {\bibinfo {title} {{Qubit State Monitoring by Measurement of Three
  Complementary Observables}},}\ }\href {\doibase
  10.1103/PhysRevLett.105.100506} {\bibfield  {journal} {\bibinfo  {journal}
  {Physical Review Letters}\ }\textbf {\bibinfo {volume} {105}}~(\bibinfo
  {number} {10}),\ \bibinfo {pages} {100506}}\BibitemShut {NoStop}%
\bibitem [{\citenamefont {Ruskov}\ \emph {et~al.}(2007)\citenamefont {Ruskov},
  \citenamefont {Mizel},\ and\ \citenamefont {Korotkov}}]{Ruskov2007}%
  \BibitemOpen
  \bibfield  {author} {\bibinfo {author} {\bibfnamefont {R.}~\bibnamefont
  {Ruskov}}, \bibinfo {author} {\bibfnamefont {A.}~\bibnamefont {Mizel}}, \
  and\ \bibinfo {author} {\bibfnamefont {A.~N.}\ \bibnamefont {Korotkov}}}
  (\bibinfo {year} {2007}),\ \bibfield  {title} {\enquote {\bibinfo {title}
  {{Crossover of phase qubit dynamics in the presence of a negative-result weak
  measurement}},}\ }\href {\doibase 10.1103/PhysRevB.75.220501} {\bibfield
  {journal} {\bibinfo  {journal} {Physical Review B}\ }\textbf {\bibinfo
  {volume} {75}}~(\bibinfo {number} {22}),\ \bibinfo {pages}
  {220501}}\BibitemShut {NoStop}%
\bibitem [{\citenamefont {Sank}\ \emph {et~al.}(2016)\citenamefont {Sank},
  \citenamefont {Chen}, \citenamefont {Khezri}, \citenamefont {Kelly},
  \citenamefont {Barends}, \citenamefont {Campbell}, \citenamefont {Chen},
  \citenamefont {Chiaro}, \citenamefont {Dunsworth}, \citenamefont {Fowler},
  \citenamefont {Jeffrey}, \citenamefont {Lucero}, \citenamefont {Megrant},
  \citenamefont {Mutus}, \citenamefont {Neeley}, \citenamefont {Neill},
  \citenamefont {O'Malley}, \citenamefont {Quintana}, \citenamefont {Roushan},
  \citenamefont {Vainsencher}, \citenamefont {White}, \citenamefont {Wenner},
  \citenamefont {Korotkov},\ and\ \citenamefont
  {Martinis}}]{Sank2016-T1vsNbar}%
  \BibitemOpen
  \bibfield  {author} {\bibinfo {author} {\bibfnamefont {D.}~\bibnamefont
  {Sank}}, \bibinfo {author} {\bibfnamefont {Z.}~\bibnamefont {Chen}}, \bibinfo
  {author} {\bibfnamefont {M.}~\bibnamefont {Khezri}}, \bibinfo {author}
  {\bibfnamefont {J.}~\bibnamefont {Kelly}}, \bibinfo {author} {\bibfnamefont
  {R.}~\bibnamefont {Barends}}, \bibinfo {author} {\bibfnamefont
  {B.}~\bibnamefont {Campbell}}, \bibinfo {author} {\bibfnamefont
  {Y.}~\bibnamefont {Chen}}, \bibinfo {author} {\bibfnamefont {B.}~\bibnamefont
  {Chiaro}}, \bibinfo {author} {\bibfnamefont {A.}~\bibnamefont {Dunsworth}},
  \bibinfo {author} {\bibfnamefont {A.}~\bibnamefont {Fowler}}, \bibinfo
  {author} {\bibfnamefont {E.}~\bibnamefont {Jeffrey}}, \bibinfo {author}
  {\bibfnamefont {E.}~\bibnamefont {Lucero}}, \bibinfo {author} {\bibfnamefont
  {A.}~\bibnamefont {Megrant}}, \bibinfo {author} {\bibfnamefont
  {J.}~\bibnamefont {Mutus}}, \bibinfo {author} {\bibfnamefont
  {M.}~\bibnamefont {Neeley}}, \bibinfo {author} {\bibfnamefont
  {C.}~\bibnamefont {Neill}}, \bibinfo {author} {\bibfnamefont {P.~J.~J.}\
  \bibnamefont {O'Malley}}, \bibinfo {author} {\bibfnamefont {C.}~\bibnamefont
  {Quintana}}, \bibinfo {author} {\bibfnamefont {P.}~\bibnamefont {Roushan}},
  \bibinfo {author} {\bibfnamefont {A.}~\bibnamefont {Vainsencher}}, \bibinfo
  {author} {\bibfnamefont {T.}~\bibnamefont {White}}, \bibinfo {author}
  {\bibfnamefont {J.}~\bibnamefont {Wenner}}, \bibinfo {author} {\bibfnamefont
  {A.~N.}\ \bibnamefont {Korotkov}}, \ and\ \bibinfo {author} {\bibfnamefont
  {J.~M.}\ \bibnamefont {Martinis}}} (\bibinfo {year} {2016}),\ \bibfield
  {title} {\enquote {\bibinfo {title} {{Measurement-Induced State Transitions
  in a Superconducting Qubit: Beyond the Rotating Wave Approximation}},}\
  }\href {\doibase 10.1103/PhysRevLett.117.190503} {\bibfield  {journal}
  {\bibinfo  {journal} {Physical Review Letters}\ }\textbf {\bibinfo {volume}
  {117}}~(\bibinfo {number} {19}),\ \bibinfo {pages} {190503}}\BibitemShut
  {NoStop}%
\bibitem [{\citenamefont {Sauter}\ \emph {et~al.}(1986)\citenamefont {Sauter},
  \citenamefont {Neuhauser}, \citenamefont {Blatt},\ and\ \citenamefont
  {Toschek}}]{Sauter1986}%
  \BibitemOpen
  \bibfield  {author} {\bibinfo {author} {\bibfnamefont {T.}~\bibnamefont
  {Sauter}}, \bibinfo {author} {\bibfnamefont {W.}~\bibnamefont {Neuhauser}},
  \bibinfo {author} {\bibfnamefont {R.}~\bibnamefont {Blatt}}, \ and\ \bibinfo
  {author} {\bibfnamefont {P.~E.}\ \bibnamefont {Toschek}}} (\bibinfo {year}
  {1986}),\ \bibfield  {title} {\enquote {\bibinfo {title} {{Observation of
  Quantum Jumps}},}\ }\href {\doibase 10.1103/PhysRevLett.57.1696} {\bibfield
  {journal} {\bibinfo  {journal} {Physical Review Letters}\ }\textbf {\bibinfo
  {volume} {57}}~(\bibinfo {number} {14}),\ \bibinfo {pages}
  {1696--1698}}\BibitemShut {NoStop}%
\bibitem [{\citenamefont {Sayrin}\ \emph {et~al.}(2011)\citenamefont {Sayrin},
  \citenamefont {Dotsenko}, \citenamefont {Zhou}, \citenamefont {Peaudecerf},
  \citenamefont {Rybarczyk}, \citenamefont {Gleyzes}, \citenamefont {Rouchon},
  \citenamefont {Mirrahimi}, \citenamefont {Amini}, \citenamefont {Brune},
  \citenamefont {Raimond},\ and\ \citenamefont {Haroche}}]{Sayrin2001}%
  \BibitemOpen
  \bibfield  {author} {\bibinfo {author} {\bibfnamefont {C.}~\bibnamefont
  {Sayrin}}, \bibinfo {author} {\bibfnamefont {I.}~\bibnamefont {Dotsenko}},
  \bibinfo {author} {\bibfnamefont {X.}~\bibnamefont {Zhou}}, \bibinfo {author}
  {\bibfnamefont {B.}~\bibnamefont {Peaudecerf}}, \bibinfo {author}
  {\bibfnamefont {T.}~\bibnamefont {Rybarczyk}}, \bibinfo {author}
  {\bibfnamefont {S.}~\bibnamefont {Gleyzes}}, \bibinfo {author} {\bibfnamefont
  {P.}~\bibnamefont {Rouchon}}, \bibinfo {author} {\bibfnamefont
  {M.}~\bibnamefont {Mirrahimi}}, \bibinfo {author} {\bibfnamefont
  {H.}~\bibnamefont {Amini}}, \bibinfo {author} {\bibfnamefont
  {M.}~\bibnamefont {Brune}}, \bibinfo {author} {\bibfnamefont {J.-M.}\
  \bibnamefont {Raimond}}, \ and\ \bibinfo {author} {\bibfnamefont
  {S.}~\bibnamefont {Haroche}}} (\bibinfo {year} {2011}),\ \bibfield  {title}
  {\enquote {\bibinfo {title} {{Real-time quantum feedback prepares and
  stabilizes photon number states}},}\ }\href {\doibase 10.1038/nature10376}
  {\bibfield  {journal} {\bibinfo  {journal} {Nature}\ }\textbf {\bibinfo
  {volume} {477}}~(\bibinfo {number} {7362}),\ \bibinfo {pages} {73--77}},\
  \Eprint {http://arxiv.org/abs/1107.4027} {arXiv:1107.4027} \BibitemShut
  {NoStop}%
\bibitem [{\citenamefont {Schreier}\ \emph {et~al.}(2008)\citenamefont
  {Schreier}, \citenamefont {Houck}, \citenamefont {Koch}, \citenamefont
  {Schuster}, \citenamefont {Johnson}, \citenamefont {Chow}, \citenamefont
  {Gambetta}, \citenamefont {Majer}, \citenamefont {Frunzio}, \citenamefont
  {Devoret}, \citenamefont {Girvin},\ and\ \citenamefont
  {Schoelkopf}}]{Schreier2008-transmon}%
  \BibitemOpen
  \bibfield  {author} {\bibinfo {author} {\bibfnamefont {J.~A.}\ \bibnamefont
  {Schreier}}, \bibinfo {author} {\bibfnamefont {A.~A.}\ \bibnamefont {Houck}},
  \bibinfo {author} {\bibfnamefont {J.}~\bibnamefont {Koch}}, \bibinfo {author}
  {\bibfnamefont {D.~I.}\ \bibnamefont {Schuster}}, \bibinfo {author}
  {\bibfnamefont {B.~R.}\ \bibnamefont {Johnson}}, \bibinfo {author}
  {\bibfnamefont {J.~M.}\ \bibnamefont {Chow}}, \bibinfo {author}
  {\bibfnamefont {J.~M.}\ \bibnamefont {Gambetta}}, \bibinfo {author}
  {\bibfnamefont {J.}~\bibnamefont {Majer}}, \bibinfo {author} {\bibfnamefont
  {L.}~\bibnamefont {Frunzio}}, \bibinfo {author} {\bibfnamefont {M.~H.}\
  \bibnamefont {Devoret}}, \bibinfo {author} {\bibfnamefont {S.~M.}\
  \bibnamefont {Girvin}}, \ and\ \bibinfo {author} {\bibfnamefont {R.~J.}\
  \bibnamefont {Schoelkopf}}} (\bibinfo {year} {2008}),\ \bibfield  {title}
  {\enquote {\bibinfo {title} {{Suppressing charge noise decoherence in
  superconducting charge qubits}},}\ }\href {\doibase
  10.1103/PhysRevB.77.180502} {\bibfield  {journal} {\bibinfo  {journal}
  {Physical Review B}\ }\textbf {\bibinfo {volume} {77}}~(\bibinfo {number}
  {18}),\ \bibinfo {pages} {180502}}\BibitemShut {NoStop}%
\bibitem [{\citenamefont
  {Schr{\"{o}}dinger}(1935)}]{schrodinger1935gegenwartige}%
  \BibitemOpen
  \bibfield  {author} {\bibinfo {author} {\bibfnamefont {E.}~\bibnamefont
  {Schr{\"{o}}dinger}}} (\bibinfo {year} {1935}),\ \bibfield  {title} {\enquote
  {\bibinfo {title} {{Die gegenw{\"{a}}rtige Situation in der Quantenmechanik
  I, II, III, Naturwiss 23, 807, 823, 844}},}\ }\href@noop {} {\bibinfo
  {journal} {English translation: Quantum Theory and Measurement (Translation:
  Wheeler, JA, Zurek, WH (eds.)) (Princeton University Press, Princeton,
  1983)}\ }\BibitemShut {NoStop}%
\bibitem [{\citenamefont {Schr{\"{o}}dinger}(1952)}]{Schrodinger1952}%
  \BibitemOpen
\bibfield  {journal} {  }\bibfield  {author} {\bibinfo {author} {\bibfnamefont
  {E.}~\bibnamefont {Schr{\"{o}}dinger}}} (\bibinfo {year} {1952}),\ \bibfield
  {title} {\enquote {\bibinfo {title} {{Are There Quantum Jumps?}}}\ }\href
  {https://www.jstor.org/stable/685552?seq=1{\#}page{\_}scan{\_}tab{\_}contents
  https://philpapers.org/rec/SCHATQ-3} {\bibfield  {journal} {\bibinfo
  {journal} {British Journal for the Philosophy of Science}\ }\textbf {\bibinfo
  {volume} {3}}~(\bibinfo {number} {10}),\ \bibinfo {pages} {109 and
  233}}\BibitemShut {NoStop}%
\bibitem [{\citenamefont {Schuster}\ \emph {et~al.}(2007)\citenamefont
  {Schuster}, \citenamefont {Houck}, \citenamefont {Schreier}, \citenamefont
  {Wallraff}, \citenamefont {Gambetta}, \citenamefont {Blais}, \citenamefont
  {Frunzio}, \citenamefont {Majer}, \citenamefont {Johnson}, \citenamefont
  {Devoret}, \citenamefont {Girvin},\ and\ \citenamefont
  {Schoelkopf}}]{Schuster2007}%
  \BibitemOpen
  \bibfield  {author} {\bibinfo {author} {\bibfnamefont {D.~I.}\ \bibnamefont
  {Schuster}}, \bibinfo {author} {\bibfnamefont {A.~A.}\ \bibnamefont {Houck}},
  \bibinfo {author} {\bibfnamefont {J.~A.}\ \bibnamefont {Schreier}}, \bibinfo
  {author} {\bibfnamefont {A.}~\bibnamefont {Wallraff}}, \bibinfo {author}
  {\bibfnamefont {J.~M.}\ \bibnamefont {Gambetta}}, \bibinfo {author}
  {\bibfnamefont {A.}~\bibnamefont {Blais}}, \bibinfo {author} {\bibfnamefont
  {L.}~\bibnamefont {Frunzio}}, \bibinfo {author} {\bibfnamefont
  {J.}~\bibnamefont {Majer}}, \bibinfo {author} {\bibfnamefont
  {B.}~\bibnamefont {Johnson}}, \bibinfo {author} {\bibfnamefont {M.~H.}\
  \bibnamefont {Devoret}}, \bibinfo {author} {\bibfnamefont {S.~M.}\
  \bibnamefont {Girvin}}, \ and\ \bibinfo {author} {\bibfnamefont {R.~J.}\
  \bibnamefont {Schoelkopf}}} (\bibinfo {year} {2007}),\ \bibfield  {title}
  {\enquote {\bibinfo {title} {{Resolving photon number states in a
  superconducting circuit}},}\ }\href {\doibase 10.1038/nature05461} {\bibfield
   {journal} {\bibinfo  {journal} {Nature}\ }\textbf {\bibinfo {volume}
  {445}}~(\bibinfo {number} {7127}),\ \bibinfo {pages} {515--518}}\BibitemShut
  {NoStop}%
\bibitem [{\citenamefont {Schuster}\ \emph {et~al.}(2005)\citenamefont
  {Schuster}, \citenamefont {Wallraff}, \citenamefont {Blais}, \citenamefont
  {Frunzio}, \citenamefont {Huang}, \citenamefont {Majer}, \citenamefont
  {Girvin},\ and\ \citenamefont {Schoelkopf}}]{Schuster2005-ACStark}%
  \BibitemOpen
  \bibfield  {author} {\bibinfo {author} {\bibfnamefont {D.~I.}\ \bibnamefont
  {Schuster}}, \bibinfo {author} {\bibfnamefont {A.}~\bibnamefont {Wallraff}},
  \bibinfo {author} {\bibfnamefont {A.}~\bibnamefont {Blais}}, \bibinfo
  {author} {\bibfnamefont {L.}~\bibnamefont {Frunzio}}, \bibinfo {author}
  {\bibfnamefont {R.-S.}\ \bibnamefont {Huang}}, \bibinfo {author}
  {\bibfnamefont {J.}~\bibnamefont {Majer}}, \bibinfo {author} {\bibfnamefont
  {S.~M.}\ \bibnamefont {Girvin}}, \ and\ \bibinfo {author} {\bibfnamefont
  {R.~J.}\ \bibnamefont {Schoelkopf}}} (\bibinfo {year} {2005}),\ \bibfield
  {title} {\enquote {\bibinfo {title} {{ac Stark Shift and Dephasing of a
  Superconducting Qubit Strongly Coupled to a Cavity Field}},}\ }\href
  {\doibase 10.1103/PhysRevLett.94.123602} {\bibfield  {journal} {\bibinfo
  {journal} {Physical Review Letters}\ }\textbf {\bibinfo {volume}
  {94}}~(\bibinfo {number} {12}),\ \bibinfo {pages} {123602}}\BibitemShut
  {NoStop}%
\bibitem [{\citenamefont {Sears}\ \emph {et~al.}(2012)\citenamefont {Sears},
  \citenamefont {Petrenko}, \citenamefont {Catelani}, \citenamefont {Sun},
  \citenamefont {Paik}, \citenamefont {Kirchmair}, \citenamefont {Frunzio},
  \citenamefont {Glazman}, \citenamefont {Girvin},\ and\ \citenamefont
  {Schoelkopf}}]{Sears2012-PhotonShot}%
  \BibitemOpen
  \bibfield  {author} {\bibinfo {author} {\bibfnamefont {A.~P.}\ \bibnamefont
  {Sears}}, \bibinfo {author} {\bibfnamefont {A.}~\bibnamefont {Petrenko}},
  \bibinfo {author} {\bibfnamefont {G.}~\bibnamefont {Catelani}}, \bibinfo
  {author} {\bibfnamefont {L.}~\bibnamefont {Sun}}, \bibinfo {author}
  {\bibfnamefont {H.}~\bibnamefont {Paik}}, \bibinfo {author} {\bibfnamefont
  {G.}~\bibnamefont {Kirchmair}}, \bibinfo {author} {\bibfnamefont
  {L.}~\bibnamefont {Frunzio}}, \bibinfo {author} {\bibfnamefont {L.~I.}\
  \bibnamefont {Glazman}}, \bibinfo {author} {\bibfnamefont {S.~M.}\
  \bibnamefont {Girvin}}, \ and\ \bibinfo {author} {\bibfnamefont {R.~J.}\
  \bibnamefont {Schoelkopf}}} (\bibinfo {year} {2012}),\ \bibfield  {title}
  {\enquote {\bibinfo {title} {{Photon shot noise dephasing in the
  strong-dispersive limit of circuit QED}},}\ }\href {\doibase
  10.1103/PhysRevB.86.180504} {\bibfield  {journal} {\bibinfo  {journal}
  {Physical Review B}\ }\textbf {\bibinfo {volume} {86}}~(\bibinfo {number}
  {18}),\ \bibinfo {pages} {180504}}\BibitemShut {NoStop}%
\bibitem [{\citenamefont {Serniak}\ \emph {et~al.}(2018)\citenamefont
  {Serniak}, \citenamefont {Hays}, \citenamefont {de~Lange}, \citenamefont
  {Diamond}, \citenamefont {Shankar}, \citenamefont {Burkhart}, \citenamefont
  {Frunzio}, \citenamefont {Houzet},\ and\ \citenamefont
  {Devoret}}]{Serniak2018}%
  \BibitemOpen
  \bibfield  {author} {\bibinfo {author} {\bibfnamefont {K.}~\bibnamefont
  {Serniak}}, \bibinfo {author} {\bibfnamefont {M.}~\bibnamefont {Hays}},
  \bibinfo {author} {\bibfnamefont {G.}~\bibnamefont {de~Lange}}, \bibinfo
  {author} {\bibfnamefont {S.}~\bibnamefont {Diamond}}, \bibinfo {author}
  {\bibfnamefont {S.}~\bibnamefont {Shankar}}, \bibinfo {author} {\bibfnamefont
  {L.~D.}\ \bibnamefont {Burkhart}}, \bibinfo {author} {\bibfnamefont
  {L.}~\bibnamefont {Frunzio}}, \bibinfo {author} {\bibfnamefont
  {M.}~\bibnamefont {Houzet}}, \ and\ \bibinfo {author} {\bibfnamefont {M.~H.}\
  \bibnamefont {Devoret}}} (\bibinfo {year} {2018}),\ \bibfield  {title}
  {\enquote {\bibinfo {title} {{Hot Nonequilibrium Quasiparticles in Transmon
  Qubits}},}\ }\href {\doibase 10.1103/PhysRevLett.121.157701} {\bibfield
  {journal} {\bibinfo  {journal} {Physical Review Letters}\ }\textbf {\bibinfo
  {volume} {121}}~(\bibinfo {number} {15}),\ \bibinfo {pages} {157701}},\
  \Eprint {http://arxiv.org/abs/1803.00476} {arXiv:1803.00476} \BibitemShut
  {NoStop}%
\bibitem [{\citenamefont {Serniak}\ \emph {et~al.}(2015)\citenamefont
  {Serniak}, \citenamefont {Minev}, \citenamefont {Pop}, \citenamefont
  {Frunzio}, \citenamefont {Schoelkopf},\ and\ \citenamefont
  {Devoret}}]{Serniak2015-APSMM}%
  \BibitemOpen
  \bibfield  {author} {\bibinfo {author} {\bibfnamefont {K.}~\bibnamefont
  {Serniak}}, \bibinfo {author} {\bibfnamefont {Z.}~\bibnamefont {Minev}},
  \bibinfo {author} {\bibfnamefont {I.}~\bibnamefont {Pop}}, \bibinfo {author}
  {\bibfnamefont {L.}~\bibnamefont {Frunzio}}, \bibinfo {author} {\bibfnamefont
  {R.}~\bibnamefont {Schoelkopf}}, \ and\ \bibinfo {author} {\bibfnamefont
  {M.}~\bibnamefont {Devoret}}} (\bibinfo {year} {2015}),\ \bibfield  {title}
  {\enquote {\bibinfo {title} {{Fabrication of transmon qubits embedded in
  superconducting whispering gallery mode resonators}},}\ }in\ \href@noop {}
  {\emph {\bibinfo {booktitle} {APS March Meeting Abstracts}}},\ Vol.\ \bibinfo
  {volume} {2015},\ p.\ \bibinfo {pages} {Y39.008}\BibitemShut {NoStop}%
\bibitem [{\citenamefont {Siddiqi}\ \emph {et~al.}(2006)\citenamefont
  {Siddiqi}, \citenamefont {Vijay}, \citenamefont {Metcalfe}, \citenamefont
  {Boaknin}, \citenamefont {Frunzio}, \citenamefont {Schoelkopf},\ and\
  \citenamefont {Devoret}}]{Siddiqi2006}%
  \BibitemOpen
  \bibfield  {author} {\bibinfo {author} {\bibfnamefont {I.}~\bibnamefont
  {Siddiqi}}, \bibinfo {author} {\bibfnamefont {R.}~\bibnamefont {Vijay}},
  \bibinfo {author} {\bibfnamefont {M.}~\bibnamefont {Metcalfe}}, \bibinfo
  {author} {\bibfnamefont {E.}~\bibnamefont {Boaknin}}, \bibinfo {author}
  {\bibfnamefont {L.}~\bibnamefont {Frunzio}}, \bibinfo {author} {\bibfnamefont
  {R.~J.}\ \bibnamefont {Schoelkopf}}, \ and\ \bibinfo {author} {\bibfnamefont
  {M.~H.}\ \bibnamefont {Devoret}}} (\bibinfo {year} {2006}),\ \bibfield
  {title} {\enquote {\bibinfo {title} {{Dispersive measurements of
  superconducting qubit coherence with a fast latching readout}},}\ }\href
  {\doibase 10.1103/PhysRevB.73.054510} {\bibfield  {journal} {\bibinfo
  {journal} {Physical Review B}\ }\textbf {\bibinfo {volume} {73}}~(\bibinfo
  {number} {5}),\ \bibinfo {pages} {054510}}\BibitemShut {NoStop}%
\bibitem [{\citenamefont {Slichter}\ \emph {et~al.}(2016)\citenamefont
  {Slichter}, \citenamefont {M{\"{u}}ller}, \citenamefont {Vijay},
  \citenamefont {Weber}, \citenamefont {Blais},\ and\ \citenamefont
  {Siddiqi}}]{Slichter2016-T1vsNbar}%
  \BibitemOpen
  \bibfield  {author} {\bibinfo {author} {\bibfnamefont {D.~H.}\ \bibnamefont
  {Slichter}}, \bibinfo {author} {\bibfnamefont {C.}~\bibnamefont
  {M{\"{u}}ller}}, \bibinfo {author} {\bibfnamefont {R.}~\bibnamefont {Vijay}},
  \bibinfo {author} {\bibfnamefont {S.~J.}\ \bibnamefont {Weber}}, \bibinfo
  {author} {\bibfnamefont {A.}~\bibnamefont {Blais}}, \ and\ \bibinfo {author}
  {\bibfnamefont {I.}~\bibnamefont {Siddiqi}}} (\bibinfo {year} {2016}),\
  \bibfield  {title} {\enquote {\bibinfo {title} {{Quantum Zeno effect in the
  strong measurement regime of circuit quantum electrodynamics}},}\ }\href
  {http://stacks.iop.org/1367-2630/18/i=5/a=053031} {\bibfield  {journal}
  {\bibinfo  {journal} {New Journal of Physics}\ }\textbf {\bibinfo {volume}
  {18}}~(\bibinfo {number} {5}),\ \bibinfo {pages} {53031}}\BibitemShut
  {NoStop}%
\bibitem [{\citenamefont {Slichter}\ \emph {et~al.}(2012)\citenamefont
  {Slichter}, \citenamefont {Vijay}, \citenamefont {Weber}, \citenamefont
  {Boutin}, \citenamefont {Boissonneault}, \citenamefont {Gambetta},
  \citenamefont {Blais},\ and\ \citenamefont {Siddiqi}}]{Slichter2012}%
  \BibitemOpen
  \bibfield  {author} {\bibinfo {author} {\bibfnamefont {D.~H.}\ \bibnamefont
  {Slichter}}, \bibinfo {author} {\bibfnamefont {R.}~\bibnamefont {Vijay}},
  \bibinfo {author} {\bibfnamefont {S.~J.}\ \bibnamefont {Weber}}, \bibinfo
  {author} {\bibfnamefont {S.}~\bibnamefont {Boutin}}, \bibinfo {author}
  {\bibfnamefont {M.}~\bibnamefont {Boissonneault}}, \bibinfo {author}
  {\bibfnamefont {J.~M.}\ \bibnamefont {Gambetta}}, \bibinfo {author}
  {\bibfnamefont {A.}~\bibnamefont {Blais}}, \ and\ \bibinfo {author}
  {\bibfnamefont {I.}~\bibnamefont {Siddiqi}}} (\bibinfo {year} {2012}),\
  \bibfield  {title} {\enquote {\bibinfo {title} {{Measurement-Induced Qubit
  State Mixing in Circuit QED from Up-Converted Dephasing Noise}},}\ }\href
  {\doibase 10.1103/PhysRevLett.109.153601} {\bibfield  {journal} {\bibinfo
  {journal} {Physical Review Letters}\ }\textbf {\bibinfo {volume}
  {109}}~(\bibinfo {number} {15}),\ \bibinfo {pages} {153601}}\BibitemShut
  {NoStop}%
\bibitem [{\citenamefont {Smith}\ \emph {et~al.}(2016)\citenamefont {Smith},
  \citenamefont {Kou}, \citenamefont {Vool}, \citenamefont {Pop}, \citenamefont
  {Frunzio}, \citenamefont {Schoelkopf},\ and\ \citenamefont
  {Devoret}}]{Smith2016}%
  \BibitemOpen
  \bibfield  {author} {\bibinfo {author} {\bibfnamefont {W.~C.}\ \bibnamefont
  {Smith}}, \bibinfo {author} {\bibfnamefont {A.}~\bibnamefont {Kou}}, \bibinfo
  {author} {\bibfnamefont {U.}~\bibnamefont {Vool}}, \bibinfo {author}
  {\bibfnamefont {I.~M.}\ \bibnamefont {Pop}}, \bibinfo {author} {\bibfnamefont
  {L.}~\bibnamefont {Frunzio}}, \bibinfo {author} {\bibfnamefont {R.~J.}\
  \bibnamefont {Schoelkopf}}, \ and\ \bibinfo {author} {\bibfnamefont {M.~H.}\
  \bibnamefont {Devoret}}} (\bibinfo {year} {2016}),\ \bibfield  {title}
  {\enquote {\bibinfo {title} {{Quantization of inductively shunted
  superconducting circuits}},}\ }\href {\doibase 10.1103/PhysRevB.94.144507}
  {\bibfield  {journal} {\bibinfo  {journal} {Physical Review B}\ }\textbf
  {\bibinfo {volume} {94}}~(\bibinfo {number} {14}),\ \bibinfo {pages}
  {144507}},\ \Eprint {http://arxiv.org/abs/1602.01793} {arXiv:1602.01793}
  \BibitemShut {NoStop}%
\bibitem [{\citenamefont {Solgun}\ \emph {et~al.}(2014)\citenamefont {Solgun},
  \citenamefont {Abraham},\ and\ \citenamefont {DiVincenzo}}]{Solgun2014}%
  \BibitemOpen
  \bibfield  {author} {\bibinfo {author} {\bibfnamefont {F.}~\bibnamefont
  {Solgun}}, \bibinfo {author} {\bibfnamefont {D.~W.}\ \bibnamefont {Abraham}},
  \ and\ \bibinfo {author} {\bibfnamefont {D.~P.}\ \bibnamefont {DiVincenzo}}}
  (\bibinfo {year} {2014}),\ \bibfield  {title} {\enquote {\bibinfo {title}
  {{Blackbox quantization of superconducting circuits using exact impedance
  synthesis}},}\ }\href {\doibase 10.1103/PhysRevB.90.134504} {\bibfield
  {journal} {\bibinfo  {journal} {Physical Review B}\ }\textbf {\bibinfo
  {volume} {90}}~(\bibinfo {number} {13}),\ \bibinfo {pages}
  {134504}}\BibitemShut {NoStop}%
\bibitem [{\citenamefont {Solgun}\ and\ \citenamefont
  {DiVincenzo}(2015)}]{Solgun2015}%
  \BibitemOpen
  \bibfield  {author} {\bibinfo {author} {\bibfnamefont {F.}~\bibnamefont
  {Solgun}}, \ and\ \bibinfo {author} {\bibfnamefont {D.~P.}\ \bibnamefont
  {DiVincenzo}}} (\bibinfo {year} {2015}),\ \bibfield  {title} {\enquote
  {\bibinfo {title} {{Multiport impedance quantization}},}\ }\href {\doibase
  10.1016/j.aop.2015.07.005} {\bibfield  {journal} {\bibinfo  {journal} {Annals
  of Physics}\ }\textbf {\bibinfo {volume} {361}},\ \bibinfo {pages}
  {605--669}},\ \Eprint {http://arxiv.org/abs/1505.04116} {arXiv:1505.04116}
  \BibitemShut {NoStop}%
\bibitem [{\citenamefont {Srinivasan}\ \emph {et~al.}(2011)\citenamefont
  {Srinivasan}, \citenamefont {Hoffman}, \citenamefont {Gambetta},\ and\
  \citenamefont {Houck}}]{Srinivasan2011}%
  \BibitemOpen
  \bibfield  {author} {\bibinfo {author} {\bibfnamefont {S.~J.}\ \bibnamefont
  {Srinivasan}}, \bibinfo {author} {\bibfnamefont {A.~J.}\ \bibnamefont
  {Hoffman}}, \bibinfo {author} {\bibfnamefont {J.~M.}\ \bibnamefont
  {Gambetta}}, \ and\ \bibinfo {author} {\bibfnamefont {A.~A.}\ \bibnamefont
  {Houck}}} (\bibinfo {year} {2011}),\ \bibfield  {title} {\enquote {\bibinfo
  {title} {{Tunable Coupling in Circuit Quantum Electrodynamics Using a
  Superconducting Charge Qubit with a V-Shaped Energy Level Diagram}},}\ }\href
  {\doibase 10.1103/PhysRevLett.106.083601} {\bibfield  {journal} {\bibinfo
  {journal} {Physical Review Letters}\ }\textbf {\bibinfo {volume}
  {106}}~(\bibinfo {number} {8}),\ \bibinfo {pages} {083601}},\ \Eprint
  {http://arxiv.org/abs/1011.4317} {arXiv:1011.4317} \BibitemShut {NoStop}%
\bibitem [{\citenamefont {Steck}(2017)}]{SteckQuantumNotes}%
  \BibitemOpen
  \bibfield  {author} {\bibinfo {author} {\bibfnamefont {D.~A.}\ \bibnamefont
  {Steck}}} (\bibinfo {year} {2017}),\ \href
  {http://atomoptics-nas.uoregon.edu/{~}dsteck/teaching/quantum-optics/}
  {\enquote {\bibinfo {title} {{Quantum and Atom Optics}},}\ }\bibinfo
  {howpublished} {available online at http://steck.us/teaching}\BibitemShut
  {NoStop}%
\bibitem [{\citenamefont {Sun}\ \emph {et~al.}(2013)\citenamefont {Sun},
  \citenamefont {Petrenko}, \citenamefont {Leghtas}, \citenamefont {Vlastakis},
  \citenamefont {Kirchmair}, \citenamefont {Sliwa}, \citenamefont {Narla},
  \citenamefont {Hatridge}, \citenamefont {Shankar}, \citenamefont {Blumoff},
  \citenamefont {Frunzio}, \citenamefont {Mirrahimi}, \citenamefont {Devoret},\
  and\ \citenamefont {Schoelkopf}}]{Sun2013}%
  \BibitemOpen
  \bibfield  {author} {\bibinfo {author} {\bibfnamefont {L.}~\bibnamefont
  {Sun}}, \bibinfo {author} {\bibfnamefont {A.}~\bibnamefont {Petrenko}},
  \bibinfo {author} {\bibfnamefont {Z.}~\bibnamefont {Leghtas}}, \bibinfo
  {author} {\bibfnamefont {B.}~\bibnamefont {Vlastakis}}, \bibinfo {author}
  {\bibfnamefont {G.}~\bibnamefont {Kirchmair}}, \bibinfo {author}
  {\bibfnamefont {K.~M.}\ \bibnamefont {Sliwa}}, \bibinfo {author}
  {\bibfnamefont {A.}~\bibnamefont {Narla}}, \bibinfo {author} {\bibfnamefont
  {M.}~\bibnamefont {Hatridge}}, \bibinfo {author} {\bibfnamefont
  {S.}~\bibnamefont {Shankar}}, \bibinfo {author} {\bibfnamefont
  {J.}~\bibnamefont {Blumoff}}, \bibinfo {author} {\bibfnamefont
  {L.}~\bibnamefont {Frunzio}}, \bibinfo {author} {\bibfnamefont
  {M.}~\bibnamefont {Mirrahimi}}, \bibinfo {author} {\bibfnamefont {M.~H.}\
  \bibnamefont {Devoret}}, \ and\ \bibinfo {author} {\bibfnamefont {R.~J.}\
  \bibnamefont {Schoelkopf}}} (\bibinfo {year} {2013}),\ \bibfield  {title}
  {\enquote {\bibinfo {title} {{Tracking Photon Jumps with Repeated Quantum
  Non-Demolition Parity Measurements}},}\ }\href {\doibase 10.1038/nature13436}
  {\bibfield  {journal} {\bibinfo  {journal} {Nature}\ }\textbf {\bibinfo
  {volume} {511}}~(\bibinfo {number} {7510}),\ \bibinfo {pages} {444--448}},\
  \Eprint {http://arxiv.org/abs/1311.2534} {arXiv:1311.2534} \BibitemShut
  {NoStop}%
\bibitem [{\citenamefont {Szangolies}(2015)}]{Szangolies2015-book}%
  \BibitemOpen
  \bibfield  {author} {\bibinfo {author} {\bibfnamefont {J.}~\bibnamefont
  {Szangolies}}} (\bibinfo {year} {2015}),\ \href {\doibase
  10.1007/978-3-658-09200-9} {\emph {\bibinfo {title} {{Testing Quantum
  Contextuality}}}}\ (\bibinfo  {publisher} {Springer Fachmedien Wiesbaden},\
  \bibinfo {address} {Wiesbaden})\BibitemShut {NoStop}%
\bibitem [{\citenamefont {Tan}\ \emph {et~al.}(2017)\citenamefont {Tan},
  \citenamefont {Foroozani}, \citenamefont {Naghiloo}, \citenamefont
  {Kiilerich}, \citenamefont {M{\o}lmer},\ and\ \citenamefont
  {Murch}}]{Tan2017}%
  \BibitemOpen
  \bibfield  {author} {\bibinfo {author} {\bibfnamefont {D.}~\bibnamefont
  {Tan}}, \bibinfo {author} {\bibfnamefont {N.}~\bibnamefont {Foroozani}},
  \bibinfo {author} {\bibfnamefont {M.}~\bibnamefont {Naghiloo}}, \bibinfo
  {author} {\bibfnamefont {A.~H.}\ \bibnamefont {Kiilerich}}, \bibinfo {author}
  {\bibfnamefont {K.}~\bibnamefont {M{\o}lmer}}, \ and\ \bibinfo {author}
  {\bibfnamefont {K.~W.}\ \bibnamefont {Murch}}} (\bibinfo {year} {2017}),\
  \bibfield  {title} {\enquote {\bibinfo {title} {{Homodyne monitoring of
  postselected decay}},}\ }\href {\doibase 10.1103/PhysRevA.96.022104}
  {\bibfield  {journal} {\bibinfo  {journal} {Physical Review A}\ }\textbf
  {\bibinfo {volume} {96}}~(\bibinfo {number} {2}),\ \bibinfo {pages}
  {022104}},\ \Eprint {http://arxiv.org/abs/1705.04287} {arXiv:1705.04287}
  \BibitemShut {NoStop}%
\bibitem [{\citenamefont {Tilloy}\ \emph {et~al.}(2015)\citenamefont {Tilloy},
  \citenamefont {Bauer},\ and\ \citenamefont {Bernard}}]{Tilloy2015}%
  \BibitemOpen
  \bibfield  {author} {\bibinfo {author} {\bibfnamefont {A.}~\bibnamefont
  {Tilloy}}, \bibinfo {author} {\bibfnamefont {M.}~\bibnamefont {Bauer}}, \
  and\ \bibinfo {author} {\bibfnamefont {D.}~\bibnamefont {Bernard}}} (\bibinfo
  {year} {2015}),\ \bibfield  {title} {\enquote {\bibinfo {title} {{Spikes in
  quantum trajectories}},}\ }\href {\doibase 10.1103/PhysRevA.92.052111}
  {\bibfield  {journal} {\bibinfo  {journal} {Physical Review A}\ }\textbf
  {\bibinfo {volume} {92}}~(\bibinfo {number} {5}),\ \bibinfo {pages}
  {052111}},\ \Eprint {http://arxiv.org/abs/1510.01232} {arXiv:1510.01232}
  \BibitemShut {NoStop}%
\bibitem [{\citenamefont {Touzard}\ \emph {et~al.}(2018)\citenamefont
  {Touzard}, \citenamefont {Grimm}, \citenamefont {Leghtas}, \citenamefont
  {Mundhada}, \citenamefont {Reinhold}, \citenamefont {Axline}, \citenamefont
  {Reagor}, \citenamefont {Chou}, \citenamefont {Blumoff}, \citenamefont
  {Sliwa}, \citenamefont {Shankar}, \citenamefont {Frunzio}, \citenamefont
  {Schoelkopf}, \citenamefont {Mirrahimi},\ and\ \citenamefont
  {Devoret}}]{Touzard2017}%
  \BibitemOpen
  \bibfield  {author} {\bibinfo {author} {\bibfnamefont {S.}~\bibnamefont
  {Touzard}}, \bibinfo {author} {\bibfnamefont {A.}~\bibnamefont {Grimm}},
  \bibinfo {author} {\bibfnamefont {Z.}~\bibnamefont {Leghtas}}, \bibinfo
  {author} {\bibfnamefont {S.~O.}\ \bibnamefont {Mundhada}}, \bibinfo {author}
  {\bibfnamefont {P.}~\bibnamefont {Reinhold}}, \bibinfo {author}
  {\bibfnamefont {C.}~\bibnamefont {Axline}}, \bibinfo {author} {\bibfnamefont
  {M.}~\bibnamefont {Reagor}}, \bibinfo {author} {\bibfnamefont
  {K.}~\bibnamefont {Chou}}, \bibinfo {author} {\bibfnamefont {J.}~\bibnamefont
  {Blumoff}}, \bibinfo {author} {\bibfnamefont {K.~M.}\ \bibnamefont {Sliwa}},
  \bibinfo {author} {\bibfnamefont {S.}~\bibnamefont {Shankar}}, \bibinfo
  {author} {\bibfnamefont {L.}~\bibnamefont {Frunzio}}, \bibinfo {author}
  {\bibfnamefont {R.~J.}\ \bibnamefont {Schoelkopf}}, \bibinfo {author}
  {\bibfnamefont {M.}~\bibnamefont {Mirrahimi}}, \ and\ \bibinfo {author}
  {\bibfnamefont {M.~H.}\ \bibnamefont {Devoret}}} (\bibinfo {year} {2018}),\
  \bibfield  {title} {\enquote {\bibinfo {title} {{Coherent Oscillations inside
  a Quantum Manifold Stabilized by Dissipation}},}\ }\href {\doibase
  10.1103/PhysRevX.8.021005} {\bibfield  {journal} {\bibinfo  {journal}
  {Physical Review X}\ }\textbf {\bibinfo {volume} {8}}~(\bibinfo {number}
  {2}),\ 10.1103/PhysRevX.8.021005},\ \Eprint {http://arxiv.org/abs/1705.02401}
  {arXiv:1705.02401} \BibitemShut {NoStop}%
\bibitem [{\citenamefont {Valenzuela}\ \emph {et~al.}(2006)\citenamefont
  {Valenzuela}, \citenamefont {Oliver}, \citenamefont {Berns}, \citenamefont
  {Berggren}, \citenamefont {Levitov},\ and\ \citenamefont
  {Orlando}}]{Valenzuela2006}%
  \BibitemOpen
  \bibfield  {author} {\bibinfo {author} {\bibfnamefont {S.~O.}\ \bibnamefont
  {Valenzuela}}, \bibinfo {author} {\bibfnamefont {W.~D.}\ \bibnamefont
  {Oliver}}, \bibinfo {author} {\bibfnamefont {D.~M.}\ \bibnamefont {Berns}},
  \bibinfo {author} {\bibfnamefont {K.~K.}\ \bibnamefont {Berggren}}, \bibinfo
  {author} {\bibfnamefont {L.~S.}\ \bibnamefont {Levitov}}, \ and\ \bibinfo
  {author} {\bibfnamefont {T.~P.}\ \bibnamefont {Orlando}}} (\bibinfo {year}
  {2006}),\ \bibfield  {title} {\enquote {\bibinfo {title} {{Microwave-Induced
  Cooling of a Superconducting Qubit}},}\ }\href {\doibase
  10.1126/science.1134008} {\bibfield  {journal} {\bibinfo  {journal}
  {Science}\ }\textbf {\bibinfo {volume} {314}}~(\bibinfo {number} {5805}),\
  \bibinfo {pages} {1589--1592}}\BibitemShut {NoStop}%
\bibitem [{\citenamefont {Ventura}\ and\ \citenamefont
  {Risegari}(2010)}]{ventura2010book}%
  \BibitemOpen
  \bibfield  {author} {\bibinfo {author} {\bibfnamefont {G.}~\bibnamefont
  {Ventura}}, \ and\ \bibinfo {author} {\bibfnamefont {L.}~\bibnamefont
  {Risegari}}} (\bibinfo {year} {2010}),\ \href
  {https://books.google.com/books?id=4kvzBRUuGDkC} {\emph {\bibinfo {title}
  {{The Art of Cryogenics: Low-Temperature Experimental Techniques}}}}\
  (\bibinfo  {publisher} {Elsevier Science})\BibitemShut {NoStop}%
\bibitem [{\citenamefont {Verney}\ \emph {et~al.}(2019)\citenamefont {Verney},
  \citenamefont {Lescanne}, \citenamefont {Devoret}, \citenamefont {Leghtas},\
  and\ \citenamefont {Mirrahimi}}]{Verney2018}%
  \BibitemOpen
  \bibfield  {author} {\bibinfo {author} {\bibfnamefont {L.}~\bibnamefont
  {Verney}}, \bibinfo {author} {\bibfnamefont {R.}~\bibnamefont {Lescanne}},
  \bibinfo {author} {\bibfnamefont {M.~H.}\ \bibnamefont {Devoret}}, \bibinfo
  {author} {\bibfnamefont {Z.}~\bibnamefont {Leghtas}}, \ and\ \bibinfo
  {author} {\bibfnamefont {M.}~\bibnamefont {Mirrahimi}}} (\bibinfo {year}
  {2019}),\ \bibfield  {title} {\enquote {\bibinfo {title} {{Structural
  Instability of Driven Josephson Circuits Prevented by an Inductive Shunt}},}\
  }\href {\doibase 10.1103/PhysRevApplied.11.024003} {\bibfield  {journal}
  {\bibinfo  {journal} {Physical Review Applied}\ }\textbf {\bibinfo {volume}
  {11}}~(\bibinfo {number} {2}),\ \bibinfo {pages} {024003}},\ \Eprint
  {http://arxiv.org/abs/1805.07542} {arXiv:1805.07542} \BibitemShut {NoStop}%
\bibitem [{\citenamefont {Vijay}\ \emph {et~al.}(2012)\citenamefont {Vijay},
  \citenamefont {Macklin}, \citenamefont {Slichter}, \citenamefont {Weber},
  \citenamefont {Murch}, \citenamefont {Naik}, \citenamefont {Korotkov},\ and\
  \citenamefont {Siddiqi}}]{Vijay2012}%
  \BibitemOpen
  \bibfield  {author} {\bibinfo {author} {\bibfnamefont {R.}~\bibnamefont
  {Vijay}}, \bibinfo {author} {\bibfnamefont {C.}~\bibnamefont {Macklin}},
  \bibinfo {author} {\bibfnamefont {D.~H.}\ \bibnamefont {Slichter}}, \bibinfo
  {author} {\bibfnamefont {S.~J.}\ \bibnamefont {Weber}}, \bibinfo {author}
  {\bibfnamefont {K.~W.}\ \bibnamefont {Murch}}, \bibinfo {author}
  {\bibfnamefont {R.}~\bibnamefont {Naik}}, \bibinfo {author} {\bibfnamefont
  {A.~N.}\ \bibnamefont {Korotkov}}, \ and\ \bibinfo {author} {\bibfnamefont
  {I.}~\bibnamefont {Siddiqi}}} (\bibinfo {year} {2012}),\ \bibfield  {title}
  {\enquote {\bibinfo {title} {{Stabilizing Rabi oscillations in a
  superconducting qubit using quantum feedback}},}\ }\href {\doibase
  10.1038/nature11505} {\bibfield  {journal} {\bibinfo  {journal} {Nature}\
  }\textbf {\bibinfo {volume} {490}}~(\bibinfo {number} {7418}),\ \bibinfo
  {pages} {77--80}},\ \Eprint {http://arxiv.org/abs/1205.5591}
  {arXiv:1205.5591} \BibitemShut {NoStop}%
\bibitem [{\citenamefont {Vijay}\ \emph {et~al.}(2011)\citenamefont {Vijay},
  \citenamefont {Slichter},\ and\ \citenamefont {Siddiqi}}]{Vijay2011}%
  \BibitemOpen
  \bibfield  {author} {\bibinfo {author} {\bibfnamefont {R.}~\bibnamefont
  {Vijay}}, \bibinfo {author} {\bibfnamefont {D.~H.}\ \bibnamefont {Slichter}},
  \ and\ \bibinfo {author} {\bibfnamefont {I.}~\bibnamefont {Siddiqi}}}
  (\bibinfo {year} {2011}),\ \bibfield  {title} {\enquote {\bibinfo {title}
  {{Observation of Quantum Jumps in a Superconducting Artificial Atom}},}\
  }\href {\doibase 10.1103/PhysRevLett.106.110502} {\bibfield  {journal}
  {\bibinfo  {journal} {Physical Review Letters}\ }\textbf {\bibinfo {volume}
  {106}}~(\bibinfo {number} {11}),\ \bibinfo {pages} {110502}},\ \Eprint
  {http://arxiv.org/abs/1009.2969} {arXiv:1009.2969} \BibitemShut {NoStop}%
\bibitem [{\citenamefont {de~Visser}\ \emph {et~al.}(2011)\citenamefont
  {de~Visser}, \citenamefont {Baselmans}, \citenamefont {Diener}, \citenamefont
  {Yates}, \citenamefont {Endo},\ and\ \citenamefont
  {Klapwijk}}]{deVisser2011-qp}%
  \BibitemOpen
  \bibfield  {author} {\bibinfo {author} {\bibfnamefont {P.~J.}\ \bibnamefont
  {de~Visser}}, \bibinfo {author} {\bibfnamefont {J.~J.~A.}\ \bibnamefont
  {Baselmans}}, \bibinfo {author} {\bibfnamefont {P.}~\bibnamefont {Diener}},
  \bibinfo {author} {\bibfnamefont {S.~J.~C.}\ \bibnamefont {Yates}}, \bibinfo
  {author} {\bibfnamefont {A.}~\bibnamefont {Endo}}, \ and\ \bibinfo {author}
  {\bibfnamefont {T.~M.}\ \bibnamefont {Klapwijk}}} (\bibinfo {year} {2011}),\
  \bibfield  {title} {\enquote {\bibinfo {title} {{Number Fluctuations of
  Sparse Quasiparticles in a Superconductor}},}\ }\href {\doibase
  10.1103/PhysRevLett.106.167004} {\bibfield  {journal} {\bibinfo  {journal}
  {Physical Review Letters}\ }\textbf {\bibinfo {volume} {106}}~(\bibinfo
  {number} {16}),\ \bibinfo {pages} {167004}}\BibitemShut {NoStop}%
\bibitem [{\citenamefont {Volz}\ \emph {et~al.}(2011)\citenamefont {Volz},
  \citenamefont {Gehr}, \citenamefont {Dubois}, \citenamefont {Esteve},\ and\
  \citenamefont {Reichel}}]{Volz2011}%
  \BibitemOpen
  \bibfield  {author} {\bibinfo {author} {\bibfnamefont {J.}~\bibnamefont
  {Volz}}, \bibinfo {author} {\bibfnamefont {R.}~\bibnamefont {Gehr}}, \bibinfo
  {author} {\bibfnamefont {G.}~\bibnamefont {Dubois}}, \bibinfo {author}
  {\bibfnamefont {J.}~\bibnamefont {Esteve}}, \ and\ \bibinfo {author}
  {\bibfnamefont {J.}~\bibnamefont {Reichel}}} (\bibinfo {year} {2011}),\
  \bibfield  {title} {\enquote {\bibinfo {title} {{Measurement of the internal
  state of a single atom without energy exchange}},}\ }\href
  {http://dx.doi.org/10.1038/nature10225} {\bibfield  {journal} {\bibinfo
  {journal} {Nature}\ }\textbf {\bibinfo {volume} {475}}~(\bibinfo {number}
  {7355}),\ \bibinfo {pages} {210--213}}\BibitemShut {NoStop}%
\bibitem [{\citenamefont {Vool}\ and\ \citenamefont
  {Devoret}(2017)}]{Vool2017}%
  \BibitemOpen
  \bibfield  {author} {\bibinfo {author} {\bibfnamefont {U.}~\bibnamefont
  {Vool}}, \ and\ \bibinfo {author} {\bibfnamefont {M.}~\bibnamefont
  {Devoret}}} (\bibinfo {year} {2017}),\ \href {\doibase 10.1002/cta.2359}
  {\enquote {\bibinfo {title} {{Introduction to quantum electromagnetic
  circuits}},}\ }\Eprint {http://arxiv.org/abs/1610.03438} {arXiv:1610.03438}
  \BibitemShut {NoStop}%
\bibitem [{\citenamefont {Vool}\ \emph {et~al.}(2014)\citenamefont {Vool},
  \citenamefont {Pop}, \citenamefont {Sliwa}, \citenamefont {Abdo},
  \citenamefont {Wang}, \citenamefont {Brecht}, \citenamefont {Gao},
  \citenamefont {Shankar}, \citenamefont {Hatridge}, \citenamefont {Catelani},
  \citenamefont {Mirrahimi}, \citenamefont {Frunzio}, \citenamefont
  {Schoelkopf}, \citenamefont {Glazman},\ and\ \citenamefont
  {Devoret}}]{Vool2014-qp}%
  \BibitemOpen
  \bibfield  {author} {\bibinfo {author} {\bibfnamefont {U.}~\bibnamefont
  {Vool}}, \bibinfo {author} {\bibfnamefont {I.~M.}\ \bibnamefont {Pop}},
  \bibinfo {author} {\bibfnamefont {K.}~\bibnamefont {Sliwa}}, \bibinfo
  {author} {\bibfnamefont {B.}~\bibnamefont {Abdo}}, \bibinfo {author}
  {\bibfnamefont {C.}~\bibnamefont {Wang}}, \bibinfo {author} {\bibfnamefont
  {T.}~\bibnamefont {Brecht}}, \bibinfo {author} {\bibfnamefont {Y.~Y.}\
  \bibnamefont {Gao}}, \bibinfo {author} {\bibfnamefont {S.}~\bibnamefont
  {Shankar}}, \bibinfo {author} {\bibfnamefont {M.}~\bibnamefont {Hatridge}},
  \bibinfo {author} {\bibfnamefont {G.}~\bibnamefont {Catelani}}, \bibinfo
  {author} {\bibfnamefont {M.}~\bibnamefont {Mirrahimi}}, \bibinfo {author}
  {\bibfnamefont {L.}~\bibnamefont {Frunzio}}, \bibinfo {author} {\bibfnamefont
  {R.~J.}\ \bibnamefont {Schoelkopf}}, \bibinfo {author} {\bibfnamefont
  {L.~I.}\ \bibnamefont {Glazman}}, \ and\ \bibinfo {author} {\bibfnamefont
  {M.~H.}\ \bibnamefont {Devoret}}} (\bibinfo {year} {2014}),\ \bibfield
  {title} {\enquote {\bibinfo {title} {{Non-Poissonian Quantum Jumps of a
  Fluxonium Qubit due to Quasiparticle Excitations}},}\ }\href {\doibase
  10.1103/PhysRevLett.113.247001} {\bibfield  {journal} {\bibinfo  {journal}
  {Physical Review Letters}\ }\textbf {\bibinfo {volume} {113}}~(\bibinfo
  {number} {24}),\ \bibinfo {pages} {247001}}\BibitemShut {NoStop}%
\bibitem [{\citenamefont {Wallraff}\ \emph {et~al.}(2004)\citenamefont
  {Wallraff}, \citenamefont {Schuster}, \citenamefont {Blais}, \citenamefont
  {Frunzio}, \citenamefont {Huang}, \citenamefont {Majer}, \citenamefont
  {Kumar}, \citenamefont {Girvin},\ and\ \citenamefont
  {Schoelkopf}}]{Wallraff2004}%
  \BibitemOpen
  \bibfield  {author} {\bibinfo {author} {\bibfnamefont {A.}~\bibnamefont
  {Wallraff}}, \bibinfo {author} {\bibfnamefont {D.~I.}\ \bibnamefont
  {Schuster}}, \bibinfo {author} {\bibfnamefont {A.}~\bibnamefont {Blais}},
  \bibinfo {author} {\bibfnamefont {L.}~\bibnamefont {Frunzio}}, \bibinfo
  {author} {\bibfnamefont {R.-S.}\ \bibnamefont {Huang}}, \bibinfo {author}
  {\bibfnamefont {J.}~\bibnamefont {Majer}}, \bibinfo {author} {\bibfnamefont
  {S.}~\bibnamefont {Kumar}}, \bibinfo {author} {\bibfnamefont {S.~M.}\
  \bibnamefont {Girvin}}, \ and\ \bibinfo {author} {\bibfnamefont {R.~J.}\
  \bibnamefont {Schoelkopf}}} (\bibinfo {year} {2004}),\ \bibfield  {title}
  {\enquote {\bibinfo {title} {{Strong coupling of a single photon to a
  superconducting qubit using circuit quantum electrodynamics}},}\ }\href
  {\doibase 10.1038/nature02851} {\bibfield  {journal} {\bibinfo  {journal}
  {Nature}\ }\textbf {\bibinfo {volume} {431}}~(\bibinfo {number} {7005}),\
  \bibinfo {pages} {162--167}},\ \Eprint {http://arxiv.org/abs/0407325}
  {arXiv:0407325 [cond-mat]} \BibitemShut {NoStop}%
\bibitem [{\citenamefont {Wallraff}\ \emph {et~al.}(2005)\citenamefont
  {Wallraff}, \citenamefont {Schuster}, \citenamefont {Blais}, \citenamefont
  {Frunzio}, \citenamefont {Majer}, \citenamefont {Devoret}, \citenamefont
  {Girvin},\ and\ \citenamefont {Schoelkopf}}]{Wallraff2005-readout}%
  \BibitemOpen
  \bibfield  {author} {\bibinfo {author} {\bibfnamefont {A.}~\bibnamefont
  {Wallraff}}, \bibinfo {author} {\bibfnamefont {D.~I.}\ \bibnamefont
  {Schuster}}, \bibinfo {author} {\bibfnamefont {A.}~\bibnamefont {Blais}},
  \bibinfo {author} {\bibfnamefont {L.}~\bibnamefont {Frunzio}}, \bibinfo
  {author} {\bibfnamefont {J.}~\bibnamefont {Majer}}, \bibinfo {author}
  {\bibfnamefont {M.~H.}\ \bibnamefont {Devoret}}, \bibinfo {author}
  {\bibfnamefont {S.~M.}\ \bibnamefont {Girvin}}, \ and\ \bibinfo {author}
  {\bibfnamefont {R.~J.}\ \bibnamefont {Schoelkopf}}} (\bibinfo {year}
  {2005}),\ \bibfield  {title} {\enquote {\bibinfo {title} {{Approaching Unit
  Visibility for Control of a Superconducting Qubit with Dispersive
  Readout}},}\ }\href {\doibase 10.1103/PhysRevLett.95.060501} {\bibfield
  {journal} {\bibinfo  {journal} {Physical Review Letters}\ }\textbf {\bibinfo
  {volume} {95}}~(\bibinfo {number} {6}),\ \bibinfo {pages}
  {060501}}\BibitemShut {NoStop}%
\bibitem [{\citenamefont {Walter}\ \emph {et~al.}(2017)\citenamefont {Walter},
  \citenamefont {Kurpiers}, \citenamefont {Gasparinetti}, \citenamefont
  {Magnard}, \citenamefont {Poto{\v{c}}nik}, \citenamefont {Salath{\'{e}}},
  \citenamefont {Pechal}, \citenamefont {Mondal}, \citenamefont {Oppliger},
  \citenamefont {Eichler},\ and\ \citenamefont {Wallraff}}]{Walter2017}%
  \BibitemOpen
  \bibfield  {author} {\bibinfo {author} {\bibfnamefont {T.}~\bibnamefont
  {Walter}}, \bibinfo {author} {\bibfnamefont {P.}~\bibnamefont {Kurpiers}},
  \bibinfo {author} {\bibfnamefont {S.}~\bibnamefont {Gasparinetti}}, \bibinfo
  {author} {\bibfnamefont {P.}~\bibnamefont {Magnard}}, \bibinfo {author}
  {\bibfnamefont {A.}~\bibnamefont {Poto{\v{c}}nik}}, \bibinfo {author}
  {\bibfnamefont {Y.}~\bibnamefont {Salath{\'{e}}}}, \bibinfo {author}
  {\bibfnamefont {M.}~\bibnamefont {Pechal}}, \bibinfo {author} {\bibfnamefont
  {M.}~\bibnamefont {Mondal}}, \bibinfo {author} {\bibfnamefont
  {M.}~\bibnamefont {Oppliger}}, \bibinfo {author} {\bibfnamefont
  {C.}~\bibnamefont {Eichler}}, \ and\ \bibinfo {author} {\bibfnamefont
  {A.}~\bibnamefont {Wallraff}}} (\bibinfo {year} {2017}),\ \bibfield  {title}
  {\enquote {\bibinfo {title} {{Rapid High-Fidelity Single-Shot Dispersive
  Readout of Superconducting Qubits}},}\ }\href {\doibase
  10.1103/PhysRevApplied.7.054020} {\bibfield  {journal} {\bibinfo  {journal}
  {Physical Review Applied}\ }\textbf {\bibinfo {volume} {7}}~(\bibinfo
  {number} {5}),\ \bibinfo {pages} {054020}}\BibitemShut {NoStop}%
\bibitem [{\citenamefont {Wang}\ \emph {et~al.}(2015)\citenamefont {Wang},
  \citenamefont {Axline}, \citenamefont {Gao}, \citenamefont {Brecht},
  \citenamefont {Chu}, \citenamefont {Frunzio}, \citenamefont {Devoret},\ and\
  \citenamefont {Schoelkopf}}]{Wang2015}%
  \BibitemOpen
  \bibfield  {author} {\bibinfo {author} {\bibfnamefont {C.}~\bibnamefont
  {Wang}}, \bibinfo {author} {\bibfnamefont {C.}~\bibnamefont {Axline}},
  \bibinfo {author} {\bibfnamefont {Y.~Y.}\ \bibnamefont {Gao}}, \bibinfo
  {author} {\bibfnamefont {T.}~\bibnamefont {Brecht}}, \bibinfo {author}
  {\bibfnamefont {Y.}~\bibnamefont {Chu}}, \bibinfo {author} {\bibfnamefont
  {L.}~\bibnamefont {Frunzio}}, \bibinfo {author} {\bibfnamefont {M.~H.}\
  \bibnamefont {Devoret}}, \ and\ \bibinfo {author} {\bibfnamefont {R.~J.}\
  \bibnamefont {Schoelkopf}}} (\bibinfo {year} {2015}),\ \bibfield  {title}
  {\enquote {\bibinfo {title} {{Surface participation and dielectric loss in
  superconducting qubits}},}\ }\href {\doibase 10.1063/1.4934486} {\bibfield
  {journal} {\bibinfo  {journal} {Applied Physics Letters}\ }\textbf {\bibinfo
  {volume} {107}}~(\bibinfo {number} {16}),\ \bibinfo {pages}
  {162601}}\BibitemShut {NoStop}%
\bibitem [{\citenamefont {Wang}\ \emph {et~al.}(2014)\citenamefont {Wang},
  \citenamefont {Gao}, \citenamefont {Pop}, \citenamefont {Vool}, \citenamefont
  {Axline}, \citenamefont {Brecht}, \citenamefont {Heeres}, \citenamefont
  {Frunzio}, \citenamefont {Devoret}, \citenamefont {Catelani}, \citenamefont
  {Glazman},\ and\ \citenamefont {Schoelkopf}}]{Wang2014}%
  \BibitemOpen
  \bibfield  {author} {\bibinfo {author} {\bibfnamefont {C.}~\bibnamefont
  {Wang}}, \bibinfo {author} {\bibfnamefont {Y.~Y.}\ \bibnamefont {Gao}},
  \bibinfo {author} {\bibfnamefont {I.~M.}\ \bibnamefont {Pop}}, \bibinfo
  {author} {\bibfnamefont {U.}~\bibnamefont {Vool}}, \bibinfo {author}
  {\bibfnamefont {C.}~\bibnamefont {Axline}}, \bibinfo {author} {\bibfnamefont
  {T.}~\bibnamefont {Brecht}}, \bibinfo {author} {\bibfnamefont {R.~W.}\
  \bibnamefont {Heeres}}, \bibinfo {author} {\bibfnamefont {L.}~\bibnamefont
  {Frunzio}}, \bibinfo {author} {\bibfnamefont {M.~H.}\ \bibnamefont
  {Devoret}}, \bibinfo {author} {\bibfnamefont {G.}~\bibnamefont {Catelani}},
  \bibinfo {author} {\bibfnamefont {L.~I.}\ \bibnamefont {Glazman}}, \ and\
  \bibinfo {author} {\bibfnamefont {R.~J.}\ \bibnamefont {Schoelkopf}}}
  (\bibinfo {year} {2014}),\ \bibfield  {title} {\enquote {\bibinfo {title}
  {{Measurement and control of quasiparticle dynamics in a superconducting
  qubit}},}\ }\href {\doibase 10.1038/ncomms6836} {\bibfield  {journal}
  {\bibinfo  {journal} {Nature Communications}\ }\textbf {\bibinfo {volume}
  {5}},\ \bibinfo {pages} {5836}}\BibitemShut {NoStop}%
\bibitem [{\citenamefont {Wang}\ \emph {et~al.}(2018)\citenamefont {Wang},
  \citenamefont {Shankar}, \citenamefont {Minev}, \citenamefont
  {Campagne-Ibarcq}, \citenamefont {Narla},\ and\ \citenamefont
  {Devoret}}]{Wang2018-cav-atten-aps}%
  \BibitemOpen
  \bibfield  {author} {\bibinfo {author} {\bibfnamefont {Z.}~\bibnamefont
  {Wang}}, \bibinfo {author} {\bibfnamefont {S.}~\bibnamefont {Shankar}},
  \bibinfo {author} {\bibfnamefont {Z.}~\bibnamefont {Minev}}, \bibinfo
  {author} {\bibfnamefont {P.}~\bibnamefont {Campagne-Ibarcq}}, \bibinfo
  {author} {\bibfnamefont {A.}~\bibnamefont {Narla}}, \ and\ \bibinfo {author}
  {\bibfnamefont {M.}~\bibnamefont {Devoret}}} (\bibinfo {year} {2018}),\
  \bibfield  {title} {\enquote {\bibinfo {title} {{Reducing qubit decoherence
  in 3D circuit quantum electrodynamics with cold cavity attenuators}},}\ }in\
  \href@noop {} {\emph {\bibinfo {booktitle} {APS March Meeting Abstracts}}},\
  \bibinfo {series} {APS Meeting Abstracts}, Vol.\ \bibinfo {volume} {2018},\
  p.\ \bibinfo {pages} {X33.009}\BibitemShut {NoStop}%
\bibitem [{\citenamefont {Wang}\ \emph {et~al.}(2019)\citenamefont {Wang},
  \citenamefont {Shankar}, \citenamefont {Minev}, \citenamefont
  {Campagne-Ibarcq}, \citenamefont {Narla},\ and\ \citenamefont
  {Devoret}}]{Wang2019-cav-atten}%
  \BibitemOpen
  \bibfield  {author} {\bibinfo {author} {\bibfnamefont {Z.}~\bibnamefont
  {Wang}}, \bibinfo {author} {\bibfnamefont {S.}~\bibnamefont {Shankar}},
  \bibinfo {author} {\bibfnamefont {Z.}~\bibnamefont {Minev}}, \bibinfo
  {author} {\bibfnamefont {P.}~\bibnamefont {Campagne-Ibarcq}}, \bibinfo
  {author} {\bibfnamefont {A.}~\bibnamefont {Narla}}, \ and\ \bibinfo {author}
  {\bibfnamefont {M.~H.}\ \bibnamefont {Devoret}}} (\bibinfo {year} {2019}),\
  \bibfield  {title} {\enquote {\bibinfo {title} {{Cavity Attenuators for
  Superconducting Qubits}},}\ }\href {\doibase
  10.1103/PhysRevApplied.11.014031} {\bibfield  {journal} {\bibinfo  {journal}
  {Physical Review Applied}\ }\textbf {\bibinfo {volume} {11}}~(\bibinfo
  {number} {1}),\ \bibinfo {pages} {014031}},\ \Eprint
  {http://arxiv.org/abs//arxiv.org/pdf/1807.04849}
  {arXiv:/arxiv.org/pdf/1807.04849 [https:]} \BibitemShut {NoStop}%
\bibitem [{\citenamefont {Weber}\ \emph {et~al.}(2014)\citenamefont {Weber},
  \citenamefont {Chantasri}, \citenamefont {Dressel}, \citenamefont {Jordan},
  \citenamefont {Murch},\ and\ \citenamefont {Siddiqi}}]{Weber2014}%
  \BibitemOpen
  \bibfield  {author} {\bibinfo {author} {\bibfnamefont {S.~J.}\ \bibnamefont
  {Weber}}, \bibinfo {author} {\bibfnamefont {A.}~\bibnamefont {Chantasri}},
  \bibinfo {author} {\bibfnamefont {J.}~\bibnamefont {Dressel}}, \bibinfo
  {author} {\bibfnamefont {A.~N.}\ \bibnamefont {Jordan}}, \bibinfo {author}
  {\bibfnamefont {K.~W.}\ \bibnamefont {Murch}}, \ and\ \bibinfo {author}
  {\bibfnamefont {I.}~\bibnamefont {Siddiqi}}} (\bibinfo {year} {2014}),\
  \bibfield  {title} {\enquote {\bibinfo {title} {{Mapping the optimal route
  between two quantum states}},}\ }\href {\doibase 10.1038/nature13559}
  {\bibfield  {journal} {\bibinfo  {journal} {Nature}\ }\textbf {\bibinfo
  {volume} {511}}~(\bibinfo {number} {7511}),\ \bibinfo {pages} {570--573}},\
  \Eprint {http://arxiv.org/abs/1403.4992} {arXiv:1403.4992} \BibitemShut
  {NoStop}%
\bibitem [{\citenamefont {White}\ \emph {et~al.}(2016)\citenamefont {White},
  \citenamefont {Mutus}, \citenamefont {Dressel}, \citenamefont {Kelly},
  \citenamefont {Barends}, \citenamefont {Jeffrey}, \citenamefont {Sank},
  \citenamefont {Megrant}, \citenamefont {Campbell}, \citenamefont {Chen},
  \citenamefont {Chen}, \citenamefont {Chiaro}, \citenamefont {Dunsworth},
  \citenamefont {Hoi}, \citenamefont {Neill}, \citenamefont {O'Malley},
  \citenamefont {Roushan}, \citenamefont {Vainsencher}, \citenamefont {Wenner},
  \citenamefont {Korotkov},\ and\ \citenamefont {Martinis}}]{White2016}%
  \BibitemOpen
  \bibfield  {author} {\bibinfo {author} {\bibfnamefont {T.~C.}\ \bibnamefont
  {White}}, \bibinfo {author} {\bibfnamefont {J.~Y.}\ \bibnamefont {Mutus}},
  \bibinfo {author} {\bibfnamefont {J.}~\bibnamefont {Dressel}}, \bibinfo
  {author} {\bibfnamefont {J.}~\bibnamefont {Kelly}}, \bibinfo {author}
  {\bibfnamefont {R.}~\bibnamefont {Barends}}, \bibinfo {author} {\bibfnamefont
  {E.}~\bibnamefont {Jeffrey}}, \bibinfo {author} {\bibfnamefont
  {D.}~\bibnamefont {Sank}}, \bibinfo {author} {\bibfnamefont {A.}~\bibnamefont
  {Megrant}}, \bibinfo {author} {\bibfnamefont {B.}~\bibnamefont {Campbell}},
  \bibinfo {author} {\bibfnamefont {Y.}~\bibnamefont {Chen}}, \bibinfo {author}
  {\bibfnamefont {Z.}~\bibnamefont {Chen}}, \bibinfo {author} {\bibfnamefont
  {B.}~\bibnamefont {Chiaro}}, \bibinfo {author} {\bibfnamefont
  {A.}~\bibnamefont {Dunsworth}}, \bibinfo {author} {\bibfnamefont {I.-C.}\
  \bibnamefont {Hoi}}, \bibinfo {author} {\bibfnamefont {C.}~\bibnamefont
  {Neill}}, \bibinfo {author} {\bibfnamefont {P.~J.~J.}\ \bibnamefont
  {O'Malley}}, \bibinfo {author} {\bibfnamefont {P.}~\bibnamefont {Roushan}},
  \bibinfo {author} {\bibfnamefont {A.}~\bibnamefont {Vainsencher}}, \bibinfo
  {author} {\bibfnamefont {J.}~\bibnamefont {Wenner}}, \bibinfo {author}
  {\bibfnamefont {A.~N.}\ \bibnamefont {Korotkov}}, \ and\ \bibinfo {author}
  {\bibfnamefont {J.~M.}\ \bibnamefont {Martinis}}} (\bibinfo {year} {2016}),\
  \bibfield  {title} {\enquote {\bibinfo {title} {{Preserving entanglement
  during weak measurement demonstrated with a violation of the
  Bell-Leggett-Garg inequality}},}\ }\href {\doibase 10.1038/npjqi.2015.22}
  {\bibfield  {journal} {\bibinfo  {journal} {npj Quantum Information}\
  }\textbf {\bibinfo {volume} {2}}~(\bibinfo {number} {1}),\ \bibinfo {pages}
  {15022}}\BibitemShut {NoStop}%
\bibitem [{\citenamefont {Wiseman}(1995)}]{Wiseman1995-adaptive}%
  \BibitemOpen
  \bibfield  {author} {\bibinfo {author} {\bibfnamefont {H.~M.}\ \bibnamefont
  {Wiseman}}} (\bibinfo {year} {1995}),\ \bibfield  {title} {\enquote {\bibinfo
  {title} {{Adaptive Phase Measurements of Optical Modes: Going Beyond the
  Marginal Q Distribution}},}\ }\href {\doibase 10.1103/PhysRevLett.75.4587}
  {\bibfield  {journal} {\bibinfo  {journal} {Physical Review Letters}\
  }\textbf {\bibinfo {volume} {75}}~(\bibinfo {number} {25}),\ \bibinfo {pages}
  {4587--4590}}\BibitemShut {NoStop}%
\bibitem [{\citenamefont {Wiseman}\ \emph {et~al.}(2007)\citenamefont
  {Wiseman}, \citenamefont {Jones},\ and\ \citenamefont
  {Doherty}}]{Wiseman2007-steering}%
  \BibitemOpen
  \bibfield  {author} {\bibinfo {author} {\bibfnamefont {H.~M.}\ \bibnamefont
  {Wiseman}}, \bibinfo {author} {\bibfnamefont {S.~J.}\ \bibnamefont {Jones}},
  \ and\ \bibinfo {author} {\bibfnamefont {A.~C.}\ \bibnamefont {Doherty}}}
  (\bibinfo {year} {2007}),\ \bibfield  {title} {\enquote {\bibinfo {title}
  {{Steering, Entanglement, Nonlocality, and the Einstein-Podolsky-Rosen
  Paradox}},}\ }\href {\doibase 10.1103/PhysRevLett.98.140402} {\bibfield
  {journal} {\bibinfo  {journal} {Physical Review Letters}\ }\textbf {\bibinfo
  {volume} {98}}~(\bibinfo {number} {14}),\ \bibinfo {pages}
  {140402}}\BibitemShut {NoStop}%
\bibitem [{\citenamefont {Wiseman}\ and\ \citenamefont
  {Milburn}(2010)}]{wiseman2010book}%
  \BibitemOpen
  \bibfield  {author} {\bibinfo {author} {\bibfnamefont {H.~M.}\ \bibnamefont
  {Wiseman}}, \ and\ \bibinfo {author} {\bibfnamefont {G.~J.}\ \bibnamefont
  {Milburn}}} (\bibinfo {year} {2010}),\ \href
  {https://books.google.com/books?id=ZNjvHaH8qA4C} {\emph {\bibinfo {title}
  {{Quantum Measurement and Control}}}}\ (\bibinfo  {publisher} {Cambridge
  University Press})\BibitemShut {NoStop}%
\bibitem [{\citenamefont {Yan}\ \emph {et~al.}(2016)\citenamefont {Yan},
  \citenamefont {Gustavsson}, \citenamefont {Kamal}, \citenamefont {Birenbaum},
  \citenamefont {Sears}, \citenamefont {Hover}, \citenamefont {Gudmundsen},
  \citenamefont {Rosenberg}, \citenamefont {Samach}, \citenamefont {Weber},
  \citenamefont {Yoder}, \citenamefont {Orlando}, \citenamefont {Clarke},
  \citenamefont {Kerman},\ and\ \citenamefont {Oliver}}]{FYan2016}%
  \BibitemOpen
  \bibfield  {author} {\bibinfo {author} {\bibfnamefont {F.}~\bibnamefont
  {Yan}}, \bibinfo {author} {\bibfnamefont {S.}~\bibnamefont {Gustavsson}},
  \bibinfo {author} {\bibfnamefont {A.}~\bibnamefont {Kamal}}, \bibinfo
  {author} {\bibfnamefont {J.}~\bibnamefont {Birenbaum}}, \bibinfo {author}
  {\bibfnamefont {A.~P.}\ \bibnamefont {Sears}}, \bibinfo {author}
  {\bibfnamefont {D.}~\bibnamefont {Hover}}, \bibinfo {author} {\bibfnamefont
  {T.~J.}\ \bibnamefont {Gudmundsen}}, \bibinfo {author} {\bibfnamefont
  {D.}~\bibnamefont {Rosenberg}}, \bibinfo {author} {\bibfnamefont
  {G.}~\bibnamefont {Samach}}, \bibinfo {author} {\bibfnamefont
  {S.}~\bibnamefont {Weber}}, \bibinfo {author} {\bibfnamefont {J.~L.}\
  \bibnamefont {Yoder}}, \bibinfo {author} {\bibfnamefont {T.~P.}\ \bibnamefont
  {Orlando}}, \bibinfo {author} {\bibfnamefont {J.}~\bibnamefont {Clarke}},
  \bibinfo {author} {\bibfnamefont {A.~J.}\ \bibnamefont {Kerman}}, \ and\
  \bibinfo {author} {\bibfnamefont {W.~D.}\ \bibnamefont {Oliver}}} (\bibinfo
  {year} {2016}),\ \bibfield  {title} {\enquote {\bibinfo {title} {{The flux
  qubit revisited to enhance coherence and reproducibility}},}\ }\href
  {\doibase 10.1038/ncomms12964} {\bibfield  {journal} {\bibinfo  {journal}
  {Nature Communications}\ }\textbf {\bibinfo {volume} {7}},\ \bibinfo {pages}
  {12964}}\BibitemShut {NoStop}%
\bibitem [{\citenamefont {Yeh}\ \emph {et~al.}(2017)\citenamefont {Yeh},
  \citenamefont {LeFebvre}, \citenamefont {Premaratne}, \citenamefont
  {Wellstood},\ and\ \citenamefont {Palmer}}]{Yeh2017Atten}%
  \BibitemOpen
  \bibfield  {author} {\bibinfo {author} {\bibfnamefont {J.-H.}\ \bibnamefont
  {Yeh}}, \bibinfo {author} {\bibfnamefont {J.}~\bibnamefont {LeFebvre}},
  \bibinfo {author} {\bibfnamefont {S.}~\bibnamefont {Premaratne}}, \bibinfo
  {author} {\bibfnamefont {F.~C.}\ \bibnamefont {Wellstood}}, \ and\ \bibinfo
  {author} {\bibfnamefont {B.~S.}\ \bibnamefont {Palmer}}} (\bibinfo {year}
  {2017}),\ \bibfield  {title} {\enquote {\bibinfo {title} {{Microwave
  attenuators for use with quantum devices below 100 mK}},}\ }\href {\doibase
  10.1063/1.4984894} {\bibfield  {journal} {\bibinfo  {journal} {Journal of
  Applied Physics}\ }\textbf {\bibinfo {volume} {121}}~(\bibinfo {number}
  {22}),\ \bibinfo {pages} {224501}}\BibitemShut {NoStop}%
\bibitem [{\citenamefont {Yu}\ and\ \citenamefont {Oh}(2012)}]{Yu2012}%
  \BibitemOpen
  \bibfield  {author} {\bibinfo {author} {\bibfnamefont {S.}~\bibnamefont
  {Yu}}, \ and\ \bibinfo {author} {\bibfnamefont {C.~H.}\ \bibnamefont {Oh}}}
  (\bibinfo {year} {2012}),\ \bibfield  {title} {\enquote {\bibinfo {title}
  {{State-Independent Proof of Kochen-Specker Theorem with 13 Rays}},}\ }\href
  {\doibase 10.1103/PhysRevLett.108.030402} {\bibfield  {journal} {\bibinfo
  {journal} {Physical Review Letters}\ }\textbf {\bibinfo {volume}
  {108}}~(\bibinfo {number} {3}),\ \bibinfo {pages} {030402}},\ \Eprint
  {http://arxiv.org/abs/1109.4396} {arXiv:1109.4396} \BibitemShut {NoStop}%
\bibitem [{\citenamefont {Yurke}\ and\ \citenamefont
  {Denker}(1984)}]{Yurke1984}%
  \BibitemOpen
  \bibfield  {author} {\bibinfo {author} {\bibfnamefont {B.}~\bibnamefont
  {Yurke}}, \ and\ \bibinfo {author} {\bibfnamefont {J.~S.}\ \bibnamefont
  {Denker}}} (\bibinfo {year} {1984}),\ \bibfield  {title} {\enquote {\bibinfo
  {title} {{Quantum network theory}},}\ }\href {\doibase
  10.1103/PhysRevA.29.1419} {\bibfield  {journal} {\bibinfo  {journal}
  {Physical Review A}\ }\textbf {\bibinfo {volume} {29}}~(\bibinfo {number}
  {3}),\ \bibinfo {pages} {1419--1437}}\BibitemShut {NoStop}%
\bibitem [{\citenamefont {Zhang}\ \emph {et~al.}(2017)\citenamefont {Zhang},
  \citenamefont {Liu}, \citenamefont {Raftery},\ and\ \citenamefont
  {Houck}}]{Zhang2017}%
  \BibitemOpen
  \bibfield  {author} {\bibinfo {author} {\bibfnamefont {G.}~\bibnamefont
  {Zhang}}, \bibinfo {author} {\bibfnamefont {Y.}~\bibnamefont {Liu}}, \bibinfo
  {author} {\bibfnamefont {J.~J.}\ \bibnamefont {Raftery}}, \ and\ \bibinfo
  {author} {\bibfnamefont {A.~A.}\ \bibnamefont {Houck}}} (\bibinfo {year}
  {2017}),\ \bibfield  {title} {\enquote {\bibinfo {title} {{Suppression of
  photon shot noise dephasing in a tunable coupling superconducting qubit}},}\
  }\href {\doibase 10.1038/s41534-016-0002-2} {\bibfield  {journal} {\bibinfo
  {journal} {npj Quantum Information}\ }\textbf {\bibinfo {volume}
  {3}}~(\bibinfo {number} {1}),\ \bibinfo {pages} {1}}\BibitemShut {NoStop}%
\bibitem [{\citenamefont {Zmuidzinas}(2012)}]{Zmuidzinas2012}%
  \BibitemOpen
  \bibfield  {author} {\bibinfo {author} {\bibfnamefont {J.}~\bibnamefont
  {Zmuidzinas}}} (\bibinfo {year} {2012}),\ \bibfield  {title} {\enquote
  {\bibinfo {title} {{Superconducting Microresonators: Physics and
  Applications}},}\ }\href {\doibase 10.1146/annurev-conmatphys-020911-125022}
  {\bibfield  {journal} {\bibinfo  {journal} {Annual Review of Condensed Matter
  Physics}\ }\textbf {\bibinfo {volume} {3}}~(\bibinfo {number} {1}),\ \bibinfo
  {pages} {169--214}}\BibitemShut {NoStop}%
\end{thebibliography}%

\medskip
\appendix

\end{document}